\documentclass[12pt]{article}
\pdfoutput=1 
\expandafter\let\csname equation*\endcsname\relax
\expandafter\let\csname endequation*\endcsname\relax
\usepackage{amsmath,amssymb,amsthm}
\usepackage{txfonts}
\usepackage{graphicx}
\usepackage{bxeepic}
\usepackage[pdfencoding=auto,bookmarks=true,bookmarksnumbered=true,colorlinks=true]{hyperref}
\usepackage{here}
\theoremstyle{plain}
\newtheorem{thm}{Theorem}[section]

\newtheorem{ex}[thm]{Example}
\newtheorem{rem}[thm]{Remark}
\renewcommand{\P}{\mathbb{P}}

\newcommand{\oF}{\overline{F}}
\newcommand{\of}{\overline{f}}
\newcommand{\og}{\overline{g}}
\newcommand{\oy}{\overline{y}}

\newcommand{\ug}{\underline{g}}
\newcommand{\uG}{\underline{G}}
\newcommand{\hypotilde}[2]{\vrule depth #1 pt width 0pt%
{\smash{{\mathop{#2} \limits_{\displaystyle\widetilde{}}}}}}
\newcommand{\Z}{\mathbb{Z}}

\newcommand{\kk}{\kappa}
\newcommand{\esan}{E_3}
\newcommand{\eni}{E_2}

\makeatletter
\@addtoreset{equation}{section} 

\makeatother
\begin{document}
\title{Geometric Aspects of Painlev\'e Equations}
\author{Kenji {\sc Kajiwara}\\[2mm]
Institute of Mathematics for Industry, Kyushu University\\
744 Motooka, Fukuoka 819-0395, Japan\\
e-mail: kaji@imi.kyushu-u.ac.jp \\[4mm]
Masatoshi {\sc Noumi} \\
Department of Mathematics, Kobe University\\
Rokko, Kobe 657-8501, Japan\\
e-mail: noumi@math.kobe-u.ac.jp \\[4mm]
Yasuhiko  {\sc Yamada}\\[2mm]
Department of Mathematics, Kobe University\\
Rokko, Kobe 657-8501, Japan\\
e-mail: yamaday@math.kobe-u.ac.jp 
}
\maketitle
\begin{abstract}
In this paper a comprehensive review is given on the current status of achievements in the geometric
aspects of the Painlev\'e equations, with a particular emphasis on the discrete Painlev\'e
equations. The theory is controlled by the geometry of certain rational surfaces called the spaces
of initial values, which are characterized by eight point configuration on
$\mathbb{P}^1\times\mathbb{P}^1$ and classified according to the degeneration of points. We give a
systematic description of the equations and their various properties, such as affine Weyl group
symmetries, hypergeometric solutions and Lax pairs under this framework, by using the language of
Picard lattice and root systems. We also provide with a collection of basic data; equations, point
configurations/root data, Weyl group representations, Lax pairs, and hypergeometric solutions of all
possible cases.
\end{abstract}
\newpage
\setcounter{tocdepth}{2}
\tableofcontents
\section{Introduction}
Today the Painlev\'e equations, both continuous and discrete, are well-established subjects in
mathematics and mathematical physics \cite{Painleve_100,Forrester:book}. In the geometric approach
to the Painlev\'e equations, initiated by Okamoto \cite{Okamoto:SIV} and subsequently extended by
Sakai \cite{Sakai:SIV} to the discrete cases, the theory is controlled by the geometry of certain
rational surfaces called the spaces of initial values. This framework gives a systematic description
of the equations, symmetries, special solutions, Lax pairs and so forth. The aim of this paper is to
provide a comprehensive review on the current status of achievements in the geometric aspects of the
Painlev\'e equations so as to serve it as a foundation of future researches in mathematics and
mathematical physics. It also contains some materials which have been newly developed to complete a
unified description. We put a particular emphasis on studying the discrete Painlev\'e equations of
the second order.

Historically, the Painlev\'e differential equations were discovered by Painlev\'e and
Gambier \cite{Gambier,Ince,Painleve_1,Painleve_2} in the efforts for finding new transcendental
functions defined by ``good'' nonlinear ordinary differential equations. They imposed the condition
that the solutions should admit only poles as movable singular points, which is now referred to as
the {\em Painlev\'e property}.  Then R. Fuchs \cite{Fuchs_1,Fuchs_2} formulated them as the
monodromy preserving deformations of linear ordinary differential equations, and subsequently
Schlesinger \cite{Schlesinger} and Garnier \cite{Garnier_1,Garnier_2} investigated their
generalizations. Almost sixty years later, the Painlev\'e equations were found to describe the
correlation function of the Ising model by
Wu-McCoy-Tracy-Barouch \cite{Wu-McCoy-Tracy-Barouch}. Inspired by this discovery, the theory of
holonomic quantum field theory \cite{Jimbo-Miwa-Mori-Sato,Sato-Miwa-Jimbo} and a general theory of
the monodromy preserving deformation of linear ordinary differential equations was established by
Sato's group in Kyoto \cite{J-M:Monodromy2,J-M:Monodromy3,J-M:Monodromy1}.

The study of geometric aspects of the Painlev\'e equations, which is a main topic of this paper, has
been initiated by Okamoto's pioneering work \cite{Okamoto:SIV}. For each Painlev\'e equation, he
constructed the {\em space of initial values} which parametrizes all the solutions. Takano further
found that the Painlev\'e equations themselves can be reproduced uniquely from the space of initial
values \cite{Takano2,Takano1}. These works are the basis of Sakai's approach for discrete Painlev\'e
equations which will be mentioned below.

On the other hand, discrete integrable systems, started in 70's by the pioneering works of
Ablowitz-Ladik and Hirota, have been regarded as equally or more important as the continuous
integrable systems. The discrete Painlev\'e equations also appear in
\cite{J-M:Monodromy2,J-M:Monodromy3,J-M:Monodromy1}, and attracted attention by the discoveries of
the scaling limit to the Painlev\'e differential equations in the context of two-dimensional quantum
gravity \cite{Brezin-Kazakov,Douglas-Shenker,Gross-Migdal}. In accordance with these studies,
Grammaticos, Ramani and Papageorgiou introduced the concept of {\em singularity confinement} as a
discrete counterpart of the Painlev\'e property and proposed to use it as an integrability detector
for discrete systems \cite{GRP:sc}. Then Ramani, Grammaticos and Hietarinta applied this idea to
obtain non-autonomous version of the two-dimensional integrable mappings known as the
Quispel-Roberts-Thompson (QRT) mappings \cite{QRT1,QRT2} and succeeded in constructing discrete
Painlev\'e equations systematically \cite{GRH:dP}.

Subsequently, discrete Painlev\'e equations as well as their generalizations have been studied from
various points of view, such as B\"acklund transformations, Lax pairs, particular solutions, $\tau$
functions and so on. For a review of those developments, we refer to
\cite{GR:review2004,TTGR:review2004}. In the meanwhile, underlying mathematical structures have
gradually been clarified. Jimbo and Sakai constructed a $q$-difference analogue of Painlev\'e VI
equation in the spirit of deformation theory of linear $q$-difference equation
\cite{Jimbo-Sakai:qp6}. A universal symmetry structure behind the continuous and discrete Painlev\'e
equations has been revealed in terms of the birational representations of affine Weyl groups which
is also applicable to higher dimensional Painlev\'e type equations
\cite{KNY:qp4,KNY:AmAn,KNY:qKP,Noumi:book,NY_CMP1998,Noumi-Yamada:p4}.

In the efforts for finding a unified framework for the Painlev\'e type equations, Sakai proposed a
class of second order discrete Painlev\'e equations arising from Cremona transformations of rational
surfaces obtained as nine-point blow-ups of $\mathbb{P}^2$ \cite{Sakai:SIV}. Those rational surfaces
are regarded as the spaces of initial values for discrete and continuous Painlev\'e equations, and
are classified into 22 cases according to the configuration of nine points. The master equation of
those Painlev\'e equations, the {\em elliptic Painlev\'e equation}, is obtained from the most
generic configuration; it provides with the geometric construction of discrete Painlev\'e equation
with affine Weyl group symmetry of type $E_8^{(1)}$ as proposed by Ohta, Ramani and Grammaticos
\cite{ORG:e8}. Other equations are obtained from the degenerate configurations.

On the basis of this geometric approach, above-mentioned various aspects of the Painlev\'e equations
can be investigated in a unified manner in the language of Picard lattice and root systems. For
example, the hypergeometric seed solutions to all possible discrete Painlev\'e equations have been
constructed in \cite{K:hyper_additive,KMNOY:hyper,KMNOY:hyper2}. Lax pairs for discrete
Painlev\'e equations have been constructed through their characterization in terms of the point
configuration \cite{NTY,Yamada:Lax,Yamada:Pade}. Geometric approach is effective particularly in the
study of Painlev\'e equations with high symmetry such as $E$ type.  \par\medskip

The plan of this review is as follows. In Section 2, we give an overview of various aspects of the
Painlev\'e equations to be discussed in this review. Taking the examples of the Painlev\'e IV
equation (P$_{\rm IV}$) and a discrete Painlev\'e II equation (dP$_{\rm II}$) which arises as a
B\"acklund transformation of P$_{\rm IV}$, we introduce basic objects in the theory of Painlev\'e
equations, such as affine Weyl group symmetry, Lax pairs, hypergeometric solutions, $\tau$ functions
and the space of initial values in the sense of Okamoto and Sakai.

The space of initial values is, roughly speaking, a surface on which the solutions of the Painlev\'e
equation in question are parametrized. For each Painlev\'e equation, this surface is characterized
by a pair of affine root systems which represent the symmetry type and the surface type. Many
properties of Painlev\'e equations as presented in Section 2 are systematically controlled by
geometry of the surface. In Section 3, we provide with general frameworks of the root systems, Weyl
groups and the Picard lattice relevant to the Painlev\'e equations. These devices will be utilized
throughout subsequent sections as fundamental and powerful tools for studying the Painlev\'e
equations.

One of the common properties of the Painlev\'e equations is that the space of initial values is
obtained from $\mathbb{P}^1\times\mathbb{P}^1$, the product of two copies of the Riemann sphere, by
blowing up at eight points. Therefore the configuration of the eight points, which possibly includes
infinitely near points, provides with the most fundamental data of the equation.  With the items
obtained in Section 3 in hand, we demonstrate how to associate a point configuration on
$\mathbb{P}^1\times\mathbb{P}^1$ to a given discrete equation in Section 4.  This provides us a
practical method for determining whether it is a discrete Painlev\'e equation in Sakai's class, and
if so, identifying the type of the equation by its surface type and symmetry type.

If the configuration of eight points in $\mathbb{P}^1\times\mathbb{P}^1$ is generic, the
corresponding space of initial values has the largest symmetry of type $E_8^{(1)}$. Other
configurations can be regarded as the degenerate cases, among all possible 22 configurations
classified by Sakai. In Section 5, we describe how to construct the equations and relevant
characteristic features from the point configuration. In particular, we formulate a representation
of affine Weyl group of type $E_8^{(1)}$ from the configuration of generic eight points in
$\mathbb{P}^1\times\mathbb{P}^1$, as well as the formalism of $\tau$ functions. We then derive a new
explicit form of the three equations of type $E_8^{(1)}$, which are the elliptic, $q$- and
difference Painlev\'e equations, from a translation of the root lattice. We also give an example
demonstrating how to construct the birational representation of the affine Weyl group for a given
degenerate point configuration.

Most of the Painlev\'e equations admit a class of particular solutions expressible in terms of the
special functions of hypergeometric type for special values of parameters which correspond to
reflection hyperplanes in the parameter space. In Section 6, we demonstrate how to construct the
hypergeometric solutions by decoupling a given equation to the Riccati equation and by linearizing
it. Then we give an intrinsic formulation of this procedure by the geometric language of point
configurations. The list of hypergeometric solutions associated with possible point configurations
will be given in Section 8.

It is a common feature of nonlinear integrable systems that they arise as the compatibility
condition of certain systems of linear equations which is called a Lax pair. In Section 7 we give a
geometric formulation of the Lax pairs for Painlev\'e equations in terms of associated point
configurations.

Section 8 is a comprehensive collection of data for all Painlev\'e equations which can be obtained
by various methods discussed in this review.  For each case, we provide with explicit forms of
equations, point configurations/root data, Weyl group representations, Lax pairs and hypergeometric
seed solutions.

\par\medskip

In this review, we present a general geometric framework as well as algebraic tools for studying the
Painlev\'e equations, confining ourselves to the second order equations. Even in the second order
equations, there are various discrete Painlev\'e equations which are not directly investigated in
this paper, e.g., equations arising from the translations with different directions or length in the
root lattice \cite{KNT:projective,RG:infinite,Takenawa:qp5}.  Also, we do not deal with higher order
or multi-variable generalizations, which are now actively studied in relation with soliton equations
\cite{Fuji-Suzuki:D,Fuji-Suzuki:A,KNY:AmAn,KNY:qKP,Noumi-Yamada:p5,Suzuki-Fuji:A,Tsuda:UC,Tsuda:UC_monodromy},
geometry of space of initial values \cite{Masuda:q-Sasano,Sasano}, geometry of flag varieties
\cite{NY:Poisson}, or general theory of monodromy preserving deformations
\cite{Kawakami-Nakamura-Sakai}. The Painlev\'e equations are believed to define new transcendental
functions, and it was rigorously proved for the Painlev\'e differential equations (see, for example
\cite{Umemura:P1,Umemura:Sugaku_expositions}). Similar investigations for discrete Painlev\'e
equations have been done in \cite{Nishioka:qp2}. Asymptotic analysis for the solutions of Painlev\'e
equations are also an important subject in view of applications
\cite{Deift:book,Duistermaat-Joshi,Fokas:book,Joshi:quicksilver,Kawai-Takei:book}. Recently,
applications to various areas of physics and mathematical sciences, including probability theory and
combinatorics, have been actively studied based on the random matrix theory
\cite{Forrester:book}. There are also other interesting relationships to various areas, such as
discrete differential geometry, integrable models of quantum physics and lattice models,
ultradiscrete systems, quivers and cluster algebras. We hope that the materials provided in this
review will be utilized for further developments of the theory of Painlev\'e equations and related
areas.

%
\section{Overview of the Painlev\'e equations}\label{section:overview}
This section is an overview of various aspects of the Painlev\'e equations to be discussed
in this review. Taking the examples of P$_{\rm IV}$ and dP$_{\rm II}$, we introduce basic objects in
the theory of Painlev\'e equations, such as affine Weyl group symmetry, Lax pairs, hypergeometric
solutions, $\tau$ functions and the space of initial values.
%
\subsection{Hamilton system and symmetry of P$_{\rm IV}$}\label{subsubsec:P4_intro}
Let us consider P$_{\rm IV}$
\begin{equation}\label{eqn:p4}
q'' = \frac{1}{2q}\left(q'\right)^2 + \frac{3}{2}q^3 +  2 tq^2 
+ \left(a_2-a_0+\frac{t^2}{2}\right)q - \frac{a_1^2}{2 q},
\end{equation}
which can be rewritten as the non-autonomous Hamiltonian system as
\begin{equation}\label{eqn:p4_hamilton}
\left\{
\begin{array}{l}\medskip
{\displaystyle  q' = \frac{\partial H_{\rm IV}}{\partial p}
= -a_1 + 2pq - q^2 - q t},\\
{\displaystyle p' = -\frac{\partial H_{\rm IV}}{\partial q}
=  a_2 - p^2 + 2 pq + pt},
\end{array}
\right.
\end{equation}
where
\begin{equation}\label{eqn:p4_hamiltonian}
 H_{\rm IV}=-a_1 p - a_2 q + pq(p-q-t),
\end{equation}
$t$ is an independent variable and $a_i$ ($i=0,1,2$) are parameters such that
\begin{equation}\label{eqn:p4_params}
{}'=\frac{d}{dt},\quad a_i{}'=0\quad (i=0,1,2), \quad  a_0 + a_1 + a_2 = 1.
\end{equation}

We introduce 
\begin{equation}\label{eqn:p4_pq_f}
 f_0 = -(p-q-t),\quad f_1=-q,\quad f_2=p.
\end{equation}
Then $f_i$ ($i=0,1,2$) satisfy the following equation
\begin{equation}\label{eqn:p4_sym}
\begin{split}
& f_0' =  f_0(f_1 - f_2 ) + a_0\\
& f_1' =  f_1(f_2 - f_0) + a_1 ,\\
& f_2' =  f_2(f_0 - f_1 ) + a_2,
\end{split} \qquad f_0 + f_1 + f_2 = t,
\end{equation}
which is called the {\em symmetric form} of P$_{\rm IV}$ \cite{Adler,Noumi:book,Noumi-Yamada:p4}. 

The following transformations $s_i$ ($i=0,1,2$) and $\pi$ on variables $p$, $q$ and $a_i$ ($i=0,1,2$) commute
with the differentiation and are called the {\em B\"acklund transformations}:
\begin{equation}\label{eqn:p4_BT_pq_a}
\renewcommand{\arraystretch}{1.8}
\begin{tabular}{|c|c|c|c|c|c|}
\hline
&$p$ &$q$ &$a_0$ &$a_1$ &$a_2$ \\
\hline
$s_0$&$ p+\dfrac{a_0}{p-q-t}$ &$ q+\dfrac{a_0}{p-q-t}$ & $-a_0$ &$a_0+a_1$ & $a_0+a_2$\\
\hline
$s_1$& $p-\dfrac{a_1}{q}$ & $q$ & $a_0+a_1$ & $-a_1$ & $a_1+a_2$\\
\hline
$s_2$ & $p$ & $q+\dfrac{a_2}{p}$ & $a_0+a_2$ & $a_1+a_2$ & $-a_2$\\
\hline
$\pi$ & $-p+q+t$ & $-p$ & $a_1$ & $a_2$ & $a_0$\\
\hline
\end{tabular} 
\end{equation}
\renewcommand{\arraystretch}{1}
For instance, $s_0$ is defined by the variable transformation that replaces $p$, $q$, $a_0$, $a_1$, $a_2$ by
\begin{equation}
 \begin{split}
&s_0(p)=p+\dfrac{a_0}{p-q-t}, \quad s_0(q)=q+\dfrac{a_0}{p-q-t}, \\
&s_0(a_0)=-a_0,\quad s_0(a_1)=a_0+a_1,\quad s_0(a_2)=a_0+a_2,  
 \end{split}
\end{equation}
respectively. Commutativity with the differentiation, $w(f')=(w(f))'$ for
$w=s_0,s_1,s_2,\pi$ and $f=p,q,a_0,a_1,a_2$, can be verified by direct calculations. Composition of those
transformations are computed, for example, as
\begin{equation}\label{eqn:p4_BT_composition}
s_2s_1(p)=s_2\left(p-\frac{a_1}{q}\right)=s_2(p)-\frac{s_2(a_1)}{s_2(q)}=p-\frac{a_1+a_2}{q+\frac{a_2}{p}}.
\end{equation}
These transformations satisfy the fundamental relations
\begin{equation}
 s_i^2=1,\quad (s_is_{i+1})^3=1,\quad \pi s_i = s_{i+1}\pi\quad (i\in\mathbb{Z}/3\mathbb{Z}),\quad
\pi^3=1,
\end{equation}
and form the extended affine Weyl group of type $A_2^{(1)}$ (we will give a general account of
the affine Weyl groups in Section \ref{section:root}). We define the {\em translation} $T$ by
\begin{equation}\label{eqn:translation_A2}
 T = s_2 s_0\pi^{-1}.
\end{equation}
Then the action of $T$ is given by
\begin{equation}
\begin{tabular}{|c|c|c|c|c|c|}
\hline
&$p$ &$q$ &$a_0$ &$a_1$ &$a_2$ \\
\hline
$T$& $\overline{p}$&$\overline{q}$ & $a_0$ &$a_1-1$ & $a_2+1$\\
\hline
\end{tabular} 
\end{equation}
where
\begin{equation}\label{eqn:dP_A2}
\begin{split}
&\overline{q}=p-q-t-\dfrac{a_2}{p},\\
& \overline{p}=\overline{q}-p+t + \dfrac{a_1-1}{\overline{q}} =
-\dfrac{a_2}{p}-q + \dfrac{a_1-1}{p-q-t-\dfrac{a_2}{p}}.
\end{split}
\end{equation}
When the iteration of this transformation is viewed as a discrete dynamical system,
\eqref{eqn:dP_A2} is identified as a discrete analogue of the Painlev\'e II equation (dP$_{\rm
II}$) \cite{NY_CMP1998}.  With the notation $T^n(q)=q_n$, $T^n(p)=p_n$ ($n\in\mathbb{Z}$),
\eqref{eqn:dP_A2} is interpreted as a difference equation with respect to $n$:
\begin{equation}\label{eqn:dP_A2-2}
\begin{split}
&q_{n+1} = p_n - q_{n}-t-\dfrac{a_2+n}{p_n},\\
&p_{n+1} = q_{n+1} - p_n + t + \dfrac{a_1-(n+1)}{q_{n+1}}.
\end{split}
\end{equation}
\begin{rem}\label{rem:convention_action}\rm
There are two possible ways to compute the compositions of the B\"acklund transformations.  The
composition defined by substitution of symbols as \eqref{eqn:p4_BT_composition} is interpreted in
terms of automorphisms of the field of rational functions
$K=\mathbb{C}(p,q,a_0,a_1,a_2)$. We call this convention the {\em symbolical composition}, since it is
convenient for symbolic computations. The other way is to regard the B\"acklund transformations as
the transformations of five variables $(p,q,a_0,a_1,a_2)$. 
We define $F_{s_1}$ and $F_{s_2}$, for instance, by
\begin{align}
& F_{s_1}(p,q,a_0,a_1,a_2)=\Big(p-\frac{a_1}{q},q,a_0+a_1,-a_1,a_1+a_2\Big),\\
& F_{s_2}(p,q,a_0,a_1,a_2)=\Big(p,q+\frac{a_2}{p},a_0+a_2,a_1+a_2,-a_2\Big).
\end{align}
In this convention, the composition of $F_{s_1}F_{s_2}$, for example, is calculated as follows.
\begin{align}
&F_{s_2}(p,q,a_0,a_1,a_2)
=\Big(p,q+\frac{a_2}{p},a_0+a_2,a_1+a_2,-a_2\Big)= (\tilde{p},\tilde{q},\tilde{a}_0,\tilde{a}_1,\tilde{a}_2),\\
& F_{s_1}(\tilde{p},\tilde{q},\tilde{a}_0,\tilde{a}_1,\tilde{a}_2)=
\Big(\tilde{p}-\frac{\tilde{a_1}}{\tilde{q}},\tilde{q},
\tilde{a}_0+\tilde{a}_1,-\tilde{a}_1,\tilde{a}_1+\tilde{a}_2\Big).
\end{align}
Eliminating $\tilde{p},\tilde{q},\tilde{a}_0,\tilde{a}_1,\tilde{a}_2$, we obtain
\begin{equation}
 F_{s_1}F_{s_2}(p,q,a_0,a_1,a_2)=
\Big(p-\frac{a_1+a_2}{q+\frac{a_2}{p}},q+\frac{a_2}{p},
a_0+a_1+2a_2,-a_1-a_2,a_1\Big).
\end{equation}
Therefore we have $F_{s_1}F_{s_2}=F_{s_2s_1}$, where $F_{s_2s_1}$ is the birational transformation
corresponding to $s_2s_1$. As we will demonstrate later, this convention is convenient for
numerical computations. We call this convention the {\em numerical composition}.
Note that in these two ways of computation, the order of composition is
opposite to each other.  This is a general phenomena as is shown schematically
\begin{equation}
 F\circ G(x) = F(G(x)) = F(x)\bigr|_{x\to G(x)} 
  = \left(x\bigr|_{x\to F(x)}\right)\bigr|_{x\to G(x)}.
\end{equation}
In order to see the difference of two conventions, the following simple example may be useful.
If we introduce the mappings $f, g:\ \mathbb{C}\to \mathbb{C}$ by
\begin{equation}
 f:\ x\mapsto x+1,\quad g:\ x\mapsto x^2,
\end{equation}
then the composition of mappings (numerical composition) is computed as
\begin{equation}
f(g(x))=f(x^2)=x^2+1,\quad
g(f(x))=g(x+1)=(x+1)^2.
\end{equation}
On the other hand, if we introduce the automorphisms $f,g: \mathbb{C}(x)\to\mathbb{C}(x)$ by
the substitutions
\begin{equation}
 f:\ x\mapsto x+1,\quad g:\ x\mapsto x^2,
\end{equation}
of the variable $x$, then the composition of substitutions (symbolical composition) implies
\begin{equation}
f(g(x))=f(x^2)=(x+1)^2,\quad
g(f(x))=g(x+1)=x^2+1.
\end{equation}
We usually adopt the convention of symbolical composition unless otherwise stated.
\end{rem}
\subsection{Lax pair}\label{subsec:P4_Lax}
P$_{\rm IV}$ \eqref{eqn:p4} can be expressed as the compatibility condition of the following
system of linear differential equations for $\psi=\psi(x,t)$:
\begin{align}
&\label{eqn:p4_lax_L} 
\psi_{xx}+ \left(\frac{1 - a_1}{x} - x - t-\frac{1}{x-q}\right) \psi_x
+ \left(-a_2 - \frac{H_{\rm IV}}{x}+ \frac{ pq}{x(x-q)}\right)\psi=0, \\
&\label{eqn:p4_lax_B}  \psi' = -\frac{pq}{x-q}\psi + \frac{x}{x-q}\psi_x,
\end{align}
where $'=\frac{\partial}{\partial t}$ and $(q,p)=(q(t),p(t))$.
We call \eqref{eqn:p4_lax_L} and \eqref{eqn:p4_lax_B} the {\em auxiliary linear problem} (or the
{\em Lax pair}) of P$_{\rm IV}$. In general, consider the following system of differential equations
\begin{equation}\label{eqn:linear_diffeq_21}
 \begin{split}
&  \psi_{xx} + u\psi_x + v\psi=0,\\
&  \psi' = a\psi + b\psi_x,
 \end{split}
\end{equation}
where $u$, $v$, $a$, $b$ are functions of $x$, $t$.  From these equations we can compute 
$(\psi_{xx})'= (-u\psi_x-v\psi)'$ and $(\psi')_{xx}=(a\psi+b\psi_x)_{xx}$ 
in the form of linear combinations of $\psi$ and $\psi_x$, assuming that $(\psi_x)'=(\psi')_x$.  Hence we obtain
$(\psi_{xx})' - (\psi')_{xx}= P\psi + Q\psi_x$, where $P$ and $Q$ are expressed in terms of $a$, $b$, $u$ and
$v$. We say that the system \eqref{eqn:linear_diffeq_21} is {\em compatible} if the coefficients $P$
and $Q$ are zero, which is a natural requirement for \eqref{eqn:linear_diffeq_21} to have two
linearly independent solutions. This implies the following equations for $u$, $v$, $a$, $b$:
\begin{equation}\label{eqn:compatibility_linear21}
u' = -2a_{x} - b_{xx} + b_xu + bu_x,\quad
v' = -a_{xx} -a_xu + 2b_xv + bv_x,
\end{equation}
which is called the {\em compatibility condition} of \eqref{eqn:linear_diffeq_21}. In case of
\eqref{eqn:p4_lax_L} and \eqref{eqn:p4_lax_B}, substituting the coefficients of the system
\eqref{eqn:p4_lax_L} and \eqref{eqn:p4_lax_B} into \eqref{eqn:compatibility_linear21}, and requiring
that \eqref{eqn:compatibility_linear21} holds for arbitrary $x$, we obtain the differential equation
in $t$ which is nothing but P$_{\rm IV}$.

dP$_{\rm II}$ \eqref{eqn:dP_A2} arises as the compatibility condition of
\eqref{eqn:p4_lax_L} and the following differential-difference equation
\begin{equation}\label{eqn:dp2_lax_B}
 \overline{\psi} =  \frac{p}{x-q}\psi -\frac{1}{x-q}\psi_x.
\end{equation}
We note that \eqref{eqn:dp2_lax_B} is known as a {\em Schlesinger transformation} for the system
\eqref{eqn:p4_lax_L} and \eqref{eqn:p4_lax_B} \cite{J-M:Monodromy2,J-M:Monodromy3,J-M:Monodromy1}. Consider the following
system of differential-difference equations,
\begin{equation}\label{eqn:linear_diffrential_differenceeq_21}
 \begin{split}
&  \psi_{xx} + u\psi_x + v\psi=0,\\
&  \overline{\psi} = a\psi + b\psi_x.
 \end{split}
\end{equation}
Then the discussion similar to the case of P$_{\rm IV}$ shows that the condition
\begin{equation}
 \overline{(\psi_{xx})} = \Big(\overline{\psi}\Big)_{xx},
\end{equation}
gives
\begin{equation}\label{eqn:compatibility_dlinear21}
\begin{split}
&(a_x - bv)\overline{u} + a\overline{v} - (2b_x+a-bu)v - bv_x + a_{xx} =0,\\
& (a + b_x- bu)\overline{u} + b\overline{v} - (a+2b_x)u  - bu_x + bu^2  - bv + 2a_x + b_{xx} =0,
\end{split}
\end{equation}
which is the compatibility condition of the system \eqref{eqn:linear_diffrential_differenceeq_21}.
Substituting the coefficients of the system \eqref{eqn:p4_lax_L} and \eqref{eqn:dp2_lax_B} into
\eqref{eqn:compatibility_dlinear21}, and requiring that \eqref{eqn:compatibility_dlinear21} holds
for arbitrary $x$, we obtain dP$_{\rm II}$.  In this sense, \eqref{eqn:p4_lax_L} and
\eqref{eqn:dp2_lax_B} can be regarded as the Lax pair of dP$_{\rm II}$ \eqref{eqn:dP_A2}.

\subsection{Hypergeometric solutions}\label{subsec:hyper_p4}
P$_{\rm IV}$ admits a class of special solutions expressible in terms of hypergeometric type
functions.  For instance, putting $a_i=0$ in \eqref{eqn:p4_sym}, we find that \eqref{eqn:p4_sym}
admits a specialization $f_i=0$. When $a_0=0$ setting $f_0=-p+q+t=0$ we have the Riccati equation
\begin{equation}\label{eqn:p4_riccati}
q' = q^2 + tq - a_1.
\end{equation}
Equation \eqref{eqn:p4_riccati} is linearized by putting $q = -\frac{w'}{w}$
\begin{equation}\label{eqn:p4_hermite}
w'' -  tw' - a_1w = 0.
\end{equation}
Let $H_\lambda(t)$ denote the Hermite function defined by \cite{Abramowitz-Stegun}
\begin{equation}
 H_\lambda(t)=2^{\frac{\lambda}{2}}\sqrt{\pi}
\left[\frac{1}{\Gamma\left(\frac{1-\lambda}{2}\right)}\,{}_1F_1\left(\frac{-\lambda}{2},\frac{1}{2};\frac{t^2}{2}\right) 
- 
\frac{\sqrt{2}t}{\Gamma\left(\frac{-\lambda}{2}\right)}\,{}_1F_1\left(\frac{1-\lambda}{2},\frac{3}{2};\frac{t^2}{2}\right)
\right].
\end{equation}
$H_\lambda(t)$ satisfies the differential equation
\begin{equation}
 H_\lambda''(t) - tH'_\lambda(t) + \lambda H_\lambda(t) =0 ,
\end{equation}
and the contiguity relation
\begin{equation}\label{eqn:hermite_contiguity}
 H'_\lambda(t) = \lambda H_{\lambda-1}(t).
\end{equation}
Note that if $\lambda=n\in\mathbb{N}$, $H_n(t)$ is the Hermite polynomial
\begin{equation}
 H_n(t) = (-1)^ne^{\frac{t^2}{2}}\left(\frac{d}{dt}\right)^ne^{-\frac{t^2}{2}}.
\end{equation}
Therefore $H_{-a_1}(t)$ solves \eqref{eqn:p4_hermite} and corresponding $q$ is given by
\begin{equation}\label{eqn:p4_q_hermite}
 q=-\frac{H_{-a_1}'(t)}{H_{-a_1}(t)}.
\end{equation}
By taking this solution as a {\em seed} we can apply the B\"acklund transformations to obtain the
solutions expressible by the rational functions of the Hermite functions (actually the ratio of
determinants of them). We call this class of the special solutions the {\em hypergeometric
solutions} to P$_{\rm IV}$.

Let us apply the same specialization $a_0=0$ to the dP$_{\rm II}$ \eqref{eqn:dP_A2}.  Then we see
that it admits the specialization $f_0=-p+q+t=0$
\begin{equation}\label{eqn:dp2_riccati}
\overline{q} = -\dfrac{a_2}{q+t},
\end{equation}
which is linearized by putting $q = -\frac{\underline{a}_2\,\underline{w}}{w}$ as
\begin{equation}\label{eqn:p4_parabolic_contiguity}
\overline{w} - tw + (a_2-1)\underline{w}=0.
\end{equation}
Since $H_\lambda(t)$ satisfies the recursion relation
\begin{equation}
 H_{\lambda+1}(t) - t H_{\lambda}(t) + \lambda H_{\lambda-1}(t)=0,
\end{equation}
$w=H_{a_2-1}(t)= H_{-a_1}(t)$ solves \eqref{eqn:p4_parabolic_contiguity}. Moreover,
from the contiguity relation \eqref{eqn:hermite_contiguity} $q$ is rewritten as
$q=-\frac{H_{-a_1}'(t)}{H_{-a_1}(t)}$. 

Thus, we have confirmed the existence of a solution which satisfies both P$_{\rm IV}$
\eqref{eqn:p4_hamilton} and dP$_{\rm II}$ \eqref{eqn:dP_A2} simultaneously. This fact is expected by
construction, since the dP$_{\rm II}$ flow commutes with P$_{\rm IV}$ flow.
\subsection{Biquadratic pencils and autonomous dP$_{\rm II}$}\label{subsec:QRT}
In this subsection, we consider the {\em autonomous} case.  In P$_{\rm IV}$ \eqref{eqn:p4_sym}, the
parameters are normalized in such a way that $a_0+a_1+a_2=1$. We here rescale the variables as
$(q,p,t,a_{0}, a_{1},a_2)_{\rm old}=(\delta^{-1/2} q,\delta^{-1/2} p,\delta^{-1/2} t, \delta^{-1}
a_0, \delta^{-1} a_1,\delta^{-1}a_2)$ so that $a_0+a_1+a_2=\delta$. We also introduce a new
independent variable $s=t/\delta$, hence $\frac{dt}{ds}=\delta$. Then the autonomous case is given
by taking the limit $\delta\to 0$. The resulting equation has the same form as \eqref{eqn:p4} or
\eqref{eqn:p4_hamilton}, where $'$ is understood as the differentiation with respect to $s$, and
$t'=0$.

\begin{figure}[ht]
\begin{center}
\includegraphics[scale=0.3]{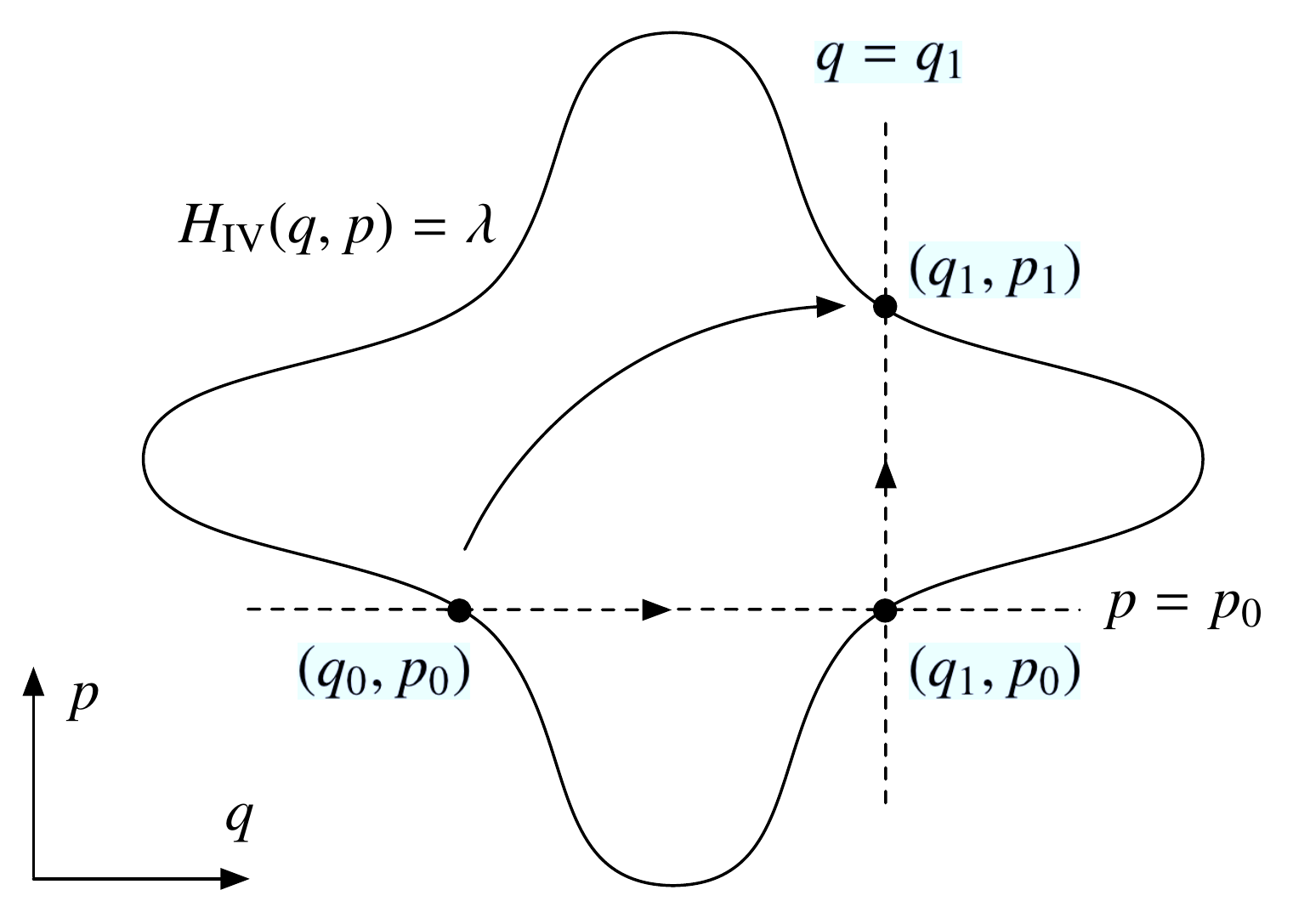}
\end{center}
\caption{QRT mapping.}\label{fig:QRT}
\end{figure}

The Hamiltonian $H_{\rm IV}$ is a conserved quantity of this autonomous P$_{\rm IV}$. The integral
curves
\begin{equation}
C_\lambda:~H_{\rm IV} =-a_1 p - a_2 q + pq(p-q-t)= \lambda,
\end{equation}
define a one-parameter family (pencil) of curves of bidegree (2,2) on $(q,p)$-plane. In terms of
this pencil of curves, the autonomous version of dP$_{\rm II}$ \eqref{eqn:dP_A2} is geometrically
reformulated as follows. For a point $(q_0,p_0)$ given, we choose the parameter $\lambda$ so that
$C_\lambda$ passes through it.  Then we have the following two points $(q_1,p_0)$ and $(q_0,p_1)$
where $q_1$ (resp. $p_1$) is determined by solving $H_{\rm IV}(q_0,p_0)=H_{\rm IV}(q_1,p_0)$
(resp. $H_{\rm IV}(q_1,p_1)=H_{\rm IV}(q_1,p_0)$) as (see Fig.\ref{fig:QRT})
 \begin{equation}\label{eqn:dP_A2_autonomous}
\begin{split}
&q_1=p_0-q_0-t-\dfrac{a_2}{p_0},\\
&p_1=q_1-p_0+t + \dfrac{a_1}{q_1},
\end{split}
\end{equation}
This procedure defines a discrete dynamical system $(q_0,p_0)\to (q_1,p_1)$ on the $(q,p)$-plane
(the autonomous dP$_{\rm II}$).  Note that the mapping $(q_0,p_0)\to (q_1,p_1)$ is composed of two
steps, the {\em horizontal flip} $(q_0,p_0)\to (q_1,p_0)$ followed by the {\em vertical flip}
$(q_1,p_0)\to (q_1,p_1)$.  In general, the class of discrete dynamical systems arising from pencils
of biquadratic curves by the above procedure is called the {\em QRT mappings}. The QRT mappings were
originally obtained in \cite{QRT1,QRT2} as reduction of discrete soliton equations; the above geometric
formulation is due to \cite{Tsuda:QRT}(see also \cite{Duistermaat:book,KMNOY:10E9}).

In general, a pencil of biquadratic curves $\lambda F(q,p)+\mu G(q,p)=0$ is characterized by the
eight points in $\mathbb{P}^1\times \mathbb{P}^1$ which are the intersection of $F(q,p)=0$ and
$G(q,p)=0$. Such a configuration of eight points is {\em special}, since any generic configuration
of eight points determines a unique biquadratic curve passing through them.  As will be discussed
later, the discrete Painlev\'e equations (non-autonomous cases) arise from the {\em non-special}
configurations of eight points in $\mathbb{P}^1\times \mathbb{P}^1$.

Historically, the QRT mappings played a crucial role in the construction and development of the
theory of discrete Painlev\'e equations. Grammaticos, Ramani and Papageorgiou \cite{GRP:sc} first
observed the {\em singularity confinement property} of the QRT mapping and noticed that it can be
regarded as a discrete analogue of the Painlev\'e property. By using the singularity confinement
property as the integrability detector, many interesting discrete Painlev\'e equations were found by
de-autonomizing the QRT mappings \cite{GR:review2004,GRH:dP}.
\subsection{$\tau$ functions}
The $\tau$ function is one of the most important objects in the theory of integrable systems.  In
the context of Painlev\'e differential equations, it is introduced as a function whose logarithmic
derivative gives the Hamiltonian
\cite{J-M:Monodromy2,J-M:Monodromy3,J-M:Monodromy1,Okamoto:tau,Okamoto:book}. One can also define
the B\"acklund transformations on $\tau$ functions in such a way that they are consistent with the
differential equation.

Consider the Hamiltonian \eqref{eqn:p4_hamiltonian} for P$_{\rm IV}$ in terms of $f_0$, $f_1$, $f_2$
\begin{equation}
 H_{\rm IV}=-a_1 f_2 + a_2 f_1 + f_2f_1f_0.
\end{equation}
Applying the B\"acklund transformation $s_1$ and $s_2$ defined in \eqref{eqn:p4_BT_pq_a}, we observe
that
\begin{equation}
 s_1(H_{\rm IV}) = H_{\rm IV} + a_1t,\quad s_2(H_{\rm IV})=H_{\rm IV}-a_2t.
\end{equation}
Hence, we slightly modify the Hamiltonian so that it is invariant with respect to $s_1$ and $s_2$:
\begin{equation}
h_0= -a_1 f_2 + a_2 f_1 + f_2f_1f_0 + \frac{a_1-a_2}{3}t.
\end{equation}
Then we see that
\begin{equation}\label{eqn:p4_BT_hamiltonian}
 s_1(h_0) = s_2(h_0)=h_0,\quad s_0(h_0) = h_0 + \frac{a_0}{f_0}.
\end{equation}
Introducing two other Hamiltonians $h_1$, $h_2$ by
\begin{align}
& h_1 = \pi(h_0) = -a_2 f_0 + a_0 f_2 + f_0f_2f_1 + \frac{a_2-a_0}{3}t,\\
& h_2 = \pi^2(h_0)= -a_0 f_1 + a_1 f_0 + f_1f_0f_2 + \frac{a_0-a_1}{3}t,
\end{align}
we have
\begin{equation}\label{eqn:p4_h2-h1}
 h_2 - h_1 = f_0  - \frac{ t}{3},\quad
 h_0 - h_2 = f_1  - \frac{ t}{3},\quad
 h_1 - h_0 = f_2  - \frac{ t}{3}.
\end{equation}
From the first equation of \eqref{eqn:p4_sym}, \eqref{eqn:p4_h2-h1} and
\eqref{eqn:p4_BT_hamiltonian} we have
\begin{equation}\label{eqn:p4_h_rel1}
\frac{f_0'}{f_0} =  f_1 - f_2  + \frac{a_0}{f_0}
=2h_0-h_1-h_2 + \frac{a_0}{f_0}
=s_0(h_0) + h_0 - h_1-h_2 .
\end{equation}
We now introduce the $\tau$ function $\tau_i$ ($i=0,1,2$) by
\begin{equation}\label{eqn:p4_tau_def}
 h_i = (\log \tau_i)' = \frac{\tau_i'}{\tau_i}.
\end{equation}
Substituting \eqref{eqn:p4_tau_def} into \eqref{eqn:p4_h_rel1}, 
we find that $f_0$ should be expressed as
\begin{equation}
 f_0 = c_0\frac{\tau_0s_0(\tau_0)}{\tau_1\tau_2},
\end{equation}
where $c_0$ is an integration constant. Similarly, we also obtain
\begin{equation}
  f_1 = c_1\frac{\tau_1s_1(\tau_1)}{\tau_2\tau_0},\quad
  f_2 = c_2\frac{\tau_2s_2(\tau_2)}{\tau_0\tau_1}.
\end{equation}
It is a subtle question how to fix the constants $c_i$, since they may depend on the parameters
$a_0$, $a_1$, $a_2$. While being aware of this point, we make the simplest possible choice by
setting $c_0=c_1=c_2=1$; namely 
\begin{equation}
  f_0 = \frac{\tau_0s_0(\tau_0)}{\tau_1\tau_2},\quad
  f_1 = \frac{\tau_1s_1(\tau_1)}{\tau_2\tau_0},\quad
  f_2 = \frac{\tau_2s_2(\tau_2)}{\tau_0\tau_1}.
\end{equation}
Hence we introduce the B\"acklund transformations on $\tau$ functions as follows:
\begingroup \renewcommand{\arraystretch}{1.8}
\begin{equation}\label{eqn:BT_on_tau}
\begin{tabular}{|c|c|c|c|}
\hline
&$\tau_0$ &$\tau_1$ &$\tau_2$ \\
\hline
$s_0$&$f_0~\dfrac{\tau_1\tau_2}{\tau_0}$  & $\tau_1$ &$\tau_2$\\
\hline
$s_1$& $\tau_0$ & $f_1~\dfrac{\tau_2\tau_0}{\tau_1}$ & $\tau_2$ \\
\hline
$s_2$ & $\tau_0$ & $\tau_1$ & $f_2~\dfrac{\tau_1\tau_0}{\tau_2}$ \\
\hline
$\pi$ & $\tau_1$ & $\tau_2$ & $\tau_0$ \\
\hline
\end{tabular} 
\end{equation}
\endgroup One can show that this definition of the B\"acklund transformations is consistent with the
differential equation \eqref{eqn:p4_sym} and \eqref{eqn:p4_tau_def}, and that they form the
extended affine Weyl group of type $A_2^{(1)}$.

\subsection{Space of initial values}
\subsubsection{Resolution of singularities by blowing-up: a simple example}\label{subsubsection:blow-up}
\begin{figure}[ht]
\begin{center}
\includegraphics[scale=0.4]{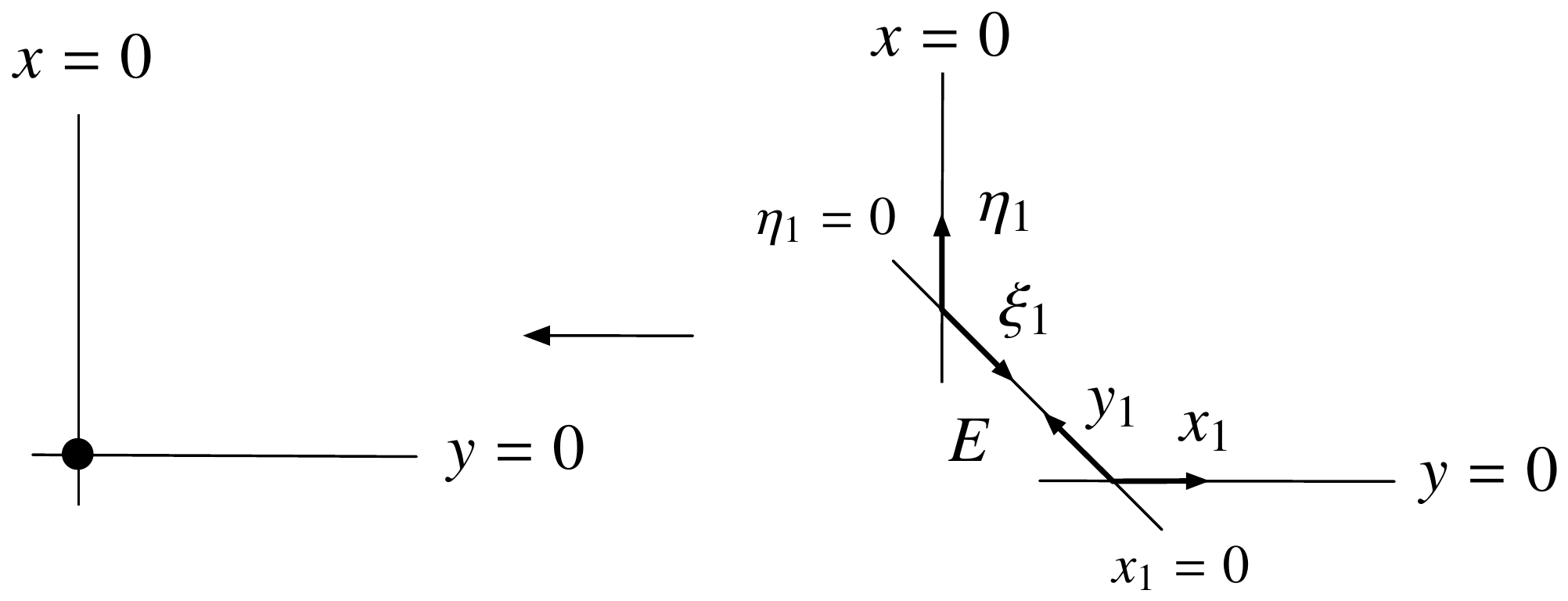}
\end{center}
\caption{A graphical representation of the process of blowing up. }\label{fig:blowup_simple_ex_diagram}
\end{figure}

Let us first consider the following simple differential equation
\begin{equation}\label{eqn:siv_simple_ex}
x'=1,\quad xy'=y.
\end{equation}
There exists a unique solution for generic initial value $(x(t_0),y(t_0))=(\xi,\eta)$ with $\xi\neq 0$. 
In case of $\xi=0$, (i) if $\eta\neq 0$ the point $(0,\eta)$ is {\em inaccessible}, namely,
there is no solution passing through this point, (ii) if $\eta=0$ there are infinitely many
solutions.  To see this, we change the variables as
\begin{equation}\label{eqn:blow-up}
 (x_1,y_1)=\left(x,\frac{y}{x}\right),\quad \mbox{namely}\quad (x,y)=(x_1,x_1y_1),
\end{equation}
 which yields a regular differential equation
\begin{equation}\label{eqn:siv_simple_ex_2}
x_1'=1,\quad y_1'=0.
\end{equation}
The general solution to \eqref{eqn:siv_simple_ex_2} is given by $(x_1,y_1)=(t-t_0,C)$, where $C$ is  an arbitrary constants.
In terms of the variables $(x,y)$, $(x,y)=(t-t_0,C(t-t_0))$ parametrizes the solutions of \eqref{eqn:siv_simple_ex} 
passing through $(x,y)=(0,0)$ at $t=t_0$.
\begin{figure}[ht]
\begin{center}
\includegraphics[scale=0.4]{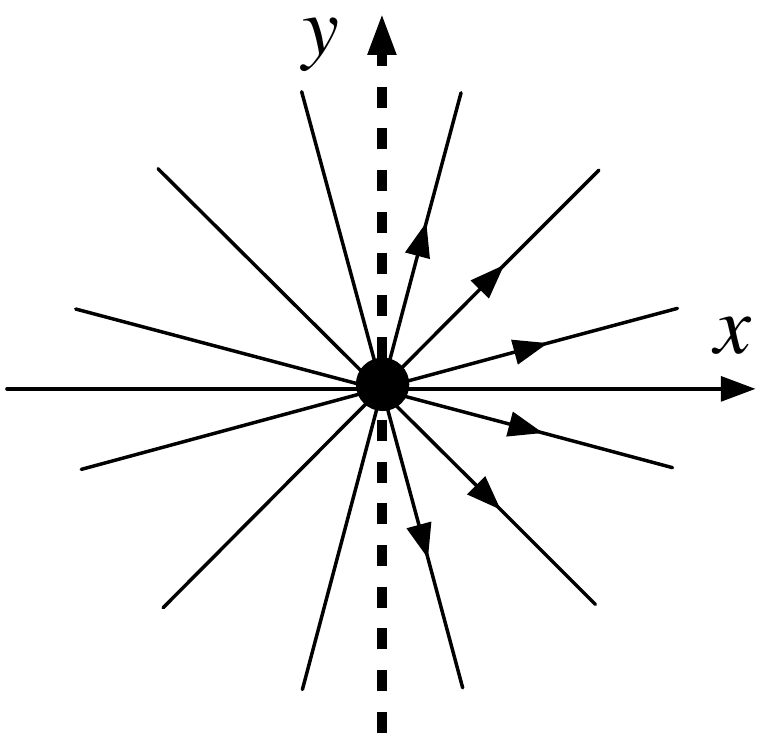}\hskip50pt
\includegraphics[scale=0.4]{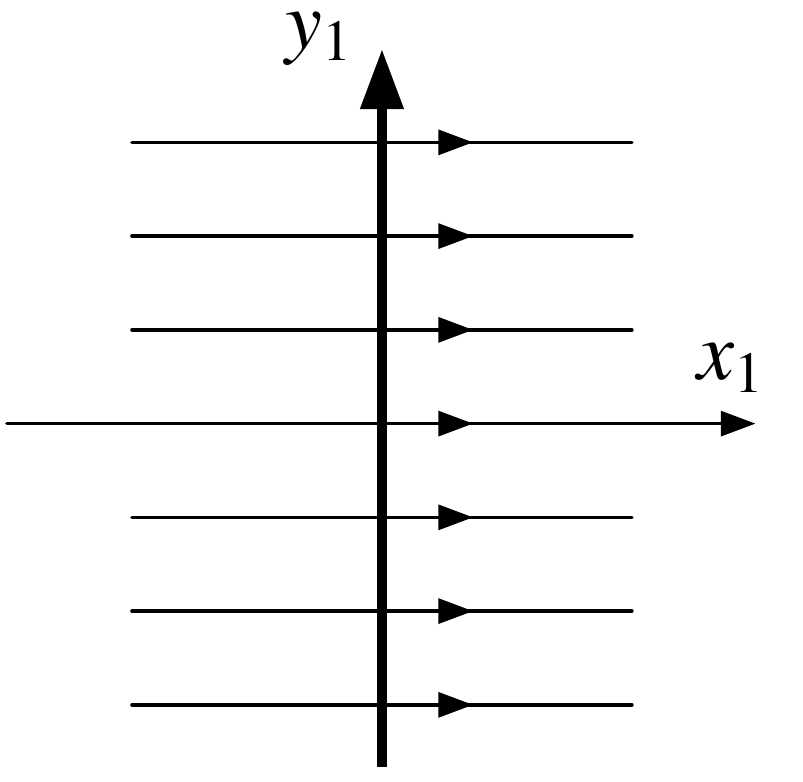}
\end{center}
\caption{A simple example of blowing-up. The line $x_1=0$ (thick line in the right figure)
corresponds to the singularity in $(x,y)$-coordinates (black circle in the left figure). The dotted
line in the left figure is inaccessible.}\label{fig:blowup_simple_ex}
\end{figure}
\noindent This means that the singularity of \eqref{eqn:siv_simple_ex} at $(x,y)=(0,0)$ is resolved,
and the infinitely many solutions passing through $(x,y)=(0,0)$ are separated by the gradient
variable $y_1=\frac{y}{x}$.  The transformation \eqref{eqn:blow-up} is called the {\em blowing up}
at $(x,y)=(0,0)$. By this transformation the point $(x,y)=(0,0)$ corresponds to the line $x_1=0$,
called the {\em exceptional line} (see Figure \ref{fig:blowup_simple_ex}). To be more precise, the
exceptional line should be considered as $\mathbb{P}^1$ including the point where the gradient
variable is $y_1=\infty$. In order to cover the whole exceptional line, we also use the variable
$(\xi_1,\eta_1)$ such that $(x,y)=(\xi_1\eta_1,\eta_1)$ as a companion to $(x_1,y_1)$. This process
of blowing up replaces the point $(x,y)=(0,0)$ by the exceptional line $E=\{x_1=0\}\cup
\{\eta_1=0\}$, which is graphically described in Figure \ref{fig:blowup_simple_ex_diagram}.
\subsubsection{Resolution of singularities of P$_{\rm IV}$}\label{subsubsection:resolution_P4}
Okamoto applied this type of procedures to each of the Painlev\'e equations to construct the {\em
space of initial values} which parametrizes the whole set of solutions. Taking the example of
P$_{\rm IV}$
\begin{equation}\label{eqn:p4_pq}
\left\{
\begin{array}{l}\medskip
{\displaystyle  q' = -a_1 + 2pq - q^2 - q t},\\
{\displaystyle  p' =  a_2 - p^2 + 2 pq + pt},
\end{array}
\right.
\end{equation}
we describe how this procedure works without getting into the details. 
As is easily seen, there is no singularities for finite $(q,p)$.  Regarding $(q,p)$ as
the inhomogeneous coordinates of $\mathbb{P}^1\times\mathbb{P}^1$, we investigate the singularities
around the points at infinity by using three sets of local coordinates (1) $(q,1/p)$, (2) $(1/q,p)$
and (3) $(1/q,1/p)$.\par\medskip

(1) We first change the dependent variables $(q,p)$ to $(q_0,p_0)=(q,1/p)$ to see the solutions that
pass through the line $p=\infty$ ($p_0=0$), which yields
\begin{equation}\label{eqn:p4_p0q0}
\left\{
\begin{array}{l}\medskip
{\displaystyle  q_0' = \frac{2q_0}{p_0} - a_1 -tq_0  - q_0^2},\\
{\displaystyle  p_0' =  1 - (t+2q_0)p_0 -a_2p_0^2}.
\end{array}
\right.
\end{equation}
We see that if $p_0=0$, $q_0\neq 0$  the point ($p_0,q_0$) is {\em inaccessible} (no solution
can pass through the point), and $(q_0,p_0)=(0,0)$ is the singular point at which we should
apply the blowing-up: $(q_0,p_0)=(q_1p_1,p_1)$. Then we obtain
\begin{equation}\label{eqn:p4_p1q1}
\left\{
\begin{array}{l}\medskip
{\displaystyle  q_1' = \frac{q_1-a_1}{p_1} + \left(a_2+q_1\right)q_1p_1  },\\
{\displaystyle  p_1' =  1 - tp_1  - \left(a_2 + 2 q_1\right)p_1^2 }.
\end{array}
\right.
\end{equation}
When $p_1=0$, the point $(q_1,p_1)=(a_1,0)$ is the only {\em accessible singularity}, where we need
another blowing-up: $(q_1,p_1)=(a_1 + q_2p_2,p_2)$. Then we obtain a regular differential equation
\begin{equation}\label{eqn:p4_p2q2}
\left\{
\begin{array}{l}\medskip
{\displaystyle  q_2' =  a_1^2 + a_1 a_2 + tq_2  + 2 (2 a_1+a_2)q_2p_2   + 3 q_2^2p_2^2 },\\
{\displaystyle  p_2' =  1 - tp_2 - (2 a_1 + a_2)p_2^2  - 2 q_2p_2^3 }.
\end{array}
\right.
\end{equation}
In this way, the singularity at $(q,p)=(0,\infty)$ has been resolved by two successive blowing-ups.

\begin{figure}[ht]
\begin{center}
\includegraphics[scale=0.33]{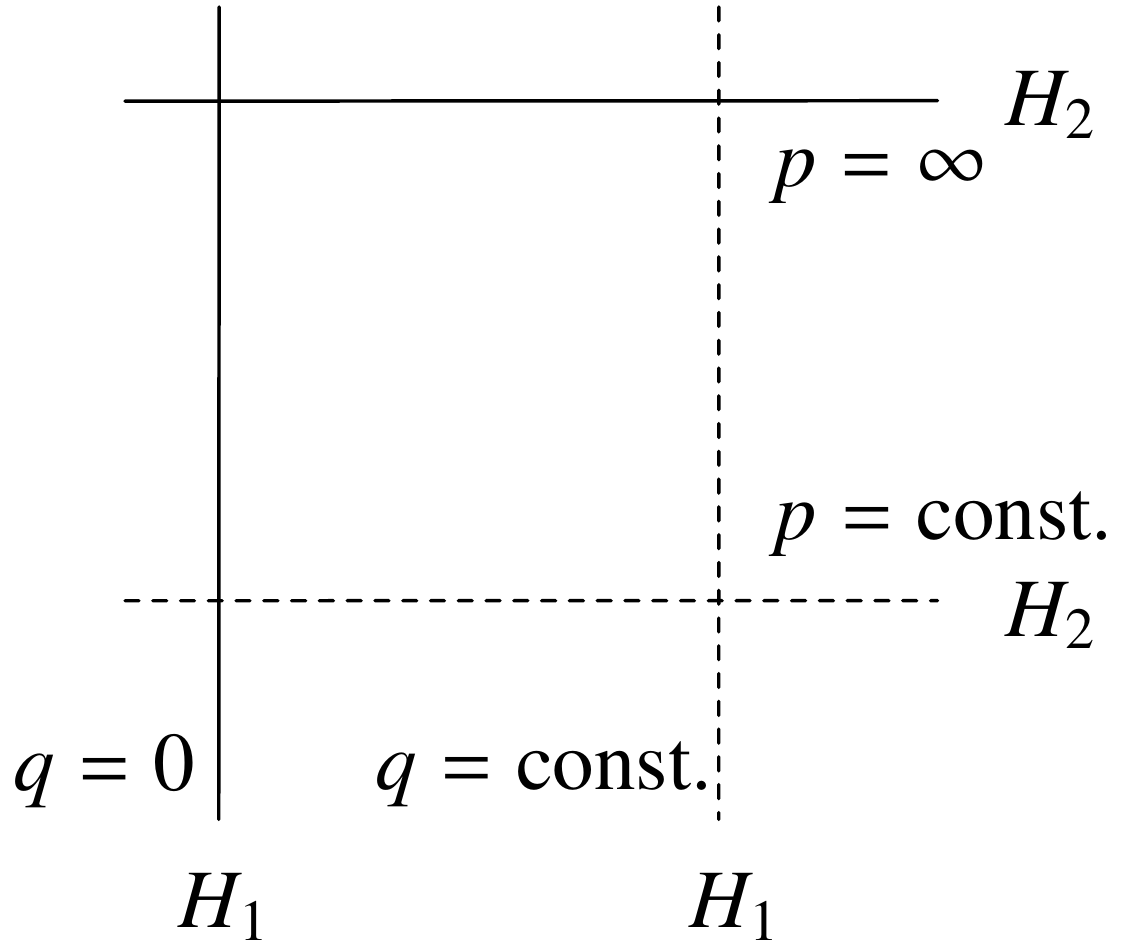}\hskip10pt\raise50pt\hbox{$\longleftarrow$}\hskip10pt
\includegraphics[scale=0.33]{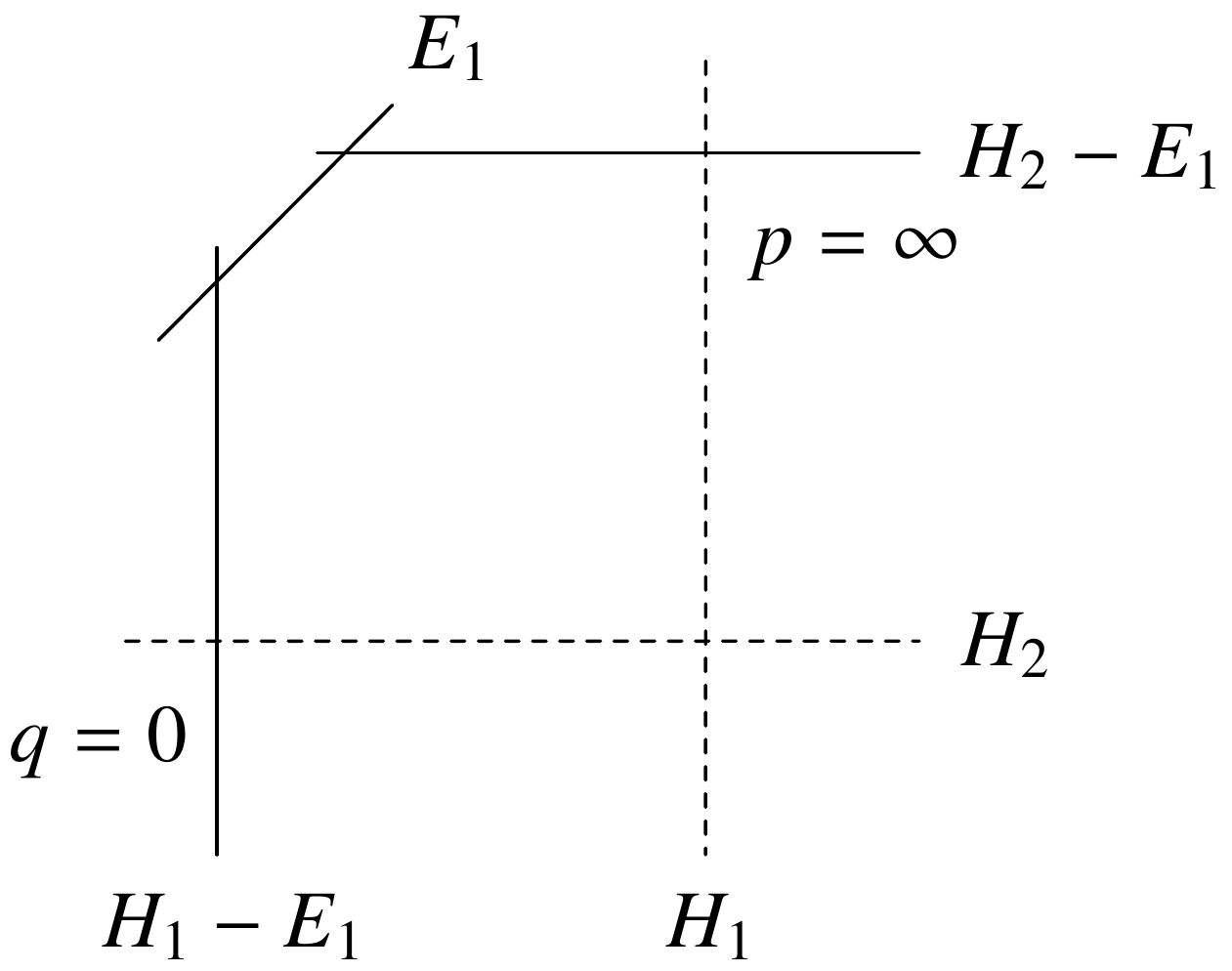}\\
\hskip10pt\raise50pt\hbox{$\longleftarrow$}\hskip10pt\includegraphics[scale=0.33]{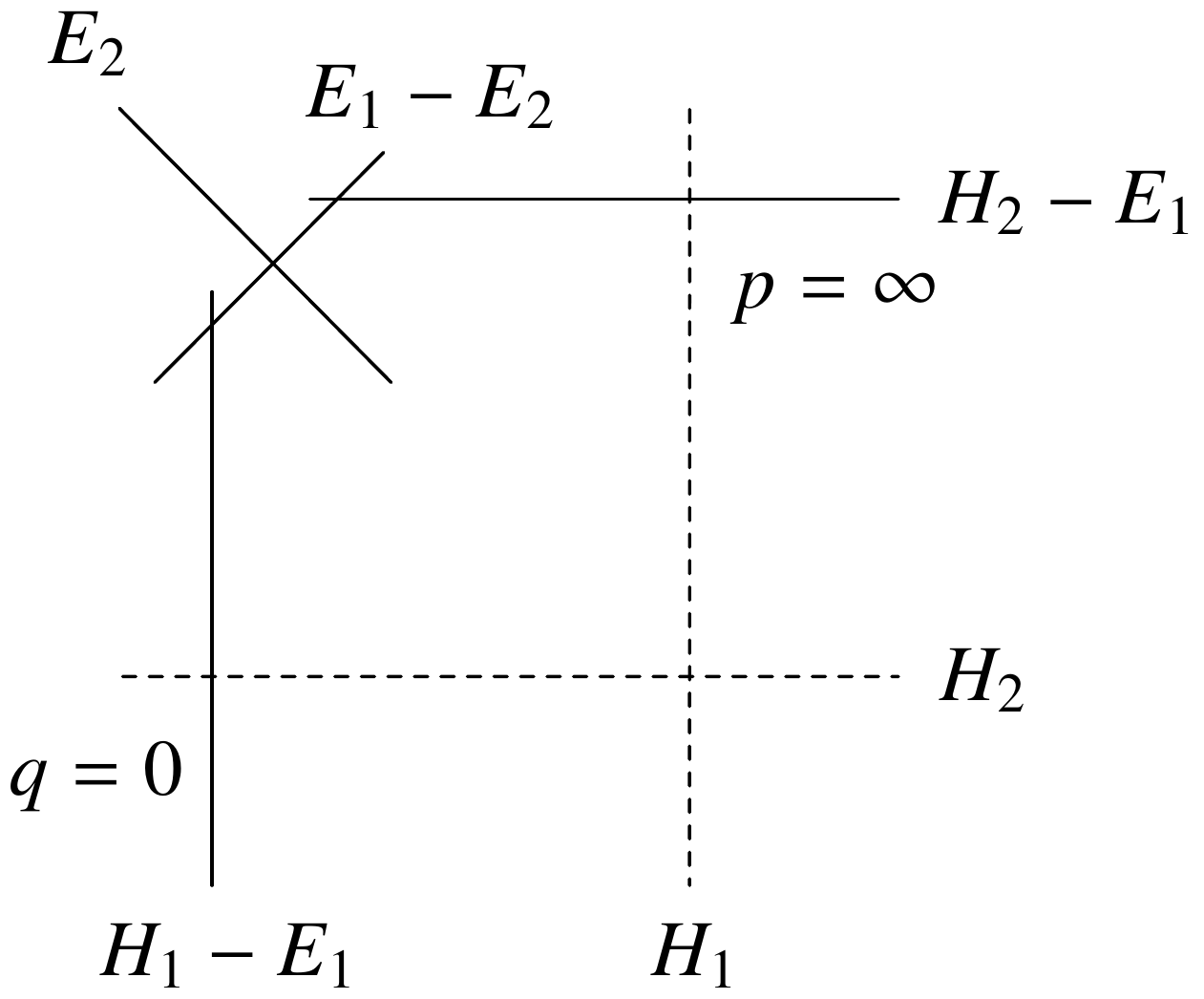}
\end{center}
\caption{Blowing-up of singularity at $(q,p)=(0,\infty)$ of P$_{\rm IV}$.} 
\label{fig:blowup_p4_012}
\end{figure}

We introduce some notations of algebraic geometry in order to book-keep this procedure.  A formal
$\mathbb{Z}$-linear combination of curves on a surface is called a {\em divisor}. In
$\mathbb{P}^1\times \mathbb{P}^1$ with inhomogeneous coordinates $(q,p)$, we denote by $H_1$ and $H_2$
the classes of divisors (curves) $q={\rm const.}$ and $p={\rm const.}$, respectively.  In the first
blowing-up, we denote by $E_1$ the exceptional divisor $p_1=0$ obtained from the singularity
$(q,p)=(0,\infty)$. In the blowing-up space, the divisor corresponding to $p=\infty$ ($p_0=0$) has
two components; one is the exceptional divisor $E_1$ and the other, called the {\em proper
transform} of $p=\infty$, is denoted by $H_2-E_1$.  In the second blowing-up space, we denote by
$E_2$ the exceptional divisor $p_2=0$ obtained from the singularity $(q_1,p_1)=(a_1,0)$, and by
$E_1-E_2$ the proper transform of $E_1$. (See Figure \ref{fig:blowup_p4_012})

(2) We next change the dependent variables $(q,p)$ to $(q_0,p_0)=(1/q,p)$ to see the solutions that pass
through the divisor $q=\infty$ ($q_0=0$) which yields the differential equation
\begin{equation}\label{eqn:p4_p0q0_2}
\left\{
\begin{array}{l}\medskip
{\displaystyle  q_0' = 1+ q_0 \left(t-2 p_0\right)+ a_1 q_0^2   },\\
{\displaystyle  p_0' =  \frac{2 p_0}{q_0} + a_2  + p_0 t - p_0^2,}
\end{array}
\right.
\end{equation}
with the only accessible singularity at $(q_0,p_0)=(0,0)$. Similarly to the previous case, this
singularity $(q,p)=(\infty,0)$ can be resolved by two successive blowing-ups: 
\begin{equation}\label{eqn:p4_blowup_var2}
\text{(i)}\quad (q_0,p_0)=(q_1,q_1p_1),\quad \text{(ii)}\quad (q_1,p_1)=(q_2,-a_2+q_2p_2).
\end{equation}
The corresponding exceptional divisors are denoted by $E_3$ and $E_4$, respectively.

\begin{figure}[ht]
\begin{center}
\includegraphics[scale=0.33]{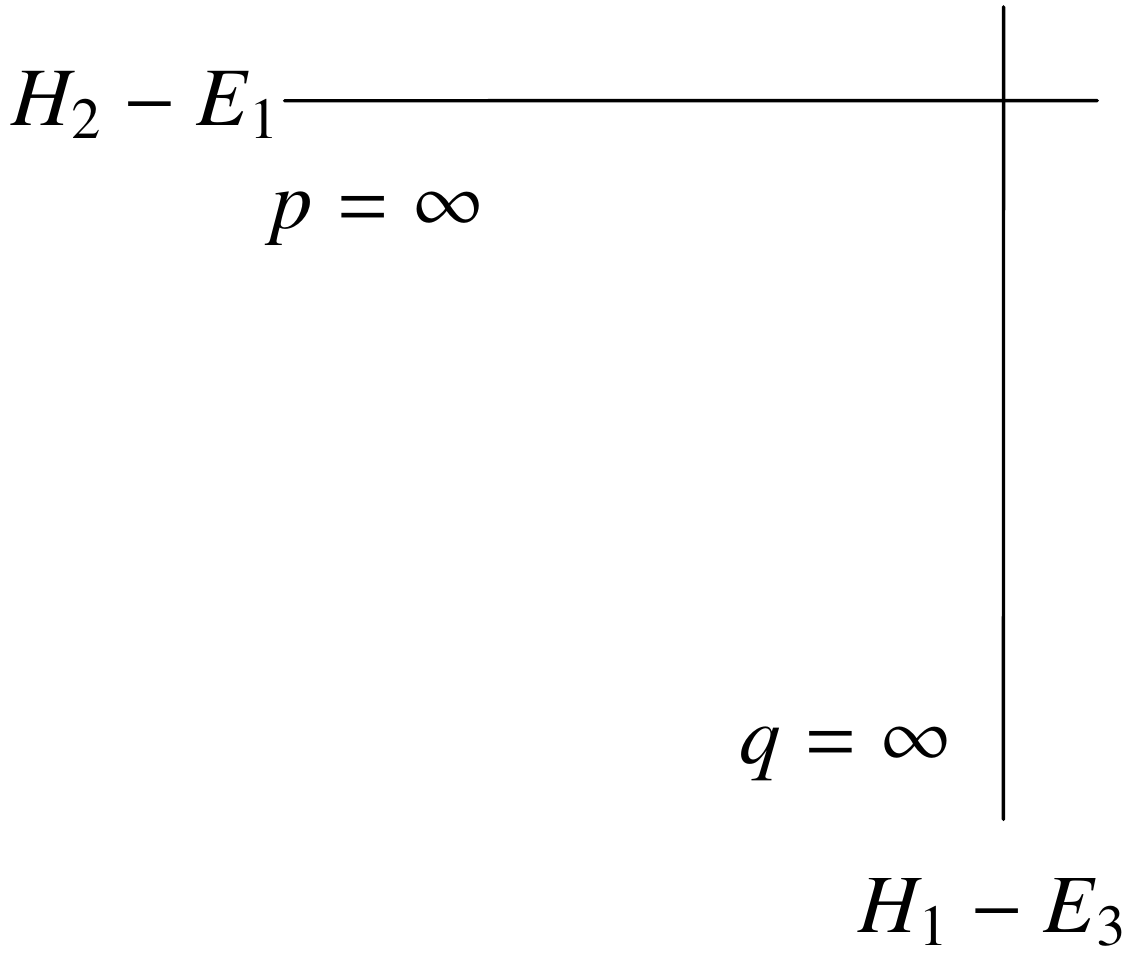}\hskip10pt
\raise40pt\hbox{$\longleftarrow$}\includegraphics[scale=0.33]{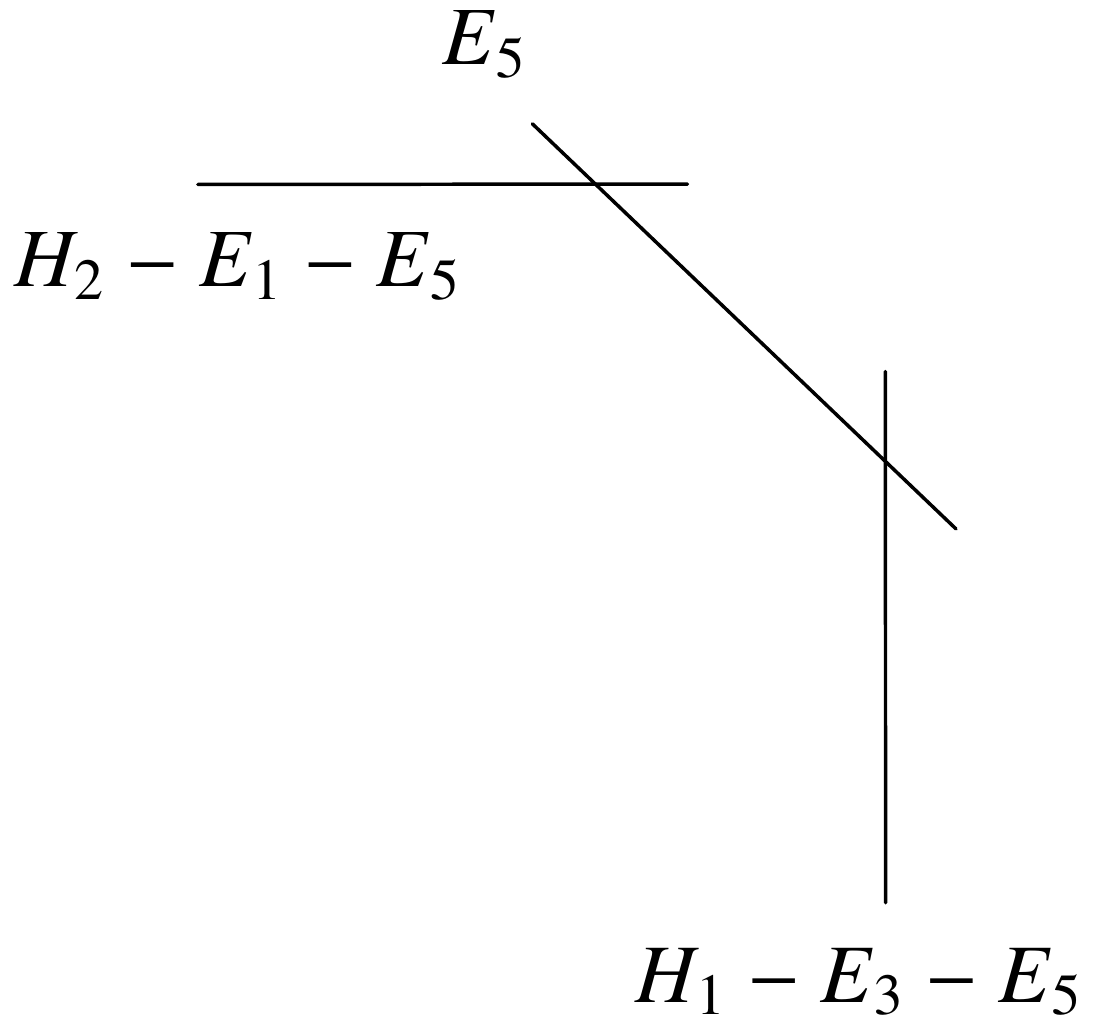}\\[5mm]
\hskip10pt\raise40pt\hbox{$\longleftarrow$}\hskip10pt\includegraphics[scale=0.33]{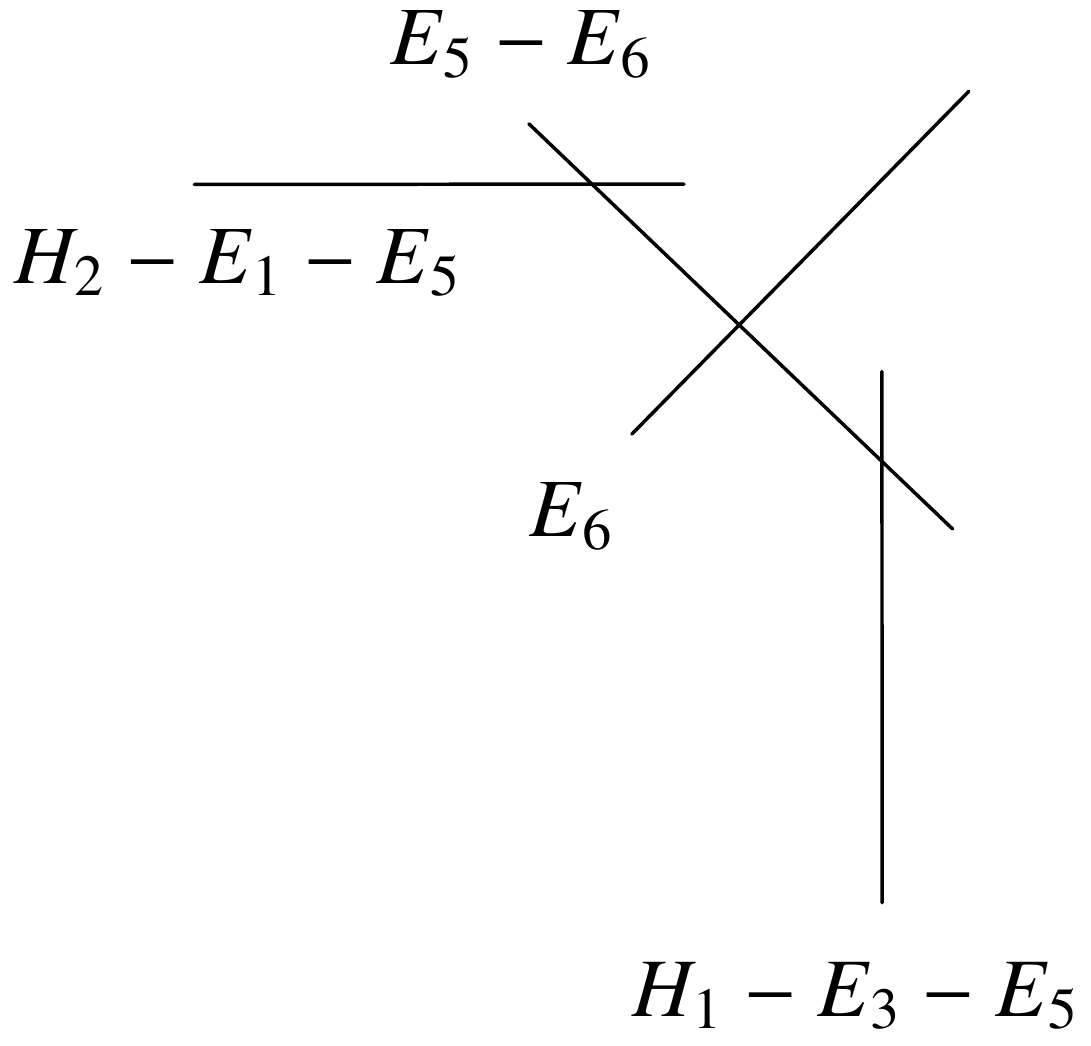}
\hskip10pt\raise40pt\hbox{$\longleftarrow$}\hskip20pt\includegraphics[scale=0.33]{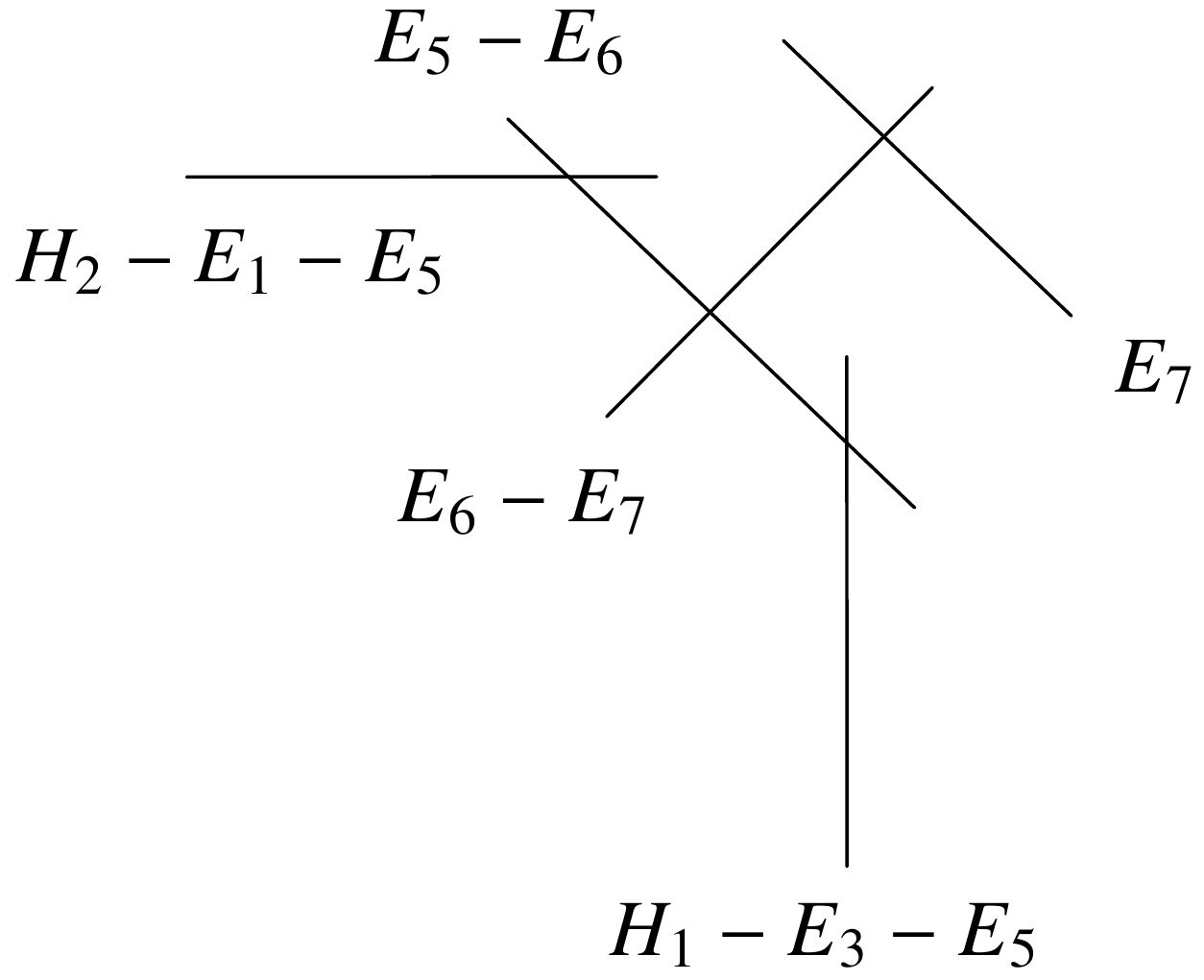}\\[5mm]
\hskip10pt\raise40pt\hbox{$\longleftarrow$}\hskip20pt\includegraphics[scale=0.33]{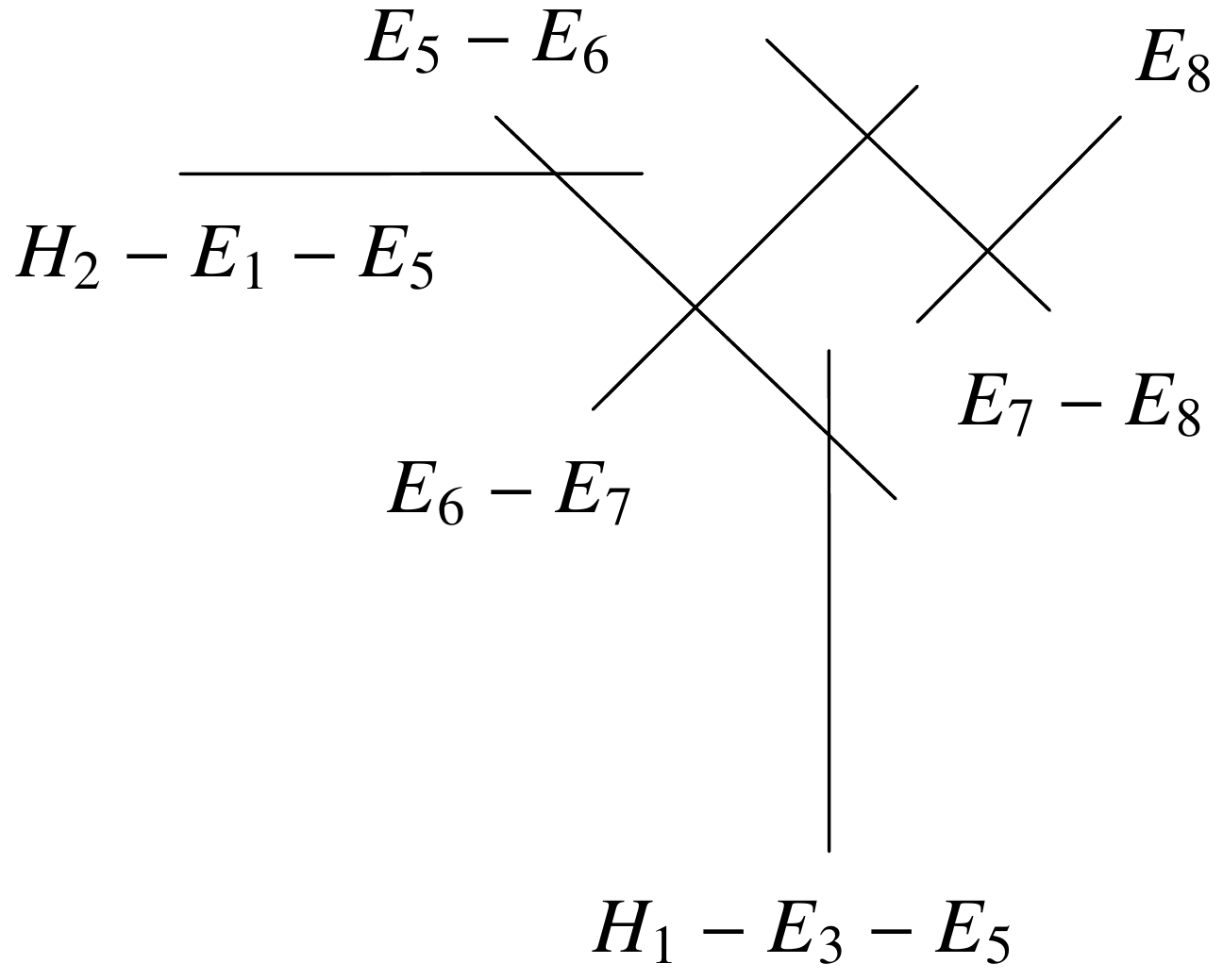}
\end{center}
\caption{Blowing-up of singularity at $(q,p)=(\infty,\infty)$ of P$_{\rm IV}$.} 
\label{fig:blowup_p4_34567}
\end{figure}

(3) We finally investigate the solution around $(q,p)=(\infty,\infty)$ by the change of coordinates
$(q,p)=(1/q_0,1/p_0)$:
\begin{equation}\label{eqn:p4_p0q0_3}
\left\{
\begin{array}{l}\medskip
{\displaystyle  q_0' =  1+ \frac{q_0 \left(p_0 t-2\right)}{p_0} + a_1 q_0^2 },\\
{\displaystyle  p_0' =  -\frac{2 p_0}{q_0} +1 - t p_0- a_2 p_0^2  }.
\end{array}
\right.
\end{equation} 
In this case, we need four successive blowing-ups to resolve the singularity at $(q_0,p_0)=(0,0)$:
\begin{equation}
 \begin{split}
\text{(i)}\quad  (q_0,p_0)&=(q_1,q_1p_1),\\
\text{(ii)}\quad (q_1,p_1)&=(q_2,1+q_2p_2),\\
\text{(iii)} \quad (q_2,p_2)&=(q_3,-t+q_3p_3),\\
 \text{(iv)}\quad (q_3,p_3)&=(q_4,a_0+t^2+q_4p_4).
 \end{split}
\end{equation}
In fact, after the fourth blowing-up, we obtain a rational differential equation with respect to
$(q_4,p_4)$, but it has no singularity at $q_4=0$ any more. The corresponding
exceptional divisors are denoted by $E_5$, $E_6$, $E_7$ and $E_8$, respectively (see Figure
\ref{fig:blowup_p4_34567}).\par\medskip

\begin{figure}[ht]
\begin{center}
\includegraphics[scale=0.35]{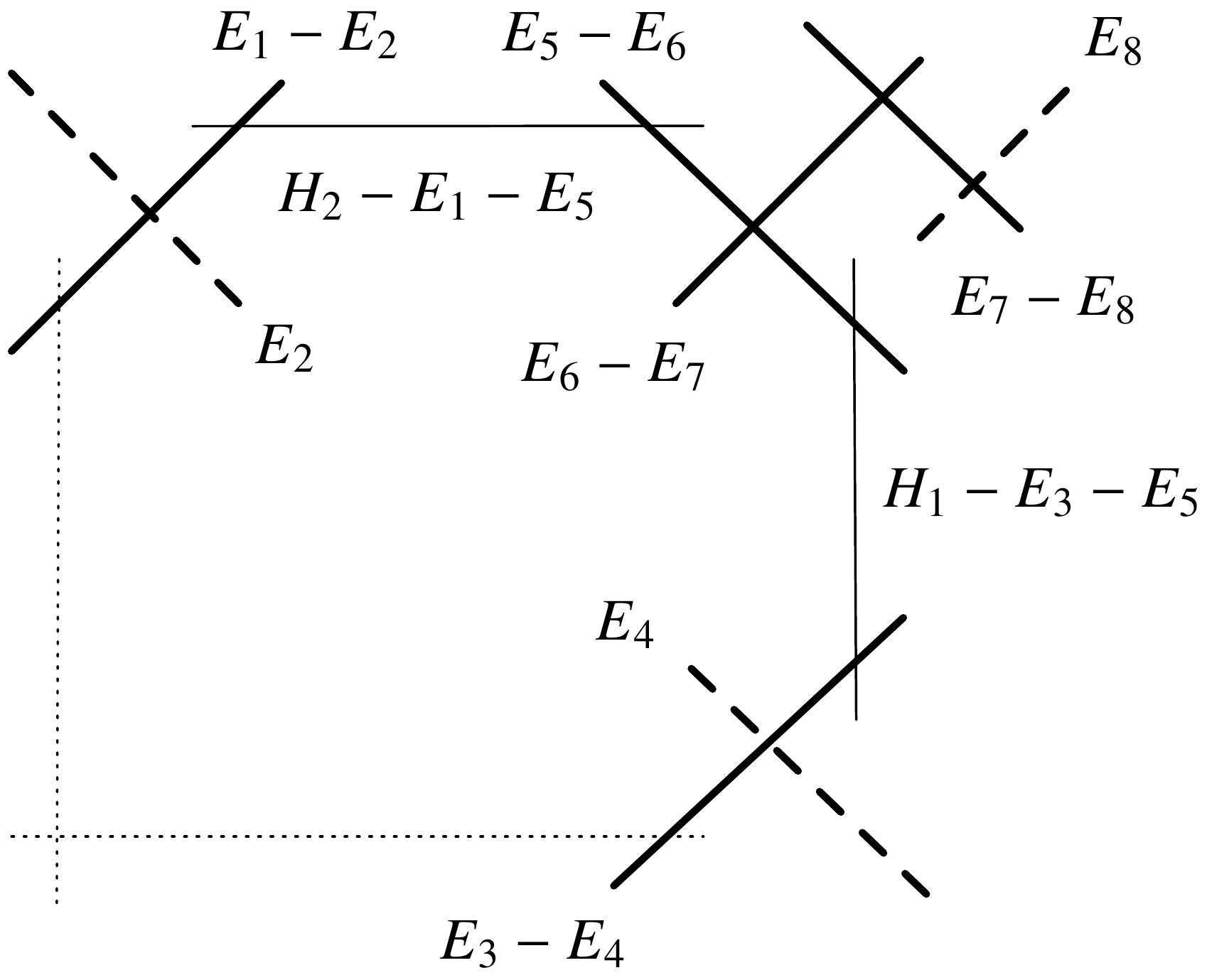}
\end{center}
\caption{Surface of P$_{\rm IV}$ obtained by eight blowing-ups. Solid lines are inaccessible divisors. Thick lines are exceptional divisors arising in the blowing-ups.} 
\label{fig:blowup_p4_final}
\end{figure}

As we have seen, all the singularities of P$_{\rm IV}$ \eqref{eqn:p4_pq} were resolved by eight
blowing-ups. The process of blowing-ups is read off graphically from Figure
\ref{fig:blowup_p4_final}. We denote by $X$ the surface obtained from
$\mathbb{P}^1\times\mathbb{P}^1$ by eight blowing-ups in this way.
This surface $X$ still contains inaccessible points of P$_{\rm IV}$ on a divisor $D$ with seven components 
\begin{equation}\label{eqn:p4_removed_divisors}
\begin{array}{c}\medskip
\delta_1= E_1-E_2,\quad  \delta_2=H_2-E_1-E_5,\quad  \delta_3=E_5-E_6,\quad  \delta_4=H_1-E_3-E_5,\\
\delta_5=E_3-E_4,\quad \delta_6=E_6-E_7,\quad  \delta_0=E_7-E_8. 
\end{array}
\end{equation}
The space of initial values of Okamoto is obtained as $X\setminus D$ by removing the inaccessible
divisor $D$ (sometimes called the {\em vertical leaves}) from $X$.  Then P$_{\rm IV}$ becomes a
regular differential equation globally defined on this surface $X\setminus D$.  In fact, P$_{\rm
IV}$ is represented as a polynomial Hamiltonian system on each chart of $X\setminus D$ constructed
above.  Conversely, it is also known that P$_{\rm IV}$ is uniquely determined by this property from
the surface $X\setminus D$ itself, as was shown by Takano et al \cite{Takano2,Takano1}. We also
remark that the B\"acklund transformations of P$_{\rm IV}$ as well as dP$_{\rm II}$ are regularized
on the space of initial values $X\setminus D$. The same story applies to the whole class of
Painlev\'e differential equations \cite{Noumi-Takano-Yamada:BT}. One of the purposes of this paper
is to pursue this philosophy in the theory of discrete Painlev\'e equations.

%
\section{Root Systems, Weyl Groups and Picard Lattice}\label{section:root}
In this Section, we give a brief introduction to powerful tools that are necessary for  systematically developing
the geometric theory of Painlev\'e equations. 
\subsection{Root systems}\label{subsection:root}
The fundamental reference to the contents of this subsection is \cite{Kac:book}.  Let
$A=(a_{ij})_{i,j\in I}$ be a generalized Cartan matrix, namely
\begin{equation}
a_{ii}=2,\quad a_{ij}\in\mathbb{Z}_{\leq 0},\quad a_{ij}=0\Longleftrightarrow a_{ji}=0\quad (i\neq j).
\end{equation}
We define the {\em Weyl group} $W(A)$ associated with the generalized Cartan matrix $A$ by the generators $s_i$
($i\in I$) and the fundamental relations
\begin{equation}\label{eqn:fundamental_relation_Weyl_group}
 \begin{array}{cll}
s_i^2=1 &\mbox{for all $i\in I$} & \\
s_is_j = s_js_i &\mbox{when}  & (a_{ij},a_{ji})=(0,0),\\
s_i s_j s_i = s_j s_i s_j&\mbox{when} 
& (a_{ij},a_{ji})=(-1,-1),\\
s_i s_j s_i s_j = s_j s_i s_j s_i&\mbox{when} 
& (a_{ij},a_{ji})=(-1,-2),\\
s_i s_j s_i s_j s_i s_j = s_j s_i s_j s_i s_j s_i
&\mbox{when}
& (a_{ij},a_{ji})=(-1,-3).
\end{array}
\end{equation}
As we will see below, this group $W(A)$ can be realized as a group generated by reflections acting on a
vector space. 

In the following, we confine ourselves to the {\em symmetrizable} cases where the matrix elements $a_{ij}$
are realized by the inner product (non-degenerated symmetric bilinear form)
$\langle\cdot,\cdot\rangle: V\times V\rightarrow \mathbb{Q}$ on a $\mathbb{Q}$-vector space $V$ as
\begin{equation}
 a_{ij}=\langle \alpha_i^{\vee},\alpha_j\rangle,\quad 
\alpha_i^{\vee}=\frac{2\alpha_i}{\langle \alpha_i,\alpha_i\rangle},
\end{equation}
in terms of a set of $\mathbb{Q}$-linearly independent vectors $\alpha_i$ ($i\in I$) such that
$\langle \alpha_i,\alpha_i\rangle\neq 0$ ; $\alpha_i$ and $\alpha_i^\vee$ are called the {\em simple
roots} and the {\em simple coroots}, respectively. The free $\mathbb{Z}$-submodules 
\begin{equation}
 Q\left(A\right)=\bigoplus_{i\in I}\mathbb{Z}\alpha_i,\quad
 Q^\vee\left(A\right)=\bigoplus_{i\in I}\mathbb{Z}\alpha_i^\vee,
\end{equation}
are called the {\em root lattice} and the {\em coroot lattice}, respectively.

For each element $\alpha\in V$ with $\langle \alpha,\alpha\rangle\neq 0$ we define the {\em
reflection} $r_\alpha: V\to V$ by
 \begin{equation}\label{eqn:roots_reflection}
 r_\alpha (\lambda) 
=   \lambda - 2\frac{\langle \alpha, \lambda\rangle}{\langle \alpha,\alpha\rangle} \;\alpha
=  \lambda - \langle \alpha^\vee, \lambda\rangle \;\alpha
,\quad \lambda\in V. 
\end{equation}
One can verify that $r_\alpha$ have the following properties:
\begin{enumerate}
 \item $(r_\alpha)^2=\text{id}_V$; $r_\alpha(\alpha)=-\alpha$ and $r_\alpha(\lambda)=\lambda$ if
       $\langle \alpha,\lambda\rangle=0$.
 \item $r_\alpha$ and $r_\beta$ commute with each other when
$\langle \alpha, \beta\rangle=0$.
 \item (braid relations) 
\begin{equation}
\begin{array}{ccl}
r_\alpha r_\beta r_\alpha = r_\beta r_\alpha r_\beta&\mbox{when} 
& (\langle \alpha^\vee, \beta\rangle,\langle \beta^\vee, \alpha\rangle)=(-1,-1),\\
r_\alpha r_\beta r_\alpha r_\beta = r_\beta r_\alpha r_\beta r_\alpha&\mbox{when} 
& (\langle \alpha^\vee, \beta\rangle,\langle \beta^\vee, \alpha\rangle)=(-1,-2),\\
r_\alpha r_\beta r_\alpha r_\beta r_\alpha r_\beta = r_\beta r_\alpha r_\beta r_\alpha r_\beta r_\alpha
&\mbox{when}
& (\langle \alpha^\vee, \beta\rangle,\langle \beta^\vee, \alpha\rangle)=(-1,-3).
\end{array}
\end{equation}
\item (isometry) $\langle r_\alpha(\lambda), r_\alpha(\mu)\rangle=\langle \lambda,\mu\rangle$
for any $\lambda,\mu\in V$.
 \item For any $\mathbb{Q}$-linear isometry $f:V\to V$, $f r_\alpha=r_{f(\alpha)} f$.
\end{enumerate}
The reflections $r_i=r_{\alpha_i}\in {\rm GL}(V)$ by the simple roots are called the {\em simple
reflections}.  From the properties (1), (2) and (3), we see that the correspondence $s_i\rightarrow
r_i$ ($i\in I$) defines a linear representation of $W(A)$ on $V$.  We remark that each simple
reflection stabilizes $Q(A)$ and $Q^\vee(A)$ and so does $W(A)$.

The generalized Cartan matrices are classified into three types according to the signatures: (i)
finite type $(+,+,\cdots)$ (ii) affine type $(0,+,+,\cdots)$ (iii) indefinite type (otherwise).  We
are particularly interested in the affine root systems of type A, D, E for which the inner product
can be renormalized so that $\langle\alpha_i,\alpha_i\rangle=2$ and hence $\alpha_i^\vee = \alpha_i$
for all $i\in I$: the corresponding generalized Cartan matrices
$A=(a_{i,j})_{i,j=0,\ldots,l}=(\langle \alpha_i,\alpha_j\rangle)_{i,j=0,\ldots,l}$ are given by
\begin{small}
\renewcommand{\arraycolsep}{1.5pt}
\allowdisplaybreaks
\begin{align}
A_1^{(1)}:\quad& \left[\begin{array}{cc}2 & -2\\-2 & 2\end{array}\right],
&&\hskip-20pt
\begin{array}{c}
\setlength{\unitlength}{1mm}
\begin{picture}(17,4)(0,0)
\put(0.7,2){$\stackrel{\alpha_0}{\circ}$}
\put(13,2){$\stackrel{\alpha_1}{\circ}$}
\put(2,2){$<$}
\put(12,2){$>$}
\put(2.7,3){\line(1,0){10.8}}
\put(2.7,2.5){\line(1,0){10.9}}
\end{picture}\\
{\displaystyle \delta=\alpha_0 + \alpha_1}
\end{array}\\
 A_l^{(1)}:\quad&
\left[\begin{array}{ccccccc}
2 &-1 & & & & &-1 \\
-1&2 &-1 & & & & \\
  &-1 &2 &-1 & & & \\
  & &\ddots &\ddots &\ddots & & \\
  & & &-1 &2 &-1 & \\
  & & & &-1 &2 &-1 \\
-1& & & & &-1 &2 
\end{array}\right],
&&\hskip-20pt
\begin{array}{c}
\setlength{\unitlength}{1mm}
\begin{picture}(30,12)(0,0)
\put(1,2){$\stackrel{\displaystyle\circ}{\scriptstyle\alpha_1}$}
\put(7,2){$\stackrel{\displaystyle\circ}{\scriptstyle\alpha_2}$}
 \put(2.7,4.7){\line(1,0){4.8}}
\dashline[0]{1}(9,4.7)(19.7,4.7)
\put(19,2){$\stackrel{\displaystyle\circ}{\scriptstyle\alpha_{l-1}}$}
\put(21.5,4.7){\line(1,0){4.8}}
\put(26,2){$\stackrel{\displaystyle\circ}{\scriptstyle\alpha_{l}}$}
\put(14,10){$\stackrel{\scriptstyle\alpha_{0}}{\displaystyle\circ}$}
\drawline(2.7,4.9)(14.6,10.6)
\drawline(15.6,10.6)(26.4,5)
\put(33,5){($l\geq 2$)}
\end{picture}\\
\delta = \alpha_0+\alpha_1+\cdots+\alpha_{l}
\end{array} \label{eqn:Cartan_A}\\
 D_l^{(1)}:\quad & 
\left[\begin{array}{ccccccc}
2 &  &-1 & & & & \\
&2 &-1 & & & & \\
-1 &-1 &2 &-1 & & & \\
  & &\ddots &\ddots &\ddots & & \\
  & & &-1 &2 &-1 &-1 \\
  & & & &-1 &2 & \\
 & & & &-1 & &2 
\end{array}\right],
&&\hskip-30pt
\begin{array}{c}
\setlength{\unitlength}{1mm}
\begin{picture}(45,20)(0,0)
\put(1,1){$\circ$}
\put(0,4){$\scriptstyle \alpha_1$}
\put(1,13){$\circ$}
\put(0,16){$\scriptstyle \alpha_0$}
\put(8,7){$\circ$}
\put(8,11){$\scriptstyle \alpha_2$}
\dashline[0]{1}(15.7,7.9)(29,7.9)
\drawline(2.4,2.3)(8.4,7.4)
\drawline(2.4,13.7)(8.4,8.4)
\drawline(9.4,7.9)(14.1,7.9)
\put(14,7){$\circ$}
\put(14,11){$\scriptstyle \alpha_3$}
\put(29,7){$\circ$}
\put(27,11){$\scriptstyle \alpha_{l-3}$}
\drawline(30.2,7.9)(35.3,7.9)
\put(35.2,7){$\circ$}
\put(33,11){$\scriptstyle \alpha_{l-2}$}
\put(42,1){$\circ$}
\put(43,4){$\scriptstyle \alpha_{l-1}$}
\put(42,13){$\circ$}
\put(43,16){$\scriptstyle \alpha_l$}
\drawline(36.3,7.4)(42.4,2.3)
\drawline(36.3,8.4)(42.4,13.7)
\put(50,8){($l\geq 4$)}
\end{picture}\\
\delta= \alpha_0+\alpha_1+2\alpha_2+\cdots +2\alpha_{l-2}+\alpha_{l-1}+\alpha_l
\end{array}\label{eqn:Cartan_D}
\\
 E_6^{(1)}: \quad& 
\left[\begin{array}{ccccccc}
2 & & & & & &-1 \\
  &2 &-1 & & & & \\
  &-1 &2 &-1 & & & \\
  & &-1 &2 &-1 & &-1 \\
  & & &-1 &2 &-1 & \\
  & & & &-1 &2 & \\
-1  & & &-1 & & &2 
\end{array}\right],
&&\hskip-30pt
\begin{array}{c}
\setlength{\unitlength}{1mm}
\begin{picture}(27,18)(0,0)
\put(1,3){$\circ$}
\put(1,1){$\scriptstyle \alpha_1$}
\drawline(2.5,4)(7.2,4)
\put(7,3){$\circ$}
\put(7,1){$\scriptstyle \alpha_2$}
\drawline(8.5,4)(13.2,4)
\put(13,3){$\circ$}
\put(13,1){$\scriptstyle \alpha_3$}
\drawline(14.5,4)(19.2,4)
\put(19,3){$\circ$}
\put(19,1){$\scriptstyle \alpha_4$}
\drawline(20.5,4)(25.2,4)
\put(25,3){$\circ$}
\put(25,1){$\scriptstyle \alpha_5$}
\put(13,9){$\circ$}
\put(9,9.5){$\scriptstyle \alpha_6$}
\drawline(13.8,10.5)(13.8,15.4)
\put(13,15){$\circ$}
\put(9,15.5){$\scriptstyle \alpha_0$}
\drawline(13.8,4.5)(13.8,9.4)
\end{picture}\\
\delta=\alpha_0+\alpha_1+2\alpha_2+3\alpha_3+2\alpha_4+\alpha_5+2\alpha_6
\end{array}\label{eqn:Cartan_E6}\\
 E_7^{(1)}:\quad&
\left[\begin{array}{cccccccc}
2 &  &  & & -1& & & \\
  &2 &-1&  & & & & \\
  &-1&2 &-1& & & & \\
  &  &-1&2 &-1 & & \\
  &  &  &-1&2 &-1 & \\
-1 & & & &-1 &2 &-1 & \\
  & & & & &-1 &2 &-1 \\
  & & & & & &-1 &2 
\end{array}\right],
&&\hskip-15pt
\begin{array}{c}
\setlength{\unitlength}{1mm}
\begin{picture}(40,12)(0,0)
\put(1,3){$\circ$}
\put(1,1){$\scriptstyle \alpha_1$}
\drawline(2.5,4)(7.2,4)
\put(7,3){$\circ$}
\put(7,1){$\scriptstyle \alpha_2$}
\drawline(8.5,4)(13.2,4)
\put(13,3){$\circ$}
\put(13,1){$\scriptstyle \alpha_3$}
\drawline(14.5,4)(19.2,4)
\put(19,3){$\circ$}
\put(19,1){$\scriptstyle \alpha_4$}
\drawline(20.5,4)(25.2,4)
\put(25,3){$\circ$}
\put(25,1){$\scriptstyle \alpha_5$}
\drawline(26.5,4)(31.2,4)
\put(31,3){$\circ$}
\put(31,1){$\scriptstyle \alpha_6$}
\drawline(32.5,4)(37.2,4)
\put(37,3){$\circ$}
\put(37,1){$\scriptstyle \alpha_7$}
\put(19,9){$\circ$}
\put(15,9.5){$\scriptstyle \alpha_0$}
\drawline(19.8,4.5)(19.8,9.4)
\end{picture}\\
\delta = 2\alpha_0+\alpha_1+2\alpha_2+3\alpha_3+4\alpha_4+3\alpha_5+2\alpha_6+\alpha_7
\end{array}\label{eqn:Cartan_E7}\\
 E_8^{(1)}:\quad&
\left[\begin{array}{ccccccccc}
2 &  & &-1 & & & & &  \\
  &2 &-1 & & & & & & \\
  &-1 &2 &-1 & & & & &\\
-1  & &-1 &2 &-1 & & & &\\
  & & &-1 &2 &-1 & & & \\
  & & & &-1 &2 &-1 & & \\
  & & & &   &-1&2 &-1 & \\
  & & & &   &  &-1 &2 &-1\\
  & & & &   &  &  &-1 &2
\end{array}\right],
&&
\begin{array}{c}
\setlength{\unitlength}{1mm}
\begin{picture}(45,12)(0,0)
\put(1,3){$\circ$}
\put(1,1){$\scriptstyle \alpha_1$}
\drawline(2.5,4)(7.2,4)
\put(7,3){$\circ$}
\put(7,1){$\scriptstyle \alpha_2$}
\drawline(8.5,4)(13.2,4)
\put(13,3){$\circ$}
\put(13,1){$\scriptstyle \alpha_3$}
\drawline(14.5,4)(19.2,4)
\put(19,3){$\circ$}
\put(19,1){$\scriptstyle \alpha_4$}
\drawline(20.5,4)(25.2,4)
\put(25,3){$\circ$}
\put(25,1){$\scriptstyle \alpha_5$}
\drawline(26.5,4)(31.2,4)
\put(31,3){$\circ$}
\put(31,1){$\scriptstyle \alpha_6$}
\drawline(32.5,4)(37.2,4)
\put(37,3){$\circ$}
\put(37,1){$\scriptstyle \alpha_7$}
\drawline(38.5,4)(43.2,4)
\put(43,3){$\circ$}
\put(43,1){$\scriptstyle \alpha_8$}
\put(13,9){$\circ$}
\put(9,9.5){$\scriptstyle \alpha_0$}
\drawline(13.9,4.5)(13.9,9.4)
\end{picture}\\
\begin{array}{c}
\delta= 3\alpha_0 + 2\alpha_1 + 4\alpha_2 + 6\alpha_3 \\
\hspace*{30pt} + 5\alpha_4 + 4\alpha_5 + 3\alpha_6 + 2\alpha_7 + \alpha_8. 
\end{array}
\end{array}\label{eqn:Cartan_E8}
\end{align}
\end{small}
In each case, the vector $\delta$ defined above carries a characteristic property $\langle
\alpha_i,\delta\rangle=0$ ($i=0,\cdots,l$) and called the {\em null root}. Writing $\delta=
\sum_{j=0}^ln_j\alpha_j$, we have $\langle \alpha_i,\delta\rangle = \sum_{j=0}^l \langle \alpha_i,
\alpha_j\rangle n_j=0$. This means that ${}^t[n_0,\cdots,n_l]\in\mathbb{Z}^{l+1}$ is the eigenvector
of the zero eigenvalue of $A$. 

The diagram illustrated in each case is called the {\em Dynkin diagram}, from which one can
recover the off-diagonal entries of the generalized Cartan matrix by the following rule:
\begin{equation}
 \begin{tabular}{|c|c|}
\hline
 ($a_{ij}$, $a_{ji}$) &\hskip3pt $\alpha_i$\hskip25pt $\alpha_j$\\
\hline
($0,0$) & 
\setlength{\unitlength}{1mm}
\begin{picture}(13,3)(0,0) 
\put(0,0){$\circ$}
\put(12,0){$\circ$}
\end{picture}\\
\hline
($-1,-1$) & 
\setlength{\unitlength}{1mm}
\begin{picture}(13,3)(0,0) 
\put(0,0){$\circ$}
\drawline(1,1)(12,1)
\put(12,0){$\circ$}
\end{picture}\\
\hline
($-2,-2$) & 
\setlength{\unitlength}{1mm}
\begin{picture}(13,3)(0,0) 
\put(0,0){$\circ$}
\drawline(1,1.0)(12,1.0)
\drawline(1,0.5)(12,0.5)
\put(12,0){$\circ$}
\put(1,0){$<$}
\put(10.1,0){$>$}
\end{picture}\\
\hline
 \end{tabular}
\end{equation}

We call a permutation $\sigma$ of index set $I$ such that $ a_{\sigma(i)\sigma(j)}=a_{ij}$ ($i,j\in
I$) a {\em Dynkin diagram automorphism}. By abuse of terminology, the phrase of {\em Dynkin diagram
automorphism} is used for a wider class of objects that are induced from such a permutation
$\sigma$. For a group $G$ of Dynkin diagram automorphisms, one can define an {\em extended Weyl
group} $\widetilde{W}(A)$ as the group generated by the Weyl group $W(A)$ together with $\pi_\sigma$
($\sigma\in G$) such that $\pi_\sigma s_i = s_{\sigma(i)}\pi_\sigma$.
%
\subsection{Kac translation}\label{subsec:Kac}
An important feature of these affine root systems is that the associated Weyl groups are of infinite
order and include the {\em translations}. Denoting $V_0=\{\lambda\in V\mid
\langle\delta,\lambda\rangle=0\}$, for each $\alpha\in V_0$ we define the {\em Kac translation}
$T_\alpha: V\to V$ by
\begin{equation}
 T_\alpha(\lambda)= \lambda + \langle \delta,\lambda\rangle \alpha - \left(\frac{1}{2}\langle \alpha,\alpha\rangle\langle\delta,\lambda\rangle + \langle \alpha,\lambda \rangle\right)\delta,\quad \lambda\in V.
\label{eqn:Kac_translation}
\end{equation}
One can verify that the linear transformation $T_\alpha$ has the following properties:
\begin{enumerate}
 \item For any $\alpha,\beta\in V_0$, $T_\alpha T_\beta=T_{\alpha+\beta}$
       and hence $T_\alpha T_\beta = T_\beta T_\alpha $.
 \item For any $w \in W(A)$, $wT_\alpha=T_{w(\alpha)}w$.
 \item If $\alpha\in V_0$ and $\langle\alpha,\alpha \rangle\neq 0$, 
then $T_\alpha=r_{\delta-\alpha^\vee}r_{\alpha^\vee}$.
 \item For $\beta\in V_0$, $T_\alpha(\beta)= \beta - \langle \alpha,\beta\rangle \delta$.
 \item (isometry) For any $\lambda,\mu\in V$, $\langle
       T_\alpha(\lambda),T_\alpha(\mu)\rangle=\langle\lambda,\mu\rangle$.
\end{enumerate}
As was shown in \eqref{eqn:dP_A2}, if we have a suitable birational representation of an affine Weyl
group, discrete Painlev\'e equations arise from its translations. Here we describe how to
construct such translations as compositions of simple reflections and Dynkin diagram automorphisms.
\begin{ex}\label{example:A2}\rm
Let us consider the root system of type $A_2^{(1)}$ which is generated by three simple roots $\alpha_i$
($i=0,1,2$). The inner product is characterized by the generalized Cartan matrix
\eqref{eqn:Cartan_A} with $l=2$. The corresponding Weyl group $W(A_2^{(1)})$ is generated by three
reflections $r_{i}=r_{\alpha_i}$ ($i=0,1,2$). Their action on the simple roots $\alpha_i$
($i=0,1,2$) is computed by using \eqref{eqn:roots_reflection} as
\begin{equation}\label{eqn:A2_reflection}
\begin{tabular}{|c|c|c|c|}
\hline
&$\alpha_0$ &$\alpha_1$ &$\alpha_2$ \\
\hline
$r_0$& $-\alpha_0$ &$\alpha_0+\alpha_1$ & $\alpha_0+\alpha_2$\\
\hline
$r_1$& $\alpha_0+\alpha_1$ & $-\alpha_1$ & $\alpha_1+\alpha_2$\\
\hline
$r_2$& $\alpha_0+\alpha_2$ & $\alpha_1+\alpha_2$ & $-\alpha_2$\\
\hline
\end{tabular}
\end{equation}
Though the above formulae are derived as linear transformations of the vectors $\alpha_i \in V$,
they can be interpreted also as a linear substitutions of variables in the field of rational functions
$\mathbb{C}(\alpha_0,\alpha_1,\alpha_2)$.
The action of the translation $T_{\alpha_1}$ on the simple roots is
obtained from \eqref{eqn:Kac_translation}
\begin{equation}\label{eqn:T1}
T_{\alpha_1}(\alpha_0) = \alpha_0+\delta,\quad
T_{\alpha_1}(\alpha_1) = \alpha_1-2\delta,\quad
T_{\alpha_1}(\alpha_2) = \alpha_2+\delta.
\end{equation}
The translation by root vectors can be expressed as a product of simple reflections. An explicit
expression for $T_{\alpha_1}$ is obtained in the following manner. We write \eqref{eqn:T1} as
\begin{equation}
T_{\alpha_1} [\alpha_0,\alpha_1,\alpha_2]
=[\alpha_0 + \delta, \alpha_1-2\delta,\alpha_2+\delta]
= [2\alpha_0 + \alpha_1 + \alpha_2, -2\alpha_0-\alpha_1-2\alpha_2,\alpha_0+\alpha_1+2\alpha_2].
\label{eqn:T1_on_roots}
\end{equation}
In general, suppose that we have a substitution $f$ such that
\begin{equation}
f[\alpha_0,\alpha_1,\alpha_2]=[f(\alpha_0),f(\alpha_1),f(\alpha_2)]=[\beta_0,\beta_1,\beta_2],
\end{equation}
where $\beta_i$ ($i=0,1,2$) are linear combinations of $\alpha_i$ ($i=0,1,2$). Then, if $\beta_1$
has negative coefficient for instance, we consider $f r_1$ to obtain
\begin{equation}
\begin{split}
&f r_1[\alpha_0,\alpha_1,\alpha_2]
=f[r_1(\alpha_0),r_1(\alpha_1),r_1(\alpha_2)]\\
=&f[\alpha_0+\alpha_1,-\alpha_1,\alpha_2+\alpha_1]
=[\beta_0+\beta_1,-\beta_1,\beta_2+\beta_1]. 
\end{split}
\end{equation}
(We adopt the convention of symbolical compositions in the sense of Remark
\ref{rem:convention_action} unless otherwise stated.) We continue this procedure until all the
coefficients become positive.  If negative signs appear at two or more positions, one can apply the
procedure at any one of them. The results may give different but equivalent {\em reduced} (shortest)
decompositions up to the fundamental relations.  In case of \eqref{eqn:T1_on_roots}, the procedure goes as follows
\begin{align*}
& T_{\alpha_1}r_1 [\alpha_0,\alpha_1,\alpha_2] =T_{\alpha_1}[\alpha_0+\alpha_1,-\alpha_1,\alpha_2+\alpha_1]
=[-\alpha_2,2\alpha_0+\alpha_1+2\alpha_2,-\alpha_0],\\
& T_{\alpha_1}r_1r_0[\alpha_0,\alpha_1,\alpha_2]=T_{\alpha_1}r_1[-\alpha_0,\alpha_1+\alpha_0,\alpha_2+\alpha_0]
=[\alpha_2,2\alpha_0+\alpha_1+\alpha_2,-\alpha_0-\alpha_2],\\
&T_{\alpha_1}r_1r_0r_2[\alpha_0,\alpha_1,\alpha_2]
=T_{\alpha_1}r_1r_0[\alpha_0+\alpha_2,\alpha_1+\alpha_2,-\alpha_2]
=[-\alpha_0,\alpha_0+\alpha_1,\alpha_0+\alpha_2],\\
&T_{\alpha_1}r_1r_0r_2r_0[\alpha_0,\alpha_1,\alpha_2]
=T_{\alpha_1}r_1r_0r_2[-\alpha_0,\alpha_1+\alpha_0,\alpha_2+\alpha_0]
=[\alpha_0,\alpha_1,\alpha_2].
\end{align*}
Therefore $T_{\alpha_1}r_1r_0r_2r_0={\rm id}$ i.e. $T_{\alpha_1}=r_0r_2r_0r_1$. It is known that
this procedure terminates after a finite number of steps (see, for example, \cite{Kac:book} Lemma
3.11).The same procedure also applies to more general transformation such as
$S[\alpha_0,\alpha_1,\alpha_2]=[\alpha_0+m_0\delta,\alpha_1+m_1\delta,\alpha_2+m_2\delta]$ with
$m_0+m_1+m_2=0$. For example, in the case of $(m_0,m_1,m_2)=(1,-1,0)$, namely,
$S[\alpha_0,\alpha_1,\alpha_2]=[\alpha_0+\delta,\alpha_1-\delta,\alpha_2]$, we have:
\begin{align*}
&S[\alpha_0,\alpha_1,\alpha_2]=[2\alpha_0+\alpha_1+\alpha_2,-\alpha_0-\alpha_2,\alpha_2],\\
&Sr_1[\alpha_0,\alpha_1,\alpha_2]=[\alpha_0+\alpha_1,\alpha_0+\alpha_2,-\alpha_0],\\
&Sr_1r_2[\alpha_0,\alpha_1,\alpha_2]=[\alpha_1,\alpha_2,\alpha_0].
\end{align*}
Hence, $Sr_1r_2=\pi$, where $\pi(\alpha_i)=\alpha_{i+1}$ ($i\in\mathbb{Z}/3\mathbb{Z}$) is an
automorphism of the Dynkin diagram of type $A_2^{(1)}$. This example shows that the above procedure
terminates possibly with a permutation of the simple roots corresponding to a Dynkin diagram
automorphism. This phenomena occurs if we consider finer translations than those by root vectors.
\end{ex}
\begin{ex}\label{example:D5}\rm
\newcommand{\DynkinD}[6]{
\renewcommand{\arraycolsep}{1.5pt}
\begin{array}{ccccccc}
& & #5& & #6& & \\
& & | & & | & & \\
#1&\text{\textemdash}&#2&\text{\textemdash}&#3&\text{\textemdash}&#4
\end{array}
}
In the procedure explained in Example \ref{example:A2} one can use arbitrary positive numbers in
place of the symbols $\alpha_i$. For this purpose, the convention of numerical composition in
Remark \ref{rem:convention_action} can be applied more effectively than that of symbolic
composition, which reduces the complexity of computations drastically. We show such computations
in case of the root system of type $D_5^{(1)}$ whose Cartan matrix is \eqref{eqn:Cartan_D} with
$l=5$. The Dynkin diagram and the null root are given by
\begin{equation}
\DynkinD{\alpha_1}{\alpha_2}{\alpha_3}{\alpha_4}{\alpha_0}{\alpha_5}  \qquad
\delta=\alpha_0 +\alpha_1 + 2\alpha_2+2\alpha_3+\alpha_4+\alpha_5,
\end{equation}
respectively. The corresponding Weyl group $W(D_5^{(1)})$ is generated by six reflections
$r_{i}=r_{\alpha_i}$ ($i=0,1,\ldots,5$).  Their actions on the simple roots $\alpha_i$ are computed
by using \eqref{eqn:roots_reflection} as
\begin{equation}\label{eqn:D5_reflection}
r_i(\alpha_j) = \left\{
\begin{array}{cl}
-\alpha_i& j=i,\\
\alpha_i+\alpha_j & \stackrel{\alpha_i}{\circ}\text{\textemdash}\stackrel{\alpha_j}{\circ},\\
\alpha_j & \mbox{otherwise}.
\end{array}\right.
\end{equation}
In the convention of numerical composition on the parameter space, the map $F_2=F_{r_2}: \mathbb{C}^6
\to \mathbb{C}^6$, for instance, is expressed graphically as
\begin{equation}
\DynkinD{\alpha_1}{\alpha_2}{\alpha_3}{\alpha_4}{\alpha_0}{\alpha_5}
\stackrel{F_2}{\longrightarrow}
\DynkinD{\alpha_1\!+\!\alpha_2}{-\alpha_2}{\alpha_3\!+\!\alpha_2}{\alpha_4}{\alpha_0\!+\!\alpha_2}{\alpha_5} 
\end{equation}
We consider a translation $S[\alpha_0,\alpha_1,\alpha_2,\alpha_3,\alpha_4,\alpha_5]=
[\alpha_0-\delta,\alpha_1+\delta,\alpha_2,\alpha_3,\alpha_4,\alpha_5]$, namely
$F_s(\alpha_0,\alpha_1,\alpha_2,\alpha_3,\alpha_4,\alpha_5)=
(\alpha_0-\delta,\alpha_1+\delta,\alpha_2,\alpha_3,\alpha_4,\alpha_5)$, and 
try to represent it
by simple reflections.
Taking an initial value
$(\alpha_0,\alpha_1,\alpha_2,\alpha_3,\alpha_4,\alpha_5)=(10,1,2,3,4,5)$ with $\delta=30$ for instance, we start from
\begin{equation}
\DynkinD{1}{2}{3}{4}{10}{5}\stackrel{F_S}{\longrightarrow}
\DynkinD{31}{2}{3}{4}{-20}{5}
\end{equation}
Then the procedure goes as follows:
\begin{equation}
\begin{split}
& \DynkinD{31}{2}{3}{4}{-20}{5}\stackrel{F_0}{\longrightarrow}
\DynkinD{31}{-18}{3}{4}{20}{5}\stackrel{F_2}{\longrightarrow}
\DynkinD{13}{18}{-15}{4}{2}{5}\stackrel{F_3}{\longrightarrow}\\
& \DynkinD{13}{3}{15}{-11}{2}{-10}\stackrel{F_4}{\longrightarrow}
\DynkinD{13}{3}{4}{11}{2}{-10}\stackrel{F_5}{\longrightarrow}
\DynkinD{13}{3}{-6}{11}{2}{10}\stackrel{F_3}{\longrightarrow}\\
&\DynkinD{13}{-3}{6}{5}{2}{4}\stackrel{F_2}{\longrightarrow}
\DynkinD{10}{3}{3}{5}{-1}{4}\stackrel{F_0}{\longrightarrow}
\DynkinD{10}{2}{3}{5}{1}{4}
\end{split}
\end{equation}
As a result, we obtain $F_0F_2F_3F_5F_4F_3F_2F_0F_S=F_{\pi}$, i.e. $S=\pi r_0r_2r_3r_5r_4r_3r_2r_0$,
where $\pi$ is a Dynkin diagram automorphism
$\{\alpha_0 \leftrightarrow \alpha_1, \ \alpha_4 \leftrightarrow \alpha_5\}$.
\end{ex}
The procedure demonstrated in Example \ref{example:A2} and Example \ref{example:D5} provides a
practical method for expressing a given translation $\alpha_i\mapsto \alpha_i+m_i\delta$,
$m_i\in\mathbb{Z}$, such that $\sum_{i}n_im_i=0$ ($\delta=\sum_i n_i\alpha_i$) in terms of the
product of simple reflections and Dynkin diagram automorphisms.
\begin{rem}\rm
In the case of the affine Weyl groups of type A, D, E, it is known that the translation $T_\alpha$
by any element $\alpha\in Q(A)$ can be expressed as the product of simple reflections. Moreover, the
affine Weyl group is decomposed into semi-direct product of the corresponding finite Weyl group
and the group of translations \cite{Kac:book}.
\end{rem}
%
\subsection{Picard lattice}\label{subsection:Picard}
Here we introduce another fundamental tool for the geometry of Painlev\'e equations, the {\em Picard
lattice} associated with the eight point blowing-up of $\mathbb{P}^1\times\mathbb{P}^1$, which
corresponds to the root system of type $E_8^{(1)}$. This tool enables us to manipulate rational maps
by the language of linear algebra.  We denote by
\begin{equation}
  \Lambda = \Z H_1 \oplus \Z H_2 \oplus \Z E_1 \oplus\cdots\oplus \Z E_8,
\end{equation}
the free $\Z$-module of rank 10 generated by the symbols $H_1$, $H_2$ and $E_1,\ldots, E_8$.  We
introduce the symmetric bilinear form $\Lambda\times \Lambda\to \Z$: $(\lambda,\mu)\mapsto
\lambda\cdot \mu$ such that
\begin{equation}\label{eqn:E_8_intersection_form}
\begin{split}
&H_1\cdot H_2=1,\quad  H_1\cdot H_1 = H_2\cdot H_2=0,\\
& E_i\cdot E_j=-\delta_{ij}, \quad H_k\cdot E_j=0\quad (i,j=1,\ldots,8;\ k=1,2) .
\end{split}
\end{equation}
For any surface $X$ obtained from $\mathbb{P}^1\times\mathbb{P}^1$ by eight blowing-ups, $\Lambda$
is identified with the {\em Picard lattice} $\text{Pic}~X$, in which $H_1$ and $H_2$ represent the divisor
classes corresponding to the $q=\text{const.}$ and $p=\text{const.}$, respectively, and
$E_1,\ldots,E_8$ are the exceptional divisors. We consider the 10-dimensional $\mathbb{Q}$-vector space
\begin{equation}
V=\Lambda\otimes \mathbb{Q}  = \mathbb{Q}H_1\oplus\mathbb{Q}H_2\oplus\mathbb{Q}E_1\oplus\cdots \oplus\mathbb{Q}E_8,
\end{equation}
and take the inner product $\langle \cdot,\cdot\rangle:V\times V\to
\mathbb{Q}$ such that $\langle\lambda,\mu\rangle=-\lambda\cdot\mu$ for each
$\lambda,\mu\in\Lambda$. We define the simple roots $\alpha_0,\alpha_1,\ldots,\alpha_8$ by
\begin{equation}\label{eqn:simple_roots_E8}
\begin{array}{c}\smallskip
\alpha_0 = E_1-E_2,\  \alpha_1= H_1-H_2,\  \alpha_2=H_2-E_1-E_2,\  \alpha_3=E_2-E_3, \ \alpha_4=E_3-E_4,\\ 
\alpha_5 = E_4-E_5,\  \alpha_6 = E_5 - E_6,\  \alpha_7 = E_6-E_7,\  \alpha_8=E_7-E_8.
\end{array}
\end{equation}
Note that $\langle\alpha_i, \alpha_i\rangle=-\alpha_i\cdot \alpha_i=2$ and hence
$\alpha_i^\vee=\alpha_i$ for $i=0,1,\ldots,8$. One can verify by direct calculation that $(\langle
\alpha_i,\alpha_j\rangle)_{i,j=0,\ldots,8}=(-\alpha_i\cdot\alpha_j)_{i,j=0,\ldots,8}$ gives the
generalized Cartan matrix of type $E_8^{(1)}$ \eqref{eqn:Cartan_E6}. The null root is given by
\begin{equation}\label{eqn:E8_null_anti_canonical_divisor}
\delta= 3\alpha_0 + 2\alpha_1 + 4\alpha_2 + 6\alpha_3 + 5\alpha_4 + 4\alpha_5 + 3\alpha_6 + 2\alpha_7 + \alpha_8
=2H_1 + 2H_2 - E_1 - \cdots - E_8.
\end{equation}
In particular, $\alpha_i\cdot \delta=0$ and $\delta\cdot\delta=0$.  Then we obtain 
the root lattice
\begin{equation}
Q\left(E_8^{(1)}\right) = \Z\alpha_0 \oplus \Z \alpha_1 \oplus\cdots\oplus \Z\alpha_8\subset \Lambda \subset V,
\end{equation}
and the action of the
affine Weyl group $W(E_8^{(1)})=\langle r_0,\ldots,r_8\rangle$ ($r_i=r_{\alpha_i}$) on $V$ according
to the construction in Section \ref{subsection:root}.  Note also that the lattices $\Lambda$ and 
$Q\left(E_8^{(1)}\right)$ are stabilized by $W(E_8^{(1)})$.

The bilinear form $\lambda\cdot \mu$ on the Picard lattice $\Lambda$ is interpreted as the
intersection form of divisors $\lambda$, $\mu$. In general,
\begin{equation}\label{eqn:Lamda}
 \lambda = d_1H_1 + d_2H_2 - m_1E_1 - \cdots - m_8E_8 \in \Lambda,\quad d_i,\ m_i\in\Z,
\end{equation}
corresponds to the class of curves on $\P^1\times \P^1$ of bidegree $(d_1,d_2)$ passing through the
blowing-up points ${\rm P}_i$ with multiplicity $\geq m_i$ ($i=1,\ldots,8$). For example, the null root
$\delta$ given in \eqref{eqn:E8_null_anti_canonical_divisor}, called the {\em anti-canonical
divisor}, corresponds to the class of curves of bidegree $(2,2)$ passing through the eight points.

\begin{figure}[ht]
\begin{center}
\includegraphics[scale=0.35]{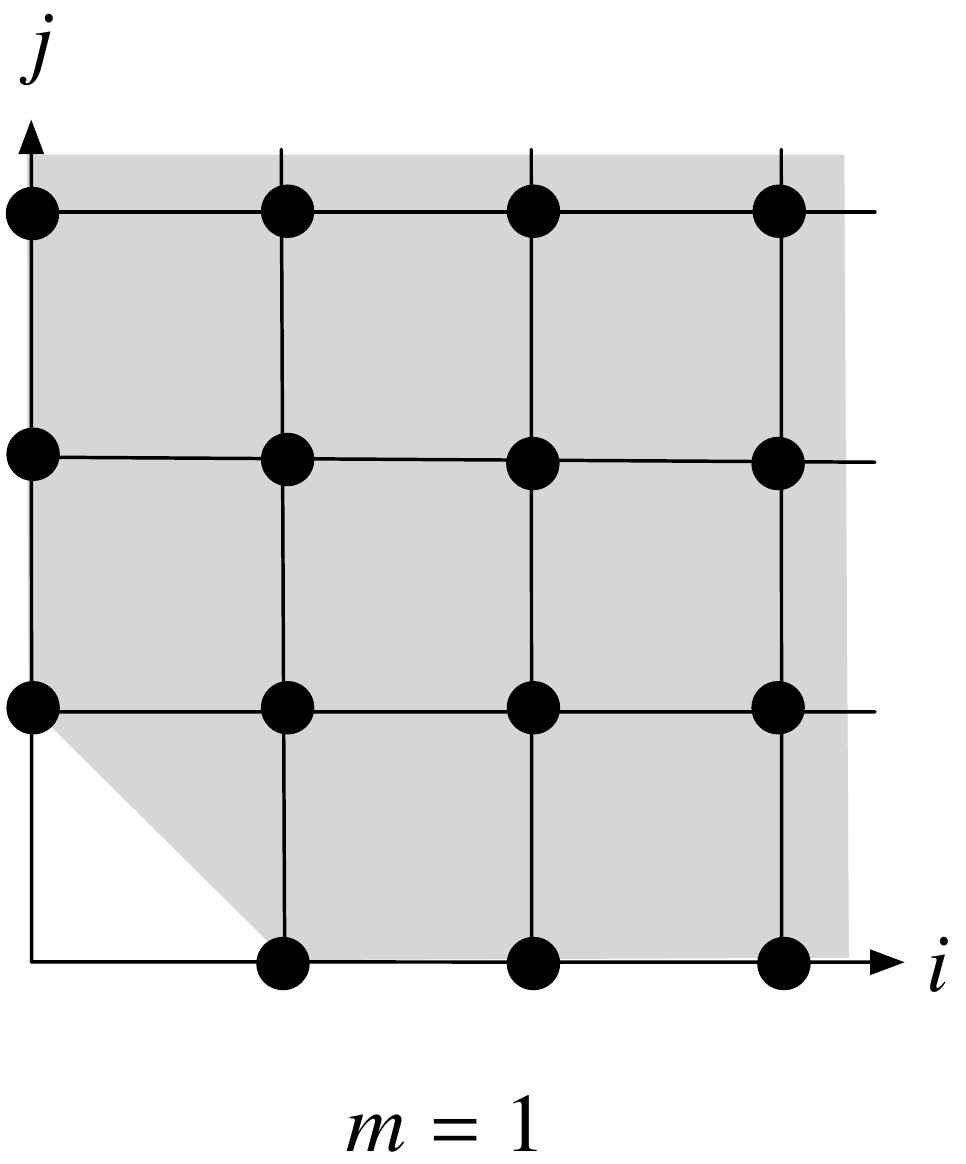}\hskip20pt
\includegraphics[scale=0.35]{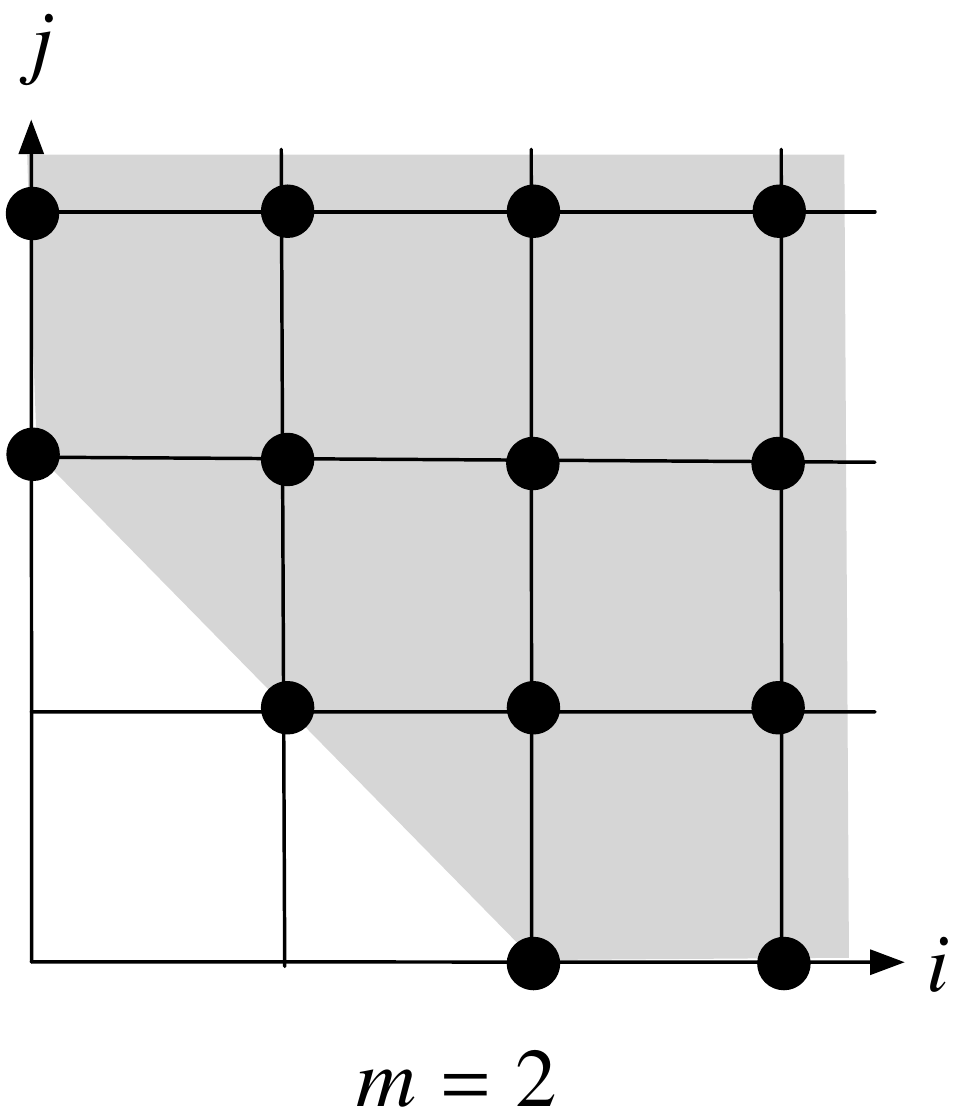}\hskip20pt
\includegraphics[scale=0.35]{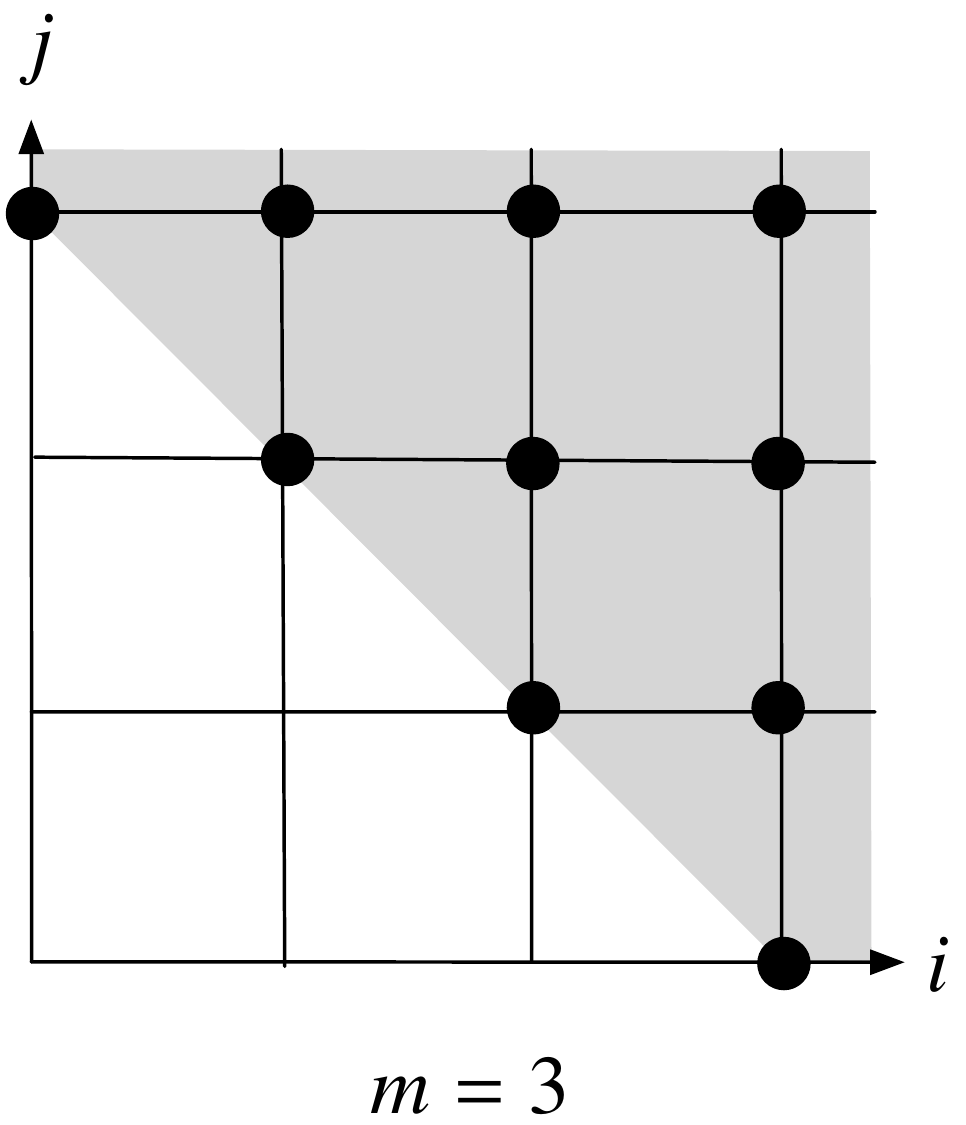}
\end{center}
\caption{Newton polygon and multiplicity} 
\label{fig:Newton_polygon_example}
\end{figure}

We remark that the multiplicity $m$ of a curve $\phi(x,y)=0$ at $(x,y)=(0,0)$ can be read from the
{\em Newton polygon} which displays the exponents of $x^iy^j$ of $\phi$ with nonzero coefficients as
in Figure \ref{fig:Newton_polygon_example}. In order to see the multiplicity at $(x,y)=(a,b)$, we
change variables to $(\xi,\eta)=(x-a,y-b)$ and construct the Newton polygon from the coefficients of
$\xi^i\eta^j$. For the multiplicity at $(\infty,b)$ and $(a,\infty)$, we take the coordinates
$(\xi,\eta)=(\frac{1}{x},y-b)$ and $(\xi,\eta)=(x-a,\frac{1}{y})$ respectively. Accordingly, for a
given Newton polygon with respect to the coordinate $(x,y)$, the multiplicities at $(x,y)=(0,0),
(\infty,0)$, $(0,\infty)$, $(\infty,\infty)$ can be read off from the bottom-left, bottom-right,
top-left, and top-right corners respectively.

For generic configurations of eight points, it is classically known that the {\em dimension}
$d(\lambda)$ of the family and the {\em genus} $g(\lambda)$ of generic curves of the class $\lambda$
\eqref{eqn:Lamda} are given by
\begin{equation}
d(\lambda) = (d_1+1)(d_2+1)-\sum_{i=1}^8\frac{m_i(m_i+1)}{2}-1,\quad
g(\lambda) = (d_1-1)(d_2-1)-\sum_{i=1}^8\frac{m_i(m_i-1)}{2},
\end{equation}
and hence it follows that 
\begin{equation}\label{eqn:dim_divisor}
2d(\lambda) =\lambda\cdot(\lambda+\delta),\quad 2g(\lambda)-2 = \lambda\cdot(\lambda-\delta),
\end{equation}
respectively. 

Noticing that each exceptional divisor $E_i$ ($i=1,\cdots,8$) satisfies $ E_i\cdot E_i = -1$ and
$E_i\cdot \delta=1$, we define the subset $M\subset\Lambda$ by
\begin{equation}\label{eqn:-1_curve}
M=\{\lambda\in\Lambda\mid \lambda\cdot \lambda = -1,\ \lambda\cdot \delta=1\}.
\end{equation}
Then $d(\lambda)=0$ and $g(\lambda)=0$ for $\lambda\in M$. 
The elements of $M$ are sometimes called the {\em exceptional classes}. Table \ref{tab:elements_M}
shows some typical elements of $M$.
\begin{table}[ht]
\begin{center}
\begin{tabular}{|c|c|}
\hline
$\lambda\in M$ & geometric meaning\\
\hline
$E_i$ & exceptional curve\\
\hline
$H_i-E_j$ & line passing through P$_j$\\
\hline
$H_1+H_2-E_i-E_j-E_k$ & curve of bidegree (1,1) passing through P$_i$, P$_j$, P$_k$\\
\hline
\end{tabular}
\caption{Typical elements of $M$.} \label{tab:elements_M}
\end{center}
\end{table}
\begin{ex}\label{ex:M}\rm
 We denote by $(f,g)$ the inhomogeneous coordinates of $\mathbb{P}^1\times\mathbb{P}^1$ and consider
 eight blowing-up points ${\rm P}_i=(f_i,g_i)$ $(i=1,\ldots,8)$. For $\lambda=H_1-E_j,\ H_2-E_j$ and
 $H_1+H_2-E_i-E_j-E_k$, the defining equations of corresponding curves are given as 
\begin{equation}
 f-f_j=0,\quad g-g_j=0,\quad
\left|\begin{array}{cccc}
1& f & g& fg\\
1& f_i & g_i& f_ig_i\\
1& f_j & g_j& f_jg_j\\
1& f_k & g_k& f_kg_k
\end{array}\right| = 0,
\end{equation}
respectively.
\end{ex}

We also define the subset $R\subset\Lambda$ by
\begin{equation}\label{eqn:def_R}
R=\{\alpha\in\Lambda\mid\alpha\cdot \alpha=-2,\ \alpha\cdot\delta=0\}.
\end{equation}
We simply call the elements of $R$ the {\em roots} (real roots of type $E_8^{(1)}$).  Note that if
$\alpha\in R$ we have $d(\alpha)=-1$ and $g(\alpha)=0$.  Table \ref{tab:elements_R} shows some
typical elements of $R$.
\begin{table}[ht]
\begin{center}
\begin{tabular}{|c|c|}
\hline
$\lambda\in R$ & geometric meaning\\
\hline
$E_i-E_j$ & exceptional curve passing through P$_j$\\
\hline
$H_i-E_j-E_k$ & line passing through P$_j$ and P$_k$\\
\hline
$H_1+H_2-E_i-E_j-E_k-E_l$ & curve of bidegree (1,1) passing through P$_i$, P$_j$, P$_k$, P$_l$\\
\hline
\end{tabular}
\caption{Typical elements of $R$.} \label{tab:elements_R}
\end{center}
\end{table}
\begin{ex}\label{ex:R}\rm
The meaning of the dimensionality $d(\alpha)=-1$ is that, for the existence of a curve of the class $\alpha$,
an extra condition is required on the configuration of points. For example, for $\lambda=H_1-E_j-E_k$,
$H_2-E_j-E_k$ and $H_1+H_2-E_i-E_j-E_k-E_l$, such conditions are given by
\begin{equation}
  f_k-f_j=0,\quad g_k-g_j=0,\quad
\left|\begin{array}{cccc}
1& f_l & g_l& f_lg_l\\
1& f_i & g_i& f_ig_i\\
1& f_j & g_j& f_jg_j\\
1& f_k & g_k& f_kg_k
\end{array}\right| = 0,
\end{equation}
respectively.
\end{ex}

\begin{rem}\label{rem:exceptional}\rm
It is known that the set $M$ defined by \eqref{eqn:-1_curve} is generated by $W(E_8^{(1)})$ from one
of the exceptional divisor $E_1,\ldots,E_8$. Also, the set $R$ is generated by $W(E_8^{(1)})$ from
one of the simple roots $\alpha_0,\ldots,\alpha_8$. 
\end{rem}

\begin{rem}\label{rem:rational_fn_Picard_lattice}\rm
By abuse of the terminology, we say that a polynomial $\phi(x,y)$ belongs to the class $\lambda\in
\Lambda$ when the curve $\phi=0$ is of the class $\lambda$. Then the dimension of the vector space
of polynomials belonging to the class $\lambda$ is given by $d(\lambda)+1$.  Note also that if two
polynomials $\phi(x,y)$ and $\psi(x,y)$ belong to the class $\lambda$ and $\mu$ respectively, then
the product $\phi(x,y)\psi(x,y)$ belongs to the class $\lambda+\mu$.  Moreover, we say that a ratio
of such polynomials is a rational function of the class $\lambda$. For example, the polynomials of
the class $H_1$ are expressed as $ax+b$, and the rational functions of the class $H_1$ as
$(ax+b)/(cx+d)$.
\end{rem}

\begin{rem}\rm
The Painlev\'e equations are sometimes discussed in the framework of nine point configuration on
$\mathbb{P}^2$ \cite{KMNOY:point_configuration,Sakai:SIV} as well as eight point configuration on
$\mathbb{P}^1\times\mathbb{P}^1$. We here give a correspondence between these two
formulations. Consider a birational mapping between $\mathbb{P}^2$ and
$\mathbb{P}^1\times\mathbb{P}^1$ given by
\begin{equation}
(X_0:X_1:X_2) \to (X_0:X_1) \times (X_0:X_2),
\end{equation}
where $(X_0:X_1:\ldots:X_i)$ is a homogeneous coordinate of $\mathbb{P}^i$.  Let $X$ be a blow-up of
$\mathbb{P}^2$ at two points $(0:1:0)$ and $(0:0:1)$, and let $Y$ be a blow-up of
$\mathbb{P}^1\times\mathbb{P}^1$ at $(0:1)\times (0:1)$, then the birational mapping gives a
biholomorphic mapping between $X$ and $Y$ (see Figure \ref{fig:P2_and_P1P1}).
\begin{figure}[ht]
{\footnotesize
\setlength{\unitlength}{0.6mm}
\begin{picture}(200,120)(-30,-10)
\put(-10,10){\line(1,0){40}}\put(-20,10){${\cal E}_0$}
\put(0,0){\line(0,1){40}}\put(-20,40){${\cal E}_0$}
\put(-10,40){\line(1,-1){40}}\put(0,-5){${\cal E}_0$}
\put(20,30){$\swarrow$}
\put(50,60){\line(1,0){40}}\put(15,58){${\cal E}_0-{\cal E}_1=H_2$}
\put(60,50){\line(0,1){40}}\put(36,43){${\cal E}_0-{\cal E}_2=H_1$}
\put(70,90){\line(1,-1){20}}\put(65,95){${\cal E}_0-{\cal E}_1-{\cal
E}_2=E_1$}
\put(50,85){\line(1,0){32}}\put(15,83){${\cal E}_2=H_2-E_1$}
\put(85,50){\line(0,1){32}}\put(75,43){${\cal E}_1=H_1-E_1$}
\put(110,30){$\searrow$}
\put(130,10){\line(1,0){45}}\put(120,8){$H_2$}
\put(130,35){\line(1,0){45}}\put(120,33){$H_2$}
\put(140,0){\line(0,1){45}}\put(138,-7){$H_1$}
\put(165,0){\line(0,1){45}}\put(163,-7){$H_1$}
\end{picture}
}
\caption{Correspondence of $\mathbb{P}^2$ and $\mathbb{P}^1\times\mathbb{P}^1$.} \label{fig:P2_and_P1P1}
\end{figure}
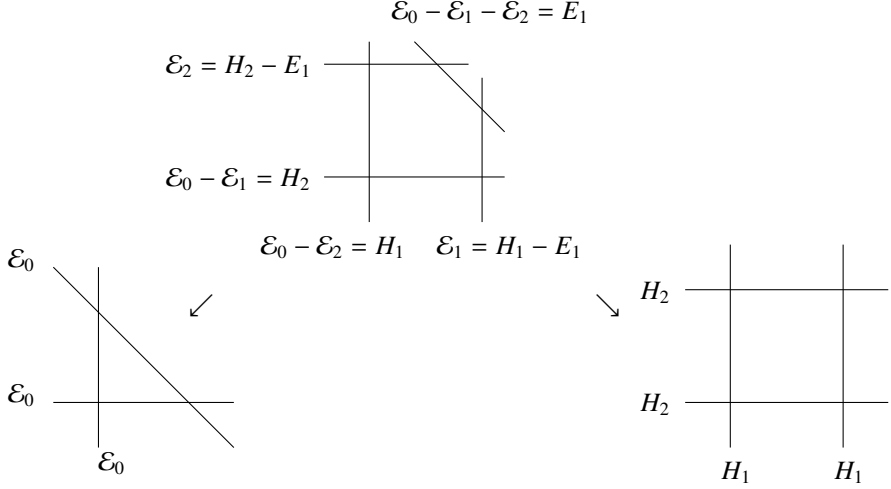
The Picard lattice of $X$ and $Y$ are given as
\begin{equation}
\text{Pic}(X)=\Z {\cal E}_0 \oplus \Z {\cal E}_1 \oplus \Z {\cal E}_2,
\quad
\text{Pic}(Y)=\Z H_1 \oplus \Z H_2 \oplus \Z E_1,
\end{equation}
respectively, where non-vanishing intersections are given by ${\cal E}_0^2=1$, 
${\cal E}_1^2={\cal E}_2^2=-1$ and $H_1\cdot H_2=1$, $E_1^2=-1$ respectively.
An isomorphism of these lattices are given by
\begin{equation}\label{eq: picP2picP1P1}
{\cal E}_0=H_1+H_2-E_1, \quad
{\cal E}_1=H_1-E_1, \quad
{\cal E}_2=H_2-E_1.
\end{equation}
Further blowing up the spaces $X$ and $Y$, we have an isomorphism between $(n+1)$-point blow-up of
$\mathbb{P}^2$ and $n$-point blow-up of $\mathbb{P}^1\times \mathbb{P}^1$. In particular, one can
use nine point blow-up of $\mathbb{P}^2$ instead of eight point blow-up of $\mathbb{P}^1 \times
\mathbb{P}^1$ to describe the Painlev\'e equations.  In the $\mathbb{P}^2$ formulation, the Picard
lattice is given as
\begin{equation}
\Lambda=\Z {\cal E}_0 \oplus \Z {\cal E}_1 \oplus \cdots \oplus \Z {\cal
E}_9,
\end{equation}
with ${\cal E}_0^2=1$ and ${\cal E}_1^2=\cdots={\cal E}_9^2=-1$. Then the simple roots of
$E_8^{(1)}$ are
\begin{equation}
\alpha_0={\cal E}_0-{\cal E}_1-{\cal E}_2-{\cal E}_3, \quad
\alpha_i={\cal E}_{i}-{\cal E}_{i+1} \ (i=1, \ldots, 8).
\end{equation}
These corresponds to the roots in eq.(3.28) in $\text{Pic}(\mathbb{P}^1\times \mathbb{P}^1)$ by the
isomorphism eq.(\ref{eq: picP2picP1P1}) extended by ${\cal E}_{i+1}=E_i$ $(i=2,\ldots,8)$.
\end{rem}

\subsection{Surface type and symmetry type}\label{subsection:P4_surface_type}
It has been observed by Okamoto for all the Painlev\'e differential equations that there is a remarkable
complementary relation between the surface type and the symmetry type in the common root lattice of
type $E_8^{(1)}$, as in the following list (see Table \ref{tab:list_P}).  The origin of this
phenomena is described by the geometry of the space of initial values, as we will demonstrated below
by taking P$_{\rm IV}$ as an example
 \cite{Ohyama-Kawamuko-Sakai-Okamoto,Okamoto:SIV,Okamoto:p24,Okamoto:p6,Okamoto:p5,Okamoto:p3,Sakai:SIV}.
\begin{table}[ht]
\begin{center}
\begin{tabular}{|c|c|c|c|ccc|c|c|}
\hline
 &P$_{\rm VI}$ &  P$_{\rm V}$ & P$_{\rm IV}$ &  &P$_{\rm III}$ &  &P$_{\rm II}$ &P$_{\rm I}$ \\
 \hline
\raise3pt\hbox{\rule{0pt}{2.2ex}} 
Surface type & $D_4^{(1)}$ & $D_5^{(1)}$ & $E_6^{(1)}$ & $D_6^{(1)}$ & $D_7^{(1)}$ & $D_8^{(1)}$ & $E_7^{(1)}$ & $E_8^{(1)}$\\[1mm]
 \hline
\raise3pt\hbox{\rule{0pt}{2.2ex}}
Symmetry type& $D_4^{(1)}$ & $A_3^{(1)}$ & $A_2^{(1)}$ & $(2A_1)^{(1)}$ & $A_1^{(1)}$ & $A_0^{(1)}$ & $A_1^{(1)}$ & $A_0^{(1)}$\\[1mm]
\hline
\end{tabular} 
\end{center}
\caption{Table of the surface type and the symmetry type of each Painlev\'e equation.}\label{tab:list_P}
\end{table}

In the example of P$_{\rm IV}$ shown in Section \ref{subsubsection:resolution_P4}, recall that the
divisor of inaccessible points \eqref{eqn:p4_removed_divisors} has seven components $\delta_i$
($i=0,\ldots,6$):
\begin{equation}\label{eqn:p4_removed_divisors_2}
\begin{array}{c}\medskip
\delta_1= E_1-E_2,\quad  \delta_2=H_2-E_1-E_5,\quad  \delta_3=E_5-E_6,\quad  \delta_4=H_1-E_3-E_5,\\
\delta_5=E_3-E_4,\quad \delta_6=E_6-E_7,\quad  \delta_0=E_7-E_8. 
\end{array}
\end{equation}
 Regarded as elements of $\Lambda$, they satisfy
\begin{equation} \label{eqn:p4_delta_decomposition}
 \delta_0+\delta_1+2\delta_2+3\delta_3+2\delta_4+\delta_5+2\delta_6=\delta,\quad 
\delta_i\cdot\delta_i=-2,\quad \delta_i\cdot \delta=0\quad (i=0,\ldots,6).
\end{equation}
Furthermore, the intersection numbers among $\delta_i$ are given by the following $7\times 7$ matrix
\begin{equation}\label{eqn:p4_surface_Cartan}
\setlength{\arraycolsep}{2pt}
-\left[\delta_i\cdot\delta_j\right]_{i,j=0,\ldots,6}=
\left[\begin{array}{ccccccc}
2 & & & & & &-1 \\
  &2 &-1 & & & & \\
  &-1 &2 &-1 & & & \\
  & &-1 &2 &-1 & &-1 \\
  & & &-1 &2 &-1 & \\
  & & & &-1 &2 & \\
-1  & & &-1 & & &2 
\end{array}\right],\quad
 \begin{array}{c}
 \delta_0\\
|\\
\delta_6\\
|\\
\delta_1\ \text{\textemdash}\ \delta_2\ \text{\textemdash}\  \delta_3\ \text{\textemdash}\  \delta_4\ \text{\textemdash}\ \delta_5
 \end{array}
\end{equation}
which is the Cartan matrix of type $E_6^{(1)}$. In view of the decomposition
\eqref{eqn:p4_delta_decomposition} of $\delta$ in terms of $\delta_0,\ldots,\delta_6$, the space of
initial values of P$_{\rm IV}$ is said to have the {\em surface type} of $E_6^{(1)}$.

As we mentioned before, the symmetry of P$_{\rm IV}$ is described by the extended affine Weyl
group of type $A_2^{(1)}$. This fact is explained geometrically as follows. Consider the elements
$\alpha\in R$ such that
\begin{equation}\label{eqn:p4_alpha}
 \alpha\cdot \alpha=-2,\quad \delta_i\cdot \alpha=0\quad (i=0,\ldots,6).
\end{equation}
For such an element $\alpha$, the reflection $r_\alpha$ satisfies $r_\alpha(\delta_i)=\delta_i$
($i=0,\ldots,6$); namely, such reflections $r_\alpha$ preserve the surface type. Hence they are
expected to generate the symmetry of P$_{\rm IV}$. In fact, we can take three fundamental elements
satisfying \eqref{eqn:p4_alpha} as
\begin{equation}\label{eqn:p4_alpha_by_E8}
\alpha_0 = H_1 + H_2 - E_5 - E_6 - E_7 - E_8,\quad
\alpha_1 = H_1 - E_1 - E_2,\quad
\alpha_2 = H_2 - E_3 - E_4.
\end{equation}
Furthermore, the intersection numbers among $\alpha_0$, $\alpha_1$,
$\alpha_2$ are given by the Cartan matrix of type $A_2^{(1)}$
\begin{equation}
 -\left[\alpha_i\cdot \alpha_j\right]_{i,j=0,1,2}
=\left[
\begin{array}{ccc}
 2& -1& -1\\
-1 & 2 & -1\\
-1 & -1 &2
\end{array}
\right],\qquad
\setlength{\unitlength}{0.8mm}
\begin{picture}(27,12)(4,5)
\put(5,0){$\alpha_1$}
\put(25,0){$\alpha_2$}
\put(15,15){$\alpha_0$}
\put(10,1){\line(1,0){12.5}}
\put(24,4){\line(-2,3){6}}
\put(8,4){\line(2,3){6}}
\end{picture} 
\end{equation}
and we have
\begin{equation}
 \alpha_0 + \alpha_1 + \alpha_2 = \delta.
\end{equation} 
As we will see later, the actions of the three simple reflections $r_{\alpha_i}$ ($i=0,1,2$) on the
Picard lattice correspond to the B\"acklund transformations $s_i$ ($i=0,1,2$) for P$_{\rm IV}$.
\begin{rem}\rm 
 We use the common symbols $\alpha_i$ ($i=0,1,2,\ldots,l$) and $\delta_j$ ($j=0,\ldots, 8-l$) for
the simple roots for any pair of root systems representing symmetry and surface types, respectively.
Therefore $\alpha_i$ ($i=0,1,2$) of type $A_2^{(1)}$ used above are different from the simple roots
of type $E_8^{(1)}$ in Section \ref{subsection:Picard}.
\end{rem}
Then we see that the two sublattices
\begin{equation}
L_1=\Z\delta_0 \oplus \Z\delta_1\oplus\cdots\oplus \Z\delta_6,\quad 
L_2= \Z \alpha_0 \oplus \Z \alpha_1 \oplus \Z \alpha_2,
\end{equation}
are orthogonal complements in the root lattice $Q\left(E_8^{(1)}\right)$ to each other, namely,
$(L_1)^\perp = L_2$, $(L_2)^\perp = L_1$. In fact, if $\lambda=d_1H_1+d_2H_2-\sum_{i=1}^8 m_iE_i\in
\Lambda$, the condition $\delta_i\cdot\lambda=0$  ($i=0,\ldots,6$) is equivalent to
\begin{equation}
  m_1=m_2,\  d_1=m_1+m_5,\  m_5=m_6,\  d_2=m_3+m_5,\  m_3=m_4,\  m_6=m_7,\  m_7=m_8. 
\end{equation}
Hence, $\lambda\in (L_1)^\perp$ if and only if
\begin{align*}
\lambda &= (m_1+m_5)H_1+(m_3+m_5)H_2-m_1(E_1+E_2) - m_3(E_3+E_4) - m_5(E_5+E_6+E_7+E_8)\\
&=m_1(H_1-E_1-E_2) + m_3 (H_2-E_3-E_4) + m_5 (H_1+H_2-E_5-E_6-E_7-E_8)\\
&= m_1 \alpha_1 + m_3\alpha_2 + m_5 \alpha_0,
\end{align*}
which shows that $(L_1)^\perp = L_2$. One can verify $(L_2)^\perp = L_1$ in a similar manner.
\begin{rem}\rm\hfill
\begin{enumerate}
 \item $Q(E_8^{(1)})$ and $\Z \delta$ are orthogonal complements to each other in $\Lambda$.
 \item The elements of $Q(E_8^{(1)})$ are not necessarily expressible as $\Z$-linear combinations of $\delta_i$
and $\alpha_i$; they are expressible as linear combinations with coefficients in
$\frac{1}{3}Z$.  This means that $Q(E_8^{(1)})\supsetneq L_1+ L_2$, while $\mathbb{Q}$-vector space
$\mathbb{Q}\otimes Q(E_8^{(1)})$ is generated by $L_1$ and $L_2$.
\end{enumerate}
\end{rem}
\subsection{Example of P$_{\rm IV}$}\label{subsec:P4_Picard}
Recall that the symmetry of P$_{\rm IV}$ is given by the affine Weyl group
$\widetilde{W}(A_2^{(1)})=\langle s_0,s_1,s_2,\pi\rangle$ (see the last paragraph of Section
\ref{subsection:root}). Here we will describe the symmetry in terms of the Picard lattice.

The action of the reflection $r_i=r_{\alpha_i}$ corresponding to $s_i$ ($i=0,1,2$) on the basis of
$\Lambda$ is computed by using \eqref{eqn:E_8_intersection_form}, \eqref{eqn:roots_reflection} and
\eqref{eqn:p4_alpha_by_E8} as (trivial one is omitted)
\begin{align}
& r_0:\quad
\left\{
\begin{array}{c}\medskip
H_1\to 2 H_1+H_2-E_5-E_6-E_7-E_8,\quad H_2\to H_1+2 H_2-E_5-E_6-E_7-E_8,\\
\medskip
E_5\to H_1+H_2-E_6-E_7-E_8,\quad E_6\to H_1+H_2-E_5-E_7-E_8,\\
 E_7\to H_1+H_2-E_5-E_6-E_8,\quad E_8\to H_1+H_2-E_5-E_6-E_7,
\end{array}\right.\\[2mm]
& r_1:\quad H_2\to H_1+H_2-E_1-E_2,\quad E_1\to H_1-E_2,\quad E_2\to H_1-E_1,\\[2mm]
&r_2:\quad H_1\to H_1+H_2-E_3-E_4,\quad E_3\to H_2-E_4,\quad E_4\to H_2-E_3.
\end{align}
On the other hand, the B\"acklund transformation $\pi$ corresponds to the action of the diagram
automorphism $\pi$ on $E_6^{(1)}\oplus A_2^{(1)}$
\begin{center}
\includegraphics[scale=0.4]{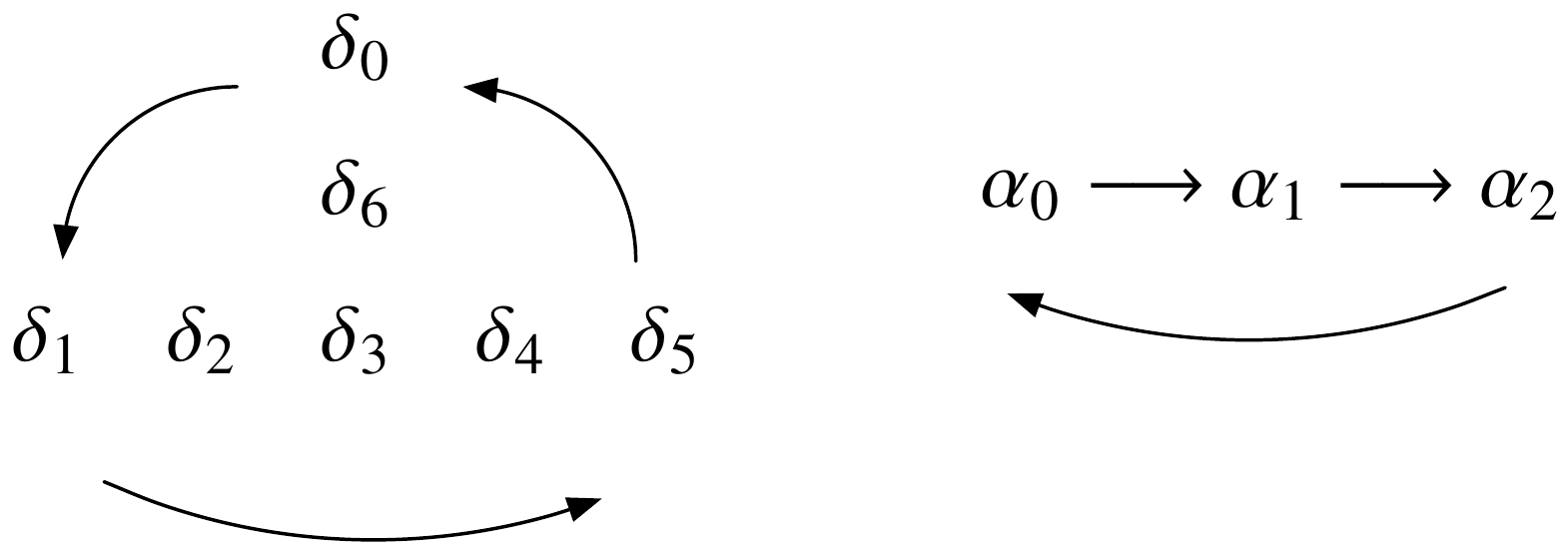}
\end{center}
which is realized by 
\begin{align}
&\pi:\ 
\left\{
\begin{array}{c}\medskip
 E_1\to E_3,\ E_2\to E_4,\ E_3\to E_7,\ E_4\to E_8,\ E_5\to H_2-E_6,\ E_6\to H_2-E_5,\\
E_7\to E_1,\ E_8\to E_2,\ H_1\to H_2,\ H_2\to H_1 + H_2 - E_5-E_6 .
\end{array}\right.
\end{align}
Then $r_0$, $r_1$, $r_2$ and $\pi$ gives the representation of $\widetilde{W}(A_2^{(1)})$ on the
lattice $\Lambda$. The relation between the B\"acklund transformation and the above action on the
lattice is stated as follows. The curve obtained from the polynomial factor arising from
$w(\tau_0)$, $w(\tau_1)$, $w(\tau_2)$, $w\in\widetilde{W}(A_2^{(1)})$ coincides with the curves in
the exceptional class $w(E_8)$, $w(E_2)$, $w(E_4)$, respectively, under the identification $s_i =
r_i$, where $r_i=r_{\alpha_i}$ ($i=0,1,2$).

\begin{figure}[ht]
\begin{center}
\includegraphics[scale=0.35]{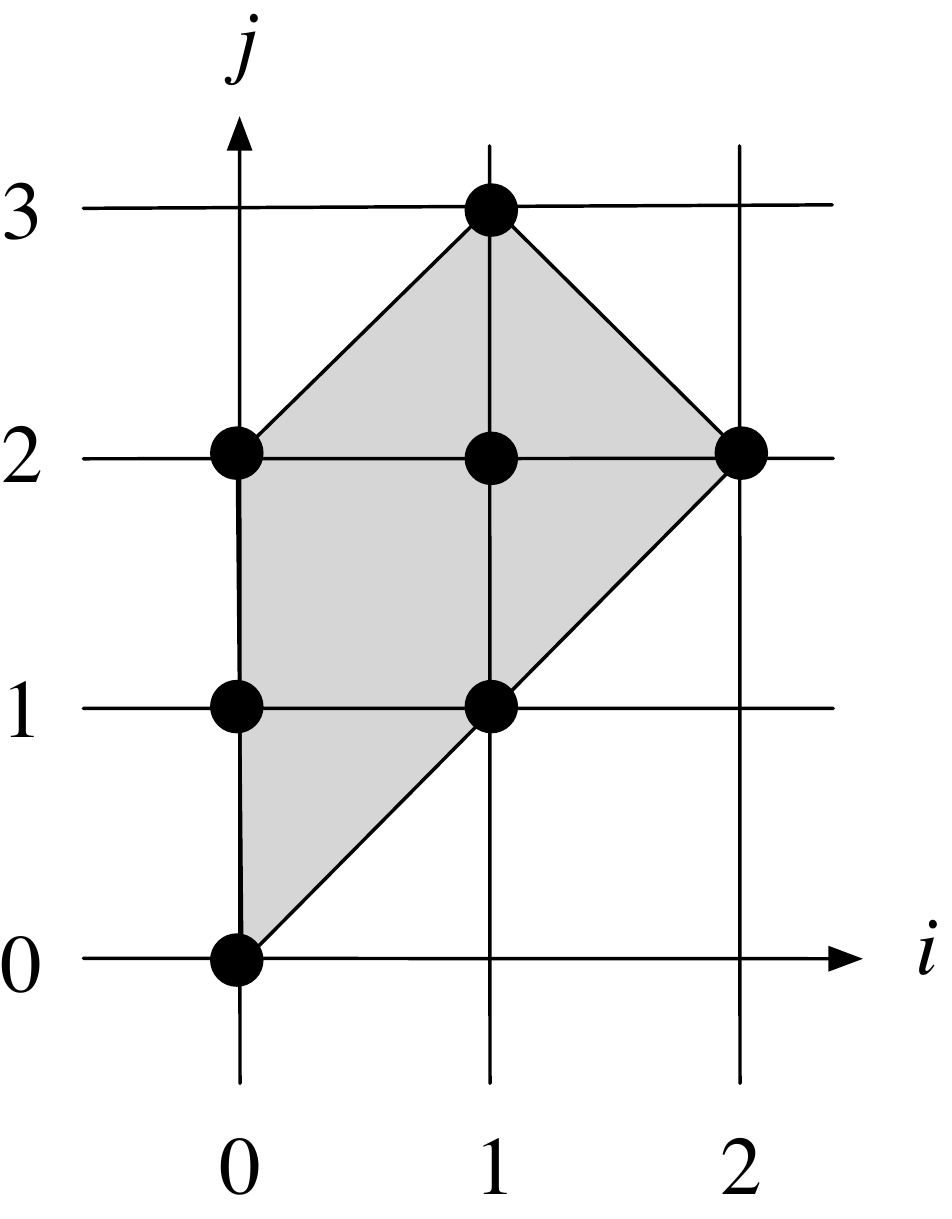}
\end{center}
\caption{Newton polygon for $\phi$.} \label{fig:Newton_polygon}
\end{figure}

We demonstrate this relation by taking an example of $s_2s_1s_0(\tau_0)$.  Noticing
\eqref{eqn:p4_pq_f}, we have by using \eqref{eqn:BT_on_tau}
\begin{align}
s_2s_1s_0(\tau_0) &= s_2s_1\left(f_0\frac{\tau_1\tau_2}{\tau_0}\right)
= s_2\left(\left(f_0 - \frac{a_1}{f_1}\right)f_1\frac{\tau_0\tau_2}{\tau_1}\frac{\tau_2}{\tau_0}\right)
=\cdots=\phi~\frac{\tau_0^2\tau_1}{\tau_2^2},\nonumber\\
\phi&= (f_0f_2-a_2)(f_1f_2+a_2) - (a_1+a_2)f_2^2\nonumber\\
& =- a_1 p^2-a_2^2 - 2 a_2 q p - a_2 p t+ qp^3 - q^2 p^2 - qp^2 t.
\end{align}
The curve $\phi=0$ corresponds to the divisor class $\lambda=d_1H_1 + d_2H_2-\sum_{i=1}^8
m_iE_i\in\Lambda$ as
\begin{equation}\label{eqn:p4_lambda_s2s1s0tau0}
\lambda = 2H_1+3H_2-E_1-E_2-2E_3-2E_4-E_5-E_6-E_7.
\end{equation}
In fact, $\phi$ is the curve of bidegree $(2,3)$ which implies $d_1=2$ and $d_2=3$. We compute
the multiplicity $m_3$, $m_4$ at $E_3$, $E_4$, respectively, around $(q,p)=(\infty,0)$ for
instance. To this end, we apply the variable change \eqref{eqn:p4_blowup_var2} where the divisors
$E_3$ and $E_4$ are defined.  First we change variable as $(q,p)=(\frac{1}{q_0},p_0)$ and then
$(q_0, p_0) = (q_1, q_1p_1)$ to obtain
\begin{align}
q_0^{d_1}\phi(\frac{1}{q_0},p_0)
&= - \left(a_2q_0 + p_0\right)^2
- q_0p_0t\left(a_2 q_0 + p_0\right) + q_0p_0^2\left( - a_1 q_0   + p_0  t\right)\nonumber\\
&= q_1^2\left\{- \left(a_2 + p_1\right)^2
- q_1p_1t\left(a_2 + p_1\right) + q_1^2p_1^2\left( - a_1   + p_1  t\right)\right\}.\label{eqn:phi_2}
\end{align}
This shows that the multiplicity $m_3$ at $(q_0,p_0)=(0,0)$ is $2$.
To see the multiplicity $m_4$, we change variables as $(q_1, p_1) = (q_2, -a_2 + q_2 p_2)$. Then 
the second factor of the right hand side of \eqref{eqn:phi_2} gives
\begin{equation}
q_2^2\left[
p_2
- p_2(-a_2 + q_2 p_2)t + (-a_2 + q_2 p_2)^2\left\{- a_1   +   (-a_2 + q_2 p_2)t\right\}
\right],
\end{equation}
which implies that $m_4=2$. Repeating the similar calculations at
$(q,p)=(\infty,0),(\infty,\infty)$, we see that $\phi$ belongs to the class $\lambda$
\eqref{eqn:p4_lambda_s2s1s0tau0}.  We note that the multiplicities $m_1=1$, $m_3=2$, $m_5=1$ can
also be simply read off from the pattern of the Newton polygon for $\phi$ (Figure \ref{fig:Newton_polygon}) 
at the corners $(0,\infty)$, $(\infty,0)$, $(\infty,\infty)$ respectively.

On the other hand, one can verify that $r_0r_1r_2(E_8)=2H_1+3H_2-E_1-E_2-2E_3-2E_4-E_5-E_6-E_7$ by
direct computation. Therefore, we have confirmed the correspondence between the B\"acklund
transformations and affine Weyl group actions on the Picard lattice. For the proof of this
correspondence, we refer to \cite{KMNOY:point_configuration}.  We remark that $\lambda$ corresponding
to the $\tau$ functions always belongs to the exceptional class $M$ \eqref{eqn:-1_curve} with
$d(\lambda)=0$. This implies that the curve is uniquely determined from $\lambda$.

This idea can be applied for analyzing the iteration of $T$ on $f$-variables or
equivalently $(q,p)$-variables. Since $f_i = \frac{\tau_is_i(\tau_i)}{\tau_{i+1}\tau_{i-1}}$
$(i\in\mathbb{Z}/3\mathbb{Z})$, we have
\begin{equation}
 w(f_i) = \frac{w(\tau_i)\, ws_i(\tau_i)}{w(\tau_{i+1})\, w(\tau_{i-1})},
\end{equation}
for general $w\in\widetilde{W}(A_2^{(1)})$.  As seen in the above example and will be clarified in
the later sections in the general setting, the polynomial factors of the rational function obtained
as $w(f_i)$ can be controlled by the language of the Picard lattice. This observation also suggests
that the polynomial factors of the iterations $T^n(f_i)$ will provide us the information of the
underlying point configuration, which will be utilized in Section \ref{subsec:point_configuration2}.
\section{Detecting Point Configurations in Discrete Painlev\'e Equations}\label{sec:Points_from_dP}
In this Section we demonstrate how to associate a configuration of singular points on
$\mathbb{P}^1\times\mathbb{P}^1$ to a given discrete equation.  This provides us a method for
determining whether it is a discrete Painlev\'e equation, and if so, identifying the type of the equation by
its surface type and symmetry type.
\subsection{Point configuration for $q$-P$(E_6^{(1)}):$ an example}\label{subsubsection:qpe6}
We start with an example of a discrete Painlev\'e equation from which we can easily read off the
point configuration. Fixing a nonzero constant $q$, let us consider the following rational mapping
$T$ of $\left(\mathbb{C^\times}\right)^{10}\times (\mathbb{P}^1\times\mathbb{P}^1)$ 
 \cite{MSY:Riccati,GRTT:special_fn,Witte-Ormerod}:
\begin{equation}
 T:\ (\kappa_1,\kappa_2,v_1,\ldots,v_8,f,g)\mapsto \left(\frac{\kappa_1}{q},\kappa_2 q,v_1,\ldots,v_8,\overline{f},\overline{\mathstrut g}\right), 
\label{eqn:q-e6_ev}
\end{equation}
where $\of$ and $\og$ are rational functions of $f$ and $g$ determined by
\begin{equation}\label{eqn:q-e6}
\left\{
\begin{array}{c}\medskip
{\displaystyle 
\frac{(fg-1)(\of g-1)}{f\of} 
= \frac{\left(g-\frac{1}{v_1}\right)\left(g-\frac{1}{v_2}\right)
\left(g-\frac{1}{v_3}\right)\left(g-\frac{1}{v_4}\right)}
{\left(g-\frac{v_5}{\kappa_2}\right)\left(g-\frac{v_6}{\kappa_2}\right)}} ,\\
{\displaystyle 
\frac{(\of g-1)(\of \overline{\mathstrut g} -1)}{g\og} 
= \frac{(\of -v_1)(\of -v_2)(\of -v_3)(\of -v_4)}
{\left(\of - \frac{\kappa_1}{qv_7}\right)\left(\of - \frac{\kappa_1}{qv_8}\right)}}.
\end{array}\right. 
\end{equation}
Here the variables $\kappa_i$ ($i=1,2$) and $v_i$ ($i=1,\ldots,8$) are subject to the relation
\begin{equation}
  q = \frac{\kappa_1^2\kappa_2^2}{\prod_{i=1}^8 v_i}. \label{eqn:q-e6_q}
\end{equation}
With the notation $T^n(f)=f_n$, $T^n(g)=g_n$ ($n\in\mathbb{Z}$), \eqref{eqn:q-e6_ev}--\eqref{eqn:q-e6_q} can be 
interpreted as a difference equation with respect to $n$:
\begin{equation}\label{eqn:q-e6_2}
\left\{
\begin{array}{c}\medskip
{\displaystyle 
\frac{(f_ng_n-1)(f_{n+1} g_n-1)}{f_nf_{n+1}} 
= \frac{\left(g_n-\frac{1}{v_1}\right)\left(g_n-\frac{1}{v_2}\right)
\left(g_n-\frac{1}{v_3}\right)\left(g_n-\frac{1}{v_4}\right)}
{\left(g_n-\frac{v_5}{\kappa_2q^n}\right)\left(g_n-\frac{v_6}{\kappa_2q^n}\right)}} ,\\
{\displaystyle 
\frac{(f_{n+1} g_n-1)(f_{n+1} g_{n+1} -1)}{g_ng_{n+1}} 
= \frac{(f_{n+1} -v_1)(f_{n+1} -v_2)(f_{n+1} -v_3)(f_{n+1} -v_4)}
{\left(f_{n+1} - \frac{\kappa_1}{v_7q^{n+1}}\right)\left(f_{n+1} - \frac{\kappa_1}{v_8q^{n+1}}\right)}}.
\end{array}\right. 
\end{equation}
We also use the notation $T(x)=\overline{x}$, $T^{-1}(x)=\underline{x}$ for the mapping $T$ and its
inverse, respectively. We call the discrete Painlev\'e equation \eqref{eqn:q-e6} or \eqref{eqn:q-e6_2} 
$q$-P$(E_6^{(1)})$, since it has the affine Weyl group symmetry of type $E_6^{(1)}$ as will be shown later.

Since $\of$, $\overline{g}$ are rational functions of $f$ and $g$, there may be some points
$(f,g)\in\mathbb{P}^1\times\mathbb{P}^1$ where their images 
$(\of,\overline{\mathstrut g})\in\mathbb{P}^1\times\mathbb{P}^1$ cannot be determined uniquely.  To investigate such points,
called the {\em points of indeterminacy}, we assume that $\kappa_1$, $\kappa_2$, $v_1,\ldots,v_8$
are generic and rewrite \eqref{eqn:q-e6} as
\begin{equation}
\left\{
\begin{array}{c}\medskip
{\displaystyle 
\frac{\of g-1}{\of} 
= \frac{f\left(g-\frac{1}{v_1}\right)\left(g-\frac{1}{v_2}\right)
\left(g-\frac{1}{v_3}\right)\left(g-\frac{1}{v_4}\right)}
{\left(fg-1\right)\left(g-\frac{v_5}{\kappa_2}\right)\left(g-\frac{v_6}{\kappa_2}\right)}} ,\\
{\displaystyle 
\frac{\of \overline{\mathstrut g} -1}{\og} 
= \frac{g(\of -v_1)(\of -v_2)(\of -v_3)(\of -v_4)}
{\left(\of g-1\right)\left(\of - \frac{\kappa_1}{qv_7}\right)\left(\of - \frac{\kappa_1}{qv_8}\right)}}.
\end{array}\right. \label{eqn:q-e6_3}
\end{equation}
Note that the second equation is equivalent to
\begin{equation}
\frac{f \ug-1}{\ug} 
= \frac{g (f -v_1)(f -v_2)(f -v_3)(f -v_4)}
{(f g -1)\left(f - \frac{\kappa_1}{v_7}\right)\left(f - \frac{\kappa_1}{v_8}\right)}.\label{eqn:q-e6_40}
\end{equation}
From the first equation in \eqref{eqn:q-e6_3} we see that $\of$ is uniquely determined from $(f,g)$ unless 
\begin{equation}
 (f,g)=\left(0,\frac{v_5}{\kappa_2}\right),\ \left(0,\frac{v_6}{\kappa_2}\right),\ \left(v_1, \frac{1}{v_1}\right),\
\left(v_2, \frac{1}{v_2}\right), \ \left(v_3, \frac{1}{v_3}\right), \ \left(v_4, \frac{1}{v_4}\right).
\end{equation}
Also, \eqref{eqn:q-e6_40} implies that $\ug$ is uniquely determined from $(f,g)$ unless 
\begin{equation}
  (f,g)=\left(\frac{\kappa_1}{v_7},0\right),\ \left(\frac{\kappa_1}{v_8},0\right),\ \left(v_1, \frac{1}{v_1}\right),\
\left(v_2, \frac{1}{v_2}\right), \ \left(v_3, \frac{1}{v_3}\right), \ \left(v_4, \frac{1}{v_4}\right).
\end{equation}
In this way, we find that the eight points
\begin{equation}\label{eqn:8points_q-e6}
{\rm P}_i:\ \left(v_i, \frac{1}{v_i}\right)\ (i=1,2,3,4),\quad \left(0,\frac{v_i}{\kappa_2}\right)\ (i=5,6),\quad 
\left(\frac{\kappa_1}{v_i},0\right)\ (i=7,8),
\end{equation}
are the points of indeterminacy of the mapping.

\begin{center}
\begin{figure}[ht]
\begin{center}
\includegraphics[scale=0.4,clip,viewport=0 0 350 280]{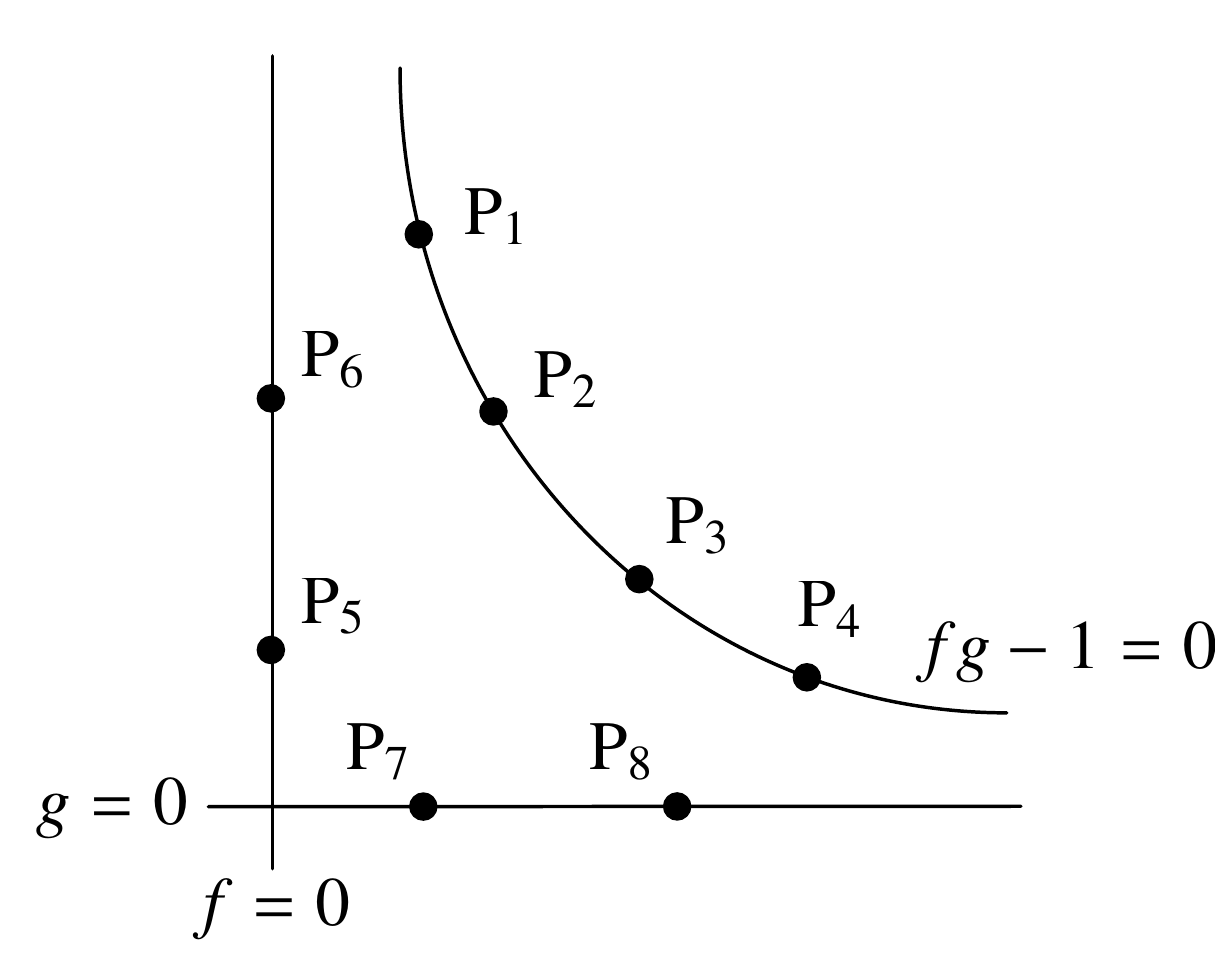}
\hskip20pt\includegraphics[scale=0.4,clip,viewport=0 0 370 300]{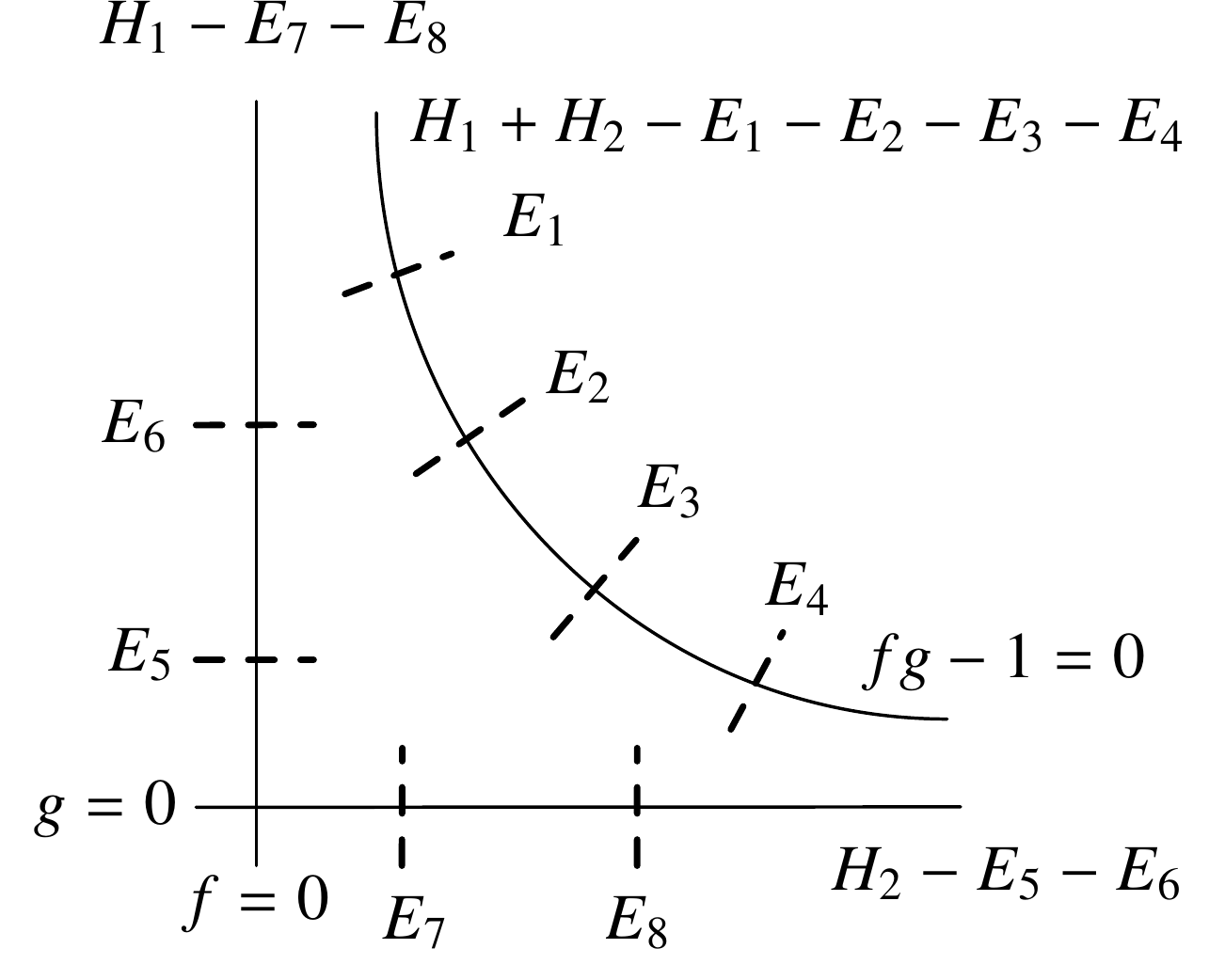}
\end{center}
\caption{Surface of the mapping \eqref{eqn:q-e6_ev} and
\eqref{eqn:q-e6}. Left: Configuration of the points of
indeterminacy. Right: Configuration of divisors. Solid lines are
inaccessible divisors. Thick lines are exceptional divisors arising in
the blow-ups. } \label{fig:qPE6_surface}
\end{figure}
\end{center}

In order to investigate the behaviour of the mapping around the points of indeterminacy, we next
apply the blowing-up at these points. Around the point of indeterminacy $(f,g)=(v_1,
\frac{1}{v_1})$ for example, we introduce new variables $(f_1,g_1)$ by setting
\begin{equation}
 f=v_1+f_1,\quad g = \frac{1}{v_1} + f_1g_1.
\end{equation}
Then the indeterminate factor in \eqref{eqn:q-e6} becomes
\begin{equation}
 \frac{g-\frac{1}{v_1}}{fg-1} = \frac{g_1}{v_1g_1+\frac{1}{v_1}+f_1g_1},
\end{equation}
which is regular at $f_1=0$, namely, $(f,g)=(v_1, \frac{1}{v_1})$ and $\of$ is determined uniquely in terms of $(f_1,g_1)$.
In this way, we see that the rational mapping $T$ is promoted to a regular mapping on the surface $X$ obtained from 
$\mathbb{P}^1\times\mathbb{P}^1$ by blowing-up at the eight points of indeterminacy. This process 
is illustrated in Fig.\ref{fig:qPE6_surface}.
The eight points of indeterminacy are on the curve
\begin{equation}
C:\quad fg(fg-1)=0,
\end{equation}
and the points on $D=C\setminus \{{\rm P}_1,\ldots,{\rm P}_8\}$ are inaccessible in a similar sense
in Section \ref{subsubsection:blow-up}, in other words, $D$ itself is stable by $T$\footnote{The
divisor $D$ is {\em inaccessible} in the sense that it cannot be reached from outside irrespective
of parameters. For special values of parameters, it may happen that some other divisors are
inaccessible, such as {\em invariant divisors} on which hypergeometric solutions exist. See Section
\ref{sec:hyper}.}.  This can be
verified by computing the image of the points on $D$:
\begin{enumerate}
 \item ${\displaystyle (f,0)\longmapsto (\of,\overline{\mathstrut g})=\left(\frac{\kappa_1^2}{qv_7v_8}\frac{1}{f}, 
\frac{qv_7v_8}{\kappa_1^2}f\right)}$,\quad $\of\overline{\mathstrut g}=1$,
 \item ${\displaystyle (0,g)\longmapsto  (\of,\overline{\mathstrut g})=\left(\frac{1}{g},0\right)}$,\quad $\og=0$,
 \item ${\displaystyle \left(f,\frac{1}{f}\right)\longmapsto  
(\of,\overline{\mathstrut g})=\left(0,\frac{v_5v_6}{q\kappa_2^2}f\right)}$,\quad $\of=0$,
\end{enumerate}
which implies that if $P=(f,g)\in D$, then $\overline{P}=(\of,\og)\in D$.  Accordingly, if $P\in X\setminus D$
then $\overline{P}\in X\setminus D$.

The inaccessible curve (divisor) $D$ has three components $f=0$, $g=0$ and $fg-1=0$.  In the terminology of
the Picard lattice, the divisor $D$ and its three components correspond to
\begin{equation}
\begin{split}
& \delta=2H_1 + 2H_2 - E_1 -E_2-E_3-E_4-E_5-E_6 -E_7 - E_8,  \\
&  \delta_1 = H_1 - E_7-E_8,\quad \delta_2 = H_2 - E_5-E_6,\quad \delta_0 = H_1+H_2-E_1-E_2-E_3-E_4,
\end{split}
\end{equation}
respectively, so that $\delta=\delta_0+\delta_1+\delta_2$. Then the intersection numbers among
$\delta_0$, $\delta_1$ and $\delta_2$ are calculated in a similar manner to Section
\ref{subsection:Picard} as
 \begin{equation}\label{eqn:qp_e6_surface_Cartan}
-\left[\delta_i\cdot\delta_j\right]_{i,j=0,\ldots,2}=
\left[
\begin{array}{ccc}
 2& -1& -1\\
-1 & 2 & -1\\
-1 & -1 &2
\end{array}
\right],
\end{equation}
which is the Cartan matrix of type $A_2^{(1)}$. In this sense, the surface $X$ associated with the
difference equation \eqref{eqn:q-e6} is of type $A_2^{(1)}$. As explained in Section
\ref{subsection:P4_surface_type}, the symmetry type of \eqref{eqn:q-e6} is given by the
orthogonal complement in $\Lambda$, which is $E_6^{(1)}$.
%
\subsection{Point configuration for a discrete Painlev\'e I equation: second example}\label{subsec:point_configuration1}
When we are given a discrete equation to study, we first try appropriate integrability
criteria, such as the singularity confinement test or degree growth criterion (the algebraic
entropy) \cite{Bellon-Viallet:entropy,GR:review2004}.  If the test suggests the equation to be possibly integrable, we may
next try to detect the point configuration for further investigation. As an example of such cases,
we consider the difference equation
\begin{equation}
 x_{n+1} + x_{n-1} = \frac{n\delta + a_0}{x_n} + b. \label{eqn:dp1-alternate}
\end{equation}
In this case, we need more elaborate investigation compared with the previous example. 
\begin{rem}\rm
Equation \eqref{eqn:dp1-alternate} is derived in \cite{RG:coales} and it is shown to have a
continuous limit to the Painlev\'e I equation. Generalizing the equation through the singularity
confinement test \cite{GPR:dressing1997}, \eqref{eqn:dp1-alternate} is interpreted as a B\"acklund
transformation of the Painlev\'e V equation (P$_{\rm V}$)(see, for example
\cite{Ohta_RIMS:1999,Tokihiro-Grammaticos-Ramani:2002}), which will be also demonstrated later.
\end{rem}
 
Introducing the four variables $f$, $g$, $\overline{f}$, $\overline{\mathstrut g}$ by
$(f,g)=(x_n,x_{n+1})$, $(\overline{f},\overline{\mathstrut g})=(x_{n+1},x_{n+2})$, we consider the rational mapping
\begin{equation}\label{eqn:dp1-alternate-1}
\begin{split}
&F:\ (a,b,f,g)\mapsto (\overline{a},b,\of,\overline{\mathstrut g}), \\
&\of = g,\quad \og = -f + \frac{a}{g} + b,\quad \overline{a}=a+\delta.
\end{split} 
\end{equation}
We also consider the inverse $G=F^{-1}$ of the mapping $F$:
\begin{equation}\label{eqn:dp1-alternate-1:inverse}
\begin{split}
&G:\ (\overline{a},b,\overline{f},\overline{\mathstrut g})\mapsto (a,b,f,g), \\
&f=-\og+\frac{\overline{a}-\delta}{\of}+b,\quad g=\of,\quad a=\overline{a}-\delta.
\end{split} 
\end{equation}
Denoting by $X$ and $X'$ two copies of $\mathbb{P}^1\times \mathbb{P}^1$ with inhomogeneous coordinates
$(f,g)$ and $(\overline{f},\overline{\mathstrut g})$ respectively, we attempt to regularize two rational mappings
$F:X\to X'$ and $G:X'\to X$.

\begin{figure}[ht]
\begin{center}
\includegraphics[scale=0.5]{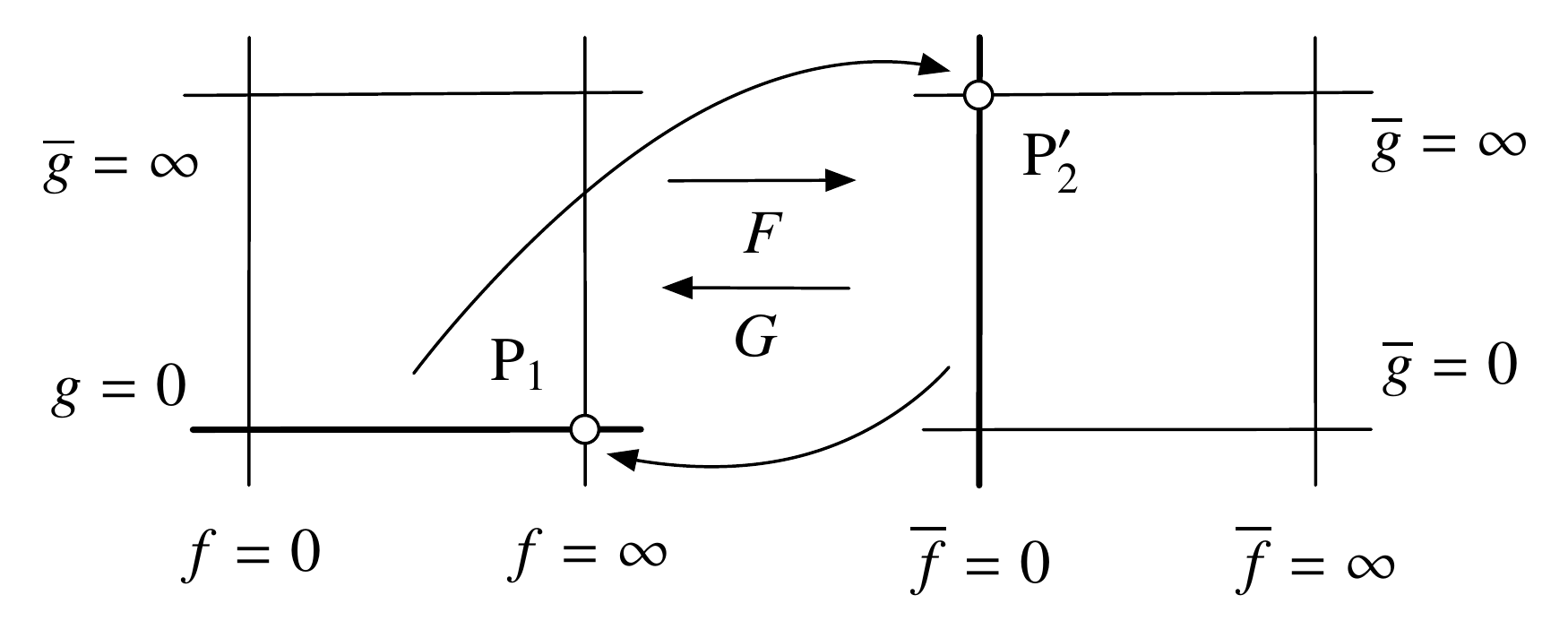}
\end{center}
\caption{Points of indeterminacy of $F$ and $G$}\label{fig:F_and_G}
\end{figure}

The points of indeterminacy of the rational mappings \eqref{eqn:dp1-alternate-1} and
\eqref{eqn:dp1-alternate-1:inverse} can be read off easily. In fact, ${\rm P}_1:(f,g)=(\infty,0)$
and ${\rm P}'_2:(\overline{f},\overline{\mathstrut g})=(0,\infty)$ are the points of indeterminacy of $F$ and
$G$ respectively.  Then we have regular mappings $F: X\setminus {\rm P}_1\to X'$ and 
$G: X'\setminus {\rm P}'_2\to X$.  We observe from \eqref{eqn:dp1-alternate-1} and
\eqref{eqn:dp1-alternate-1:inverse} that $F$ maps the line $\{g=0\}\setminus {\rm P}_1$ to 
${\rm P}'_2$, and similarly $G$ does the line $\{\overline{f}=0\}\setminus {\rm P}'_2$ to ${\rm P}_1$ 
(See Figure \ref{fig:F_and_G}). Therefore $F$ induces a biregular mapping from $X\setminus \{g=0\}$ to
$X'\setminus \{\overline{f}=0\}$. To summarize, we have the following correspondence:
\begin{equation}
\begin{array}{ccc}\medskip
X\setminus \{g=0\} &\stackrel{\sim}{\leftrightarrow} & X'\setminus \{\overline{f}=0\}\\
\medskip
\{g=0\}\setminus {\rm P}_1 &\rightarrow  & {\rm P}'_2\\
{\rm P}_1&  \leftarrow         & \{\overline{f}=0\}\setminus {\rm P}'_2
\end{array}
\end{equation}

\begin{rem}\rm 
A formal method to find the points of indeterminacy of the rational mapping \eqref{eqn:dp1-alternate-1} and
\eqref{eqn:dp1-alternate-1:inverse} is to introduce the homogeneous coordinates of
$\mathbb{P}^1\times\mathbb{P}^1$ as $f=\frac{x_1}{x_0}$, $g=\frac{y_1}{y_0}$, which yields
\begin{align}
F:\ &\frac{\overline{x}_1}{\overline{x}_2}=\frac{y_1}{y_0},\quad
 \frac{\overline{y}_1}{\overline{y}_0}= \frac{ax_0y_0 + bx_0y_1-x_1y_1}{x_0y_1},\label{eqn:dp1-alternate-2} \\
G:\ 
&\frac{x_1}{x_0}=\frac{a\overline{x}_0\overline{y}_0+b \overline{x}_1\overline{y}_0-\overline{x}_1\overline{y}_1}
{\overline{x}_1\overline{y}_0},\quad
\frac{y_1}{y_0}=\frac{\overline{x}_1}{\overline{x}_0}.
\end{align}
The numerator and the denominator of the right hand side of the second equation of
\eqref{eqn:dp1-alternate-2} have a common zero when $x_0=0$, $y_1=0$, which implies that the point
of indeterminacy of $F$ is ${\rm P}_1:\, (f,g)=(\infty,0)$. Similarly, the point of indeterminacy of
$G$ is ${\rm P}'_2:\, (\overline{f},\overline{\mathstrut g})=(0,\infty)$.  
\end{rem}

To resolve the indeterminacy, we blow up $X$ at ${\rm P}_1$ and $X'$ at ${\rm P}'_2$ by introducing
new variables $(f_1,g_1)$, $(\overline{f}_2,\overline{\mathstrut g}_2)$ and their companions
$(\varphi_1,\psi_1)$, $(\overline{\mathstrut \varphi}_2,\overline{\psi}_2)$ as
\begin{equation}
(f,g) = \Bigl(\frac{1}{f_1},f_1g_1\Bigr) = \Bigl(\frac{1}{\varphi_1\psi_1},\psi_1\Bigr),\quad
(\overline{f},\overline{\mathstrut g}) = \Bigl(\overline{f}_2\overline{\mathstrut g}_2,\frac{1}{\overline{g}_2}\Bigr)
= \Bigl(\overline{\varphi}_2,\frac{1}{\overline{\mathstrut \varphi}_2\overline{\psi}_2}\Bigr).
\end{equation}
We denote by $X_{(1)}$ the surface obtained from $X$ by blowing up at ${\rm P}_1$ 
and by $E_1=\{f_1=0\}\cup\{\psi_1=0\}$ the corresponding exceptional curve. Similarly, 
we denote by $X'_{(2)}$  and $E'_2=\{\overline{g}_2=0\}\cup\{\overline{\varphi}_2=0\}$ 
the surface obtained from $X'$ by blowing up at ${\rm P}'_2$
and the corresponding exceptional curve respectively. In terms of the local coordinates
$(f_1,g_1)$ and $(\overline{f}_2,\overline{\mathstrut g}_2)$ we have
\begin{equation}
F:\ \left\{
\begin{array}{l}
{\displaystyle \overline{f}_2=a+bg-fg=a+bf_1g_1-g_1,}
\\
{\displaystyle \overline{g}_2=\frac{g}{a+bg-fg}=\frac{f_1g_1}{a+bf_1g_1-g_1}.}
\end{array}\right.\label{eqn:dp1-alternate-3}
\end{equation}
This mapping $F$ still has an indeterminacy at the point ${\rm P}_3:(f_1,g_1)=(0,a)$ on $E_1$.  We
see that $F$ maps all the points $(f,g)=(f,0)$ $(f\neq\infty)$ on $\{g=0\}\setminus E_1$ to 
${\rm P}'_4: (\overline{f}_2,\overline{\mathstrut g}_2)=(a,0)$ on $E'_2$. The points $(f_1,g_1)=(0,g_1)$ on $E_1$
except ${\rm P}_3$ ($g_1=a$) are mapped to the points $(\overline{f}_2,\overline{\mathstrut g}_2)=(a-g_1,0)$
except ${\rm Q}'_2: (\overline{f}_2,\overline{\mathstrut g}_2)=(0,0)$. It will be shown later that ${\rm Q}'_2$
is actually a regular point.

To investigate the mapping $G$, solving \eqref{eqn:dp1-alternate-3} in terms of $(f_1,g_1)$ we have
\begin{equation}\label{eqn:dp1-alternate-4}
G:\ f_1=\frac{\overline{f}}{a+b\overline{f}-\overline{f}\overline{\mathstrut g}}
=\frac{\overline{f}_2\overline{\mathstrut g}_2}{a+b\overline{f}_2\overline{\mathstrut g}_2-\overline{f}_2},
\quad
g_1=a+b\overline{f}-\overline{f}\overline{\mathstrut g}
=a + b\overline{f}_2\overline{\mathstrut g}_2-\overline{f}_2,
\end{equation}
from which we see that the mapping $G$ has an indeterminacy at ${\rm P}'_4:
(\overline{f}_2,\overline{\mathstrut g}_2)=(a,0)$ on $E'_2$. We observe that $G$ maps all the points
$(\overline{f},\overline{\mathstrut g})=(0,\overline{g})$ $(\overline{g}\neq\infty)$ on
$\{\overline{f}=0\}\setminus E'_2$ to ${\rm P}_3$ on $E_1$. The points
$(\overline{f}_2,\overline{\mathstrut g}_2)=(\overline{f}_2,0)$ on $E'_2$ except ${\rm P}'_4$
($\overline{f}_2=a$) are mapped to the points $(f_1,g_1)=(0,a-\overline{f}_2)$ except ${\rm Q}_1:
(f_1,g_1)=(0,0)$.

\begin{figure}[ht]
\begin{center}
 \includegraphics[scale=0.5]{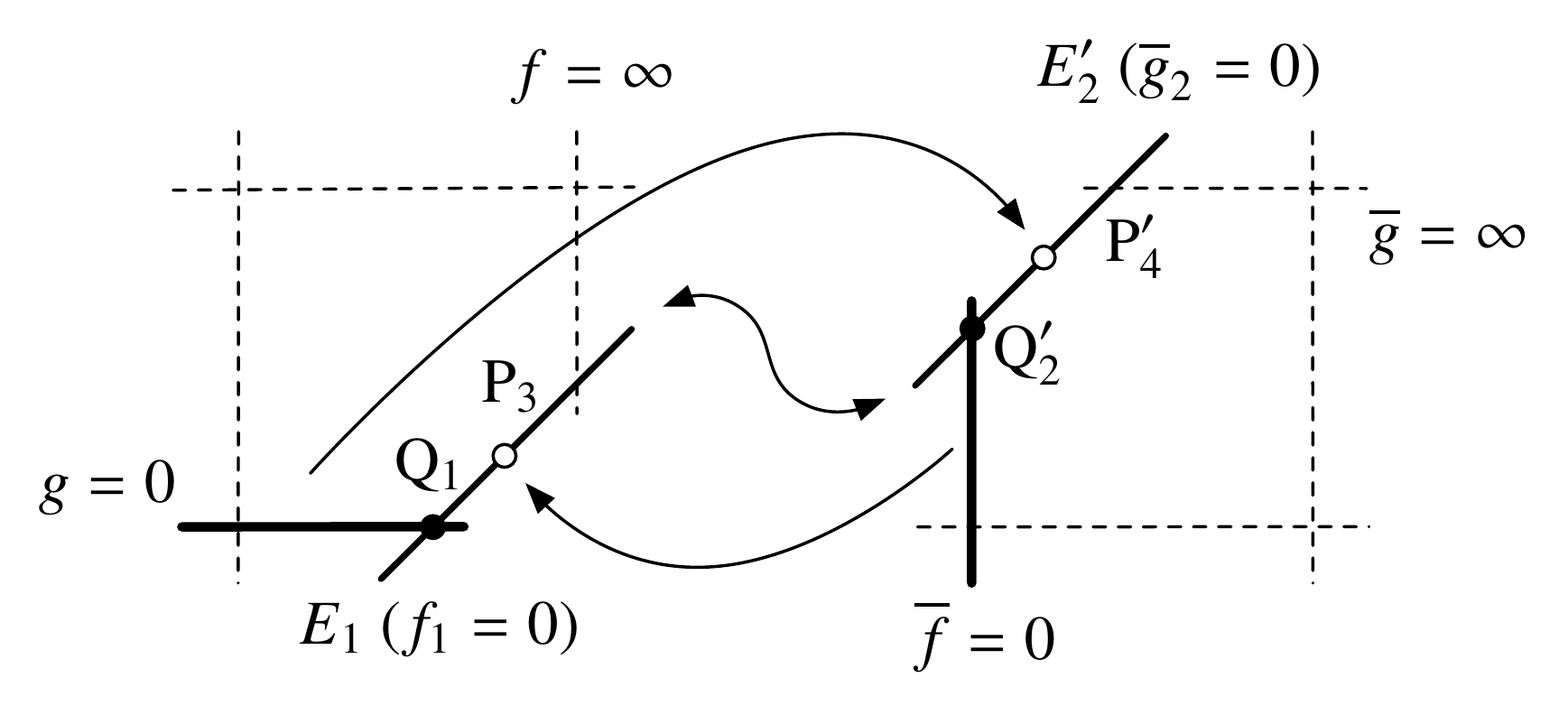}
\end{center}
\caption{The second step of regularization.}\label{fig:2nd_step}
\end{figure}

In summary, at this stage the above investigation shows the following correspondence between $X_{(1)}$ and $X'_{(2)}$
(see Figure \ref{fig:2nd_step}):
\begin{equation}
\begin{array}{ccc}
{\displaystyle  X_{(1)}} &  & {\displaystyle X'_{(2)} }\\
\hline\\[-2mm]
{\displaystyle \{g=0\}\setminus E_1}& {\displaystyle \to} & {\rm P}'_4\\
{\displaystyle E_1 \setminus {\rm P}_3}& {\displaystyle \stackrel{\sim}{\to}} & {\displaystyle E'_2\setminus {\rm Q}'_2}\\
{\rm P}_3& {\displaystyle \leftarrow} & {\displaystyle \{\overline{f}=0\}\setminus E'_2}\\
{\displaystyle E_1\setminus {\rm Q}_1}& {\displaystyle \stackrel{\sim}{\leftarrow}} & {\displaystyle E'_2 \setminus {\rm P}'_4}
\end{array}
\end{equation}
We remark that $E_1\setminus ({\rm P}_3\cup {\rm Q}_1)$ is mapped bijectively to $E'_2\setminus ({\rm P}'_4\cup {\rm Q}'_2)$.

\begin{figure}[ht]
\begin{center}
\includegraphics[scale=0.5]{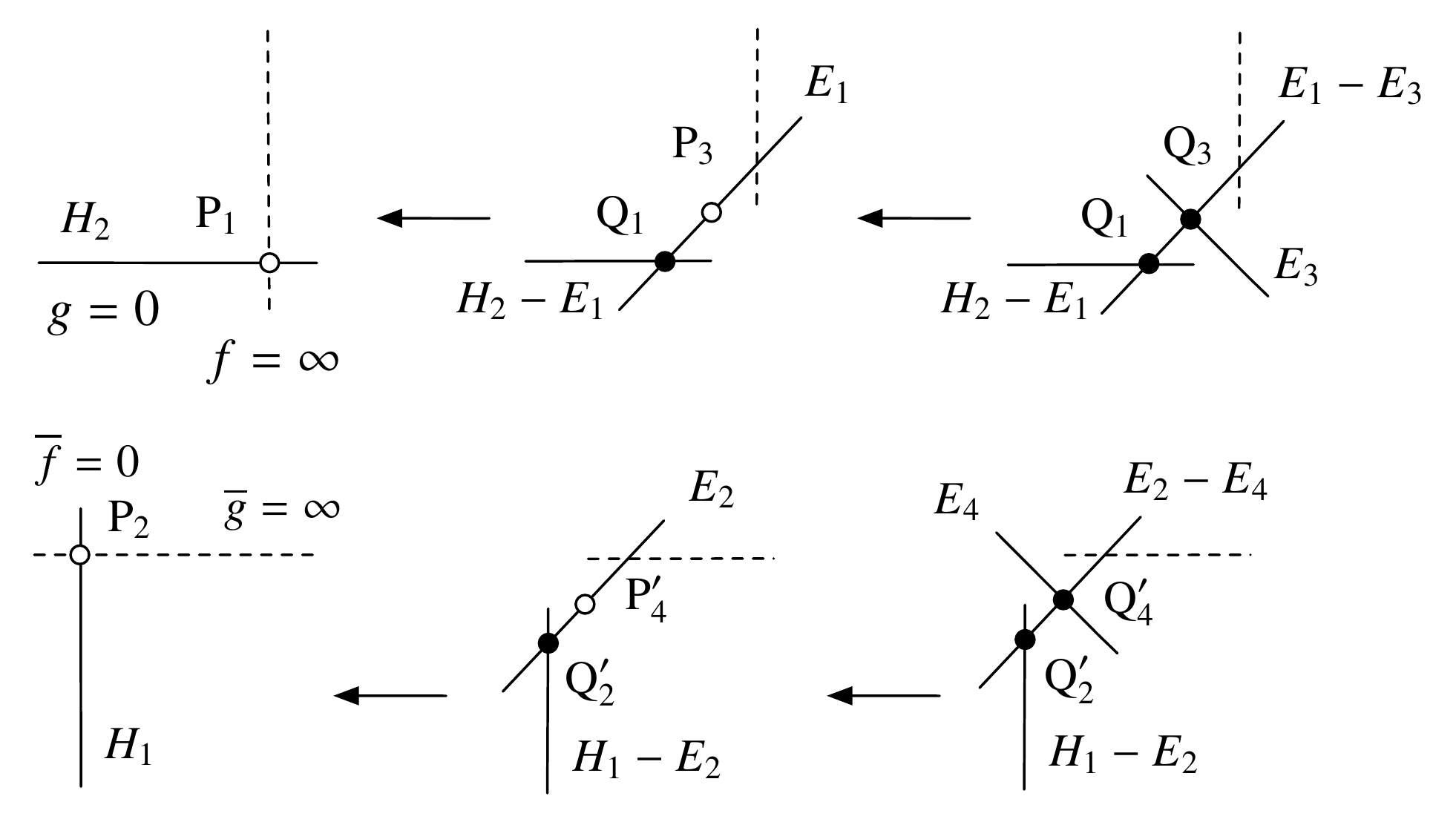}
\end{center}
\caption{Process of blowing up.} \label{fig:A3_process_blowup}
\end{figure}

\begin{figure}[ht]
\begin{center}
\includegraphics[scale=0.5]{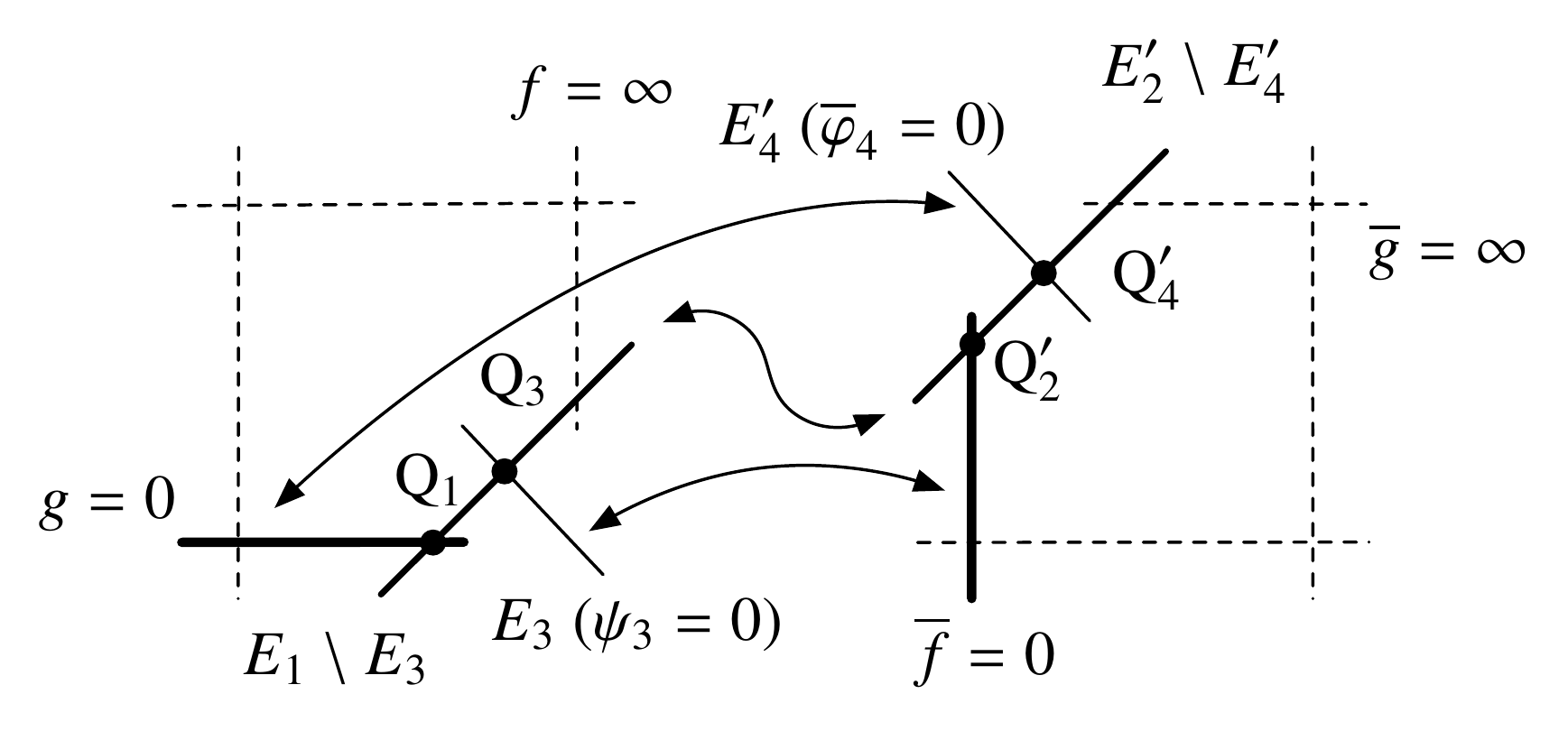}
\end{center}
\caption{The third step of regularization.}\label{fig:3rd_step}
\end{figure}

To resolve the above indeterminacies, we take the blow-up $X_{(13)}$ of $X_{(1)}$ at ${\rm P}_3$
by introducing the variables $(f_3,g_3)$ and its companion $(\varphi_3,\psi_3)$ by
\begin{equation}\label{eqn:f3g3}
(f_1,g_1) = (f_3,a+f_3g_3) = (\varphi_3\psi_3, a+\psi_3),
\end{equation}
and denote by $E_3=\{f_3=0\}\cup\{\psi_3=0\}$ the corresponding exceptional curve.  Similarly, we consider the
blow-up $X'_{(24)}$ of $X'_{(2)}$ at ${\rm P}'_4$ by introducing the variables
$(\overline{f}_4,\overline{\mathstrut g}_4)$ and $(\overline{\varphi_4},\overline{\psi}_4)$ by
\begin{equation}\label{eqn:f4g4}
 (\overline{f}_2,\overline{\mathstrut g}_2) = (a+\overline{f}_4\overline{\mathstrut g}_4,\overline{\mathstrut g}_4)
= (a+\overline{\mathstrut \varphi}_4,\overline{\mathstrut \varphi}_4\overline{\psi}_4),
\end{equation}
and denote by $E'_4=\{\overline{g}_4=0\}\cup\{\overline{\varphi}_4=0\}$ the corresponding exceptional curve.  
The process of blowing-up so far is summarized in Figure \ref{fig:A3_process_blowup}.
We first investigate the image of $\{g=0\}\setminus E_1$ in terms of the local coordinates
$(\overline{\mathstrut \varphi}_4,\overline{\psi}_4)$. We have from \eqref{eqn:dp1-alternate-3} and \eqref{eqn:f4g4}
\begin{equation}
F:\  \overline{\varphi}_4 = (b-f)g,\quad \overline{\psi}_4 = \frac{1}{(b-f)(a+bg-fg)},
\end{equation}
which implies that $F$ maps all the points $(f,g)=(f,0)$ $(f\neq\infty)$ on $\{g=0\}\setminus E_1$
to $(\overline{\mathstrut \varphi}_4,\overline{\psi}_4)=(0,1/a(b-f))$ 
on $E'_4\setminus {\rm Q}'_4$ where ${\rm Q}'_4:(\overline{\mathstrut \varphi}_4,\overline{\psi}_4)=(0,0)$. 
We next investigate the image of $\{\overline{f}=0\}\setminus E_2'$ by $G$ in terms of 
the local coordinates $(\varphi_3,\psi_3)$. We have from \eqref{eqn:dp1-alternate-4} and \eqref{eqn:f3g3}
\begin{equation}
 G:\ \varphi_3 = \frac{1}{(b-\overline{g})(a+b\overline{f}-\overline{f}\overline{\mathstrut g})},
\quad
\psi_3=\overline{f}(b-\overline{\mathstrut g}),
\end{equation}
which implies that $G$ maps all the points $(\overline{f},\overline{\mathstrut g})=(0,\overline{g})$
$(\overline{g}\neq\infty)$ on $\{\overline{f}=0\}\setminus E'_2$ to
$(\varphi_3,\psi_3)=(1/a(b-\overline{g}),0)$ on $E_3\setminus {\rm Q}_3$ where ${\rm Q}_3:(\varphi_3,\psi_3)=(0,0)$. 
Now the correspondence between $X_{(13)}$ and $X'_{(24)}$ is given by
\begin{equation}
\begin{array}{ccc}
{\displaystyle  X_{(13)}} &  & {\displaystyle X'_{(24)} }\\
\hline\\[-2mm]
{\displaystyle \{g=0\}\setminus E_1}& {\displaystyle \stackrel{\sim}{\to}} & {\displaystyle E'_4\setminus {\rm Q}'_4}\\
{\displaystyle E_1 \setminus (E_3\cup {\rm Q}_1)}& {\displaystyle \stackrel{\sim}{\to}} & 
{\displaystyle E'_2\setminus (E'_4 \cup {\rm Q}'_2)}\\
E_3\setminus {\rm Q}_3& {\displaystyle \stackrel{\sim}{\leftarrow}} & {\displaystyle \{\overline{f}=0\}\setminus E'_2}
\end{array}
\end{equation}
We investigate the mapping around ${\rm Q}_3$. 
We write the image of ${\rm Q}_3$ in terms of the
local coordinates $(\overline{f}_2,\overline{\mathstrut g}_2)$ on $E'_2$. We have from
\eqref{eqn:dp1-alternate-3} and \eqref{eqn:f3g3}
\begin{equation}
F:\ \overline{f}_2  = (b\varphi_3(a+\psi_3)-1)\psi_3,\quad \overline{g}_2 = \frac{\varphi_3(a+\psi_3)}{b\varphi_3(a+\psi_3)-1},
\end{equation}
which implies that $F$ is regular at ${\rm Q}_3$ and it is mapped to ${\rm Q}'_2$. 
We can verify by similar computations that ${\rm Q}_1$ is mapped regularly to ${\rm Q}'_4$ as
well. The correspondence between $X_{(13)}$ and $X'_{(24)}$ is now described as (see Figure \ref{fig:3rd_step})
\begin{equation}\label{eqn:correspondence_X13_X24}
\begin{array}{ccc}
{\displaystyle   X_{(13)}} &  & {\displaystyle X'_{(24)} }\\
\hline\\[-2mm]
{\displaystyle \left(\{g=0\}\setminus E_1\right)\cup {\rm Q}_1}& {\displaystyle \stackrel{\sim}{\leftrightarrow}} & E'_4\\
{\displaystyle \left(E_1 \setminus E_3\right)}\cup {\rm Q}_3& {\displaystyle \stackrel{\sim}{\leftrightarrow}} 
& {\displaystyle (E'_2\setminus E'_4)\cup {\rm Q}'_4}\\
{\displaystyle E_3} & {\displaystyle \stackrel{\sim}{\leftrightarrow}} & 
{\displaystyle \left(\{\overline{f}=0\}\setminus E'_2\right)\cup {\rm Q}'_2}\\
\end{array}
\end{equation}
Therefore $F$ induces a biregular mapping $X_{(13)}\stackrel{\sim}{\to} X'_{(24)}$.

In order to consider the iteration of $F$, we identify $X'$ with $X$ everytime after we apply $F:X\to X'$. 
Accordingly, we need to blow up $X_{(13)}$ successively at
\begin{equation}
\begin{array}{ll}
{\displaystyle {\rm P}_2:\ \ (f,g)=(0,\infty)},& {\displaystyle (f,g)=(f_2g_2,\frac{1}{g_2}),}\\
{\displaystyle {\rm P}_4:\ \ (f_2,g_2)=(a-\delta,0),}&{\displaystyle (f_2,g_2)=(a-\delta+f_4g_4,g_4),}
\end{array}
\end{equation}
where ${\rm P}_2$ and ${\rm P}_4$ are copies of ${\rm P}'_2\in X'$ and ${\rm P}'_4\in X'_{(2)}$
respectively. For simplicity, we show one of the two coordinate systems only for each blow-up.  Note
that the parameter $a$ is downshifted to $a-\delta$ in ${\rm P}_4$ when identifying $X'{}$ with
$X$. Tracing the orbit of ${\rm P}_2$ by $F:X_{(13)}\to X'{}_{(24)}$ and applying the same
procedure, we find two additional points ${\rm P}_5$, ${\rm P}_6$ where we need to apply blow-ups:
\begin{equation}\label{eqn:P2_orbit}
\begin{array}{lll}
{\displaystyle F({\rm P}_2)={\rm P}'_5:} & {\displaystyle {\rm P}_5:\ (f,g)=(\infty,b),}&{\displaystyle (f,g)=(\frac{1}{f_5},b+f_5g_5),}\\
{\displaystyle F({\rm P}_5)={\rm P}'_6:} & {\displaystyle {\rm P}_6:\ (f,g)=(b,\infty),}&{\displaystyle (f,g)=(b+f_6g_6,\frac{1}{g_6}),}\\
{\displaystyle F({\rm P}_6)={\rm P}'_1:}& {\displaystyle {\rm P}_1:\ (f,g)=(\infty,0).} & 
\end{array}
\end{equation}
Writing the final step in terms of the local coordinates $(f_6,g_6)$ on $E_6$ and
$(\overline{f}_1,\overline{\mathstrut g}_1)$ on $E'_1$ as
\begin{equation}
 F:\ \overline{f}_1 = g_6,\quad \overline{g}_1=a-f_6,
\end{equation}
we can check that $E_6$ is mapped to $E'_1$, which guarantees the regularity of
the iterated mapping on this orbit.

We next trace the orbit of ${\rm P}_4$ on the exceptional curve $E_2=\{g_2=0\}$ to find the two points
${\rm P}_7$ on $E_5=\{f_5=0\}$ and ${\rm P}_8$ on $E_6=\{g_6=0\}$:
\begin{equation}\label{eqn:P4_orbit}
\begin{array}{lll}\smallskip
{\displaystyle F({\rm P}_4)={\rm P}'_7:} &  {\displaystyle {\rm P}_7:\ (f_5,g_5)=(0,\delta),}&{\displaystyle (f_5,g_5)=(f_7,\delta+f_7g_7),}\\
\smallskip
{\displaystyle F({\rm P}_7)={\rm P}'_8:}& {\displaystyle {\rm P}_8:\ (f_6,g_6)=(-\delta,0), }&{\displaystyle (f_6,g_6)=(-\delta+f_8g_8,g_8),}
\\
{\displaystyle F({\rm P}_8)={\rm P}'_3:}& {\displaystyle {\rm P}_3:\ (f_1,g_1)=(0,a).}
\end{array}
\end{equation}
We can verify the regularity of the iterated mapping by checking that $E_8$ is mapped to $E'_3$. We
thus obtain an eight-point blow-up $X_{(12563478)}$ of $X$ by four blow-ups at ${\rm P}_1$, ${\rm
P}_2$, ${\rm P}_5$, ${\rm P}_6$ followed by four blow-ups at ${\rm P}_3$, ${\rm P}_4$, ${\rm
P}_7$, ${\rm P}_8$, on which $F$ induces a biregular mapping.  
\begin{figure}[ht]
\begin{center}
\includegraphics[scale=0.4]{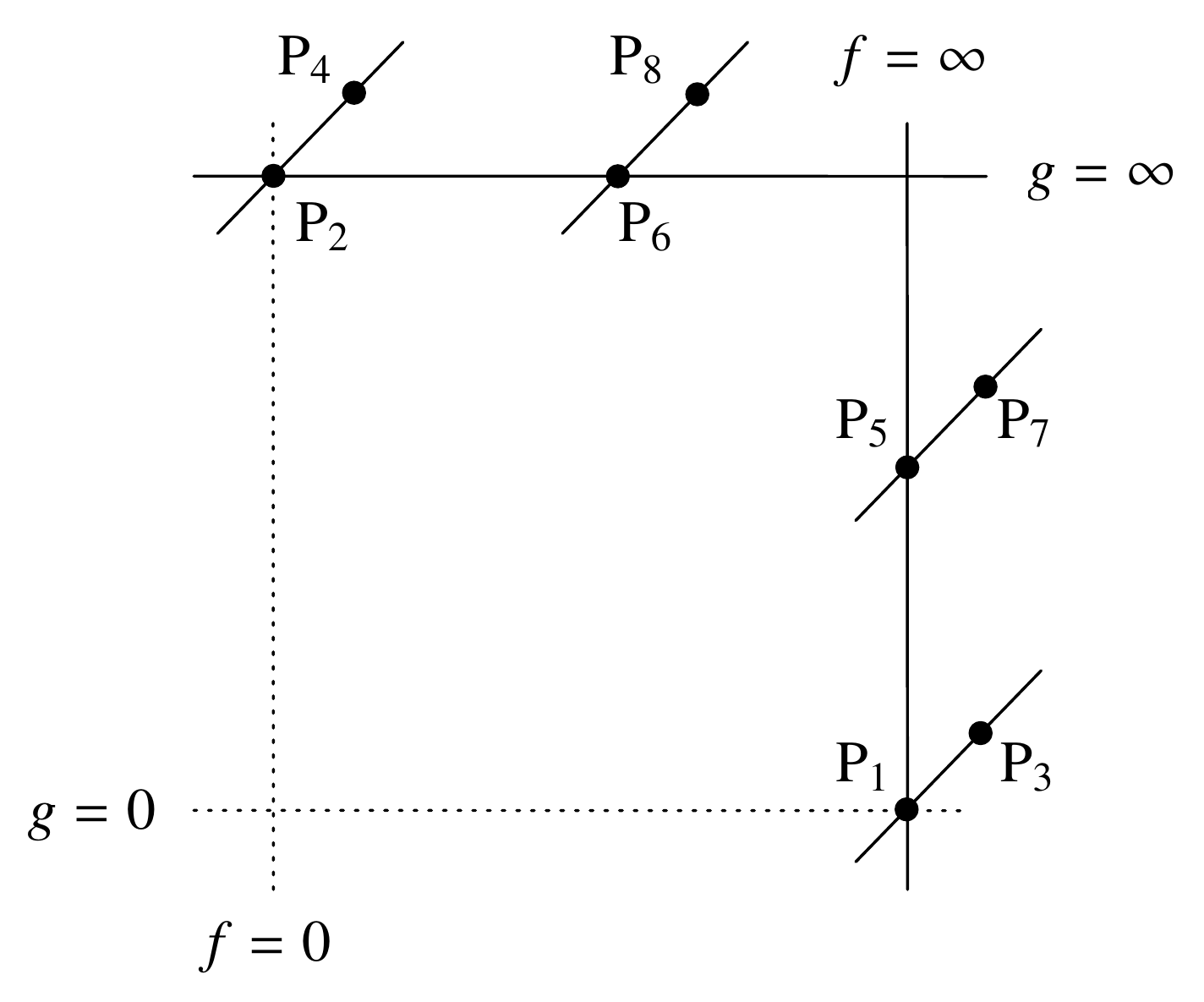}\hskip30pt \includegraphics[scale=0.4]{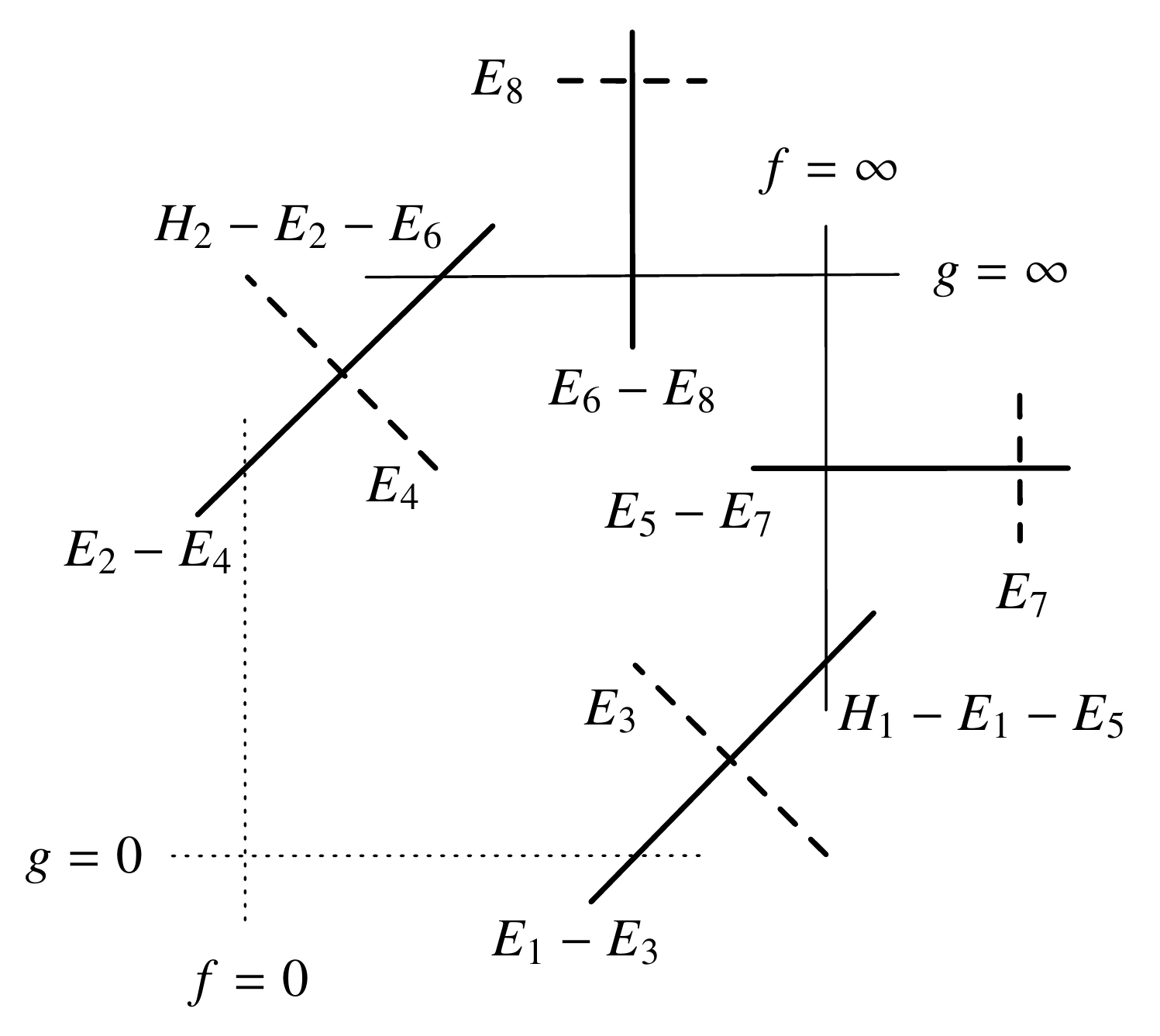}\\
\end{center}
\caption{Surface of the mapping \eqref{eqn:dp1-alternate-1}. Left: Configuration of the points of indeterminacy. Right: Configuration of 
divisors. Solid lines are inaccessible divisors. Thick lines are exceptional divisors arising in the blow-ups.}
\label{fig:dA3_surface}
\end{figure}
%
\begin{rem}\rm \hfill
\begin{enumerate}
 \item In this example, only ${\rm P}_1$ and ${\rm P}_2$ arise as the indeterminacies of $F$ and $F^{-1}$
respectively.  The other points appear as the indeterminacies of iterate $F^4$. Such a phenomenon
occurs when the mapping is finer than translations of the underlying root lattice (``{\em
projective reduction}''), as shown below \cite{KNT:projective,Takenawa:qp5}.
\item Here we have constructed the space of initial values by successive blow-ups. Some examples
to which we need to apply blow-downs are investigated in \cite{Carstea-Takenawa}.
\end{enumerate}
\end{rem}

Let us identify the underlying root lattice from the configuration of the eight points ${\rm
P}_1,\ldots,{\rm P}_8$: ${\rm P}_1$ and ${\rm P}_5$ are on the line $f=\infty$.  ${\rm P}_2$ and
${\rm P}_6$ are on the line $g=\infty$. ${\rm P}_3$, ${\rm P}_4$, ${\rm P}_7$, ${\rm P}_8$ are
infinitely near point of ${\rm P}_1$, ${\rm P}_2$, ${\rm P}_5$, ${\rm P}_6$, respectively.  We
then identify the surface type of $X_{(12563478)}$ as follows. We note that the above
set-theoretical computations tracing the orbits of points are dual to our convention of the symbolic
computations on the Picard lattice (recall the difference of the symbolical composition and the
numerical composition). Hence the action of discrete time evolution $T$ on the Picard lattice is
obtained as follows from the set-theoretical action of $G$. Denoting by $H_1$ and $H_2$ the divisors
corresponding to the lines $f={\rm const.}$ and $g={\rm const.}$ respectively, we have from
\eqref{eqn:correspondence_X13_X24}
\begin{align}
T:\ E_4 \to H_2-E_1  ,\quad E_2-E_4 \to E_1-E_3,\quad H_1-E_2\to E_3 .
\end{align}
We also have from \eqref{eqn:P2_orbit} and \eqref{eqn:P4_orbit} 
\begin{equation}
T:\ E_1\to E_6\to E_5\to E_2,\quad E_3\to E_8\to E_7\to E_4.
\end{equation}
It is obvious from \eqref{eqn:dp1-alternate-1} that $\{f=\infty\}\setminus {\rm P}_1$ is mapped to
$\{g=\infty\}\setminus {\rm P}_2$, which implies 
\begin{equation}
T:\ H_2-E_2 \to H_1-E_1 .
\end{equation}
Combining these formulae we obtain 
\begin{equation}\label{eqn:T_Picard}
T:\quad
\begin{array}{l}
H_1\to H_2,\quad H_2\to H_1+H_2-E_1-E_3,\quad E_2\to H_2-E_3,\\ 
E_4\to H_2-E_1,\quad E_{\{135768\}}\to E_{\{682457\}} . 
\end{array}
\end{equation}

In terms of the Picard lattice, the surface type is described by the following inaccessible root vectors, as
illustrated in Fig.\ref{fig:dA3_surface}. 
\begin{equation}
\begin{split}
& \delta_0 = E_2-E_4,\quad \delta_1 = E_6-E_8,\quad \delta_2 = H_2-E_2-E_6,\\
& \delta_3 = H_1-E_1-E_5, \quad \delta_4 = E_1-E_3,\quad \delta_5 = E_5-E_7.
\end{split}
\end{equation}
In fact, $T$ induces the following transformations on the root vectors
\begin{equation}
 \delta_0\to \delta_4\to \delta_1\to \delta_5\to \delta_0,\quad \delta_2\leftrightarrow\delta_3.
\end{equation}
Note also that the other divisors in Figure \ref{fig:dA3_surface} are accessible which is 
obvious from the tracing of the orbit of indeterminacies. We remark that they generate
the following infinite orbit of divisors by successive applications of $T$:
\begin{equation}
\cdots\to H_1-E_2 \to E_3\to E_8\to E_7\to E_4 \to H_2-E_1\to \cdots.
\end{equation}

The intersection numbers among those components of inaccessible divisors are computed
in a similar manner to Section \ref{subsection:Picard} as
 \begin{equation}\label{eqn:dp1_surface_Cartan}
-\left[\delta_i\cdot\delta_j\right]_{i,j=0,\ldots,5}=
\left[\begin{array}{cccccc}
2 &  &-1&  &  & \\
  &2 &-1&  &  & \\
-1&-1&2 &-1&  & \\
  &  &-1&2 &-1&-1 \\
  &  &  &-1&2 &  \\
  &  &  &-1&  &2 
\end{array}\right],
\end{equation}
which is the Cartan matrix of type $D_5^{(1)}$. It is also expressed as the Dynkin diagram by
\begin{equation}\label{eqn:Dynkin_D_5}
 \begin{array}{ccccccc}
        &                  & \delta_0&                 & \delta_5&                 &       \\
        &                  &   |    &                  &   |    &                  &       \\
\delta_1&\text{\textemdash}&\delta_2&\text{\textemdash}&\delta_3&\text{\textemdash}&\delta_4
 \end{array}
\end{equation}
The corresponding surface is the type $D_5^{(1)}$. Note that the basis of the orthogonal complement
in the Picard lattice is given by
\begin{equation}
\alpha_0 = H_2-E_5-E_7, \quad \alpha_1 = H_1-E_2-E_4,\quad \alpha_2 = H_2-E_1-E_3,\quad \alpha_3 = H_1-E_6-E_8,
\end{equation}
whose intersection numbers are given by
\begin{equation}
-\left[\alpha_i\cdot \alpha_j\right]_{i,j=0,\ldots,3}=
\left[\begin{array}{cccc}
2 &-1&  &-1 \\
-1&2 &-1& \\
  &-1&2 &-1\\
-1&  &-1&2 \\
\end{array}\right].
\end{equation}
This is the Cartan matrix of type $A_3^{(1)}$ and its Dynkin diagram is given by
\begin{equation}
\begin{array}{ccc}
\alpha_0&\text{\textemdash}&\alpha_3\\
    |   &                  &   |    \\
\alpha_1&\text{\textemdash}&\alpha_2
\end{array}
\end{equation}
which implies that the symmetry is type $A_3^{(1)}$. We remark that the action of $T$ on
$\alpha_i$ ($i=0,1,2,3$) is given by
\begin{equation}
\alpha_0\to \alpha_1 + \alpha_2,\quad \alpha_1\to -\alpha_2,\quad \alpha_2\to \alpha_3+\alpha_2,\quad \alpha_3\to \alpha_0.
\end{equation}
The action of the simple reflections $s_i$ ($i=0,1,2,3$) on $\alpha_i$ ($i=0,1,2,3$) can be computed
in the similar manner to Example \ref{example:A2} as
\begin{equation}
 s_i(\alpha_i) = -\alpha_i,\quad s_i(\alpha_{i\pm 1})=\alpha_{i\pm 1}+\alpha_i\quad (i\in\mathbb{Z}/4\mathbb{Z}),
\end{equation}
and the Dynkin diagram automorphism $\pi$
\begin{equation}
 \pi(\alpha_i) = \alpha_{i+1}\quad (i\in\mathbb{Z}/4\mathbb{Z}).
\end{equation}
\begin{rem}\rm\hfill
 \begin{enumerate}
  \item Applying the similar analysis as above to the difference equation \eqref{eqn:dP_A2} or
	\eqref{eqn:dP_A2-2}, one obtains the surface of type $E_6^{(1)}$ which is exactly the same
	as that constructed from P$_{\rm IV}$ in Sections
	\ref{subsubsection:resolution_P4} and \ref{subsection:P4_surface_type}. Namely, the
	differential equation P$_{\rm IV}$ \eqref{eqn:p4_pq} and the difference equation 
\eqref{eqn:dP_A2} or \eqref{eqn:dP_A2-2} share the same surface. 
  \item We can introducing the variables $F_i$ in the similar manner to the case of P$_{\rm IV}$ (the case
of symmetry type $A_2^{(1)}$) such that
\begin{equation}
 s_i(F_{i\pm 1})=F_{i\pm 1} \pm \frac{\alpha_i}{F_i},\ \pi(F_i) = F_{i+1}\ (i\in\mathbb{Z}/4\mathbb{Z}),
\ F_0 + F_2 = k_0,\ F_1+F_3=k_1,
\end{equation}
where $k_0$ and $k_1$ are constants. Then we can verify that
\begin{equation}
 T=\pi s_1,
\end{equation}
under the identification $(F_1,F_2)=(f,g)$, $(\alpha_0,\alpha_1,\alpha_2,\alpha_3)=(\delta,\delta-a,a,\delta)$,
$(k_0,k_1)=(b,b)$.
Also, it is known that this action of the affine Weyl group of type $A_3^{(1)}$ admits a continuous
flow which is nothing but the P$_{\rm V}$. Actually, putting 
$(F_0,F_1,F_2,F_3)=(\frac{p+t}{\sqrt{t}},\sqrt{t}q,-\frac{p}{\sqrt{t}}, \sqrt{t}(1-q))$ under the
normalization $\delta=1$ and $k_0=k_1=\sqrt{t}$, the variables $q$ and $p$
obey the canonical equations with the Hamiltonian
\begin{equation}
 tH=q(q-1)p(p+t) - (\alpha_1+\alpha_3)qp + \alpha_1p + \alpha_2tq,
\end{equation}
which is equivalent to P$_{\rm V}$ 
\begin{equation} \label{eqn:P5}
 \frac{d^2y}{dt^2} = \left(\frac{1}{2y}+\frac{1}{y-1}\right)\left(\frac{dy}{dt}\right)^2 - \frac{1}{t}\frac{dy}{dt}
+ \frac{(y-1)^2}{2t^2}\left(\alpha_1^2 y-\frac{\alpha_3^2}{y}\right)
-(\alpha_2-\alpha_0)\frac{y}{t} - \frac{y}{2}\frac{y+1}{y-1},
\end{equation}
for the variable $y=1-\frac{1}{q}$ \cite{Masuda:hyper,Noumi-Takano-Yamada:BT,Noumi-Yamada:p5,Tsuda-Okamoto-Sakai:folding}.
Therefore, the analysis in this section reveals that the difference equation
\eqref{eqn:dp1-alternate} describes a B\"acklund transformation of P$_{\rm V}$. Further, Takano's
coordinates for the space of initial values of P$_{\rm V}$ correspond to $(f,g)$,
$(f_i,g_i)$ $(i=3,4,7,8)$ \cite{Takano2,Takano1}.
 \end{enumerate}
\end{rem}
%
\subsection{Practical method for finding point configuration}\label{subsec:point_configuration2}
The point configuration for a difference equation of Painlev\'e type, as constructed in Section
\ref{subsec:point_configuration1} for \eqref{eqn:dp1-alternate}, can be obtained by an alternative
``short-cut'' method which is suitable for computer algebra. Instead of repeating blow-ups, we
may iterate the mapping, in other words, the discrete time evolution. Sufficient number of
iterations will give all the point of indeterminacy.

To find the points of indeterminacy of the rational mapping \eqref{eqn:dp1-alternate-1}, we iterate the substitution
\begin{equation}
 T(f) = g,\quad T(g) = -f + \frac{a}{g} + b,\quad T(a)=a+\delta.
\end{equation}
Computing $T^2(g)$, $T^3(g)$, $\ldots$, we have
\begin{equation}
T(g)=\frac{F_1}{g}, \quad
T^2(g)=\frac{F_2}{F_1}, \quad
T^3(g)=\frac{F_3}{g F_2},\ \ldots,\ T^{n}(g)=\frac{F_nF_{n-4}}{F_{n-1}F_{n-3}},\ \ldots,
\end{equation}
where
\begin{equation}
\begin{split}
&F_1=a+b g-f g,\\
&F_2=a b+b^2 g-b f g-b g^2+f g^2+\delta  g,\\
&F_3=-a^2 b+a^2 g-a b^2 g+2 a b f g+2 a b g^2-2 a f g^2+a \delta  g+b^2 f
 g^2-b f^2 g^2\\
&\hskip20pt -b f g^3+2 b \delta  g^2+f^2 g^3-\delta  f g^2,\quad\cdots.
\end{split}
\end{equation}
We observe two points $(f,g)=(b,\infty), (\infty,0)$ appearing as the common zeros of the
polynomials $F_1, F_2$, which give the points of indeterminacy of $T^2(g)$. These points correspond
to ${\rm P}_6$ and ${\rm P}_1$ respectively in Section \ref{subsec:point_configuration1}. To see
infinitely near points to ${\rm P}_6:\ (f,g)=(b, \infty)$, we introduce a parameter $u$ by $(f,g)=(b+u \epsilon,
\frac{1}{\epsilon})$. Then for $\epsilon \rightarrow 0$ we have
\begin{equation}
F_2=\frac{u+\delta}{\epsilon}+O(\epsilon^0),\quad
F_3=\frac{b(u+\delta)}{\epsilon^2}+O(\epsilon^{-1}),
\end{equation}
and hence
\begin{equation}
 T^3(g)=\frac{F_3}{gF_2} = \frac{b(u+\delta)+O(\epsilon^{1})}{u+\delta+O(\epsilon^{1})},
\end{equation}
which indicates another point of indeterminacy at $u=-\delta$, namely 
\begin{equation}\label{eqn:pt_infinitely_near}
{\rm P}_8:\ (f,g)=(b-\delta \epsilon+O(\epsilon^2), \frac{1}{\epsilon}).
\end{equation}
Similarly, near the point ${\rm P}_1:\ (f,g)=(\infty,0)$ we have an infinitely near singular point
\begin{equation}
{\rm P}_3:\ (f,g)=(\frac{1}{\epsilon}, a \epsilon+O(\epsilon^2)).
\end{equation}
Analyzing $T^{-1}(f), T^{-2}(f), \cdots$ in a similar way, we have four more points:
\begin{equation}
\begin{split}
&{\rm P}_5:\ (f,g)=(\infty,b), \quad {\rm P}_7:\ (f,g)=(\frac{1}{\epsilon}, b+\delta \epsilon+O(\epsilon^2)),\\
&{\rm P}_2:\ (f,g)=(0,\infty), \quad {\rm P}_4:\ (f,g)=( (a-\delta) \epsilon+O(\epsilon^2),\frac{1}{\epsilon}). 
\end{split}
\end{equation}
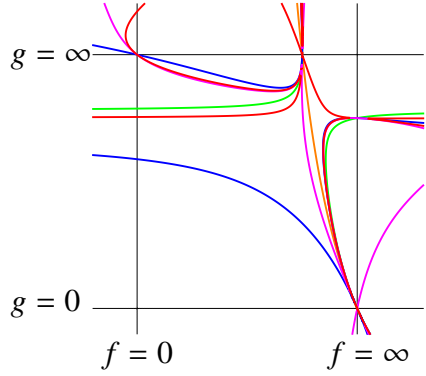
\begin{figure}[ht]
\begin{center}
\setlength{\unitlength}{1mm}
\begin{picture}(60,45)(0,0)
\put(10,5){\includegraphics[scale=0.5]{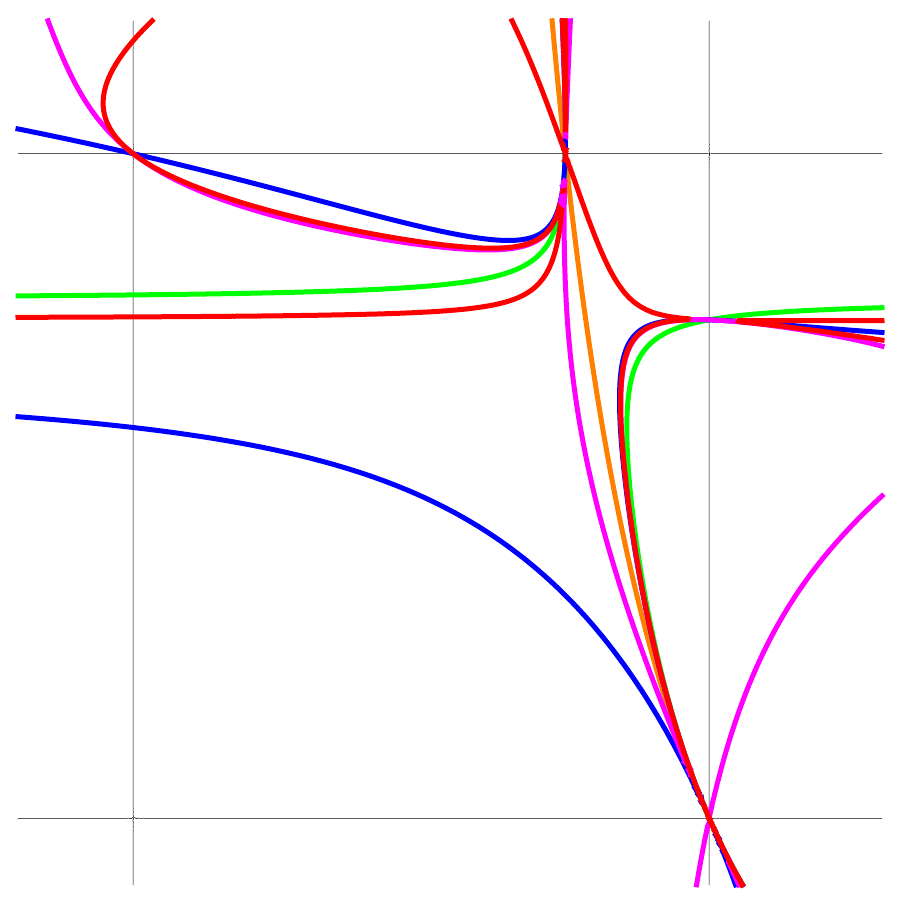} }
\put(0,9){$g=0$}
\put(12,2){$f=0$}
\put(42,2){$f=\infty$}
\put(0,42){$g=\infty$}
\end{picture}
\caption{Curves generated from the iterates of
$T$ with $a=1$, $b=3$, $\delta=2$.  The graph is shown in the $(x,y)$ coordinates where
$f=\frac{x}{1-x}$, $g=\frac{y}{1-y}$ to show the points of infinity. Orange: $F_1=0$, green:
$F_2=0$, blue: $F_3=0$, magenta:$F_4=0$, red: $F_5=0$. }\label{fig:dA3_curves}
\end{center}
\end{figure}

Let us give a brief account of the reason why this method works effectively for finding the point
configuration.  As we have suggested in Section \ref{subsec:P4_Picard}, the polynomial factors such
as $F_n$ are controlled by the Picard lattice. Therefore the singularities (self-intersections) of a
polynomial $F_n$ with sufficient degree, or the common zeros of $F_n$'s would give the points in the
configuration. Figure \ref{fig:dA3_curves} illustrates the behavior of curves $F_n$
$(n=1,\ldots,5)$, from which we observe the four common zeros. Also, we can identify the four
infinitely near points by the common tangent at each of the four common zeros.

\section{Discrete Painlev\'e Equation from Point Configuration}\label{sec:dP}
If the configuration of eight points in $\mathbb{P}^1\times\mathbb{P}^1$ is generic, the
corresponding space of initial values has the largest symmetry of type $E_8^{(1)}$, and other
configurations can be regarded as the degenerate cases. In this Section, we describe how to
construct the equations and relevant characteristic features from the point configuration. In
particular, we formulate a representation of affine Weyl group of type $E_8^{(1)}$ from the
configuration of generic eight points, as well as the formalism of $\tau$ functions. We then derive
a new explicit form of the three equations of type $E_8^{(1)}$, which are the elliptic, $q$- and
difference Painlev\'e equations, from a translation of the root lattice. We also give an example
demonstrating how to construct the birational representation of the affine Weyl group for a given
degenerate point configuration.
\subsection{Configuration of points on $\P^1\times \P^1$}\label{subsec:8points}
Suppose that generic $n$ points ${\rm P}_i\ (x_i,y_i)$ ($i=1,\ldots,n$) in $\P^1\times \P^1$ are
given ($n \geq 4$).  Since $\P^1\times \P^1$ admits the action of ${\rm PGL}(2)^2$ given by the
linear fractional transformations $(x,y)\mapsto \left(\frac{ax+b}{cx+d},\
\frac{a'y+b'}{c'y+d'}\right)$, we choose the inhomogeneous coordinates
\begin{equation}
(f,g)=\left(\frac{(x-x_2)(x_3-x_1)}{(x-x_1)(x_3-x_2)},\ \frac{(y-y_2)(y_3-y_1)}{(y-y_1)(y_3-y_2)}\right),
\end{equation}
of $\P^1\times \P^1$ so that ${\rm P_1}\ (\infty,\infty)$, ${\rm P}_2\ (0,0)$, ${\rm P}_3\ (1,1)$
and ${\rm P}_i\ (f_i,g_i)$ ($i=4,\ldots,n$).  We define birational actions $s_0,\ldots,s_n$ on the
field of rational functions in variables $f_i$, $g_i$ ($i=4,\ldots,n$) by
\begin{equation}\label{eqn:E8_fg}
\begin{split}
 s_0:\quad & f_i\ \rightarrow\  \frac{1}{f_i},\quad  g_i\ \rightarrow\  \frac{1}{g_i},\\
 s_1:\quad & f_i\ \rightarrow\  g_i,\quad  g_i\ \rightarrow\  f_i,\\
 s_2:\quad & f_i\ \rightarrow\  \frac{f_i}{g_i},\quad  g_i\ \rightarrow\  \frac{1}{g_i},\\
 s_3:\quad & f_i\ \rightarrow\  1-f_i,\quad  g_i\ \rightarrow\  1-g_i,\\
 s_4:\quad & 
{\displaystyle f_4\ \rightarrow\  \frac{1}{f_4}},\quad
{\displaystyle g_4\ \rightarrow\  \frac{1}{g_4}},\quad
{\displaystyle f_i\ \rightarrow\ \frac{f_i}{f_4}}, \quad
{\displaystyle g_i\ \rightarrow\ \frac{g_i}{g_4}\quad (i\geq 5)}, \\
 s_j:\quad & f_{j-1}\ \leftrightarrow\  f_j,\quad  g_{j-1}\ \leftrightarrow\  g_j\qquad\qquad (j=5,\ldots,n).
\end{split}
\end{equation}
Then one can verify directly that $s_i$ ($i=0,\ldots,n$) satisfy the fundamental relations specified
by \eqref{eqn:fundamental_relation_Weyl_group} corresponding to the following Dynkin diagram:
\begin{equation}
\begin{array}{cccccccccccccc}
&&&& s_0\\
&&&& \vert\\
s_1&-&s_2&-&s_3&-&s_4&-&\cdots&-&s_{n},\\
\end{array}
\end{equation}
 (e.g. $E_8^{(1)}$ \eqref{eqn:Cartan_E8} for $n=8$).  We remark that the variables $(f_i,g_i)$ ($i=4,\ldots,n$) are regarded as the inhomogeneous coordinates of the 
{\em configuration space}
\begin{equation}
\left\{
\left[\begin{array}{cccc}x_1& x_2&\cdots &x_n \\y_1& y_2&\cdots &y_n \\\end{array}\right]
\right\}\Big/\ {\rm PGL}(2)^2,
\end{equation} 
of generic $n$ points in $\P^1\times \P^1$. The above transformations have simple interpretations
on this configuration space except $s_2$. The action of
$\mathfrak{S}_n=\langle s_0,s_3,\ldots,s_n\rangle$ comes from the permutation of eight
points in $(x,y)$ coordinates, and $s_1$ from the exchange of two coordinates of each point. 

We remark that in the formulation of point configurations in $\mathbb{P}^2$, those transformations
have different geometric meaning; $s_0$ is the standard Cremona transformation and other transformations
correspond to exchange of points or coordinates \cite{KMNOY:point_configuration,Sakai:SIV}.

In the context of the discrete Painlev\'e equations, we consider the case $n=9$, and the coordinates
$(f_i,g_i)$ of the points ${\rm P}_i$ ($i=1,\ldots,8$) play the role of parameters (or independent
variables). The ninth point $(f_9,g_9)=(f,g)$ plays different role (the dependent variable) and we
do not use the action $s_9$.

Calculating actions for some elements $w \in W(E_8^{(1)})$, we observe that $w(f)$ and $w(g)$ are
rational functions in $f$, $g$ with the factorized form
\begin{equation}\label{eqn:factor_wfg}
 w(f)={\rm const.}\frac{P_{w(E_2)}P_{w(H_1-E_2)}}{P_{w(E_1)}P_{w(H_1-E_1)}}, 
\quad  w(g)={\rm const.}\frac{P_{w(E_2)}P_{w(H_2-E_2)}}{P_{w(E_1)}P_{w(H_2-E_1)}},
\end{equation}
where $P_\lambda$ is the polynomial in $f, g$ corresponding to the exceptional element $\lambda \in M
\subset \Lambda=\Z H_1 \oplus \Z H_2 \oplus \Z E_1 \oplus \cdots \oplus\Z E_8$ \eqref{eqn:-1_curve},
which is unique up to normalization constants.
\begin{ex}\label{example:cocycle}\rm
For $w=s_4s_3s_2s_1s_6s_5s_4s_3s_0s_7s_6s_5s_4 s_3s_2$, we have:
\begin{equation}
\begin{split}
w(f)&=\frac{\left(g_7-g\right) \left(f_6 g_4-f g_4-f_4  g_6+f g_6+f_4 g-f_6 g\right)}
{\left(g_6-g\right)  \left(f_7 g_4-f g_4-f_4 g_7+f g_7+f_4 g-f_7  g\right)},\\
w(g)&=\frac{\left(g_7-g_4\right) \left(f_6 g_4 - f  g_4 - f_4 g_6 + f g_6 + f_4 g - f_6  g\right)}
{\left(g_6-g_4\right) \left(f_7 g_4 - f g_4 - f_4 g_7 + f g_7 + f_4 g - f_7 g\right)},
\end{split}
\end{equation}
whose factors are identified as
\begin{equation}
\begin{split}
&P_{w(E_1)}=P_{H_1+H_2 -E_1-E_4-E_7}=f_7 g_4-f g_4-f_4  g_7+f g_7+f_4 g-f_7 g,\\
&P_{w(E_2)}=P_{H_1+H_2 -E_1-E_4-E_6}=f_6 g_4-f g_4-f_4  g_6+f g_6+f_4 g-f_6 g,\\
&P_{w(H_1-E_1)}=P_{H_2-E_6}= g-g_6,\quad 
P_{w(H_1-E_2)}=P_{H_2-E_7}=g-g_7,\\
&P_{w(H_2-E_1)}=P_{E_7}=1,\quad
P_{w(H_2-E_2)}=P_{E_6}=1.
\end{split}
\end{equation}
In fact, one can check that $P_{w(E_1)}=0$ is a bidegree (1,1) curve which passes through ${\rm
P}_1(\infty,\infty)$, ${\rm P}_{4}(f_4,g_4)$ and ${\rm P}_{7}(f_7,g_7)$ with multiplicity $1$.
\end{ex}
For any polynomial $P(f,g)$ belonging to the class $\lambda$, one can show that $s_k(P(f,g))$ is a
polynomial belonging to the class $s_k(\lambda)$ up to multiplication by a monomial factor in $f$,
$g$ for each $k=0,\ldots,8$.  If $\lambda=d_1H_1 + d_2H_2-\sum_{i=1}^8m_iE_i$ then $P(f,g)$ is
expressed in the form
\begin{equation}\label{eqn:factor_1}
 P(f,g)=\sum_{(i,j)\in S}
c_{ij}f^ig^j,\quad S=\left\{(i,j)~\Bigr|~
\begin{array}{c}
0\leq i\leq d_1,\ 0\leq j\leq d_2, \\
 m_2\leq i+j\leq d_1+d_2-m_1
\end{array}
\right\}
\end{equation}
Applying $s_2$, for example, we obtain
\begin{align*}
 s_2(P(f,g))
&=\sum_{(i,j)\in S}
s_2(c_{ij})\left(\frac{f}{g}\right)^i\left(\frac{1}{g}\right)^j
=g^{-d_1-d_2+m_1}
\sum_{(i,j)\in S}
s_2(c_{ij})f^i g^{d_1+d_2-m_1-i-j}\\
&=g^{-d_1-d_2+m_1}Q(f,g),
\end{align*}
where
\begin{equation}\label{eqn:factor_2}
\begin{split}
& Q(f,g)=\sum_{(i,j)\in T}s_2(c_{i\ d_1+d_2-m_1-i-j})f^i g^{j},\\[2mm]
&T=\left\{(i,j)~\Bigr|~
\begin{array}{c}0\leq i\leq d_1,\ 0\leq j\leq d_1+d_2-m_1-m_2\\ d_1-m_1\leq i+j\leq 
d_1+d_2-m_1\end{array}\right\} .
\end{split}
\end{equation}
Hence $Q(f,g)$ is a polynomial of bidegree $(d_1,d_1+d_2-m_1-m_2)$ having zeros at
${\rm P}_1(\infty,\infty)$, ${\rm P}_2(0,0)$ with multiplicity $d_1-m_2$ and $d_1-m_1$, respectively.
Multiplicities of other zeros remain the same. This implies that  $Q(f,g)$ belongs to the class
\begin{align*}
&  d_1H_1+(d_1+d_2-m_1-m_2)H_2 - (d_1-m_2)E_1 - (d_1-m_1)E_2-m_3E_3-\cdots-m_8E_8\\
=&d_1(H_1+H_2-E_1-E_2) + d_2H_2 - m_1(H_2-E_2)-m_2(H_2-E_1)-m_3E_3-\cdots-m_8E_8
 = s_2(\lambda).
\end{align*}
One can verify the case of other $s_k$ in a similar manner.  The factorized formulae
\eqref{eqn:factor_wfg} for $w=s_{k_r}\cdots s_{k_2}s_{k_1}$ can be obtained by applying $s_{k_1}$,
$s_{k_2}$, $\ldots$ successively to \eqref{eqn:factor_wfg} for $w={\rm id}$ 
with $P_{{E}_1}=1$, $P_{{E}_2}=1$, $P_{H_1-{E}_1}=1$, $P_{{H}_1-{E}_2}=f$, $P_{{H}_2-{E}_1}=1$,
$P_{{H}_2-{ E}_2}=g$.  In Section \ref{subsec:tau}, we will introduce the $\tau$ functions, which play
an essential role in the complete description of those polynomials.
%
\subsection{Parametrization of the eight points and the curve}
Recall that the affine root system of type $E_8^{(1)}$ is realized by the simple roots
\begin{equation}\label{eqn:alpha_E8}
\begin{array}{c}
\alpha_0 = E_1-E_2,\quad \alpha_1= H_1-H_2,\quad \alpha_2=H_2-E_1-E_2,\quad 
\alpha_3=E_2-E_3,\quad \alpha_4=E_3-E_4,\\
\alpha_5 = E_4-E_5,\quad \alpha_6 = E_5 - E_6,\quad \alpha_7 = E_6-E_7,\quad \alpha_8=E_7-E_8.
\end{array}
\end{equation}
in the Picard lattice
$\Lambda=\Z H_1 \oplus \Z H_2 \oplus \Z E_1 \oplus \cdots \oplus\Z E_8$
and that the affine Weyl group $W(E_8^{(1)})=\langle s_0,\ldots,s_8\rangle$ has a natural linear
action \eqref{eqn:roots_reflection} on $\Lambda$  through the simple reflections $r_{\alpha_i}$
($i=0,\ldots,8$). The action of $s_j$ on the basis of $\Lambda$ is computed explicitly by
using \eqref{eqn:E_8_intersection_form}, \eqref{eqn:roots_reflection} and
\eqref{eqn:alpha_E8} as follows:
\begin{equation}\label{eqn:E8_linear}
 \begin{split}
  s_0:\quad& E_1\ \leftrightarrow\ E_2,\\
  s_1:\quad& H_1\ \leftrightarrow\ H_2,\\
  s_2:\quad& E_1\ \rightarrow\ H_2-E_2,\quad E_2\ \rightarrow\ H_2-E_1,\quad H_1\ \rightarrow\ H_1+H_2-E_1-E_2,\\
  s_j:\quad& E_{j-1}\ \leftrightarrow\ E_{j}\quad (j=3,\ldots,8).
 \end{split}
\end{equation}
Also, we introduce the set of parameters $h_1$, $h_2$, $e_1,\ldots,e_8$ and define the actions on them as
\begin{equation}\label{eqn:E8_linear_parameters}
 \begin{split}
  s_0:\quad& e_1\ \leftrightarrow\ e_2,\\
  s_1:\quad& h_1\ \leftrightarrow\ h_2,\\
  s_2:\quad& e_1\ \rightarrow\ h_2-e_2,\quad e_2\ \rightarrow\ h_2-e_1,\quad h_1\ \rightarrow\ h_1+h_2-e_1-e_2,\\
  s_j:\quad& e_{j-1}\ \leftrightarrow\ e_{j}\quad (j=3,\ldots,8)
 \end{split}
\end{equation}
by identifying $H_i$ with $h_i$ ($i=1,2$) and $E_i$ with $e_i$ ($i=1,\ldots,8$),
respectively. The parameter
\begin{equation}\label{eqn:delta_par}
 \delta = 2h_1 + 2h_2 - e_1-\cdots - e_8,
\end{equation}
corresponding to the null root, denoted by the same symbol, is $W(E_8^{(1)})$-invariant and will
play the role of the step-size for difference equations. The set of parameters $\kappa_1$,
$\kappa_2$, $v_1,\ldots,v_8$ used in Section \ref{subsubsection:qpe6} and $h_1$, $h_2$,
$e_1,\ldots,e_8$ are related with each other (see Remark \ref{rem:kappa-var}).

In order to establish a connection between the two representations \eqref{eqn:E8_fg} and
\eqref{eqn:E8_linear}, \eqref{eqn:E8_linear_parameters} of $W(E_8^{(1)})$, we introduce a
parametrization of the eight points by the parameters $h_1$, $h_2$, $e_1,\ldots,e_8$. 
For notational convenience, we introduce a function
\begin{equation}\label{eqn:phi_kappa}
 \varphi_\kappa(s,t) = [s-t][\kappa-s-t],
\end{equation}
supposing that $[u]$ is an odd function in $u\in\mathbb{C}$, namely $[-u]=-[u]$ and $[0]=0$;
$ \varphi_\kappa(s,t)$ satisfy the relations
\begin{equation}\label{eqn:varphi_relations}
 \varphi_{\kappa}(s,t)=-\varphi_{\kappa}(t,s)=\varphi_{\kappa}(\kappa-s,t),\quad \varphi_\kappa(t,t)=0.
\end{equation}
We also require {\it the Riemann relation}
\begin{equation}\label{eqn:Riemann_phi}
\varphi_\kappa(a,b)\varphi_\kappa(c,u)+ \varphi_\kappa(b,c)\varphi_\kappa(a,u) 
+  \varphi_\kappa(c,a) \varphi_\kappa(b,u) =0,
\end{equation}
which is equivalent to 
\begin{equation}\label{eqn:Riemann_relation}
 [u+a][u-a][b+c][b-c] 
+[u+b][u-b][c+a][c-a] 
+[u+c][u-c][a+b][a-b] =0.
\end{equation}
See Remark \ref{rem:[u]} for concrete examples of the function $[u]$.

We introduce the parametrization of 8 points as
\begin{equation}\label{eqn:E8_fg_params}
\begin{split}
&(f_i,g_i)  = \left(\frac{F(e_i)}{F(e_3)}, \frac{G(e_i)}{G(e_3)}\right)\quad (i=1,\ldots,8),\\[2mm]
&F(u)=\frac{\varphi_{h_1}(e_2,u)}{\varphi_{h_1}(e_1,u)}=\frac{[e_2-u][h_1-e_2-u]}{[e_1-u][h_1-e_1-u]},\\
&G(u)=\frac{\varphi_{h_2}(e_2,u)}{\varphi_{h_2}(e_1,u)}=\frac{[e_2-u][h_2-e_2-u]}{[e_1-u][h_2-e_1-u]}.
\end{split}
\end{equation}
By this parametrization, the representations \eqref{eqn:E8_fg} and \eqref{eqn:E8_linear_parameters}
are compatible.  For example, since
\begin{align*}
s_2(F(u))&=s_2\left(\frac{[e_2-u][h_1-e_2-u]}{[e_1-u][h_1-e_1-u]}\right)
=\frac{[s_2(e_2)-u][s_2(h_1)-s_2(e_2)-u]}{[s_2(e_1)-u][s_2(h_1)-s_2(e_1)-u]}\\
&=\frac{[h_2-e_1-u][h_1-e_2-u]}{[h_2-e_2-u][h_1-e_1-u]}
=\frac{F(u)}{G(u)}, 
\end{align*}
provided that $s_2(u)=u$, we verify the consistency of the action of $s_2$ on $f_i$\,:
\begin{align*}
s_2(f_i)&=s_2\left(\frac{F(e_i)}{F(e_3)}\right)
=\frac{F(e_i)}{F(e_3)}\frac{G(e_3)}{G(e_i)}=\frac{f_i}{g_i}. 
\end{align*}
Also, for the action of $s_3$, we have
\begin{align*}
 s_3(f_i)&=
s_3\left(\frac{F(e_i)}{F(e_3)}\right)
=
s_3\left(\frac{\varphi_{h_1}(e_2,e_i)}{\varphi_{h_1}(e_1,e_i)}\frac{\varphi_{h_1}(e_1,e_3)}{\varphi_{h_1}(e_2,e_3)}\right)
=
\frac{\varphi_{h_1}(e_3,e_i)}{\varphi_{h_1}(e_1,e_i)}\frac{\varphi_{h_1}(e_1,e_2)}{\varphi_{h_1}(e_3,e_2)},
\end{align*}
hence the condition $s_3(f_i)=1-f_i$ is equivalent to
\begin{displaymath}
\frac{\varphi_{h_1}(e_3,e_i)}{\varphi_{h_1}(e_1,e_i)}\frac{\varphi_{h_1}(e_1,e_2)}{\varphi_{h_1}(e_3,e_2)}
=1-\frac{\varphi_{h_1}(e_2,e_i)}{\varphi_{h_1}(e_1,e_i)}\frac{\varphi_{h_1}(e_1,e_3)}{\varphi_{h_1}(e_2,e_3)}, 
\end{displaymath}
which is nothing but the Riemann relation \eqref{eqn:Riemann_phi}.
%
\begin{rem}\label{rem:[u]}\rm
It is known that there are three classes of functions satisfying the Riemann relation
\eqref{eqn:Riemann_relation}:
\begin{enumerate}
\setcounter{enumi}{-1}
 \item rational function $[u]=u$,
 \item trigonometric function $[u]=\sin \frac{\pi u}{\omega}$,
 \item elliptic function $[u]=\sigma(u;\omega_1,\omega_2)$, where $\sigma(u;\omega_1,\omega_2)$ is
       the Weierstrass sigma function or an odd theta function.
\end{enumerate} 

For given generic eight points in $\P^1\times \P^1$, there exists a unique
bidegree (2,2) curve
\begin{equation}\label{eqn:C0}
 C_0:\quad p(f,g)=\sum_{i,j=0}^2 c_{ij}f^ig^j=0,
\end{equation}
passing through them. In fact, since we have the nine coefficients $c_{ij}$, the eight linear
equations $p(f_i,g_i)=0$ ($i=1,\ldots,8$) uniquely determines the curve $C_0$.  If the curve $C_0$
is non-singular, it is an elliptic curve and otherwise a rational curve.  The three classes of the
function $[u]$, (2) elliptic, (1) trigonometric and (0) rational, correspond to the cases where the
curve $C_0$ is (2) smooth, (1) nodal and (0) cuspidal, respectively.

In the elliptic case, the curve $C_0$ can be identified with the complex torus $\mathbb{C}/\Omega$,
$\Omega=\mathbb{Z}\omega_1 + \mathbb{Z}\omega_2$, and the rational function on $C_0$ are expressed
by elliptic functions ($\Omega$--periodic meromorphic functions on $\mathbb{C}$).  Since $C_0$ is a
$(2,2)$ curve, the lines $f={\rm const.}$ and $g={\rm const.}$ intersect with $C_0$ at two points,
respectively. Therefore, in terms of the coordinate $u$ of the complex torus $\mathbb{C}/\Omega$, it
is known that the coordinates $(f,g)$ of a point on $C_0$ can be parametrized by elliptic functions
of order 2 \cite{Mumford,Whittaker-Watson}:
\begin{equation}
\begin{split}
& (f,g) = \left(c\frac{\sigma(u-\alpha)\sigma(u-\beta)}{\sigma(u-\gamma)\sigma(u-\delta)},
 c'\frac{\sigma(u-\alpha')\sigma(u-\beta')}{\sigma(u-\gamma')\sigma(u-\delta')}\right),\\[2mm]
& \alpha+\beta=\gamma+\delta,\quad \alpha'+\beta'=\gamma'+\delta', 
\end{split}
\end{equation}
where $\sigma(u)=\sigma(u;\omega_1,\omega_2)$ is the Weierstrass sigma function or a theta
function. From this we obtain the parametrization of the curve $C_0$
\begin{equation}
C_0:\quad (f,g)=\left(\frac{F(u)}{F(e_3)}, \frac{G(u)}{G(e_3)}\right),\quad u\in\mathbb{C},
\end{equation}
through the renormalization by the action of ${\rm PGL}(2)^2$ and the translation of $u$.
The trigonometric and rational cases are understood as degenerations of the elliptic case explained above.
\end{rem}
%

Through the Parametrization of the eight points \eqref{eqn:E8_fg_params}, we obtain from
\eqref{eqn:E8_fg} and \eqref{eqn:E8_linear_parameters} the following representation of
$W(E_8^{(1)})$ on the variables $h_1$, $h_2$, $e_1,\ldots,e_8$ and $(f,g)=(f_9,g_9)$:
\begin{equation}\label{eqn:E8_parameters_and_fg}
\begin{array}{lll}
{\displaystyle s_0:\quad} &  {\displaystyle e_1\ \leftrightarrow\ e_2,}&
{\displaystyle f\ \rightarrow\  \frac{1}{f},\quad  g\ \rightarrow\  \frac{1}{g},}\\
{\displaystyle s_1:\quad}& {\displaystyle h_1\ \leftrightarrow\ h_2,}
&{\displaystyle f\ \leftrightarrow\  g,}\\[2mm]
{\displaystyle s_2:\quad}&\hspace{-5pt}
\left\{\begin{array}{l}
{\displaystyle e_1\ \rightarrow\ h_2-e_2,}\\[2mm] 
{\displaystyle e_2\ \rightarrow\ h_2-e_1,}  \\[2mm]
{\displaystyle  h_1\ \rightarrow\ h_1+h_2-e_1-e_2,}
\end{array}\right.
& {\displaystyle  f\ \rightarrow\  \frac{f}{g},\quad  g\ \rightarrow\  \frac{1}{g},}\\[4mm]
{\displaystyle s_3:\quad } & {\displaystyle e_{2}\ \leftrightarrow\ e_{3},}
&{\displaystyle f\ \rightarrow\  1-f,\quad  g\ \rightarrow\  1-g,}\\[2mm]
{\displaystyle s_4:\quad} &{\displaystyle  e_{3}\ \leftrightarrow\ e_{4},}
&\left\{\begin{array}{l}
{\displaystyle f\ \rightarrow\  \frac{\varphi_{h_1}(e_1,e_4)}{\varphi_{h_1}(e_2,e_4)}
\frac{\varphi_{h_1}(e_2,e_3)}{\varphi_{h_1}(e_1,e_3)}f,}\\[4mm]
{\displaystyle g\ \rightarrow\  \frac{\varphi_{h_2}(e_1,e_4)}{\varphi_{h_2}(e_2,e_4)}
\frac{\varphi_{h_2}(e_2,e_3)}{\varphi_{h_2}(e_1,e_3)}g,}
\end{array}\right. \\[1cm]
{\displaystyle  s_j:\quad}& {\displaystyle e_{j-1}\ \leftrightarrow\ e_{j}}, &
{\displaystyle f\ \rightarrow\ f,\quad g\ \rightarrow\ g}\qquad (j=5,\ldots,8).
\end{array}
 \end{equation}
In this representation, the variables $h_1$, $h_2$, $e_1,\ldots,e_8$ play the role of independent
variables and parameters for our elliptic Painlev\'e equation, while $f$ and $g$ are the dependent
variables.

We remark that the ninth variables $(f,g)$ are free variables independent of the curve $C_0$
\eqref{eqn:C0}. However, one may specialize the point $(f,g)$ onto $C_0$ and parametrized
it as
\begin{equation}\label{eqn:E8_fg_spesialization}
f\,\Bigr|_{C_0}
=\frac{\varphi_{h_1}(e_1,e_3)}{\varphi_{h_1}(e_2,e_3)}\frac{\varphi_{h_1}(e_2,u)}{\varphi_{h_1}(e_1,u)},\quad
g\,\Bigr|_{C_0}
=\frac{\varphi_{h_2}(e_1,e_3)}{\varphi_{h_2}(e_2,e_3)}\frac{\varphi_{h_2}(e_2,u)}{\varphi_{h_2}(e_1,u)}.
\end{equation}
This expression is consistent with the action of $W(E_8^{(1)})$ if we regard $u$ as a constant (a
$W(E_8^{(1)})$-invariant parameter). Namely we have for any $w\in W(E_8^{(1)})$, 
$w(f)\bigr|_{C_0} = w\left(f\bigr|_{C_0}\right)$, $w(g)\bigr|_{C_0} = w\left(g\bigr|_{C_0}\right)$,
or, more generally,
\begin{equation}\label{eqn:specialization_fg_C0}
 w(F)\Bigr|_{C_0} = w\left(F\Bigr|_{C_0}\right),
\end{equation}
for any rational function $F=F(f,g)$. This implies that the points on the curve $C_0$ are confined to
$C_0$ by the action of $W(E_8^{(1)})$. 

Conversely, the condition \eqref{eqn:specialization_fg_C0} can be used to determine the action of $w
\in W(E_8^{(1)})$ on $f, g$. From \eqref{eqn:factor_wfg}, we see that $w(f)$ and $w(g)$ are rational
functions of the class $w(H_1)$ and $w(H_2)$, respectively (see Remark
\ref{rem:rational_fn_Picard_lattice}).  Since the dimension of polynomials of the class $w(H_i)$
($i=1,2$) is $2$, $w(f)$ and $w(g)$ are expressed in the form
\begin{equation}\label{eqn:wxy}
 w(f) = \frac{a_1A_1(f,g)+b_1B_1(f,g)}{c_1A_1(f,g)+d_1B_1(f,g)},\quad
 w(g) = \frac{a_2A_2(f,g)+b_2B_2(f,g)}{c_2A_2(f,g)+d_2B_2(f,g)},
\end{equation}
where $\{A_i(f,g), B_i(f,g)\}$ is an arbitrary basis of the vector space of polynomials in the class
$w(H_i)$ ($i=1,2$).  Then the condition \eqref{eqn:specialization_fg_C0}, which is an identity with
respect to the parameter $u$ of $C_0$, gives enough information to determine the ratios of
coefficients $a_i:b_i:c_i:d_i$ ($i=1,2$) uniquely. We note that this method can be applied also for
any inhomogeneous coordinates on $\P^1 \times \P^1$.
%
\begin{rem}\rm\label{rem:w(f)}
The fact that $w(f)$ and $w(g)$ can be expressed as the rational functions of the class $w(H_1)$ and
$w(H_2)$ is also expected in the degenerate cases that will be discussed later. We will use this
property as a guiding principle for constructing the birational actions of affine Weyl groups on
variables $f$ and $g$.
\end{rem}

Similarly to the elliptic case described here, one can (and we will) find a compatible
parametrization of the eight points in terms of parameters $h_i$, $e_i$ for each configuration. A
mathematical basis for such parametrization is the {\it period mapping}, which is also a key tool of
Sakai's theory (see \cite{Sakai:SIV} Section 5).
%
\subsection{$\tau$ functions}\label{subsec:tau}
We now introduce the $\tau$ functions and lift the representation \eqref{eqn:E8_parameters_and_fg}
of $W(E_8^{(1)})$ further to the level of them. For this purpose, we define the $\tau$ functions as
a family of variables $\tau(\lambda)$ parametrized by the exceptional classes $\lambda$ in $M$
\eqref{eqn:-1_curve}.  The action of the affine Weyl group $W(E_8^{(1)})$ on $M$ induces a natural
action of $w\in W(E_8^{(1)})$ on the $\tau$ variables as $w(\tau(\lambda))=\tau(w(\lambda))$.  In
order to obtain a representation of $W(E_8^{(1)})$ on a finite number of $\tau$ variables, we impose
appropriate algebraic (bilinear) relations among the $\tau$ variables which are consistent with
\eqref{eqn:E8_parameters_and_fg}. In view of \eqref{eqn:E8_fg_spesialization}, we represent the
variables $f$, $g$ in terms of the $\tau$ variables as
\begin{equation}\label{eqn:E8_fg_tau}
 f =\frac{\varphi_{h_1}(e_1,e_3)}{\varphi_{h_1}(e_2,e_3)}
\frac{\tau(E_2)\tau(H_1-E_2)}{\tau(E_1)\tau(H_1-E_1)},\quad
 g =\frac{\varphi_{h_2}(e_1,e_3)}{\varphi_{h_2}(e_2,e_3)}
\frac{\tau(E_2)\tau(H_2-E_2)}{\tau(E_1)\tau(H_2-E_1)}, 
\end{equation}
It is easy to check the action \eqref{eqn:E8_parameters_and_fg} of $W(E_8^{(1)})$ on $f$, $g$ is consistent if and
only if the $\tau$ variables satisfy the bilinear relations
\begin{equation}
\begin{split}\label{eqn:E8_bl1}
&   \varphi_{h_1}(e_i,e_j)\tau(E_k)\tau(H_1-E_k)
+  \varphi_{h_1}(e_j,e_k)\tau(E_i)\tau(H_1-E_i)\\
&\hskip60pt +  \varphi_{h_1}(e_k,e_i)\tau(E_j)\tau(H_1-E_j)=0,\\
&   \varphi_{h_2}(e_i,e_j)\tau(E_k)\tau(H_2-E_k)
+  \varphi_{h_2}(e_j,e_k)\tau(E_i)\tau(H_2-E_i)\\
&\hskip60pt +  \varphi_{h_2}(e_k,e_i)\tau(E_j)\tau(H_2-E_j)=0.
\end{split}
\end{equation}
%
\begin{rem}\rm
We have infinitely many bilinear relations by applying the elements of $W(E_8^{(1)})$ to \eqref{eqn:E8_bl1}.
In general, such bilinear relation can be expressed as
\begin{equation} \label{eqn:E8_bl_general}
 [\widetilde{b}-c][b-c]\tau(A)\tau(\widetilde A) + 
 [\widetilde{c}-a][c-a]\tau(B)\tau(\widetilde B) + 
 [\widetilde{a}-b][a-b]\tau(C)\tau(\widetilde C) =0.
\end{equation}
Here, $A$, $B$, $C$, $\widetilde A$, $\widetilde B$, $\widetilde{C}$ are the vertices of a regular octahedron in
$M$ whose edge length is $\sqrt{2}$; $A$, $B$, $C$ and $\widetilde A$, $\widetilde B$, $\widetilde{C}$ are
antipodal to each other, respectively (see Figure \ref{fig:bilinear_octahedron}). Moreover, $a$,
$b$, $c$, $\tilde a$, $\tilde b$, $\tilde{c}$ are the complex parameters associated with $A$, $B$,
$C$, $\widetilde A$, $\widetilde B$, $\widetilde{C}$, respectively.  We remark that such an infinite family of
bilinear relations is essentially equivalent to the bilinear equations on the root lattice of type
$E_8$ obtained by Ohta, Ramani and Grammaticos \cite{ORG:e8}.
\begin{figure}[ht]
\begin{center}
\includegraphics[scale=0.4]{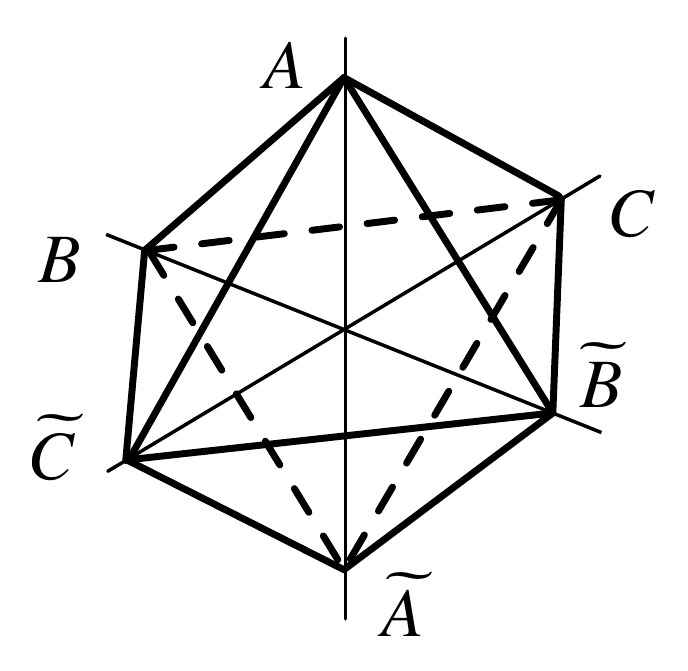}
\end{center}
\caption{$A$, $B$, $C$, $\widetilde A$, $\widetilde B$, $\widetilde{C}$ form an octahedron in $M$.}
\label{fig:bilinear_octahedron}
\end{figure}
\end{rem}
%
As we have seen before, the representation of $W(E_8^{(1)})$ on $f$, $g$ can be specialized
consistently to the curve $C_0$ as \eqref{eqn:E8_fg_spesialization}. On the level of the $\tau$
functions, this specialization is realized by
\begin{equation}\label{eqn:tau_specialize1}
\tau(E_i)\,\Big|_{C_0} = [e_i-u],\quad
\tau(H_k-E_i)\,\Big|_{C_0}=[h_k-e_i-u]\quad 
(i=1,\ldots,8,\ k=1,2,\ u\in\mathbb{C}), 
\end{equation}
and for the general exceptional class $\lambda=d_1H_1+d_2H_2-\sum_{i=1}^8 m_iE_i\in M$
\begin{equation}\label{eqn:tau_specialize2}
 \tau(\lambda)\,\Big|_{C_0}=\Bigl[d_1h_1+d_2h_2-{\textstyle\sum\limits_{i=1}^8} m_ie_i-u\Bigr]\quad
(u\in\mathbb{C}).
\end{equation}

The representation of $W(E_8^{(1)})$ on the $\tau$ variables with the bilinear relations
\eqref{eqn:E8_bl_general} can be expressed in a closed form in terms of the twelve $\tau$ variables
$\tau(E_i)$ ($i=1,\ldots,8$) and $\tau(H_k-E_1)$, $\tau(H_k-E_2)$ ($k=1,2$) as follows:
\begin{equation}\label{eqn:E8_tau}
 \begin{split}
  s_0:\quad& \tau(E_1)\ \leftrightarrow\ \tau(E_2),\quad \tau(H_k-E_1)\ \leftrightarrow\ \tau(H_k-E_2)\quad (k=1,2),\\
  s_1:\quad& \tau(H_1-E_i)\ \leftrightarrow\ \tau(H_2-E_i),\\
  s_2:\quad& \tau(E_1)\ \leftrightarrow\ \tau(H_2-E_2),\quad \tau(E_2)\ \leftrightarrow\ \tau(H_2-E_1),\\
  s_3:\quad& \tau(E_2)\ \leftrightarrow\ \tau(E_3),\\
& \tau(H_k-E_2)\ \rightarrow\ 
\frac{\varphi_{h_1}(e_3,e_2)\tau(E_1)\tau(H_k-E_1)+\varphi_{h_1}(e_1,e_3)\tau(E_2)\tau(H_k-E_2)}
{\varphi_{h_1}(e_1,e_2)\tau(E_3)}, \\[2mm]
  s_j:\quad& \tau(E_{j-1})\ \leftrightarrow\ \tau(E_{j})\quad (j=4,\ldots,8).
 \end{split}
\end{equation}
Note that the $s_3$ action on $\tau(H_k-E_2)$ ($k=1,2$) comes from the bilinear relations.  This
representation is equivalently rewritten as
\begin{equation}\label{eqn:E8_homogeneous}
 \begin{split}
  s_0:\quad& \xi_1\ \leftrightarrow\ \xi_2,\quad \eta_1\ \leftrightarrow\ \eta_2,
\quad \tau_1\ \leftrightarrow\ \tau_2,\\
  s_1:\quad& \xi_i\ \leftrightarrow\ \eta_i \quad (i=1,2),\\
  s_2:\quad& \xi_1\ \rightarrow\ \frac{\xi_1\eta_2}{\tau_1\tau_2},\quad 
\xi_2\ \rightarrow\ \frac{\xi_2\eta_1}{\tau_1\tau_2},\quad \eta_1\ \leftrightarrow\ \eta_2,\quad 
\tau_1\ \rightarrow\ \frac{\eta_2}{\tau_2},\quad \tau_2\ \rightarrow\ \frac{\eta_1}{\tau_1},\\
  s_3:\quad& \hspace{-5pt}
\left\{\begin{array}{l}
{\displaystyle \xi_2\ \rightarrow\ 
\frac{\varphi_{h_1}(e_2,e_3)}{\varphi_{h_1}(e_2,e_1)}\xi_1
 +  \frac{\varphi_{h_1}(e_1,e_3)}{\varphi_{h_1}(e_1,e_2)}\xi_2,} \\[4mm]
{\displaystyle \eta_2\ \rightarrow\ 
\frac{\varphi_{h_2}(e_2,e_3)}{\varphi_{h_2}(e_2,e_1)}\eta_1
 +  \frac{\varphi_{h_2}(e_1,e_3)}{\varphi_{h_2}(e_1,e_2)}\eta_2,} \\[4mm]
{\displaystyle \tau_2\ \leftrightarrow\ \tau_3,}
\end{array}\right.\\[2mm]
  s_j:\quad& \tau_{j-1}\ \leftrightarrow\ \tau_{j}\quad (j=4,\ldots,8),
 \end{split}
\end{equation}
in terms of the twelve variables
\begin{equation}\label{eqn:xieta12only}
\xi_i=\tau(E_i)\tau(H_1-E_i),\quad \eta_i=\tau(E_i)\tau(H_2-E_i)\quad (i=1,2),\quad
\tau_i=\tau(E_i)\quad (i=1,\ldots,8) .
\end{equation}
The variables $\xi_i$, $\eta_i$ can be regarded as homogeneous coordinates of $\P^1\times \P^1$, and
the representation \eqref{eqn:E8_parameters_and_fg} is recovered by
\begin{equation}\label{eqn:P1P1_homogeneous}
 f = \frac{\varphi_{h_1}(e_1,e_3)}{\varphi_{h_1}(e_2,e_3)}\frac{\xi_2}{\xi_1},
\quad
 g = \frac{\varphi_{h_2}(e_1,e_3)}{\varphi_{h_2}(e_2,e_3)}\frac{\eta_2}{\eta_1}.
\end{equation}
On the level of homogeneous coordinates $(\xi,\eta)=(\xi_1:\xi_2,\eta_1:\eta_2)$, the
parametrization of the curve $C_0$ and the eight points ${\rm P}_1,\ldots,{\rm P}_8$ are given by
\begin{equation}\label{eqn:specialization_xi_eta}
 \xi_i\,\Big|_{C_0} =\varphi_{h_1}(e_i,u),\quad
 \eta_i\,\Big|_{C_0} = \varphi_{h_1}(e_i,u)\quad (i=1,2).
\end{equation}
\begin{equation}\label{eqn:points_homogeneous}
{\rm P}_j:\ (\xi_1:\xi_2,\eta_1:\eta_2)
=\left(\varphi_{h_1}(e_1,e_j):\varphi_{h_1}(e_2,e_j),\varphi_{h_2}(e_1,e_j):\varphi_{h_2}(e_2,e_j)\right).
\end{equation}

In the sequel we denote by $\mathcal{K}$ the field of meromorphic function in the parameters,
in $h_1,h_2,e_1,\ldots,e_8\in\mathbb{C}$.
For each $\lambda=d_1H_1+d_2H_2-\sum\limits_{i=1}^8 m_iE_i\in \Lambda$ 
we introduce the $\mathcal{K}$--vector space $L(\lambda)$ of functions of the form
\begin{equation}
P(\xi,\eta)\prod_{i=1}^8 \tau_i^{-m_i}.
\end{equation}
Here, $P(\xi,\eta)\in\mathcal{K}[\xi,\eta]$ is a homogeneous polynomial in $\xi=(\xi_1,\xi_2)$,
$\eta=(\eta_1,\eta_2)$ of bidegree $(d_1,d_2)$, i.e., $P(s\xi_1,s\xi_2,t\eta_1,t\eta_2)=s^{d_1}t^{d_2}
P(\xi_1,\xi_2,\eta_1,\eta_2)$, having zeros at the eight points ${\rm P}_i$ with
multiplicity $\geq m_i$ ($i=1,\ldots,8$). For example,
\begin{equation}
\begin{split}
&L(H_1)=\mathcal{K}\xi_1\oplus \mathcal{K}\xi_2
= \mathcal{K}\tau(E_1)\tau(H_1-E_1)\oplus\mathcal{K}\tau(E_2)\tau(H_1-E_2),\\
&L(H_2)=\mathcal{K}\eta_1\oplus \mathcal{K}\eta_2
= \mathcal{K}\tau(E_1)\tau(H_2-E_1)\oplus\mathcal{K}\tau(E_2)\tau(H_2-E_2),\\
&L(E_i) = \mathcal{K}\tau_i\ (i=1,\ldots,8),\\
&L(H_1-E_i)=\mathcal{K}\xi_i\tau_i^{-1},\quad
L(H_2-E_i)=\mathcal{K}\eta_i\tau_i^{-1}\ (i=1,2).
\end{split}
\end{equation}
By a computation similar to that from \eqref{eqn:factor_1} to \eqref{eqn:factor_2}, one can show
that for each $k=0,\ldots,8$, $s_k$ transforms $L(\lambda)$ bijectively to $L(s_k(\lambda))$ for any
$\lambda\in \Lambda$. Hence, each $w\in W(E_8^{(1)})$ induces an isomorphism
$L(\lambda)\xrightarrow{\raisebox{-0.35 em}{\smash{\ensuremath{\sim}}}} L(w(\lambda))$ for any
$\lambda\in\Lambda$. Note that for each exceptional class $\lambda\in M$ there exists an element
$w\in W(E_8^{(1)})$ such that $w(E_1)=\lambda$ (see Remark \ref{rem:exceptional}), which induces an
isomorphism $w:\ L(E_1)\xrightarrow{\raisebox{-0.35 em}{\smash{\ensuremath{\sim}}}} L(\lambda)$.
Since $L(E_1)=\mathcal{K}\tau_1$, we see that $L(\lambda)$ is one-dimensional and
$L(\lambda)=\mathcal{K}w(\tau_1)=\mathcal{K}\tau(\lambda)$. This implies that for each
$\lambda=d_1H_1+d_2H_2-\sum_{i=1}^8 m_iE_i\in M$ the $\tau$ function $\tau(\lambda)$ can be
expressed in the form
\begin{equation}\label{eqn:tau_lambda}
 \tau(\lambda) = \phi_\lambda(\xi,\eta)\prod_{i=1}^8 \tau_i^{-m_i},
\end{equation}
where $\phi_\lambda(\xi,\eta)$ is a homogeneous polynomial of bidegree $(d_1,d_2)$ such that the
curve $\phi_\lambda(\xi,\eta)=0$ passes through the eight points ${\rm P}_i$ with multiplicity $m_i$
($i=1,\ldots,8$).  Such a homogeneous polynomial is determined uniquely up to constant multiple.  By
specializing \eqref{eqn:tau_lambda} to the curve $C_0$ according to \eqref{eqn:tau_specialize1} and
\eqref{eqn:tau_specialize2}, we obtain
\begin{equation}
\Bigl[d_1h_1+d_2h_2-{\textstyle\sum\limits_{i=1}^8} m_ie_i-u\Bigr]=
\phi_\lambda(\xi,\eta)\Bigr|_{C_0}\prod_{i=1}^8 [e_i-u]^{-m_i}.
\end{equation}
Consequently our polynomial $\phi_\lambda(\xi,\eta)$ is normalized in such a way that
\begin{equation}
 \phi_\lambda(\xi,\eta)\,\Bigr|_{C_0}= \Bigl[d_1h_1+d_2h_2-{\textstyle\sum\limits_{i=1}^8} m_ie_i-u\Bigr]
\prod_{i=1}^8 [e_i-u]^{m_i},
\end{equation}
when specialized to the curve $C_0$. By applying $w$ to \eqref{eqn:E8_fg_tau}, we have
\begin{equation}
\begin{split}
 w(f)&= w\left(\frac{\varphi_{h_1}(e_1,e_3)}{\varphi_{h_1}(e_2,e_3)}\right)
\frac{\phi_{w(E_2)}(\xi,\eta)\phi_{w(H_1-E_2)}(\xi,\eta)}
{\phi_{w(E_1)}(\xi,\eta)\phi_{w(H_1-E_1)}(\xi,\eta)},\\ 
 w(g)&=  w\left(\frac{\varphi_{h_2}(e_1,e_3)}{\varphi_{h_2}(e_2,e_3)}\right)
\frac{\phi_{w(E_2)}(\xi,\eta)\phi_{w(H_2-E_2)}(\xi,\eta)}
{\phi_{w(E_1)}(\xi,\eta)\phi_{w(H_2-E_1)}(\xi,\eta)},
\end{split} 
\end{equation}
which give the precise form of $w(f)$ and $w(g)$ observed in \eqref{eqn:factor_wfg}.
%
\begin{rem}\rm
There exists a simple geometric meaning of the bilinear relations \eqref{eqn:E8_bl1}.  For instance,
the functions of the form $\tau(E_i)\tau(H_1-E_i)=\phi_{H_1-E_i}(\xi,\eta)$ ($i=1,\ldots,8$) belongs
to the same vector space $L(H_1)$ of dimension 2. As a result, there exists a linear relation among
any three of such functions, for instance,
\begin{align}
  c_1\tau(E_1)\tau(H_1-E_1)
 +  c_2\tau(E_2)\tau(H_1-E_2)
 + c_3\tau(E_3)\tau(H_1-E_3)=0.
\end{align}
The coefficients can be easily recovered by specializing to $C_0$ according to \eqref{eqn:tau_specialize1}
and putting $u=e_1,e_2,e_3$.
In general, the bilinear relations arise in such a situation that elements of a
two-dimensional vector space such as $L(H_1)$ can be obtained in several ways as products of elements associated to a exceptional class. This is a universal
structure to yield bilinear relations among the $\tau$ functions that can be observed also in the
degenerate cases.
\end{rem}
\subsection{$\mathfrak{S}_8$-invariant coordinates}

We define convenient coordinates $\xi(t), \eta(t)$ on $\P^1\times\P^1$ in which the 8 points can be
treated in symmetric manner.  First, we extend the coordinates $\xi_i, \eta_i$ ($i=1,2$) in
\eqref{eqn:xieta12only} to
\begin{equation}\label{eqn:xi_eta_by_tau}
\xi_i=\tau(E_i)\tau(H_1-E_i), \quad \eta_i=\tau(E_i)\tau(H_2-E_i), \quad (i=1,\ldots,8).
\end{equation}
With these notations the bilinear relations for the $\tau$ functions \eqref{eqn:E8_bl1} are rewritten as
\begin{equation}\label{eqn:E8_bl3} 
\begin{split}
&\varphi_{h_1}(e_i,e_j)\xi_k+\varphi_{h_1}(e_j,e_k)\xi_i+\varphi_{h_1}(e_k,e_i)\xi_j=0,\\
&\varphi_{h_2}(e_i,e_j)\eta_k+\varphi_{h_2}(e_j,e_k)\eta_i+\varphi_{h_2}(e_k,e_i)\eta_j=0,
\end{split}
\end{equation}
for any $i,j,k\in\{1,\ldots,8\}$. The specialization of the coordinates $\xi_i$, $\eta_i$ to the curve $C_0$ is given by
\begin{equation}\label{eqn:special_xi_eta}
 \xi_i\,\Bigl|_{C_0}=\varphi_{h_1}(e_i,u),\quad 
 \eta_i\,\Bigl|_{C_0}=\varphi_{h_2}(e_i,u)\quad (i=1,\ldots,8),
\end{equation}
so that $\xi_i$, $\eta_i$ have zero at $u=e_i$. 

For any $W(E_8^{(1)})$-invariant parameter $t$, we define a more general homogeneous coordinates 
$\xi(t)\in L(H_1)=\mathcal{K}\xi_1\oplus\mathcal{K}\xi_2$ and 
$\eta(t)\in L(H_2)=\mathcal{K}\eta_1\oplus\mathcal{K}\eta_2$, 
that have a zero at the point $u=t$ on $C_0$ as
\begin{equation}\label{eqn:def_xi_a}
\xi(t)=\frac{\varphi_{h_1}(t,e_2)\xi_1+\varphi_{h_1}(e_1,t)\xi_2}{\varphi_{h_1}(e_1,e_2)}, \quad
\eta(t)=\frac{\varphi_{h_2}(t,e_2)\eta_1+\varphi_{h_2}(e_1,t)\eta_2}{\varphi_{h_2}(e_1,e_2)}.
\end{equation}
Then, from \eqref{eqn:Riemann_phi} and \eqref{eqn:E8_bl1},  we have
\begin{equation}\label{eqn:special_xi_a}
\begin{split}
&\xi(e_i)=\xi_i \quad (i=1,\ldots,8), \quad \xi(t)\,\Bigl|_{C_0}=\varphi_{h_1}(t,u),\\
&\eta(e_i)=\eta_i\quad (i=1,\dots,8),\quad \eta(t)\,\Bigl|_{C_0}=\varphi_{h_2}(t,u).
\end{split} 
\end{equation}
We remark that the Riemann relation \eqref{eqn:Riemann_phi} also implies the three-term relations
\begin{equation}
\begin{split}\label{eqn:Riemann_xi}
\varphi_{h_1}(a,b)\xi(c)+  \varphi_{h_1}(b,c)\xi(a) +  \varphi_{h_1}(c,a)\xi(b)  =0,\\ 
\varphi_{h_2}(a,b)\eta(c)+  \varphi_{h_2}(b,c)\eta(a) +  \varphi_{h_2}(c,a)\eta(b)  =0. 
\end{split}
\end{equation}

By the construction given above, we have $L(H_1)=\mathcal{K}\xi(a)\oplus\mathcal{K}\xi(b)$ and
$L(H_2)=\mathcal{K}\eta(a)\oplus\mathcal{K}\eta(b)$ for generic $a$, $b$.  The coordinates used by
Ohta-Ramani-Grammaticos \cite{ORG:e8} correspond to the following ones
\begin{equation}\label{eqn:ORG_coordinates}
 Z = \frac{\xi(b)}{\xi(a)},\quad W=\frac{\eta(b)}{\eta(a)},
\end{equation}
where $a$ and $b$ are arbitrary fixed parameters (invariant under the Weyl group actions). 

Then, in the inhomogeneous coordinates $(Z,W)$, we see from \eqref{eqn:Riemann_phi} and
\eqref{eqn:points_homogeneous} that the original eight points ${\rm P}_i$ ($i=1,\ldots,8$) are
parametrized as
\begin{equation}\label{eqn:8points_par_ORG0}
{\rm P}_i:\ (Z_i,W_i)=\left(\varphi(e_i),\psi(e_i)\right),\quad (i=1,\ldots,8),
\end{equation}
where
\begin{equation}\label{eqn:8points_par_ORG}
\varphi(u)=\frac{\varphi_{h_1}(b,u)}{\varphi_{h_1}(a,u)}, \quad
\psi(u)=\frac{\varphi_{h_2}(b,u)}{\varphi_{h_2}(a,u)}.
\end{equation}
We note that $\varphi(u)$ and $\psi(u)$ are elliptic functions of order $2$ and satisfy
\begin{equation}\label{eqn:symmetry_xi_eta}
 \varphi(h_1-u)=\varphi(u),\quad \psi(h_2-u)=\psi(u),
\end{equation}
which follows from \eqref{eqn:phi_kappa} and \eqref{eqn:8points_par_ORG}.

Advantage of the coordinates $(Z,W)$ is that the action of $s_0,\ldots,s_8\in W(E_8^{(1)})$ can be
described simply \cite{MSY:Riccati}. In fact, they are invariant under the permutations
$\mathfrak{S}_8=\langle s_0,s_3,\ldots,s_8\rangle$.  The following actions
\begin{equation}\label{eqn:s1_on_Z}
 s_1(Z)=W,\quad s_1(W)=Z, \quad s_2(W)=W,
\end{equation}
are also obvious from \eqref{eqn:varphi_relations}, \eqref{eqn:E8_parameters_and_fg} and
\eqref{eqn:points_homogeneous}. And we also have
\begin{equation}\label{eqn:s2_on_Z}
s_2\left(\frac{Z-Z_2}{Z-Z_1}\right)=\frac{Z-Z_2}{Z-Z_1}\ \frac{W-W_1}{W-W_2},
\end{equation}
which follows from
\begin{equation}
s_2\left(\dfrac{\xi_2}{\xi_1}\right)=\dfrac{\tau(H_1-E_2)\tau(H_2-E_1)}{\tau(H_1-E_1)\tau(H_2-E_2)}
=\dfrac{\xi_2}{\xi_1}\dfrac{\eta_1}{\eta_2},
\end{equation}
and
\begin{equation}\label{eqn:Z_minus_varphi}
Z-\varphi(t)=\dfrac{\varphi_{h_1}(a,b)}{\varphi_{h_1}(a,t)}\dfrac{\xi(t)}{\xi(a)}, \quad
W-\psi(t)=\dfrac{\varphi_{h_2}(a,b)}{\varphi_{h_2}(a,t)}\dfrac{\eta(t)}{\eta(a)}.
\end{equation}

We also remark that the coordinates $(Z,W)$ are expressed in terms of the $\tau$ functions as follows.
From \eqref{eqn:xi_eta_by_tau}, \eqref{eqn:special_xi_a}, \eqref{eqn:ORG_coordinates}, 
\eqref{eqn:8points_par_ORG0}, \eqref{eqn:8points_par_ORG} and \eqref{eqn:Z_minus_varphi} we have
\begin{equation}\label{eqn:WZ_by_tau}
 \frac{\varphi_{h_1}(a,e_i)}{\varphi_{h_1}(a,e_j)}\frac{Z-Z_i}{Z-Z_j} 
= \frac{\tau(E_i)\tau(H_1-E_i)}{\tau(E_j)\tau(H_1-E_j)},\quad
 \frac{\varphi_{h_2}(a,e_i)}{\varphi_{h_2}(a,e_j)}\frac{W-W_i}{W-W_j} 
= \frac{\tau(E_i)\tau(H_2-E_i)}{\tau(E_j)\tau(H_2-E_j)}.
\end{equation}

We finally give a geometric description of the action of $W(E_8^{(1)})$ on $\xi(t)$ and $\eta(t)$.
Since $\xi(t)\in L(H_1)$, we have $w(\xi(t))\in L(w(H_1))$ for each $w\in W(E_8^{(1)})$. 
If $w(H_1)=d_1H_1+d_2H_2-\sum\limits_{i=1}^8 m_iE_i$, $w(\xi(t))$ is expressed in the form
\begin{equation}\label{eqn:x_xi}
 w(\xi(t)) = P_{w}(\xi,\eta;t)\prod_{i=1}^8 \tau_i^{-m_i},
\end{equation}
where $P_{w}(\xi,\eta;t)$ is a homogeneous polynomial in $\xi$, $\eta$ with parameter $t$ of
bidegree $(d_1,d_2)$ having zeros at ${\rm P}_i$ with multiplicity $\geq m_i$ ($i=1,\ldots,8$).  Also, by
\eqref{eqn:tau_specialize1}, \eqref{eqn:special_xi_a} and \eqref{eqn:x_xi} the specialization of
$P_{w}(\xi,\eta;t)$ to $C_0$ is given by
\begin{equation}\label{eqn:P_normalization}
 P_w(\xi,\eta;t)\Bigr|_{C_0} = [w(h_1)-t-u][t-u]\prod_{i=1}^8 [e_i-u]^{m_i},
\end{equation}
where we used $w\bigl(\xi(t)\bigr)\bigr|_{C_0} = w\left(\xi(t)\bigr|_{C_0}\right)$
\eqref{eqn:specialization_fg_C0}. In particular, $P_{w}(\xi,\eta;t)$ has a zero at $u=t$ on
$C_0$. The polynomial $P_{w}(\xi,\eta;t)$ of the class $w(H_1)$ is determined uniquely by the
normalization condition \eqref{eqn:P_normalization}. The action of $w$ on $\eta(t)$ can be described
in a similar way.
\subsection{Elliptic Painlev\'e equation}\label{subsec:eP}
On the basis of the materials provided in the preceding sections, we now write down the elliptic
Painlev\'e equation explicitly. 

By choosing $w$ to be a Kac transformation $T_\alpha$ associated with $\alpha\in Q(E_8^{(1)})$
\eqref{eqn:Kac_translation}, we obtain the elliptic Painlev\'e equation of the direction $\alpha$.
Note that for any roots $\alpha, \beta\in R$ there exists an element $w\in W(E_8^{(1)})$ such that
$\beta=w(\alpha)$ and hence $T_\beta = wT_\alpha w^{-1}$. This means that the elliptic Painlev\'e
equations of two different directions $\alpha, \beta\in R$ are transformed to each other by certain
birational transformations ({\em B\"acklund transformations}).

We fix the simple root $\alpha_1=H_1-H_2$, and investigate the elliptic Painvlev\'e equation in this
direction. We see from \eqref{eqn:Kac_translation} that the Kac translation $T_{\alpha_1}$ acts
on the Picard lattice as
\begin{equation}\label{eqn:T1_lattice}
\begin{split}
& T_{\alpha_1}(H_1)=H_1 - 2(H_1-H_2) + \delta= H_1 + 4H_2 - E_1-\cdots-E_8,\\
& T_{\alpha_1}(H_2)=H_2 - 2(H_1-H_2)+3\delta =4H_1+9H_2-3E_1-\cdots-3E_8,\\
&T_{\alpha_1}(E_i)=E_i - (H_1-H_2) + \delta\quad (i=1,\ldots,8).
\end{split}
\end{equation}
This means that $T_{\alpha_1}(x)$ and $T_{\alpha_1}(y)$ are rational function of bidegree (1,4) and
(4,9), respectively. However, we could use $T_{\alpha_1}^{-1}(y)=T_{-\alpha_1}(y)$, since we have
\begin{equation}\label{eqn:T1_inverse}
T_{\alpha_1}^{-1}(H_2)=H_2 - 2(H_2-H_1)+\delta=4H_1+H_2-E_1-\cdots-E_8,
\end{equation}
which implies that $T_{\alpha_1}^{-1}(y)$ is a rational function of bidegree (4,1).  We remark that
the Kac translation $T_{\alpha_1}$ can be expressed as a product of two involutions
\begin{equation}\label{eqn:T1_w2w1}
 T_{\alpha_1} = w_2w_1,
\end{equation}
where each of $w_i$ ($i=1,2$) is given as a product of eight commuting reflections as
\begin{equation}\label{eqn:w1w2}
\begin{split}
& w_1=r_{E_7-E_8}r_{H_1-E_7-E_8}r_{E_5-E_6}r_{H_1-E_5-E_6}
r_{E_3-E_4}r_{H_1-E_3-E_4}r_{E_1-E_2}r_{H_1-E_1-E_2},\\
& w_2=r_{E_7-E_8}r_{H_2-E_7-E_8}r_{E_5-E_6}r_{H_2-E_5-E_6}
r_{E_3-E_4}r_{H_2-E_3-E_4}r_{E_1-E_2}r_{H_2-E_1-E_2}.
\end{split}
\end{equation}
Notice from \eqref{eqn:roots_reflection}, \eqref{eqn:E_8_intersection_form} and
\eqref{eqn:simple_roots_E8} that
\begin{equation}\label{eqn:root_reflection_E8}
\begin{split}
r_{H_1-E_i-E_j}&:\ E_i\to H_1-E_j,\ E_j\to H_1-E_i,\ H_2\to H_1+H_2-E_i-E_j,\\ 
r_{H_2-E_i-E_j}&:\ E_i\to H_2-E_j,\ E_j\to H_2-E_i,\ H_1\to H_1+H_2-E_i-E_j,\\ 
r_{E_i-E_j}&:\ E_i\ \leftrightarrow\ E_j.
\end{split}
\end{equation}
Then we find that
\begin{equation}
\begin{split}
r_{E_i-E_j} r_{H_1-E_i-E_j}&:\ E_i\to H_1-E_i,\ E_j\to H_1-E_j,\ H_2\to H_1+H_2-E_i-E_j,\\  
r_{E_i-E_j} r_{H_2-E_i-E_j}&:\ E_i\to H_2-E_i,\ E_j\to H_2-E_j,\ H_1\to H_1+H_2-E_i-E_j,
\end{split}
\end{equation}
and hence these $w_i$ ($i=1,2$) act on the Picard lattice as follows:
\begin{equation}\label{eqn:action_w1w2_Picard}
\begin{split}
 w_1(H_1)=H_1,\ w_1(H_2)=4H_1 + H_2 - E_1-\cdots-E_8,\ w_1(E_i) = H_1-E_i\ (i=1,\ldots,8),\\
 w_2(H_1)=H_1 + 4H_2 - E_1-\cdots-E_8,\ w_2(H_2)=H_2,\ w_2(E_i) = H_2-E_i\ (i=1,\ldots,8).
\end{split}
\end{equation}
One can immediately verify that the product $w_2w_1$ actually gives $T_{\alpha_1}$ by 
comparing this with \eqref{eqn:T1_lattice}. 
\begin{rem}\label{rem:flips}\rm
The elliptic Painlev\'e equation in the direction of $\alpha_1$ can be regarded as a non-autonomous
version of the QRT mapping (see Section \ref{subsec:QRT}). The two involutions $w_1$ and $w_2$
correspond to the vertical and horizontal flips, respectively. In fact, noticing from
\eqref{eqn:alpha_E8} that $s_2=r_{H_2-E_1-E_2}$ and applying permutations $\mathfrak{S}_8=\langle
s_0,s_3,\ldots,s_8\rangle$ on \eqref{eqn:s2_on_Z}, we see that
\begin{equation}\label{eqn:s2_gen_on_Z}
r_{H_2-E_i-E_j}\left(\frac{Z-Z_i}{Z-Z_j}\right)=\frac{Z-Z_i}{Z-Z_j}\ \frac{W-W_j}{W-W_i},\quad
r_{H_2-E_i-E_j}(W)=W.
\end{equation}
This implies that $w_2(W)=W$, namely $w_2$ corresponds to the horizontal flip.  Similarly, we see that
$r_{H_1-E_i-E_j}$ leaves $Z$ invariant and thus $w_1$ corresponds to the vertical flip.
\end{rem}

As to the description of \eqref{eqn:wxy} for $w=T_{\alpha_1}$, several expressions are
known in the literature \cite{Murata:FE,MSY:Riccati,ORG:e8}. Here we derive a new explicit
expression based on the representation of the affine Weyl group $W(E_8^{(1)})$ discussed 
in the previous sections.

In view of \eqref{eqn:T1_lattice}, \eqref{eqn:T1_inverse}, let us compute $T_{\alpha_1}(\xi(t))$ and
$T^{-1}_{\alpha_1}(\eta(t))$ with a $W(E_8^{(1)})$-invariant parameter $t$.  Note that we have from
\eqref{eqn:xieta12only}, \eqref{eqn:def_xi_a} and \eqref{eqn:action_w1w2_Picard}
\begin{equation}\label{eqn:flip_xi_eta}
\begin{split}
&w_1(\xi(t))=\xi(t),\quad w_2(\eta(t))=\eta(t),\\
& T_{\alpha_1}(\xi(t))=w_2w_1(\xi(t)) = w_2(\xi(t)),\quad
 T_{\alpha_1}^{-1}(\eta(t))=w_1w_2(\eta(t)) = w_1(\eta(t)). 
\end{split}
\end{equation}
For notational convenience, we write
\begin{equation}
 \overline{F} = T_{\alpha_1}(F),\quad  \underline{F} = T_{\alpha_1}^{-1}(F),
\end{equation}
for any function $F$. Applying formula \eqref{eqn:x_xi} with $w=T_{\alpha_1}$, we have from
\eqref{eqn:T1_lattice}
\begin{equation}
 T_{\alpha_1}\left(\xi(t)\right) = 
\overline{\xi}(t)=P(\xi,\eta;t)(\tau_1\cdots\tau_8)^{-1}\in L(H_1+4H_2-E_1-\cdots-E_8),
\end{equation}
where $P(\xi,\eta;t)=P_{T_{\alpha_1}}(\xi,\eta;t)$. In order to obtain an explicit and compact
expression for $P(\xi,\eta;t)$, we use the linear functions (see \eqref{eqn:def_xi_a})
\begin{equation}
\xi_j = \xi(e_j),\quad \eta_j=\eta(e_j)\quad (j=1,\ldots,9),
\end{equation}
that have a zero at ${\rm P}_j$, where we set $e_9=t$ regarding $t$ as the coordinate of ${\rm P}_9$ on $C_0$.
Noting that $P(\xi,\eta;t)$ is a homogeneous polynomial of bidegree $(1,4)$, we use $\eta_5$, $\eta_6$,
$\eta_7$, $\eta_8$, $\eta_9$ to form the basis
\begin{align}
\eta_5\cdots\widehat{\eta_k}\cdots\eta_9\qquad (k=5,\ldots,9),
\end{align}
of five polynomials for the homogeneous polynomials of degree $4$ in $\eta$.  We remark that instead
of ${\rm P}_5, {\rm P}_6, {\rm P}_7, {\rm P}_8$, one may use any four points among ${\rm
P}_1,\ldots, {\rm P}_8$.  Since the polynomial $P(\xi,\eta;t)$ vanishes at ${\rm P}_5,\ldots, {\rm
P}_9$, it takes the form
\begin{align}\label{eqn:P}
P(\xi,\eta;t)&=\sum_{k=5}^{9}c_k(t) \xi_k
\eta_5\cdots\widehat{\eta_k}\cdots\eta_9.
\end{align}
The coefficient $c_k(t)$ is determined by the normalization condition \eqref{eqn:P_normalization}
\begin{equation}\label{eqn:P_normalization2}
 P(\xi,\eta;t)\Bigr|_{C_0} = [T_{\alpha_1}(h_1)-t-u][t-u]\prod_{i=1}^8 [e_i-u]
=\prod_{i=0}^9 [e_i-u],
\end{equation}
by specializing at $u=h_2-e_k$, and we have
\begin{equation}\label{eqn:c_elliptic}
\begin{split}
 c_k(t) &= 
\dfrac{\prod\limits_{i=0}^{4}[h_2-e_i-e_k]}
{[h_2-h_1]\prod\limits_{{5\le j\le 9}\atop{j\ne k}}^{}[e_j-e_k]}\quad (k=5,\ldots,9) ,
\end{split}
\end{equation}
where $e_9=t$ and $e_0=\delta-h_1+2h_2-t$.  A similar formula for the polynomial
\begin{equation}
 T_{\alpha_1}^{-1}(\eta(t))=\underline{\eta}(t)
=Q(\xi,\eta,t)(\tau_1\cdots\tau_8)^{-1}\in L(4H_1+H_2-E_1-\cdots-E_8),
\end{equation}
can be obtained by applying the simple reflection $s_1=s_{\alpha_1}$ to $T_{\alpha_1}(\xi(t))$
as (see \eqref{eqn:E8_parameters_and_fg}, \eqref{eqn:E8_homogeneous})
\begin{align}\label{eqn:Q}
Q(\xi,\eta;t)&=\sum_{k=5}^{9}d_k(t) \eta_k
\xi_5\cdots\widehat{\xi_k}\cdots\xi_9, \quad
d_k(t)=c_k(t)|_{h_1 \leftrightarrow h_2}.
\end{align}

Using the polynomials $P,Q$ in \eqref{eqn:P} and \eqref{eqn:Q}, the {\em elliptic Painlev\'e
equation} $e$-P$(E_8^{(1)})$ is expressed as
\begin{equation}\label{eqn:elliptic_Painleve}
T_{\alpha_1}\left(\frac{\xi(t)}{\xi(s)}\right) = \frac{P(\xi,\eta;t)}{P(\xi,\eta;s)},\quad
T_{\alpha_1}^{-1}\left(\frac{\eta(t)}{\eta(s)}\right) = \frac{Q(\xi,\eta;t)}{Q(\xi,\eta;s)}.
\end{equation}
Equation \eqref{eqn:elliptic_Painleve} gives a general form of the elliptic Painlev\'e equation in
the direction of the root $\alpha_1$,
\begin{equation}\label{eqn:T1_pars}
\begin{split}
&T_{\alpha_1}(h_1)=h_1 - 2(h_1-h_2) + \delta,\quad 
 T_{\alpha_1}(h_2)=h_2 - 2(h_1-h_2)+3\delta ,\\
&T_{\alpha_1}(e_i)=e_i - (h_1-h_2) + \delta,\\
&T_{\alpha_1}^{-1}(h_1)=h_1 + 2(h_1-h_2) + 3\delta,\quad 
 T_{\alpha_1}^{-1}(h_2)=h_2 + 2(h_1-h_2)+\delta ,\\
&T_{\alpha_1}^{-1}(e_i)=e_i + (h_1-h_2) + \delta\quad (i=1,\ldots,8). 
\end{split}
\end{equation}

It is possible to derive a simple expression of the elliptic Painlev\'e equation in terms of the
coordinates $(Z,W)$. From \eqref{eqn:8points_par_ORG}, \eqref{eqn:symmetry_xi_eta} and
\eqref{eqn:Z_minus_varphi}, one can introduce two parameters $t_W$ and $s_W=h_2-t_W$ such that
$W=\psi(t_W)=\psi(s_W)$.  By noting that $h_2$ and $W$ are $w_2$-invariant from \eqref{eqn:w1w2},
Remark \ref{rem:flips} and \eqref{eqn:flip_xi_eta}, these parameters are also chosen to be
$w_2$-invariant.  Since \eqref{eqn:Z_minus_varphi} implies $\eta(t_W)=\eta(s_W)=0$, we have from
\eqref{eqn:P}
\begin{equation}
 P(\xi,\eta;t_W) = c_9(t_W)\xi(t_W)\eta_4\cdots\eta_8,\quad
 P(\xi,\eta;s_W) = c_9(s_W)\xi(s_W)\eta_4\cdots\eta_8.
\end{equation}
The right hand side of \eqref{eqn:elliptic_Painleve} is thus drastically simplified as
\begin{equation}\label{eqn:eP_xi}
w_2\left(\dfrac{\xi(t_W)}{\xi(s_W)}\right)
=\dfrac{\xi(t_W)}{\xi(s_W)}\prod_{i=1}^8 \dfrac{[e_i-s_W]}{[e_i-t_W]}. 
\end{equation}
Similarly, by introducing $w_1$-invariant parameters $t_Z$ and $s_Z=h_1-t_Z$ such that 
$Z=\phi(t_Z)=\phi(s_Z)$, we have
\begin{equation}\label{eqn:eP_eta}
w_1\left(\dfrac{\eta(t_Z)}{\eta(s_Z)}\right)
=\dfrac{\eta(t_Z)}{\eta(s_Z)}\prod_{i=1}^8 \dfrac{[e_i-s_Z]}{[e_i-t_Z]}. 
\end{equation}
Using the relations \eqref{eqn:Z_minus_varphi}, one can rewrite \eqref{eqn:eP_xi} and
\eqref{eqn:eP_eta} in terms of $(Z,W)$ coordinates as
\begin{equation}\label{eqn:elliptic_Painleve_3}
\begin{split}
&\frac{[\overline{h}_1-a-t_W]}{[\overline{h}_1-a-s_W]}
\frac{\overline{Z}-\overline{\varphi}(t_W)}{\overline{Z}-\overline{\varphi}(s_W)}
=\frac{[h_1-a-t_W]}{[h_1-a-s_W]}
\frac{Z-\varphi(t_W)}{Z-\varphi(s_W)}\prod_{i=1}^8\frac{[e_i-s_W]}{[e_i-t_W]},\\
& \frac{[\underline{h}_2-a-t_Z]}{[\underline{h}_2-a-s_Z]}
\frac{\underline{W}-\underline{\psi}(t_Z)}{\underline{W}-\underline{\psi}(s_Z)}
=\frac{[h_2-a-t_Z]}{[h_2-a-s_Z]}
\frac{W-\psi(t_Z)}{W-\psi(s_Z)}\prod_{i=1}^8\frac{[e_i-s_Z]}{[e_i-t_Z]},
\end{split}
\end{equation}
which is an alternative form of the elliptic Painlev\'e equation $e$-P$(E_8^{(1)})$.
%
\begin{rem}\rm
In a similar way one can show that 
\begin{equation}\label{eqn:s2_theta}
r_{H_2-E_i-E_j}\left(\frac{[h_1-a-t_W]}{[h_1-a-s_W]}
\frac{Z-\varphi(t_W)}{Z-\varphi(s_W)}\right)
=
\frac{[h_1-a-t_W]}{[h_1-a-s_W]}
\frac{Z-\varphi(t_W)}{Z-\varphi(s_W)}
\frac{[e_i-s_W]}{[e_i-t_W]}\frac{[e_j-s_W]}{[e_j-t_W]},
\end{equation}
with the parameters $t_W$ and $s_W=h_2-t_W$ such that $W=\psi(s_W)=\psi(t_W)$. Then it is possible
to derive the first equation in \eqref{eqn:elliptic_Painleve_3} by compositions of
\eqref{eqn:s2_theta} \cite{NTY}.
\end{rem}
\subsection{Degeneration to $q$- and d-${\rm P}(E_8^{(1)})$ cases}
\subsubsection{Degeneration to $q$-${\rm P}(E_8^{(1)})$}\label{subsec:qe8}
By choosing $[u]$ as trigonometric and rational functions, the bidegree $(2,2)$-curve $C_0$ on
$\mathbb{P}^1\times\mathbb{P}^1$ passing through the eight points ${\rm P}_1,\ldots, {\rm P}_8$
becomes nodal and cuspidal curve respectively. Those cases give rise to discrete Painlev\'e
equations $q$-${\rm P}(E_8^{(1)})$ and d-${\rm P}(E_8^{(1)})$ respectively.

For the description of those cases, it is convenient to change the parameters and parametrization
of points. We first recall the parametrization of the curve $C_0$ on
$\mathbb{P}^1\times\mathbb{P}^1$ \eqref{eqn:8points_par_ORG}:
\begin{equation}\label{eqn:8points_par_ORG2}
(Z,W)=\left(\varphi(u),\psi(u)\right),\quad
\varphi(u)=\frac{[b-u][h_1-b-u]}{[a-u][h_1-a-u]},\quad
\psi(u)=\frac{[b-u][h_2-b-u]}{[a-u][h_2-a-u]},
\end{equation}
where we take $ [x] = (e^{x/2}-e^{-x/2})/2$ in $q$-${\rm P}(E_8^{(1)})$ case. Then we have
\begin{equation}\label{eqn:curve_parametrization_qe8}
\varphi(u) = \frac{f(u)-f(b)}{f(u)-f(a)},\ 
\psi(u) = \frac{g(u)-g(b)}{g(u)-g(a)},\ f(u)=e^{u} + e^{h_1-u},\ 
g(u)=e^{u} + e^{h_2-u}.
\end{equation}
Since $\varphi(u)$ and $\psi(u)$ are fractional linear transforms of $f(u)$ and $g(u)$ respectively,
it is convenient to use $f(u)$ and $g(u)$ for parametrization of $C_0$.  

The above expression is in terms of ``additive'' parameters $h_i$ ($i=1,2$), $u$, $e_i$
($i=1,\ldots,8$). For the case of $q$-difference Painlev\'e equations, we use the same
symbols as ``multiplicative'' parameters. Then the curve $C_0$ 
and the eight points ${\rm P}_i$ ($i=1,\ldots,8$) on $C_0$ are parametrized as
\begin{equation}
(x,y) = (f(u), g(u)),\quad  f(u)  = u + \frac{h_1}{u},\quad  g(u)  = u + \frac{h_2}{u},\label{eqn:qpe8_curve_parameterization}
\end{equation} 
\begin{equation}\label{eqn:qpe8_8points}
 {\rm P}_i:\ (x_i,y_i) = (f(e_i), g(e_i))\quad (i=1,\ldots,8),
\end{equation}
respectively, and the $(2,2)$-curve $C_0$ passing through ${\rm P}_i$ is given by
\begin{equation}\label{eqn:C_0_qe8}
 (x-y)(h_2x - h_1y) + (h_1-h_2)^2=0.
\end{equation}

The inhomogeneous coordinates $(Z,W)$ and $(x,y)$ are related as
\begin{equation}\label{eqn:ZW_and_xy}
 Z = \frac{x-f(b)}{x-f(a)},\quad  W = \frac{y-g(b)}{y-g(a)}.
\end{equation}
From \eqref{eqn:E8_linear_parameters}, \eqref{eqn:s1_on_Z} and \eqref{eqn:s2_on_Z}, the Weyl group
$W(E_8^{(1)})$ acts on these multiplicative parameters as
\begin{equation}\label{eqn:E8_linear_parameters_multi}
 \begin{split}
  s_0:\quad& e_1\ \leftrightarrow\ e_2,\\
  s_1:\quad& h_1\ \leftrightarrow\ h_2,\quad x\ \leftrightarrow\ y,\\
  s_2:\quad& e_1\ \rightarrow\ \frac{h_2}{e_2},\quad e_2\ \rightarrow\ \frac{h_2}{e_1},
\quad h_1\ \rightarrow\ \frac{h_1 h_2}{e_1e_2},\quad
x\ \rightarrow \tilde{x},\\
  s_j:\quad& e_{j-1}\ \leftrightarrow\ e_{j}\quad (j=3,\ldots,8),
 \end{split}
\end{equation}
where $\tilde{x}$ is determined by
\begin{equation}
\frac{\tilde{x}-s_2(x_1)}{\tilde{x}-s_2(x_{2})} = \frac{x-x_1}{x-x_{2}}\frac{y-y_{2}}{y-y_1}.
\end{equation}
We also use $q$ given by
\begin{equation}
 q=e^\delta = \frac{h_1^2h_2^2}{e_1\cdots e_8}.
\end{equation}
as the multiplicative parameter for the null root $\delta$. The Weyl group actions
for the $\tau$ variables are the same as \eqref{eqn:E8_tau}, where 
the $s_3$ action simplifies to
\begin{equation}\label{eqn:qe8_action_tau}
\begin{array}{cl}\medskip
&{\displaystyle \tau(E_2)\ \leftrightarrow\ \tau(E_3),}\\
\medskip
 s_3:&{\displaystyle  \tau(H_1-E_2)\ \rightarrow\ 
\frac{x_3-x_2}{x_1-x_2}\frac{\tau(E_1)\tau(H_1-E_1)}{\tau(E_3)}
+  \frac{x_3-x_1}{x_2-x_1}\frac{\tau(E_2)\tau(H_1-E_2)}{\tau(E_3)}}, \\
&{\displaystyle  \tau(H_2-E_2)\ \rightarrow\ 
\frac{y_3-y_2}{y_1-y_2}\frac{\tau(E_1)\tau(H_2-E_1)}{\tau(E_3)}
+  \frac{y_3-y_1}{y_2-y_1}\frac{\tau(E_2)\tau(H_2-E_2)}{\tau(E_3)}.}
\end{array}
\end{equation}
By specializing \eqref{eqn:WZ_by_tau} to the trigonometric case and using \eqref{eqn:ZW_and_xy}, we
find that the inhomogeneous coordinates $(x,y)$ of $\mathbb{P}^1\times \mathbb{P}^1$ can be
expressed in terms of the $\tau$ variables as
\begin{equation}\label{eqn:xy_and_tau}
 \frac{x-x_j}{x-x_i}=\frac{\tau(E_j)\tau(H_1-E_j)}{\tau(E_i)\tau(H_1-E_i)},
\quad
 \frac{y-y_j}{y-y_i}=\frac{\tau(E_j)\tau(H_2-E_j)}{\tau(E_i)\tau(H_2-E_i)}\quad (i,j=1,\ldots,8).
\end{equation}

We next write down $q$-P$(E_8^{(1)})$. Under the translation
$T_{\alpha_1}$ in the direction of $\alpha_1$, the multiplicative parameters transform as (see \eqref{eqn:T1_pars})
\begin{equation}\label{eqn:T1_pars_multi}
\begin{split}
&\overline{h}_1=q\frac{h_2^2}{h_1},\quad 
\overline{h}_2= q^3\frac{h_2^3}{h_1^2} ,\quad
\overline{e}_i=qe_i\frac{h_2}{h_1},\\
&\underline{h}_1=q^3\frac{h_1^3}{h_2^2} ,\quad 
\underline{h}_2= q\frac{h_1^2}{h_2} ,\quad
\underline{e}_i=qe_i\frac{h_1}{h_2}\quad (i=1,\ldots,8). 
\end{split}
\end{equation}
Then we obtain in a similar manner to \eqref{eqn:elliptic_Painleve}
\begin{equation}\label{eqn:q_e8}
\frac{\overline{x}-\overline{f}(b)}{\overline{x}-\overline{f}(a)}
=\frac{P(x,y;b)}{P(x,y;a)}, \quad
\frac{\underline{y}-\underline{g}(b)}{\underline{y}-\underline{g}(a)}
=\frac{Q(x,y;b)}{Q(x,y;a)},
\end{equation}
where 
\begin{equation}\label{eqn:c_mult}
\begin{split}
&P(x,y;t)=\sum_{k=5}^{9}c_k (x-x_k)\prod\limits_{5\leq j\leq 9\atop j\neq k} (y-y_j),\quad
c_k= 
\dfrac{\prod\limits_{i=0}^{4}\left(1-\frac{h_2}{e_ie_k}\right)}
{\prod\limits_{{5\le j\le 9}\atop{j\ne k}}^{}\left(1-\frac{e_j}{e_k}\right)},
\end{split}
\end{equation}
with $e_9=t$, $e_0=q h_2^2/(t h_1)$, and $Q(x,y;t)=P(x,y;t)|_{x\leftrightarrow y, f \leftrightarrow
g, h_1 \leftrightarrow h_2}$.  Equation \eqref{eqn:q_e8} is the $q$-difference Painlev\'e equation
$q$-P$(E_8^{(1)})$. We also obtain an alternative form of $q$-P$(E_8^{(1)})$
\begin{equation}\label{eqn:q_e8_2}
\begin{split}
&\frac{\overline{x}-\overline{f}(t_y)}{\overline{x}-\overline{f}\Bigl(\frac{h_2}{t_y}\Bigr)}
=\frac{x-f(t_y)}{x-f\Bigl(\frac{h_2}{t_y}\Bigr)}\,\frac{t_y^8}{h_2^4}\,\prod_{i=1}^8\frac{e_i-\frac{h_2}{t_y}}{e_i-t_y},
\\
&\frac{\underline{y}-\underline{g}(t_x)}{\underline{y}-\underline{g}\Bigl(\frac{h_1}{t_x}\Bigr)}
=\frac{y-g(t_x)}{y-g\Bigl(\frac{h_1}{t_x}\Bigr)}\,\frac{t_x^8}{h_1^2}\,
\prod_{i=1}^8\frac{e_i-\frac{h_1}{t_x}}{e_i-t_x},
\end{split}
\end{equation}
with the parameters $t_y$, $t_x$ such that $y=g(t_y)$, $x=f(t_x)$ respectively,
by putting $b=t_y$, $a=\frac{h_2}{t_y}$ in the first equation in \eqref{eqn:q_e8}, and
similarly, by putting $b=t_x$, $a=\frac{h_2}{t_x}$  in the second equation. 

\subsubsection{Degeneration to d-${\rm P}(E_8^{(1)})$}
We simply take $[u]=u$. Then $\varphi(u)$ and $\psi(u)$ in \eqref{eqn:8points_par_ORG2} become
\begin{equation}\label{eqn:curve_parametrization_de8}
\varphi(u) = \frac{f(u)-f(b)}{f(u)-f(a)},\ 
\psi(u) = \frac{g(u)-g(b)}{g(u)-g(a)},\ f(u)=u(u-h_1),\
g(u)=u(u-h_2).
\end{equation}
We use the above $f(u)$ and $g(u)$ for parametrization of $C_0$ and the eight points ${\rm P}_i$
($i=1,\ldots,8$) on $C_0$:
\begin{equation}
(x,y) = (f(u), g(u)),\quad  {\rm P}_i:\ (x_i,y_i) = (f(e_i), g(e_i))\quad (i=1,\ldots,8).
\end{equation} 
Then $C_0$ is given by
\begin{equation}\label{eqn:C_0_de8}
 (x-y)^2 + (h_1-h_2)(h_2x - h_1 y) = 0.
\end{equation}
The inhomogeneous coordinates $(Z,W)$ and $(x,y)$ are related as
\begin{equation}\label{eqn:ZW_and_xy2}
 Z = \frac{x-f(b)}{x-f(a)},\quad  W = \frac{y-g(b)}{y-g(a)}.
\end{equation}
The Weyl group $W(E_8^{(1)})$ action on the parameters is the same as
\eqref{eqn:E8_linear_parameters}. From \eqref{eqn:s1_on_Z} and \eqref{eqn:s2_on_Z}, we obtain the 
action on $x$, $y$ as
\begin{equation}\label{eqn:E8_linear_variables_additive}
  s_1:\quad x\ \leftrightarrow\ y,\qquad   s_2:\quad x\ \rightarrow \tilde{x},
\end{equation}
where $\tilde{x}$ is determined by
\begin{equation}
\frac{\tilde{x}-s_2(x_1)}{\tilde{x}-s_2(x_{2})} = \frac{x-x_1}{x-x_{2}}\frac{y-y_1}{y-y_{2}}.
\end{equation}
The Weyl group actions for the $\tau$ variables has the same form as $q$-P$(E_8^{(1)})$ case, namely
they are given by \eqref{eqn:E8_tau}, \eqref{eqn:qe8_action_tau} with $x_i=e_i(e_i-h_1)$, $y_i=e_i(e_i-h_2)$.
The inhomogeneous coordinates $(x,y)$ of $\mathbb{P}^1\times \mathbb{P}^1$ can be
expressed in terms of the $\tau$ by the same formula as \eqref{eqn:xy_and_tau}.

We next write down d-P$(E_8^{(1)})$. By using \eqref{eqn:ZW_and_xy2} in
\eqref{eqn:elliptic_Painleve} (or taking $q\to 1$ limit in \eqref{eqn:q_e8}), we have
\begin{equation}\label{eqn:d_e8}
\frac{\overline{x}-\overline{f}(b)}{\overline{x}-\overline{f}(a)}
=\frac{P(x,y;b)}{P(x,y;a)}, \quad
\frac{\underline{y}-\underline{g}(b)}{\underline{y}-\underline{g}(a)}
=\frac{Q(x,y;b)}{Q(x,y;a)},
\end{equation}
where
\begin{equation}\label{eqn:c_add}
\begin{split}
&P(x,y;t)=\sum_{k=5}^{9}c_k (x-x_k)\prod\limits_{5\leq j\leq 9\atop j\neq k} (y-y_j),\quad
c_k= 
\dfrac{\prod\limits_{i=0}^{4}(h_2-e_i-e_j)}
{\prod\limits_{{5\le j\le 9}\atop{j\ne k}}^{}(e_j-e_k)},
\end{split}
\end{equation}
with $e_9=t$, $e_0=\delta+2 h_2-t -h_1$, and $Q(x,y;t)=P(x,y;t)|_{x\leftrightarrow y, f \leftrightarrow g, h_1 \leftrightarrow h_2}$.
Equation \eqref{eqn:d_e8} is the difference Painlev\'e equation d-P$(E_8^{(1)})$. We also have
an alternate expression of d-P$(E_8^{(1)})$ as
\begin{equation}\label{eqn:d_e8_2}
\begin{split}
&
\frac{\overline{x}-\overline{f}(t_y)}{\overline{x}-\overline{f}(h_2-t_y)}
=\frac{x-f(t_y)}{x-f(h_2-t_y)}\prod_{i=1}^8\frac{h_2-e_i-t_y}{e_i-t_y},
\\
&\frac{\underline{y}-\underline{g}(t_x)}{\underline{y}-\underline{g}(h_1-t_x)}
=\frac{y-g(t_x)}{y-g(h_1-t_x)}
\prod_{i=1}^8\frac{h_1-e_i-t_x}{e_i-t_x},
\end{split}
\end{equation}
with the parameters $t_y$, $t_x$ such that $y=g(t_y)$, $x=f(t_x)$ respectively.

\subsubsection{Relation to the ORG form}
Let us finally mention on the relations between $q$-P$(E_8^{(1)})$ \eqref{eqn:q_e8},
d-P$(E_8^{(1)})$ \eqref{eqn:d_e8} and the $q$-difference and difference Painlev\'e equations with
$W(E_8^{(1)})$-symmetry derived by Ohta-Ramani-Grammaticos \cite{ORG:e8}.  For $x$, $y$ satisfying
$q$-P$(E_8^{(1)})$ \eqref{eqn:q_e8}, it is possible to verify
\begin{equation}\label{eqn:ORG_qe8}
\begin{split}
&\frac{(x-y)(h_2{\overline x}-{\overline h_1} y)-(h_1-h_2)(h_2-{\overline h_1})}
{({\overline x}-y)(h_2 x-h_1 y)-({\overline h_1}-h_2)(h_2-h_1)}
=\frac{A(h_2,y)}{B(h_2,y)},\\
& \frac{(y-x)(h_1{\underline y}-{\underline h_2} x)-(h_2-h_1)(h_1-{\underline h_2})}
{({\underline y}-x)(h_1 y-h_2 x)-({\overline h_2}-h_1)(h_1-h_2)}
=\frac{A(h_1,x)}{B(h_1,x)}.
\end{split}
\end{equation}
Here $A(h,z)$ and $B(h,z)$ are given by
\begin{equation}\label{eqn:AB}
 \begin{split}
  A(h,z) &= m_0 z^4 - m_1 z^3 + \left(-3hm_0+ m_2-h^{-3}m_8\right) z^2 \\
& + \left(2hm_1 - m_3 + h^{-2}m_7\right)z + \left(h^2m_0-hm_2 + m_4-h^{-1}m_6 + h^{-2}m_8\right),\\
B(h,z) &= h^{-4}m_8z^4 - h^{-3}m_7z^3 + \left(-hm_0 + h^{-2}m_6 - 3h^{-3}m_8\right) z^2\\
& + \left(hm_1 - h^{-1}m_5 + 2h^{-2}m_7\right)z
+ \left(h^2m_0 - hm_2 + m_4 - h_2^{-1}m_6 + h^{-2}m_8\right),
 \end{split}
\end{equation}
where $m_i$ ($i=0,\ldots,8$) defined by $U(z)= z^{-4}\prod\limits_{i=1}^8 (e_i-z)=
z^{-4}\sum\limits_{i=0}^8 m_{8-i} (-z)^i$; $m_{i}$ are the $i$-th elementary symmetric
polynomials of $e_1,\ldots, e_8$.  Equation \eqref{eqn:ORG_qe8} is equivalent to the the
$q$-difference Painlev\'e equation with $W(E_8^{(1)})$-symmetry derived by
Ohta-Ramani-Grammaticos (\cite[formula (4.6a), (4.6b)]{ORG:e8}).  We note that the polynomials in the left hand sides of
\eqref{eqn:ORG_qe8} are regarded as analogues of the curve $C_0$ \eqref{eqn:C_0_qe8} passing through
the eight points ${\rm P}_i$.

Equations \eqref{eqn:ORG_qe8} can be derived as follows. Solving the first equation of
\eqref{eqn:q_e8_2} in terms of $\overline{x}$ and substituting it into the left hand side of the
first equation of \eqref{eqn:ORG_qe8}, we have
\begin{equation}
\frac{(x-y)(h_2{\overline x}-{\overline h_1} y)-(h_1-h_2)(h_2-{\overline h_1})}
{({\overline x}-y)(h_2 x-h_1 y)-({\overline h_1}-h_2)(h_2-h_1)}
= \frac{uU(u)-\frac{h_2}{u}U\left(\frac{h_2}{u}\right)} 
{uU\left(\frac{h_2}{u}\right)-\frac{h_2}{u}U\left(u\right)} ,
\end{equation}
where $u=t_y$ is a parameter such that $y=u+\frac{h_2}{u}$.  By virtue of the symmetry with respect
to interchanging $u\leftrightarrow \frac{h_2}{u}$ of the right hand side, there exist functions
$A(h,z)$ and $B(h,z)$ which are polynomials in $z$ such that
\begin{equation}
 \begin{split}
\left(u-\frac{h_2}{u}\right)A\left(h_2,u+\frac{h_2}{u}\right)=u U(u)-\frac{h_2}{u}U\left(\frac{h_2}{u}\right),\\
\left(u-\frac{h_2}{u}\right)B\left(h_2,u+\frac{h_2}{u}\right)=u U\left(\frac{h_2}{u}\right)-\frac{h_2}{u}U(u).
 \end{split}
\end{equation}
By solving $A$ and $B$ from these equations we obtain \eqref{eqn:AB}, which gives the first equation of 
\eqref{eqn:ORG_qe8}. The second equation is derived in a similar manner.

Similarly, d-P$(E_8^{(1)})$ \eqref{eqn:d_e8} is shown to be equivalent to the difference Painlev\'e
equation with $W(E_8^{(1)})$-symmetry obtained in \cite{ORG:e8}.
%
\begin{rem}\label{rem:kappa-var}\rm
It is sometimes convenient to change the parameters $(h_1,h_2,e_1,\ldots,e_8,u)$ to
$(\kappa_1,\kappa_2,v_1,\ldots,v_8,v)$ 
\begin{equation}\label{eqn:kappa-var}
\kappa_i = h_i - 2\lambda, \quad v_i = e_i-\lambda, \quad v=u-\lambda,
 \end{equation}
by introducing a new parameter $\lambda$ such that 
\begin{equation}
 s_2(\lambda) = \lambda + \frac{h_1-e_1-e_2}{4},\quad s_j(\lambda)=\lambda\quad (j\neq 2).
\end{equation}
Note that $c=\lambda-\frac{h_1+h_2}{4}$ is a central element ($W(E_8^{(1)})$-invariant).  Then
\eqref{eqn:E8_linear_parameters} and \eqref{eqn:T1_pars} yield
 \begin{equation}\label{eqn:E8_for_d}
 \begin{split}
  s_0:\quad& v_1\ \leftrightarrow\ v_2,\\
  s_1:\quad& \kappa_1\ \leftrightarrow\ \kappa_2,\\
  s_2:\quad& \kappa_1\ \leftrightarrow\ \kappa_1+2\mu,\quad 
\kappa_2\ \leftrightarrow\ \kappa_2-2\mu,\\
 & v_1\ \leftrightarrow\ v_1+3\mu,\quad v_2\ \leftrightarrow\ v_2+3\mu,\quad 
v_j\ \leftrightarrow\ v_j-\mu\quad (j=3,\ldots,8),\\
& v\ \leftrightarrow\ v-\mu,\quad  \lambda\ \leftrightarrow \lambda+\mu,\quad 
\mu = \frac{\kappa_2-v_1-v_2}{4},\\
  s_j:\quad& v_{j-1}\ \leftrightarrow\ v_{j}\quad (j=3,\ldots,8),
 \end{split}
\end{equation}
and
\begin{equation}\label{eqn:T1_pars3}
\begin{split}
& T_{\alpha_1}(\kappa_1)=\kappa_1-\delta,\quad T_{\alpha_1}(\kappa_2)=\kappa_2 + \delta,
\quad T_{\alpha_1}(\lambda)=\lambda - \kappa_1+\kappa_2+\delta,\\
& T_{\alpha_1}(v_i)=v_i\quad (i=1,\ldots,8),\quad T_{\alpha_1}(v) = v+\kappa_1-\kappa_2-\delta,
\end{split}
\end{equation}
respectively. 
It is convenient to use $(\kappa_1,\kappa_2,v_1,\ldots,v_8,v)$ to describe
$q$-P$(E_8^{(1)})$ as a difference equation, while $(h_1,h_2,e_1,\ldots,e_8,u)$ is convenient for
the description of underlying symmetry.

Similar change of parameters $(h_1,h_2,e_1,\ldots,e_8,u)$ to
$(\kappa_1,\kappa_2,v_1,\ldots,v_8,v)$ are also used in multiplicative cases:
\begin{equation}
\kappa_i = \frac{h_i}{\lambda^2}, \quad v_i = \frac{e_i}{\lambda}, \quad v=\frac{u}{\lambda}, \label{eqn:parameter_kappa}
 \end{equation}
where $\lambda$ is a parameter such that
$s_2(\lambda)=\lambda\left(\frac{\kappa_1}{v_1v_2}\right)^{\frac{1}{4}}$ and $s_j(\lambda)=\lambda$
$(j\neq 2)$.  Note that $c=\frac{\lambda}{\left(h_1h_2\right)^{\frac{1}{4}}}$ is a central element.
Then \eqref{eqn:E8_linear_parameters_multi} and \eqref{eqn:T1_pars_multi} yield
\begin{equation}\label{eqn:E8_for_q}
\begin{split}
 s_0:\quad& v_1\ \leftrightarrow\ v_2,\\
 s_1:\quad& \kappa_1\ \leftrightarrow\ \kappa_2,\\
 s_2:\quad& \kappa_1\ \leftrightarrow\ \kappa_1\mu^2,\quad 
                   \kappa_2\ \leftrightarrow\ \frac{\kappa_2}{\mu^2},\quad
                   v_1\ \leftrightarrow\ v_1\mu^3,\quad 
                   v_2\ \leftrightarrow\ v_2\mu^3,\\
               & v_j\ \leftrightarrow\ \frac{v_j}{\mu}\quad (j=3,\ldots,8),\quad
                  v\ \leftrightarrow\ \frac{v}{\mu}, \quad
\lambda\ \leftrightarrow\ \lambda\mu,\quad
                   \mu = \left(\frac{\kappa_2}{v_1v_2}\right)^{\frac{1}{4}},\\
  s_j:\quad& v_{j-1}\ \leftrightarrow\ v_{j}\quad (j=3,\ldots,8),
 \end{split}
\end{equation}
and
\begin{equation}\label{eqn:T1_pars2}
\begin{split}
& T_{\alpha_1}(\kappa_1)=\frac{\kappa_1}{q}\quad T_{\alpha_1}(\kappa_2)=q\kappa_2,\quad
T_{\alpha_1}(\lambda) = \lambda\frac{q\kappa_2}{\kappa_1},\\
& T_{\alpha_1}(v_i)=v_i\quad (i=1,\ldots,8),\quad T_{\alpha_1}(v) = v\frac{\kappa_1}{q\kappa_2} ,
\end{split}
\end{equation}
respectively. 
\end{rem}

Under the parametrization given in Remark \ref{rem:kappa-var}, we rewrite $e$-P($E_8^{(1)}$)
\eqref{eqn:elliptic_Painleve_3}, i.e.,
\begin{equation}\label{eqn:elliptic_Painleve_4}
\begin{split}
&\frac{\varphi_{\overline{h}_1}(a,t_W)\overline{Z} - \varphi_{\overline{h}_1}(b,t_W)}
{\varphi_{\overline{h}_1}(a,s_W)\overline{Z} - \varphi_{\overline{h}_1}(b,s_W)}
=\frac{\varphi_{h_1}(a,t_W) Z  - \varphi_{h_1}(b,t_W)}
{\varphi_{h_1}(a,s_W) Z  - \varphi_{h_1}(b,s_W)}\prod_{i=1}^8\frac{[e_i-s_W]}{[e_i-t_W]},\\
& 
\frac{\varphi_{\underline{h}_2}(a,t_Z)\underline{W} - \varphi_{\underline{h}_2}(b,t_Z)}
{\varphi_{\underline{h}_2}(a,s_Z)\underline{W} - \varphi_{\underline{h}_2}(b,s_Z)}
=\frac{\varphi_{h_2}(a,t_Z) W  - \varphi_{h_2}(b,t_Z)}
{\varphi_{h_2}(a,s_Z) W  - \varphi_{h_2}(b,s_Z)}\prod_{i=1}^8\frac{[e_i-s_Z]}{[e_i-t_Z]}.
\end{split}
\end{equation}
In terms of the parameters in \eqref{eqn:kappa-var}
together with
\begin{equation}
 a = \alpha+\lambda,\quad  b = \beta+\lambda,\quad 
s_W = u_W +\lambda,\quad t_W = v_W +\lambda,\quad
s_Z = u_Z +\lambda,\quad t_Z = v_Z +\lambda,
\end{equation}
so that $u_W=\kappa_2-v_W$, $u_Z=\kappa_1-v_Z$ hold, we have
\begin{equation}\label{eqn:elliptic_Painleve_5}
\begin{split}
&\frac{\varphi_{\overline{\kappa}_1}(\overline{\alpha},u_W)\overline{Z} - \varphi_{\overline{\kappa}_1}(\overline{\beta},u_W)}
{\varphi_{\overline{\kappa}_1}(\overline{\alpha},v_W)\overline{Z} - \varphi_{\overline{\kappa}_1}(\overline{\beta},v_W)}
=\frac{\varphi_{\kappa_1}(\alpha,v_W) Z  - \varphi_{\kappa_1}(\beta,v_W)}
{\varphi_{\kappa_1}(\alpha,u_W) Z  - \varphi_{\kappa_1}(\beta,u_W)}
\prod_{i=1}^8\frac{[v_i-u_W]}{[v_i-v_W]},\\
&\frac{\varphi_{\underline{\kappa}_2}(\underline{\alpha},u_Z)\underline{W} 
- \varphi_{\underline{\kappa}_2}(\underline{\beta},u_Z)}
{\varphi_{\underline{\kappa}_2}(\underline{\alpha},v_Z)\underline{W} 
- \varphi_{\underline{\kappa}_2}(\underline{\beta},v_Z)}
= \frac{\varphi_{\kappa_2}(\alpha,v_Z) W  - \varphi_{\kappa_2}(\beta,v_Z)}
{\varphi_{\kappa_2}(\alpha,u_Z) W  - \varphi_{\kappa_2}(\beta,u_Z)}
\prod_{i=1}^8\frac{[v_i-u_Z]}{[v_i-v_Z]},\\
&\hskip60pt W=\dfrac{\varphi_{\kappa_2}(\beta,v_W)}{\varphi_{\kappa_2}(\alpha,v_W)},\quad 
Z=\dfrac{\varphi_{\kappa_1}(\beta,v_Z)}{\varphi_{\kappa_1}(\alpha,v_Z)},
\end{split}
\end{equation}
where we have used the following relations
\begin{equation}\label{eqn:varphi_h_kappa}
\begin{split}
&\varphi_{h_1}(a,t_W) = \varphi_{\kappa_1}(\alpha,v_W),\quad 
\varphi_{h_1}(a,h_2-t_W) = \varphi_{\kappa_1}(\alpha,\kappa_2-v_W), \\
& \varphi_{\overline{h}_1}(a,t_W)= \varphi_{\overline{\kappa}_1}(\overline{\alpha},\kappa_2-v_W),\quad
 \varphi_{\overline{h}_1}(a,h_2-t) = \varphi_{\overline{\kappa}_1}(\overline{\alpha},v_W).
\end{split}
\end{equation}
For example, the third equation in \eqref{eqn:varphi_h_kappa} can be verified as follows.
Since $\overline{h}_1=-h_1+2h_2+\delta=-\kappa_1+2\kappa_2+\delta+2\lambda$, we have
\begin{displaymath}
\begin{split}
 \varphi_{\overline{h}_1}(a,t_W)&=[a-t_W][\overline{h}_1-a-t_W] 
= [\alpha-v_W][-\kappa_1+2\kappa_2+\delta-\alpha-v_W]\\
&=[\overline{\kappa}_1-\overline{\alpha}-(\kappa_2-v_W)][\overline{\alpha}-(\kappa_2-v_W)]
=\varphi_{\overline{\kappa}_1}(\overline{\alpha},\kappa_2-v_W),
\end{split}
\end{displaymath}
where we have used $\overline{\alpha}=\overline{a}-\overline{\lambda}=\alpha+\kappa_1-\kappa_2-\delta$.

%
\subsection{Birational representation of affine Weyl groups}\label{subsec:Weyl_group}
In this section, we discuss how to construct an explicit birational representation of the symmetry
group of the surface characterized by a given point configuration, which is generated by simple
reflections and lattice isomorphisms (Dynkin diagram automorphisms).
We demonstrate the procedure by taking the case of the symmetry type/surface type
$A_4^{(1)}/A_4^{(1)}$. 

There are several methods to construct the birational representation of the affine Weyl group
associated with a given symmetry type/surface type. One is to trace the procedures of blowing up and
blowing down according to the transformations of the Picard lattice \cite{Sakai:SIV}.  Another way
is to consider the degeneration of the birational representation of the generic
$E_8^{(1)}/A_0^{(1)}$ case to the given symmetry/surface type according to the scheme that is
explained in Section \ref{subsec:degeneration}. In this section, we explain another direct way based
on the principle in Remark \ref{rem:w(f)}.

We use the multiplicative parameters $h_1$, $h_2$, $e_1,\ldots,e_8$.  On these parameters the simple
reflections act multiplicatively in the same way as they do on the basis of the Picard lattice
$H_1$, $H_2$, $E_1,\ldots, E_8$. As for the variables $f$ and $g$, we construct the birational
action on them according to the guiding principle as mentioned in Remark \ref{rem:w(f)}; for each
element $w$ of the affine Weyl group, $w(f)$ and $w(g)$ should be rational functions in the class
$w(H_1)$ and $w(H_2)$, respectively.

\begin{figure}[th]
\begin{center}
\setlength{\unitlength}{0.72mm}
\footnotesize{
\begin{picture}(50,50)(0,0)
\put(5,10){\line(1,0){42}}
\put(-11,8){{\large $g=0$}}
\put(5,40){\line(1,0){42}}
\put(-11,38){{\large $g=\infty$}}
\put(10,5){\line(0,1){40}}
\put(3,0){{\large $f=0$}}
\put(40,5){\line(0,1){40}}
\put(33,0){{\large $f=\infty$}}
\put(20,10){\circle*{1.5}}\put(30,10){\circle*{1.5}}
\put(18,13){{\large $7$}} \put(28,13){{\large $8$}}
\put(20,40){\circle*{1.5}}\put(40,40){\circle{3}}
\put(18,43){\large{$4$}}\put(41,43){\large{$23$}}
\put(10,20){\circle*{1.5}\quad \large{$6$}}\put(10,30){\circle*{1.5}\quad \large{$5$}}
\put(40,20){\circle*{1.5}\quad {\large $1$}}\put(40,40){\circle*{1.5} ${}$}
\end{picture} }
\qquad
\includegraphics[scale=0.3]{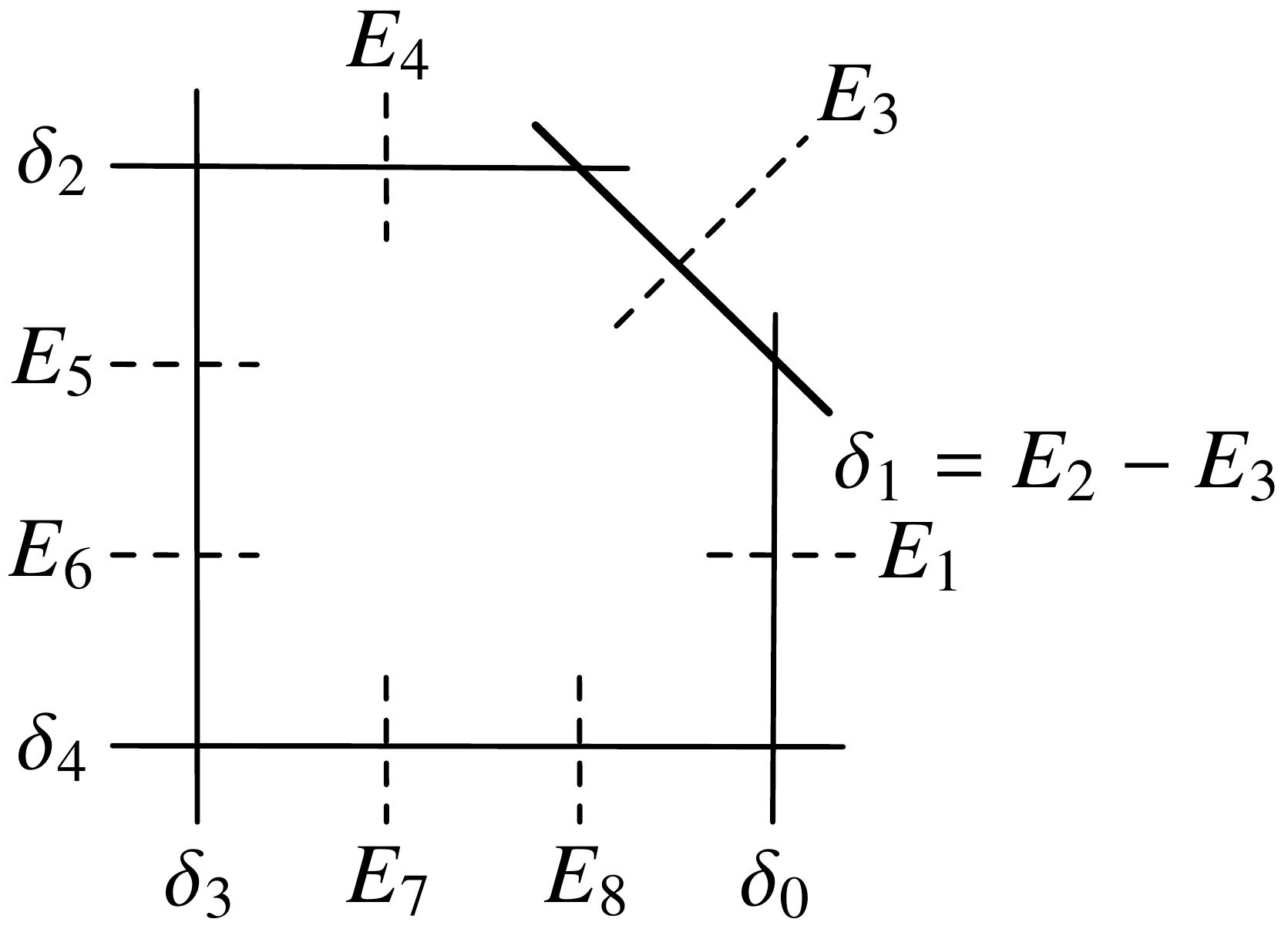}
\caption{Point configuration of symmetry/surface type $A_4^{(1)}/A_4^{(1)}$.}\label{fig:points_qP_a4}
\end{center}
\end{figure}
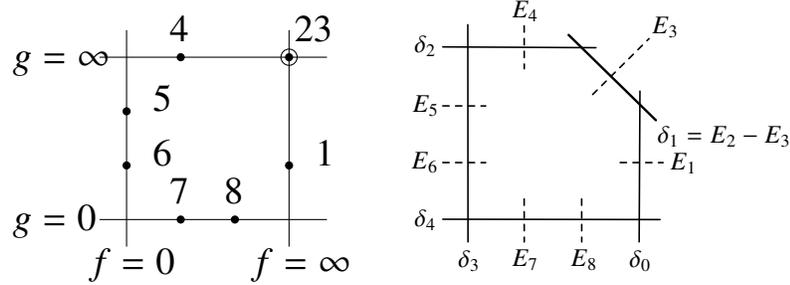
We consider the point configuration given by
\begin{equation}\label{eqn:conf_a4/a4}
\begin{split}
(f,g)=\left(\infty,\frac{1}{e_1}\right),\ 
\left(-\frac{e_2e_3}{\epsilon}, \frac{1}{\epsilon}\right)_2,\ 
\left(e_4,\infty \right),\ 
\left(0,\frac{e_i}{h_2}\right)\ (i=5,6),\ 
\left(\frac{h_1}{e_i},0\right)\ (i=7,8).
\end{split}
\end{equation}
which is illustrated graphically in Figure \ref{fig:points_qP_a4}.  The second point is the
double point at $(\infty,\infty)$ with gradient $\frac{f}{g}=-e_2e_3$ (see
\eqref{eqn:pt_infinitely_near}).  The pair of root bases $\{\alpha_i\}$, $\{\delta_i\}$ representing
the symmetry/surface types associated with the point configuration \eqref{eqn:conf_a4/a4},
and the Dynkin diagram automorphisms are given by
\begin{equation}
\begin{split}
 \setlength{\unitlength}{1.3mm}
\begin{picture}(25,20)(-7,-7)
\put(0,7){$\alpha_0$}
\drawline(-1,6)(-4,3.8)
\put(-6.657,2.163){$\alpha_1$}
\drawline(-5,1)(-4,-3.3)
\put(-4.114,-5.663){$\alpha_2$}
\drawline(-1,-5)(3.3,-5)
\put(4.114,-5.663){$\alpha_3$}
\drawline(6,-3.5)(7.5,1)
\put(6.657,2.163){$\alpha_4$}
\drawline(6,3.8)(3,6)
\end{picture}
\qquad
 \setlength{\unitlength}{1.3mm}
\begin{picture}(25,20)(-7,-7)
\put(0,7){$\delta_0$}
\drawline(-1,6)(-4,3.8)
\put(-6.657,2.163){$\delta_1$}
\drawline(-5,1)(-4,-3.3)
\put(-4.114,-5.663){$\delta_2$}
\drawline(-1,-5)(3.3,-5)
\put(4.114,-5.663){$\delta_3$}
\drawline(6,-3.5)(7.5,1)
\put(6.657,2.163){$\delta_4$}
\drawline(6,3.8)(3,6)
\end{picture}
\end{split}
\end{equation}
\begin{equation}
\begin{split}
&
\alpha_0=E_7-E_8,\ 
\alpha_1=H_1-E_4-E_7,\ 
\alpha_2=H_2-E_1-E_5,\ 
\\
&
\alpha_3=E_5-E_6,\ 
\alpha_4=H_1+H_2-E_2-E_3-E_5-E_7,\ 
\\
&
\delta_0=H_1-E_1-E_2,\ 
\delta_1=E_2-E_3,\ 
\delta_2=H_2-E_2-E_4,\ 
\\
&
\delta_3=H_1-E_5-E_6,\ 
\delta_4=H_2-E_7-E_8,\ 
\\
&
\pi_1=\pi_{42687153}r_{H_1-H_2}r_{H_2-E_2-E_7}r_{H_1-E_5-E_7},\ \pi_2=\pi_{42317856}r_{H_1-H_2}.
\end{split}
\end{equation}
\begin{displaymath}
\includegraphics[scale=0.6]{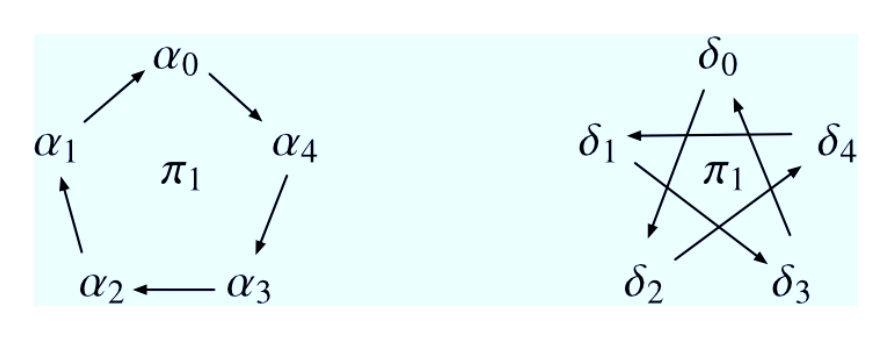} \quad
\includegraphics[scale=0.6]{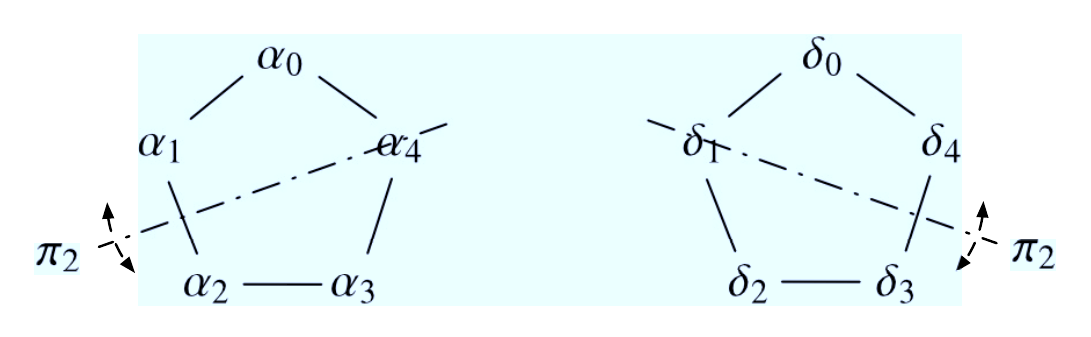} 
\end{displaymath}
Here $\pi_{i_1i_2\ldots i_8}$ is the permutation $E_1\to E_{i_1},\ E_2\to
E_{i_2},\ldots,E_{8}\to E_{i_8}$ and $r_\alpha$ is the simple reflection with respect to $\alpha$.

From the above data, the action of affine Weyl group of type $A_4^{(1)}$ 
together with the Dynkin diagram automorphisms on
parameters $e_i$ ($i=1,\ldots,8$), $h_i$ ($i=1,2$) is given as follows:
\begin{equation}
\begin{split}
s_0: &\ e_7\ \leftrightarrow\ e_8 ,\\ 
s_1: &\ e_4\ \rightarrow\ \frac{h_1}{e_7},\ 
e_7\ \rightarrow\ \frac{h_1}{e_4},\ 
h_2\ \rightarrow\ \frac{h_1h_2}{e_4e_7},\\
s_2: &\ e_1\ \rightarrow\ \frac{h_2}{e_5},\ 
e_5\ \rightarrow\ \frac{h_2}{e_1},\ 
h_1\ \rightarrow\ \frac{h_1h_2}{e_1e_5},\\
s_3: &\ e_5\ \leftrightarrow\ e_6 ,\\ 
s_4: &\ e_2\ \rightarrow\ \frac{h_1h_2}{e_3e_5e_7},\ 
e_3\ \rightarrow\ \frac{h_1h_2}{e_5e_7e_2},\ 
e_5\ \rightarrow\ \frac{h_1h_2}{e_7e_2e_3},\ 
e_7\ \rightarrow\ \frac{h_1h_2}{e_2e_3e_5},\\
& h_1\ \rightarrow \frac{h_1^2h_2}{e_2e_3e_5e_7},\ 
h_2\ \rightarrow \frac{h_1h_2^2}{e_2e_3e_5e_7},\\
\pi_1:&\ \ e_1\ \rightarrow\ e_2,\ e_2\ \rightarrow\ \frac{h_1}{e_5},\ e_3\ \rightarrow\ e_6,
\ e_4\ \rightarrow\ e_5,\\
&\  e_5\ \rightarrow\ \frac{h_2}{e_5},\ e_6\ \rightarrow\ e_1,\ 
e_7\ \rightarrow\ \frac{h_1h_2}{e_2e_5e_7},\ e_8\ \rightarrow\ e_3,\\
& h_1\ \rightarrow\ \frac{h_1h_2}{e_2e_5},\ h_2\rightarrow\ \frac{h_1h_2}{e_5e_7},\\
\pi_2:&\ \ e_1\ \leftrightarrow\ \frac{1}{e_4},\ e_2\ \leftrightarrow\ \frac{1}{e_2},\ 
e_3\ \leftrightarrow\ \frac{1}{e_3},\ e_5\ \leftrightarrow\ \frac{1}{e_7},\ 
e_6\ \leftrightarrow\ \frac{1}{e_8},
\ h_1\ \leftrightarrow\ \frac{1}{h_2}.
\end{split}
\end{equation}
Note that the Picard lattice has a trivial lattice isomorphism $H_i\rightarrow -H_i$,
$E_i\rightarrow -E_i$ which does not belong to the affine Weyl group. As to the Dynkin diagram
automorphism $\pi_2$, we need to incorporate the corresponding transformation $h_i\rightarrow h_i^{-1}$,
$e_i\rightarrow e_i^{-1}$ in constructing the birational representation. We also remark that
$\pi_2$ transforms $q=h_1^2h_2^2/(e_1\cdots e_8)$ into its reciprocal $q^{-1}$, 
while the other transformations $s_0,\ldots,s_4,\pi_1$ leave $q$ invariant.

Then the action of affine Weyl group on the variables $f$ and $g$ can be constructed as:
\begin{equation}\label{eqn:weyl_a4_fg}
 \begin{split}
s_1: &\ g\ \rightarrow \frac{e_7 \left(e_4 - f\right)}{h_1 - e_7 f}g,\\
s_2: &\ f\ \rightarrow \frac{h_2\left(1 - e_1g\right)}{e_1(e_5 - h_2g)}f,\\ 
s_4: &\ f\ \rightarrow\ \frac{h_1h_2}{e_2e_3e_7}\frac{-h_1+e_7f+e_2e_3e_7g}{-h_1e_5 + e_5e_7f + h_1h_2g}\,f,\\
&\ g\ \rightarrow\ \frac{e_5e_7}{h_2} \frac{-e_2e_3e_5 + h_2f + h_2e_2e_3g}{-h_1e_5 + e_5e_7f + h_1h_2g}\,g,\\
\pi_1:&\ f\ \rightarrow \frac{h_1}{e_5}\frac{e_5 - h_2g}{f},\ 
g\ \rightarrow\ \frac{1}{h_1h_2}\frac{-h_1e_5 + e_5e_7 f + h_1h_2g}{fg},\\
\pi_2:&\ f\ \leftrightarrow\ g.
\end{split}
\end{equation}
In the following, we discuss how to construct \eqref{eqn:weyl_a4_fg} from the above data.  For
example, we demonstrate the construction of $s_4(f)$. Noting that
$s_4(H_1)=2H_1+H_2-E_2-E_3-E_5-E_7$, we first determine a basis of polynomials belonging to the
divisor class of $s_4(H_1)$ as (see Section \ref{subsec:configuration} for treatment of the double
point)
\begin{equation}
 A(f,g) = -\frac{e_5}{h_2} + \frac{e_5e_7}{h_1h_2}f + g,\quad
 B(f,g) = f\left(-\frac{h_1}{e_7} + f + e_2e_3g\right).
\end{equation}
Then, according to our guiding principle, $s_4(f)$ should be expressed in the form
\begin{equation}
 s_4(f)=\frac{a A(f,g) + b B(f,g)}{c A(f,g) + d B(f,g)}.
\end{equation}
One can determine the coefficients by investigating the image of appropriate points or divisors. 
In this case, $H_1-E_2$, $H_1-E_5$ and $H_1-E_7$ are perpendicular to
$\alpha_4=H_1+H_2-E_2-E_3-E_5-E_7$. Hence, the corresponding lines $f=\infty$, $f=0$, and
$f=\frac{h_1}{e_7}$ are invariant with respect to the action of $s_4$. Thus we have
\begin{equation}
 s_4(f)\bigr|_{f=\infty} = \infty,\quad
 s_4(f)\bigr|_{f=0} = 0,\quad
s_4(f)\bigr|_{f=\frac{h_1}{e_7}}= \frac{h_1}{e_7},
\end{equation}
first two of which yield $d=0$ and $a=0$, respectively. Then the third equation gives 
$\frac{b}{c}e_2e_3=1$, which determines $s_4(f)$ as in \eqref{eqn:weyl_a4_fg}.
Other actions can be determined in a similar manner. One can verify that these transformations
$\langle s_0,\ldots,s_4,\pi_1,\pi_2\rangle$ satisfy the fundamental relations of the 
affine Weyl group of type $A_4^{(1)}$ and the Dynkin diagram automorphisms,
\begin{equation}
\begin{split}
& s_i^2=1,\quad (s_is_{i+1})^3=1,\quad (s_is_j)^2=1\quad (j\not\equiv i,i\pm 1\ {\rm mod}~5),\\
& \pi_1^5 = \pi_2^2 = (\pi_1\pi_2)^2=1,\quad
\pi_1s_{\{01234\}}=s_{\{40123\}}\pi_1,\quad \pi_2s_{\{01234\}}=s_{\{32104\}}\pi_2. 
\end{split}
\end{equation}
%
\section{Hypergeometric Solutions}\label{sec:hyper}
Most of the Painlev\'e equations admit a class of particular solutions expressible in terms of
hypergeometric type functions for special values of parameters which correspond to reflection
hyperplanes in the parameter space. We call this class of solutions the hypergeometric solutions.
In this Section, taking the example of $q$-P$(E_6^{(1)})$ \eqref{eqn:q-e6}, we demonstrate how to
construct the hypergeometric solutions through the Riccati equation and by linearizing it. Then we
give an intrinsic formulation of this procedure by the geometric language of point configurations.

\subsection{Hypergeometric solution to $q$-P$(E_6^{(1)}):$ an example}\label{subsec:hyper_qpe6}
We have already demonstrated in Section \ref{subsec:hyper_p4} a simple example of P$_{\rm IV}$ and a
dP$_{\rm II}$, where we constructed particular solutions by means of the Hermite functions. In
general, the simplest hypergeometric solutions can be constructed by looking for the special values
of parameters where the equation is decoupled into Riccati equations. Then one can linearize the
Riccati equation to the second order linear differential or difference equation by the standard
procedure, which may be identified with the equation for a certain hypergeometric type function.

Before proceeding to the example of $q$-P$(E_6^{(1)})$, we give general remarks on the linearization of
(discrete) Riccati equation
\begin{equation} \label{eqn:d-Riccati}
\overline{y} = \frac{ay+b}{cy+d},
\end{equation}
where the coefficients $a$, $b$, $c$ and $d$ are given functions. Substituting
\begin{equation}
 y = \frac{G}{F}
\end{equation}
into \eqref{eqn:d-Riccati}, we have
\begin{equation}
 \frac{\overline{G}}{\overline{F}} = \frac{aG+bF}{cG+dF},
\end{equation}
which may be linearized by introducing a decoupling function $h$ as
\begin{equation}
 \overline{G}=h\left(aG+bF\right),\quad \overline{F}=h\left(cG+dF\right).
\end{equation}
Then we obtain a second order linear difference equation for $F$
\begin{equation}\label{eqn:d-Riccati_linear}
 \overline{F} - \frac{h}{\underline{c}}\left( \underline{a}c + \underline{c}d\right)F
 + \frac{c}{\underline{c}}h\underline{h}\left(\underline{a}\underline{d}-\underline{bc}\right)\underline{F}=0,\quad
G=\frac{1}{c}\left(\frac{\overline{F}}{h}-dF\right).
\end{equation}
By suitable choice of decoupling function $h$, \eqref{eqn:d-Riccati_linear} is expected to reduce to
some hypergeometric equation which typically takes the form
\begin{equation}
 A(\overline{F}-F) + BF + C(\underline{F}-F)=0.
\end{equation}
Here, the coefficients $A$, $B$, $C$ are of factorized form. In view of this, setting $h=\frac{1}{d}$, namely,
\begin{equation}\label{eqn:hypergeometric_solution_general}
 y=\frac{d}{c}\frac{\overline{F}-F}{F},
\end{equation}
we see that \eqref{eqn:d-Riccati} yields the following linear difference equation
\begin{equation}\label{eqn:hyper_linear_general}
 \underline{c}\underline{d}d(\overline{F}-F)-\underline{b}\underline{c}{c}F + \underline{\Delta}c(\underline{F}-F)=0,\quad
\Delta = ad-bc.
\end{equation}
Therefore, in the context of construction of hypergeometric solutions to the discrete Painlev\'e
equation, it is a good strategy to choose the $y$ variable in such a way that the coefficients $b$, $c$, $d$ and
$\Delta=ad-bc$ are factorized.

We show how this procedure works for the case of $q$-P$(E_6^{(1)})$ \eqref{eqn:q-e6}:
\begin{equation}\label{eqn:q-e6_5}
 \begin{split}
\frac{(fg-1)(\of g-1)}{f\of} 
= \frac{\left(g-\frac{1}{v_1}\right)\left(g-\frac{1}{v_2}\right)\left(g-\frac{1}{v_3}\right)\left(g-\frac{1}{v_4}\right)}
{\left(g-\frac{v_5}{\kappa_2}\right)\left(g-\frac{v_6}{\kappa_2}\right)},\\
\frac{(f g-1)(f \ug -1)}{g\ug} 
= \frac{(f -v_1)(f -v_2)(f -v_3)(f -v_4)}
{\left(f - \frac{\kappa_1}{v_7}\right)\left(f - \frac{\kappa_1}{v_8}\right)},
 \end{split}
\end{equation}
where $\kappa_1$, $\kappa_2$, $v_1,\ldots,v_8$ are parameters with $q\prod\limits_{i=1}^8v_i
= \kappa_1^2\kappa_2^2$ introduced in Remark \ref{rem:kappa-var}, and $\overline{\phantom y}$ is the
time evolution corresponding to $T_{\alpha_1}$ such that
$(\overline{\kappa}_1,\overline{\kappa}_2,\overline{v}_1,\ldots,\overline{v}_8)
=(\frac{\kappa_1}{q}, q\kappa_2,v_1,\ldots,v_8)$.  Note that the corresponding eight
points configuration 
\begin{equation}\label{eqn:8points_q-e6-2}
{\rm P}_i:\ (f_i,g_i)
=\left(v_i, \frac{1}{v_i}\right)\ (i=1,2,3,4),\quad\left(0,\frac{v_i}{\kappa_2}\right) \ (i=5,6),\quad 
\left(\frac{\kappa_1}{v_i},0\right)\ (i=7,8),
\end{equation}
as given in \eqref{eqn:8points_q-e6}, and those points are on the reference curve $C_0:\,fg(fg-1)=0$. 

Consider the case where the parameters satisfy
\begin{equation}\label{eqn:qe6_hyper_parameters}
 \kappa_1\kappa_2 = v_1v_3v_5v_7,\quad \mbox{i.e.}\quad q^{-1}\kappa_1\kappa_2=v_2v_4v_6v_8.
\end{equation}
Then \eqref{eqn:q-e6_5} admits the following specialization:
\begin{align}\label{eqn:hyper_qpe6_decoupling1}
& \frac{fg-1}{f} = \frac{\left(g-\frac{1}{v_1}\right)\left(g-\frac{1}{v_3}\right)}{g-\frac{v_5}{\kappa_2}},\quad
 \frac{\overline{f}g-1}{\overline{f}} = \frac{\left(g-\frac{1}{v_2}\right)\left(g-\frac{1}{v_4}\right)}{g-\frac{v_6}{\kappa_2}},\\
&\label{eqn:hyper_qpe6_decoupling2}
\frac{f g-1}{g} = \frac{(f -v_1)(f -v_3)}{f - \frac{\kappa_1}{v_7}},\quad
\frac{f \underline{g}-1}{\underline{g}} = \frac{(f -v_2)(f -v_4)}{f - \frac{\kappa_1}{v_8}}.
\end{align}
This decoupling is consistent in the sense that \eqref{eqn:hyper_qpe6_decoupling1} imply the
\eqref{eqn:hyper_qpe6_decoupling2}, and {\it vice versa}.  In other words, the discrete time evolution
admits the specialization \eqref{eqn:hyper_qpe6_decoupling1} under the condition \eqref{eqn:qe6_hyper_parameters}.
In fact, under the condition
\eqref{eqn:qe6_hyper_parameters}, both of the first equations of \eqref{eqn:hyper_qpe6_decoupling1}
and \eqref{eqn:hyper_qpe6_decoupling2} imply that the point $(f,g)$ is on a $(1,1)$-curve passing
through P$_1$, P$_3$, P$_5$ and P$_7$, which can be verified directly by substituting the
coordinates of the points.  Similarly, the second equation of \eqref{eqn:hyper_qpe6_decoupling1} 
and upshift of the second equation of \eqref{eqn:hyper_qpe6_decoupling2} mean
that $(\overline{f},g)$ is on a $(1,1)$-curve passing through P$_2$, P$_4$, P$_6$ and 
${\rm P}_8'=(\frac{\kappa_1}{qv_8},0)$.  This shows the consistency of the decoupling and geometric meaning
of the constraint \eqref{eqn:qe6_hyper_parameters} as well. 

Equation \eqref{eqn:hyper_qpe6_decoupling2} is a coupled Riccati equation; we obtain a Riccati
equation with respect to $f$ by eliminating $g$ from the first equation and the upshift of the
second equation, and $g$ is determined from $f$ by a fractional linear transformation. The Riccati
equation for $f$ is given by
\begin{equation}
\begin{split}
&\hskip90pt \overline{f} = \frac{\alpha f + \beta}{\gamma f + \delta},\\
& \alpha = \kappa_1^2 - \kappa_1v_1v_7 - \kappa_1v_3v_7 + qv_2v_4v_7v_8,\quad 
   \beta = \kappa_1 (v_1v_3v_7 - qv_2v_4v_8 ),\\
&\gamma = q \kappa_1 v_8 - \kappa_1v_7 - qv_1v_7v_8 + qv_2v_7v_8 - qv_3v_7v_8 + qv_4v_7v_8,\\
&\delta =  \kappa_1^2 - q\kappa_1v_2 v_8 - q\kappa_1 v_4 v_8 + q v_1v_3v_7v_8.
\end{split}
\end{equation}
Choosing the $y$ and $F$ variables as
\begin{equation}\label{eqn:Riccati_qpe6_y_and_F}
y =  \frac{f-v_1}{f-\frac{\kappa_1}{v_8}} 
= - \frac{1-\frac{v_1v_7}{\kappa_1}}{1-\frac{v_7}{v_8}}~\frac{\overline{F}-F}{F},
\end{equation}
we have the Riccati equation
\begin{equation}
\begin{split}
&\hskip130pt \overline{y}=\frac{\zeta y + \eta}{\lambda y + \mu },\\ 
& \eta = q^2v_8^3(\kappa_1 - v_1v_7)(v_1 - v_2)(v_1 - v_4),\quad 
\lambda =\kappa_1(\kappa_1 - qv_2v_8)(\kappa_1 - qv_4v_8)(v_7 - v_8),\\
& \mu =v_8(\kappa_1 - qv_2v_8)(\kappa_1 - qv_4v_8)(\kappa_1 - v_1v_7),\\
&\Delta = qv_8^2(\kappa_1 - qv_1v_8)(\kappa_1 - qv_2v_8)(\kappa_1 - qv_4v_8)
(\kappa_1 - v_1v_7)(\kappa_1 - v_1v_8)(\kappa_1 - v_3v_7),
\end{split}
\end{equation}
which is linearized to
\begin{equation}\label{eqn:qpe6_hyper_linear}
\begin{split}
&\hskip90pt A(\overline{F}-F) + BF + C(\underline{F}-F)=0,\\
&A = \left(1-\frac{\kappa_2}{v_3v_5}\right)\left(1-\frac{qv_2v_6}{\kappa_2}\right)\left(1-\frac{qv_4v_6}{\kappa_2}\right),
\quad  
B= \left(1-\frac{v_1}{v_2}\right)\left(1-\frac{v_1}{v_4}\right)\left(1-\frac{v_7}{v_8}\right),\\
& C= q\frac{v_6}{v_5}\left(1-\frac{qv_1v_5}{\kappa_2}\right)
      \left(1-\frac{\kappa_1}{v_1}\right)\left(1-\frac{v_1v_8}{q\kappa_1}\right).
\end{split}
\end{equation}
It is known that the balanced ${}_3\phi_2$ series
$\varphi={}_3\phi_2\left[\begin{array}{c}a,b,c\\d,e\end{array};q;z\right]$, $z=\frac{de}{abc}$  ($|q|<1$, $|\frac{de}{abc}|<1$)
satisfies 
\begin{equation}\label{eqn:3term-3phi2}
\begin{split}
&\hskip50pt V_1(\overline{\varphi}-\varphi) + V_2\varphi + V_3(\underline{\varphi}-\varphi)=0,
\\ 
&V_1= \left(1-\frac{a}{e}\right)\left(1-\frac{b}{e}\right)(1-c),\quad
V_2=(1-a)(1-b)\left(1-\frac{c}{e}\right),\\
&V_3=\frac{q}{z}\left(1-\frac{e}{q}\right)\left(1-\frac{d}{c}\right)\left(1-\frac{1}{e}\right),
\end{split}
\end{equation}
where
\begin{equation}
\overline{\varphi}=\varphi\bigr|_{{c\to qc}\atop{e\to qe}},
\quad\underline{\varphi}=\varphi\bigr|_{{c\to c/q}\atop{e\to e/q}}, 
\end{equation}
Equation \eqref{eqn:3term-3phi2} is obtained from the following three-term relation (\cite[formula (2.7)]{Gupta-Ismail-Masson})
\begin{equation}
\begin{split}
 U_1\left(\widetilde{\varphi}-\varphi\right) + U_2\varphi + U_3(\lower0.5pt\hbox{$\hypotilde{0}{\varphi}$}- \varphi)=0,\quad 
\widetilde{\varphi}=\varphi\bigr|_{a\to qa},\quad\lower0.5pt\hbox{$\hypotilde{0}{\varphi}$}=\varphi\bigr|_{a\to a/q},\\
U_1 =\left(1-\frac{q}{z}\right)\left(1-a\right),\quad 
U_2=(1-b)(1-c),\quad U_3=\frac{a}{z}\left(1-\frac{d}{a}\right)\left(1-\frac{e}{a}\right),
\end{split}
\end{equation}
by applying the transformation(\cite[formula (III.10)]{Gasper-Rahman})
\begin{equation}
 {}_3\phi_2\left[\begin{array}{c}a,b,c\\d,e\end{array};q;\frac{de}{abc}\right]
= \frac{(b,de/ab,de/bd;q)_\infty}{(d,e,de/abc;q)_\infty}
 {}_3\phi_2\left[\begin{array}{c}d/b,e/b,de/abc\\de/bc,de/ab\end{array};q;b\right].
\end{equation}
Comparing \eqref{eqn:qpe6_hyper_linear} and \eqref{eqn:3term-3phi2}, we find that
\eqref{eqn:qpe6_hyper_linear} is solved by
\begin{equation}
 F =  {}_3\phi_2\left[\begin{array}{c}\frac{v_1}{v_2},\frac{v_1}{v_4},\frac{\kappa_2}{v_3v_5}\\[2mm]\frac{qv_1}{v_3},\frac{v_1v_8}{\kappa_1}\end{array};q;\frac{v_5}{v_6}\right].
\end{equation}
Also, the numerator of $y$ is expressed by
\begin{equation}
\overline{F}- F 
= 
\frac{\kappa_2\left(1-\frac{v_1}{v_2}\right)\left(1-\frac{v_1}{v_4}\right)\left(1-\frac{v_8}{v_7}\right)}
{v_3v_6\left(1-\frac{qv_1}{v_3}\right)\left(1-\frac{v_1v_8}{\kappa_1}\right)\left(1-\frac{qv_1v_8}{\kappa_1}\right)}
~
  {}_3\phi_2\left[\begin{array}{c}\frac{qv_1}{v_2},\frac{qv_1}{v_4},\frac{q\kappa_2}{v_3v_5}\\[2mm]
\frac{q^2v_1}{v_3},\frac{q^2v_1v_8}{\kappa_1}\end{array};q;\frac{v_5}{v_6}\right],
\end{equation}
so that
\begin{equation}
 y =  
\frac{\kappa_2v_8\left(1-\frac{v_1v_7}{\kappa_1}\right)\left(1-\frac{v_1}{v_2}\right)\left(1-\frac{v_1}{v_4}\right)}
{v_3v_6v_7\left(1-\frac{qv_1}{v_3}\right)\left(1-\frac{v_1v_8}{\kappa_1}\right)\left(1-\frac{qv_1v_8}{\kappa_1}\right)}
~
\frac{ {}_3\phi_2\left[\begin{array}{c}\frac{qv_1}{v_2},\frac{qv_1}{v_4},\frac{q\kappa_2}{v_3v_5}\\[2mm]
\frac{q^2v_1}{v_3},\frac{q^2v_1v_8}{\kappa_1}\end{array};q;\frac{v_5}{v_6}\right]}
{{}_3\phi_2\left[\begin{array}{c}\frac{v_1}{v_2},\frac{v_1}{v_4},\frac{\kappa_2}{v_3v_5}\\[2mm]\frac{qv_1}{v_3},\frac{v_1v_8}{\kappa_1}\end{array};q;\frac{v_5}{v_6}\right]}.
\end{equation}

\subsection{Hypergeometric solution from point configuration}\label{subsec:hyper_generic}
The construction of the hypergeometric solutions demonstrated in Section \ref{subsec:hyper_qpe6}
have the geometric background, and the procedures and quantities appeared in the construction can be
understood from the geometry of the point configuration. Let us describe the fundamental principle
of the construction according to the example of $q$-P$(E_6^{(1)})$. As mentioned before, the
condition of the parameter \eqref{eqn:qe6_hyper_parameters} implies that the points P$_1$, P$_3$,
P$_5$, P$_7$ is on a $(1,1)$--curve $C_1$, and similarly, ${\rm P}_2'$, ${\rm P}_4'$, ${\rm P}_6'$, ${\rm
P}_8'$ are on another $(1,1)$--curve $C_2$. Namely, we have
\begin{equation}\label{eqn:riccati_parameters_general}
C_1:\ d_{1357} = 0,
\quad 
C_2:\ d_{2468}'= 0,
\end{equation}
where
\begin{equation}
\begin{split}
& {\rm P}_i=(f_i,g_i),\quad {\rm P}_i'=(\overline{f}_i,g_i),\\
&d_{ijkl}=\left|
\begin{array}{cccc}
1& f_i & g_i& f_ig_i\\
1& f_j & g_j& f_jg_j\\
1& f_k & g_k& f_kg_k\\
1& f_l & g_l& f_lg_l
\end{array}\right|,\quad 
d_{ijkl}'=\left|
\begin{array}{cccc}
1& \overline{f}_i & g_i& \overline{f}_ig_i\\
1& \overline{f}_j & g_j& \overline{f}_jg_j\\
1& \overline{f}_k & g_k& \overline{f}_kg_k\\
1& \overline{f}_l & g_l& \overline{f}_lg_l
\end{array}\right|.
\end{split}
\end{equation}
Recall from Example \ref{ex:M} and Example \ref{ex:R} that 
\begin{equation}
 d_{*ijk}=
\left|\begin{array}{cccc}
1& f & g& fg\\
1& f_i & g_i& f_ig_i\\
1& f_j & g_j& f_jg_j\\
1& f_k & g_k& f_kg_k
\end{array}\right| = 0,\quad {\rm P}_*=(f,g),
\end{equation}
represents a $(1,1)$-curve passing through the three points ${\rm P}_i$, ${\rm P}_j$, ${\rm P}_k$;
it also passes through ${\rm P}_l$ if and only if $d_{ijkl}=0$.
The Riccati equation \eqref{eqn:hyper_qpe6_decoupling1} or \eqref{eqn:hyper_qpe6_decoupling2} 
can be expressed as
\begin{equation}\label{eqn:riccati_general}
 d_{*135}=0,\quad d'_{*246}=0,\quad \quad {\rm P}_*'=(\overline{f},g).
\end{equation}
Note that the $(1,1)$--curve $d_{*135}=0$ corresponds to the root $\beta=H_1+H_2-E_1-E_3-E_5-E_7\in
R\subset\Lambda$ (see \eqref{eqn:def_R}).  Consistency of those specialization with the discrete
time evolution, $T_\alpha$, $\alpha=H_1-H_2\in R$, is guaranteed by the condition $\alpha\cdot
\beta=0$.  In fact, the orthogonality $\alpha\cdot \beta=0$ implies $T_\alpha(\beta)=\beta$ and
hence the corresponding curve $C_\beta:\,d_{*135}=0$ is preserved by $T_\alpha$, i.e., ${\rm
P}=(f,g)\in C_\beta\mapsto \overline{{\rm P}}=T_\alpha({\rm P})\in C_\beta$. Since the genus of
$C_\beta$ is $0$ (see \eqref{eqn:def_R}), $C_\beta$ is isomorphic to $\mathbb{P}^1$ and thus it is
natural that the resulting dynamical system on $C_\beta$ is the linear fractional transformation,
i.e., the Riccati equation. We remark that the above argument applies to $T_\alpha$ and
$C_\beta$ for any $\alpha, \beta\in R$ provided that $\alpha\cdot \beta=0$. 

In the context of the (discrete) Painlev\'e equations and their affine Weyl group symmetries,
$C_\beta$ is the so-called the {\em invariant divisor} along the reflection hyperplane of the
B\"acklund transformation $r_\beta$ in the Umemura theory
\cite{Noumi-Okamoto,Umemura:P1,Umemura:Sugaku_expositions,Umemura:Suugaku,Umemura-Watanabe:P4,Umemura-Watanabe:P3,Watanabe:P5,Watanabe:P6}. The
condition of the parameters (\eqref{eqn:qe6_hyper_parameters} in the above example) is the defining
relation of the reflection hyperplane in the parameter space. Further, classification of invariant
divisors of Painlev\'e differential equations is a crucial step of understanding the irreducibility
of Painlev\'e transcendents.

As to the $y$ variable \eqref{eqn:Riccati_qpe6_y_and_F}, we make the choice  
\begin{equation}
 y = \frac{f-f_1}{f-f_8},
\end{equation}
in the generic context, as motivated by the expression in terms of the $\tau$ function
\eqref{eqn:xy_and_tau}.  Then the Riccati equation for $y$ takes the form
\begin{equation}
\begin{split}
&\hskip120pt  \overline{y}=\frac{ay+b}{cy+d},\\
& b = -f_{13} f_{15} g_{35}d'_{1468},\quad c=\overline{f}_{48}\overline{f}_{68}g_{46}d_{1358},\quad
d = f_{13}f_{15}\overline{f}_{48}\overline{f}_{68}g_{18}g_{35}g_{46},\\
& \Delta = ad-bc=f_{13}f_{15}f_{35}f_{18}\overline{f}_{46}\overline{f}_{48}\overline{f}_{68}\overline{f}_{18}
g_{13}g_{15}g_{35}g_{46}g_{48}g_{68},\\
&f_{ij}=f_i-f_j,\quad \overline{f}_{ij} = \overline{f}_{i}-\overline{f}_j,\quad g_{ij}=g_i-g_j,
\end{split}
\end{equation}
as verified by the direct computation from \eqref{eqn:riccati_parameters_general} and
\eqref{eqn:riccati_general}.  Accordingly, the coefficients of the linear difference equation for $F$
variable \eqref{eqn:hyper_linear_general} are expressed as factorized form. These formulae apply to
any configuration of points which does not contain infinitely near points, namely to the cases of
symmetry type $E_{8}^{(1)}$, $E_{7}^{(1)}$, $E_{6}^{(1)}$ and $D_5^{(1)}$. Otherwise, we need to
employ appropriate limiting procedures, or it may be easier to construct the hypergeometric
solutions from the equation itself directly. 

We include the basic data of the fundamental hypergeometric solution of each discrete Painlev\'e
equation in Section \ref{subsec:hyper_data}.
%
\begin{rem}\rm\hfill
\begin{enumerate}
 \item Applying B\"acklund transformations (birational transformations by elements of the affine Weyl group) 
to a known solution (seed solution), we obtain a class of solutions expressible by the
rational functions of the seed solutions. Moreover, in known examples, they are always given by the
ratio of determinants whose entries are the seed solutions. The determinant structure of the
solutions is understood as a universal property of the Painlev\'e equations \cite{KMNOY:Toda,KMO:Toda,Noumi:book,Ohyama-Kawamuko-Sakai-Okamoto,Okamoto:p24,Okamoto:p6,Okamoto:p5,Okamoto:p3,Yamada:det}. 
 \item In the class of particular solutions of hypergeometric type, the
       corresponding determinants are referred to as the {\em hypergeometric $\tau$ functions}
       \cite{HK:qA4,HKW:qp2,KK:qp3-1,KNY:qp4,KOS:dp3,KOSGR:dp2,Masuda:hyper,Masuda:qE7,Masuda:qE8,NSKGR:alt-dp2,Noumi:book,NTY,OKS:dp1,Yamada:Pade}. Historically, the discrete Painlev\'e equations
       became familiar after they were derived as the recursion relations satisfied
by the ratio of hypergeometric $\tau$ functions \cite{Brezin-Kazakov,Douglas-Shenker,Fokas-Its-Kitaev,Gross-Migdal,Periwal-Shevitz} which appeared as the partition functions of the random matrix theory \cite{Forrester:book}.
 \item There is another important class of particular solutions, called the {\em algebraic
       solutions}. Typical examples of this class are obtained by applying B\"acklund
       transformations to the simple solutions characterized by the invariance with respect to the
       Dynkin diagram automorphisms. Many of such solutions are interpreted as simple specialization
       of the Schur functions or the universal characters
       \cite{KM:p3_rational,KO:p2_rational,KO:p4_rational,KYO:dp2_rational,Masuda:p6_rational,MOK:p5_rational,Noumi-Yamada:p4,Noumi-Yamada:p5_rational,Tsuda:UC}.
 \item Recently, the general solutions to some Painlev\'e equations are found to admit explicit 
formal series solutions \cite{ILST:P6_conformal_blocks} in the context of conformal field theory.
\end{enumerate}
\end{rem}

\section{Lax Pairs}\label{sec:Lax}
It is a common feature of nonlinear integrable systems that they arise as the compatibility
condition of certain systems of linear equations. The system of linear equations is called a Lax
pair of the nonlinear equation.  As we have seen in Section \ref{subsec:P4_Lax},
\eqref{eqn:p4_lax_L} and \eqref{eqn:p4_lax_B} constitute a Lax pair of P$_{\rm IV}$. See
\cite{J-M:Monodromy2,J-M:Monodromy3,J-M:Monodromy1,Yamada-Nagoya,Ohyama-Kawamuko-Sakai-Okamoto,
Ohyama_Okumura:Lax} for Lax pairs of other Painlev\'e equations.

Lax pairs of discrete Painlev\'e equations have been discussed by many authors from various points
of view.  Earlier works are discussed in \cite{GR:review2004}, and subsequently more systematic
approach has been used in
\cite{Arinkin-Borodin,Boalch,Jimbo-Sakai:qp6,Murata:Lax,Rains,Sakai:Lax,Witte-Ormerod,Yamada:Lax,Yamada:IMRN}.
We will explain below how one can use the geometric method for constructing Lax pairs of the
discrete Painlev\'e equations according to the idea in \cite{Yamada:Lax,Yamada:IMRN}.
\subsection{Lax pair for $q$-P$(E_6^{(1)}):$ an example}\label{subsec:lax_qpe6}
In order to see the relation between the Lax pair and the point configuration, we consider a Lax
pair for $q$-P$(E_6^{(1)})$ \eqref{eqn:q-e6}
\begin{equation}\label{eqn:q-e6_4}
 \begin{split}
\frac{(fg-1)(\of g-1)}{f\of} 
= \frac{\left(g-\frac{1}{v_1}\right)\left(g-\frac{1}{v_2}\right)\left(g-\frac{1}{v_3}\right)\left(g-\frac{1}{v_4}\right)}
{\left(g-\frac{v_5}{\kappa_2}\right)\left(g-\frac{v_6}{\kappa_2}\right)},\\
\frac{(f g-1)(f \ug -1)}{g\ug} 
= \frac{(f -v_1)(f -v_2)(f -v_3)(f -v_4)}
{\left(f - \frac{\kappa_1}{v_7}\right)\left(f - \frac{\kappa_1}{v_8}\right)},
 \end{split}
\end{equation}
as an example, where $\kappa_1$, $\kappa_2$, $v_1,\ldots,v_8$ are parameters with $q\prod_{i=1}^8v_i
= \kappa_1^2\kappa_2^2$ introduced in Remark \ref{rem:kappa-var}, and $\overline{\phantom y}$ is the
time evolution corresponding to $T_{\alpha_1}$ such that
$(\overline{\kappa}_1,\overline{\kappa}_2,\overline{v}_1,\ldots,\overline{v}_8)
=(\frac{\kappa_1}{q}, q\kappa_2,v_1,\ldots,v_8)$.  Note that the corresponding eight
points configuration is given in \eqref{eqn:8points_q-e6}, and those points are on 
the reference curve $C_0:\,fg(fg-1)=0$. A Lax pair is given by \cite{Yamada:IMRN}
\begin{align}
&L_1(z) = 
\dfrac{z \prod_{i=1}^4 (g v_i-1)}{g (f g-1) (g z-1)}y(z)
-\dfrac{\prod_{i=5}^6(\frac{g \kk _2}{v_i}-1) \kk _1^2}{f g q v_7 v_8}y(z)
+\dfrac{\prod_{i=1}^4(v_i-z) }{f-z}\Big\{\frac{g}{1-g z}y(z)-y\left(\frac{z}{q}\right)\Big\}\nonumber\\
&\hskip40pt +\dfrac{\prod_{i=7}^8(\frac{\kk_1}{v_i}-q z) }{q (f-q z)}\left\{\left(\frac{1}{g}-q z\right)y(z) -y(qz)\right\}=0,
\label{eqn:Lax_qpe61}\\[2mm]
&L_2(z)= \left(1-\frac{f}{z}\right)\, \overline{y}\left(\frac{z}{q}\right)
+ y(z) -\left(\frac{1}{g}-z\right)\, y\left(\frac{z}{q}\right)=0.\label{eqn:Lax_qpe62}
\end{align}
Equation \eqref{eqn:Lax_qpe61} is a linear $q$-difference equation for $y(z)$ and
\eqref{eqn:Lax_qpe62} describes a deformation of \eqref{eqn:Lax_qpe61}. It turns out that the
compatibility condition of the linear system \eqref{eqn:Lax_qpe61}, \eqref{eqn:Lax_qpe62} gives
$q$-P$(E_6^{(1)})$. More precise meaning of the compatibility condition is described as follows.
Consider the equations $L_1(z)=0$, $L_1(qz)=0$, $L_2(z)=0$, $L_2(qz)=0$ and $L_2(q^2z)=0$. One can
eliminate four variables $y(z/q)$, $y(z)$ $y(qz)$ and $y(q^2z)$ from these five equations to obtain
a linear relation among $\overline{y}(z/q)$, $\overline{y}(z)$ and $\overline{y}(z/q)$, which should
coincide with $\overline{L}_1(z)$.  Note that one can rewrite \eqref{eqn:Lax_qpe61} and
\eqref{eqn:Lax_qpe62} in a matrix form as
\begin{align}
&Y(qz) = M_1(z)Y(z),\quad Y(z)=\left[\begin{array}{c}y(z)\\ y(z/q)\end{array}\right],\quad 
M_1(z) = \left[\begin{array}{cc} a(z)& b(z)\\1 & 0\end{array}\right],\\
&\overline{Y}(z) = M_2(z)Y(z),\quad M_2(z) = 
\left[\begin{array}{cc}a(z)\alpha(qz) + \beta(qz) & b(z)\alpha(qz) \\ \alpha(z)  &\beta(z)\end{array}\right].
\end{align}
Then the compatibility condition mentioned above is equivalently written as
\begin{equation}
 \overline{M}_1(z)M_2(z) = M_2(qz)M_1(z),
\end{equation}
which yields
\begin{equation}
\begin{split}
& \overline{a}(z) = 
 \frac{\left(\alpha(q^2z)a(qz) + \beta(q^2z)\right)b(z) - \beta(z)\overline{b}(z)}{b(z)\alpha(qz)},\quad
\overline{b}(z) = \frac{b(z)\Delta(qz)}{\Delta(z)},
\\
& \Delta(z)=\det M_1(z) = \alpha(z)\alpha(qz)b(z) - a(z)\alpha(qz)\beta(z) - \beta(z)\beta(qz). 
\end{split}
\end{equation}

More practically, one can verify the compatibility of $L_1(z)=0$ and $L_2(z)=0$ 
as follows. Eliminating $y(q z)$ and $y(z/q)$ from $L_1(z)=0$,
$L_2(z)=0$ and $L_2(qz)=0$, we have
\begin{equation}
L_3(z)=w z(z-\overline{f})\, y(z)
+g \prod_{i=1}^4(z-v_i)\,\overline{y}\left(\frac{z}{q}\right)
-(1-g z)\prod_{i=7}^8\left(\frac{\kappa_{1}}{q v_{i}}-z\right)\, \overline{y}(z)=0, 
\end{equation}
where
\begin{equation}
\begin{array}l
{\displaystyle w=\frac{v_{1} v_{2} v_{3} v_{4}
(g-\frac{v_{5}}{\kappa_{2}})(g-\frac{v_{6}}{\kappa_{2}})}{\overline{f}}
-\frac{(1-g v_{1}) (1-g v_{2}) (1-g v_{3}) (1-g v_{4})}
{g(\overline{f} g-1)}} \\ \quad
{\displaystyle =\frac{v_{1} v_{2} v_{3} v_{4}
(g-\frac{v_{5}}{\kappa_{2}})(g-\frac{v_{6}}{\kappa_{2}})}{f}
-\frac{(1-g v_{1}) (1-g v_{2}) (1-g v_{3}) (1-g v_{4})}
{g(fg-1)}}.
\end{array} 
\end{equation}
Here we have used the first equation of \eqref{eqn:q-e6_4}.
Then eliminating $y(z)$, $y(qz)$ from $L_2(qz)=0$, $L_3(z)=0$ and $L_3(qz)=0$,
we get the three-term relation between $\overline{y}(qz)$, $\overline{y}(z)$ and $\overline{y}(z/q)$
\begin{small}
\begin{equation}
\begin{split}
&\left(\frac{\prod\limits_{i=1}^4
v_i\prod\limits_{i=5}^6(g-\frac{v_i}{\kappa_2})}{q\overline{f}g}+
\frac{z\prod\limits_{i=1}^4(1-v_i g)}{(\overline{f}g-1)(1-q zg)g}\right)\,\overline{y}(z)
+\frac{\left(\frac{1}{g}-z\right)\prod\limits_{i=7}^8\left(\frac{\kappa_1}{qv_1}-z\right)\overline{y}(z)
- \prod\limits_{i=1}^4(v_i-z)\,\overline{y}\left(\frac{z}{q}\right)}{z-\overline{f}}\\[3mm]
&+
\frac{\prod\limits_{i=1}^4(v_i-q z)\,\overline{y}(z)-\left(\frac{1}{g}-qz\right)
\prod\limits_{i=7}^8\left(\frac{\kappa_1}{q u_i}-q z\right)\,\overline{y}(qz)}{q(qz-\overline{f})
\left(\frac{1}{g}-q z\right)} = 0.
 \end{split}
\end{equation} 
\end{small}
Written in terms of $\overline{g}$ by using the second equation of \eqref{eqn:q-e6_4}, this gives
exactly $\overline{L}_1(z)$.

\begin{center}
\begin{figure}[ht]
\begin{center}
\includegraphics[scale=0.5,clip,viewport=0 0 370 280]{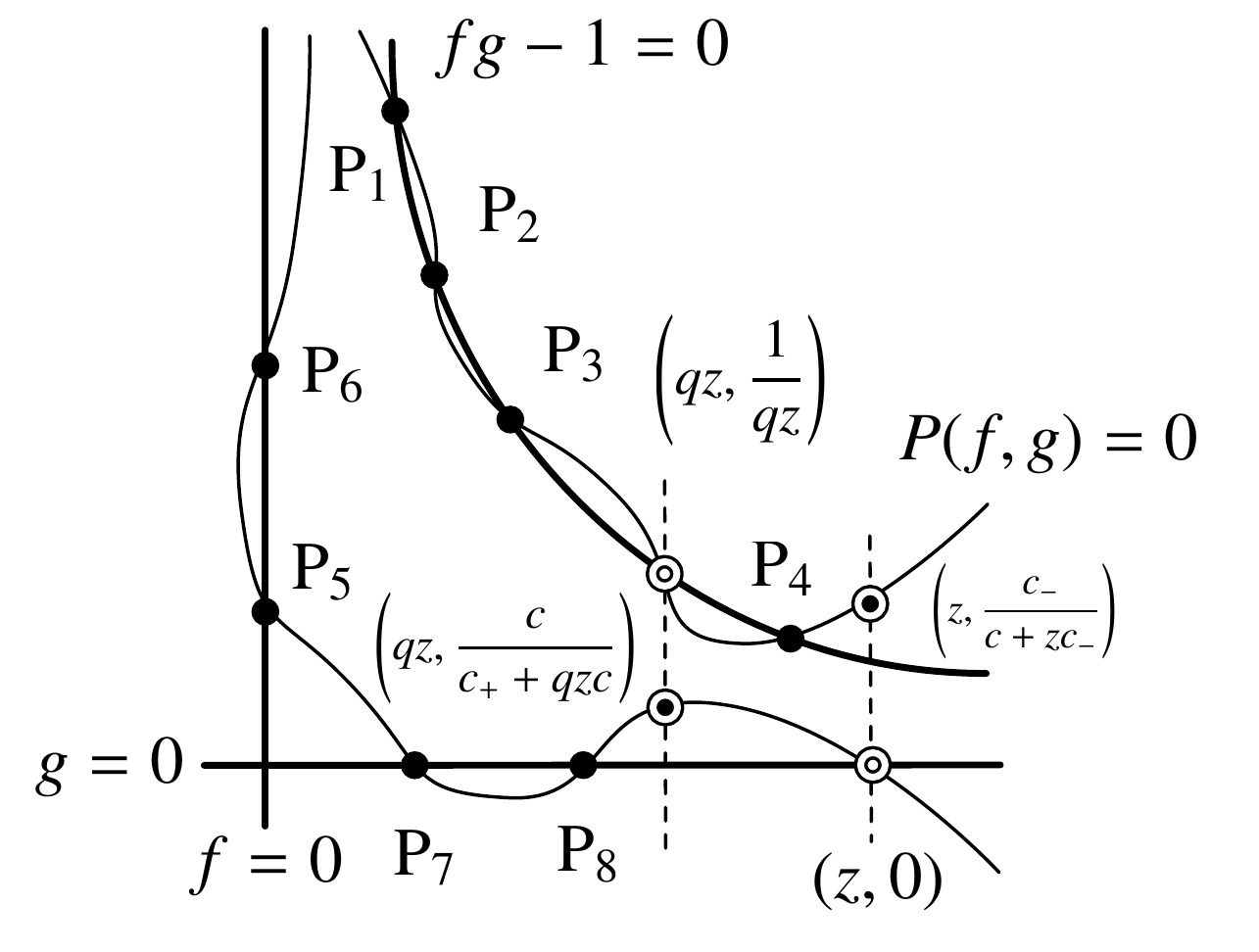}\\
\caption{Twelve points specifying the polynomial $P(f,g)$.} \label{fig:qe6_Lax_12pts}
\end{center}
\end{figure} 
\end{center}

We next discuss the geometric characterization of the difference equation \eqref{eqn:Lax_qpe61}. 
Multiplying $fg(fg-1)(f-z)(f-qz)$ to \eqref{eqn:Lax_qpe61} yields
\begin{align}
\widetilde{L}_1(z) 
&=
\dfrac{z f(f-z)(f-qz)\prod\limits_{i=1}^4 (g v_i-1)}{g z-1}y(z)
-\dfrac{(fg-1)(f-z)(f-qz)\prod\limits_{i=5}^6\left(\frac{g \kk _2}{v_i}-1\right) \kk _1^2}{ q v_7 v_8}y(z)\nonumber\\
&+fg(fg-1)(f-qz)\prod_{i=1}^4(v_i-z) \left\{\frac{g}{1-g z}y(z)-y\left(\frac{z}{q}\right)\right\}\nonumber\\
& +\dfrac{fg(fg-1)(f-z)\prod\limits_{i=7}^8
\left(\frac{\kk_1}{v_i}-q z\right) }{q }\left\{\left(\frac{1}{g}-q z\right)y(z) - y(qz)\right\}=0,
\label{eqn:Lax_qpe63}
\end{align}
from which we observe that the pole at $g=\frac{1}{z}$ cancels out and $\widetilde{L}_1(z)=P(f,g)$
becomes a polynomial in $(f,g)$ of degree $(3,2)$ with coefficients depending on $\kappa_1$,
$\kappa_2$, $v_1,\ldots,v_8$, $z$, $c_{\pm }=y(q^{\pm 1}z)$ and $c=y(z)$. We first observe that 
the $(3,2)$-curve $P(f,g)=0$ and the $(2,2)$-curve $C_0:\,fg(fg-1)=0$ intersect at the following ten points:
\begin{alignat}{3}
(f,g)=&
\left(v_i, \frac{1}{v_i}\right)\ (i=1,\ldots,4),\ && \left(0,\frac{v_i}{\kappa_2}\right)\ (i=5,6),\ 
&&\left(\frac{\kappa_1}{v_i},0\right)\ (i=7,8), \label{eqn:qe6_Lax_12pts_kappa}\\[2mm]
&\left(qz,\frac{1}{qz}\right),&& \left(z,0\right).&&\label{eqn:qe6_Lax_12pts_z}
\end{alignat}
Further, investigating the section of $P(f,g)=0$ with $f=z,\, qz$, we find that $P(f,g)$ also vanishes
at the following two points (Fig.\ref{fig:qe6_Lax_12pts}):
\begin{equation}
(f,g)=\left(z,\frac{c_{-}}{c+zc_{-}}\right),\quad
\left(qz,\frac{c}{c_{+}+ qzc}\right). \label{eqn:qe6_Lax_12pts_y}
\end{equation}
We note that $(3,2)$-curve $P(f,g)$ is uniquely determined up to a scalar multiple by the vanishing
property at these twelve points.  It is nontrivial, however, that the polynomial $P(f,g)$ becomes
linear homogeneous in $c_{\pm}$ and $c$ by the particular choice in \eqref{eqn:qe6_Lax_12pts_y};
this is the key property for constructing the Lax pair from the point configuration.

We now turn to the example of $q$-P$(E_8^{(1)})$ discussed in Section \ref{subsec:qe8} and
demonstrate how to construct a linear problem from the corresponding point configuration. Under the
parametrization in Remark \ref{rem:kappa-var}, the eight blowing-up points are given by
\begin{equation}\label{eqn:qpe8_points_kappa}
 \left(f(v_i),g(v_i)\right)_{i=1,\ldots,8}
\end{equation}
where
\begin{equation}\label{eqn:fg_symmetry}
  f(v)=v+\frac{\kappa_1}{v}=f\left(\frac{\kappa_1}{v}\right),\quad g(v)=v+\frac{\kappa_2}{v}=g\left(\frac{\kappa_2}{v}\right),
\end{equation}
and the reference curve $C_0$ passing through the eight points \eqref{eqn:qpe8_points_kappa} is written as
\begin{equation}
\varphi_{22}(f,g)= (f-g)(\kappa_2 f - \kappa_1 g) + (\kappa_1-\kappa_2)^2=0.
\end{equation}
Supposing that the linear problem is expressed by a polynomial $P(f,g)$ of bidegree $(3,2)$, we
specify a set of twelve points on $P(f,g)=0$ by which the polynomial is characterized.  Among the
twelve points, it is natural to take eight points from the blowing-up points
\eqref{eqn:qpe8_points_kappa}. There are two additional points
\begin{equation}
\left(f(z),g(z)\right),\quad \left(f(w), g(w)\right),
\end{equation}
at which the curve $P(f,g)=0$ intersects with the reference curve $C_0$. Here, the parameters $z$
and $w$ should satisfy
\begin{equation}
v_1\cdots v_8zw= \kappa_1^3\kappa_2^2,\label{eqn:qpe8_parameter_constraint}
\end{equation}
namely,
\begin{equation}
 w = \frac{q\kappa_1}{z},
\end{equation}
due to Abel's Theorem, or the relation between roots and coefficients for the Laurent polynomial
\begin{equation}
P(f,g)=P\left(v+\frac{\kappa_1}{v},v+\frac{\kappa_2}{v}\right)
=\frac{{\rm const.}}{v^5}(v-v_1)\cdots(v-v_8)(v-z)(v-w).
\end{equation}
We specify the remaining two points on the $(3,2)$-curve $P(f,g)=0$ of the form
\begin{equation}
 (f(z), g_1),\quad (f(w),g_2)=(f(z/q),g_2), \label{eqn:qpe8_Lax_additional_points1}
\end{equation}
by choosing $g_1$ and $g_2$ so that the resulting polynomial $P(f,g)$ becomes linear homogeneous in
$y(qz)$, $y(z)$, $y(z/q)$. Our choice will be made as
\begin{equation}
\frac{g_1-g\left(z\right)}{g_1-g\left(\kappa_1/z\right)}=\frac{y(qz)}{y(z)},\quad
\frac{g_2-g\left(z/q\right)}{g_2-g\left(q\kappa_1/z\right)}=\frac{y(z)}{y(z/q)}.
\label{eqn:qpe8_Lax_additional_points2}
\end{equation}
The polynomials vanishing at the above ten points (effectively nine points due to the constraint
\eqref{eqn:qpe8_parameter_constraint}) form a three-parameter family.  Since
$(f-f(z))\,\varphi_{22}(f,g)$ and $(f-f(w))\,\varphi_{22}(f,g)$ belong to this family, we can write
any member $P(f,g)$ of this family as
\begin{equation}
 P(f,g)=A(f-f(z))\,\varphi_{22}(f,g)+B\left(f-f\left(w\right)\right)\,\varphi_{22}(f,g) - CF_{32}(f,g),
\end{equation}
where $F_{32}(f,g)$ is a bidegree $(3,2)$-polynomial. It is convenient to choose $F_{32}(f,g)$ by
the condition that the curve $F_{32}(f,g)=0$ is tangent to the lines $f=f(z)$ and
$f=f(w)$. Putting $f=f(z)$ we have, as a polynomial in $g$,
\begin{align}
  P(f(z),g)&=B\left(f(z)-f\left(w\right)\right)\varphi_{22}(f(z),g) - CF_{32}(f(z),g)\nonumber\\
&=B'(g-g(z))\left(g-g\left(\kappa_1/z\right)\right) - C'(g-g(z))^2\nonumber\\
&=(g-g(z))\left\{B'\left(g-g\left(\kappa_1/z\right)\right) - C'(g-g(z))\right\}.\label{eqn:Pfz1}
\end{align}
In the second line, we have used factorization
$\varphi_{22}(f(z),g)=\kappa_1(g-g(z))(g-g(\kappa_1/z))$, which follows from
$\varphi_{22}(f(z),g(z))=0$ and $f(z)=f(\kappa_1/z)$. Similarly, we have
\begin{align}
 P\left(f\left(w\right),g\right)
&=\left(g-g\left(w\right)\right)\left\{A'\left(g-g\left(\kappa_1/w\right)\right) 
 - C''\left(g-g\left(w\right)\right)\right\}\nonumber\\
&=\left(g-g\left(q\kappa_1/z\right)\right)\left\{A'\left(g-g\left(z/q\right)\right) 
 - C''\left(g-g\left(q\kappa_1/z\right)\right)\right\}.\label{eqn:Pfz2}
\end{align}
In view of the two relations \eqref{eqn:Pfz1}, \eqref{eqn:Pfz2}, we choose the additional two points
as \eqref{eqn:qpe8_Lax_additional_points1} and \eqref{eqn:qpe8_Lax_additional_points2}.  Then from
\eqref{eqn:Pfz1} and \eqref{eqn:Pfz2} it follows that $A\propto A'\propto y(z/q)$, $B\propto
B'\propto y(qz)$ and $C\propto C'\propto C''\propto y(z)$, namely, $P(f,g)$ becomes linear
homogeneous in $y(qz)$, $y(z)$ and $y(z/q)$.

%
\subsection{Lax pair for  e-P$(E_8^{(1)})$ }\label{subsec:Lax_e-E8}
%
\subsubsection{Contiguity type relations}\label{subsubsec:contiguity}
We use the multiplicative parameters, $\kk_1,\kk_2, v_1,\ldots,v_8$, with $\kk_1^2\kk_2^2=q
\prod\limits_{i=1}^8 v_i$.  Let $[x]$ be a multiplicative odd theta function satisfying
$[x^{-1}]=-[x]$ and quasi-periodicity $[px]=-x^{-1}p^{-1/2}[x]$ with period $p$, for instance,
$[x]=x^{-\frac{1}{2}}(x,\frac{p}{x},p;p)_{\infty}$, 
where $(a_1,\ldots,a_n;q)_\infty = \prod_{\nu=1}^n\prod_{i=1}^\infty (1-a_\nu q^{i-1})$. We put
\begin{equation}\label{eqn:f_rFG}
\begin{array}{l}\smallskip
\displaystyle
f_a(z)=\left[\frac{a}{z}\right]\left[\frac{\kk_1}{az}\right],\quad
g_a(z)=\left[\frac{a}{z}\right]\left[\frac{\kk_2}{az}\right],\\\smallskip
\displaystyle
F(f,z)=f_a(z)f-f_b(z)=f_a(z)\{f-f(z)\},\quad f(z)=\frac{f_b(z)}{f_a(z)},\\
\displaystyle
G(g,z)=g_a(z)g-g_b(z)=g_a(z)\{g-g(z)\},\quad g(z)=\frac{g_b(z)}{g_a(z)}.
\end{array}
\end{equation}
Note that for any $a$ and $z$ we have
\begin{equation}\label{eqn:rel_fg}
\begin{split}
&f_a(z) = -f_z(a)=f_a\left(\frac{\kappa_1}{z}\right), \quad g_a(z) = -g_z(a)=g_a\left(\frac{\kappa_2}{z}\right), \\
&F(f,z)=F\left(f,\frac{\kappa_1}{z}\right), \quad G(f,z)=G\left(f,\frac{\kappa_2}{z}\right).
\end{split}
\end{equation}
By the Riemann relation: $g_a(b)g_c(x)+g_b(c)g_a(x)+g_c(a)g_b(x)=0$, we have
\begin{equation}\label{eqn:g-g_G}
g(x)-g(y)=\dfrac{g_a(b)g_x(y)}{g_a(x)g_a(y)}, \quad
G(g(x),y)=\dfrac{g_a(b)g_x(y)}{g_a(x)}. 
\end{equation}
The time evolution $T$ is given by $T: \kk_1\mapsto \frac{\kk_1}{q}, \kk_2 \mapsto q\kk_2$. 
For any functions or variables $X$, we use the notations $\overline{X}=T(X)$,  $\underline{X}=T^{-1}(X)$, e.g.
\begin{equation}
\oF(f,z)=\overline{f_a}(z)f-\overline{f_b}(z), \quad
\overline{f_a}(z)
=\left[\frac{\overline{a}}{z}\right]\left[\frac{\overline{\kappa}_1}{\overline{a}z}\right]
=\left[\frac{\overline{a}}{z}\right]\left[\frac{\kk_1}{q\overline{a}z}\right]. 
\end{equation}
We note that in the following argument we do not need to specify the discrete time evolution of $a$
and $b$.

The most fundamental object in the scalar Lax formulation is the linear difference equation $L_1(z)$
among $y(zq)$, $y(z)$ and $y(z/q)$. The explicit form of the equation $L_1(z)$ is, however, rather
complicated (see (\ref{L1-w-eq}) below). So, it is convenient to start with the following contiguity
type equations:
\begin{equation}
L_2(z):G\left(g,\frac{\kk_1}{z}\right)\,y(qz)-G(g,z)\,y(z)-\left[\frac{\kk_1}{z^2}\right]F(f,z)\,{\overline{y}}(z)=0,
\end{equation}
\begin{equation}
L_3(z):G\left(g,\frac{\kk_1}{q z}\right)U(z)\,{\overline{y}}(z)
-G(g,z)U\left(\frac{\kk_1}{q z}\right)\,{\overline{y}}(qz)
-\left[\frac{\kk_1}{q z^2}\right]w \oF(\of,z)\,y(qz)=0,
\end{equation}
where $\displaystyle U(z)=\prod_{i=1}^8\left[\frac{v_i}{z}\right]$; $\of=\of(f,g)$ and $w=w(f,g)$
are variables independent of $z$ to be determined later (see \eqref{eqn:e8_w} and
\eqref{eqn:w-in-fg} below).
\begin{rem}\rm
The linear equations $L_2, L_3$ given above are equivalent to those in \cite{NTY}. Here, the
coefficients are simplified by a gauge transformation. As a price for this, the function $y(z)$ is
no longer periodic but quasi-periodic: $y(pz)=\lambda z^{\alpha} y(z)$ where $\overline{\lambda}=p^2
\frac{\kk_2}{\kk_1^3}\lambda$, $q^\alpha=\frac{\kk_2^2}{\kk_1^2}$.
\end{rem}

As the necessary condition for the compatibility of these equations, one can easily derive 
$e$-P$(E_8^{(1)})$ as follows. If we put $g=g(z)$, i.e. $G(g,z)=0$ in $L_2(z)$ and $L_3(z)$, we
have 
\begin{equation}
 \begin{split}
G\left(g,\frac{\kk_1}{z}\right)\,y(qz) &= \left[\frac{\kk_1}{z^2}\right]F(f,z)\,{\overline{y}}(z),\\[2mm]
G\left(g,\frac{\kk_1}{q z}\right)U(z)\,{\overline{y}}(z)
 &= \left[\frac{\kk_1}{qz^2}\right]w \oF(\of,z)\,y(qz),
 \end{split}
\end{equation}
from which we obtain
\begin{equation}\label{compati-w}
\dfrac{wF(f,z)\oF(\of,z)}{U(z)}
= \dfrac{G\left(g,\frac{\kk_1}{z}\right)G\left(g,\frac{\kk_1}{q z}\right)}
{\left[\frac{\kk_1}{z^2}\right]\left[\frac{\kk_1}{q z^2}\right]}
= \left[\frac{\kk_1}{\kk_2}\right]\left[\frac{\kk_1}{q \kk_2}\right]\left(\dfrac{g_a(b)}{g_a(z)}\right)^2,
\end{equation}
for $g=g(z)$. Here we have used \eqref{eqn:g-g_G} to derive the second equality. Since
$g(z)=g(\frac{\kk_2}{z})$, we get another relation by replacing $z$ with $\frac{\kk_2}{z}$. Then by
taking the ratio of these two expressions, we have
\begin{equation}\label{compati-f}
\dfrac{F(f,\frac{\kk_2}{z})\oF(\of,\frac{\kk_2}{z})U(z)}
{F(f,z) \oF(\of,z) U(\frac{\kk_2}{z})}=1, \quad {\rm for} \ g=g(z).
\end{equation}
On the other hand, putting $f=f(z)$, i.e. $F(f,z)=0$ in $L_2(z)$ and $\underline{L_3}(z)$, we have 
\begin{equation}\label{compati-g}
\dfrac{G(g,\frac{\kk_1}{z})\uG(\ug,\frac{\kk_1}{z})U(z)}
{G(g,z)\uG(\ug,z) U(\frac{\kk_1}{z})}=1, \quad {\rm for} \ f=f(z). 
\end{equation}
Equations \eqref{compati-f} and \eqref{compati-g} are equivalent to $e$-P$(E_8^{(1)})$
\eqref{eqn:elliptic_Painleve_5}, which we will verify in the rest of this Section \ref{subsubsec:contiguity}.

From the relations (\ref{compati-f}), (\ref{compati-g}) and (\ref{compati-w}), the variables $\of$,
$\ug$ and $w$ are determined as rational functions in $(f,g)$.  For instance, substituting
\eqref{eqn:f_rFG} into \eqref{compati-f} we have
\begin{equation}\label{eqn:fb}
\frac{\overline{f_a}\left(\frac{\kappa_2}{z}\right)\of - \overline{f_b}\left(\frac{\kappa_2}{z}\right)}
{\overline{f_a}\left(z\right)\of - \overline{f_b}\left(z\right)}
=\frac{F(f,z)U\left(\frac{\kappa_2}{z}\right)}{F(f,\frac{\kappa_2}{z})U(z)}\quad \mbox{for}\ g=g(z).
\end{equation}

Solving \eqref{eqn:fb} in terms of $\overline{f}$, we obtain
\begin{equation}
 \left.\overline{f}\,\right|_{g=g(z)}
=\frac{F(f,\frac{\kk_2}{z})U(z)\overline{f_b}(\frac{\kk_2}{z})-F(f,z)U(\frac{\kk_2}{z})\overline{f_b}(z)}
{F(f,\frac{\kk_2}{z})U(z)\overline{f_a}(\frac{\kk_2}{z})-F(f,z)U(\frac{\kk_2}{z})\overline{f_a}(z)}.
\end{equation}
Note that
\begin{equation}\label{eqn:eP_FU_elliptic}
 \frac{F(f,\frac{\kk_2}{z})\overline{f_r}(\frac{\kk_2}{z})U(z)-F(f,z)\overline{f_r}(z)U(\frac{\kk_2}{z})}
{\left[\frac{\kappa_2}{z^2}\right]g_a(z)^4 },
\quad r=a,b,
\end{equation}
is an elliptic function in $z$ with poles of order 4 at $z=a, \frac{\kappa_2}{a}$, as is easily
verified by definition \eqref{eqn:f_rFG}.  This implies that \eqref{eqn:eP_FU_elliptic} is a
rational function in $g(z)$ and $f(z)$, but from the symmetry with respect to $z\leftrightarrow
\frac{\kappa_2}{z}$ this is actually a rational function only in $g(z)$.  We also remark that the
numerator is alternating with respect to $z\leftrightarrow \frac{\kappa_2}{z}$, and thus the four
zeros of $\left[\frac{\kappa_2}{z^2}\right]$ which are apparent poles of \eqref{eqn:eP_FU_elliptic}
cancel out.  Further, noticing that $g(z)$ has poles of order $1$ at $z=a, \frac{\kappa_2}{a}$, we
see that \eqref{eqn:eP_FU_elliptic} is a polynomial of degree 4 in $g=g(z)$.  By this argument, we
have\footnote{For a degree $(1,n)$ polynomial $P(f,g)$ vanishing at
$\big(f(v_i),g(v_i)\big)_{i=1}^{2n+2}$ ($\prod\limits_{i=1}^{2n+2}v_i=\kk_1 \kk_2^{n}$), we have
$P(f,g(z))={\rm
const.}\times\dfrac{F(f,\frac{\kk_2}{z})p(z)-F(f,z)p(\frac{\kk_2}{z})}{g_a(z)^{n}\left[\frac{\kk_2}{z^2}\right]}$,
$p(z)=\prod\limits_{i=1}^{2n+2}\left[\frac{v_i}{z}\right]$. }
\begin{equation}\label{eqn:e8_w}
\of=\frac{R_b(f,g)}{R_a(f,g)},\quad
f=\frac{S_b(\of,g)}{S_a(\of,g)},
\end{equation}
where $R_r(f,g)$ ($r=a,b$) are polynomials of bidegree $(1,4)$ in $(f,g)$.  The second equation of
\eqref{eqn:e8_w} is obtained in a similar manner, starting from \eqref{eqn:fb} and solving it in
terms of $f$. Here, $S_r(\of,g)$ are bidegree $(1,4)$ in $(\of,g)$.  Also, it should be noted that
$R_a(f,g)$ and $R_b(f,g)$ vanish at the eight points $(f(v_i),g(v_i))_{i=1,\ldots,8}$, as is easily
seen by setting $f=f(z)$, $z=v_i$ $(i=1,\ldots,8)$ in \eqref{eqn:eP_FU_elliptic}.  Similarly, it
follows from \eqref{compati-g} that $\ug$ is expressed as a rational function in $(f,g)$ of bidegree
$(1,4)$ having the same eight points as points of indeterminacy. These facts will be used in the
next section.

We now choose the normalization of $R_a, R_b$ as 
\begin{equation}\label{eq:of-normalization}
R_{r}(f,g)\big|_{g=g(z)}
=\frac{F(f,\frac{\kk_2}{z})\overline{f_r}(\frac{\kk_2}{z})U(z)-F(f,z)\overline{f_r}(z)U(\frac{\kk_2}{z})}{\overline{f_a}(b)\overline{f_z}(\frac{\kk_2}{z})g_a(z)^4 },
\quad r=a,b.
\end{equation}
Then from \eqref{compati-w} it turns out that $w$ can be expressed in the form
\begin{equation}\label{eqn:w-in-fg}
w=\frac{R_a(f,g)}{\varphi(f,g)}=\frac{S_a(\of,g)}{\psi(\of,g)},
\end{equation}
where $\varphi(f,g)=0$ 
is the reference curve $C_0$ of bidegree
$(2,2)$ parametrized by $(f,g)=(f(z),g(z))$ (see Remark \ref{rem:[u]}).
Similarly, $\psi(\of,g)=0$ is a bidegree $(2,2)$ curve parametrized by
$(\of,g)=(\of(z),g(z))$. Here we normalize them as
\begin{equation}
\varphi(f,g)\,\Big|_{g=g(z)}
=
\frac{F(f,z)F(f,\frac{\kk_2}{z})}
{\left[\frac{\kappa_1}{\kappa_2}\right]\left[\frac{\kappa_1}{q\kappa_2}\right]g_a(b)^2g_a(z)^2},\quad 
\psi(\of,g)\,\Big|_{g=g(z)}
=
\frac{\oF(\of,z)\oF(\of,\frac{\kk_2}{z})}
{\left[\frac{\kappa_1}{\kappa_2}\right]\left[\frac{q\kappa_1}{\kappa_2}\right]g_a(b)^2g_a(z)^2}.
\end{equation}
From (\ref{eq:of-normalization}) we have
\begin{align}
& R_a(f,g)\oF(\of,z)\big{|}_{g=g(z)}
=\left.R_a(f,g)\left(\overline{f_a}(z)\frac{R_b(f,g)}{R_a(f,g)}-\overline{f_b}(z)\right)\right|_{g=g(z)}\nonumber\\
&=\left.\overline{f_a}(z)R_b(f,g)-\overline{f_b}(z)R_a(f,g)\right|_{g=g(z)}\nonumber\\
&=\frac{F\left(f,\frac{\kappa_2}{z}\right)U(z)}{\overline{f_a}(b)\overline{f_z}\left(\frac{\kappa_2}{z}\right)g_a(z)^4}
\left\{
\overline{f_a}(z)\overline{f_b}\left(\frac{\kappa_2}{z}\right) - \overline{f_a}\left(\frac{\kappa_2}{z}\right)\overline{f_b}(z) \right\}\nonumber\\
&=\frac{F(f,\frac{\kk_2}{z})U(z)}{g_a(z)^4},\label{eqn:RFbar_by_FUg}
\end{align}
where we have used the Riemann relation and \eqref{eqn:rel_fg} in the last equality.
Hence
\begin{equation}
 \left.\frac{R_a(f,g)}{\varphi(f,g)}\right|_{g=g(z)}
=
\frac{\left[\frac{\kappa_1}{\kappa_2}\right]\left[\frac{\kappa_1}{q\kappa_2}\right]g_a(b)^2U(z)}{F(f,z)\oF(\of,z)g_a(z)^2}= 
w\,\bigl|_{g=g(z)}\quad \mbox{for}\ g=g(z),
\end{equation}
as required by (\ref{compati-w}).  The second relation of (\ref{eqn:w-in-fg}) can be derived in a
similar way. From the expression of $\of$ and $w$ in \eqref{eqn:e8_w}, \eqref{eqn:w-in-fg},
$w\varphi=R_a(f,g)$ and $w\varphi\of=R_b(f,g)$ are both polynomials of bidegree $(1,4)$. 
%
\subsubsection{Sufficiency for compatibility}
The relations (\ref{compati-f}), (\ref{compati-g}) and (\ref{compati-w}) are not only necessary but
also sufficient for the compatibility in the sense of Section \ref{subsec:lax_qpe6}.  To see this, we construct
the $L_1(z)$ equation by eliminating $\overline{y}(z)$ and $\overline{y}(z/q)$ from $L_2(z),
L_2(z/q)$ and $L_3(z/q)$:
\begin{equation}\label{L1-w-eq}
\begin{split}
L_1(z)=&
\dfrac{\left[\frac{q\kk_1}{z^2}\right]w \oF(\of,\frac{z}{q})}{G\left(g,\frac{z}{q}\right)G\left(g,\frac{\kk_1}{z}\right)}\,y(z)
+ \dfrac{U\left(\frac{z}{q}\right)}{\left[\frac{q^2\kk_1}{z^2}\right]F\left(f,\frac{z}{q}\right)}
\left\{y\left(\frac{z}{q} \right) - \dfrac{G\left(g,\frac{q\kk_1}{z}\right)}{G\left(g,\frac{z}{q}\right)}\,y(z)\right\}\\
&+ \dfrac{U\left(\frac{\kk_1}{z}\right)}{\left[\frac{\kk_1}{z^2}\right]F(f,z)}
\left\{y(qz) - \dfrac{G(g,z)}{G\left(g,\frac{\kk_1}{z}\right)}\,y(z) \right\}=0.
\end{split}
\end{equation}
Here, the variables $w, \of$ in (\ref{L1-w-eq}) should be viewed as functions of $(f,g)$
which are determined above in \eqref{eqn:e8_w} and \eqref{eqn:w-in-fg}.  Similarly, by eliminating ${y}(z)$
and ${y}(qz)$ from $L_3(z), L_3(z/q)$ and $L_2(z)$, we obtain
\begin{equation}\label{L1p-w-eq}
\begin{split}
L_4(z):\quad
&
  \dfrac{\left[\frac{\kk_1}{z^2}\right]w F(f,z)}{G(g,z)G\left(g,\frac{\kk_1}{z}\right)}\,\oy(z)
+ \dfrac{U\left(\frac{z}{q}\right)}{\left[\frac{q\kk_1}{z^2}\right]\oF\left(\of,\frac{z}{q}\right)}
\left\{\oy\left(\frac{z}{q}\right) - \dfrac{G\left(g,\frac{z}{q}\right)}{G\left(g,\frac{\kk_1}{z}\right)}
\dfrac{U\left(\frac{\kk_1}{z}\right)}{U\left(\frac{z}{q}\right)}\,\oy(z) \right\}\\
&+\dfrac{U\left(\frac{\kk_1}{qz}\right)}{\left[\frac{\kk_1}{qz^2}\right]\oF\left(\of,z\right)}
\left\{\oy(qz) - \dfrac{G\left(g,\frac{\kk_1}{qz}\right)}{G(g,z)}\dfrac{U(z)}{U\left(\frac{\kk_1}{qz}\right)}\,\oy(z)\right\}=0. 
\end{split}
\end{equation}
\begin{figure}[h]
\begin{center}\setlength{\unitlength}{1.3mm}
\begin{picture}(50,28)(-5,-3)
\put(-1,23){$\oy(\frac{z}{q})$}
\put(20,23){$\oy(z)$}
\put(41,23){$\oy(qz)$}
\put(-1,-3){$y(\frac{z}{q})$}
\put(20,-3){$y(z)$}
\put(41,-3){$y(qz)$}
\put(0,0){\line(1,0){20}}
\put(0,0){\line(0,1){20}}
\put(0,20){\line(1,-1){20}}
\put(1,21){\line(1,0){20}}
\put(21,1){\line(0,1){20}}
\put(1,21){\line(1,-1){20}}
\put(3,4){$L_2(\frac{z}{q})$}
\put(12,14){$L_3(\frac{z}{q})$}
\put(22,0){\line(1,0){20}}
\put(22,0){\line(0,1){20}}
\put(22,20){\line(1,-1){20}}
\put(24,4){$L_2(z)$}
\put(23,21){\line(1,0){20}}
\put(23,21){\line(1,-1){20}}
\put(43,1){\line(0,1){20}}
\put(32,14){$L_3(z)$}
\end{picture}
\end{center}
\end{figure}

The compatibility means $L_1(z) \propto T^{-1}({L_4(z)})$.  We prove this compatibility assuming
\eqref{compati-w} and $e$-P$(E_8^{(1)})$ \eqref{compati-f}, \eqref{compati-g}.  A convenient way is
to use the geometric characterizations of $L_1(z)$ and $L_4(z)$ as rational functions of $f$, $g$
\cite{Yamada:Lax}.  If we multiply the rational function $L_1(z)$ by a factor
$F(f,z)F(f,\frac{z}{q})\varphi(f,g)$,
\begin{equation}
 \widetilde{L}_1(z) = F(f,z)F\left(f,\frac{z}{q}\right)\varphi(f,g) L_1(z),
\end{equation}
becomes a polynomial  of bidegree $(3,4)$ divided by the factor
$G\left(g,\frac{z}{q}\right)G\left(g,\frac{\kk_1}{z}\right)$. It is actually a polynomial of bidegree $(3,2)$ 
since the residues at $g=\frac{z}{q}, \frac{\kappa_1}{z}$ vanish by \eqref{compati-w}, which we denote by $P_{32}(f,g)$.
This polynomial is characterized by the vanishing condition at the following 12 points:
\begin{equation}\label{points-cond-L1}
\big(f(v),g(v)\big)_{v=v_1, \ldots, v_8, z, \frac{q\kk_1}{z}}\ , \quad 
\big(f(x),\gamma_x\big)_{x=z,\frac{z}{q}}\ , \quad \mbox{where}\quad
\dfrac{G(\gamma_x,\frac{\kk_1}{x})}{G(\gamma_x,x)}=\dfrac{y(x)}{y(qx)}.
\end{equation}
It is directly seen that $P_{32}(f,g)$ vanishes at the last two points by noticing
$F(f(z),z)=G(g(z),z)=0$.  Moreover, one can verify that $P_{32}(f,g)$ vanishes at the first eight
points as follows. We have
\begin{equation}\label{eqn:qpe8_tildeL_1}
\begin{split}
P_{32}(f,g)=& 
\dfrac{\left[\frac{q\kk_1}{z^2}\right]F(f,z)F\left(f,\frac{z}{q}\right)A(f,g,z)}{G\left(g,\frac{z}{q}\right)G\left(g,\frac{\kk_1}{z}\right)}\,y(z)
+ \dfrac{F(f,z)\varphi(f,g)U\left(\frac{z}{q}\right)}{\left[\frac{q^2\kk_1}{z^2}\right]}\
\left\{y\left(\frac{z}{q} \right) - \dfrac{G\left(g,\frac{q\kk_1}{z}\right)}{G\left(g,\frac{z}{q}\right)}\,y(z)\right\}\\
&+ \dfrac{F\left(f,\frac{z}{q}\right)\varphi(f,g)U\left(\frac{\kk_1}{z}\right)}{\left[\frac{\kk_1}{z^2}\right]}
\left\{y(qz) - \dfrac{G(g,z)}{G\left(g,\frac{\kk_1}{z}\right)}\,y(z) \right\}.
\end{split}
\end{equation}
Here we have from \eqref{eqn:e8_w} and \eqref{eqn:w-in-fg}
\begin{equation}
A(f,g,z)= \varphi(f,g) w \oF\left(\of,\frac{z}{q}\right) 
= \overline{f_a}\left(\frac{z}{q}\right)R_b(f,g) - \overline{f_b}\left(\frac{z}{q}\right)R_a(f,g),
\end{equation}
which vanishes at the eight points $(f(v_i), g(v_i))_{i=1,\ldots,8}$ as mentioned in the end of
Section \ref{subsubsec:contiguity}. It is also obvious that the second and third terms of
\eqref{eqn:qpe8_tildeL_1}, and thus $P_{32}(f,g)$, vanish at those eight points. It is also clear
that $P_{32}(f,g)$ vanishes at $\big(f(v),g(v)\big)_{v=z, \frac{q\kk_1}{z}}$ because of the factors
$F(f,z)$, $F\left(f,\frac{z}{q}\right)=F\left(f,\frac{q\kappa_1}{z}\right)$ (see \eqref{eqn:rel_fg})
in the first term (vanishing of the second and third term is obvious due to $\varphi(f,g)$).

Similarly, one can show that $L_4(z)=0$ is a curve of bidegree $(3,2)$ in $(\of,g)$ passing through
\begin{equation}
\big(\of(v),g(v)\big)_{v=v_1, \ldots, v_8, \frac{z}{q}, \frac{\kk_1}{qz}}, \quad
\big(\of(x),\gamma'_x\big)_{x=z,\frac{z}{q}}, \quad\mbox{where}\quad
\dfrac{G(\gamma'_x,x)}{G(\gamma'_x,\frac{\kk_1}{qx})}\dfrac{U(\frac{\kk_1}{qx})}{U(x)}=\dfrac{\oy(x)}{\oy(qx)},
\end{equation}
and hence, $T^{-1}(L_4(z))=0$ is a bidegree $(3,2)$ curve in $(f,\ug)$ passing through
\begin{equation}\label{points-cond-L1p}
\big(f(v),\ug(v)\big)_{v=v_1, \ldots, v_8, \frac{z}{q}, \frac{\kk_1}{z}}, \quad
\big(f(x),\gamma''_x\big)_{x=z,\frac{z}{q}}, \quad\mbox{where}\quad 
\dfrac{\uG(\gamma''_x,x)}{\uG(\gamma''_x,\frac{\kk_1}{x})}\dfrac{U(\frac{\kk_1}{x})}{U(x)}=\dfrac{y(x)}{y(qx)}.
\end{equation}
We denote by $Q_{32}(f,\ug)$ a polynomial in $(f,\ug)$ of bidegree $(3,2)$ defining the curve
$T^{-1}(L_4(z))=0$. 

Our remaining task is to express $Q_{32}(f,\ug)$ as a rational function in $(f,g)$ and to compare it
with $P_{32}(f,g)$. To this end, we express $\ug$ by $g$ as $\ug(f,g)
=\dfrac{B_{41}(f,g)}{A_{41}(f,g)}$, where $A_{41}(f,g)$, $B_{41}(f,g)$ are polynomials of bidegree
$(4,1)$ vanishing at the eight points $(f(v_i),g(v_i))_{i=1,\ldots,8}$, as explained above.
We have 
\begin{equation}
 Q_{32}\left(f,\frac{B_{41}(f,g)}{A_{41}(f,g)}\right) = \frac{P_{11,2}(f,g)}{A_{41}(f,g)^2},
\end{equation}
by the degree counting, where $P_{11,2}(f,g)$ is a polynomial of bidegree $(11,2)$. Writing $Q_{32}(f,\ug)$ as
\begin{equation}
 Q_{32}(f,\ug)=\alpha_3(f)\ug^2 + \beta_3(f)\ug + \gamma_3(f),
\end{equation}
we see that $P_{11,2}(f,g)$ is expressed as
\begin{equation}
 P_{11,2}(f,g) = \alpha_3(f)B_{41}(f,g)^2 + \beta_3(f)B_{41}(f,g)A_{41}(f,g) + \gamma_3(f)A_{41}(f,g)^2.
\end{equation}
Hence we see that $P_{11,2}(f,g)$ has zeros with multiplicity $2$ at the eight points
$(f(v_i),g(v_i))_{i=1,\ldots,8}$. We also note that if we set $z=v_i$ in (\ref{compati-g}) then the
numerator of the left hand side vanishes because of $U(z)$, and thus $\uG(\ug,v_i)$ in the
denominator must vanish. This implies that in case of $f=f(v_i)$ it follows that $\ug=\ug(v_i)$
regardless of the generic value of $g$ ($i=1,\ldots,8$). Then we have
\begin{equation}
 Q_{32}(f,\ug)\Big|_{f=f(v_i)} = Q_{32}(f(v_i),\ug(v_i))=0,
\end{equation}
which means that $Q_{32}(f,\ug)$ is divisible by $\prod\limits_{i=1}^8(f-f(v_i))$. Hence we have
the factorization
\begin{equation}
Q_{32}\left(f,\dfrac{B_{41}(f,g)}{A_{41}(f,g)}\right)
=\dfrac{\prod\limits_{i=1}^8(f-f(v_i))}{A_{41}(f,g)^2}\hat{P}_{32}(f,g).
\end{equation}
where $\hat{P}_{32}(f,g)$ is a certain polynomial of bidegree $(3,2)$ and linear in $y(z)$, $y(qz)$,
$y\left(\frac{z}{q}\right)$.  Our goal is to show that $\hat{P}_{32}(f,g)$ is actually proportional
to $P_{32}(f,g)$.  Noticing that P$_{11,2}(f,g)$ has zeros with multiplicity $2$ at
$(f(v_i),g(v_i))_{i=1,\ldots,8}$, it follows that $\hat{P}_{32}(f,g)$ also vanishes at those
points. One can also observe from \eqref{compati-g} that $\ug = \ug\left(\frac{\kappa_1}{v}\right)$
for $(f,g)=(f(v),g(v))$ when $v=z,\frac{z}{q}$, and $\ug = \gamma''_x$ for $(f,g)=(f(x),\gamma_x)$
when $x=z,\frac{z}{q}$. This implies that $Q_{32}(f,\ug)$ and thus $\hat{P}_{32}(f,g)$ vanish at the
remaining four points as desired. Hence we have shown that $\hat{P}_{32}(f,g)$ is equal to
$P_{32}(f,g)$ up to constant multiple.

\subsubsection{Case of $q$-P$(E_8^{(1)})$}

We continuously use the parameters, $\kk_1,\kk_2, v_1,\ldots,v_8$ with $\kk_1^2\kk_2^2=q
\prod\limits_{i=1}^8 v_i$, and the time evolution $T: \kk_1\mapsto \frac{\kk_1}{q}, \kk_2 \mapsto q \kk_2$.
In $q$-$E_8$ case, we put $f(z)=z+\frac{\kk_1}{z}$, $g(z)=z+\frac{\kk_2}{z}$,
$U(z)=z^{-4}\prod\limits_{i=1}^8(z-v_i)$.

We start with the contiguity type equations:
\begin{equation}\label{eqn:qpe8_L2}
L_2(z):\ \left\{g-g\left(\frac{\kk_1}{z}\right)\right\}\,y(qz)
-\left\{g-g(z)\right\}\,y(z)
-\left(z-\frac{\kk_1}{z}\right)\left\{f-f(z)\right\}\,{\overline{y}}(z)=0,
\end{equation}
\begin{equation}
L_3(z):\ \left\{g-g\left(\frac{\kk_1}{q z}\right)\right\}U(z)\,{\overline{y}}(z)
-\left\{g-g(z)\right\}U\left(\frac{\kk_1}{q z}\right)\,{\overline{y}}(qz)
-w \left(z-\frac{\kk_1}{q z}\right)\left\{\of-\of(z)\right\},y(qz)=0.
\end{equation}
{} From the compatibility of these equations, the $q$-P$(E_8^{(1)})$ Painlev\'e equation is derived as follows.
Putting $g=g(z)$ in $L_2(z)$ and $L_3(z)$, we have
\begin{equation}
 \frac{w\{f-f(z)\}\{\of-\of(z)\}}{U(z)}=\frac{\{g-g(\frac{\kk_1}{z})\}
\{g-g(\frac{\kk_1}{q z})\}}
{(z-\frac{\kk_1}{z})(z-\frac{\kk_1}{q z})}=\frac{(\kk_1-\kk_2)(\kk_1-q \kk_2)}{\kk_1^2 },\ 
\mbox{for}\  g=g(z).
\end{equation}
This relation holds also if $z$ is replaced by $\frac{\kappa_2}{z}$, and taking the ratio of these
two expressions, we have
\begin{equation}\label{q-compati-f}
\dfrac{\{f-f\Bigl(\frac{\kk_2}{z}\Bigr)\}\{\of-\of\Bigl(\frac{\kk_2}{z}\Bigr)\}U(z)}
{\{f-f(z)\} \{\of-\of(z)\} U\Bigl(\frac{\kk_2}{z}\Bigr)}=1, \quad {\rm for} \ g=g(z),
\end{equation}
along with
\begin{equation}
w=\frac{(\kk_1-\kk_2)q\kk_2}{\kk_1^2\left(z-\frac{\kk_2}{z}\right)}
\left[\frac{U(z)}{f-f(z)}-\frac{U\left(\frac{\kk_2}{z}\right)}{f-f\left(\frac{\kk_2}{z}\right)}\right]
=\frac{(\kk_1-q\kk_2)\kk_2}{\kk_1^2\left(z - \frac{\kk_2}{z}\right)}
\left[\frac{U(z)}{\of-\of(z)}-\frac{U\left(\frac{\kk_2}{z}\right)}{\of-\of\left(\frac{\kk_2}{z}\right)}\right].
\end{equation}
On the other hand, putting $f=f(z)$ in $L_2(z)$ and $\underline{L_3}(z)$, we have 
\begin{equation}\label{q-compati-g}
\dfrac{\left\{g-g\left(\frac{\kk_1}{z}\right)\right\}\left\{\ug-\ug\left(\frac{\kk_1}{z}\right)\right\}U(z)}
{\left\{g-g(z)\right\} \left\{\ug-\ug(z)\right\} U\left(\frac{\kk_1}{z}\right)}=1, \quad {\rm for} \ f=f(z). 
\end{equation}
These relations determine the variables $\of, \ug, w$ as rational functions in $(f,g)$.\footnote{ A
bidegree $(1,n)$ polynomial $P(f,g)$ vanishing at $\big(f(v_i),g(v_i)\big)_{i=1}^{2n+2}$
($\prod_{i=1}^{2n+2}v_i=\kk_1 \kk_2^{n}$) is given by $P(f,g(v))=\frac{{\rm const.}\big[
\big\{f-f(\frac{\kk_2}{v})\big\}p(v)-\big\{f-f(v)\big\}p(\frac{\kk_2}{v})\big]}{v-\frac{\kk2}{v}}$,
$p(v)=v^{-n}\prod_{i=1}^{2n+2}(v-v_i)$. }

In a similar way to the elliptic case, we have
\begin{equation}\label{eqn:qpe8_L1}
\begin{split}
L_1(z):\quad
&\dfrac{w\left(\frac{z}{q}-\frac{\kk_1}{z}\right)\left\{\of-\of\left(\frac{z}{q}\right)\right\}}
{\left\{g - g\left(\frac{z}{q}\right)\right\}\left\{g-g\left(\frac{\kk_1}{z}\right)\right\}}\,y(z)
+ \dfrac{U\left(\frac{z}{q}\right)}{\left(\frac{z}{q}-\frac{q\kk_1}{z}\right)\left\{f-f(\frac{z}{q})\right\}}
\left[y\left(\frac{z}{q}\right) - \dfrac{g - g\left(\frac{q\kk_1}{z}\right)}{g - g\left(\frac{z}{q}\right)}\,y(z) \right]\\
&+\dfrac{U\left(\frac{\kk_1}{z}\right)}{\left(z-\frac{\kk_1}{z}\right)\left\{f-f(z)\right\}}
\left[y(qz) - \dfrac{g-g(z)}{g-g\left(\frac{\kk_1}{z}\right)}y(z)\right]=0,\\[6mm]
L_4(z):\quad
&\dfrac{w\left(z-\frac{\kk_1}{z}\right)\left\{f-f(z)\right\}}{\{g-g(z)\}\left\{g-g\left(\frac{\kk_1}{z}\right)\right\}}\,\oy(z)\\
&+ \dfrac{1}{\left(\frac{z}{q}-\frac{\kk_1}{z}\right)\left\{\of-\of\left(\frac{z}{q}\right)\right\}}
\left[
U\left(\frac{z}{q}\right)\,\oy\left(\frac{z}{q}\right)
- \dfrac{g-g\left(\frac{z}{q}\right)}{g-g\left(\frac{\kk_1}{z}\right)}U\left(\frac{\kk_1}{z}\right)\,\oy(z)
\right]\\
&+\dfrac{1}{\left(z-\frac{\kk_1}{qz}\right)\left\{\of-\of(z)\right\}}
\left[U\left(\frac{\kk_1}{qz}\right)\,\oy(qz) - \dfrac{g-g\left(\frac{\kk_1}{qz}\right)}{g-g(z)}U(z)\,\oy(z)\right]=0.
\end{split}
\end{equation}
The compatibility $L_1(z) \propto T^{-1}({L_4(z)})$ is confirmed by using the the following geometric characterizations.\\
$L_1(z)=0$: 
the curve of bidegree $(3,2)$ in $(f,g)$ passing through 
\begin{equation}
\big(f(v),g(v)\big)_{v=v_1, \ldots, v_8, z, \frac{q\kk_1}{z}}, \quad 
\big(f(x),\gamma_x\big)_{x=z,\frac{z}{q}}, \quad
\dfrac{\gamma_x-g\Bigl(\frac{\kk_1}{x}\Bigr)}{\gamma_x-g(x)}=\dfrac{y(x)}{y(qx)}. 
\end{equation}
$L_4(z)=0$:  the curve of bidegree $(3,2)$ in $(\of,g)$ passing through
\begin{equation}
 \big(\of(v),g(v)\big)_{v=v_1, \ldots, v_8, \frac{z}{q}, \frac{\kk_1}{qz}}, \quad
\big(\of(x),\gamma'_x\big)_{x=z,\frac{z}{q}}, \quad
\dfrac{\gamma'_x-g(x)}{\gamma'_x-g\Bigl(\frac{\kk_1}{qx}\Bigr)}
\dfrac{U\Bigl(\frac{\kk_1}{qx}\Bigr)}{U(x)}=\dfrac{\oy(x)}{\oy(qx)}.
\end{equation}
$T^{-1}(L_4(z))=0$: the curve of bidegree $(3,2)$ in $(f,\ug)$ passing through
\begin{equation}
\big(f(v),\ug(v)\big)_{v=v_1, \ldots, v_8, \frac{z}{q}, \frac{\kk_1}{z}}, \quad
\big(f(x),\gamma''_x\big)_{x=z,\frac{z}{q}}, \quad
\dfrac{\gamma''_x-\ug(x)}{\gamma''_x-\ug\Bigl(\frac{\kk_1}{x}\Bigl)}\dfrac{U\Bigl(\frac{\kk_1}{x}\Bigr)}{U(x)}
=\dfrac{y(x)}{y(qx)}. 
\end{equation}

\section{Basic Data for Discrete Painlev\'e Equations}
In this section we provide with basic data for all discrete Painlev\'e equations of QRT type:
equations, point configurations/root data, Weyl group representations, Lax
pairs and hypergeometric solutions.
\begin{rem}\rm
 In the following, we also use the symbols $E_3^{(1)} = (A_2 + A_1)^{(1)}$ and $E_2^{(1)} = (A_1 +
\underset{|\alpha|^2=14}{A_1})^{(1)}$ to simplify the notation.  In these cases, the labels $(a)$
and $(b)$ are used to discriminate two inequivalent equations associated with different realizations
of the same symmetry/surface type. These may be realized as equations with respect to two different
directions (see \cite{KNY:qp4} for $q$-P$_{\rm III}$ and $q$-P$_{\rm IV}$ of the case
$E_3^{(1)}/A_5^{(1)}$). In this paper, however, we formulate them in terms of different point
configurations in order to represent the equations in standard forms as in the literature. In Table
\ref{tab:configuration}, we omit the case of $A_0^{(1)}/E_8^{(1)}$ which allows P$_{\rm I}$:
$y''=6y^3+t$ as a continuous flow, since the surface cannot be realized by eight point blowing-up of
$\mathbb{P}^1\times\mathbb{P}^1$ and there is no discrete symmetry. We also omit the case of
$A_0^{(1)}/A_8^{(1)}$ which has no discrete flow.
\end{rem}
\begin{table}[H]
\begingroup
\renewcommand{\arraystretch}{1.5}
\noindent\begin{tabular}{|c||c|c|c|c|c|}
\hline
& & & & & \\[-10pt]
\parbox{1.6cm}{\centering\footnotesize Symmetry\\/ Surface} &\parbox{1.2cm}{\centering\footnotesize Equation} 
&\parbox{1.8cm}{\footnotesize Point\\ Configuration\\ / Root Data}  
&\parbox{2.1cm}{\centering\footnotesize Weyl Group\\ Representations} 
& \parbox{1.5cm}{\centering\footnotesize Lax Pair}&\parbox{2.2cm}{\centering\footnotesize Hypergeometric\\ Solution}\\[-10pt]
& & & & & \\
\hline
\hline
\parbox{1.3cm}{\hskip-3mm{\footnotesize elliptic}\\$E_8^{(1)}/A_0^{(1)}$} &Sec. \ref{subsubsec:equation_e-E8} 
&Sec. \ref{subsubsec:conf_e-E8}  
& \eqref{eqn:E8_parameters_and_fg} & Sec. \ref{subsec:Lax_e-E8}& Sec. \ref{subsubsec:hyper_e-E8}\\
\hline
\hline
\parbox{1.6cm}{\hskip-4mm{\footnotesize multiplicative}\\ $E_8^{(1)}/A_0^{(1)}$} &Sec. \ref{subsubsec:equation_q-E8} 
&Sec. \ref{subsubsec:conf_q-E8} 
&Sec. \ref{subsubsec:Weyl_q-E8} &Sec. \ref{subsubsec:Lax_q-E8} & Sec. \ref{subsubsec:hyper_q-E8}\\
\hline
$E_7^{(1)}/A_1^{(1)}$ &Sec. \ref{subsubsec:equation_q-E7} 
&Sec. \ref{subsubsec:conf_q-E7} 
&Sec. \ref{subsubsec:Weyl_q-E7} &Sec. \ref{subsubsec:Lax_q-E7} & Sec. \ref{subsubsec:hyper_q-E7}\\
\hline
$E_6^{(1)}/A_2^{(1)}$ &Sec. \ref{subsubsec:equation_q-E6} 
&Sec. \ref{subsubsec:conf_q-E6} 
&Sec. \ref{subsubsec:Weyl_q-E6} &Sec. \ref{subsubsec:Lax_q-E6} & Sec. \ref{subsubsec:hyper_q-E6}\\
\hline
$D_5^{(1)}/A_3^{(1)}$ &Sec. \ref{subsubsec:equation_q-D5} 
&Sec. \ref{subsubsec:conf_q-D5} 
&Sec. \ref{subsubsec:Weyl_q-D5} &Sec. \ref{subsubsec:Lax_q-D5} & Sec. \ref{subsubsec:hyper_q-D5}\\
\hline
$A_4^{(1)}/A_4^{(1)}$ &Sec. \ref{subsubsec:equation_q-A4} 
&Sec. \ref{subsubsec:conf_q-A4} 
&Sec. \ref{subsubsec:Weyl_q-A4} &Sec. \ref{subsubsec:Lax_q-A4} & Sec. \ref{subsubsec:hyper_q-A4}\\
\hline
$E_3^{(1)}/A_5^{(1)}(a)$ &Sec. \ref{subsubsec:equation_q-E3a} 
&Sec. \ref{subsubsec:conf_q-E3a} 
&Sec. \ref{subsubsec:Weyl_q-E3a} &Sec. \ref{subsubsec:Lax_q-E3a} & Sec. \ref{subsubsec:hyper_q-E3a}\\
\hline
$E_3^{(1)}/A_5^{(1)}(b)$ &Sec. \ref{subsubsec:equation_q-E3b} 
&Sec. \ref{subsubsec:conf_q-E3b} 
&Sec. \ref{subsubsec:Weyl_q-E3b} &Sec. \ref{subsubsec:Lax_q-E3b} & Sec. \ref{subsubsec:hyper_q-E3b}\\
\hline
$E_2^{(1)}/A_6^{(1)}(a)$ &Sec. \ref{subsubsec:equation_q-E2a} 
&Sec. \ref{subsubsec:conf_q-E2a} 
&Sec. \ref{subsubsec:Weyl_q-E2a} &Sec. \ref{subsubsec:Lax_q-E2a} & Sec. \ref{subsubsec:hyper_q-E2a}\\
\hline
$E_2^{(1)}/A_6^{(1)}(b)$ &Sec. \ref{subsubsec:equation_q-E2b} 
&Sec. \ref{subsubsec:conf_q-E2b} 
&Sec. \ref{subsubsec:Weyl_q-E2b} &Sec. \ref{subsubsec:Lax_q-E2b} & None\\
\hline
$\underset{|\alpha|^2=8}{A_1^{(1)}}/A_7^{(1)}$ &Sec. \ref{subsubsec:equation_q-A1a} 
&Sec. \ref{subsubsec:conf_q-A1a} 
&Sec. \ref{subsubsec:Weyl_q-A1a} &Sec. \ref{subsubsec:Lax_q-A1a} &None\\
\hline
$A_1^{(1)}/A_7^{(1)}$ &Sec. \ref{subsubsec:equation_q-A1b} 
&Sec. \ref{subsubsec:conf_q-A1b} 
&Sec. \ref{subsubsec:Weyl_q-A1b} &Sec. \ref{subsubsec:Lax_q-A1b} &None\\
\hline
\hline
\parbox{1.3cm}{\hskip-3mm{\footnotesize additive}\\$E_8^{(1)}/A_0^{(1)}$} &Sec. \ref{subsubsec:equation_d-E8} 
&Sec. \ref{subsubsec:conf_d-E8} 
&Sec. \ref{subsubsec:Weyl_d-E8} &Sec. \ref{subsubsec:Lax_d-E8} &Sec. \ref{subsubsec:hyper_d-E8}\\
\hline
$E_7^{(1)}/A_1^{(1)}$ &Sec. \ref{subsubsec:equation_d-E7} 
&Sec. \ref{subsubsec:conf_d-E7} 
&Sec. \ref{subsubsec:Weyl_d-E7} &Sec. \ref{subsubsec:Lax_d-E7} &Sec. \ref{subsubsec:hyper_d-E7}\\
\hline
$E_6^{(1)}/A_2^{(1)}$ &Sec. \ref{subsubsec:equation_d-E6} 
&Sec. \ref{subsubsec:conf_d-E6} 
&Sec. \ref{subsubsec:Weyl_d-E6} &Sec. \ref{subsubsec:Lax_d-E6} &Sec. \ref{subsubsec:hyper_d-E6}\\
\hline
$D_4^{(1)}/D_4^{(1)}$ &Sec. \ref{subsubsec:equation_d-D4} 
&Sec. \ref{subsubsec:conf_d-D4} 
&Sec. \ref{subsubsec:Weyl_d-D4} &Sec. \ref{subsubsec:Lax_d-D4} &Sec. \ref{subsubsec:hyper_d-D4}\\
\hline
$A_3^{(1)}/D_5^{(1)}$ &Sec. \ref{subsubsec:equation_d-A3} 
&Sec. \ref{subsubsec:conf_d-A3} 
&Sec. \ref{subsubsec:Weyl_d-A3} &Sec. \ref{subsubsec:Lax_d-A3} &Sec. \ref{subsubsec:hyper_d-A3}\\
\hline
$2A_1^{(1)}/D_6^{(1)}$ &Sec. \ref{subsubsec:equation_d-2A1} 
&Sec. \ref{subsubsec:conf_d-2A1} 
&Sec. \ref{subsubsec:Weyl_d-2A1} &Sec. \ref{subsubsec:Lax_d-2A1} &Sec. \ref{subsubsec:hyper_d-2A1}\\
\hline
$A_2^{(1)}/E_6^{(1)}$ &Sec. \ref{subsubsec:equation_d-A2} 
&Sec. \ref{subsubsec:conf_d-A2} 
&Sec. \ref{subsubsec:Weyl_d-A2} &Sec. \ref{subsubsec:Lax_d-A2} &Sec. \ref{subsubsec:hyper_d-A2}\\
\hline
$\underset{|\alpha|^2=4}{A_1^{(1)}}/D_7^{(1)}$ &Sec. \ref{subsubsec:equation_d-A1'} 
&Sec. \ref{subsubsec:conf_d-A1'} 
&Sec. \ref{subsubsec:Weyl_d-A1'} &Sec. \ref{subsubsec:Lax_d-A1'} &None\\
\hline
$A_1^{(1)}/E_7^{(1)}$ &Sec. \ref{subsubsec:equation_d-A1} 
&Sec. \ref{subsubsec:conf_d-A1} 
&Sec. \ref{subsubsec:Weyl_d-A1} &Sec. \ref{subsubsec:Lax_d-A1} & None\\
\hline
$A_0^{(1)}/D_8^{(1)}$ &Sec. \ref{subsubsec:equation_d-A0} 
&Sec. \ref{subsubsec:conf_d-A0D8}
&Sec. \ref{subsubsec:Weyl_d-A0} &Sec. \ref{subsubsec:Lax_d-A0D8} & None\\
\hline
\end{tabular}
\endgroup 
\caption{List of data associated with possible point configurations.}\label{tab:configuration}
\end{table}
%
\subsection{Discrete Painlev\'e equations}\label{subsec:dPs}
In this subsection, for each point configuration in the Table \ref{tab:configuration} we give an
explicit form of the discrete Painlev\'e equation with respect to the QRT direction.  We use the
symbols [$e$,$q$,d]-P({\it symmetry/surface type}) for reference to the (discrete) Painlev\'e
equations; the first symbol represents the type of time evolution, $e$: elliptic, $q$:
multiplicative ($q$-difference), d: additive (difference), none: continuous (differential).

Basically we use below the parameters $\kappa_i$ ($i=1,2$) and $v_i$ ($i=1,\ldots,8$),
and $f$, $g$ denote dependent variables. Also $\overline{\phantom{F}}$ is the time evolution such that
\begin{equation}\label{eqn:T1_pars4}
\overline{\kappa}_1=\frac{\kappa_1}{q},\quad \overline{\kappa}_2=q\kappa_2,\quad
\overline{v}_i=v_i\ (i=1,\ldots,8),\quad \kappa_1^2\kappa_2^2=q\prod_{i=1}^8v_i,
\end{equation}
for elliptic and multiplicative cases and
\begin{equation}\label{eqn:T1_pars5}
\overline{\kappa}_1=\kappa_1 - \delta,\quad \overline{\kappa_2}=\kappa_2 + \delta,\quad
\overline{v}_i=v_i\ (i=1,\ldots,8),\quad 2(\kappa_1 + \kappa_2) =\delta + \sum_{i=1}^8v_i
\end{equation}
for additive cases. Relation to the parameters $h_i$ ($i=1,2$) and $e_i$ ($i=1,\ldots,8$) used in Section \ref{sec:dP}
is given in Remark \ref{rem:kappa-var}.

\subsubsection{$e$-P$(E_8^{(1)}/A_0^{(1)})$}\label{subsubsec:equation_e-E8}
\begin{equation}
\begin{split}
\frac{\Bigl\{f - f\Bigl(\dfrac{\kappa_2}{t}\Bigr)\Bigr\}
\Bigl\{\of - \overline{f}\Bigl(\dfrac{\kappa_2}{t}\Bigr)\Bigr\}}
{\Bigl\{f - f(t)\Bigr\}\Bigl\{\of - \overline{f}(t)\Bigr\}}
=\frac{f_a(t)\overline{f_a}(t)}{f_a\Bigl(\dfrac{\kappa_2}{t}\Bigr)\overline{f_a}\Bigl(\dfrac{\kappa_2}{t}\Bigr)}
\frac{U\Bigl(\dfrac{\kappa_2}{t}\Bigr)}{U(t)},\\
 \frac{\Bigl\{g - g\Bigl(\dfrac{\kappa_1}{s}\Bigr)\Bigr\}
\Bigl\{\ug - \underline{g}\Bigl(\dfrac{\kappa_1}{s}\Bigr)\Bigr\}}
{\Bigl\{g - g(s)\Bigr\}\Bigl\{\ug - \underline{g}(s)\Bigr\}}
=\frac{g_a(s)\underline{g_a}\Bigl(s\Bigr)}
{g_a\Bigl(\dfrac{\kappa_1}{s}\Bigr)\underline{g_a}\Bigl(\dfrac{\kappa_1}{s}\Bigr)}
\frac{U\Bigl(\dfrac{\kappa_1}{s}\Bigr)}{U(s)},
\end{split}
\end{equation}
where $t$ and $s$ are the variables such that $g=g(t)$, $f=f(s)$,
\begin{equation}
\begin{split}
& f_a(z)=\left[\frac{a}{z}\right]\left[\frac{\kappa_1}{az}\right],\quad
 g_a(z)=\left[\frac{a}{z}\right]\left[\frac{\kappa_2}{az}\right],\\
&f(z)=\frac{f_b(z)}{f_a(z)},\quad g(z)=\frac{g_b(z)}{g_a(z)},\quad U(z)=\prod_{i=1}^8\left[\frac{v_i}{z}\right],
\end{split}
\end{equation}
$[z]$ is the multiplicative theta function given in Section \ref{subsubsec:contiguity},
and $a$, $b$ are arbitrary.\\

\subsubsection{$q$-P$(E_8^{(1)}/A_0^{(1)})$}\label{subsubsec:equation_q-E8}
\begin{equation}
\begin{array}{ll}\bigskip
{\displaystyle 
\dfrac{\Bigl\{f-f\Bigl(\dfrac{\kappa_2}{t}\Bigr)\Bigr\}\Bigl\{\overline{f}-\overline{f}\Bigl(\dfrac{\kappa_2}{t}\Bigr)\Bigr\}}
{\{f-f(t)\} \{\overline{f}-\overline{f}(t)\} }
=\frac{U\Bigl(\dfrac{\kappa_2}{t}\Bigr)}{U(t)},} &{\displaystyle g=g(t)},\\
{\displaystyle \dfrac{\Bigl\{g-g\Bigl(\dfrac{\kappa_1}{s}\Bigr)\Bigr\}
\Bigl\{\underline{g}-\underline{g}\Bigl(\dfrac{\kappa_1}{s}\Bigr)\Bigr\}}
{\{g-g(s)\} \{\underline{g}-\underline{g}(s)\} }=\frac{U\Bigl(\dfrac{\kappa_1}{s}\Bigr)}{U(s)},} &{\displaystyle f=f(s). }
\end{array}
\end{equation}
where $t$ and $s$ are the variables such that $g=g(t)$, $f=f(s)$,
\begin{equation}
U(z)=\frac{1}{z^{4}}\prod_{i=1}^8(z-v_i),\quad f(z)=z+\frac{\kappa_1}{z},\quad g(z)=z+\frac{\kappa_2}{z}.
\end{equation}
%
\subsubsection{$q$-P$(E_7^{(1)}/A_1^{(1)})$}\label{subsubsec:equation_q-E7}
\begin{equation}
\begin{split}
\dfrac{\Bigl(f g-\dfrac{\kappa_1}{\kappa_2}\Bigr) \Bigl(\overline{f} g -\dfrac{\kappa_1}{q\kappa_2}\Bigr)}
{(f g-1)(\overline{f} g-1) }
=\frac{\prod\limits_{i=5}^8\Bigl(g-\dfrac{v_i}{\kappa_2}\Bigr)}{\prod\limits_{i=1}^4\Bigl(g-\dfrac{1}{v_i}\Bigr)}, \\[2mm]
\dfrac{\Bigl(f g-\dfrac{\kappa_1}{\kappa_2}\Bigr) \Bigl(f \underline{g} -\dfrac{q\kappa_1}{\kappa_2}\Bigr)}
{(f g-1) (f \underline{g}-1) }
= \frac{\prod\limits_{i=5}^8 \Bigl(f-\dfrac{\kappa_1}{v_i}\Bigr)}{\prod\limits_{i=1}^4(f-v_i)}.
\end{split}
\end{equation}
%
\subsubsection{$q$-P$(E_6^{(1)}/A_2^{(1)})$}\label{subsubsec:equation_q-E6}
\begin{equation}
\frac{(f g-1) (\overline{f}g-1) }{f \overline{f} }
= \dfrac{\prod\limits_{i=1}^4\Bigl(g -\dfrac{1}{v_i}\Bigr)}
{\prod\limits_{i=5}^6\Bigl(g-\dfrac{v_i}{\kappa_2}\Bigr)},
\quad
\frac{(fg-1) (f \underline{g}-1) }{g \underline{g} }
=\dfrac{\prod\limits_{i=1}^4(f-v_i)}{\prod\limits_{i=7}^8\Bigl(f -\dfrac{\kappa_1}{v_i}\Bigr)}. 
\end{equation}
%
\subsubsection{$q$-P$(D_5^{(1)}/A_3^{(1)})$}\label{subsubsec:equation_q-D5}
\begin{equation}\label{eqn:q-D5}
f \overline{f} 
=v_3 v_4~ \frac{\prod\limits_{i=5}^6\Bigl(g -\dfrac{v_i}{\kappa_2}\Bigr)}{\prod\limits_{i=1}^2\Bigl(g-\dfrac{1}{v_i}\Bigr)},
\quad
g \underline{g}
= \frac{1}{v_1v_2}\,\frac{\prod\limits_{i=7}^8\Bigl(f-\dfrac{\kappa_1}{v_i}\Bigr)}{ \prod\limits_{i=3}^4(f-v_i)}.
\end{equation}
%
\subsubsection{$q$-P$(A_4^{(1)}/A_4^{(1)})$}\label{subsubsec:equation_q-A4}
\begin{equation}
f \overline{f}
=-v_2v_3v_4~\frac{\prod\limits_{i=5}^6\Bigl(g-\dfrac{v_i}{\kappa_2}\Bigr)}{g-\dfrac{1}{v_1}}, \quad
g \underline{g}
=-\frac{1}{ v_1v_2v_3}~\frac{\prod\limits_{i=7}^8\Bigl(f-\dfrac{\kappa_1}{v_i}\Bigr)}{f-v_4}. 
\end{equation}
%
\subsubsection{$q$-P$(\esan^{(1)}/A_5^{(1)};a)$}\label{subsubsec:equation_q-E3a}
\begin{equation}
f \overline{f} 
= -v_2v_3v_4~ \frac{\prod\limits_{i=5}^6\Bigl(g-\dfrac{v_i}{\kappa_2}\Bigr)}{g}, \quad
g \underline{g} 
= \frac{\kappa_1}{v_1v_2v_3v_8}~\frac{f-\dfrac{\kappa_1}{v_7}}{f-v_4 }. 
\end{equation}
%
\subsubsection{$q$-P$(\eni^{(1)}/A_6^{(1)};a)$}\label{subsubsec:equation_q-E2a}
\begin{equation}
f \overline{f} = -v_2v_3v_4\Bigl(g-\dfrac{v_5}{\kappa_2}\Bigr), \quad
g \underline{g}  = \frac{\kappa_1}{v_1v_2v_3v_8}~\frac{f}{f-v_4 }. 
\end{equation}
%
\subsubsection{$q$-P$(\underset{|\alpha|^2=8}{A_1^{(1)}}/A_7^{(1)})$}\label{subsubsec:equation_q-A1a}
\begin{equation}\label{eqn:q-A1a}
f \overline{f} = -v_2v_3v_4 g, \quad
g \underline{g} = \frac{\kappa_1}{v_1v_2v_3v_8}~\frac{f}{f-v_4 }. 
\end{equation}
%
\subsubsection{$q$-P$(\esan^{(1)}/A_5^{(1)};b)$}\label{subsubsec:equation_q-E3b}
\begin{equation}
f \overline{f}  = -v_2v_3v_4~\frac{g\Bigl(g-\frac{v_5}{\kappa_2}\Bigr)}{g-\dfrac{1}{v_1}}, \quad
g \underline{g}   = -\frac{1}{v_1v_2v_3}~\frac{f \Bigl(f-\frac{\kappa_1}{v_8}\Bigr)}{f-v_4}.
\end{equation}
%
\subsubsection{$q$-P$(\eni^{(1)}/A_6^{(1)};b)$}\label{subsubsec:equation_q-E2b}
\begin{equation}\label{eqn:q-E2b}
f \overline{f} = -v_2v_3v_4~\frac{g^2}{g-\dfrac{1}{v_1}}, \quad
g \underline{g} = -\frac{1}{v_1v_2v_3}~\frac{f \Bigl(f-\dfrac{\kappa_1}{v_8}\Bigr)}{f-v_4}.
\end{equation}
%
\subsubsection{$q$-P$(A_1^{(1)}/A_7^{(1)})$}\label{subsubsec:equation_q-A1b}
\begin{equation}\label{eqn:q-A1b}
f \overline{f} = v_1v_2v_3v_4g^2, \quad
g \underline{g}= - v_1v_2v_3~\frac{f \Bigl(f-\dfrac{\kappa_1}{v_8}\Bigr)}{f-v_4}.
\end{equation}

The cases $\esan^{(1)}(a)$, $\esan^{(1)}(b)$ have the same symmetry/surface type
$(A_2+A_1)^{(1)}$/$A_5^{(1)}$, while the time evolutions $T$ belong to the inequivalent directions:
$A_1^{(1)}$ for $(a)$: $q$-P$_{\rm IV}$ and $A_2^{(1)}$ for $(b)$: $q$-P$_{\rm III}$
\cite{K:qp3-2,KK:qp3-1,KNY:qp4,Tsuda:qp3_qp4}. Similarly, the cases $\eni^{(1)}(a),\eni^{(1)}(b)$
have the same symmetry/surface type $(A_1+\underset{|\alpha|^2=14}{A_1})^{(1)}/A_6^{(1)}$ with
different directions $\underset{|\alpha|^2=14}{A_1^{(1)}}$ for $(a)$: $q$-P$_{\rm II}$
\cite{HKW:qp2,Nishioka:qp2} and $A_1^{(1)}$ for $(b)$.

%
\subsubsection{d-P$(E_8^{(1)}/A_0^{(1)})$}\label{subsubsec:equation_d-E8}
\begin{equation}
\begin{split}
&
\frac{(f-f(\kappa_2-t))(\overline{f}-\overline{f}(\kappa_2-t))}{(f-f(t))(\overline{f}-\overline{f}(t))}
= 
\frac{U(\kappa_2-t)}{U(t)},
\quad g=g(t),\\
&\frac{(g-g(\kappa_1-s))(\underline{g}-\underline{g}(\kappa_1-s))}{(g-g(s))(\underline{g}-\underline{g}(s))}
= \frac{U(\kappa_1-s)}{U(s)},
\quad f=f(s),
\end{split}
\end{equation}
where $t$ and $s$ are the variables such that $g=g(t)$, $f=f(s)$,
\begin{equation}
U(z)=\prod_{i=1}^8 (z-v_i),\quad f(z)=z(z-\kappa_1),\quad g(z)=z(z-\kappa_2).
\end{equation}
%
\subsubsection{d-P$(E_7^{(1)}/A_1^{(1)})$} \label{subsubsec:equation_d-E7}
\begin{equation}
\begin{split}
\frac{(f+g-\kappa_1+\kappa_2)(\overline{f} + g - \kappa_1+\kappa_2+\delta)}
{(f+g)(\of+g)}=\frac{\prod\limits_{i=5}^8(g+\kappa_2-v_i)}{\prod\limits_{i=1}^4(g+v_i)},\\
\frac{(f+g-\kappa_1+\kappa_2)(f + \underline{g} - \kappa_1+\kappa_2-\delta)}{(f+g)(f+\underline{g})}
= \frac{\prod\limits_{i=5}^8(f-\kappa_1 + v_i)}{\prod\limits_{i=1}^4(f-v_i)}.
\end{split}
\end{equation}
%
\subsubsection{d-P$(E_6^{(1)}/A_2^{(1)})$}\label{subsubsec:equation_d-E6}
\begin{equation}
(f+g)(\overline{f}+g) = \frac{\prod\limits_{i=1}^4(g+v_i)}{\prod\limits_{i=5}^6(g+\kappa_2-v_i)},\quad
(f+g)(f+\underline{g}) = \frac{\prod\limits_{i=1}^4(f-v_i)}{\prod\limits_{i=5}^6(f-\kappa_1 + v_i)},
\end{equation}

The following cases admit the Painlev\'e equations as the continuous flows commuting with the
discrete time evolutions.  For these equations we use the parameters $a_0, a_1,\ldots$ corresponding
to the simple roots, as can be found in the literature
\cite{Noumi:book,Tsuda-Okamoto-Sakai:folding}.  We use two types of inhomogeneous coordinates $(q,p)$ and
$(f,g)=(q,qp)$ of $\mathbb{P}^1\times\mathbb{P}^1$ depending on the situations.
%
\subsubsection{d-P$(D_4^{(1)}/D_4^{(1)})$ and P$(D_4^{(1)}/D_4^{(1)})$ (P$_{\rm VI}$)} \label{subsubsec:equation_d-D4}
\noindent(i) Discrete Painlev\'e equation
\begin{equation}\label{eqn:d-D4}
\begin{split}
&\overline{a_0}=a_0-1, \quad
\overline{a_2}=a_2+1, \quad
\overline{a_3}=a_3-1, \quad a_0+a_1+2a_2+a_3+a_4=1,\\
&\overline{f}f=\frac{tg(g-a_4)}{(g+a_2)(g+a_1+a_2)}, \quad
g + \underline{g} = a_0 + a_3 + a_4 + \frac{a_3}{f-1} + \frac{ta_0}{f-t}.  
\end{split}
\end{equation}
\noindent(ii) Painlev\'e differential equation:  P$_{\rm VI}$ ($y=q$)
 \begin{equation}
\begin{split}
& \frac{d^2y}{dt^2}=\frac{1}{2}\left(\frac{1}{y}+\frac{1}{y-1}+\frac{1}{y-t}\right)\left(\frac{dy}{dt}\right)^2
-\left(\frac{1}{t}+\frac{1}{t-1}+\frac{1}{y-t}\right)\frac{dy}{dt}\\ 
& \qquad +\frac{y(y-1)(y-t)}{t^2(t^2-1)}
\left\{\alpha+\beta\frac{t}{y^2}+\gamma\frac{t-1}{(y-1)^2}+\delta\frac{t(t-1)}{(y-t)^2}\right\},\\
& \alpha = \frac{a_1^2}{2},\quad \beta=-\frac{a_4^2}{2},\quad \gamma=\frac{\alpha_3^2}{2},\quad
\delta = -\frac{a_0^2-1}{2}.
\end{split}
\end{equation}
%
\subsubsection{d-P$(A_3^{(1)}/D_5^{(1)})$ and P$(A_3^{(1)}/D_5^{(1)})$ (P$_{\rm V}$)}\label{subsubsec:equation_d-A3}
\noindent(i) Discrete Painlev\'e equation
\begin{equation}\label{eqn:d-A3}
\begin{split}
 &\overline{a_1}=a_1-1, \quad
\overline{a_2}= a_2+1, \quad
\overline{a_3}=a_3-1, \quad a_0+a_1+a_2+a_3=1,\\
&\overline{q} + q =1-\frac{a_2}{p}-\frac{a_0}{p+t}, \quad
p + \underline{p} = -t+\frac{a_1}{q}+\frac{a_3}{q-1}. 
\end{split}
\end{equation}

\noindent(ii) Painlev\'e differential equation: P$_{\rm V}$ ($y=1-\frac{1}{q}$)
\begin{equation} \label{eqn:P5_2}
\begin{split}
& \frac{d^2y}{dt^2} = \left(\frac{1}{2y}+\frac{1}{y-1}\right)\left(\frac{dy}{dt}\right)^2 - \frac{1}{t}\frac{dy}{dt}
+ \frac{(y-1)^2}{t^2}\left(\alpha_1 y + \frac{\beta}{y}\right)
+\gamma\frac{y}{t} + \delta\frac{y(y+1)}{y-1},\\
&\qquad \alpha=\frac{a_1^2}{2}, \quad \beta=-\frac{a_3^2}{2},\quad \gamma=a_0-a_2,\quad \delta=-\frac{1}{2}.\\
\end{split}
\end{equation}
%
\subsubsection{d-P$(A_2^{(1)}/E_6^{(1)})$ and P$(A_2^{(1)}/E_6^{(1)})$ (P$_{\rm IV}$)}\label{subsubsec:equation_d-A2}
\noindent(i) Discrete Painlev\'e equation
\begin{equation}\label{eqn:d-A2}
\begin{split}
&\overline{a_1}=a_1-1, \quad
\overline{a_2}= a_2+1, \quad a_0+a_1+a_2=1,\\
&\overline{q} + q = p - t-\frac{a_2}{p}, \quad
p + \underline{p}  = q+t+\frac{a_1}{q}. 
\end{split}
\end{equation}
\noindent(ii) Painlev\'e differential equation: P$_{\rm IV}$ ($y=q$)
\begin{equation}\label{eqn:P4_2}
\begin{split}
\frac{d^2y}{dt^2} &= \frac{1}{2y}\left(\frac{dy}{dt}\right)^2 + \frac{3}{2} y^2 + 4ty^2 + 2(t^2-\alpha) + \frac{\beta}{y}, \\
& \alpha = a_0-a_2,\quad \beta = -2a_1^2.
\end{split}
\end{equation}

%
\subsubsection{d-P$(A_1^{(1)}/E_7^{(1)})$ and P$(A_1^{(1)}/E_7^{(1)})$ (P$_{\rm II}$)}\label{subsubsec:equation_d-A1}
\noindent(i) Discrete Painlev\'e equation
\begin{equation}\label{eqn:d-A1}
\begin{split}
&\overline{a_1}= a_1+1, \quad a_0+a_1=1,\\
&\overline{q} + q = -\frac{a_1}{p}, \quad
p + \underline{p}= 2\, q^2+t.
\end{split}
\end{equation}
\noindent(ii) Painlev\'e differential equation: P$_{\rm II}$ ($y=q$)
\begin{equation}
\begin{split}
& \frac{d^2y}{dt^2} = 2y^3 + ty + \alpha, \\
& \alpha = a_1 - \frac{1}{2}.
\end{split}
\end{equation}

%
\subsubsection{d-P$((2A_1)^{(1)}/D_6^{(1)})$ and P$((2A_1)^{(1)}/D_6^{(1)})$ (P$_{\rm III}^{D_6^{(1)}}$)}\label{subsubsec:equation_d-2A1}
\noindent(i) Discrete Painlev\'e equation
\begin{equation}\label{eqn:d-2A1}
\begin{split}
&\overline{a_0}=a_0+1, \quad
\overline{a_1}= a_1+1, \quad a_0+a_1=1, \\
&\overline{q}+q = -\frac{a_0}{p}-\frac{a_1}{p-1}, \quad
p + \underline{p}=1 + \frac{1-a_0-a_1}{q}-\frac{t}{q^2}.
\end{split}
\end{equation}
\noindent(ii) Painlev\'e differential equation: P$_{\rm III}^{D_6^{(1)}}$ ($y=\frac{q}{s}$, $s^2=t$)
\begin{equation}
 \begin{split}
&  \frac{d^2y}{ds^2} = \frac{1}{y}\left(\frac{dy}{ds}\right)^2 -\frac{1}{s}\frac{dy}{ds} 
+ \frac{1}{s}(\alpha y^2+\beta) + \gamma y^3 + \frac{\delta}{y},\\
&\alpha = 4(1+2a_0-2a_1),\quad \beta = -4(1 + a_0 - a_1),\quad \gamma=4,\quad \delta=-4.
 \end{split}
\end{equation}
%
\subsubsection{d-P$(\underset{|\alpha|^2=4}{A_1^{(1)}}/D_7^{(1)})$ and P$(\underset{|\alpha|^2=4}{A_1^{(1)}}/D_7^{(1)})$ (P$_{\rm III}^{D_7^{(1)}}$)}\label{subsubsec:equation_d-A1'}
\noindent(i) Discrete Painlev\'e equation
\begin{equation}
\begin{split}
&\overline{a_1}= a_1+2, \\
&\overline{q}=-q-\frac{a_1}{p}-\frac{1}{p^2}, \quad
\overline{p}=-p-\frac{a_1+1}{\overline{q}}-\frac{t^2}{\overline{q}^2}.  
\end{split}
\end{equation}

\noindent(ii) Painlev\'e differential equation: P$_{\rm III}^{D_7^{(1)}}$ ($y=\frac{q}{s}$, $s^2=t$)
\begin{equation}
\begin{split}
&\frac{d^2y}{ds^2} = \frac{1}{y}\left(\frac{dy}{ds}\right)^2 -\frac{1}{s}\frac{dy}{ds} 
+ \frac{1}{s}(\alpha y^2+\beta)  + \frac{\delta}{y},\\ 
& \alpha = -8,\quad \beta = -4(a_1-1),\quad \delta=-4.
\end{split}
\end{equation}
%
\subsubsection{P$({A_0^{(1)}}/D_8^{(1)})$ (P$_{\rm III}^{D_8^{(1)}}$)}\label{subsubsec:equation_d-A0}
\noindent(i) There is no discrete discrete Painlev\'e equation.\\

\noindent(ii) Painlev\'e differential equation: P$_{\rm III}^{D_8^{(1)}}$ ($y=\frac{q}{s}$, $s^2=t$)
\begin{equation}
\frac{d^2y}{ds^2} = \frac{1}{y}\left(\frac{dy}{ds}\right)^2 -\frac{1}{s}\frac{dy}{ds} 
+ \frac{1}{s}(-8y^2+8) .
\end{equation}
%
\subsection{Point configurations and root data}\label{subsec:configuration}
In this subsection, we give the list of configurations of eight points on
$\mathbb{P}^1\times\mathbb{P}^1$ relevant to the Painlev\'e equations in Section
\ref{subsec:dPs}.

We denote by $P_{ij}, P_{ijk},\ldots$ multiple points where $P_j$ is infinitely near to $P_i$, $P_k$
infinitely near to $P_{ij}$, and so on. Moreover, $(x,y)=(A(\epsilon),B(\epsilon))_n$ in $(x,y)$
coordinates represents $n$ infinitely near points around $(A(0),B(0))$. Namely, when we say a curve
$F(x,y)=0$ passes through $(A(\epsilon),B(\epsilon))_n$, it means that the first $n$ coefficients
vanish in the $\epsilon$-expansion of $F(A(\epsilon),B(\epsilon))$. We attach schematic pictures of
configurations of eight points and associated divisors for each case. In the pictures of configuration of
divisors, we use the following notations such as $i|jk=H_i-E_i-E_k$, $ij=E_i-E_j$, $i=E_i$.

We also give a realization of the root basis $\{\alpha_i\}$, configuration of the divisors of
inaccessible points $\{\delta_i\}$ and the lattice isomorphisms (Dynkin diagram automorphisms)
$\{\pi_i\}$ corresponding to each case. We denote by $\pi_{i_1i_2,\ldots,i_8}$ the permutation $E_1\to
E_{i_1}$, $E_2\to E_{i_2}$, $\ldots$, $E_8\to E_{i_8}$. 
%
\subsubsection{$e$-P$(E_8^{(1)}/A_0^{(1)})$}\label{subsubsec:conf_e-E8} Point configuration in $(f,g)$ coordinates:
\begin{equation}
{\rm P}_i: (f(v_i),g(v_i))\ (i=1,\ldots,8),
\quad  f(z)=\frac{\Bigl[\frac{b}{z}\Bigr]\Bigl[\frac{\kappa_1}{az}\Bigl]}{\Bigl[\frac{a}{z}\Bigr]\Bigl[\frac{\kappa_1}{az}\Bigr]},\quad
g(z)=\frac{\Bigl[\frac{b}{z}\Bigr]\Bigl[\frac{\kappa_2}{az}\Bigl]}{\Bigl[\frac{a}{z}\Bigr]\Bigl[\frac{\kappa_2}{az}\Bigr]}.
\end{equation}
\begin{equation}
\lower2cm\hbox{\includegraphics[width=4cm]{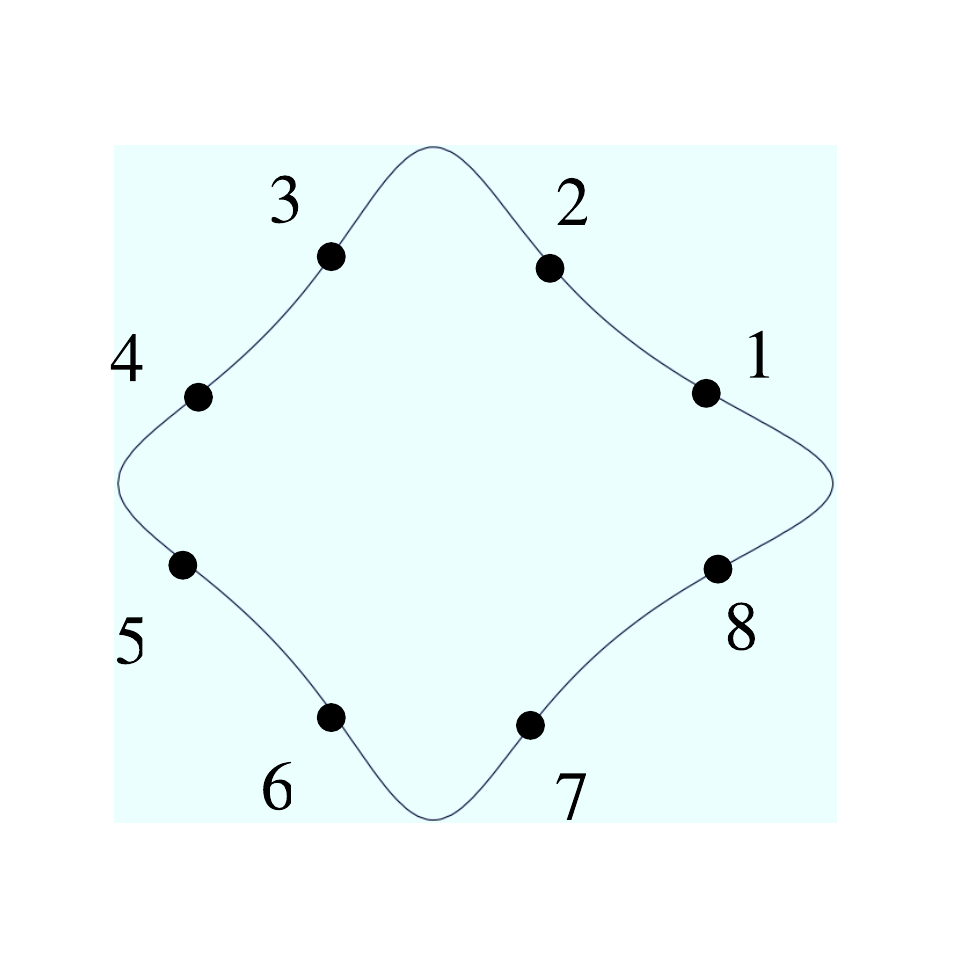}}
\hskip40pt
\lower2cm\hbox{\includegraphics[width=4cm]{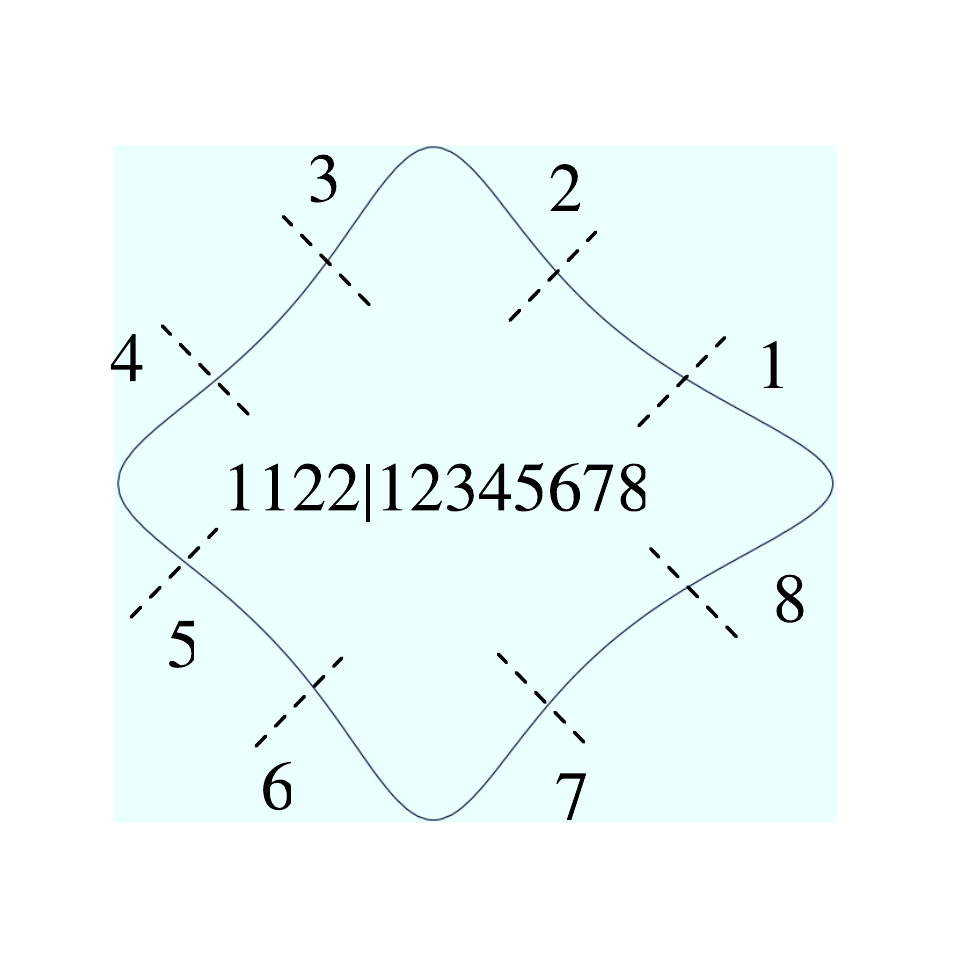}}
\end{equation}
Root data:
\begin{equation}
  \begin{array}{c}
\hskip-85pt \alpha_0\\
\hskip-85pt |\\
\alpha_1\ \text{\textemdash}\ \alpha_2\ \text{\textemdash}\  \alpha_3\ \text{\textemdash}\  \alpha_4\ \text{\textemdash}\ \alpha_5
\text{\textemdash}\ \alpha_6\text{\textemdash}\ \alpha_7\text{\textemdash}\ \alpha_8
 \end{array}
\qquad \lower5pt\hbox{\mbox{$\delta_0$}}
\end{equation}
\begin{equation}
\begin{split}
&
\alpha_0=E_1-E_2,\ 
\alpha_1=H_1-H_2,\ 
\alpha_2=H_2-E_1-E_2,\ 
\alpha_3=E_2-E_3,\ 
\\
& 
\alpha_4=E_3-E_4,\ 
\alpha_5=E_4-E_5,\ 
\alpha_6=E_5-E_6,\ 
\alpha_7=E_6-E_7,\ 
\alpha_8=E_7-E_8,
\\
&
\delta_0=2H_1+2H_2-E_1-E_2-E_3-E_4-E_5-E_6-E_7-E_8.
\\
\end{split}
\end{equation}

%
\subsubsection{$q$-P$(E_8^{(1)}/A_0^{(1)})$}\label{subsubsec:conf_q-E8} Point configuration in $(f,g)$ coordinates:
\begin{equation}
{\rm P}_i: (f(v_i),g(v_i))\ (i=1,\ldots,8),\quad f(z) = z+\frac{\kappa_1}{z},\quad g(z) = z+\frac{\kappa_2}{z}.
\end{equation}
\begin{equation}
\lower2cm\hbox{\includegraphics[width=4cm]{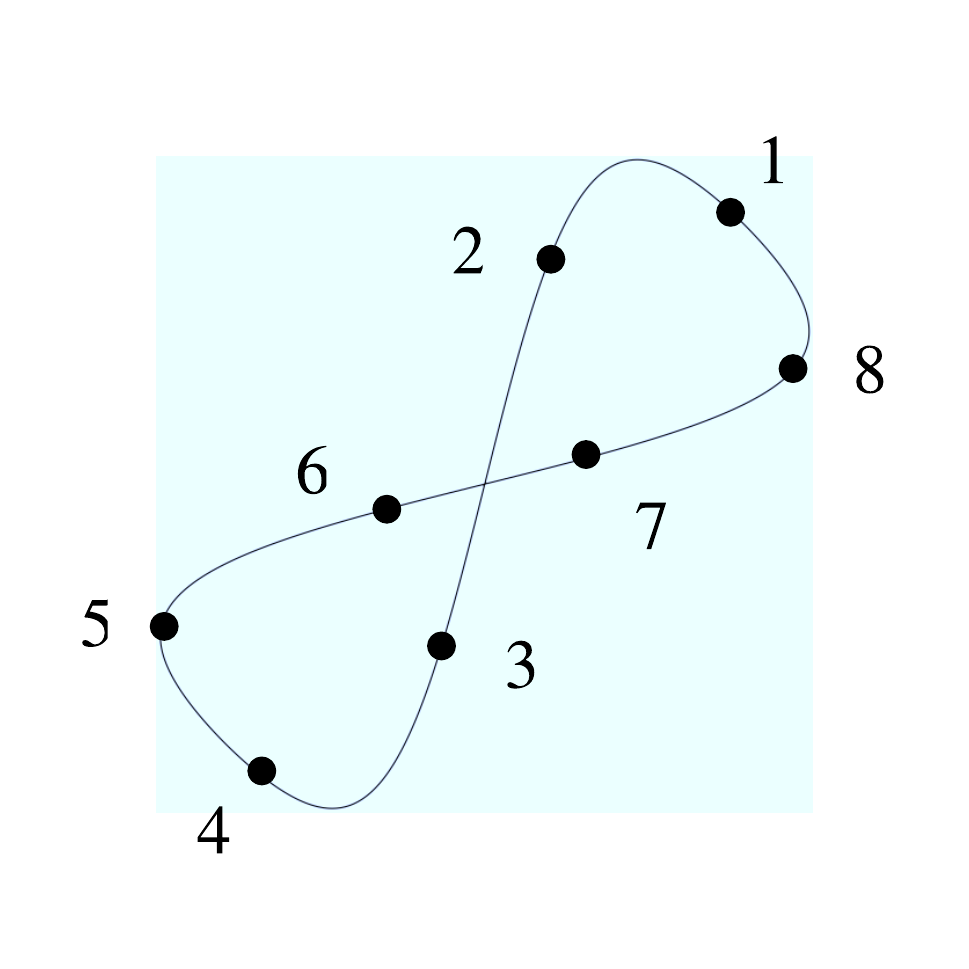}}
\hskip40pt
\lower2cm\hbox{\includegraphics[width=4cm]{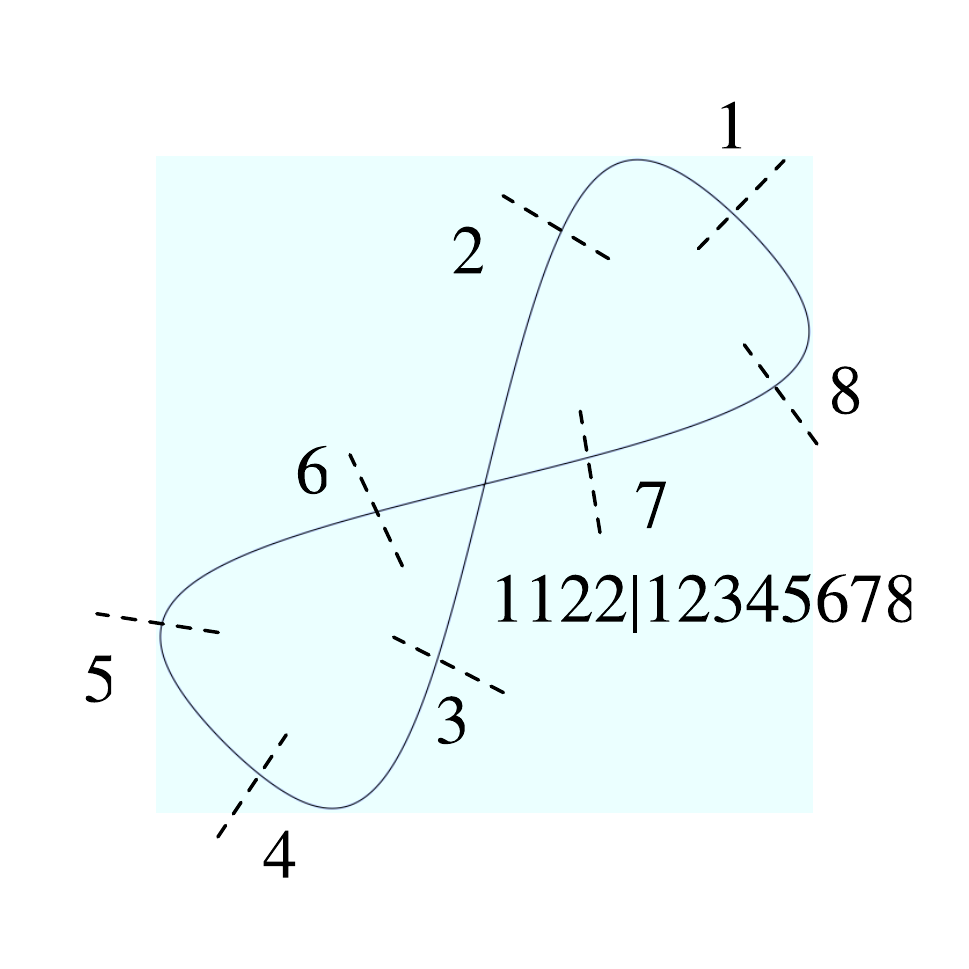}}
\end{equation}
Root data: same as Section \ref{subsubsec:conf_e-E8}.
%
\subsubsection{$q$-P$(E_7^{(1)}/A_1^{(1)})$}\label{subsubsec:conf_q-E7} Point configuration in $(f,g)$ coordinates:
\begin{equation}
{\rm P}_i:\left(v_i,\frac{1}{v_i}\right)\ (i=1,\ldots,4), 
\quad\left(\frac{\kappa_1}{v_i} ,\frac{v_i}{\kappa _2}\right)\ (i=5,\ldots,8).
\end{equation}
\setlength{\unitlength}{0.6mm}
\begin{equation}
{\lower20pt\hbox{\footnotesize
\begin{picture}(50,50)(0,0)
\qbezier(5,45)(5,6)(45,5)
\put(35,9){$5$}\put(35,6){\circle*{2}}
\put(25,12){$6$}\put(25,9){\circle*{2}}
\put(12,22){$7$}\put(12,19){\circle*{2}}
\put(10,32){$8$}\put(7,30){\circle*{2}}
\qbezier(15,48)(17,17)(48,15)
\put(40,19){$1$}\put(40,16){\circle*{2}}
\put(30,23){$2$}\put(30,20){\circle*{2}}
\put(23,30){$3$}\put(20,29.5){\circle*{2}}
\put(18,41){$4$}\put(15.5,42){\circle*{2}}
\end{picture}\hskip40pt
\begin{picture}(50,50)(0,0)
\put(37,9){$5$}\dashline[40]{2}(35,3)(35,11)
\put(28,12){$6$}\dashline[40]{2}(22,7)(29,14)
\put(12,22){$7$}\dashline[40]{2}(9,14)(16,21)
\put(10,32){$8$}\dashline[40]{2}(3,30)(12,30)

\put(42,19){$1$}\dashline[40]{2}(40,13)(40,22)
\put(34,23){$2$}\dashline[40]{2}(25,17)(32,24)
\put(23,30){$3$}\dashline[40]{2}(17,29.5)(24,36.5)
\put(18,44){$4$}\dashline[40]{2}(12,42)(21,42)
\put(50,13){$12|1234$}
\put(48,3){$12|5678$}

\thicklines
\qbezier(5,45)(5,6)(45,5)
\qbezier(15,48)(17,17)(48,15)
\end{picture}
} }
\end{equation}
Root data:
\begin{equation}
  \begin{array}{c}
 \alpha_0\\
|\\
\alpha_1\ \text{\textemdash}\ \alpha_2\ \text{\textemdash}\  \alpha_3\ \text{\textemdash}\  \alpha_4\ \text{\textemdash}\ \alpha_5
\text{\textemdash}\ \alpha_6\text{\textemdash}\ \alpha_7
 \end{array}
\qquad
\setlength{\unitlength}{0.8mm}
\begin{picture}(27,18)(0,7)
\put(5,5){$\delta_0$}
\put(25,5){$\delta_1$}
\put(10,6.3){\line(1,0){12.5}}
\put(10,5.8){\line(1,0){12.5}}
\put(9.4,5.1){$<$}
\put(21.3,5.1){$>$}
\end{picture} 
\end{equation}
\begin{equation}
\begin{split}
&
\alpha_0=H_1-H_2,\ 
\alpha_1=E_3-E_4,\ 
\alpha_2=E_2-E_3,\ 
\alpha_3=E_1-E_2,
\\
& 
\alpha_4=H_2-E_1-E_5,\ 
\alpha_5=E_5-E_6,\ 
\alpha_6=E_6-E_7,\ 
\alpha_7=E_7-E_8,
\\
&
\delta_0=H_1+H_2-E_1-E_2-E_3-E_4,\ 
\delta_1=H_1+H_2-E_5-E_6-E_7-E_8,
\\
&
\pi=\pi_{56781234}.
\end{split}
\end{equation}
\begin{displaymath}
\includegraphics[scale=0.7]{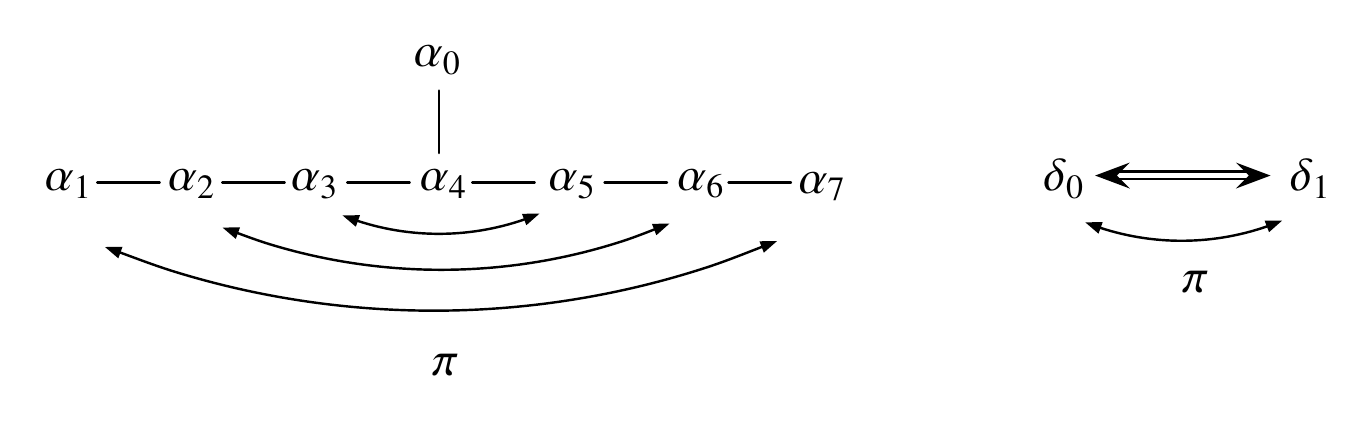} 
\end{displaymath}

%
\subsubsection{$q$-P$(E_6^{(1)}/A_2^{(1)})$}\label{subsubsec:conf_q-E6} Point configuration in $(f,g)$ coordinates:
\begin{equation}
{\rm P}_i: \left(v_i,\frac{1}{v_i}\right)\ (i=1,\ldots,4), \ 
\left(0 ,\frac{v_i}{\kappa _2}\right)\ (i=5,6), \ 
\left(\frac{\kappa_1}{v_i},0 \right)\ (i=7,8).
\end{equation}
\setlength{\unitlength}{0.6mm}
\begin{equation}
{\lower20pt\hbox{\footnotesize
\begin{picture}(50,50)(0,0)
\put(0,10){\line(1,0){50}}
\put(10,0){\line(0,1){50}}
\put(18,13){$7$}\put(20,10){\circle*{2}}
\put(28,13){$8$}\put(30,10){\circle*{2}}
\put(12,22){$6$}\put(10,20){\circle*{2}}
\put(12,32){$5$}\put(10,30){\circle*{2}}
\qbezier(15,48)(17,17)(48,15)
\put(40,19){$1$}\put(40,16){\circle*{2}}
\put(30,23){$2$}\put(30,20){\circle*{2}}
\put(23,30){$3$}\put(20,29.5){\circle*{2}}
\put(18,41){$4$}\put(15.5,42){\circle*{2}}
\end{picture}\hskip40pt
\begin{picture}(50,50)(0,0)
\put(16,13){$7$}\dashline[40]{2}(20,5)(20,15)
\put(26,13){$8$}\dashline[40]{2}(30,5)(30,15)
\put(12,24){$6$}\dashline[40]{2}(5,20)(15,20)
\put(12,34){$5$}\dashline[40]{2}(5,30)(15,30)

\put(42,19){$1$}\dashline[40]{2}(40,13)(40,22)
\put(34,23){$2$}\dashline[40]{2}(27,19)(34,26)
\put(23,30){$3$}\dashline[40]{2}(17,29.5)(24,36.5)
\put(18,44){$4$}\dashline[40]{2}(12,42)(21,42)
\put(12,0){$1|56$}
\put(0,4){$2|78$}
\put(50,15){$12|1234$}

\thicklines
\put(0,10){\line(1,0){50}}
\put(10,0){\line(0,1){50}}
\qbezier(15,48)(17,17)(48,15)
\end{picture}}}
\end{equation}
Root data:
\begin{equation}
  \begin{array}{c}
 \alpha_0\\
|\\
\alpha_6\\
|\\
\alpha_1\ \text{\textemdash}\ \alpha_2\ \text{\textemdash}\  \alpha_3\ \text{\textemdash}\  \alpha_4\ \text{\textemdash}\ \alpha_5
 \end{array}
\qquad
\setlength{\unitlength}{0.8mm}
\begin{picture}(27,18)(0,7)
\put(5,0){$\delta_1$}
\put(25,0){$\delta_2$}
\put(15,15){$\delta_0$}
\put(10,1){\line(1,0){12.5}}
\put(24,4){\line(-2,3){6}}
\put(8,4){\line(2,3){6}}
\end{picture} 
\end{equation}
\begin{equation}
\begin{minipage}{0.88\textwidth}
\begin{displaymath}
\begin{split}
&
\alpha_0=E_7-E_8,\ 
\alpha_1=E_6-E_5,\ 
\alpha_2=H_2-E_1-E_6,\ 
\\
&
\alpha_3=E_1-E_2,\ 
\alpha_4=E_2-E_3,\ 
\alpha_5=E_3-E_4,\ 
\alpha_6=H_1-E_1-E_7,
\\
&
\delta_0=H_1+H_2-E_1-E_2-E_3-E_4,\ 
\delta_1=H_1-E_5-E_6,\ 
\delta_2=H_2-E_7-E_8,\ 
\\
&
\pi_1=\pi_{12654378}r_{H_2-E_1-E_2},\
\pi_2=\pi_{12348765}r_{H_1-H_2}.
\end{split} 
\end{displaymath}
\end{minipage}
\end{equation}
\begin{displaymath}
\includegraphics[scale=0.7]{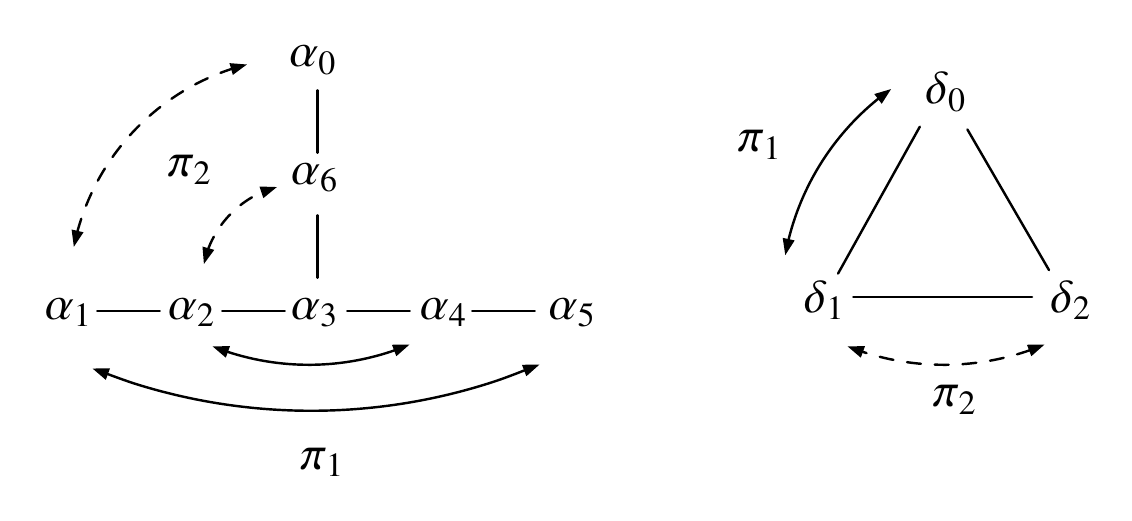} 
\end{displaymath}

In the following multiplicative cases, the eight points lie on the lines $f=0,\infty$,
$g=0,\infty$. We also show the schematic diagram of the configuration of eight points.
%
\subsubsection{$q$-P$(D_5^{(1)}/A_3^{(1)})$}\label{subsubsec:conf_q-D5} Point configuration in $(f,g)$ coordinates:
\begin{equation}
{\rm P}_i: \left(\infty ,\frac{1}{v_i}\right)\ (i=1,2), \ 
\left(v_i,\infty \right)\ (i=3,4), \ 
\left(0,\frac{v_i}{\kappa _2}\right)\ (i=5,6) , \ 
\left(\frac{\kappa _1}{v_i},0\right)\ (i=7,8).\\
\end{equation}
\begin{equation}
{\lower20pt\hbox{\footnotesize
\begin{picture}(50,50)(0,0)
\put(5,10){\line(1,0){42}}
\put(5,40){\line(1,0){42}}
\put(10,5){\line(0,1){42}}
\put(40,5){\line(0,1){42}}
\put(5,0){$f=0$}\put(35,0){$f=\infty$}\put(-7,12){$g=0$}\put(-7,35){$g=\infty$}
\put(18,13){$7$} \put(20,10){\circle*{2}}
\put(28,13){$8$}\put(30,10){\circle*{2}}
\put(18,43){$4$}\put(20,40){\circle*{2}}
\put(28,43){$3$}\put(30,40){\circle*{2}}
\put(42,32){$2$}\put(40,30){\circle*{2}}
\put(28,13){$8$}\put(30,10){\circle*{2}}
\put(12,22){$6$}\put(10,20){\circle*{2}}
\put(12,32){$5$}\put(10,30){\circle*{2}}
\put(42,22){$1$}\put(40,20){\circle*{2}}
\end{picture}
\hskip40pt 
\begin{picture}(50,50)(0,0)
\thicklines
\put(5,10){\line(1,0){45}}
\put(5,40){\line(1,0){45}}
\put(10,5){\line(0,1){45}}
\put(40,5){\line(0,1){45}}
\put(-5,38){$2|34$}
\put(-5,8){$2|78$}
\put(5,0){$1|56$}
\put(35,0){$1|12$}

\thinlines
\put(36,22){$1$}\dashline[40]{2}(35,20)(50,20) 
\put(36,32){$2$}\dashline[40]{2}(35,30)(50,30) 
\put(33,42){$3$}\dashline[40]{2.0}(30,35)(30,50) 
\put(23,42){$4$}\dashline[40]{2.0}(20,35)(20,50) 
\put(12,32){$5$}\dashline[40]{2}(0,30)(15,30) 
\put(12,22){$6$}\dashline[40]{2}(0,20)(15,20) 
\put(16,13){$7$}\dashline[40]{2}(20,5)(20,15) 
\put(26,13){$8$}\dashline[40]{2}(30,5)(30,15) 
\end{picture}
}}
\end{equation}
Root data:
\begin{equation}
\begin{split}
\setlength{\unitlength}{1mm}
\begin{picture}(25,20)(0,0)
\put(0,0){$\alpha_1$}
\put(0,14.7){$\alpha_0$}
\put(6.5,7){$\alpha_2$}
\drawline(2.4,2.1)(5.8,6.0)
\drawline(2.4,12.9)(5.8,9.0)
\drawline(10.4,7.5)(14.1,7.5)
\put(15,7){$\alpha_3$}
\drawline(18,9.1)(21.4,13.0)
\put(21.5,14.7){$\alpha_4$}
\drawline(18,6.0)(21.4,2.1)
\put(21.5,0){$\alpha_5$}
\end{picture}
\qquad\qquad
\begin{picture}(25,20)(0,0)
\put(0,0){$\delta_1$}
\put(0,14){$\delta_0$}
\put(14,0){$\delta_2$}
\put(14,14){$\delta_3$}
\drawline(1,3)(1,12)
\drawline(15,3)(15,12)
\drawline(3,1)(12,1)
\drawline(3,15)(12,15)
\end{picture}
\end{split}
\end{equation}
\begin{equation}
\begin{split}
&
\alpha_0=E_7-E_8,\ 
\alpha_1=E_3-E_4,\ 
\alpha_2=H_1-E_3-E_7,\ 
\\
&
\alpha_3=H_2-E_1-E_5,\ 
\alpha_4=E_1-E_2,\ 
\alpha_5=E_5-E_6,\ 
\\
&
\delta_0=H_1-E_1-E_2,\ 
\delta_1=H_2-E_3-E_4,\ 
\delta_2=H_1-E_5-E_6,\ 
\delta_3=H_2-E_7-E_8,\ 
\\
&
\pi_1=\pi_{12785634},\
\pi_2=\pi_{78563412}r_{H_1-H_2}.
\end{split}
\end{equation}
\begin{displaymath}
\includegraphics[scale=0.75]{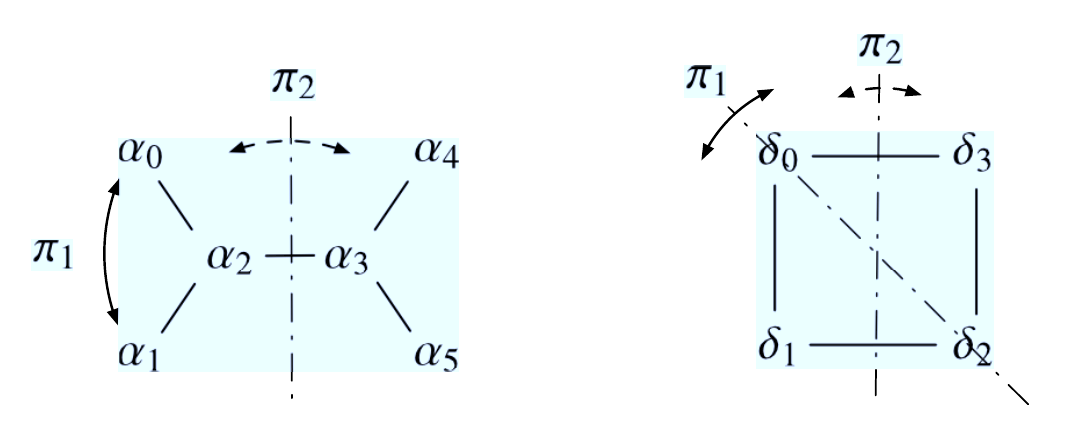} 
\end{displaymath}
%
\subsubsection{$q$-P$(A_4^{(1)}/A_4^{(1)})$}\label{subsubsec:conf_q-A4}
Point configuration in $(f,g)$ coordinates:
\begin{equation}
{\rm P}_1:\left(\infty ,\frac{1}{v_1}\right),\ 
{\rm P}_{23}:\left(-\frac{v_2v_3}{\epsilon}, \frac{1}{\epsilon}\right)_2,\ 
{\rm P}_4: \left(v_4,\infty \right), {\rm P}_i: \left(0,\frac{v_i}{\kappa_2}\right)\ (i=5,6),\ 
\left(\frac{\kappa_1}{v_i},0\right)\ (i=7,8).
\end{equation}
\begin{equation}
{\lower40pt\hbox{\footnotesize
\begin{picture}(50,50)(0,0)
\put(5,10){\line(1,0){42}}
\put(5,40){\line(1,0){42}}
\put(10,5){\line(0,1){42}}
\put(40,5){\line(0,1){42}}
\put(5,0){$f=0$}\put(35,0){$f=\infty$}\put(-7,14){$g=0$}\put(-7,35){$g=\infty$}
\put(20,10){\circle*{2}}\put(30,10){\circle*{2}}
\put(18,13){$7$} \put(28,13){$8$}
\put(20,40){\circle*{2}}\put(40,40){\circle{3}}
\put(18,43){$4$}\put(41,43){$23$}
\put(10,20){\circle*{2}}\put(12,22){$6$}
\put(10,30){\circle*{2}}\put(12,32){$5$}
\put(40,20){\circle*{2}}\put(42,22){$1$}
\put(40,40){\circle*{1.5} ${}$}
\end{picture}
\hskip40pt 
\begin{picture}(50,50)(0,0)
\thicklines
\put(5,10){\line(1,0){45}}
\put(5,40){\line(1,0){30}}
\put(10,5){\line(0,1){45}}
\put(40,5){\line(0,1){30}}
\put(28,42){\line(1,-1){15}} 
\put(-8,38){$2|24$}
\put(-8,8){$2|78$}
\put(5,0){$1|56$}
\put(35,0){$1|12$}
\put(44,25){$23$}

\thinlines
\put(36,22){$1$}\dashline[40]{2}(35,20)(50,20) 
\put(23,42){$4$}\dashline[40]{2.0}(20,35)(20,50) 
\put(12,32){$5$}\dashline[40]{2}(0,30)(15,30) 
\put(12,22){$6$}\dashline[40]{2}(0,20)(15,20) 
\put(16,13){$7$}\dashline[40]{2}(20,5)(20,15) 
\put(26,13){$8$}\dashline[40]{2}(30,5)(30,15) 

\put(29,33){$3$}\dashline[30]{2.0}(30,30)(42,42)
\end{picture}
}}
\end{equation}
Root data:
\begin{equation}
\begin{split}
 \setlength{\unitlength}{1.3mm}
\begin{picture}(25,20)(-7,-7)
\put(0,7){$\alpha_0$}
\drawline(-1,6)(-4,3.8)
\put(-6.657,2.163){$\alpha_1$}
\drawline(-5,1)(-4,-3.3)
\put(-4.114,-5.663){$\alpha_2$}
\drawline(-1,-5)(3.3,-5)
\put(4.114,-5.663){$\alpha_3$}
\drawline(6,-3.5)(7.5,1)
\put(6.657,2.163){$\alpha_4$}
\drawline(6,3.8)(3,6)
\end{picture}
\qquad
 \setlength{\unitlength}{1.3mm}
\begin{picture}(25,20)(-7,-7)
\put(0,7){$\delta_0$}
\drawline(-1,6)(-4,3.8)
\put(-6.657,2.163){$\delta_1$}
\drawline(-5,1)(-4,-3.3)
\put(-4.114,-5.663){$\delta_2$}
\drawline(-1,-5)(3.3,-5)
\put(4.114,-5.663){$\delta_3$}
\drawline(6,-3.5)(7.5,1)
\put(6.657,2.163){$\delta_4$}
\drawline(6,3.8)(3,6)
\end{picture}
\end{split}
\end{equation}
\begin{equation}
\begin{split}
&
\alpha_0=E_7-E_8,\ 
\alpha_1=H_1-E_4-E_7,\ 
\alpha_2=H_2-E_1-E_5,\ 
\\
&
\alpha_3=E_5-E_6,\ 
\alpha_4=H_1+H_2-E_2-E_3-E_5-E_7,\ 
\\
&
\delta_0=H_1-E_1-E_2,\ 
\delta_1=E_2-E_3,\ 
\delta_2=H_2-E_2-E_4,\ 
\\
&
\delta_3=H_1-E_5-E_6,\ 
\delta_4=H_2-E_7-E_8,\ 
\\
&
\pi_1=\pi_{42687153}r_{H_1-H_2}r_{H_2-E_2-E_7}r_{H_1-E_5-E_7},\
\pi_2=\pi_{42317856}r_{H_1-H_2}.
\end{split}
\end{equation}
\begin{displaymath}
\includegraphics[scale=0.7]{Dynkin_Auto_A4-eps-converted-to.pdf} \quad
\includegraphics[scale=0.7]{Dynkin_Auto_A4_2-eps-converted-to.pdf} 
\end{displaymath}
%
\subsubsection{$q$-P$(E_3^{(1)}/A_5^{(1)};a)$}\label{subsubsec:conf_q-E3a}
Point configuration in $(f,g)$ coordinates:
\begin{equation}
\begin{split}
&{\rm P}_{18}:\left(-\frac{\kappa_1}{\epsilon v_1v_8},\epsilon\right)_2,\quad
{\rm P}_{23}:\left(-\frac{v_2v_3}{\epsilon}, \frac{1}{\epsilon}\right)_2,\quad
{\rm P}_4: \left(v_4,\infty \right),\\
&{\rm P}_{i}:\left(0,\frac{v_i}{\kappa_2}\right)\ (i=5,6),\quad
{\rm P}_7: \left(\frac{\kappa_1}{v_7},0\right).
\end{split}
\end{equation}
\begin{equation}
{\lower40pt\hbox{\footnotesize
\begin{picture}(50,50)(0,0)
\put(5,10){\line(1,0){42}}
\put(5,40){\line(1,0){42}}
\put(10,5){\line(0,1){42}}
\put(40,5){\line(0,1){42}}
\put(5,0){$f=0$}\put(35,0){$f=\infty$}\put(-7,14){$g=0$}\put(-7,35){$g=\infty$}
\put(20,10){\circle*{2}}
\put(40,40){\circle*{1.5} ${}$}\put(40,40){\circle{3}}
\put(18,13){$7$}\put(41,13){$18$}\put(40,10){\circle*{1.5}}
\put(20,40){\circle*{2}}
\put(18,43){$4$}\put(41,43){$23$}
\put(10,20){\circle*{2} $6$}\put(10,30){\circle*{2} $5$}
\put(40,10){\circle{3} ${}$}
\end{picture}
\hskip40pt 
\begin{picture}(50,50)(0,0)
\thicklines
\put(5,10){\line(1,0){30}} 
\put(5,40){\line(1,0){30}}
\put(10,5){\line(0,1){45}}
\put(40,15){\line(0,1){20}}
\put(28,42){\line(1,-1){15}}
\put(28,8){\line(1,1){15}}  
\put(-7,38){$2|24$}
\put(5,0){$1|56$}
\put(42,13){$1|12$}
\put(44,25){$23$}
\put(23,3){$18$}
\put(-7,8){$2|17$}
\thinlines
\put(23,42){$4$}\dashline[40]{2.0}(20,35)(20,50) 
\put(12,32){$5$}\dashline[40]{2}(0,30)(15,30) 
\put(12,22){$6$}\dashline[40]{2}(0,20)(15,20) 
\put(16,13){$7$}\dashline[40]{2}(20,5)(20,15) 

\put(29,33){$3$}\dashline[30]{2.0}(30,30)(42,42)
\put(29,13){$8$}\dashline[30]{2.0}(30,20)(42,8)
\end{picture}
}}
\end{equation}
Root data:
\begin{equation}
\begin{split}
 \setlength{\unitlength}{0.8mm}
\begin{picture}(27,18)(0,-4)
\put(-2,0){$\alpha_1$}
\put(21,0){$\alpha_2$}
\put(10,15){$\alpha_0$}
\put(5,1){\line(1,0){12.5}}
\put(19,4){\line(-2,3){6}}
\put(3,4){\line(2,3){6}}
\end{picture} \ 
 \setlength{\unitlength}{0.8mm}
\begin{picture}(27,18)(0,-7)
\put(-2,5){$\alpha_3$}
\put(21,5){$\alpha_4$}
\put(5,6.3){\line(1,0){12.5}}
\put(5,5.8){\line(1,0){12.5}}
\put(4.4,5.1){$<$}
\put(16.3,5.1){$>$}
\end{picture} 
\qquad
 \setlength{\unitlength}{1.4mm}
\begin{picture}(25,20)(-7,-7)
\put(0,7.0){$\delta_0$}
\drawline(-1,6.5)(-4,5)
\put(- 6.06,3.5){$\delta_1$}
\drawline(-5.5,2.5)(-5.5,-1.5)
\put(- 6.06,-3.5){$\delta_2$}
\drawline(-4,-4.5)(-1,-5.5)
\put(0, -7.0){$\delta_3$}
\drawline(2,-5.5)(5,-4)
\put(6.06, -3.5){$\delta_4$}
\drawline(6.5,2.5)(6.5,-1.5)
\put(6.06,3.5){$\delta_5$}
\drawline(5.5,5)(2,6.5)
\end{picture}
\end{split}
\end{equation}
\begin{equation}\label{eqn:root_E3A5a}
\begin{split}
&
\alpha_0=H_1+H_2-E_2-E_3-E_6-E_7,\ 
\alpha_1=H_1+H_2-E_1-E_4-E_6-E_8,\\
&
\alpha_2=E_6-E_5,\ 
\alpha_3=\delta-\alpha_4,\ 
\alpha_4=H_1-E_4-E_7,\ 
\\
&
\delta_0=H_1-E_1-E_2,\ 
\delta_1=E_2-E_3,\ 
\delta_2=H_2-E_2-E_4,\ 
\\
&
\delta_3=H_1-E_5-E_6,\ 
\delta_4=H_2-E_1-E_7,\ 
\delta_5=E_1-E_8,\ 
\\
&
\pi_1=\pi_{12348675}r_{H_2-E_4-E_6}r_{H_1-E_1-E_6},\ 
\pi_2=\pi_{36457182}r_{H_1-H_2}r_{H_1-E_2-E_8}r_{H_2-E_1-E_6}.
\end{split}
\end{equation}
\begin{displaymath}
\includegraphics[scale=0.6]{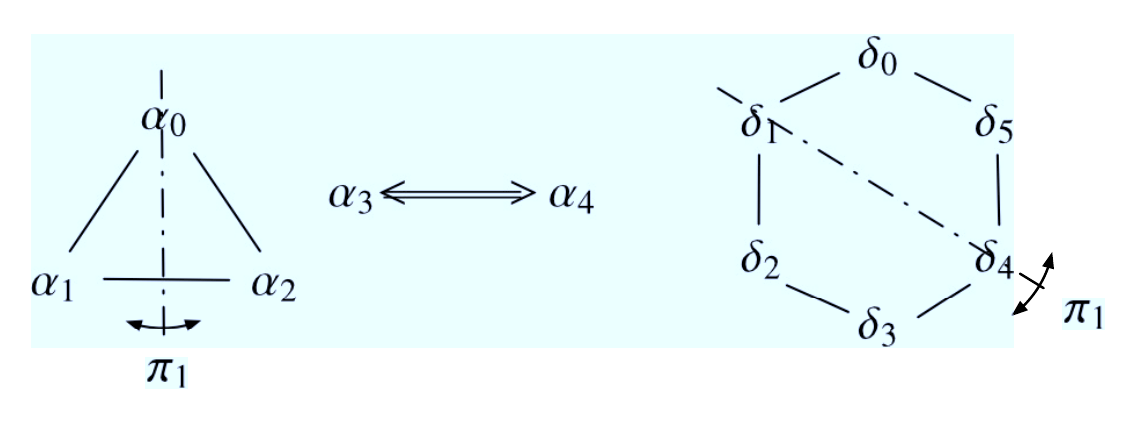}\quad
\includegraphics[scale=0.6]{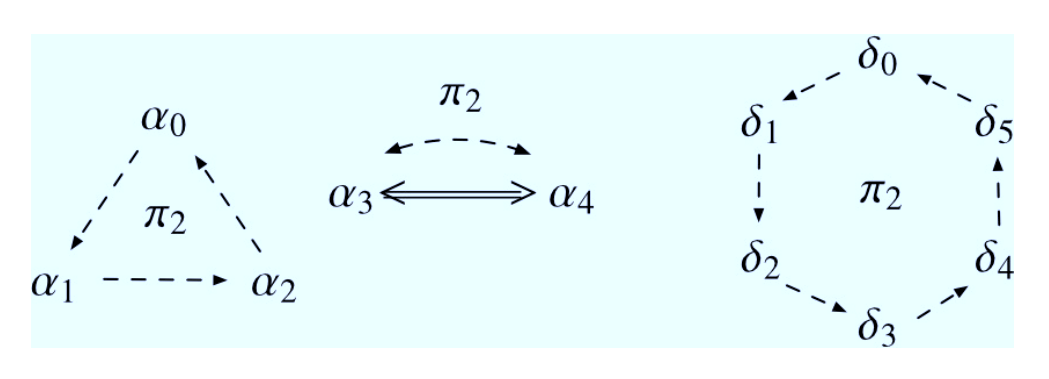}
\end{displaymath}
%
\subsubsection{$q$-P$(E_2^{(1)}/A_6^{(1)};a)$}\label{subsubsec:conf_q-E2a}
Point configuration in $(f,g)$ coordinates:
\begin{equation}
\begin{split}
&{\rm P}_{18}:\left(-\frac{\kappa_1}{\epsilon v_1v_8},\epsilon\right)_2,\quad
{\rm P}_{23}:\left(-\frac{v_2v_3}{\epsilon}, \frac{1}{\epsilon}\right)_2,\quad
{\rm P}_4: \left(v_4,\infty \right),\\
&{\rm P}_{5}:\left(0,\frac{v_5}{\kappa_2}\right),\quad
{\rm P}_{67}: \left(-\frac{\kappa_1\kappa_2}{v_6v_7}\epsilon ,\epsilon\right).
\end{split}
\end{equation}
\begin{equation}
{\lower40pt\hbox{\footnotesize
\begin{picture}(50,50)(0,0)
\put(5,10){\line(1,0){42}}
\put(5,40){\line(1,0){42}}
\put(10,5){\line(0,1){42}}
\put(40,5){\line(0,1){42}}
\put(5,0){$f=0$}\put(35,0){$f=\infty$}\put(-7,14){$g=0$}\put(-7,35){$g=\infty$}
\put(10,10){\circle{3}}\put(40,10){\circle*{1.5}}
\put(12,13){$67$}\put(41,13){$18$}
\put(20,40){\circle*{2}}\put(40,40){\circle{3}}
\put(18,43){$4$}\put(41,43){$23$}
\put(10,10){\circle*{1.5} ${}$}\put(10,30){\circle*{2} $5$}
\put(40,10){\circle{3} ${}$}\put(40,40){\circle*{1.5} ${}$}
\end{picture}
\hskip40pt 
\begin{picture}(50,50)(0,0)
\thicklines
\put(15,10){\line(1,0){20}} 
\put(5,40){\line(1,0){30}}
\put(10,15){\line(0,1){35}}
\put(40,15){\line(0,1){20}}
\put(28,42){\line(1,-1){15}}
\put(28,8){\line(1,1){15}}  
\put(7,23){\line(1,-1){15}}
\put(-7,38){$2|24$}
\put(-3,12){$1|56$}
\put(-1,22){$67$}
\put(42,13){$1|12$}
\put(44,25){$23$}
\put(25,3){$18$}
\put(10,3){$2|16$}

\thinlines
\put(23,42){$4$}\dashline[40]{2.0}(20,35)(20,50) 
\put(12,32){$5$}\dashline[40]{2}(0,30)(15,30) 

\put(29,33){$3$}\dashline[30]{2.0}(30,30)(42,42)
\put(29,13){$8$}\dashline[30]{2.0}(30,20)(42,8)

\put(15,18){$7$}\dashline[30]{2.0}(20,20)(8,8)
\end{picture}
}}
\end{equation}
Root data:
\begin{equation}
\begin{split}
  \setlength{\unitlength}{0.8mm}
\begin{picture}(40,18)(0,-6)
\put(-2,5){$\alpha_0$}
\put(21,5){$\alpha_1$}
\put(35,5){$\alpha_2$}
\put(5,6.3){\line(1,0){12.5}}
\put(5,5.8){\line(1,0){12.5}}
\put(4.05,5.05){$<$}
\put(15.95,5.05){$>$}
\end{picture} 
\qquad
 \setlength{\unitlength}{1.4mm}
\begin{picture}(15,16)(-7,-6)
\put(0,7.0){$\delta_0$}
\put(- 5.47282,4.36443){$\delta_1$}
\put( - 6.8245, - 1.55765){$\delta_2$}
\put( - 3.03719, - 6.30678){$\delta_3$}
\put(3.03719, - 6.30678){$\delta_4$}
\put(6.8245, - 1.55765){$\delta_5$}
\put(5.47282,4.36443){$\delta_6$}
\drawline(-0.8,7.0)(-3.5,5.4)
\drawline(-5.3,3.5)(-6,0.5)
\drawline(-5.8,-2.5)(-3.5,-5.2)
\drawline(-1,-5.8)(2.5,-5.8)
\drawline(4.8,-5.2)(6.8,-2)
\drawline(7.1,0.5)(6.4,3.5)
\drawline(5.2,5.4)(2.5,7.0)
\end{picture} 
\end{split}
\end{equation}
\begin{equation}\label{eqn:root_E2A6a}
\begin{split}
&
\alpha_0=H_1+H_2-E_2-E_3-E_6-E_7,\ 
\alpha_1=H_1+H_2-E_1-E_4-E_5-E_8,\ 
\\
&
\alpha_2=H_1+3H_2-E_1-2E_2- 2E_3+E_4-3E_5-E_8,
\\
&
\delta_0=H_1-E_1-E_2,\ 
\delta_1=E_2-E_3,\ 
\delta_2=H_2-E_2-E_4,\ 
\\
&
\delta_3=H_1-E_5-E_6,\ 
\delta_4=E_6-E_7,\ 
\delta_5=H_2-E_1-E_6,\ 
\delta_6=E_1-E_8,\ 
\\
 &
\pi_1 =  \pi_{65432187}r_{H_2-E_2-E_5},\\
&\pi_2=\pi_{53281674}r_{H_1-H_2}r_{H_1-E_1-E_3}r_{H_2-E_2-E_5}.
\end{split}
\end{equation}
\begin{displaymath}
\includegraphics[scale=0.4]{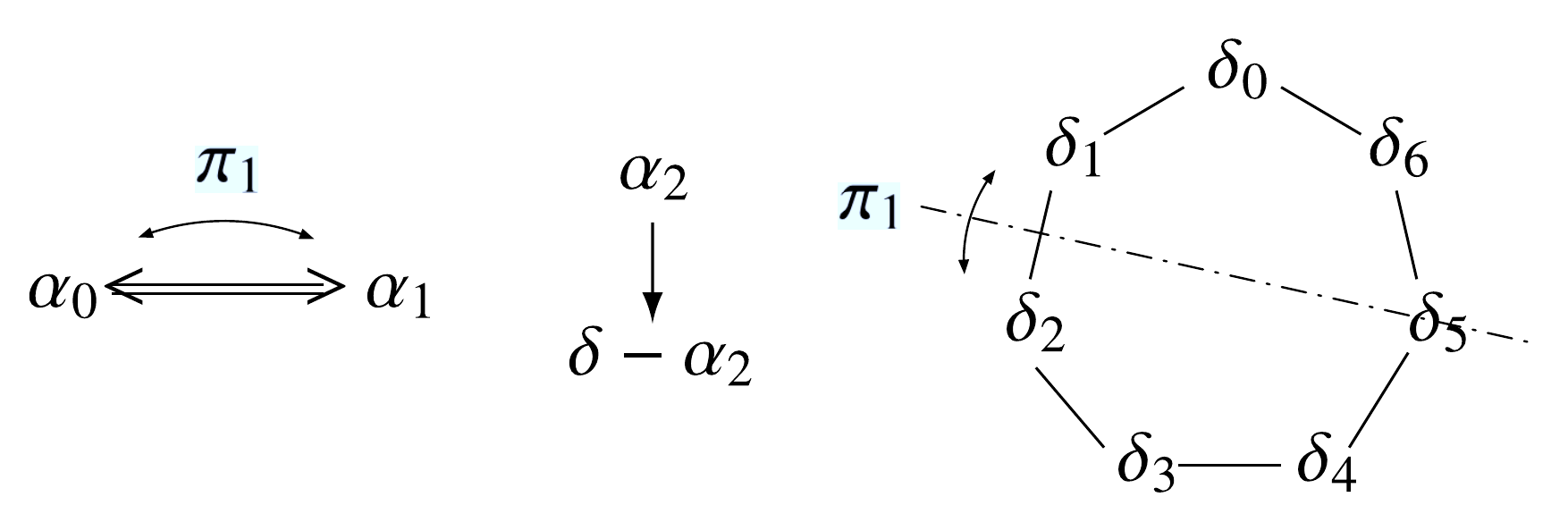}\qquad
\lower10pt\hbox{\includegraphics[scale=0.4]{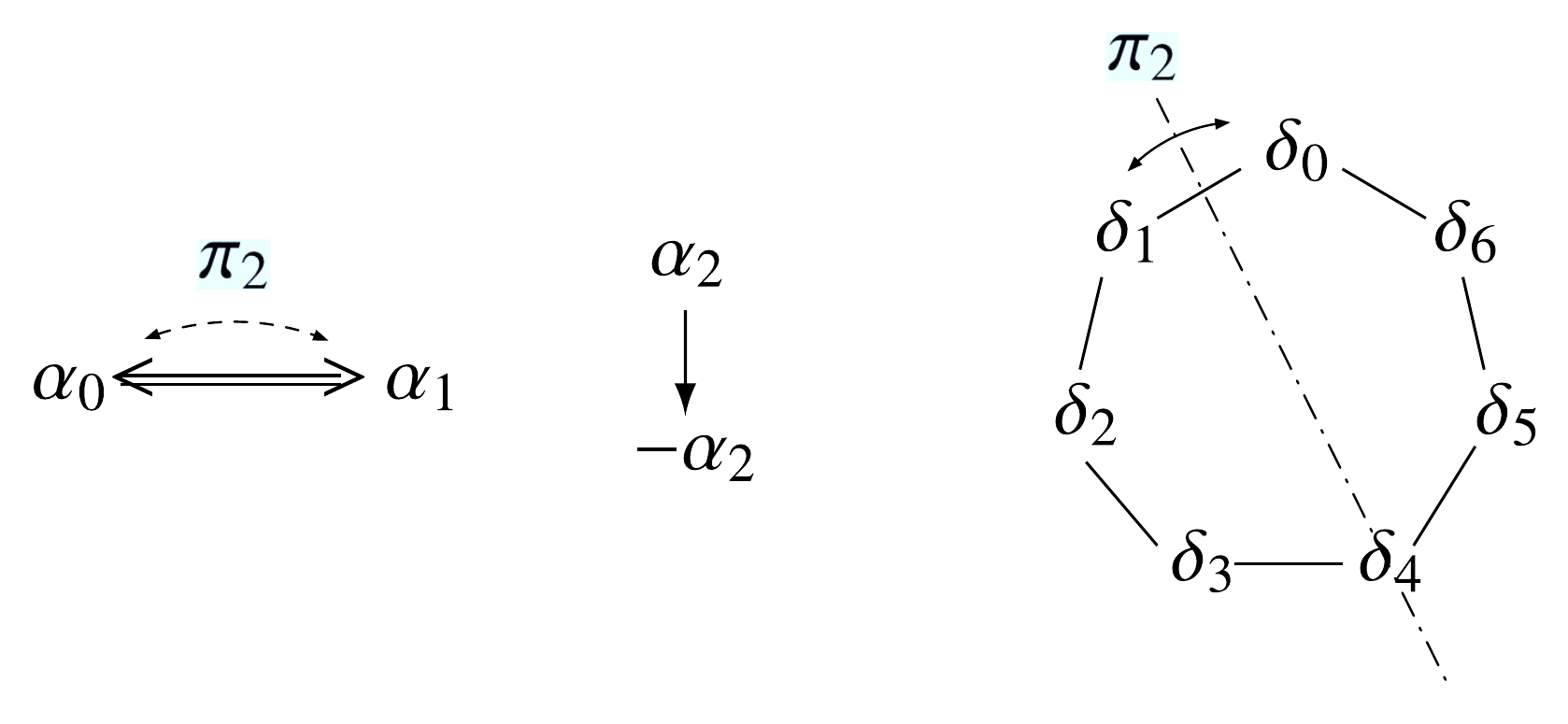}}
\end{displaymath}
%
\subsubsection{$q$-P$(\underset{|\alpha|^2=8}{A_1^{(1)}}/A_7^{(1)})$}\label{subsubsec:conf_q-A1a}
Point configuration in $(f,g)$ coordinates:
\begin{equation}
\begin{split}
&{\rm P}_{18}:\left(-\frac{\kappa_1}{\epsilon v_1v_8},\epsilon\right)_2,\quad
{\rm P}_{23}:\left(-\frac{v_2v_3}{\epsilon}, \frac{1}{\epsilon}\right)_2,\quad
{\rm P}_4:\left(v_4,\infty \right),\quad
{\rm P}_{567}:\Bigl(\frac{\kappa_1\kappa_2^2}{v_5v_6v_7}\epsilon^2,\epsilon\Bigr)_3.
\end{split}
\end{equation}
\begin{equation}
{\lower40pt\hbox{\footnotesize
\begin{picture}(50,50)(0,0)
\put(5,10){\line(1,0){42}}
\put(5,40){\line(1,0){42}}
\put(10,5){\line(0,1){42}}
\put(40,5){\line(0,1){42}}
\put(5,0){$f=0$}\put(35,0){$f=\infty$}\put(-7,14){$g=0$}\put(-7,35){$g=\infty$}
\put(10,10){\circle{3}}\put(40,10){\circle*{1.5}}
\put(12,13){$567$}\put(41,13){$18$}
\put(20,40){\circle*{2}}\put(40,40){\circle{3}}
\put(18,43){$4$}\put(41,43){$23$}
\put(10,10){\circle*{1.5}}\put(10,10){\circle{5}}
\put(40,10){\circle{3} ${}$}\put(40,40){\circle*{1.5} ${}$}
\end{picture}
\hskip40pt 
\begin{picture}(50,50)(0,0)
\thicklines
\put(15,10){\line(1,0){20}} 
\put(5,40){\line(1,0){30}}
\put(10,15){\line(0,1){35}}
\put(40,15){\line(0,1){20}}
\put(28,42){\line(1,-1){15}}
\put(28,8){\line(1,1){15}}  
\put(18,7){\line(0,1){15}}
\put(7,18){\line(1,0){15}}
\put(-7,38){$2|24$}
\put(0,10){$1|56$}
\put(0,16){$67$}
\put(42,13){$1|12$}
\put(44,25){$23$}
\put(25,1){$18$}
\put(17,1){$56$}
\put(5,2){$2|15$}

\thinlines
\put(23,42){$4$}\dashline[40]{2.0}(20,35)(20,50) 

\put(29,33){$3$}\dashline[30]{2.0}(30,30)(42,42)
\put(29,13){$8$}\dashline[30]{2.0}(30,20)(42,8)

\put(16,25){$7$}\dashline[30]{2.0}(15,23)(15,14)
\end{picture}
}}
\end{equation}
Root data:
\begin{equation}
\begin{split}
   \setlength{\unitlength}{0.8mm}
\begin{picture}(25,18)(0,-6)
\put(-2,5){$\alpha_0$}
\put(21,5){$\alpha_1$}
\put(5,6.3){\line(1,0){12.5}}
\put(5,5.8){\line(1,0){12.5}}
\put(4.1,5.1){$<$}
\put(16.25,5.1){$>$}
\end{picture} 
\qquad
 \setlength{\unitlength}{1.4mm}
\begin{picture}(15,16)(-7,-6)
\put(0,7.0){$\delta_0$}
\put(-4.95,4.95){$\delta_1$}
\put(- 7.0,0){$\delta_2$}
\put(- 4.95, - 4.95){$\delta_3$}
\put( 0, -7.0){$\delta_4$}
\put(4.95, - 4.95){$\delta_5$}
\put(7.0, 0){$\delta_6$}
\put(4.95, 4.95){$\delta_7$}
\drawline(-0.8,7.0)(-3,6)
\drawline(-5,4)(-6,2)
\drawline(-6,-1)(-5,-3)
\drawline(-2.5,-5.2)(-0.8,-6.2)
\drawline(2.5,-6.2)(4.5,-5.2)
\drawline(6.5,-3)(7.5,-1)
\drawline(6.5,4)(7.5,2)
\drawline(4.5,6)(2.3,7.0)
\end{picture}
\end{split}
\end{equation}
\begin{equation}
 \begin{split}
& \alpha_0 = H_1 + 2H_2 -2E_2 - 2E_3 + E_4 - E_5 - E_6 - E_7,\\
& \alpha_1 = H_1-E_1+E_2+E_3-2E_4-E_8,\\
& \delta_0 = H_1 - E_1 - E_2,\ \delta_1 = E_2 - E_3,\ \delta_2 = H_2 - E_2 - E_4,\ \delta_3 = H_1 - E_5 - E_6,\\
& \delta_4 = E_6 - E_7,\ \delta_5 = E_5 - E_6,\ \delta_6 = H_2 - E_1 - E_5,\ \delta_7 = E_1 - E_8,\\
& \pi_1 = \pi_{56723184}r_{H_1-E_4-E_5}r_{H_2-E_1-E_4},\ \pi_2 = \pi_{65432187}r_{H_2-E_2-E_5}.
 \end{split}
\end{equation}
\begin{displaymath}
\includegraphics[scale=0.4]{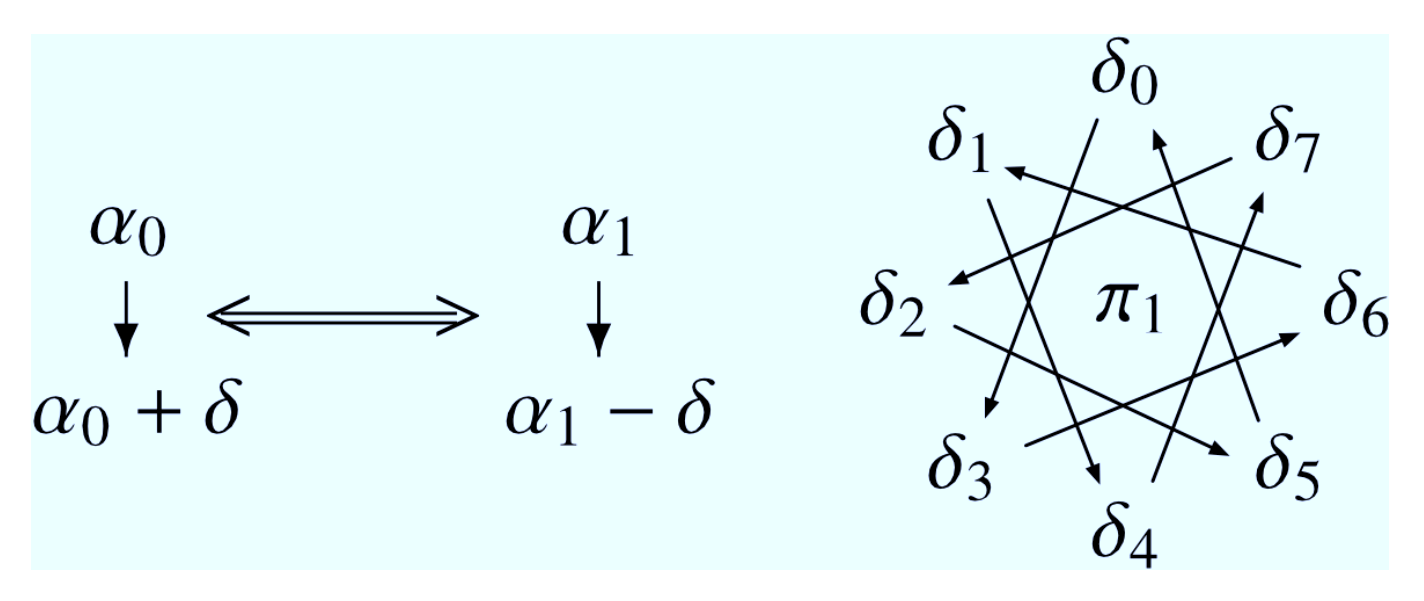}\qquad
\includegraphics[scale=0.4]{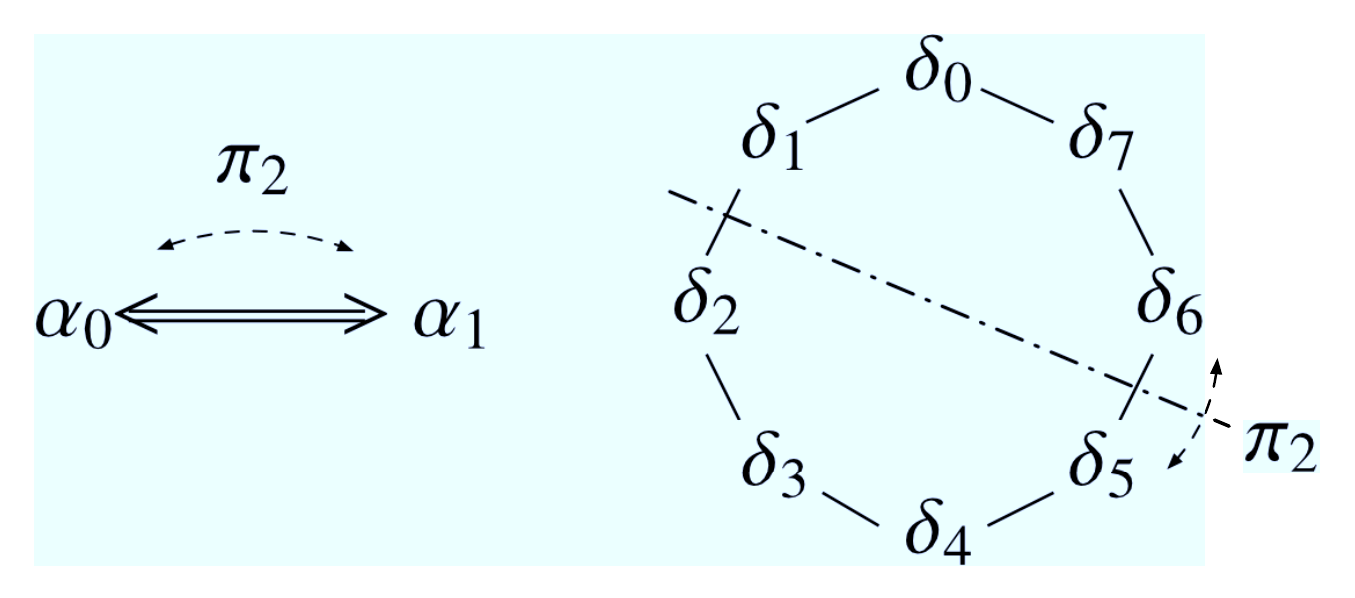}
\end{displaymath}
%
\subsubsection{$q$-P$(E_3^{(1)}/A_5^{(1)};b)$}\label{subsubsec:conf_q-E3b}
Point configuration in $(f,g)$ coordinates:
\begin{equation}
\begin{split}
&{\rm P}_1:\left(\infty ,\frac{1}{v_1}\right),\quad
{\rm P}_{23}:\left(-\frac{v_2v_3}{\epsilon}, \frac{1}{\epsilon}\right)_2,\quad
{\rm P}_4:\left(v_4,\infty \right),\\
&{\rm P}_5: \left(0,\frac{v_5}{\kappa_2}\right),\quad 
{\rm P}_{67}:\Bigl(-\frac{\kappa_1\kappa_2}{v_6v_7}\epsilon,\epsilon\Bigr)_2,\quad
{\rm P}_8:\left(\frac{\kappa_1}{v_8},0\right). 
\end{split}
\end{equation}
\begin{equation}
{\lower40pt\hbox{\footnotesize
\begin{picture}(50,50)(0,0)
\put(5,10){\line(1,0){42}}
\put(5,40){\line(1,0){42}}
\put(10,5){\line(0,1){42}}
\put(40,5){\line(0,1){42}}
\put(5,0){$f=0$}\put(35,0){$f=\infty$}\put(-7,14){$g=0$}\put(-7,35){$g=\infty$}
\put(10,10){\circle{3}}\put(30,10){\circle*{2}}
\put(12,13){$67$}\put(28,13){$8$}
\put(20,40){\circle*{2}}\put(40,40){\circle{3}}
\put(18,43){$4$}\put(41,43){$23$}
\put(10,10){\circle*{1.5} ${}$}\put(10,30){\circle*{2} $5$}
\put(40,20){\circle*{2} $1$}\put(40,40){\circle*{1.5} ${}$}
\end{picture}
\hskip40pt 
\begin{picture}(50,50)(0,0)
\thicklines
\put(15,10){\line(1,0){35}}
\put(5,40){\line(1,0){30}}
\put(10,15){\line(0,1){35}}
\put(40,5){\line(0,1){30}}
\put(28,42){\line(1,-1){15}} 
\put(8,22){\line(1,-1){15}} 
\put(-7,38){$2|24$}
\put(10,3){$2|68$}
\put(-2,11){$1|56$}
\put(35,0){$1|12$}
\put(44,25){$23$}
\put(2,22){$67$}

\thinlines
\put(36,22){$1$}\dashline[40]{2}(35,20)(50,20) 
\put(23,42){$4$}\dashline[40]{2.0}(20,35)(20,50) 
\put(12,32){$5$}\dashline[40]{2}(0,30)(15,30) 
\put(26,13){$8$}\dashline[40]{2}(30,5)(30,15) 

\put(29,33){$3$}\dashline[30]{2.0}(30,30)(42,42)
\put(14,18){$7$}\dashline[30]{2.0}(10,10)(22,22)
\end{picture}
}}
\end{equation}
Root data:
\begin{equation}
\begin{split}
 \setlength{\unitlength}{0.8mm}
\begin{picture}(27,18)(0,-4)
\put(0,0){$\alpha_1$}
\put(20,0){$\alpha_2$}
\put(10,15){$\alpha_0$}
\put(5,1){\line(1,0){12.5}}
\put(19,4){\line(-2,3){6}}
\put(3,4){\line(2,3){6}}
\end{picture} \ 
 \setlength{\unitlength}{0.8mm}
\begin{picture}(27,18)(0,-7)
\put(-2,5){$\alpha_3$}
\put(21,5){$\alpha_4$}
\put(5,6.3){\line(1,0){12.5}}
\put(5,5.8){\line(1,0){12.5}}
\put(4.4,5.1){$<$}
\put(16.3,5.1){$>$}
\end{picture} 
\qquad
 \setlength{\unitlength}{1.4mm}
\begin{picture}(25,20)(-7,-7)
\put(0,7.0){$\delta_0$}
\drawline(-1,6.5)(-4,5)
\put(- 6.06,3.5){$\delta_1$}
\drawline(-5.5,2.5)(-5.5,-1.5)
\put(- 6.06,-3.5){$\delta_2$}
\drawline(-4,-4.5)(-1,-5.5)
\put(0, -7.0){$\delta_3$}
\drawline(2,-5.5)(5,-4)
\put(6.06, -3.5){$\delta_4$}
\drawline(6.5,2.5)(6.5,-1.5)
\put(6.06,3.5){$\delta_5$}
\drawline(5.5,5)(2,6.5)
\end{picture}
\end{split}
\end{equation}
\begin{equation}\label{eqn:root_E3A5b}
\begin{split}
&
\alpha_0=H_1+H_2-E_2-E_3-E_6-E_7,\ 
\alpha_1=H_1-E_4-E_8,\ 
\alpha_2=H_2-E_1-E_5,\ 
\\
&
\alpha_3=H_1+H_2-E_2-E_3-E_5-E_8,\ 
\alpha_4=H_1+H_2-E_1-E_4-E_6-E_7,\ 
\\
&
\delta_0=H_1-E_1-E_2,\ 
\delta_1=E_2-E_3,\ 
\delta_2=H_2-E_2-E_4,\ 
\\
&
\delta_3=H_1-E_5-E_6,\ 
\delta_4=E_6-E_7,\ 
\delta_5=H_2-E_6-E_8,\ 
\\
&
\pi_1=\pi_{42318675}r_{H_1-H_2},\ 
\pi_2=\pi_{36457281}r_{H_1-H_2}r_{H_1-E_2-E_6}.
\end{split}
\end{equation}
\begin{displaymath}
\includegraphics[scale=0.7]{Dynkin_Auto_A2A1_1-eps-converted-to.pdf}\quad
\includegraphics[scale=0.7]{Dynkin_Auto_A2A1_2-eps-converted-to.pdf}
\end{displaymath}
\begin{rem}\rm
 The two realizations of root systems $q$-P$(E_3^{(1)}/A_5^{(1)};a)$ and $q$-P$(E_3^{(1)}/A_5^{(1)};b)$
are transformed with each other by the reflection $r_{H_2-E_1-E_6}$.
\end{rem}
%
\subsubsection{$q$-P$(E_2^{(1)}/A_6^{(1)};b)$}\label{subsubsec:conf_q-E2b}
Point configuration in $(f,g)$ coordinates:
\begin{equation}
\begin{split}
&{\rm P}_1:\left(\infty ,\frac{1}{v_1}\right),\ 
{\rm P}_{23}:\left(-\frac{v_2v_3}{\epsilon}, \frac{1}{\epsilon}\right)_2,\ 
{\rm P}_4:\left(v_4,\infty \right),\ 
{\rm P}_{567}:\Bigl(\frac{\kappa_1\kappa_2^2}{v_5v_6v_7}\epsilon^2,\epsilon\Bigr)_3,\ 
{\rm P}_8:\left(\frac{\kappa_1}{v_8},0\right) .
\end{split}
\end{equation}
\begin{equation}
{\lower40pt\hbox{\footnotesize
\begin{picture}(50,50)(0,0)
\put(5,10){\line(1,0){42}}
\put(5,40){\line(1,0){42}}
\put(10,5){\line(0,1){42}}
\put(40,5){\line(0,1){42}}
\put(5,0){$f=0$}\put(35,0){$f=\infty$}\put(-7,14){$g=0$}\put(-7,35){$g=\infty$}
\put(10,10){\circle{3}}\put(30,10){\circle*{2}}
\put(12,13){$567$}\put(28,13){$8$}
\put(20,40){\circle*{2}}\put(40,40){\circle{3}}
\put(18,43){$4$}\put(41,43){$23$}
\put(10,10){\circle*{1.5} ${}$}\put(10,10){\circle{5} ${}$}
\put(40,20){\circle*{2} $1$}\put(40,40){\circle*{1.5} ${}$}
\end{picture}
\hskip40pt 
\begin{picture}(50,50)(0,0)
\thicklines
\put(15,10){\line(1,0){35}}
\put(5,40){\line(1,0){30}}
\put(10,15){\line(0,1){35}}
\put(40,5){\line(0,1){30}}
\put(28,42){\line(1,-1){15}} 
\put(18,7){\line(0,1){15}}
\put(7,18){\line(1,0){15}}
\put(-7,38){$2|24$}
\put(4,3){$2|58$}
\put(2,10){$1|56$}
\put(35,0){$1|12$}
\put(44,25){$23$}
\put(2,20){$67$}
\put(18,1){$56$}

\thinlines
\put(36,22){$1$}\dashline[40]{2}(35,20)(50,20) 
\put(23,42){$4$}\dashline[40]{2.0}(20,35)(20,50) 
\put(26,13){$8$}\dashline[40]{2}(30,5)(30,15) 

\put(29,33){$3$}\dashline[30]{2.0}(30,30)(42,42)
\put(13,25){$7$}\dashline[30]{2.0}(15,23)(15,14)

\end{picture}
}}
\end{equation}
Root data:
\begin{equation}
\begin{split}
  \setlength{\unitlength}{0.8mm}
\begin{picture}(40,18)(0,-6)
\put(-2,5){$\alpha_0$}
\put(21,5){$\alpha_1$}
\put(35,5){$\alpha_2$}
\put(5,6.3){\line(1,0){12.5}}
\put(5,5.8){\line(1,0){12.5}}
\put(4.05,5.05){$<$}
\put(15.95,5.05){$>$}
\end{picture} 
\qquad
 \setlength{\unitlength}{1.4mm}
\begin{picture}(15,16)(-7,-6)
\put(0,7.0){$\delta_0$}
\put(- 5.47282,4.36443){$\delta_1$}
\put( - 6.8245, - 1.55765){$\delta_2$}
\put( - 3.03719, - 6.30678){$\delta_3$}
\put(3.03719, - 6.30678){$\delta_4$}
\put(6.8245, - 1.55765){$\delta_5$}
\put(5.47282,4.36443){$\delta_6$}
\drawline(-0.8,7.0)(-3.5,5.4)
\drawline(-5.3,3.5)(-6,0.5)
\drawline(-5.8,-2.5)(-3.5,-5.2)
\drawline(-1,-5.8)(2.5,-5.8)
\drawline(4.8,-5.2)(6.8,-2)
\drawline(7.1,0.5)(6.4,3.5)
\drawline(5.2,5.4)(2.5,7.0)
\end{picture} 
\end{split}
\end{equation}
\begin{equation}\label{eqn:root_E2A6b}
\begin{split}
&
\alpha_0=H_1+2H_2-E_1-E_2-E_3-E_5-E_6-E_7,\ 
\alpha_1=H_1-E_4-E_8,\ 
\\
&
\alpha_2=H_1+2E_1-2E_2- 2E_3+E_4-E_8,
\\
&
\delta_0=H_1-E_1-E_2,\ 
\delta_1=E_2-E_3,\ 
\delta_2=H_2-E_2-E_4,\ 
\\
&
\delta_3=H_1-E_5-E_6,\ 
\delta_4=E_6-E_7,\ 
\delta_5=E_5-E_6,\ 
\delta_6=H_2-E_5-E_8,\ 
\\
&
\pi_1=\pi_{21435678}r_{H_2-E_5-E_6}r_{H_2-E_1-E_2},\\
&
\pi_2=\pi_{35182674}r_{H_1-E_2-E_5}.
\end{split}
\end{equation}
\begin{displaymath}
\includegraphics[scale=0.4]{Dynkin_Auto_A1A1_1_revised-eps-converted-to.pdf}\qquad
\lower10pt\hbox{\includegraphics[scale=0.4]{Dynkin_Auto_A1A1_2_revised-eps-converted-to.pdf}}
\end{displaymath}
\begin{rem}\label{rem:a_and_b}\rm
The two realizations of root systems $q$-P$(E_2^{(1)}/A_6^{(1)};a)$ and
$q$-P$(E_2^{(1)}/A_6^{(1)};b)$ are transformed with each other by the reflection $r_{H_2-E_1-E_5}$.
\end{rem}
%
\subsubsection{$q$-P$(A_1^{(1)}/A_7^{(1)})$}\label{subsubsec:conf_q-A1b}
Point configuration in $(f,g)$ coordinates:
\begin{equation}
\begin{split}
&{\rm P}_{123}:\left(\frac{v_1v_2v_3}{\epsilon^2}, \frac{1}{\epsilon}\right)_3,\quad
{\rm P}_4:\left(v_4,\infty \right),\ 
{\rm P}_{567}:\Bigl(\frac{\kappa_1\kappa_2^2}{v_5v_6v_7}\epsilon^2,\epsilon\Bigr)_3,\ 
{\rm P}_8:\left(\frac{\kappa_1}{v_8},0\right).
\end{split}
\end{equation}
\begin{equation}
{\lower40pt\hbox{\footnotesize
\begin{picture}(50,50)(0,0)
\put(5,10){\line(1,0){42}}
\put(5,40){\line(1,0){42}}
\put(10,5){\line(0,1){42}}
\put(40,5){\line(0,1){42}}
\put(5,0){$f=0$}\put(35,0){$f=\infty$}\put(-7,14){$g=0$}\put(-7,35){$g=\infty$}
\put(10,10){\circle{3}}\put(30,10){\circle*{2}}
\put(12,13){$567$}\put(28,13){$8$}
\put(20,40){\circle*{2}}\put(40,40){\circle{3}}
\put(18,43){$4$}\put(41,43){$123$}
\put(10,10){\circle*{1.5} ${}$}\put(10,10){\circle{5} ${}$}
\put(40,40){\circle{5} ${}$}\put(40,40){\circle*{1.5} ${}$}
\end{picture}
\hskip40pt 
\begin{picture}(50,50)(0,0)
\thicklines
\put(15,10){\line(1,0){35}}
\put(5,40){\line(1,0){30}}
\put(10,15){\line(0,1){35}}
\put(40,5){\line(0,1){30}}
\put(18,7){\line(0,1){15}}
\put(7,18){\line(1,0){15}}
\put(30,27){\line(0,1){15}}
\put(27,30){\line(1,0){15}}

\put(-7,38){$2|14$}
\put(4,3){$2|58$}
\put(2,10){$1|56$}
\put(35,0){$1|12$}
\put(44,28){$23$}
\put(28,45){$12$}
\put(2,20){$67$}
\put(17,1){$56$}

\thinlines
\put(23,42){$4$}\dashline[40]{2.0}(20,35)(20,50) 
\put(26,13){$8$}\dashline[40]{2}(30,5)(30,15) 

\put(34,20){$3$}\dashline[30]{2.0}(35,25)(35,35)
\put(16,25){$7$}\dashline[30]{2.0}(15,23)(15,14)
\end{picture}
}}
\end{equation}
Root data:
\begin{equation}
\begin{split}
   \setlength{\unitlength}{0.8mm}
\begin{picture}(25,18)(0,-6)
\put(-2,5){$\alpha_0$}
\put(21,5){$\alpha_1$}
\put(5,6.3){\line(1,0){12.5}}
\put(5,5.8){\line(1,0){12.5}}
\put(4.1,5.1){$<$}
\put(16.25,5.1){$>$}
\end{picture} 
\qquad
 \setlength{\unitlength}{1.4mm}
\begin{picture}(15,16)(-7,-6)
\put(0,7.0){$\delta_0$}
\put(-4.95,4.95){$\delta_1$}
\put(- 7.0,0){$\delta_2$}
\put(- 4.95, - 4.95){$\delta_3$}
\put( 0, -7.0){$\delta_4$}
\put(4.95, - 4.95){$\delta_5$}
\put(7.0, 0){$\delta_6$}
\put(4.95, 4.95){$\delta_7$}
\drawline(-0.8,7.0)(-3,6)
\drawline(-5,4)(-6,2)
\drawline(-6,-1)(-5,-3)
\drawline(-2.5,-5.2)(-0.8,-6.2)
\drawline(2.5,-6.2)(4.5,-5.2)
\drawline(6.5,-3)(7.5,-1)
\drawline(6.5,4)(7.5,2)
\drawline(4.5,6)(2.3,7.0)
\end{picture}
\end{split}
\end{equation}
\begin{equation}
\begin{split}
&
\alpha_0=H_1+2H_2-E_1-E_2-E_3-E_5-E_6-E_7,\ 
\alpha_1=H_1-E_4-E_8\ 
\\
&
\delta_0=H_1-E_1-E_2,\ 
\delta_1=E_2-E_3,\ 
\delta_2=E_1-E_2,\ 
\delta_3=H_2-E_1-E_4,\ 
\\
&
\delta_4=H_1-E_5-E_6,\ 
\delta_5=E_6-E_7,\ 
\delta_6=E_5-E_6,\ 
\delta_7=H_2-E_5-E_8,\ 
\\
&
\pi_1=\pi_{52381674}r_{H_1-E_1-E_5},\ 
\pi_2=\pi_{25476183}r_{H_1-H_2}r_{H_1-E_2-E_6}r_{H_2-E_1-E_5} .
\end{split}
\end{equation}
\begin{displaymath}
\includegraphics[scale=0.4]{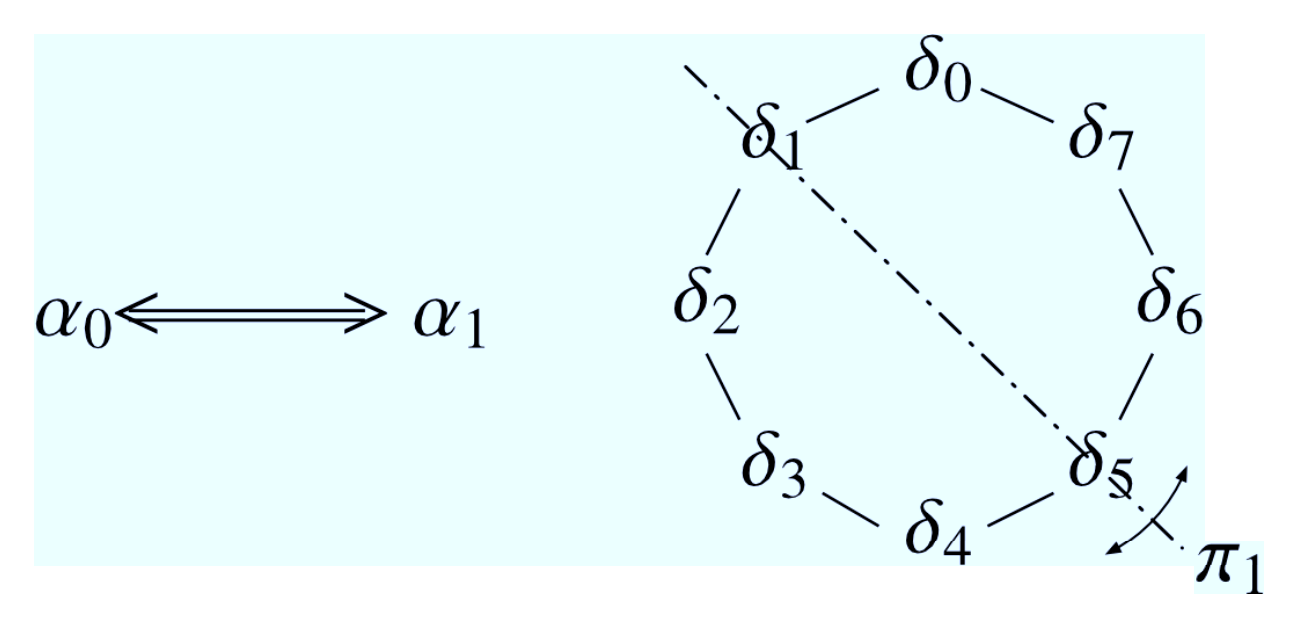}\qquad
\raise5pt\hbox{\includegraphics[scale=0.4]{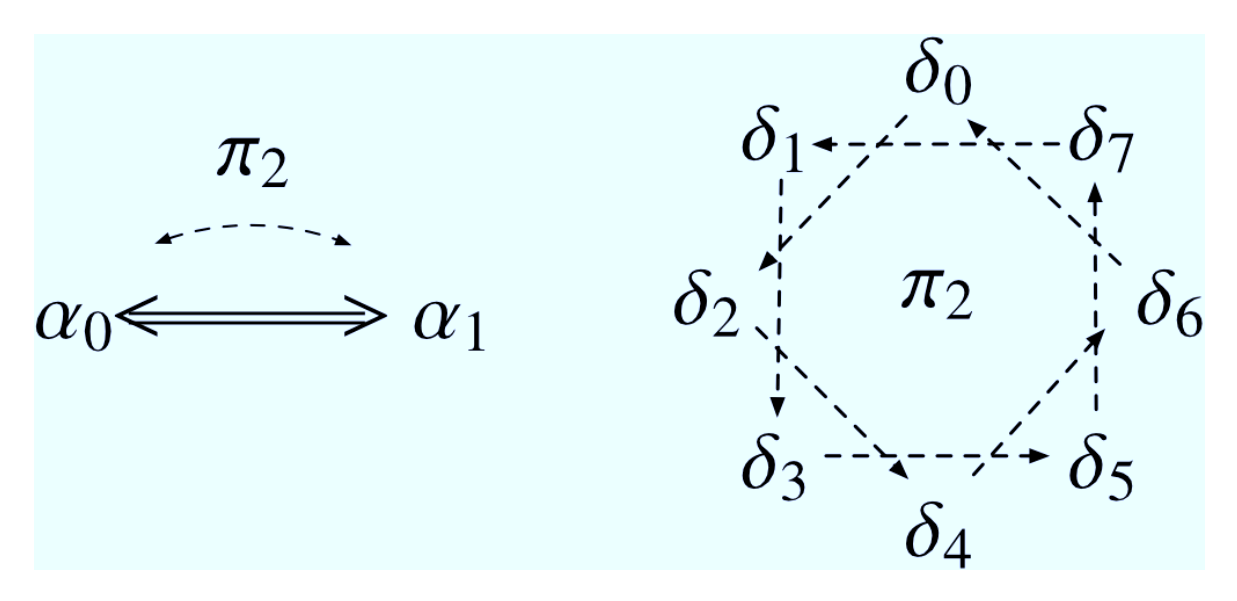}}
\end{displaymath}
%
\subsubsection{$q$-P$(A_0^{(1)}/A_8^{(1)})$}\label{subsubsec:conf_q-A0A8}
Point configuration in $(f,g)$ coordinates:
\begin{equation}
\begin{split}
&{\rm P}_{18}:\left(-\frac{\kappa_1}{\epsilon v_1v_8},\epsilon\right)_2,\quad
{\rm P}_{234}:\left(\frac{1}{\epsilon}, \frac{1}{v_2v_3v_4\epsilon^2}\right)_3,\quad
{\rm P}_{567}:\Bigl(\frac{\kappa_1\kappa_2^2}{v_5v_6v_7}\epsilon^2,\epsilon\Bigr)_3.
\end{split}
\end{equation}
\begin{equation}
{\lower40pt\hbox{\footnotesize
\begin{picture}(50,50)(0,0)
\put(5,10){\line(1,0){42}}
\put(5,40){\line(1,0){42}}
\put(10,5){\line(0,1){42}}
\put(40,5){\line(0,1){42}}
\put(5,0){$f=0$}\put(35,0){$f=\infty$}\put(-7,14){$g=0$}\put(-7,35){$g=\infty$}
\put(10,10){\circle{3}}\put(40,10){\circle*{1.5}}
\put(12,13){$567$}\put(41,13){$18$}
\put(40,40){\circle{3}}\put(40,40){\circle{5}}
\put(41,43){$234$}
\put(10,10){\circle*{1.5} ${}$}\put(10,10){\circle{5} ${}$}
\put(40,10){\circle{3} ${}$}\put(40,40){\circle*{1.5} ${}$}
\end{picture}
\hskip40pt 
\begin{picture}(50,50)(0,0)
\thicklines
\put(15,10){\line(1,0){20}} 
\put(5,40){\line(1,0){30}}
\put(10,15){\line(0,1){35}}
\put(40,15){\line(0,1){20}}
\put(28,8){\line(1,1){15}}  
\put(18,7){\line(0,1){15}}
\put(7,18){\line(1,0){15}}
\put(30,27){\line(0,1){15}}
\put(27,30){\line(1,0){15}}
\put(-7,38){$2|23$}
\put(0,10){$1|56$}
\put(0,18){$67$}
\put(42,13){$1|12$}
\put(44,28){$23$}
\put(25,1){$18$}
\put(17,1){$56$}
\put(5,4){$2|15$}
\put(28,45){$34$}

\thinlines
\put(29,13){$8$}\dashline[30]{2.0}(30,20)(42,8)
\put(16,25){$7$}\dashline[30]{2.0}(15,23)(15,14)
\put(20,34){$4$}\dashline[30]{2.0}(25,35)(35,35)
\end{picture}
}}
\end{equation}

Root data:
\begin{equation}
\begin{split}
  \setlength{\unitlength}{0.8mm}
\begin{picture}(18,18)(0,-10)
\put(10,5){$\alpha_0$}
\end{picture} 
\qquad
 \setlength{\unitlength}{1.4mm}
\begin{picture}(25,20)(-7,-10)
\put(0,7.0){$\delta_0$}
\put( - 4.5,5.36){$\delta_1$}
\put( - 6.89,1.22){$\delta_2$}
\put( - 6.06, - 3.5){$\delta_3$}
\put( - 2.39, - 6.58){$\delta_4$}
\put(2.39, - 6.58){$\delta_5$}
\put(6.06, - 3.5){$\delta_6$}
\put(6.89,1.22){$\delta_7$}
\put(4.5,5.36){$\delta_8$}
\drawline(-0.8,7.3)(-2.5,6.5)
\drawline(-4.4,4.7)(-5.5,3)
\drawline(-6,0.5)(-5.7,-1.6)
\drawline(-4.5,-4.2)(-2.8,-5.5)
\drawline(-0.2,-6.2)(2,-6.2)
\drawline(4.2,-5.5)(6.2,-4)
\drawline(7,-1.7)(7.5,0.5)
\drawline(7.2,3)(6.3,4.5)
\drawline(4,6.5)(2.2,7.3)
\end{picture} 
\end{split}
\end{equation}
\begin{equation}
\begin{split}
&
\alpha_0=2H_1+2H_2-E_1-E_2-E_3-E_4-E_5-E_6-E_7-E_8,\ 
\\
&
\delta_0=H_1-E_1-E_2,\ 
\delta_1=E_2-E_3,\ 
\delta_2=E_3-E_4,\ 
\delta_3=H_2-E_2-E_3,\ 
\\
&
\delta_4=H_1-E_5-E_6,\ 
\delta_5=E_6-E_7,\ 
\delta_6=E_5-E_6,\ 
\delta_7=H_2-E_1-E_5,\ 
\delta_8=E_1-E_8,
\\
&
\pi_1=\pi_{35182674}r_{H_1-E_2-E_5},\ 
\pi_2=\pi_{32675184}r_{H_1-H_2}r_{H_2-E_2-E_5} .
\end{split}
\end{equation}
\begin{displaymath}
\includegraphics[scale=0.4]{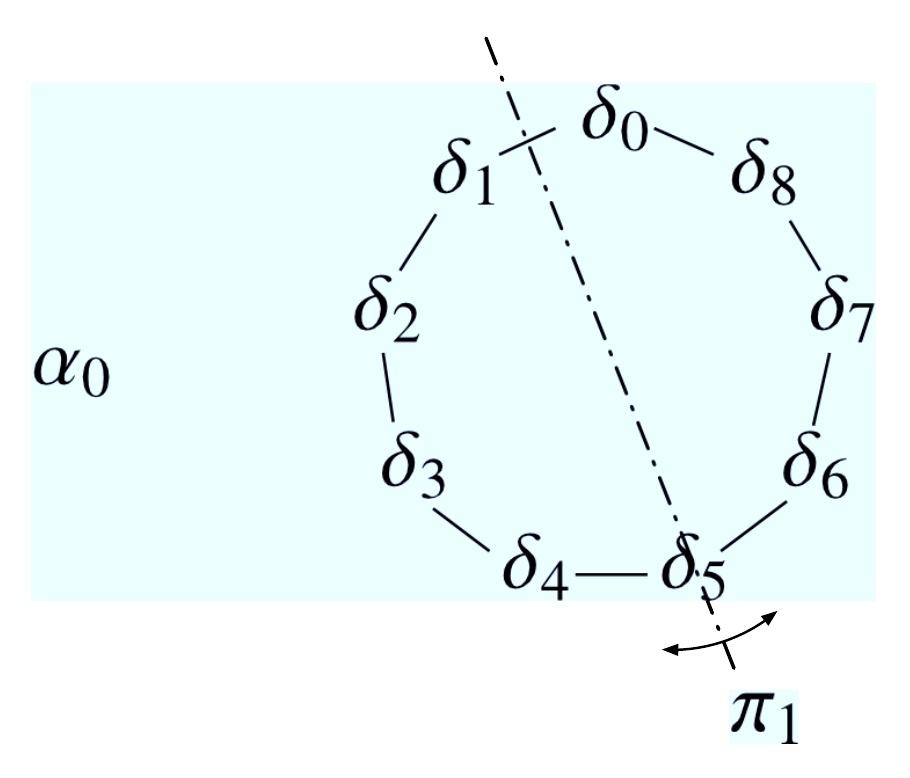}\qquad
\raise15pt\hbox{
\includegraphics[scale=0.4]{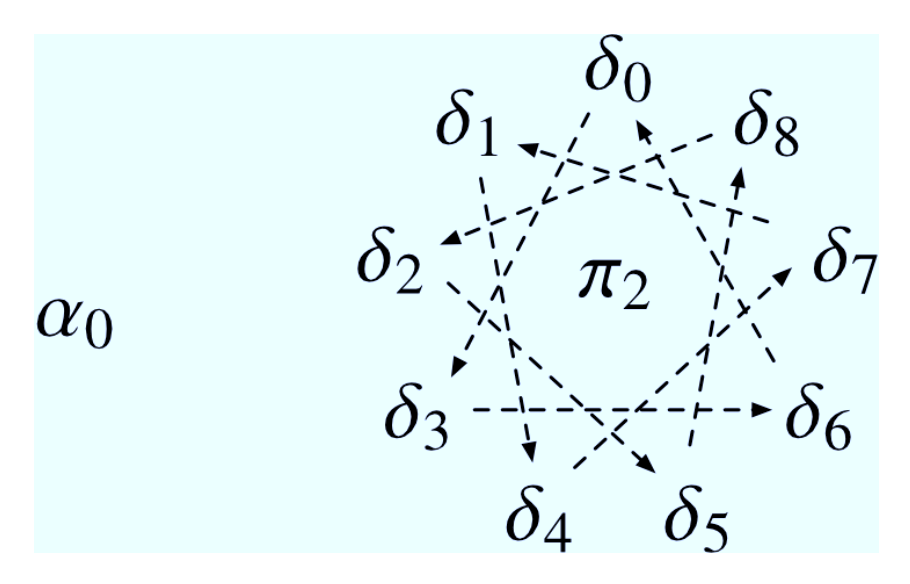}
}
\end{displaymath}

\subsubsection{d-P$(E_8^{(1)}/A_0^{(1)})$}\label{subsubsec:conf_d-E8}
Point configuration in $(f,g)$ coordinates:
\begin{equation}
{\rm P}_i: (f(v_i),g(v_i))\ (i=1,\ldots,8),\quad  f(z)=z(z-\kappa_1),\quad g(z)=z(z-\kappa_2).
\end{equation}
\begin{equation}
\lower2cm\hbox{\includegraphics[width=4cm]{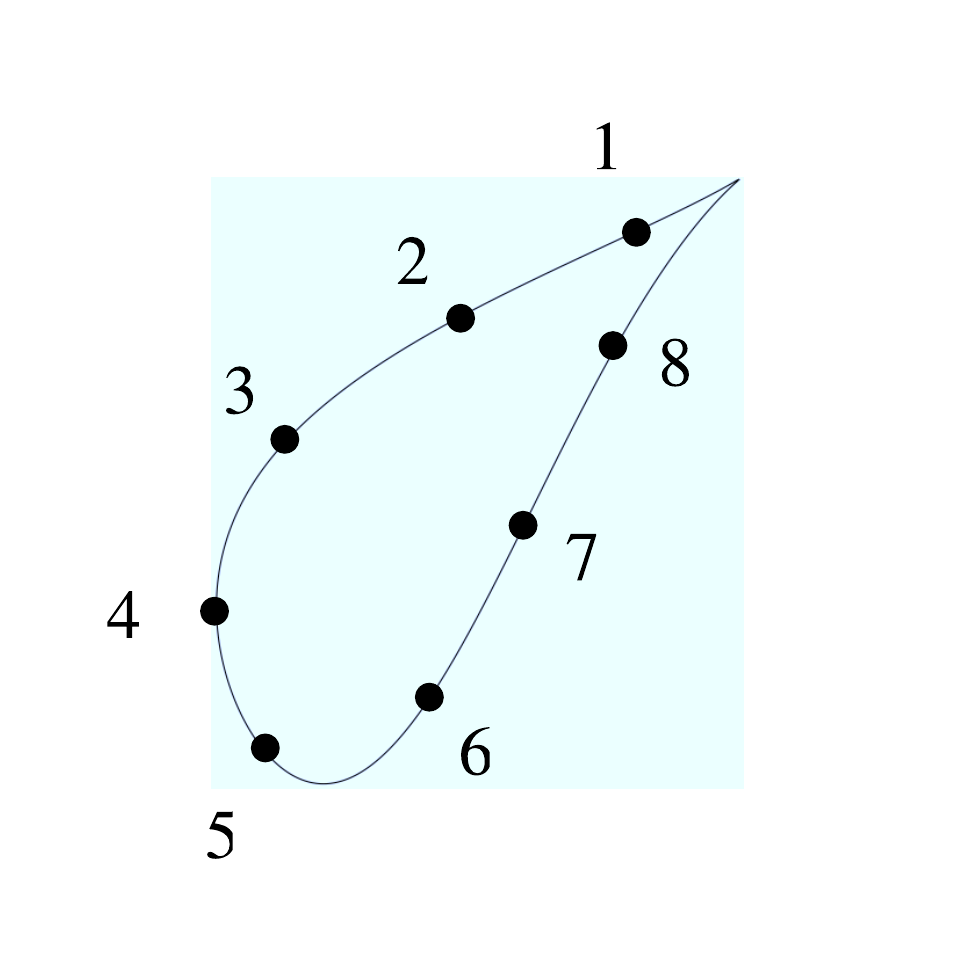}}
\hskip40pt
\lower2cm\hbox{\includegraphics[width=4cm]{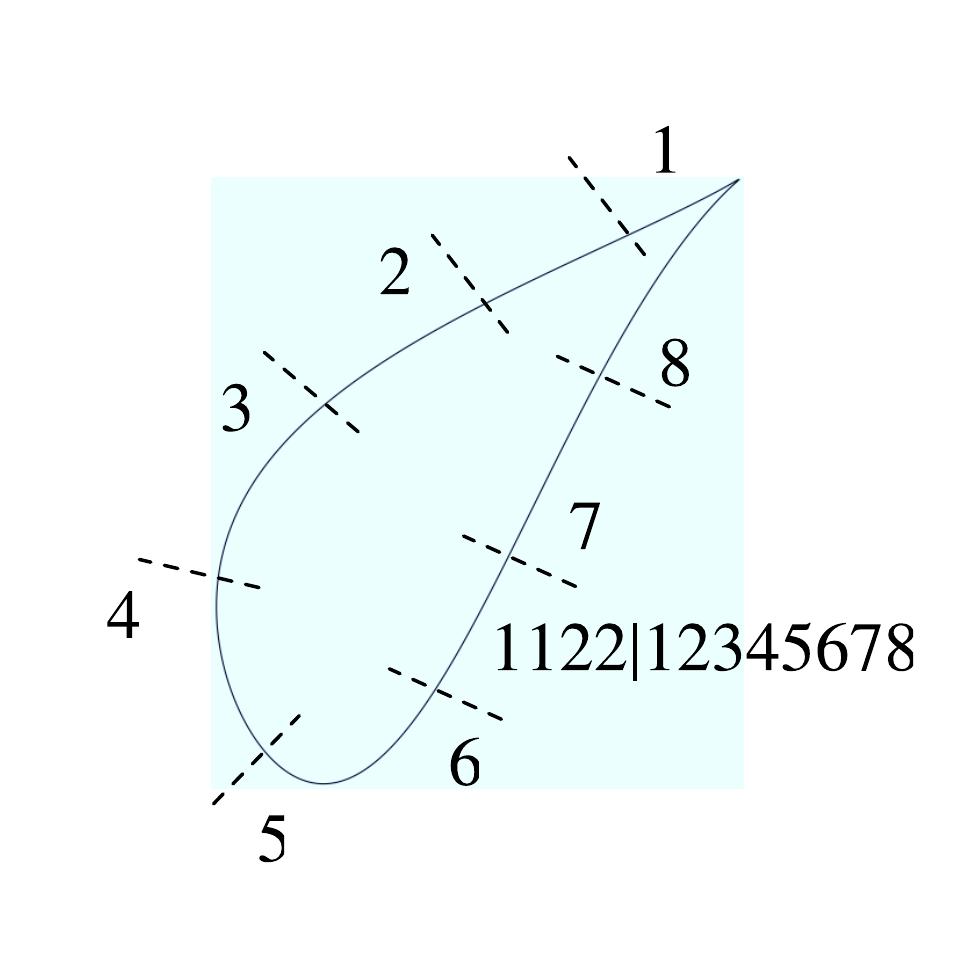}}
\end{equation}
Root data: same as Section \ref{subsubsec:conf_e-E8}.
%
\subsubsection{d-P$(E_7^{(1)}/A_1^{(1)})$} \label{subsubsec:conf_d-E7}
Point configuration in $(f,g)$ coordinates:
\begin{equation}
{\rm P}_i: (v_i,-v_i)\ (i=1,\ldots,4),\quad (\kappa_1-v_i, v_i-\kappa_2)\ (i=5,\ldots,8).
\end{equation}
\begin{equation}
{\lower40pt\hbox{\footnotesize
\begin{picture}(50,50)(0,0)
\put(10,25){$8$}\put(12,22){\circle*{2}}
\put(18,33){$7$}\put(20,30){\circle*{2}}
\put(23,38){$6$}\put(25,35){\circle*{2}}
\put(31,46){$5$}\put(33,43){\circle*{2}}

\put(40,26){$1$}\put(40,33){\circle*{2}}
\put(35,21){$2$}\put(35,28){\circle*{2}}
\put(28,16){$3$}\put(25,18){\circle*{2}}
\put(23,11){$4$}\put(20,13){\circle*{2}}

\put(10,20){\line(1,1){25}}
\put(10,3){\line(1,1){35}}
\end{picture}\hskip40pt
\begin{picture}(50,50)(0,0)
\put(10,25){$8$}\dashline[40]{2}(7,27)(14,20)
\put(18,33){$7$}\dashline[40]{2}(15,35)(22,28)
\put(23,38){$6$}\dashline[40]{2}(20,40)(27,33)
\put(31,46){$5$}\dashline[40]{2}(28,48)(35,41)

\put(40,26){$1$}\dashline[40]{2}(38,35)(45,28)
\put(35,21){$2$}\dashline[40]{2}(32,30)(39,23)
\put(28,16){$3$}\dashline[40]{2}(23,20)(30,13)
\put(23,11){$4$}\dashline[40]{2}(18,15)(25,8)
\put(12,0){$12|1234$}
\put(0,13){$12|5678$}

\thicklines
\put(10,20){\line(1,1){25}}
\put(10,3){\line(1,1){35}}
\end{picture}
}}
\end{equation}
Root data: same as Section \ref{subsubsec:conf_q-E7}.
%
\subsubsection{d-P$(E_6^{(1)}/A_2^{(1)})$} \label{subsubsec:conf_d-E6}
Point configuration in $(f,g)$ coordinates:
\begin{equation}
{\rm P}_i: (v_i,-v_i)\ (i=1,\ldots,4),\quad (\infty,v_i-\kappa_2)\ (i=5,6),\quad (\kappa_1-v_i,\infty)\ (i=7,8).
\end{equation}
\begin{equation}
{\lower40pt\hbox{\footnotesize
\begin{picture}(50,50)(0,0)
\put(0,30){\line(1,0){50}}
\put(30,0){\line(0,1){50}}
\put(8,24){$7$}\put(10,30){\circle*{2}}
\put(18,24){$8$}\put(20,30){\circle*{2}}
\put(25,9){$6$}\put(30,10){\circle*{2}}
\put(25,19){$5$}\put(30,20){\circle*{2}}
\put(13,47){\line(1,-1){35}}
\put(15,38){$1$}\put(20,40){\circle*{2}}
\put(20,33){$2$}\put(25,35){\circle*{2}}
\put(38,24){$3$}\put(35,25){\circle*{2}}
\put(43,19){$4$}\put(40,20){\circle*{2}}
\end{picture}\hskip40pt
\begin{picture}(50,50)(0,0)
\put(5,25){$7$}\dashline[40]{2}(10,25)(10,35)
\put(15,25){$8$}\dashline[40]{2}(20,25)(20,35)
\put(25,5){$6$}\dashline[40]{2}(25,10)(35,10)
\put(25,15){$5$}\dashline[40]{2}(25,20)(35,20)

\put(19,44){$1$}\dashline[40]{2}(16,36)(24,44)
\put(24,39){$2$}\dashline[40]{2}(21,31)(29,39)
\put(38,24){$3$}\dashline[40]{2}(31,21)(39,29)
\put(43,19){$4$}\dashline[40]{2}(36,16)(44,24)
\put(32,45){$1|56$}
\put(43,32){$2|78$}
\put(43,5){$12|1234$}

\thicklines
\put(0,30){\line(1,0){50}}
\put(30,0){\line(0,1){50}}
\put(13,47){\line(1,-1){35}}
\end{picture}
}}
\end{equation}
Root data: same as Section \ref{subsubsec:conf_q-E6}.

\subsubsection{d-P$(D_4^{(1)}/D_4^{(1)})$}\label{subsubsec:conf_d-D4}
Point configuration in $(f,g)$ coordinates:
\begin{equation}
\begin{split}
&{\rm P}_1:(\infty,-a_2),\ {\rm P}_2:(\infty,-a_1-a_2),\ {\rm P}_{34}\left(t(1+a_0\epsilon),\frac{1}{\epsilon}\right)_2,\\
&{\rm P}_5:(0,0),\ {\rm P}_6:(0,a_4),\ {\rm P}_{78}:\left(1+a_3\epsilon,\frac{1}{\epsilon}\right)_2,\\
& a_0 + a_1 + 2a_2 + a_3 +a_4 = 1.
\end{split}
\end{equation}
\vspace*{4mm}
\begin{equation}
{\lower50pt\hbox{\footnotesize
\begin{picture}(50,50)(0,-5)
\put(5,1){$f=0$}\put(37,1){$f=\infty$}\put(41,42){$g=\infty$}
\put(10,5){\line(0,1){45}}
\put(40,5){\line(0,1){45}}
\put(20,40){\circle{3}}\put(20,40){\circle*{1.5}}\put(18,44){$78$}
\put(30,40){\circle{3}}\put(30,40){\circle*{1.5}}\put(28,44){$34$}
\put(12,20){$6$}\put(10,20){\circle*{2}}
\put(12,30){$5$}\put(10,30){\circle*{2}}
\put(42,20){$1$}\put(40,20){\circle*{2}}
\put(42,30){$2$}\put(40,30){\circle*{2}}
\put(0,40){\line(1,0){50}}
\end{picture}
\hskip40pt
\begin{picture}(50,50)(0,-5)
\thicklines
\put(5,0){$1|56$}\put(35,0){$1|12$}\put(-5,40){$2|37$}
\put(10,5){\line(0,1){45}}
\put(40,5){\line(0,1){45}}
\put(5,40){\line(1,0){45}}
\put(18,31){$78$}\put(20,35){\line(0,1){15}}
\put(28,31){$34$}\put(30,35){\line(0,1){15}}

\thinlines
\put(16,47){$8$}\dashline[40]{2.0}(13,45)(23,45)
\put(31,47){$4$}\dashline[40]{2.0}(27,45)(37,45)
\put(42,22){$1$}\dashline[40]{2}(35,20)(50,20) 
\put(42,32){$2$}\dashline[40]{2}(35,30)(50,30) 
\put(6,22){$5$}\dashline[40]{2}(0,20)(15,20) 
\put(6,32){$6$}\dashline[40]{2}(0,30)(15,30) 
\end{picture}
}}
\end{equation}
Root data:
\begin{equation}
\begin{split}
\setlength{\unitlength}{1mm}
\begin{picture}(25,20)(0,0)
\put(0,0){$\alpha_1$}
\put(0,14.7){$\alpha_0$}
\put(6.5,7){$\alpha_2$}
\drawline(2.4,2.1)(5.8,6.0)
\drawline(2.4,12.9)(5.8,9.0)
\put(13,14.7){$\alpha_3$}
\drawline(9.5,6.0)(12.9,2.1)
\put(13,0){$\alpha_4$}
\drawline(9.5,9.0)(12.9,12.9)
\end{picture}
\qquad
\begin{picture}(25,20)(0,0)
\put(0,0){$\delta_1$}
\put(0,14.7){$\delta_0$}
\put(6.5,7){$\delta_2$}
\drawline(2.4,2.1)(5.8,6.0)
\drawline(2.4,12.9)(5.8,9.0)
\put(13,14.7){$\delta_3$}
\drawline(9.5,6.0)(12.9,2.1)
\put(13,0){$\delta_4$}
\drawline(9.5,9.0)(12.9,12.9)
\end{picture}
\end{split}
\end{equation}
\begin{equation}\label{eqn:root_data_D4/A4}
\begin{split}
&
\alpha_0=H_1-E_3-E_4,\ 
\alpha_1=E_1-E_2,\ 
\alpha_2=H_2-E_1-E_5,\ 
\\
& \alpha_3=H_1-E_7-E_8,\ 
\alpha_4=E_5-E_6 \\
&
\delta_0=E_3-E_4,\ 
\delta_1=H_1-E_1-E_2,\ 
\delta_2=H_2-E_3-E_7,\ 
\\
&\delta_3=E_7-E_8,\ 
\delta_4=H_1-E_5-E_6, \\
&
\pi_1=\pi_{12345876}r_{H_1-E_5-E_7},\ 
\pi_2=\pi_{12785634},\ 
\pi_3=\pi_{56341278}.
\end{split}
\end{equation}
\begin{displaymath}
\includegraphics[scale=0.35]{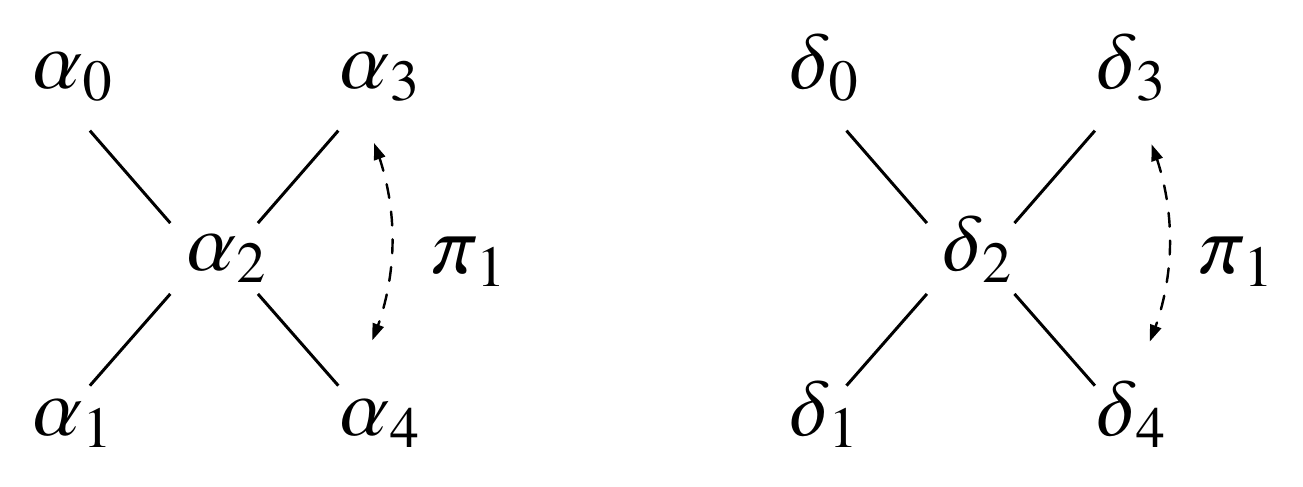}\qquad
\includegraphics[scale=0.35]{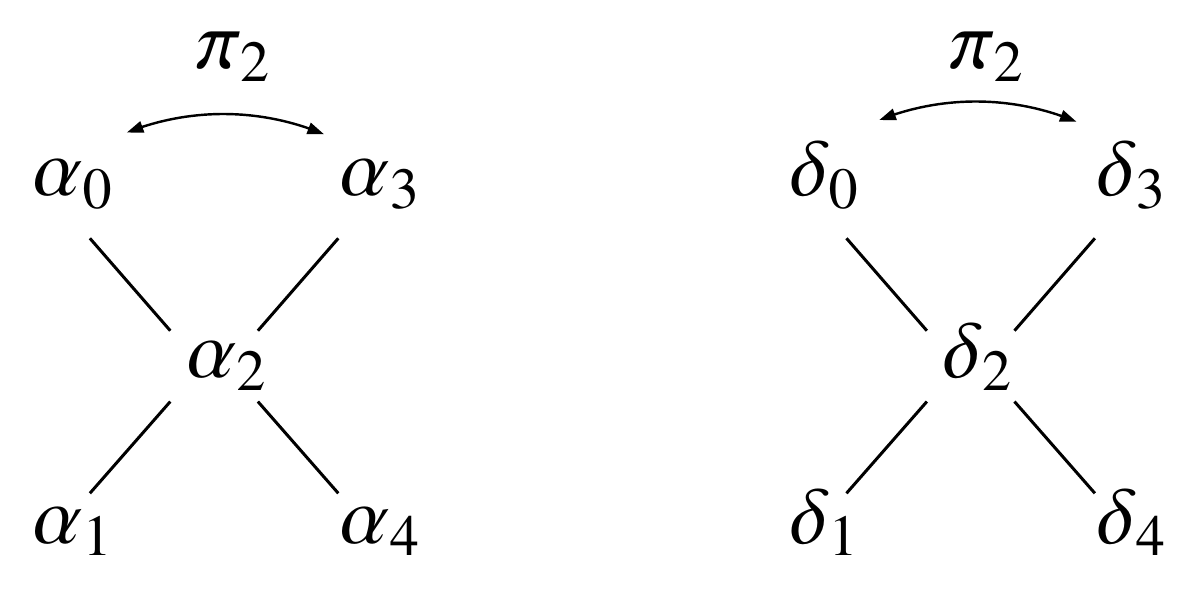}\qquad
\lower8pt\hbox{\includegraphics[scale=0.3]{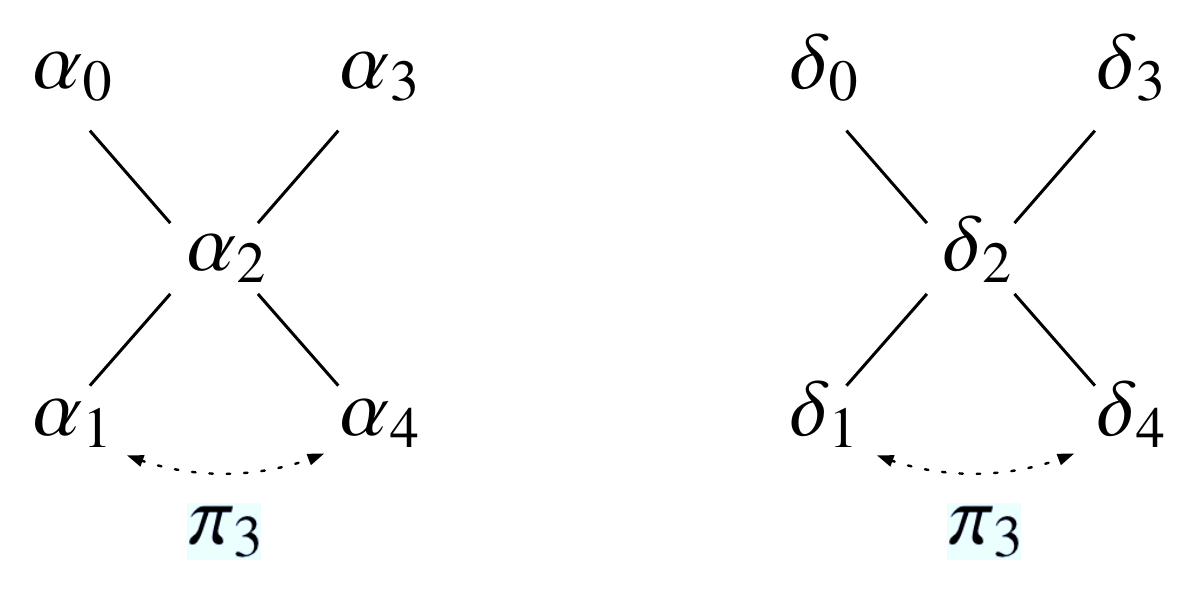}}
\end{displaymath}

\subsubsection{d-P$(A_3^{(1)}/D_5^{(1)})$}\label{subsubsec:conf_d-A3}
Point configuration in $(q,p)$ coordinates:
\begin{equation}
\begin{split}
&{\rm P}_{12}: \left(\frac{1}{\epsilon},-t+(a_1+a_2+a_3-1)\epsilon\right)_2,\quad
{\rm P}_{34}: \left(\frac{1}{\epsilon},-a_2\epsilon\right)_2, \\
&{\rm P}_{56}: \left(a_1\epsilon,\frac{1}{\epsilon}\right)_2,\ 
{\rm P}_{78}: \left(1+a_3\epsilon,\frac{1}{\epsilon}\right)_2.
\end{split}
\end{equation}
\begin{equation}
{\lower40pt\hbox{\footnotesize
\begin{picture}(50,50)(0,-5)
\put(20,40){\circle{3}}\put(20,40){\circle*{1.5}}\put(18,44){$78$}
\put(30,40){\circle{3}}\put(30,40){\circle*{1.5}}\put(28,44){$56$}
\put(40,30){\circle{3}}\put(40,30){\circle*{1.5}}\put(42,28){$34$}
\put(40,20){\circle{3}}\put(40,20){\circle*{1.5}}\put(42,18){$12$}

\put(23,8){$q=\infty$}\put(2,33){$p=\infty$}
\put(5,40){\line(1,0){45}}
\put(40,5){\line(0,1){45}}
\end{picture}
\hskip40pt
\begin{picture}(50,50)(0,-5)
\thicklines
\put(36,0){${1|13}$}\put(40,5){\line(0,1){45}}
\put(-7,38){${2|57}$}\put(5,40){\line(1,0){45}}
\put(17,29){${78}$}\put(20,35){\line(0,1){15}}
\put(26,29){${56}$}\put(30,35){\line(0,1){15}}
\put(33,15){${12}$}\put(35,20){\line(1,0){15}}
\put(33,25){${34}$}\put(35,30){\line(1,0){15}}

\thinlines
\put(16,47){${8}$}\dashline[40]{2.0}(13,45)(23,45)
\put(31,47){${6}$}\dashline[40]{2.0}(27,45)(37,45)
\put(47,32){${2}$}\dashline[40]{2}(45,38)(45,27) 
\put(47,22){${4}$}\dashline[40]{2}(45,12)(45,23) 
\end{picture}
}}
\end{equation}
Root data:
\begin{equation}
\begin{split}
\setlength{\unitlength}{1mm}
\begin{picture}(25,20)(0,0)
\put(0,0){$\alpha_1$}
\put(0,14){$\alpha_0$}
\put(14,0){$\alpha_2$}
\put(14,14){$\alpha_3$}
\drawline(1,3)(1,12)
\drawline(15,3)(15,12)
\drawline(4,1)(12,1)
\drawline(4,15)(12,15)
\end{picture}
\qquad
\begin{picture}(25,20)(0,0)
\put(0,0){$\delta_1$}
\put(0,14.7){$\delta_0$}
\put(6.5,7){$\delta_2$}
\drawline(2.4,2.1)(5.8,6.0)
\drawline(2.4,12.9)(5.8,9.0)
\drawline(10.4,7.5)(14.1,7.5)
\put(15,7){$\delta_3$}
\drawline(18,9.1)(21.4,13.0)
\put(21.5,14.7){$\delta_4$}
\drawline(18,6.0)(21.4,2.1)
\put(21.5,0){$\delta_5$}
\end{picture}
\end{split}
\end{equation}
\begin{equation}
 \begin{split}
&
\alpha_0=H_2-E_1-E_2,\ 
\alpha_1=H_1-E_5-E_6,\ 
\alpha_2=H_2-E_3-E_4,\ 
\alpha_3=H_1-E_7-E_8
\\
&
\delta_0=E_1-E_2,\ 
\delta_1=E_3-E_4,\ 
\delta_2=H_1-E_1-E_3,\ 
\\
&
\delta_3=H_2-E_5-E_7,\ 
\delta_4=E_5-E_6,\ 
\delta_5=E_7-E_8 ,
\\
&
\pi_1=\pi_{78563412}r_{H_1-H_2},\ 
\pi_2=\pi_{34125678}
\end{split}
\end{equation}
\begin{displaymath}
\includegraphics[scale=0.4]{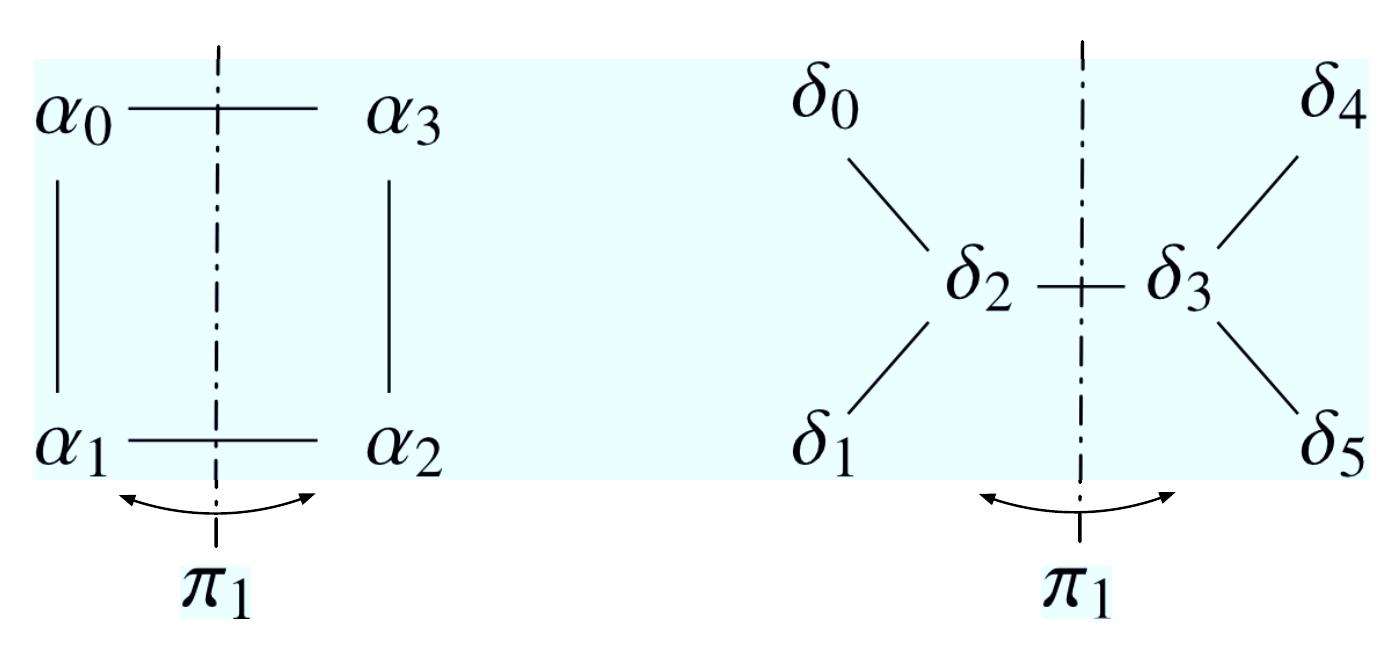}\qquad
\raise10pt\hbox{\includegraphics[scale=0.4]{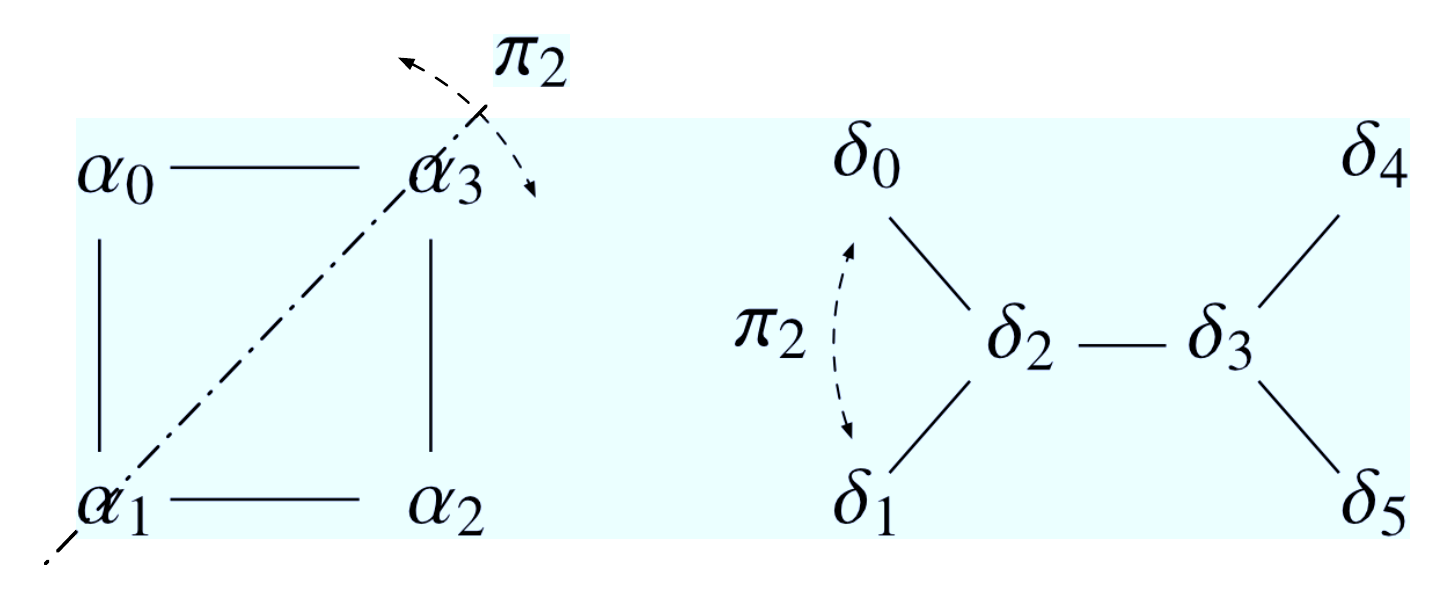}}
\end{displaymath}

\subsubsection{d-P$((2A_1)^{(1)}/D_6^{(1)})$} \label{subsubsec:conf_d-2A1}
Point configuration in $(q,p)$ coordinates:
\begin{equation}
\begin{split}
&{\rm P}_{12}:\left(\frac{1}{\epsilon},1-a_1\epsilon\right)_2,\ 
{\rm P}_{34}:\left(\frac{1}{\epsilon},-a_2\epsilon\right)_2,\ 
{\rm P}_{5678}:\left(\epsilon, -\frac{t}{\epsilon^2} + \frac{1-a_1-a_2}{\epsilon}\right)_4. 
\end{split}
\end{equation}
\vspace*{2mm}
\begin{equation}
{\lower40pt\hbox{\footnotesize
\begin{picture}(50,50)(0,-5)
\put(20,40){\circle{7}}\put(20,40){\circle{5}}\put(20,40){\circle{3}}\put(20,40){\circle*{1.5}}\put(24,42){$5678$}
\put(40,30){\circle{3}}\put(40,30){\circle*{1.5}}\put(42,28){$34$}
\put(40,20){\circle{3}}\put(40,20){\circle*{1.5}}\put(42,18){$12$}
\put(23,8){$q=\infty$}\put(2,33){$p=\infty$}
\put(5,40){\line(1,0){45}}
\put(40,5){\line(0,1){45}}
\end{picture}
\hskip40pt
\begin{picture}(50,50)(0,-5)
\thicklines
\put(35,0){${}$}\put(40,5){\line(0,1){45}}
\put(-5,40){${}$}\put(5,40){\line(1,0){45}}
\put(18,29){${67}$}\put(20,35){\line(0,1){15}}

\put(33,15){${12}$}\put(35,20){\line(1,0){15}}
\put(33,25){${34}$}\put(35,30){\line(1,0){15}}
\put(24,42){${56}$}\put(13,43){\line(1,0){10}}
\put(24,47){${78}$}\put(13,47){\line(1,0){10}}
\thinlines
\put(47,32){${2}$}\dashline[40]{2}(45,38)(45,27) 
\put(47,22){${4}$}\dashline[40]{2}(45,12)(45,23) 
\put(11,49){${8}$}\dashline[40]{2}(16,45)(16,52) 

\put(36,0){${1|13}$}\put(40,5){\line(0,1){45}}
\put(-7,38){${2|56}$}\put(5,40){\line(1,0){45}}
\end{picture}
}}
\end{equation}
Root data:
\begin{equation}
\begin{split}
  \setlength{\unitlength}{0.8mm}
\begin{picture}(50,18)(0,-3)
\put(0,5){$\alpha_0$}
\put(20,5){$\alpha_1$}
\put(5,6.3){\line(1,0){12.5}}
\put(5,5.8){\line(1,0){12.5}}
\put(4.1,5.1){$<$}
\put(16.25,5.1){$>$}
\put(30,5){$\alpha_2$}
\put(50,5){$\alpha_3$}
\put(35,6.3){\line(1,0){12.5}}
\put(35,5.8){\line(1,0){12.5}}
\put(34.1,5.1){$<$}
\put(46.25,5.1){$>$}
\end{picture}
\qquad
  \setlength{\unitlength}{1mm}
 \begin{picture}(35,20)(0,0)
\put(0,0){$\delta_1$}
\put(0,14.7){$\delta_0$}
\put(6.5,7){$\delta_2$}
\drawline(2.4,2.1)(5.8,6.0)
\drawline(2.4,12.9)(5.8,9.0)
\drawline(10.4,7.5)(14.1,7.5)
\put(15,7){$\delta_3$}
\drawline(18.9,7.5)(22.6,7.5)
\put(23.5,7){$\delta_4$}
\drawline(26.5,9.1)(29.9,13.0)
\put(29.9,14.7){$\delta_6$}
\drawline(26.5,6.0)(29.9,2.1)
\put(30,0){$\delta_5$}
\end{picture} 
\end{split}
\end{equation}
\begin{equation}
\begin{split}
&
\alpha_0=2H_1+H_2-E_3-E_4-E_5-E_6-E_7-E_8,\\
&
\alpha_1=H_2-E_1-E_2,\ 
\alpha_2=H_2-E_3-E_4,\\
&
\alpha_3=2H_1+H_2-E_1-E_2-E_5-E_6-E_7-E_8,\\
&
\delta_0=E_1-E_2,\ 
\delta_1=E_3-E_4,\ 
\delta_2=H_1-E_1-E_3,\ 
\\
&
\delta_3=H_2-E_5-E_6,\ 
\delta_4=E_6-E_7,\ 
\delta_5=E_5-E_6,\
\delta_6=E_7-E_8 ,\\
&
\pi_1=\pi_{34125678},\ 
\pi_2=\pi_{78563412}
r_{H_1-E_5-E_6}r_{H_1-E_3-E_4}.
\end{split}
\end{equation}
\begin{displaymath}
\includegraphics[scale=0.35]{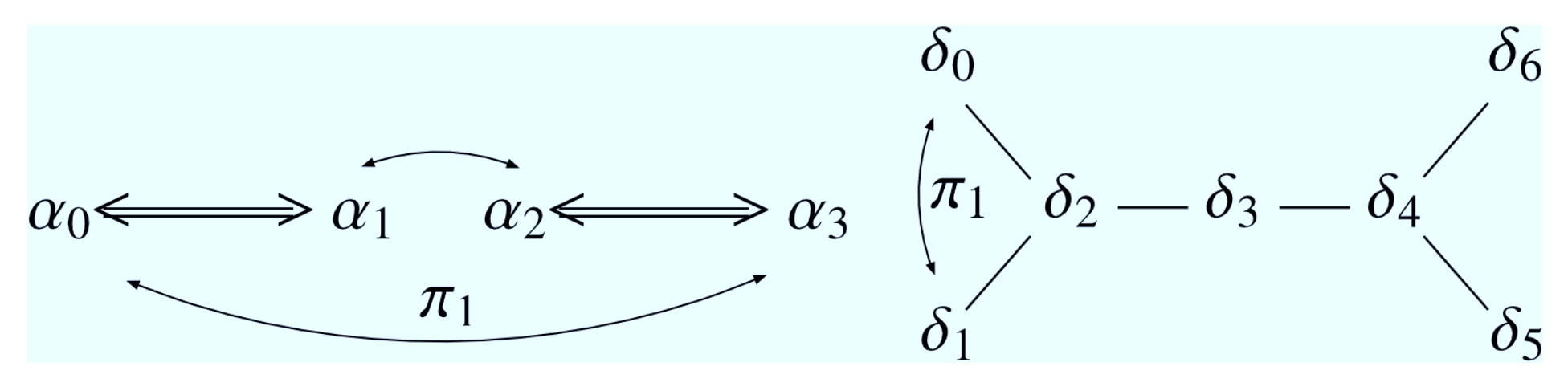}\quad
\includegraphics[scale=0.35]{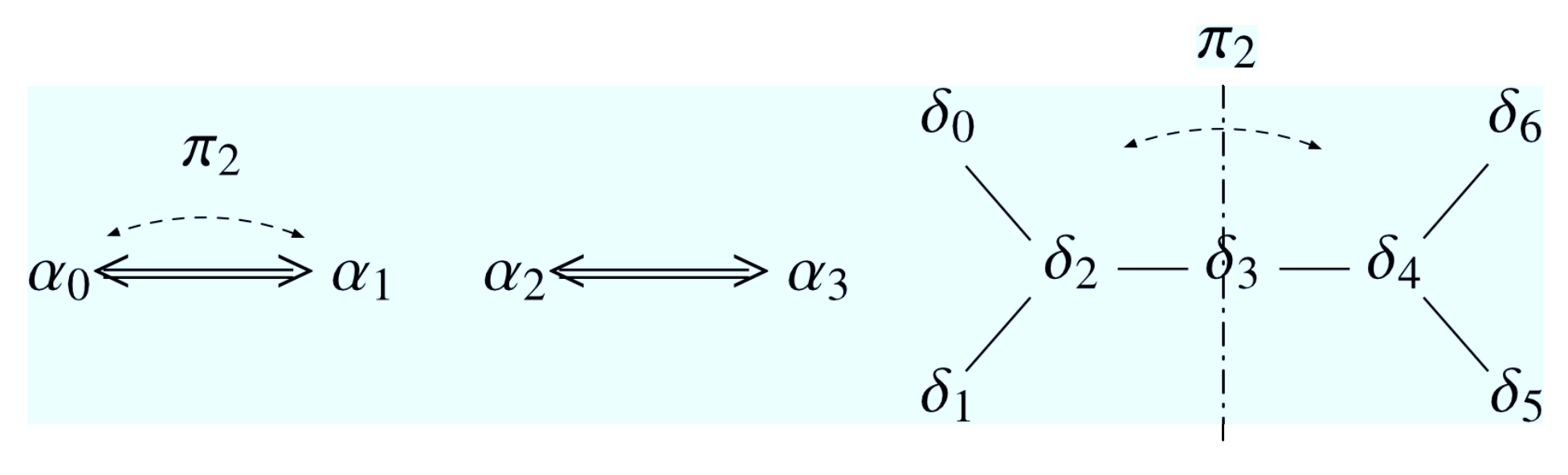}
\end{displaymath}
%
\subsubsection{d-P$(\underset{|\alpha|^2=4}{A_1^{(1)}}/D_7^{(1)})$} \label{subsubsec:conf_d-A1'}
Point configuration in $(q,p)$ coordinates:
\begin{equation}
\begin{split}
&{\rm P}_{1234}:\left(-\frac{1}{\epsilon^2},\epsilon+\frac{a_1}{2}\epsilon^2\right)_4,\ 
{\rm P}_{5678}:\left(\epsilon,-\frac{t}{\epsilon^2}+\frac{1-a_1}{\epsilon}\right)_4,\ 
\end{split}
\end{equation}
\vspace*{3mm}
\begin{equation}
{\lower30pt\hbox{\footnotesize
\begin{picture}(50,50)(0,-5)
\put(20,40){\circle{7}}\put(20,40){\circle{5}}\put(20,40){\circle{3}}\put(20,40){\circle*{1.5}}\put(24,42){$5678$}
\put(40,20){\circle{7}}\put(40,20){\circle{5}}\put(40,20){\circle{3}}\put(40,20){\circle*{1.5}}\put(42,24){$1234$}
\put(23,8){$q=\infty$}\put(2,33){$p=\infty$}
\put(5,40){\line(1,0){45}}
\put(40,5){\line(0,1){45}}
\end{picture}
\hskip40pt
\begin{picture}(50,50)(0,-5)
\thicklines
\put(35,0){${}$}\put(40,5){\line(0,1){45}}
\put(-5,40){${}$}\put(5,40){\line(1,0){45}}
\put(18,31){${}$}\put(20,35){\line(0,1){15}}

\put(33,22){${23}$}\put(35,20){\line(1,0){15}}
\put(24,42){${56}$}\put(13,43){\line(1,0){10}}
\put(24,47){${78}$}\put(13,47){\line(1,0){10}}
\put(40,24){${12}$}\put(43,12){\line(0,1){10}}
\put(48,24){${34}$}\put(47,12){\line(0,1){10}}

\thinlines
\put(11,49){${8}$}\dashline[40]{2}(16,45)(16,52) 
\put(48,11){${4}$}\dashline[40]{2}(45,16)(52,16) 
\put(36,0){${1|12}$}\put(40,5){\line(0,1){45}}
\put(-7,38){${2|56}$}\put(5,40){\line(1,0){45}}
\end{picture}
}}
\end{equation}
Root data:
\begin{equation}
\begin{split}
  \setlength{\unitlength}{0.8mm}
\begin{picture}(25,18)(0,-3)
\put(-2,5){$\alpha_0$}
\put(21,5){$\alpha_1$}
\put(5,6.3){\line(1,0){12.5}}
\put(5,5.8){\line(1,0){12.5}}
\put(4.1,5.1){$<$}
\put(16.25,5.1){$>$}
\end{picture}
\qquad
 \setlength{\unitlength}{1mm}
 \begin{picture}(35,20)(0,0)
\put(0,0){$\delta_1$}
\put(0,14.7){$\delta_0$}
\put(6.5,7){$\delta_2$}
\drawline(2.4,2.1)(5.8,6.0)
\drawline(2.4,12.9)(5.8,9.0)
\drawline(10.4,7.5)(14.1,7.5)
\put(15,7){$\delta_3$}
\drawline(18.9,7.5)(22.6,7.5)
\put(23.5,7){$\delta_4$}
\drawline(27.4,7.5)(30.1,7.5)
\put(31.5,7){$\delta_5$}
\drawline(35,9.1)(38.4,13.0)
\put(38.4,14.7){$\delta_6$}
\drawline(35,6.0)(38.4,2.1)
\put(38.5,0){$\delta_7$}
\end{picture} 
\end{split}
\end{equation}
\begin{equation}
\begin{split}
&
\alpha_0=2H_1-E_5-E_6-E_7-E_8,\ 
\alpha_1=2H_2-E_1-E_2-E_3-E_4
\\
&
\delta_0=E_1-E_2,\ 
\delta_1=E_3-E_4,\ 
\delta_2=E_2-E_3,\ 
\delta_3=H_1-E_1-E_2,\ 
\\
&
\delta_4=H_2-E_5-E_6,\ 
\delta_5=E_6-E_7,\ 
\delta_6=E_5-E_6,\
\delta_7=E_7-E_8 ,\\
&
\pi_1=\pi_{56781234}r_{H_1-H_2},\ 
\pi_2=\pi_{34125678}
r_{H_2-E_3-E_4}r_{H_2-E_1-E_2}.
\end{split}
\end{equation}
\begin{displaymath}
\includegraphics[scale=0.35]{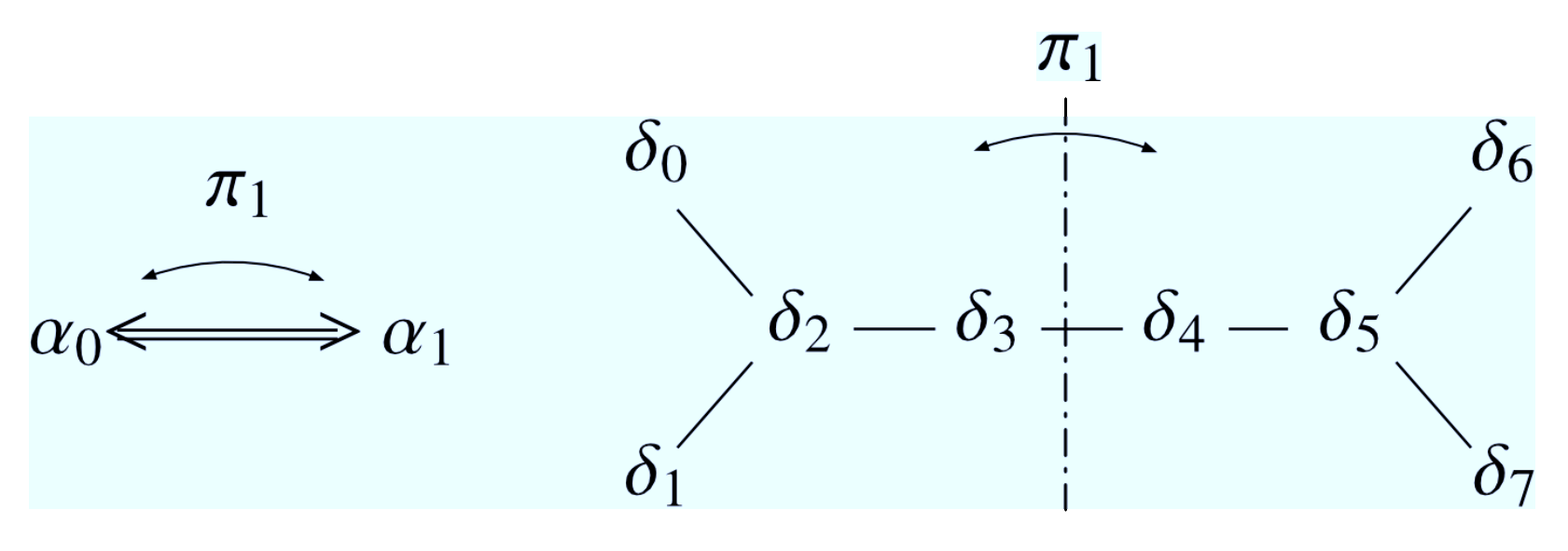}\quad
\includegraphics[scale=0.35]{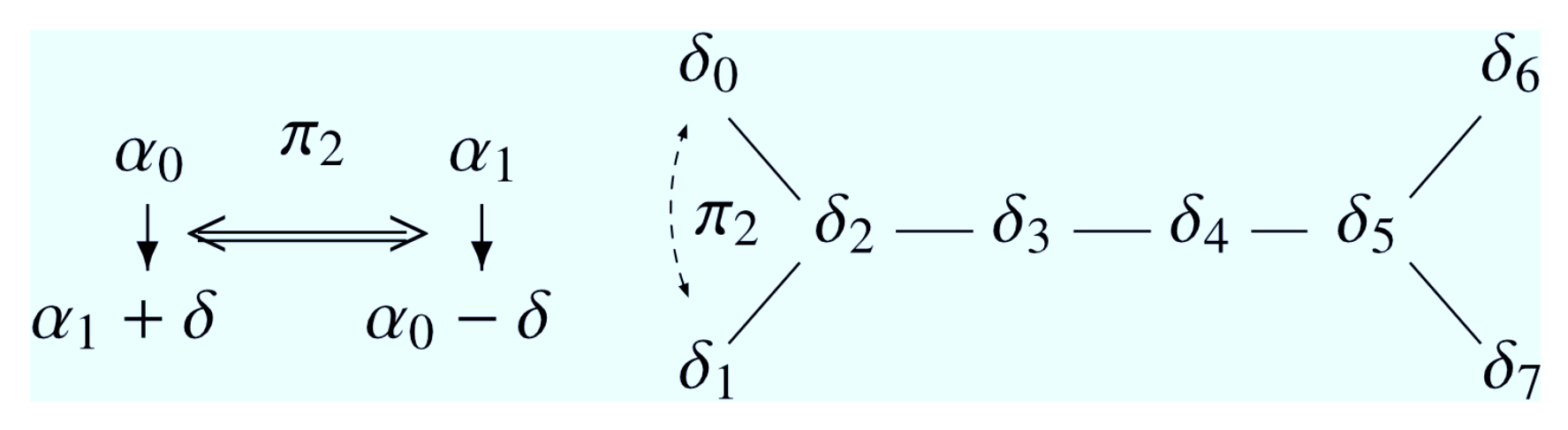}
\end{displaymath}

\subsubsection{d-P$(A_0^{(1)}/D_8^{(1)})$ }\label{subsubsec:conf_d-A0D8} 
Point configuration in $(f,g)=(q,qp)$ coordinates:
\begin{equation}
\begin{split}
&{\rm P}_{1234}:\left(-\frac{1}{\epsilon^2},-\frac{1}{\epsilon}-\frac{1}{2}\right)_4,\ 
{\rm P}_{5678}:\left(-t\epsilon^2,\frac{1}{\epsilon}\right)_4.
\end{split}
\end{equation}
\vspace*{3mm}
\begin{equation}
{\lower40pt\hbox{\footnotesize
\begin{picture}(50,50)(0,-5)
\put(36,2){$g=\infty$}\put(18,35){$f=\infty$}\put(6,2){$g=0$}
\put(10,5){\line(0,1){45}}
\put(40,5){\line(0,1){45}}
\put(0,40){\line(1,0){50}}

\put(10,40){\circle{7}}\put(10,40){\circle{5}}\put(10,40){\circle{3}}\put(10,40){\circle*{1.5}}\put(14,42){$5678$}
\put(40,40){\circle{7}}\put(40,40){\circle{5}}\put(40,40){\circle{3}}\put(40,40){\circle*{1.5}}\put(44,42){$1234$}
\end{picture}
\hskip40pt
\begin{picture}(50,50)(0,-5)
\thicklines
\put(36,0){${1|12}$}\put(40,5){\line(0,1){30}}
\put(22,34){${2|15}$}\put(15,40){\line(1,0){20}}
\put(6,0){${1|56}$}\put(10,5){\line(0,1){30}}

\put(16,25){$56$}\put(18,30){\line(0,1){15}}
\put(-5,30){$67$}\put(2,32){\line(1,0){18}}
\put(2,25){$78$}\put(4,30){\line(0,1){15}}

\put(50,30){$23$}\put(30,32){\line(1,0){18}}
\put(30,25){$12$}\put(32,30){\line(0,1){15}}
\put(44,25){$34$}\put(46,30){\line(0,1){15}}

\thinlines
\put(0,42){$8$}\dashline[40]{2}(0,40)(8,40) 
\put(48,42){$4$}\dashline[40]{2}(42,40)(50,40) 
\end{picture}
}}
\end{equation}
Root data:
\begin{equation}
\begin{split}
  \setlength{\unitlength}{0.8mm}
\begin{picture}(15,18)(0,-3)
\put(5,5){$\alpha_0$}
\end{picture}
\qquad
 \setlength{\unitlength}{1mm}
 \begin{picture}(45,20)(0,0)
\put(0,0){$\delta_1$}
\put(0,14.7){$\delta_0$}
\put(6.5,7){$\delta_2$}%
\drawline(2.4,2.1)(5.8,6.0)
\drawline(2.4,12.9)(5.8,9.0)
\drawline(10.4,7.5)(14.1,7.5)
\put(15,7){$\delta_3$}
\drawline(18.9,7.5)(22.6,7.5)
\put(23.5,7){$\delta_4$}
\drawline(27.4,7.5)(30.1,7.5)
\put(31.5,7){$\delta_5$}
\drawline(35.9,7.5)(38.6,7.5)
\put(40,7){$\delta_6$}
\drawline(43.5,9.1)(46.9,13.0)
\put(46.9,14.7){$\delta_7$}
\drawline(43.5,6.0)(46.9,2.1)
\put(47,0){$\delta_8$}
\end{picture} 
\end{split}
\end{equation}
\begin{equation}
\begin{split}
&
\alpha_0=2H_1+2H_2-E_1-E_2-E_3-E_4-
E_5-E_6-E_7-E_8
\\
&
\delta_0=H_1-E_1-E_2,\ 
\delta_1=E_3-E_4,\ 
\delta_2=E_2-E_3,\ 
\delta_3=E_1-E_2,\\
&
\delta_4=H_2-E_1-E_5,\ 
\delta_5=E_5-E_6,\ 
\delta_6=E_6-E_7,\
\delta_7=H_1-E_5-E_6,\
\delta_8=E_7-E_8 ,\\
&
\pi=\pi_{56781234}.
\end{split}
\end{equation}
\begin{displaymath}
\includegraphics[scale=0.4]{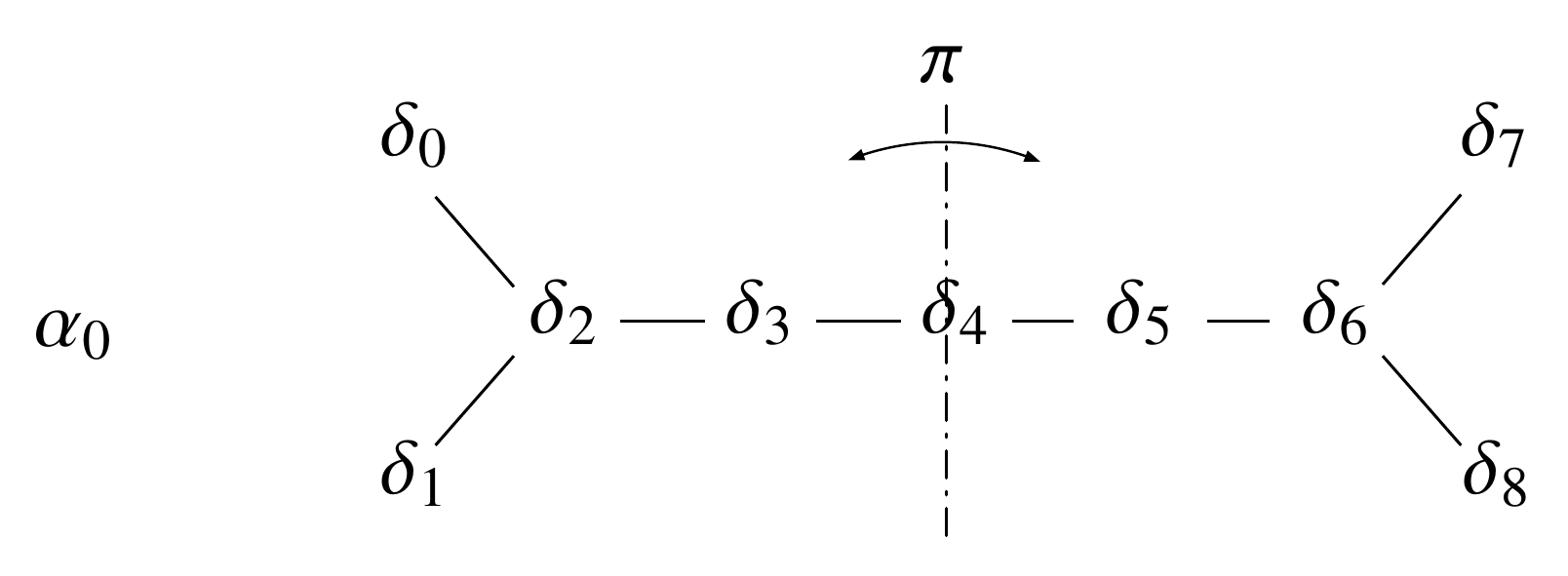}
\end{displaymath}
%
\subsubsection{d-P$(A_2^{(1)}/E_6^{(1)})$ }\label{subsubsec:conf_d-A2}
Point configuration in $(q,p)$ coordinates:
\begin{equation}
\begin{split}
&{\rm P}_{12}:\left(\frac{1}{\epsilon},-a_2\epsilon\right)_2,\ 
{\rm P}_{34}:\left(a_1\epsilon,\frac{1}{\epsilon}\right)_2,\ 
{\rm P}_{5678}: \left(\frac{1}{\epsilon}, \frac{1}{\epsilon} + t + (a_1+a_2-1)\epsilon\right)_4.
\end{split}
\end{equation}
\vspace*{5mm}
\begin{equation}
{\lower40pt\hbox{\footnotesize
\begin{picture}(50,50)(0,-5)
\put(23,8){$q=\infty$}\put(2,33){$p=\infty$}
\put(20,40){\circle{3}}\put(20,40){\circle*{1.5}}\put(18,43){$34$}
\put(40,40){\circle{7}}\put(40,40){\circle{5}}\put(40,40){\circle{3}}\put(40,40){\circle*{1.5}}\put(42,44){$5678$}
\put(40,20){\circle{3}}\put(40,20){\circle*{1.5}}\put(42,18){$12$}
\put(5,40){\line(1,0){45}}
\put(40,5){\line(0,1){45}}
\end{picture}
\hskip40pt
\begin{picture}(50,50)(0,-5)
\thicklines
\put(36,0){${1|15}$}\put(40,5){\line(0,1){30}}
\put(-7,38){${2|35}$}\put(5,40){\line(1,0){30}}
\put(18,30){${34}$}\put(20,35){\line(0,1){15}}

\put(28,18){${12}$}\put(35,20){\line(1,0){15}}
\put(44,26){${56}$}\put(28,42){\line(1,-1){15}} 
\put(25,25){${67}$}\put(30,30){\line(1,1){15}}
\put(50,30){$78$}\put(36,46){\line(1,-1){12}} 

\thinlines
\put(16,47){${4}$}\dashline[40]{2.0}(13,45)(23,45)
\put(47,15){${2}$}\dashline[40]{2}(45,12)(45,23) 
\put(46,40){${8}$}\dashline[40]{2}(43,33)(52,42) 
\end{picture}
}}
\end{equation}
Root data:
\begin{equation}
\begin{split}
 \setlength{\unitlength}{0.8mm}
\lower20pt\hbox{
\begin{picture}(27,18)(0,0)
\put(0,0){$\alpha_1$}
\put(20,0){$\alpha_2$}
\put(9,15){$\alpha_0$}
\put(5,1){\line(1,0){12.5}}
\put(19,4){\line(-2,3){6}}
\put(3,4){\line(2,3){6}}
\end{picture} 
}
\qquad
\begin{array}{c}
\delta_0\\
|\\
\delta_6\\
|\\
\delta_1\ \text{\textemdash}\ \delta_2\ \text{\textemdash}\  \delta_3\ \text{\textemdash}\  \delta_4\ \text{\textemdash}\ \delta_5
 \end{array}
\end{split}
\end{equation}
\begin{equation}
\begin{split}
&
\alpha_0=H_1+H_2-E_5-E_6-E_7-E_8,\ 
\alpha_1=H_1-E_3-E_4,\ 
\alpha_2=H_2-E_1-E_2
\\
&
\delta_0=E_7-E_8,\ 
\delta_1=E_1-E_2,\ 
\delta_2=H_1-E_1-E_5,\ 
\delta_3=E_5-E_6,\\
&
\delta_4=H_2-E_3-E_5,\ 
\delta_5=E_3-E_4,\ 
\delta_6=E_6-E_7,
\\
&
\pi_1=\pi_{34125678}r_{H_1-H_2},\ 
\pi_2=\pi_{78345612}r_{H_2-E_5-E_6}.
\end{split}
\end{equation}
\begin{displaymath}
\includegraphics[scale=0.4]{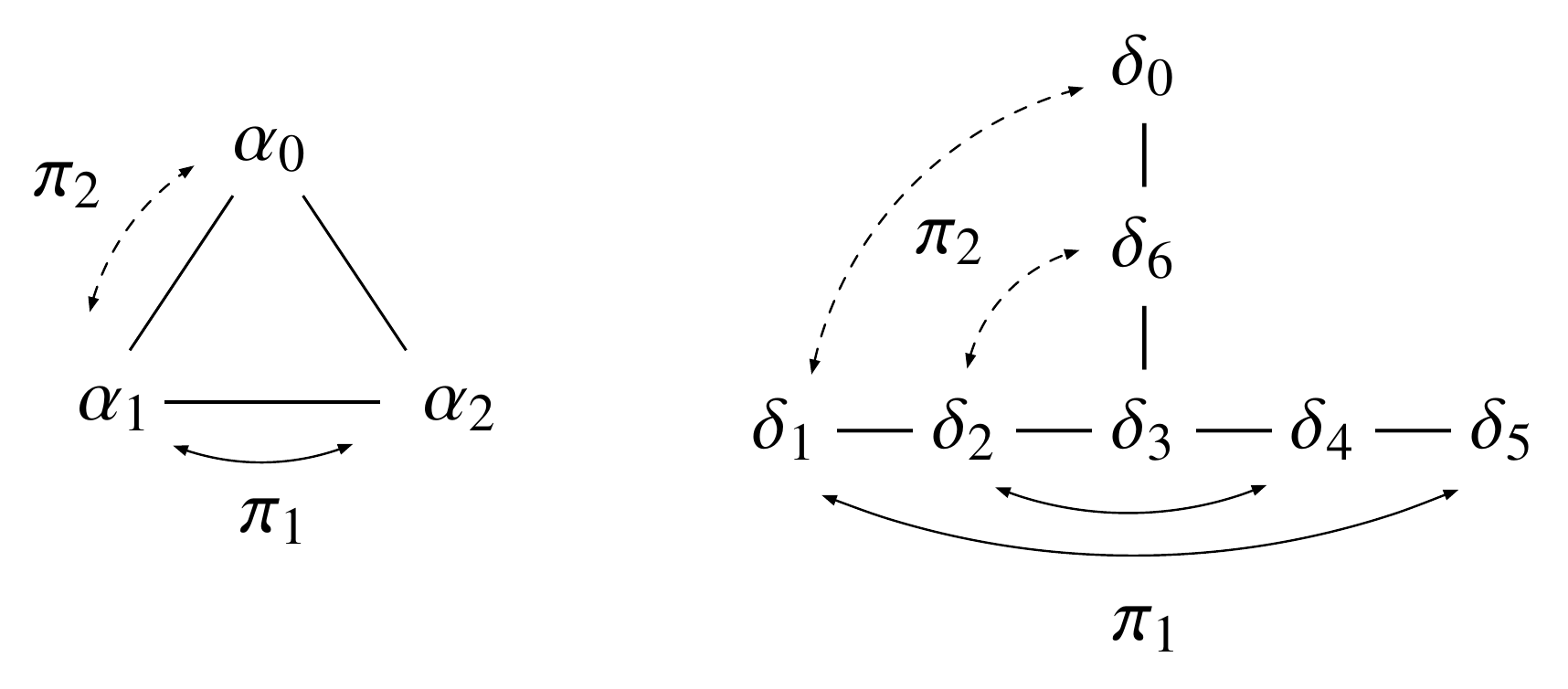}
\end{displaymath}
%

\subsubsection{d-P$(A_1^{(1)}/E_7^{(1)})$} \label{subsubsec:conf_d-A1}
Point configuration in $(q,p)$ coordinates:
\begin{equation}
\begin{split}
&{\rm P}_{12}: \left(\frac{1}{\epsilon},-a_1\epsilon\right)_2,\ 
{\rm P}_{345678}: \left(\frac{1}{\epsilon},\frac{2}{\epsilon^2}+t+(a_1-1)\epsilon\right)_6.
\end{split}
\end{equation}
\vspace*{3mm}
\begin{equation}
{\lower40pt\hbox{\footnotesize
\begin{picture}(50,50)(0,0)
\put(42,45){$345678$}
\put(40,40){\circle{9}}\put(40,40){\circle{7.5}}\put(40,40){\circle{6}}\put(40,40){\circle{4.5}}\put(40,40){\circle{3}}\put(40,40){\circle*{1.5}}
\put(40,20){\circle{3}}\put(40,20){\circle*{1.5}}\put(42,18){$12$}
\put(23,8){$q=\infty$}\put(2,33){$p=\infty$}
\put(5,40){\line(1,0){45}}
\put(40,5){\line(0,1){45}}
\end{picture}
\hskip40pt 
\begin{picture}(50,50)(0,0)
\thicklines
\put(-7,38){${2|34}$}\put(5,40){\line(1,0){30}}
\put(36,0){${1|13}$}\put(40,5){\line(0,1){30}}
\put(25,21){$45$}\put(30,27){\line(0,1){25}}
\put(44,28.5){${34}$}\put(27,30){\line(1,0){15}}
\put(30,14){${12}$}\put(35,20){\line(1,0){15}}
\put(19,43){$56$}\put(27,45){\line(1,0){15}}
\put(39,38){$67$}\put(38,42){\line(0,1){10}}
\put(31,51){$78$}\put(35,49){\line(1,0){10}}

\thinlines
\put(47,14){${2}$}\dashline[40]{2}(45,12)(45,23) 
\put(45,44){${8}$}\dashline[40]{2}(43,46)(43,52) 
\end{picture}
}}
\end{equation}
Root data:
\begin{equation}
\setlength{\unitlength}{0.8mm}
\begin{picture}(27,18)(0,7)
\put(3,5){$\alpha_0$}
\put(26,5){$\alpha_1$}
\put(10,6.3){\line(1,0){12.5}}
\put(10,5.8){\line(1,0){12.5}}
\put(9.4,5.1){$<$}
\put(21.3,5.1){$>$}
\end{picture} 
\qquad
  \begin{array}{c}
 \delta_0\\
|\\
\delta_1\ \text{\textemdash}\ \delta_2\ \text{\textemdash}\  \delta_3\ \text{\textemdash}\  \delta_4\ \text{\textemdash}\ \delta_5
\text{\textemdash}\ \delta_6\text{\textemdash}\ \delta_7
 \end{array}
\end{equation}
\begin{equation}
\begin{split}
&
\alpha_0=2H_1+H_2-E_3-E_4-E_5-E_6-E_7-E_8,\ 
\alpha_1=H_2-E_1-E_2
\\
&
\delta_0=H_2-E_3-E_4,\ 
\delta_1=E_1-E_2,\ 
\delta_2=H_1-E_1-E_3,\ 
\delta_3=E_3-E_4,\ 
\\
&
\delta_4=E_4-E_5,\ 
\delta_5=E_5-E_6,\ 
\delta_6=E_6-E_7,\ 
\delta_7=E_7-E_8,\\
&
\pi=\pi_{78563412}
r_{H_1-E_5-E_6}r_{H_1-E_3-E_4}.
\end{split}
\end{equation}
\begin{displaymath}
\includegraphics[scale=0.4]{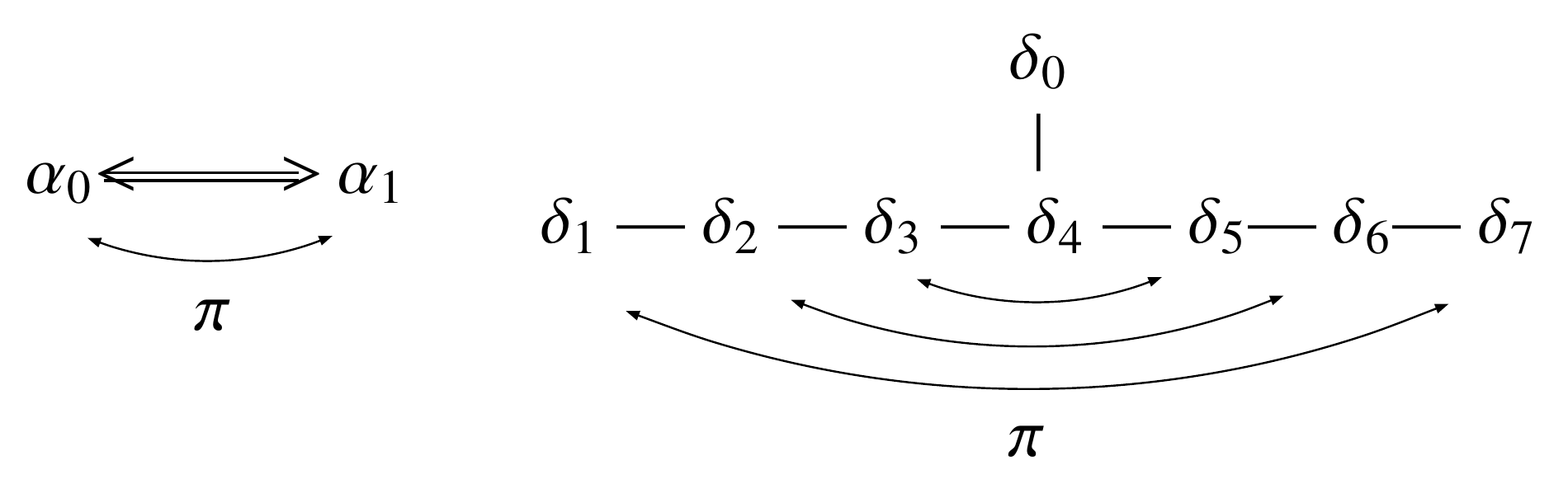}
\end{displaymath}

\subsubsection{d-P($A_0^{(1)}/E_8^{(1)}$)}\label{subsubsec:conf_d-A0E8}\quad 
This case cannot be realized by the configuration of eight points on
$\mathbb{P}^1\times\mathbb{P}^1$. It is necessary to consider the configuration of nine points on
$\mathbb{P}^2$ \cite{KMNOY:point_configuration,Sakai:SIV}.\\
Root data:
\begin{equation}
\lower5pt\hbox{\mbox{$\alpha_0$}}\qquad
  \begin{array}{c}
\hskip-78pt \delta_0\\
\hskip-78pt |\\
\delta_1\ \text{\textemdash}\ \delta_2\ \text{\textemdash}\  \delta_3\ \text{\textemdash}\  \delta_4\ \text{\textemdash}\ \delta_5
\text{\textemdash}\ \delta_6\text{\textemdash}\ \delta_7\text{\textemdash}\ \delta_8
 \end{array}
\end{equation}
\begin{equation}
\begin{split}
&
\alpha_0=2H_1+2H_2-E_1-E_2-E_3-E_4-E_5-E_6-E_7-E_8
\\
&
\delta_0=E_1-E_2,\ 
\delta_1=H_1-H_2,\ 
\delta_2=H_2-E_1-E_2,\ 
\delta_3=E_2-E_3,\ 
\\
&
\delta_4=E_3-E_4,\ 
\delta_5=E_4-E_5,\ 
\delta_6=E_5-E_6,\ 
\delta_7=E_6-E_7,\ 
\delta_8=E_7-E_8.
\end{split}
\end{equation}

\subsection{Degeneration of point configurations}\label{subsec:degeneration}
In this subsection, we describe the procedure of degeneration of the point configurations and the
corresponding discrete Painlev\'e equations given in Section \ref{subsec:configuration}.

We first show the multiplicative cases according to the following diagram.
%
\begingroup \renewcommand{\arraycolsep}{3pt}
\begin{equation}\label{eqn:degeneration_of_qP}
\begin{array}{cccccccccccccccccccccc}
E_8^{(1)}&\rightarrow &E_7^{(1)} &\rightarrow &E_6^{(1)} &\rightarrow 
&D_5^{(1)} &\overset{2/3}{\longrightarrow} &A_{4}^{(1)}
&\overset{6/7}{\longrightarrow}&\esan^{(1)}(b)
&\overset{5/7}{\longrightarrow}&\eni^{(1)}(b)
&\overset{1/3}{\longrightarrow}&A_{1}^{(1)}\\
& & & & & &&&&\overset{8/1}{\searrow}&&
\overset{8/1}{\searrow}&&\overset{8/1}{\searrow}&&\\
& & & & & &&&&&\esan^{(1)}(a)
&\overset{6/7}{\longrightarrow}&\eni^{(1)}(a)
&\overset{5/7}{\longrightarrow}&\underset{|\alpha|^2=8}{A_{1}^{(1)}}&
\end{array}
\end{equation}
\endgroup
%
In each case, introducing a small parameter $\varepsilon$, change the variables $f$, $g$, $\kappa_i$
$(i=1,2)$ and $v_i$ $(i=1,\ldots,8)$ as indicated below and then take the limit $\varepsilon\to 0$
to obtain the lower case. 
%
\subsubsection{$q$-P$(E_8^{(1)}/A_0^{(1)})$ $\rightarrow$ $q$-P$(E_7^{(1)}/A_1^{(1)})$}
\begin{equation}
\kappa_i \to \kappa_i \varepsilon \ (i=1,2), \quad v_i \to v_i \varepsilon \ (i=5,...,8),\quad g \to \frac{1}{g}.
\end{equation}
\subsubsection{$q$-P$(E_7^{(1)}/A_1^{(1)})$ $\rightarrow$ $q$-P$(E_6^{(1)}/A_2^{(1)})$}\label{subsec:degeneration_qe7_to_qe6}
\begin{equation}
v_i  \to v_i \varepsilon \ (i=7,8),\quad \kappa_1 \to \kappa_1 \varepsilon.
\end{equation}
\subsubsection{$q$-P$(E_6^{(1)}/A_2^{(1)})$ $\rightarrow$ $q$-P$(D_5^{(1)}/A_3^{(1)})$}\label{subsec:degeneration_qe6_to_qd5}
\begin{equation}
\begin{split}
& v_i \to \frac{v_i}{\varepsilon} \ (i=1,2),\quad 
v_i \to \varepsilon v_i \ (i=3,4),\quad 
\kappa_1 \to \kappa_1 \varepsilon,\\
& \kappa_2 \to \frac{\kappa_2}{\varepsilon},\quad
f \to f\varepsilon,\quad g \to g\varepsilon .  
\end{split}
\end{equation}

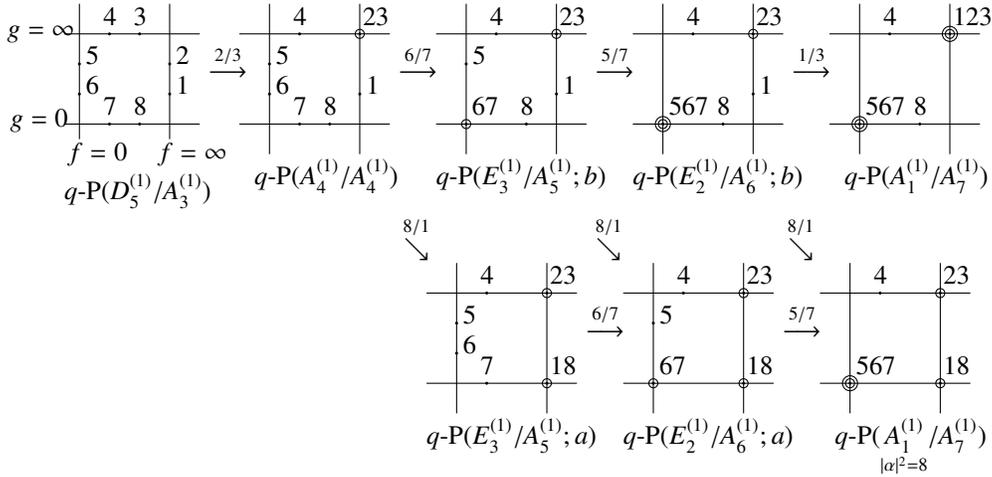
\begin{figure}[th]
\begin{center}
\setlength{\unitlength}{0.4mm}
\footnotesize{
\begin{picture}(50,50)(0,0)
\put(-13,9){$g=0$}
\put(-14,39){$g=\infty$}
\put(6,-2){$f=0$}
\put(36,-2){$f=\infty$}
\put(5,10){\line(1,0){45}}
\put(5,40){\line(1,0){45}}
\put(10,5){\line(0,1){45}}
\put(40,5){\line(0,1){45}}
\put(20,10){\circle*{1}}\put(30,10){\circle*{1}}
\put(18,13){$7$} \put(28,13){$8$}
\put(20,40){\circle*{1}}\put(30,40){\circle*{1}}
\put(18,43){$4$} \put(28,43){$3$}
\put(10,20){\circle*{1} $6$}\put(10,30){\circle*{1} $5$}
\put(40,20){\circle*{1} $1$}\put(40,30){\circle*{1} $2$}
\put(53,25){$\overset{2/3}{\longrightarrow}$}
\put(5,-15){$q$-P$(D_5^{(1)}/A_3^{(1)})$}
\end{picture}}
\quad
\footnotesize{
\begin{picture}(50,50)(0,0)
\put(0,10){\line(1,0){50}}
\put(0,40){\line(1,0){50}}
\put(10,0){\line(0,1){50}}
\put(40,0){\line(0,1){50}}
\put(20,10){\circle*{1}}\put(30,10){\circle*{1}}
\put(18,13){$7$} \put(28,13){$8$}
\put(20,40){\circle*{1}}\put(40,40){\circle{3}}
\put(18,43){$4$}\put(41,43){$23$}
\put(10,20){\circle*{1} $6$}\put(10,30){\circle*{1} $5$}
\put(40,20){\circle*{1} $1$}\put(40,40){\circle*{1} ${}$}
\put(53,25){$\overset{6/7}{\longrightarrow}$}
\put(5,-10){$q$-P$(A_4^{(1)}/A_4^{(1)})$}
\end{picture}
}
\quad
\footnotesize{
\begin{picture}(50,50)(0,0)
\put(0,10){\line(1,0){50}}
\put(0,40){\line(1,0){50}}
\put(10,0){\line(0,1){50}}
\put(40,0){\line(0,1){50}}
\put(10,10){\circle{3}}\put(30,10){\circle*{1}}
\put(12,13){$67$}\put(28,13){$8$}
\put(20,40){\circle*{1}}\put(40,40){\circle{3}}
\put(18,43){$4$}\put(41,43){$23$}
\put(10,10){\circle*{1} ${}$}\put(10,30){\circle*{1} $5$}
\put(40,20){\circle*{1} $1$}\put(40,40){\circle*{1} ${}$}
\put(53,25){$\overset{5/7}{\longrightarrow}$}
\put(0,-10){$q$-P$(E_3^{(1)}/A_5^{(1)};b)$}
\end{picture}
}
\quad
\footnotesize{
\begin{picture}(50,50)(0,0)
\put(0,10){\line(1,0){50}}
\put(0,40){\line(1,0){50}}
\put(10,0){\line(0,1){50}}
\put(40,0){\line(0,1){50}}
\put(10,10){\circle{3}}\put(30,10){\circle*{1}}
\put(12,13){$567$}\put(28,13){$8$}
\put(20,40){\circle*{1}}\put(40,40){\circle{3}}
\put(18,43){$4$}\put(41,43){$23$}
\put(10,10){\circle*{1} ${}$}\put(10,10){\circle{5} ${}$}
\put(40,20){\circle*{1} $1$}\put(40,40){\circle*{1} ${}$}
\put(53,25){$\overset{1/3}{\longrightarrow}$}
\put(0,-10){$q$-P$(E_2^{(1)}/A_6^{(1)};b)$}
\end{picture}
}
\quad
\footnotesize{
\begin{picture}(50,50)(0,0)
\put(0,10){\line(1,0){50}}
\put(0,40){\line(1,0){50}}
\put(10,0){\line(0,1){50}}
\put(40,0){\line(0,1){50}}
\put(10,10){\circle{3}}\put(30,10){\circle*{1}}
\put(12,13){$567$}\put(28,13){$8$}
\put(20,40){\circle*{1}}\put(40,40){\circle{3}}
\put(18,43){$4$}\put(41,43){$123$}
\put(10,10){\circle*{1} ${}$}\put(10,10){\circle{5} ${}$}
\put(40,40){\circle{5} ${}$}\put(40,40){\circle*{1} ${}$}
\put(5,-10){$q$-P$(A_1^{(1)}/A_7^{(1)})$}
\end{picture}
}\\[8mm]
\hskip25mm $\overset{8/1}{\searrow}$\hskip22mm
$\overset{8/1}{\searrow}$\hskip22mm$\overset{8/1}{\searrow}$
\\
\hskip48mm
\footnotesize{
\begin{picture}(50,50)(0,0)
\put(0,10){\line(1,0){50}}
\put(0,40){\line(1,0){50}}
\put(10,0){\line(0,1){50}}
\put(40,0){\line(0,1){50}}
\put(20,10){\circle*{1}}\put(40,10){\circle*{1}}
\put(18,13){$7$}\put(41,13){$18$}
\put(20,40){\circle*{1}}\put(40,40){\circle{3}}
\put(18,43){$4$}\put(41,43){$23$}
\put(10,20){\circle*{1} $6$}\put(10,30){\circle*{1} $5$}
\put(40,10){\circle{3} ${}$}\put(40,40){\circle*{1} ${}$}
\put(53,25){$\overset{6/7}{\longrightarrow}$}
\put(0,-10){$q$-P$(E_3^{(1)}/A_5^{(1)};a)$}
\end{picture}
}
\quad
\footnotesize{
\begin{picture}(50,50)(0,0)
\put(0,10){\line(1,0){50}}
\put(0,40){\line(1,0){50}}
\put(10,0){\line(0,1){50}}
\put(40,0){\line(0,1){50}}
\put(10,10){\circle{3}}\put(40,10){\circle*{1}}
\put(12,13){$67$}\put(41,13){$18$}
\put(20,40){\circle*{1}}\put(40,40){\circle{3}}
\put(18,43){$4$}\put(41,43){$23$}
\put(10,10){\circle*{1} ${}$}\put(10,30){\circle*{1} $5$}
\put(40,10){\circle{3} ${}$}\put(40,40){\circle*{1} ${}$}
\put(53,25){$\overset{5/7}{\longrightarrow}$}
\put(0,-10){$q$-P$(E_2^{(1)}/A_6^{(1)};a)$}
\end{picture}
}
\quad
\footnotesize{
\begin{picture}(50,50)(0,0)
\put(0,10){\line(1,0){50}}
\put(0,40){\line(1,0){50}}
\put(10,0){\line(0,1){50}}
\put(40,0){\line(0,1){50}}
\put(10,10){\circle{3}}\put(40,10){\circle*{1}}
\put(12,13){$567$}\put(41,13){$18$}
\put(20,40){\circle*{1}}\put(40,40){\circle{3}}
\put(18,43){$4$}\put(41,43){$23$}
\put(10,10){\circle*{1} ${}$}\put(10,10){\circle{5} ${}$}
\put(40,10){\circle{3} ${}$}\put(40,40){\circle*{1} ${}$}
\put(5,-10){$q$-P$(\underset{|\alpha|^2=8}{A_1^{(1)}}/A_7^{(1)})$}
\end{picture}}\\[5mm]
\caption{Degenerations of the point configurations from the case of $q$-P$(D_5^{(1)}/A_3^{(1)})$ case.}
\label{fig:degenerateD5A4}
\end{center}
\end{figure}

The degeneration from $q$-P$(D_5^{(1)}/A_3^{(1)})$ can be carried out by the simple substitution
\begin{equation}\label{eqn:degeneration_points_1}
v_i \to v_i \varepsilon,\quad v_j \to\frac{v_j}{\varepsilon},
\end{equation}
and the limit $\varepsilon\to 0$ which is denoted by $\overset{i/j}{\longrightarrow}$ as shown in
\eqref{eqn:degeneration_of_qP} (see also Fig.\ref{fig:degenerateD5A4}). 

%
\begin{rem}\rm
Here we remark on the relation between infinitely near points and the multiple point
\cite{KMNOY:cubic_pencil}.  For example, consider the degeneration of ${\rm P}_a:\Bigl(0,a
\varepsilon\Bigr)$ and ${\rm P}_b:\Bigl(b\varepsilon,0\Bigr)$ by the limit $\varepsilon\to 0$, as in
the degeneration of ${\rm P}_6$ and ${\rm P}_7$. The line connecting P$_a$ and P$_b$ yields
$\frac{g}{f}=-\frac{a}{b}$ in the limit $\varepsilon\to 0$. Therefore, it passes through the double
point ${\rm P}_{ab}:\Bigl(-\frac{b}{a}\epsilon,\epsilon\Bigr)_2$ in the sense of Section
\ref{subsec:configuration}.  More generally, the condition for a curve $F(f,g)=0$ to pass through
the two points ${\rm P}_a$ and ${\rm P}_b$ coincides in the limit $\varepsilon\to 0$ with the
condition that the curve passes through the double point ${\rm P}_{ab}$. This condition is written
as $F=0$ and $b\frac{\partial F}{\partial f}=a\frac{\partial F}{\partial g}$ at $(f,g)=(0,0)$.  One
can consider the degeneration of several points to a multiple point in a similar manner.
\end{rem}

\subsubsection{$q$-P($D_5^{(1)}/A_3^{(1)}$) $\rightarrow$ d-P($D_4^{(1)}/D_4^{(1)}$)}
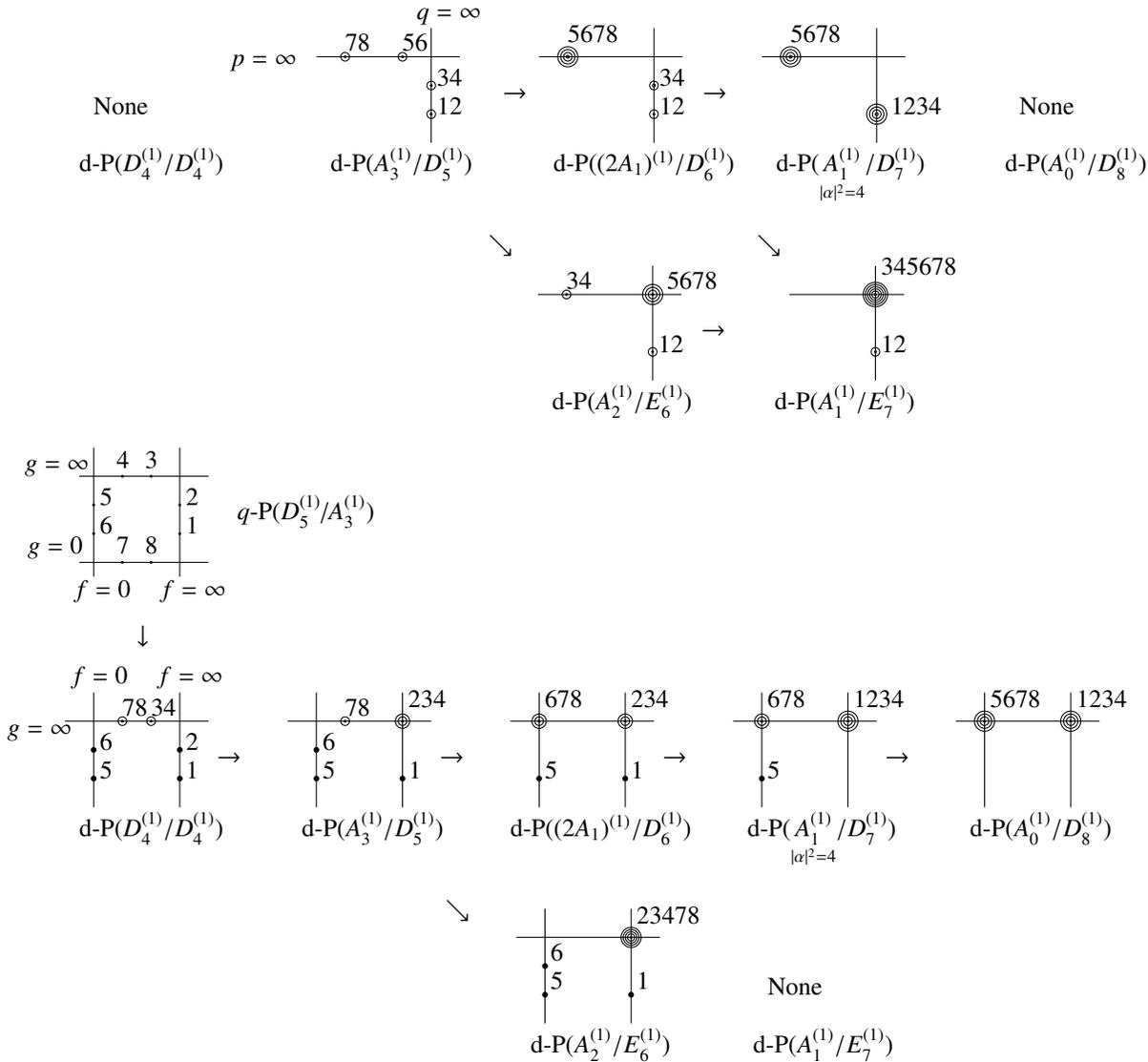
\begin{figure}[H]
{\setlength{\unitlength}{0.4mm}
\footnotesize{
\begin{picture}(70,70)(0,-5)
\put(10,20){None}
\put(5,0){d-P$(D_4^{(1)}/D_4^{(1)})$}
\end{picture}\qquad
\begin{picture}(60,50)(0,-5)
\put(-30,37){$p=\infty$}
\put(35,53){$q=\infty$}
\put(10,40){\circle*{1}}\put(10,40){\circle{3}}\put(10,42){$78$}
\put(30,40){\circle*{1}}\put(30,40){\circle{3}}\put(30,42){$56$}
\put(40,30){\circle*{1}}\put(40,30){\circle{3}}\put(42,30){$34$}
\put(40,20){\circle*{1}}\put(40,20){\circle{3}}\put(42,20){$12$}
\put(65,25){${\rightarrow}$}
\put(0,40){\line(1,0){50}}
\put(40,10){\line(0,1){40}}
\put(5,0){d-P$(A_3^{(1)}/D_5^{(1)})$}
\end{picture}\qquad
\begin{picture}(60,50)(0,-5)
\put(10,40){\circle{1}}\put(10,40){\circle{3}}\put(10,40){\circle{5}}\put(10,40){\circle{7}}\put(10,45){$5678$}
\put(40,30){\circle{1}}\put(40,30){\circle{3}}\put(42,30){$34$}
\put(40,20){\circle*{1}}\put(40,20){\circle{3}}\put(42,20){$12$}
\put(57,25){${\rightarrow}$}
\put(0,40){\line(1,0){50}}
\put(40,10){\line(0,1){40}}
\put(5,0){d-P$((2A_1)^{(1)}/D_6^{(1)})$}
\end{picture}\qquad
\begin{picture}(60,50)(0,-5)
\put(10,40){\circle{1}}\put(10,40){\circle{3}}\put(10,40){\circle{5}}\put(10,40){\circle{7}}\put(10,45){$5678$}
\put(40,20){\circle{1}}\put(40,20){\circle{3}}\put(40,20){\circle{5}}\put(40,20){\circle{7}}\put(45,20){$1234$}
\put(0,40){\line(1,0){50}}
\put(40,10){\line(0,1){40}}
\put(5,0){d-P$(\underset{|\alpha|^2=4}{A_1^{(1)}}/D_7^{(1)})$}
\end{picture}
\qquad
\begin{picture}(60,50)(0,-5)
\put(10,20){None}
\put(5,0){d-P$(A_0^{(1)}/D_8^{(1)})$}
\end{picture}\\[5mm]
\hspace*{59mm}${\searrow}$\hskip34mm${\searrow}$\\[3mm]
\hspace*{65mm}
\begin{picture}(60,50)(0,-5)
\put(10,40){\circle*{1}}\put(10,40){\circle{3}}\put(10,42){$34$}
\put(40,40){\circle*{1}}\put(40,40){\circle{3}}\put(40,40){\circle{5}}\put(40,40){\circle{7}}\put(45,42){$5678$}
\put(40,20){\circle*{1}}\put(40,20){\circle{3}}\put(42,20){$12$}
\put(57,25){${\rightarrow}$}
\put(0,40){\line(1,0){50}}
\put(40,10){\line(0,1){40}}
\put(5,0){d-P$(A_2^{(1)}/E_6^{(1)})$}
\end{picture}\qquad
\begin{picture}(60,50)(0,-5)
\put(40,40){\circle*{1}}\put(40,40){\circle{2.5}}\put(40,40){\circle{4}}\put(40,40){\circle{5.5}}\put(40,40){\circle{7}}\put(40,40){\circle{8.5}}\put(42,47){$345678$}
\put(40,20){\circle*{1}}\put(40,20){\circle{3}}\put(42,20){$12$}
\put(10,40){\line(1,0){40}}
\put(40,10){\line(0,1){40}}
\put(5,0){d-P$(A_1^{(1)}/E_7^{(1)})$}
\end{picture}
\vskip3mm
\begin{picture}(50,50)(0,0)
\put(-13,13){{\footnotesize $g=0$}}
\put(-14,42){{\footnotesize $g=\infty$}}
\put(3,-2){{\footnotesize $f=0$}}
\put(33,-2){{\footnotesize $f=\infty$}}
\put(5,10){\line(1,0){45}}
\put(5,40){\line(1,0){45}}
\put(10,5){\line(0,1){45}}
\put(40,5){\line(0,1){45}}
\put(20,10){\circle*{1}}\put(30,10){\circle*{1}}
\put(18,13){$7$}\put(28,13){$8$}
\put(20,40){\circle*{1}}\put(30,40){\circle*{1}}
\put(18,43){$4$}\put(28,43){$3$}
\put(10,20){\circle*{1} $6$}\put(10,30){\circle*{1} $5$}
\put(40,20){\circle*{1} $1$}\put(40,30){\circle*{1} $2$}
\put(60,25){$q$-P($D_5^{(1)}$/$A_3^{(1)}$)}
\end{picture}\hfill\\[3mm]
\hspace*{10mm}$\downarrow$\\[8mm]
\begin{picture}(60,50)(0,-5)
\put(-20,35){{\footnotesize $g=\infty$}}
\put(2,53){{\footnotesize $f=0$}}
\put(32,53){{\footnotesize $f=\infty$}}
\put(10,10){\line(0,1){40}}
\put(40,10){\line(0,1){40}}
\put(20,40){\circle*{1}}\put(20,40){\circle{3}}\put(20,42){$78$}
\put(30,40){\circle*{1}}\put(30,40){\circle{3}}\put(30,42){$34$}
\put(10,20){\circle*{2} $5$}\put(10,30){\circle*{2} $6$}
\put(40,20){\circle*{2} $1$}\put(40,30){\circle*{2} $2$}
\put(53,25){${\rightarrow}$}
\put(0,40){\line(1,0){50}}
\put(5,0){d-P$(D_4^{(1)}/D_4^{(1)})$}
\end{picture}\qquad
\begin{picture}(60,50)(0,-5)
\put(20,40){\circle*{1}}\put(20,40){\circle{3}}\put(20,42){$78$}
\put(40,40){\circle*{1}}\put(40,40){\circle{3}}\put(40,40){\circle{5}}\put(42,45){$234$}
\put(40,20){\circle*{2} $1$}
\put(10,20){\circle*{2} $5$}\put(10,30){\circle*{2} $6$}
\put(53,25){${\rightarrow}$}
\put(0,40){\line(1,0){50}}
\put(40,10){\line(0,1){40}}
\put(10,10){\line(0,1){40}}
\put(5,0){d-P$(A_3^{(1)}/D_5^{(1)})$}
\end{picture}\qquad
\begin{picture}(60,50)(0,-5)
\put(10,40){\circle*{1}}\put(10,40){\circle{3}}\put(10,40){\circle{5}}\put(12,45){$678$}
\put(40,40){\circle*{1}}\put(40,40){\circle{3}}\put(40,40){\circle{5}}\put(42,45){$234$}
\put(10,20){\circle*{2}}\put(12,20){$5$}
\put(40,20){\circle*{2}}\put(42,20){$1$}
\put(53,25){${\rightarrow}$}
\put(0,40){\line(1,0){50}}
\put(10,10){\line(0,1){40}}
\put(40,10){\line(0,1){40}}
\put(0,0){d-P$((2A_1)^{(1)}/D_6^{(1)})$}
\end{picture}\qquad
\begin{picture}(60,50)(0,-5)
\put(10,40){\circle*{1}}\put(10,40){\circle{3}}\put(10,40){\circle{5}}\put(12,45){$678$}
\put(40,40){\circle*{1}}\put(40,40){\circle{3}}\put(40,40){\circle{5}}\put(40,40){\circle{7}}\put(42,45){$1234$}
\put(10,20){\circle*{2}}\put(12,20){$5$}
\put(53,25){${\rightarrow}$}
\put(0,40){\line(1,0){50}}
\put(10,10){\line(0,1){40}}
\put(40,10){\line(0,1){40}}
\put(5,0){d-P$(\underset{|\alpha|^2=4}{A_1^{(1)}}/D_7^{(1)})$}
\end{picture}\qquad
\begin{picture}(60,50)(0,-5)
\put(10,40){\circle*{1}}\put(10,40){\circle{3}}\put(10,40){\circle{5}}\put(10,40){\circle{7}}\put(12,45){$5678$}
\put(40,40){\circle*{1}}\put(40,40){\circle{3}}\put(40,40){\circle{5}}\put(40,40){\circle{7}}\put(42,45){$1234$}
\put(0,40){\line(1,0){50}}
\put(10,10){\line(0,1){40}}
\put(40,10){\line(0,1){40}}
\put(5,0){d-P($A_0^{(1)}/D_8^{(1)})$}
\end{picture}
\\[5mm]
\hspace*{53mm}${\searrow}$
\\
\hspace*{62mm}
\begin{picture}(60,50)(0,-5)
\put(10,30){\circle*{2}}\put(12,32){$6$}
\put(10,20){\circle*{2}}\put(12,22){$5$}
\put(40,40){\circle*{1}}\put(40,40){\circle{2.5}}\put(40,40){\circle{4}}\put(40,40){\circle{5.5}}\put(40,40){\circle{7}}\put(42,45){$23478$}
\put(40,20){\circle*{2}}\put(42,22){$1$}
\put(0,40){\line(1,0){50}}
\put(10,10){\line(0,1){40}}
\put(40,10){\line(0,1){40}}
\put(5,0){d-P$(A_2^{(1)}/E_6^{(1)})$}
\end{picture}\qquad
\begin{picture}(50,50)(0,-5)
\put(10,20){None}
\put(5,0){d-P$(A_1^{(1)}/E_7^{(1)})$}
\end{picture}}
\caption{Degeneration of point configurations of the additive types in $\mathbb{P}^1\times
\mathbb{P}^1$ from $q$-$(D_5^{(1)}/A_3^{(1)})$ case. Top: $(q,p)$ coordinates. Bottom: $(f,g)$
coordinates. Multi-indices $i_1i_2\cdots i_k$ represent the multiple points.  ``None'' means that
the surface cannot be realized by eight point configuration on a $(2,2)$-curve in the corresponding
coordinates.  }\label{fig:degeneration_additive}}
\end{figure}

For the cases admitting the Painlev\'e differential equations as continuous flows, we have the
graphical diagram of degeneration from $q$-P$(D_5^{(1)}/A_3^{(1)})$ as shown in Figure
\ref{fig:degeneration_additive}.  We use below the root parameters as in Section \ref{subsec:dPs} 
for describing the point configurations. In the case of d-P($D_4^{(1)}/D_4^{(1)}$), the
correspondence of the parameters is given by (see \eqref{eqn:root_data_D4/A4})
\begin{equation}\label{eqn:par_correspondence_D4}
\begin{split}
& a_0 = \kappa_1-v_3-v_4,\quad a_1 = v_1-v_2,\\
& a_2 = \kappa_2 - v_1-v_5,\quad a_3 = \kappa_1-v_7-v_8,\quad a_4 = v_5-v_6.
\end{split}
\end{equation}
We will explain the degeneration limit to some additive cases in Figure
\ref{fig:degeneration_additive} starting from $q$-P($D_5^{(1)}/A_3^{(1)}$).  We only show the
degenerations relevant to the hypergeometric solutions in Section \ref{subsec:hyper_data}.

We set  
\begin{equation}\label{eqn:limit_qD5_to_dD4}
\begin{array}{c}\medskip
{\displaystyle  q\to e^\varepsilon,\quad g \to e^{-\varepsilon (g+\kappa_2-v_5)},\quad \kappa_i \to e^{\varepsilon \kappa_i}\ (i=1,2),}\\
(v_1,v_2,v_3,v_4,v_5,v_6,v_7,v_8)\rightarrow (e^{\varepsilon v_2}, e^{\varepsilon v_1},te^{\varepsilon v_3},
t^{-1}e^{\varepsilon v_7},e^{\varepsilon v_5},e^{\varepsilon v_6},e^{\varepsilon v_8},e^{\varepsilon v_4}),
\end{array}
\end{equation}
and take the limit $\varepsilon\to 0$. Actually, one can verify by direct calculation that
$q$-P($D_5^{(1)}/A_3^{(1)}$) \eqref{eqn:q-D5} gives rise to d-P($D_4^{(1)}/D_4^{(1)}$)
\eqref{eqn:d-D4} by the limiting procedure \eqref{eqn:limit_qD5_to_dD4} under the correspondence of
parameters \eqref{eqn:par_correspondence_D4}.

\begin{figure}[H]
{\setlength{\unitlength}{0.36mm}
\footnotesize{
\begin{picture}(70,65)(0,-5)
\put(15,30){None}
\put(5,-12){d-P$({D_4^{(1)}/D_4^{(1)}})$}
\end{picture}\qquad
\begin{picture}(80,65)(0,-5)
\put(-18,37){{\footnotesize $p=\infty$}}
\put(38,58){{\footnotesize $q=\infty$}}
\thicklines
\put(10,40){\line(1,0){40}}\put(4,28){$2|57$}
\put(40,10){\line(0,1){40}}\put(35,1){$1|13$}
\put(20,43){\circle*{2} $8$}
\put(30,43){\circle*{2} $6$}
\put(43,16){\circle*{2}}\put(43,19){$2$}
\put(43,30){\circle*{2}}\put(43,32){$4$}
\thinlines
\put(20,35){\line(0,1){15}}\put(17,53){$78$}
\put(30,35){\line(0,1){15}}\put(27,53){$56$}
\put(35,16){\line(1,0){15}}\put(52,14){$12$}
\put(35,30){\line(1,0){15}}\put(52,28){$34$}
\put(5,-12){d-P(${A_3^{(1)}/D_5^{(1)}})$}
\put(70,30){$\rightarrow$}
 \end{picture}\qquad
\begin{picture}(80,65)(0,-5)
\thicklines
\put(10,40){\line(1,0){40}}\put(-9,35){$2|56$}
\put(40,10){\line(0,1){40}}\put(35,1){$1|13$}
\put(18,55){\circle*{2}}\put(18,59){$8$}
\put(43,16){\circle*{2}}\put(43,19){$2$}
\put(43,30){\circle*{2}}\put(43,32){$4$}
\thinlines
\put(25,35){\line(0,1){25}}\put(20,26){$67$}
\put(15,48){\line(1,0){15}}\put(2,46){$56$}
\put(15,55){\line(1,0){15}}\put(2,55){$78$}
\put(35,16){\line(1,0){15}}\put(52,14){$12$}
\put(35,30){\line(1,0){15}}\put(52,28){$34$}
\put(5,-12){d-P$({(2A_1)^{(1)}/D_6^{(1)}})$}
\put(70,30){$\rightarrow$}
\end{picture}\qquad
\begin{picture}(70,65)(0,-5)
\thicklines
\put(10,40){\line(1,0){40}}\put(-9,35){$2|56$}
\put(40,10){\line(0,1){40}}\put(35,1){$1|12$}
\put(18,55){\circle*{2}}\put(18,59){$8$}
\put(55,30){\circle*{2} $4$}
\thinlines
\put(25,35){\line(0,1){25}}\put(16,27){$67$}
\put(15,48){\line(1,0){15}}\put(2,46){$56$}
\put(15,55){\line(1,0){15}}\put(2,55){$78$}
\put(35,25){\line(1,0){25}}\put(25,18){$23$}
\put(45,20){\line(0,1){15}}\put(41,13){$12$}
\put(55,20){\line(0,1){15}}\put(55,13){$34$}
\put(5,-12){d-P$(\underset{|\alpha|^2=4}{A_1^{(1)}}/D_7^{(1)})$}
\end{picture}\qquad
\begin{picture}(50,65)(0,-5)
\put(15,30){None}
\put(0,-12){d-P$({A_0^{(1)}/D_8^{(1)}})$}
\end{picture}
}
\\[5mm]
\hspace*{60mm}${\searrow}$\hspace*{23mm}${\searrow}$
\\[5mm]
\hspace*{70mm}
\footnotesize{
\begin{picture}(85,70)(0,-5)
\thicklines
\put(10,40){\line(1,0){25}}\put(-9,38){$2|35$}
\put(40,10){\line(0,1){25}}\put(35,1){$1|15$}
\put(20,43){\circle*{2} $4$}
\put(43,13){\circle*{2}}\put(43,15){$2$}
\put(48,37){\circle*{2} $8$}
\thinlines
\put(20,35){\line(0,1){15}}\put(17,54){$34$}
\put(35,13){\line(1,0){15}}\put(52,11){$12$}
\put(27,43){\line(1,-1){16}}\put(45,25){$56$}
\put(37,48){\line(1,-1){16}}\put(58,28){$78$}
\put(30,30){\line(1,1){15}}\put(48,48){$67$}
\put(5,-12){d-P(${A_2^{(1)}/E_6^{(1)}})$}
\put(75,30){$\rightarrow$}
\end{picture}\qquad
\begin{picture}(75,65)(0,-5)
\thicklines
\put(15,40){\line(1,0){20}}\put(-5,38){$2|34$}
\put(40,10){\line(0,1){25}}\put(35,3){$1|13$}
\put(43,60){\circle*{2}}\put(43,63){$8$}
\put(43,15){\circle*{2}}\put(43,19){$2$}
\thinlines
\put(35,15){\line(1,0){15}}\put(50,13){$12$}
\put(30,25){\line(0,1){30}}\put(24,17){$45$}
\put(25,30){\line(1,0){20}}\put(50,28){$34$}
\put(25,50){\line(1,0){20}}\put(50,48){$56$}
\put(40,45){\line(0,1){20}}\put(32,68){$67$}
\put(35,60){\line(1,0){15}}\put(54,58){$78$}
\put(5,-12){d-P$({A_1^{(1)}/E_7^{(1)}})$}
\end{picture}
}\\
\vskip5mm
\footnotesize{
\begin{picture}(50,50)(0,0)
\thicklines
\put(-18,12){{\footnotesize $g=0$}}
\put(-20,42){{\footnotesize $g=\infty$}}
\put(0,-5){{\footnotesize $f=0$}}
\put(30,-5){{\footnotesize $f=\infty$}}
\put(5,10){\line(1,0){45}}
\put(5,40){\line(1,0){45}}
\put(10,5){\line(0,1){45}}
\put(40,5){\line(0,1){45}}
\put(20,10){\circle*{2}}\put(30,10){\circle*{2}}
\put(18,13){$7$}\put(28,13){$8$}
\put(20,40){\circle*{2}}\put(30,40){\circle*{2}}
\put(18,43){$4$}\put(28,43){$3$}
\put(10,20){\circle*{2} $6$}\put(10,30){\circle*{2} $5$}
\put(40,20){\circle*{2} $1$}\put(40,30){\circle*{2} $2$}
\put(60,25){$q$-P($D_5^{(1)}$/$A_3^{(1)}$)}
\end{picture}\hfill\\[3mm]
\hspace*{8mm}$\downarrow$\\[5mm]
\begin{picture}(65,65)(0,-5)
\thicklines
\put(-29,40){{\footnotesize $g=\infty$}}
\put(-10,54){{\footnotesize $f=0$}}
\put(40,54){{\footnotesize $f=\infty$}}
\put(0,40){\line(1,0){50}}\put(-10,28){$2|37$}
\put(10,10){\line(0,1){40}}\put(5,0){$1|56$}
\put(40,10){\line(0,1){40}}\put(35,0){$1|12$}
\put(10,20){\circle*{2} $5$}
\put(10,30){\circle*{2} $6$}
\put(40,20){\circle*{2} $1$}
\put(40,30){\circle*{2} $2$}
\put(20,43){\circle*{2} $8$}
\put(30,43){\circle*{2} $4$}
\thinlines
\put(20,30){\line(0,1){20}}\put(15,53){$78$}
\put(30,30){\line(0,1){20}}\put(29,53){$34$}
\put(0,-15){d-P$({D_4^{(1)}/D_4^{(1)}})$}
\put(60,30){$\rightarrow$}
\end{picture}\qquad
\begin{picture}(65,55)(0,-5)
\thicklines
\put(5,40){\line(1,0){30}}\put(-12,38){$2|27$}
\put(10,10){\line(0,1){40}}\put(5,0){$1|56$}
\put(40,10){\line(0,1){25}}\put(35,0){$1|12$}
\put(10,20){\circle*{2} $5$}
\put(10,30){\circle*{2} $6$}
\put(40,15){\circle*{2} $1$}
\put(38,42){\circle*{2} $4$}
\put(20,43){\circle*{2} $8$}
\thinlines
\put(20,30){\line(0,1){20}}\put(17,53){$78$}
\put(28,42){\line(1,-1){15}}\put(44,25){$23$}
\put(38,25){\line(0,1){25}}\put(35,53){$34$}
\put(0,-15){d-P$({A_3^{(1)}/D_5^{(1)}})$}
\put(60,30){$\rightarrow$}
\end{picture}\qquad
\begin{picture}(65,55)(0,-5)
\thicklines
\put(15,40){\line(1,0){20}}\put(22,34){{\tiny $2|26$}}
\put(10,10){\line(0,1){25}}\put(5,0){$1|56$}
\put(40,10){\line(0,1){25}}\put(35,0){$1|12$}
\put(10,15){\circle*{2}}\put(13,13){$5$}
\put(40,15){\circle*{2}}\put(43,13){$1$}
\put(38,42){\circle*{2} $4$}
\put(12,43){\circle*{2} $8$}
\thinlines
\put(22,42){\line(-1,-1){15}}\put(-5,22){$67$}
\put(12,30){\line(0,1){20}}\put(12,53){$78$}
\put(28,42){\line(1,-1){15}}\put(45,22){$23$}
\put(38,25){\line(0,1){25}}\put(35,53){$34$}
\put(0,-15){d-P$({(2A_1)^{(1)}/D_6^{(1)}})$}
\put(60,30){$\rightarrow$}
\end{picture}\qquad
\begin{picture}(80,55)(0,-5)
\thicklines
\put(15,40){\line(1,0){22}}\put(22,34){{\tiny $2|16$}}
\put(10,10){\line(0,1){25}}\put(5,0){$1|56$}
\put(40,10){\line(0,1){25}}\put(35,0){$1|12$}
\put(10,15){\circle*{2} $5$}
\put(12,43){\circle*{2} $8$}
\thinlines
\put(22,42){\line(-1,-1){15}}\put(-4,22){$67$}
\put(12,30){\line(0,1){20}}\put(12,53){$78$}
\put(34,25){\line(0,1){25}}\put(27,53){$12$}
\put(30,32){\line(1,0){20}}\put(50,30){$23$}
\put(45,25){\line(0,1){25}}\put(42,53){$34$}\put(45,42){\circle*{2} $4$}
\put(0,-15){d-P$(\underset{|\alpha|^2=4}{A_1^{(1)}}/D_7^{(1)})$}
\put(65,30){$\rightarrow$}
\end{picture}\qquad
\begin{picture}(80,55)(0,-5)
\thicklines
\put(13,40){\line(1,0){24}}\put(20,34){{\tiny $2|15$}}
\put(10,10){\line(0,1){25}}\put(5,0){$1|56$}
\put(40,10){\line(0,1){25}}\put(35,0){$1|12$}
\thinlines
\put(16,25){\line(0,1){25}}\put(12,53){$56$}
\put(0,32){\line(1,0){20}}\put(-10,30){$67$}
\put(4,25){\line(0,1){25}}\put(-2,53){$78$}\put(4,43){\circle*{2} $8$}
\put(34,25){\line(0,1){25}}\put(28,53){$12$}
\put(30,32){\line(1,0){20}}\put(50,30){$23$}
\put(45,25){\line(0,1){25}}\put(42,53){$34$}\put(45,42){\circle*{2} $4$}
\put(0,-15){d-P$({A_0^{(1)}/D_8^{(1)}})$}
\end{picture}
}
\\[5mm]
\hspace*{53mm}${\searrow}$\\[3mm]
\hspace*{62mm}
\footnotesize{
\begin{picture}(65,60)(0,-5)
\thicklines
\put(5,40){\line(1,0){30}}\put(-15,38){$2|23$}
\put(10,10){\line(0,1){40}}\put(5,0){$1|56$}
\put(40,10){\line(0,1){25}}\put(35,0){$1|12$}
\put(10,20){\circle*{2} $5$}
\put(10,30){\circle*{2} $6$}
\put(40,15){\circle*{2} $1$}
\thinlines
\put(40,45){\line(0,1){15}}\put(38,63){$78$}\put(40,52){\circle*{2} $8$}
\put(28,25){\line(0,1){30}}\put(25,17){$34$}
\put(25,30){\line(1,0){20}}\put(48,26){$23$}
\put(25,48){\line(1,0){20}}\put(48,46){$47$}
\put(0,-15){d-P$({A_2^{(1)}/E_6^{(1)}})$}
\end{picture}\qquad
\begin{picture}(50,65)(0,-5)
\put(15,30){None}
\put(0,-15){d-P$({A_1^{(1)}/E_7^{(1)}})$}
\end{picture}
}}\\ \caption{Point configurations corresponding to Figure \ref{fig:degeneration_additive} in the
blown-up spaces. We use the abbreviated notations as $i|jk=H_i-E_j-E_k$, $ij=E_i-E_j$,
$i=E_i$.}\label{fig:config_add_blowup}
\end{figure}
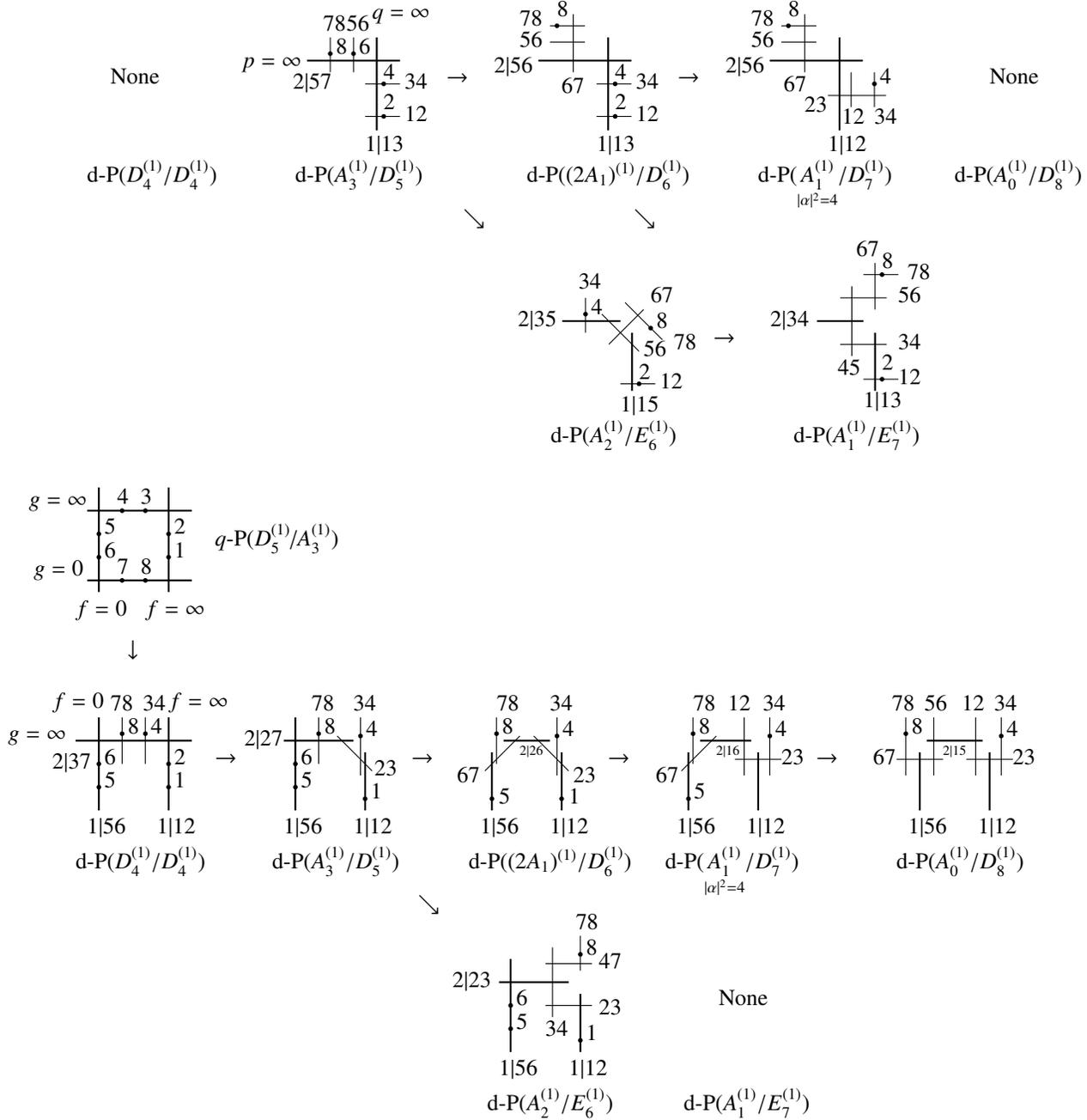
\subsubsection{d-P($D_4^{(1)}/D_4^{(1)}$) $\rightarrow$ d-P($A_3^{(1)}/D_5^{(1)}$)}
We set
\begin{equation}
\begin{split}
& f\to e^{\varepsilon tq},\quad g\to \frac{p}{t\varepsilon},\quad t\to e^{t\varepsilon},\quad a_0 \to  a_3,\\
& a_1\to -\frac{1}{\varepsilon} + a_2 ,\quad a_2\to \frac{1}{\varepsilon},\quad a_3\to a_1,\quad a_4\to -\frac{1}{\varepsilon} + a_0,
\end{split}
\end{equation}
and take the limit $\varepsilon\to 0$ to obtain \eqref{eqn:d-A3} from \eqref{eqn:d-D4}.
\subsubsection{d-P($A_3^{(1)}/D_5^{(1)}$) $\rightarrow$ d-P($A_2^{(1)}/E_6^{(1)}$)}
We set
\begin{equation}
\begin{split}
& p\to \frac{1}{\varepsilon^2}+\frac{p-t}{\varepsilon},\quad
q\to q \varepsilon,\quad
t\to \frac{t}{\varepsilon}-\frac{1}{\varepsilon^2},\\
&a_0\to a_2,\quad
a_2\to  \frac{1}{\varepsilon^2}+a_0,\quad
a_3\to -\frac{1}{\varepsilon^2},
\end{split}
\end{equation}
and take the limit $\varepsilon\to 0$ to obtain \eqref{eqn:d-A2} from \eqref{eqn:d-A3}.
\par\medskip
%
\subsection{Birational representation of affine Weyl groups}\label{subsec:birational}
In this subsection, for each symmetry/surface type we construct an explicit birational
representation of the symmetry group generated by simple reflections $s_0, s_1,\ldots,s_l$
($s_i=r_{\alpha_i}$) and lattice isomorphisms (Dynkin diagram automorphisms) $\pi_1,\ldots,\pi_r$,
with $\alpha_i$ and $\pi_i$ as listed in Section \ref{subsec:configuration}. We use the parameters
$h_1$, $h_2$, $e_1,\ldots,e_8$ instead of $\kappa_1$, $\kappa_2$, $v_1,\ldots, v_8$ (see Remark
\ref{rem:kappa-var}) for both multiplicative and additive cases.  On these parameters the simple
reflections act linearly in the same way as they do on the basis of the Picard lattice, $H_1$,
$H_2$, $E_1,\ldots, E_8$. Note that the Picard lattice has a trivial lattice isomorphism
$H_i\rightarrow -H_i$, $E_i\rightarrow -E_i$ which does not belong to the affine Weyl group. As to
the lattice automorphism $\pi_1,\ldots, \pi_r$, we need to incorporate the corresponding
transformation $h_i\rightarrow h_i^{-1}$, $e_i\rightarrow e_i^{-1}$ or $h_i\rightarrow -h_i$,
$e_i\rightarrow -e_i$ in constructing individual birational representations. We also give the
actions on $f$, $g$.

In the following, we use the following symbols for distinct $i$, $j$, $k$, $l$:
\begin{equation}
\begin{split}
s_{ij}:&\quad e_i\ \leftrightarrow\ e_j,\\
s_{H_1-H_2}:&\quad h_1\ \leftrightarrow\ h_2 \\
s_{H_1-E_i-E_j}:&\quad e_i\ \rightarrow\ \frac{h_1}{e_j},\ e_j\ \rightarrow\ \frac{h_1}{e_i},\ 
h_2\ \rightarrow\ \frac{h_1h_2}{e_ie_j},\\
s_{H_2-E_i-E_j}:&\quad e_i\ \rightarrow\ \frac{h_2}{e_j},\ e_j\ \rightarrow\ \frac{h_2}{e_i},\ 
h_1\ \rightarrow\ \frac{h_1h_2}{e_ie_j},\\
s_{H_1+H_2-E_i-E_j-E_k-E_l}:&\quad e_i\ \rightarrow\ \frac{h_1h_2}{e_je_ke_l},\ 
e_j\ \rightarrow\ \frac{h_1h_2}{e_ie_ke_l},\ 
e_k\ \rightarrow\ \frac{h_1h_2}{e_ie_je_l},\ e_l\ \leftrightarrow\ \frac{h_1h_2}{e_ie_je_k},\\
&\quad h_1\ \rightarrow\ \frac{h_1^2h_2}{e_ie_je_ke_l},\quad 
h_2\ \rightarrow\ \frac{h_1h_2^2}{e_ie_je_ke_l}.
\end{split}
\end{equation}

\subsubsection{$q$-P($E_8^{(1)}/A_0^{(1)}$)}\label{subsubsec:Weyl_q-E8}

\noindent Point configuration:
\begin{equation}
 (f_i,g_i)= \Bigl(e_i + \frac{h_1}{e_i}, e_i + \frac{h_2}{e_i}\Bigr)\quad (i=1,\ldots,8). 
\end{equation}

\noindent Generators and actions on parameters:
\begin{equation}
\begin{split}
&s_0=s_{12},\ s_1 = s_{H_1-H_2},\ s_2 =s_{H_2-E_1-E_2},\ s_3 = s_{23},\\
&s_4=s_{34},\ s_5=s_{45},\ s_6=s_{56},\ s_7=s_{67},\ s_8=s_{78}.
\end{split}
\end{equation}
Actions on $f$, $g$:
\begin{equation}
\begin{split}
 s_1: &\ f\ \leftrightarrow\ g ,\\
 s_2: &\ f\ \rightarrow \frac{1}{e_1 e_2  \left(e_1 e_2-h_2\right) f - e_1 e_2 \left(e_1  e_2-h_1\right)g 
-  e_1 e_2 \left(e_1+e_2 \right)  \left(h_1-h_2\right)}\\
&\times\left[e_1 e_2    \left(h_1-h_2\right) f g + h_2 \left(e_1+e_2\right)\left(e_1 e_2-h_1\right)f\right.\\
&\hskip40pt \left.     - h_1 \left(e_1+e_2\right)\left(e_1 e_2-h_2\right) g 
 + \left(h_1-h_2\right) \left(e_1  e_2-h_1\right) \left(e_1 e_2-h_2\right) \right].
\end{split}
\end{equation}


\subsubsection{$q$-P($E_7^{(1)}/A_1^{(1)}$)}\label{subsubsec:Weyl_q-E7}

\noindent Point configuration:
\begin{equation}
(f_i,g_i)=\Bigl(e_i,\frac{1}{e_i}\Bigr),\quad (i=1,\ldots,4),\quad
\Bigl(\frac{h_1}{e_i},\frac{e_i}{h_2}\Bigr)\quad (i=5,\ldots,8).
\end{equation}
Generators and actions on parameters:
\begin{equation}
\begin{split}
&s_0=s_{H_1-H_2},\ s_1 = s_{34},\ s_2 =s_{23},\ s_3 = s_{12},\\
&s_4=s_{H_2-E_1-E_5},\ s_5=s_{56},\ s_6=s_{67},\ s_7=s_{78},\\
&\pi:\ h_i\ \leftrightarrow\ h_i^{-1},\ 
e_{i}\ \leftrightarrow e_{\sigma(i)}^{-1},\ \sigma={\scriptsize \begin{pmatrix}12345678\\ 56781234\end{pmatrix}}.
\end{split}
\end{equation}
Actions on $f$, $g$:
\begin{equation}
\begin{split}
s_0:&\ f\ \rightarrow\ \frac{1}{g},\ g\ \rightarrow\ \frac{1}{f},\\
s_4:&\ f\ \rightarrow\ \frac{-h_2(e_1e_5 - h_1)fg - e_5(h_1-h_2)f + h_1(e_1e_5- h_2)  }{e_5\bigl\{-(e_1e_5-h_2)fg + e_1(h_1-h_2) g + (e_1e_5-h_1) \bigr\}},\\
\pi:& \ f\ \rightarrow\ \frac{f}{h_1},\ g\ \rightarrow\ h_2g.
\end{split}
\end{equation}

\subsubsection{$q$-P($E_6^{(1)}/A_2^{(1)}$)}\label{subsubsec:Weyl_q-E6}

\noindent Point configuration:
\begin{equation}
(f_i,g_i)=\Bigl(e_i,\frac{1}{e_i}\Bigr)\quad (i=1,\ldots,4),\quad
\Bigl(0,\frac{e_i}{h_2}\Bigr)\quad (i=5,6),\quad \Bigl(\frac{h_1}{e_i},0\Bigr)\quad (i=7,8). 
\end{equation}
Generators and actions on parameters:
\begin{equation}
\begin{split}
&s_0=s_{78},\ s_1 = s_{65},\ s_2 =s_{H_2-E_1-E_6},\ s_3 = s_{12},\\
& s_4=s_{23},\ s_5=s_{34},\ s_6=s_{H_1-E_1-E_7},\\
\pi_1:&\ h_1\ \rightarrow \frac{e_1e_2}{h_1h_2},\ h_2\ \rightarrow\ h_2^{-1},\ 
e_1\ \rightarrow\ \frac{e_2}{h_2},\ e_2\ \rightarrow\ \frac{e_1}{h_2},\\
&\ e_i\ \rightarrow e_{\sigma(i)}^{-1},\ \sigma={\scriptsize \begin{pmatrix}345678\\ 654378\end{pmatrix}}\ (i\neq 1,2),\\
\pi_2:&\ h_1\ \rightarrow\ h_2^{-1},\ h_2\ \rightarrow\ h_1^{-1},\ 
e_{i}\ \rightarrow e_{\sigma(i)}^{-1},\ \sigma={\scriptsize \begin{pmatrix}12345678\\ 12348765\end{pmatrix}}.
\end{split}
\end{equation}
Actions on $f$, $g$:
\begin{equation}
\begin{split}
s_2:&\ f\ \rightarrow\ \frac{h_2(e_1 g - 1)f}{- (h_2-e_1e_6) fg + e_1h_2g - e_1e_6},\\
s_6:&\ g\ \rightarrow\ \frac{e_7g(e_1-f)}{-e_7f + (e_1e_7-h_1)fg + h_1},\\
\pi_1:&\ g\ \rightarrow\ h_2g,\ f \ \rightarrow\ \frac{\frac{e_1e_2}{h_2}(1-fg)}{e_1e_2g +f -(e_1+e_2)fg},\\
\pi_2:&\ f\ \leftrightarrow\ g.
 \end{split}
\end{equation}

\subsubsection{$q$-P($D_5^{(1)}/A_3^{(1)}$)}\label{subsubsec:Weyl_q-D5}

\noindent Point configuration:
\begin{equation}
(f_i,g_i)=\Bigl(\infty,\frac{1}{e_i}\Bigr)\ (i=1,2),\ 
\Bigl(e_i,\infty\Bigr)\ (i=3,4),\ \Bigl(0,\frac{e_i}{h_2}\Bigr)\ (i=5,6),\ \Bigl(\frac{h_1}{e_i},0\Bigr)\ (i=7,8). 
\end{equation}
Generators and actions on parameters:
\begin{equation}
 \begin{split}
&  s_0 = s_{78},\ s_1=s_{34}, \ s_2=s_{H_1-E_3-E_7},\  s_3=s_{H_2-E_1-E_5},\ s_4=s_{12},\ s_5 = s_{56},\\
\pi_1:&\ h_i\ \leftrightarrow\ h_i^{-1},\ 
e_{i}\ \rightarrow e_{\sigma(i)}^{-1},\ \sigma={\scriptsize \begin{pmatrix}12345678\\ 12785634\end{pmatrix}},\\
\pi_2:&\ h_1\rightarrow \frac{1}{h_2},\ h_2 \rightarrow\ \frac{1}{h_1}, \ 
e_{i}\ \rightarrow e_{\sigma(i)}^{-1},\ \sigma={\scriptsize \begin{pmatrix}12345678\\ 78563412\end{pmatrix}}.
 \end{split}
\end{equation}
Actions on $f$, $g$:
\begin{equation}
 \begin{split}
 s_2:&\ g\ \rightarrow\ g~\frac{f-e_3}{f-\frac{h_1}{e_7}},\\
 s_3:&\ f\ \rightarrow\ f~\frac{g-\frac{1}{e_1}}{g-\frac{e_5}{h_2}},\\
\pi_1:&\ f\ \rightarrow \frac{f}{h_1},\ g\ \rightarrow\ \frac{1}{g},\\
\pi_2:&\ f\ \rightarrow \frac{1}{h_2g},\ g\ \rightarrow\ \frac{h_1}{f}.
 \end{split}
\end{equation}
%

\subsubsection{$q$-P($A_4^{(1)}/A_4^{(1)}$)}\label{subsubsec:Weyl_q-A4}
\noindent Point configuration:
\begin{equation}
(f_i,g_i)=\left(\infty ,\frac{1}{e_1}\right),\quad
\left(-\frac{e_2e_3}{\epsilon}, \frac{1}{\epsilon}\right)_2,\ 
\left(e_4,\infty \right),\ 
\left(0,\frac{e_i}{h_2}\right)\ (i=5,6),\ 
\left(\frac{h_1}{e_i},0\right)\ (i=7,8).
\end{equation}
Generators and actions on parameters:
\begin{equation}
  \begin{split}
&  s_0 = s_{78},\ s_1=s_{H_1-E_4-E_7}, \ s_2=s_{H_2-E_1-E_5}, s_3=s_{56},\ s_4=s_{H_1+H_2-E_2-E_3-E_5-E_7},\\
\pi_1:&\ e_2\ \rightarrow\ \frac{h_2}{e_5},\quad e_2\ \rightarrow\ \frac{h_1h_2}{e_2e_5e_7},\quad 
e_i\to e_{\sigma(i)},\quad \sigma={\scriptsize \begin{pmatrix}13468\\ 26513\end{pmatrix}}\ (i\neq 2,5,7),\\
&\ h_1\ \rightarrow\ \frac{h_1h_2}{e_2e_5},\quad h_2\ \rightarrow\ \frac{h_1h_2}{e_5e_7},\\
\pi_2:&\ e_i\to e^{-1}_{\sigma(i)},\quad \sigma={\scriptsize \begin{pmatrix}12345678\\ 42317856\end{pmatrix}},
\quad\ h_1\to \frac{1}{h_2},\quad h_2\to \frac{1}{h_1}.
 \end{split}
\end{equation}

\noindent Actions on $f$, $g$:
\begin{equation}
 \begin{split}
s_1: &\ g\ \rightarrow \frac{e_7 \left(e_4 - f\right)}{h_1 - e_7 f}g,\\
s_2: &\ f\ \rightarrow \frac{h_2\left(1 - e_1g\right)}{e_1(e_5 - h_2g)}f,\\ 
s_4: &\ f\ \rightarrow\ \frac{h_1h_2}{e_2e_3e_7}\frac{-h_1+e_7f+e_2e_3e_7g}{-h_1e_5 + e_5e_7f + h_1h_2g}\,f,\\
&\ g\ \rightarrow\ \frac{e_5e_7}{h_2} \frac{-e_2e_3e_5 + h_2f + h_2e_2e_3g}{-h_1e_5 + e_5e_7f + h_1h_2g}\,g,\\
\pi_1:&\ f\ \rightarrow \frac{h_1}{e_5}\frac{e_5 - h_2g}{f},\ 
g\ \rightarrow\ \frac{1}{h_1h_2}\frac{-h_1e_5 + e_5e_7 f + h_1h_2g}{fg},\\
\pi_2:&\ f\ \leftrightarrow\ g.
\end{split}
\end{equation}
%

\subsubsection{$q$-P($E_3^{(1)}/A_5^{(1)};b$)}\label{subsubsec:Weyl_q-E3b}

\noindent Point configuration:
\begin{equation}
\begin{split}
(f_i,g_i)=&\left(\infty ,\frac{1}{e_1}\right),\ 
\left(-\frac{e_2e_3}{\epsilon}, \frac{1}{\epsilon}\right)_2,\ 
\left(e_4,\infty \right),\ \left(0,\frac{e_5}{h_2}\right),\ 
\Bigl(-\frac{h_1h_2}{e_6e_7}\epsilon,\epsilon\Bigr)_2,\ 
\left(\frac{h_1}{e_8},0\right) .
\end{split}
\end{equation}
Generators and actions on parameters:
\begin{equation}
  \begin{split}
&  s_0 = s_{H_1+H_2-E_2-E_3-E_6-E_7},\ s_1=s_{H_1-E_4-E_8}, \ 
s_2=s_{H_2-E_1-E_5},\\
& s_3=s_{H_1+H_2-E_2-E_3-E_5-E_8},\ s_4=s_{H_1+H_2-E_1-E_4-E_6-E_7},\\
\pi_1:&\ h_1\ \rightarrow\ h_2^{-1},\quad h_2\ \rightarrow\ h_1^{-1},\quad e_i\to e^{-1}_{\sigma(i)},\quad
\sigma={\scriptsize \begin{pmatrix}12345678\\ 42318675\end{pmatrix}},\\
\pi_2:&\ e_2\to \frac{h_2}{e_2},\quad e_6\to \frac{h_2}{e_6},\quad 
e_i \to e^{-1}_{\sigma(i)}\ (i\neq 2,6),\quad \sigma={\scriptsize \begin{pmatrix}134578\\ 345781\end{pmatrix}},\\
&\ h_1\to h_2,\quad h_2\to\frac{h_1h_2}{e_2e_6}.
 \end{split}
\end{equation}

\noindent Actions on $f$, $g$:
\begin{equation}
 \begin{split}
s_0: &\ f\to \frac{f h_1 h_2 \left(e_2 e_3 g+f\right)}{e_2 e_3 \left(e_6 e_7 f+g h_1 h_2\right)},\quad 
g\to \frac{e_6 e_7 g \ \left(e_2 e_3 g+f\right)}{e_6 e_7 f+g h_1 h_2},\\
s_1: &\  g\to \frac{e_8 g \left(f-e_4\right)}{e_8 f-h_1},\\ 
 s_2:&\  f\to \frac{f h_2 \left(e_1 g-1\right)}{e_1 \left(g h_2-e_5\right)},\\ 
 s_3:&\  f\to \frac{f h_1 h_2 \left(e_8 \left(e_2 e_3 g+f \right)-h_1\right)}
{e_2 e_3 e_8 \left(e_5 \left(e_8 f-h_1\right)+g h_1 h_2\right)},\quad 
g\to \frac{e_5 e_8 g \left(e_2 e_3 \left(g h_2-e_5\right)+f h_2\right)}
{h_2 \left(e_5 \left(e_8 f-h_1\right)+g h_1 h_2 \right)},\\
s_4: &\  f\to -\frac{f h_1 h_2 \left(e_1 g \left(f-e_4\right)-f \right)}
{e_1 \left(e_4 \left(e_6 e_7 f+g h_1 h_2\right)-f g h_1 h_2\right)},\quad 
g\to \frac{e_6 e_7 g \left(e_1 g \left(f-e_4\right)-f \right)}
{e_6 e_7 f \left(e_1 g-1\right)-g h_1 h_2},\\
\pi_1:&\  f\ \leftrightarrow\ g,\\
\pi_2:&\  f\to g h_2,\quad g\to -\frac{e_2 g}{f}.\\
 \end{split}
\end{equation}
%

\subsubsection{$q$-P($E_3^{(1)}/A_5^{(1)};a$)}\label{subsubsec:Weyl_q-E3a}
\noindent Point configuration:
\begin{equation}
\begin{split}
(f_i,g_i)=&
\left(-\frac{h_1}{\epsilon e_1e_8},\epsilon\right)_2,\ 
\left(-\frac{e_2e_3}{\epsilon}, \frac{1}{\epsilon}\right)_2,
\left(e_4,\infty \right),\ 
\left(0,\frac{e_i}{h_2}\right)\ (i=5,6),\ 
\left(\frac{h_1}{e_7},0\right).
\end{split}
\end{equation}
Generators and actions on parameters:
\begin{equation}
  \begin{split}
&  s_0 = s_{H_1+H_2-E_2-E_3-E_6-E_7},\ s_1=s_{H_1+H_2-E_1-E_4-E_6-E_8}, \ 
s_2=s_{E_6-E_5},\\
& s_3=s_{\delta-\alpha_4},\ s_4=s_{H_1-E_4-E_7},\\
\pi_1:&\ h_1\to \frac{e_4e_6}{h_1h_2},\quad h_2\to \frac{e_1e_6}{h_1h_2},\quad e_i\to e^{-1}_{\sigma(i)},\quad
\sigma={\scriptsize \begin{pmatrix}23578\\ 23875\end{pmatrix}},\\
&\ e_1 \to \frac{e_6}{h_1},\quad e_4\to\frac{e_6}{h_2},\quad e_6\to\frac{e_1e_4e_6}{h_1h_2},\\
\pi_2:&\ e_1\to \frac{h_1h_2}{e_1e_2e_6},\quad e_2\to \frac{h_2}{e_2},\quad 
e_6\to \frac{h_1h_2}{e_2e_3e_6},\quad e_8\to \frac{h_2}{e_6},\\
&\ e_i \to e_{\sigma(i)},\quad \sigma={\scriptsize \begin{pmatrix}3457\\ 4578\end{pmatrix}},\quad
h_1\to \frac{h_1h_2^2}{e_1e_2e_3e_6},\quad h_2\to\frac{h_1h_2}{e_2e_6}.
\end{split}
\end{equation}

\noindent Actions on $f$, $g$:
\begin{equation}
 \begin{split}
s_0: &\ f\to \frac{h_1h_2 f(e_7f + e_2e_3e_7g - h_1)}{e_2 e_3e_7 \left(e_6 e_7 f -e_6h_1 + h_1 h_2g\right)},\quad 
g\to \frac{e_6 e_7 g \ \left(e_2 e_3 e_6 +h_2 f + e_2e_3 h_2g\right)}{h_2(e_6e_7 f - e_6h_1 + h_1h_2 g)},\\
s_1: &\  f\to \frac{h_2 f \left(e_1 e_8 fg -e_1 e_4 e_8 g +h_1\right)}{e_1 e_8 \left(-e_4 h_2 g+e_4 e_6+h_2 f g\right)},\quad 
g\to -\frac{e_1 e_6 e_8 g \left(e_4 h_2 g-e_4  e_6-h_2 f g\right)}{h_2 \left(e_6 h_1+e_1 e_6 e_8 fg-h_1 h_2 g\right)},\\ 
s_3:&\ f\to \frac{h_1 h_2^2 f \left(e_1 e_5 e_8 h_2 f g+\left(h_2 g-e_5\right) 
\left(e_1 e_2 e_3 e_5 e_8 g-h_1  h_2\right)\right) }{e_1^2 e_2 e_3 e_5 e_6 e_8^2 \left(e_2 e_3 \left(e_5-h_2 g\right) 
\left(e_6-h_2  g\right)+h_2^2 f g\right) }\\
  &\qquad\qquad\times
  \frac{\left(e_1 e_6 e_8 h_2 f g+\left(h_2 g-e_6\right) \left(e_1 e_2 e_3 e_6 e_8 g-h_1  h_2\right)\right)}
  {\left(h_1 \left(e_5-h_2 g\right) \left(e_6-h_2 g\right)+e_1 e_5 e_6 e_8 f g\right)},\\
  &\ g\to \frac{e_1 e_5 e_6 e_8 g \left(e_2 e_3 \left(e_5-h_2  g\right) \left(e_6-h_2 g\right)+h_2^2 fg\right)}
{h_2^2 \left(h_1 \left(e_5-h_2 g\right)  \left(e_6-h_2g\right)+e_1 e_5 e_6 e_8 fg\right)},\\
s_4: &\  g\to \frac{e_7 g \left(f-e_4\right)}{e_7 f-h_1},\\
\pi_1:&\  f\to \frac{-e_4 h_2 g+e_4 e_6+h_2 f g}{h_2  f},\quad
g\to \frac{h_2 f g}{h_2 g-e_6},\\
\pi_2:&\  f\to \frac{h_2 g \left(e_2 e_3 h_2 g-e_2 e_3  e_6+h_2 f\right)}{e_2 e_3 \left(h_2g-e_6\right)},\quad 
g\to -\frac{e_2 \left(h_2g-e_6\right)}{h_2 f}.\\
 \end{split}
\end{equation}
 \begin{rem}\rm
The two realizations of affine Weyl groups on parameters associated with $q$-P$(E_3^{(1)}/A_6^{(1)};b)$
and $q$-P$(E_3^{(1)}/A_6^{(1)};a)$ are transformed with each other by the reflection $s_{H_2-E_1-E_6}$.
Also, the actions on $(f,g)$ variables associated with the former
is transformed to that of latter by the substitution $f\to f\frac{g}{g-\frac{e_6}{h_2}}$.
Conversely, the latter is transformed to the former by $f\to f\frac{g-\frac{1}{e_1}}{g}$.
\end{rem}
%

\subsubsection{$q$-P($E_2^{(1)}/A_6^{(1)};b$)}\label{subsubsec:Weyl_q-E2b}
\noindent Point configuration:
\begin{equation}
\begin{split}
(f_i,g_i)=&\left(\infty ,\frac{1}{e_1}\right),\ 
\left(-\frac{e_2e_3}{\epsilon}, \frac{1}{\epsilon}\right)_{2},\ 
\left(e_4,\infty \right),\ 
\Bigl(\frac{h_1h_2^2}{e_5e_6e_7}\epsilon^2,\epsilon\Bigr)_3,\ 
\left(\frac{h_1}{e_8},0\right) .
\end{split}
\end{equation}
Generators and actions on parameters:
\begin{equation}
  \begin{split}
s_0 & = s_{H_1+2H_2-E_1-E_2-E_3-E_5-E_6-E_7},\ 
s_1=s_{H_1-E_4-E_8}, \\
\pi_1:&\ h_1\to \frac{e_1 e_2 e_5 e_6}{h_1 h_2^2},\ 
h_2\to \frac{1}{h_2},\ 
e_{i}\to e_{\sigma(i)}^{-1},\ \sigma={\scriptsize\begin{pmatrix} 3478\\ 4387\end{pmatrix}}, \\
&\ e_1\to \frac{e_1}{h_2},\ e_2\to \frac{e_2}{h_2},\ e_5\to \frac{e_6}{h_2},\ e_6\to \frac{e_5}{h_2},\\
\pi_2:&\ h_1\to \frac{1}{h_1},\ 
h_2\to \frac{e_2 e_5}{h_1 h_2},\ 
e_2\to \frac{e_2}{h_1},\ 
e_5\to \frac{e_5}{h_1},\ 
e_{i}\to e_{\sigma(i)}^{-1},\ \sigma={\scriptsize\begin{pmatrix} 134678\\ 318674\end{pmatrix}}.
 \end{split}
\end{equation}

\noindent Actions on $f$, $g$:
\begin{equation}
 \begin{split}
s_0: & \ s_0=\pi_1s_1\pi_1, \\
s_1: & \ g\ \to\ \frac{e_8(f-e_4)g}{e_8f-h_1},\\ 
\pi_1:&\ f\to -\frac{g^2 e_2}{f \left(g-\frac{1}{e_1}\right)},\quad
g\to g h_2,\\
\pi_2:&\ f\to \frac{f}{h_1},\quad g\to -\frac{f}{g e_2}.
\end{split}
\end{equation}
\begin{rem}\rm \hfill
 \begin{itemize}
  \item The reflection corresponding to $\alpha_2=H_2+2E_1 - 2E_2 - 2E_3 + E_4 - E_8$ ($|\alpha_2|^2=\langle \alpha_2,\alpha_2\rangle=-\alpha_2\cdot\alpha_2=14$) does not exist.
  \item We omitted the explicit formula of action of $s_0$ on $f$ and $g$, since their expressions
	are long and obtained as $\pi_1 s_1 \pi_1$. 
 \end{itemize}
\end{rem}
%

\subsubsection{$q$-P($E_2^{(1)}/A_6^{(1)};a$)}\label{subsubsec:Weyl_q-E2a}
\noindent Point configuration:
\begin{equation}
\begin{split}
(f_i,g_i)=&
\left(-\frac{h_1}{\epsilon e_1e_8},\epsilon\right)_2\ 
\left(-\frac{e_2e_3}{\epsilon}, \frac{1}{\epsilon}\right)_2,\ 
\left(e_4,\infty \right),\ 
\left(0,\frac{e_5}{h_2}\right),\ 
\Bigl(-\frac{h_1h_2}{e_6e_7}\epsilon,\epsilon\Bigr)_2.
\end{split}
\end{equation}
Generators and actions on parameters:
\begin{equation}
  \begin{split}
& s_0 = s_{H_1+H_2-E_2-E_3-E_6-E_7},\ s_1=s_{H_1+H_2-E_1-E_4-E_5-E_8}, \\
 \pi_1:&\quad e_i \to e^{-1}_{\sigma(i)},\ \sigma={\scriptsize \begin{pmatrix}134678\\ 643187\end{pmatrix}},\ 
e_2\to \frac{e_2}{h_2},\ e_5\to \frac{e_5}{h_2},\\
&\quad h_1\to \frac{e_2 e_5}{h_1  h_2},\ h_2\to \frac{1}{h_2},   \\
   \pi_2:&\quad e_i \to e^{-1}_{\sigma(i)},\ \sigma={\scriptsize \begin{pmatrix}4678\\ 8674\end{pmatrix}},\\
&\quad   e_1\to \frac{e_2}{h_2},\ e_2\to \frac{e_1 e_2 e_5}{h_1 h_2},\ e_3\to \frac{e_5}{h_2},\ e_5\to\frac{e_2 e_3 e_5}{h_1 h_2},\\
&\quad h_1\to \frac{e_1 e_2 e_3 e_5}{h_1 h_2^2},\ h_2\to \frac{e_2 e_5}{h_1 h_2}.
 \end{split}
\end{equation}

\noindent Actions on $f$, $g$:
\begin{equation}
 \begin{split}
s_0: &\quad f\to \frac{f h_1 h_2 \left(e_2 e_3 g+f\right)}{e_2 e_3 \left(e_6 e_7 f+g h_1 h_2\right)},\quad 
g\to \frac{e_6 e_7 g \left(e_2 e_3 g+f\right)}{e_6 e_7 f+g h_1 h_2},\\
s_1: &\quad f\to \frac{f h_2 \left(e_1 e_8 f g-e_1 e_4 e_8 g+h_1\right)}
{e_1 e_8 \left(-e_4 g h_2+e_4 e_5+f g h_2\right)},\quad
g\to -\frac{e_1 e_5 e_8 g \left(e_4 g h_2-e_4 e_5-f g h_2\right)}
{h_2 \left(e_1 e_5 e_8 f  g+e_5 h_1-g h_1 h_2\right)},\\ 
\pi_1:&\quad f\to -\frac{e_2 \left(g h_2-e_5\right)}{f h_2},\ g\to g h_2,\\
\pi_2:&\quad f\to \frac{e_1 e_5 g \left(e_2 e_3 g h_2-e_2 e_3 e_5+f h_2\right)}{h_1 h_2 \left(g h_2-e_5\right)},
\  g\to -\frac{f h_2}{e_2 \left(g h_2-e_5\right)}.
 \end{split}
\end{equation}
 \begin{rem}\rm
 The two realizations of affine Weyl groups on parameters associated with $q$-P$(E_2^{(1)}/A_6^{(1)};b)$
and $q$-P$(E_2^{(1)}/A_6^{(1)};a)$ are transformed with each other by the reflection $s_{H_2-E_1-E_5}$.
Also, the actions on $(f,g)$ variables associated with the former
is transformed to that of latter by the substitution $f\to f\frac{g}{g-\frac{e_5}{h_2}}$.
Conversely, the latter is transformed to the former by $f\to f\frac{g-\frac{1}{e_1}}{g}$.
\end{rem}
%

\subsubsection{$q$-P($A_1^{(1)}/A_7^{(1)}$)}\label{subsubsec:Weyl_q-A1b}
\noindent Point configuration:
\begin{equation}
\begin{split}
(f_i,g_i)=& \left(\frac{e_1e_2e_3}{\epsilon^2}, \frac{1}{\epsilon}\right)_3,\quad
\left(e_4,\infty \right),\quad
\Bigl(\frac{h_1h_2^2}{e_5e_6e_7}\epsilon^2,\epsilon\Bigr)_3,\quad
\left(\frac{h_1}{e_8},0\right) .
\end{split}
\end{equation}
Generators and actions on parameters:
\begin{equation}
  \begin{split}
s_0 & = s_{H_1+2H_2-E_1-E_2-E_3-E_5-E_6-E_7},\ 
s_1=s_{H_1-E_4-E_8}, \\
\pi_1:&\ e_{i}\to e_{\sigma(i)}^{-1}, \ 
\sigma={\scriptsize\begin{pmatrix} 234678\\ 238674\end{pmatrix}}, \quad
e_1\ \to\ \frac{e_1}{h_1}, \ 
e_5\ \to\ \frac{e_5}{h_1}, \\
&\ h_1\ \to\ h_1^{-1}, \quad
h_2\ \to\ \frac{e_1e_5}{h_1h_2},\\
\pi_2:&
\ e_{i}\to e_{\sigma(i)}, \quad \sigma={\scriptsize\begin{pmatrix} 3478 \\ 4783 \end{pmatrix}}, \quad
e_1\ \to\ \frac{h_1h_2}{e_1e_5e_6}, \quad
e_2\ \to\ \frac{h_2}{e_1}, \\
&\ e_5\ \to\ \frac{h_1h_2}{e_1e_2e_5}, \quad
e_6\ \to\ \frac{h_2}{e_5}, \quad
h_1\ \to\ \frac{h_1h_2^2}{e_1e_2e_5e_6}, \quad
h_2\ \to\ \frac{h_1h_2}{e_1e_5}.
 \end{split}
\end{equation}

\noindent Actions on $f$, $g$:
\begin{equation}
 \begin{split}
s_0: & \ f \to \frac{f (f-g^2 e_1 e_2 e_3)^2 h_1^2 h_2^4}
{e_1^2 e_2^2 e_3^2 (fe_5 e_6 e_7-g^2 h_1 h_2^2)^2},\quad
g \to \frac{g \left(g^2 e_1 e_2 e_3-f\right) e_5 e_6e_7}{g^2 h_1 h_2^2-fe_5 e_6 e_7},\\
s_1: & \ g\to \frac{g \left(f-e_4\right) e_8}{f e_8-h_1},\\ 
\pi_1:&\ f\to \frac{f}{h_1},\quad g\to -\frac{f}{g e_1},\\
\pi_2:& \ f\to \frac{g^2 h_1 h_2^2}{f e_5 e_6},\quad 
g\to -\frac{g e_1}{f}.
 \end{split}
\end{equation}
%

\subsubsection{$q$-P($\underset{|\alpha|^2=8}{A_1^{(1)}}/A_7^{(1)}$)}\label{subsubsec:Weyl_q-A1a}

\noindent Point configuration:
\begin{equation}
\begin{split}
(f_i,g_i)=&\left(-\frac{h_1}{\epsilon e_1e_8},\epsilon\right)_2,\quad 
\left(-\frac{e_2e_3}{\epsilon}, \frac{1}{\epsilon}\right)_2,\quad
\left(e_4,\infty \right),\quad
\left(\frac{h_1h_2^2}{e_5e_6e_7}\epsilon^2,\epsilon\right)_3.
\end{split}
\end{equation}
Generators and actions on parameters:
\begin{equation}
  \begin{split}
\pi_1:&\ e_{i}\to e_{\sigma(i)}, \ 
\sigma={\scriptsize\begin{pmatrix} 23678\\ 67184\end{pmatrix}}, \quad
e_1\ \to\ \frac{h_2}{e_2}, \ 
e_4\ \to\ \frac{h_1h_2}{e_2e_3e_5}, \ 
e_5\ \to\ \frac{h_1}{e_2}, \\
&\ h_1\ \to\ \frac{h_1h_2}{e_2e_5}, \quad
h_2\ \to\ \frac{h_1h_2}{e_1e_3},\\
\pi_2:&
\ e_{i}\to e_{\sigma(i)}^{-1}, \quad \sigma={\scriptsize\begin{pmatrix} 134678 \\ 643187 \end{pmatrix}}, \quad
e_2\ \to\ \frac{e_2}{h_2}, \quad
e_5\ \to\ \frac{e_5}{h_2}, \quad
e_5\ \to\ \frac{h_1h_2}{e_1e_2e_5},\\
&\  h_1\ \to\ \frac{e_2e_5}{h_1h_2}, \quad 
h_2\ \to\ h_2^{-1}.
 \end{split}
\end{equation}

\noindent Actions on $f$, $g$:
\begin{equation}
 \begin{split}
\pi_1:&\ f\ \to\ -\frac{h_1h_2 g}{e_5 f}, \quad g\ \to\ \frac{e_2e_3}{h_2(f + e_2e_3 g)},\\ 
\pi_2:& \ f\ \to\ -\frac{e_2 g}{f}, \quad g\ \to\ h_2 g.
 \end{split}
\end{equation}
\begin{rem}\rm \hfill
 \begin{itemize}
  \item The reflections corresponding to $\alpha_0=H_1+2H_2-2E_2-2E_3+E_4-E_5-E_6-E_7$ and
	$\alpha_1=H_1-E_1+E_2+E_3-2E_4-E_8$ ($|\alpha_0|^2=|\alpha_1|^2=8$) do not exist.
  \item The translation is given by $\pi_1$.
 \end{itemize}
\end{rem}

%

\subsubsection{d-P$({E_8^{(1)}}/A_0^{(1)})$}\label{subsubsec:Weyl_d-E8}
\noindent Point configuration:
\begin{equation}
 (f_i,g_i)= (e_i(e_i-h_1), e_i(e_i-h_2))\quad (i=1,\ldots,8). 
\end{equation}
Generators and actions on parameters:
\begin{equation}
\begin{split}
&s_0=s_{12},\ s_1 = s_{H_1-H_2},\ s_2 =s_{H_2-E_1-E_2},\ s_3 = s_{23},\\
&s_4=s_{34},\ s_5=s_{45},\ s_6=s_{56},\ s_7=s_{67},\ s_8=s_{78}.
\end{split}
\end{equation}
Actions on $f$, $g$:
\begin{equation}
\begin{split}
 s_1: &\ f\ \leftrightarrow\ g ,\\
 s_2: &\ f\ \rightarrow \frac{1}{(e_1+e_2-h_2) f - (e_1+e_2-h_1)g + e_1e_2(h_1-h_2)}\\
&\times\left[(h_1-h_2) f g - (e_1-h_2)(e_2-h_2)(e_1 e_2-h_1)f \right.\\
&\hskip70pt \left. + (e_1-h_1)(e_2-h_1)(e_1+e_2-h_2) g\right].
\end{split}
\end{equation}

%
\subsubsection{d-P($E_7^{(1)}/A_1^{(1)}$)}\label{subsubsec:Weyl_d-E7}
Point configuration:
\begin{equation}
(f_i,g_i)=(e_i,-e_i),\quad (i=1,\ldots,4),\quad (h_1-e_i,e_i-h_2)\quad (i=5,\ldots,8).
\end{equation}
Generators and actions on parameters:
\begin{equation}
\begin{split}
&s_0=s_{H_1-H_2},\ s_1 = s_{34},\ s_2 =s_{23},\ s_3 = s_{12},\\
&s_4=s_{H_2-E_1-E_5},\ s_5=s_{56},\ s_6=s_{67},\ s_7=s_{78},\\
&\pi:\ h_i\ \leftrightarrow\ -h_i,\ 
e_{i}\ \leftrightarrow -e_{\sigma(i)},\ \sigma={\scriptsize \begin{pmatrix}12345678\\ 56781234\end{pmatrix}}.
\end{split}
\end{equation}
Actions on $f$, $g$:
\begin{equation}
\begin{split}
s_0:&\ f\ \rightarrow\ -g,\ g\ \rightarrow\ -f,\\
s_4:&\ f\ \rightarrow\ \frac{-(h_1-h_2)fg - (e_1+e_5-h_1)(e_5-h_2)f - (e_1+e_5- h_2)(e_5-h_1)g  }{(e_1+e_5-h_2)f + (e_1+e_5-h_1)g - e_1(h_1-h_2)},\\
\pi:& \ f\ \rightarrow\ f-h_1,\ g\ \rightarrow\ g+h_2.
\end{split}
\end{equation}

%
\subsubsection{d-P($E_6^{(1)}/A_2^{(1)}$)}\label{subsubsec:Weyl_d-E6}
Point configuration:
\begin{equation}
(f_i,g_i)=(e_i,-e_i)\ (i=1,\ldots,4),\quad
(\infty,e_i-h_2)\ (i=5,6),\quad (h_1-e_i,\infty)\ (i=7,8). 
\end{equation}
Generators and actions on parameters:
\begin{equation}
\begin{split}
&s_0=s_{78},\ s_1 = s_{65},\ s_2 =s_{H_2-E_1-E_6},\ s_3 = s_{12},\\
& s_4=s_{23},\ s_5=s_{34},\ s_6=s_{H_1-E_1-E_7},\\
\pi_1:&\ h_1\ \rightarrow e_1+e_2-h_1-h_2,\ h_2\ \rightarrow\ -h_2,\ 
e_1\ \rightarrow\ e_2-h_2,\ e_2\ \rightarrow\ e_1-h_2,\\
&\ e_i\ \rightarrow -e_{\sigma(i)}\ \sigma={\scriptsize \begin{pmatrix}345678\\ 654378\end{pmatrix}}\ (i\neq 1,2),\\
\pi_2:&\ h_1\ \rightarrow\ -h_2,\ h_2\ \rightarrow\ -h_1,\ 
e_{i}\ \rightarrow -e_{\sigma(i)},\ \sigma={\scriptsize \begin{pmatrix}12345678\\ 12348765\end{pmatrix}}.
\end{split}
\end{equation}
Actions on $f$, $g$:
\begin{equation}
\begin{split}
s_2:&\ f\ \rightarrow\ \frac{g(f-e_1)-(e_6-h_2)(f+g)}{g+e_1},\\
s_6:&\ g\ \rightarrow\ \frac{f(g+e_1)+(e_7-h_1)(f+g)}{f-e_1},\\
\pi_1:&\ f\ \rightarrow\ \frac{g(-f + e_2) + e_1(g+e_2) - h_2(f+g)}{f+g},\ g\ \rightarrow\ g+h_2,\\
\pi_2:&\ f\ \leftrightarrow\ g.
 \end{split}
\end{equation}
%

\subsubsection{d-P$({D_4^{(1)}}/D_4^{(1)})$}\label{subsubsec:Weyl_d-D4}
In the following additive cases, we use the root parameters $a_i$ instead of the parameters $h_i$, $e_i$,
and variables $(q,p)$ as the dependent variables. 
\par\medskip

\noindent Point configuration:
\begin{equation}\label{eqn:conf_d-D4_fg}
 \begin{split}
 (f_i,g_i)=(q_i,q_ip_i)
=& (\infty,-a_2),\  (\infty,-a_1-a_2),\  
\left(t(1+a_0 \epsilon),\frac{1}{\epsilon}\right)_2, \\
&(0,0),\ (0,a_4),\ \left(1+a_3 \epsilon,\frac{1}{\epsilon}\right)_2.
 \end{split}
\end{equation}
Generators and actions on parameters: 
\begin{equation}
 \begin{split}
& s_0 = s_{H_1-E_3-E_4},\ s_1 = s_{E_1-E_2},\ s_2 = s_{H_2-E_1-E_5},\ s_3 = s_{H_1-E_7-E_8},\ s_4 = s_{E_5-E_6},\\
s_0 :&\ a_0\to -a_0,\ a_2\to a_0+a_2,\\
s_1 :&\ a_1\to -a_1,\ a_2\to a_1+a_2,\\
s_2 :&\ a_0\to a_0+a_2,\ a_1\to a_1+a_2,\ a_2\to -a_2,\ a_3\to a_2+a_3,\ a_4\to a_2+a_4,\\
s_3 :&\ a_2\to a_2+a_3,\ a_3\to -a_3, \\
s_4 :&\ a_2\to a_2+a_4,\ a_4\to -a_4,\\
\pi_1:&\ a_3\to a_4, a_4\to a_3,\\
\pi_2:&\ a_0\to a_3, a_3\to a_0,\\
\pi_3:&\ a_1\to a_4, a_4\to a_1.
\end{split}
\end{equation}
Actions on $p$, $q$:
\begin{equation}
 \begin{split}
s_0:&\ p\to p-\frac{a_0}{q-t}, \\
s_2:&\ q\to q+\frac{a_2}{p}, \\
s_3:&\ p\to p-\frac{a_3}{q-1}, \\
s_4:&\ p\to p-\frac{a_4}{q}, \\
\pi_1:&\ p\to -p,\ q\to 1-q, \ t\to 1-t,\\
\pi_2:&\ p\to p t,\ q\to \frac{q}{t},\ t\to \frac{1}{t}, \\
\pi_3:&\ p\to -q(qp+a_2),\ q\to \frac{1}{q}, \ t\to \frac{1}{t}.  
 \end{split}
\end{equation}
%

\subsubsection{d-P$({A_3^{(1)}}/D_5^{(1)})$}\label{subsubsec:Weyl_d-A3}
\noindent Point configuration:
\begin{equation}\label{eqn:conf_d-A3_pq}
\begin{split}
&(q_i,p_i) = \left(\frac{1}{\epsilon},-t-a_0\epsilon\right)_2,\ \left(\frac{1}{\epsilon},-a_2\epsilon\right)_2,\ 
\left(a_1\epsilon,\frac{1}{\epsilon}\right)_2,\ \left(1+a_3\epsilon,\frac{1}{\epsilon}\right)_2,\\
& a_0+a_1+a_2+a_3=1.
 \end{split}
\end{equation}

\noindent Generators and actions on parameters: 
\begin{equation}
\begin{split}
& s_0 = s_{H_2-E_1-E_2},\ s_1 = s_{H_1-E_5-E_6},\ s_2 = s_{H_2-E_3-E_4},\ s_3 = s_{H_1-E_7-E_8},\\
s_0:&\ a_0\to -a_0,\ a_1\to a_0+a_1,\ a_3\to a_3+a_0, \\
s_1:&\ a_0\to a_0+a_1,\ a_1\to -a_1,\ a_2\to a_1+a_2,\\
s_2:&\ a_1\to a_1+a_2, \ a_2\to -a_2,\ a_3\to a_2+a_3, \\
s_3:&\ a_0\to a_0+a_3,\ a_2\to a_2+a_3,\ a_3\to -a_3, \\
\pi_1:& \ a_0\to a_3,\ a_1\to a_2,\ a_2\to a_1,\ a_3\to a_0,\\
\pi_2:& \ a_0\to a_2,\ a_2\to a_0.
\end{split}
\end{equation}
\noindent Actions on $q$, $p$:
\begin{equation}
 \begin{split}
s_0:&\ q\to q+\frac{a_0}{p+t},\\
s_1:&\ p\to p-\frac{a_1}{q},\\
s_2:&\ q\to q+\frac{a_2}{p},\\
s_3:&\ p\to p-\frac{a_3}{q-1},\\
s_4:&\ q\to -\frac{p}{t},\ p\to (q-1)t,\\
\pi_1:&\ q\to -\frac{p}{t},\ p\to q t,\ t\to -t,\\
\pi_2:&\ p\to p+t,\ t\to -t.
 \end{split}
\end{equation}
%

\subsubsection{d-P$((2A_1)^{(1)}/D_6^{(1)})$}\label{subsubsec:Weyl_d-2A1}
\noindent Point configuration:
\begin{equation}\label{eqn:conf_dP2A1_pq}
\begin{split}
&(q_i,p_i)= \left(\frac{1}{\epsilon},1-a_1\epsilon\right)_2,\ \left(\frac{1}{\epsilon},-a_2\epsilon\right)_{2},\ 
\left(\epsilon,-\frac{t}{\epsilon^2}+\frac{1-a_1-a_2}{\epsilon}\right)_4.
\end{split}
\end{equation}
\noindent Generators and actions on parameters:
\begin{equation}
 \begin{split}
s_2 &= s_{H_2-E_3-E_4},\ s_1=s_{H_2-E_1-E_2},\\
s_1:&\ a_1\to -a_1, \\
s_2:&\ a_2\to -a_2, \\
\pi_1:&\ a_2\to a_1,\ a_1\to a_2, \\
\pi_2:&\ a_1\to 1-a_1.  
 \end{split}
\end{equation}

\noindent Actions on $q$, $p$:
\begin{equation}
\begin{split}
s_1:&\ q\to q + \frac{a_1}{p-1},\\
s_2:&\ q\to q + \frac{a_2}{p},\\ 
\pi_1:&\ q\to -q,\ p\to 1-p,\ t \to -t,\\
\pi_2:&\ q \to\frac{t}{q},\ p\to -\frac{q(q p+a_2)}{t}.
\end{split}
\end{equation}

%

\subsubsection{d-P$(\underset{|\alpha|^2=4}{A_{1}{}^{(1)}}/D_7^{(1)})$}\label{subsubsec:Weyl_d-A1'}

\noindent Point configuration:
\begin{equation}\label{eqn:conf_dPA1'}
\begin{split}
(q_i,p_i)=\left(-\frac{1}{\epsilon^2},\epsilon+\frac{a_1}{2}\epsilon^2\right)_4,\quad
\left(\epsilon,-\frac{t}{\epsilon^2}+\frac{1-a_1}{\epsilon}\right)_4.
\end{split}
\end{equation}

\noindent Generators and actions on parameters:
\begin{equation}
 \begin{split}
\pi_1:&\  a_1\to 1-a_1,\\
\pi_2:&\ a_1\to -a_1.
\end{split}
\end{equation}
\noindent Actions on $q$, $p$:
\begin{equation}
\begin{split}
\pi_1:&\ q\to tp,\ p\to -\frac{q}{t},\ t\to -t,\\
\pi_2:&\ q\to -q-\frac{a_1}{p}-\frac{1}{p^2},\ p\to -p,\ t\to-t.
\end{split}
\end{equation}
%

\subsubsection{d-P$(A_0^{(1)}/D_8^{(1)})$}\label{subsubsec:Weyl_d-A0}
\noindent Point configuration:
\begin{equation}\label{eqn:conf_dPA0}
\begin{split}
&(f_i,g_i)=\left(-\frac{1}{\epsilon^2},-\frac{1}{\epsilon}-\frac{1}{2}\right)_4, \quad \left(-t\epsilon^2, \frac{1}{\epsilon}\right)_4.
\end{split}
\end{equation}

\noindent The symmetry group of this case is a finite group $S_2$ generated by
\begin{equation}
\pi:\ q\to \frac{t}{q},\quad p\to -\frac{q(2qp+1)}{2 t}.
\end{equation}

%

\subsubsection{d-P$(A_2^{(1)}/E_6^{(1)})$}\label{subsubsec:Weyl_d-A2}
\noindent Point configuration:
\begin{equation}\label{eqn:conf_dPA2}
\begin{split}
&(q_i,p_i) = \left(\frac{1}{\epsilon},-a_2\epsilon\right)_2,\ 
\left(\frac{1}{\epsilon}, \frac{1}{\epsilon}+t -a_0\epsilon\right)_4,\ 
\left(a_1\epsilon,\frac{1}{\epsilon}\right)_2,\\
& a_0+a_1+a_2=\delta.
\end{split}
\end{equation}

\noindent Generators and actions on parameters:
\begin{equation}
 \begin{split}
s_0 &= s_{H_1+H_2-E_5-E_6-E_7-E_8},\quad s_1 = s_{H_1-E_3-E_4},\quad s_2 = s_{H_2-E_1-E_2},\\
s_0:&\ a_0\to -a_0,\ a_1\to  a_0+a_1,\ a_2\to  a_0+a_2, \\
s_1:&\ a_0\to a_0+a_1,\ a_1\to  -a_1,\ a_2\to  a_1+a_2, \\
s_2:&\ a_0\to a_0+a_2,\ a_1\to  a_1+a_2,\ a_2\to  -a_2, \\
\pi_1:&\ a_0\to -a_0,\ a_1\to -a_2,\ a_2\to -a_1,\\
\pi_2:&\ a_0\to -a_2,\ a_1\to -a_1,\ a_2\to -a_0.
 \end{split}
\end{equation}
\noindent Actions on $q$, $p$:
\begin{equation}
\begin{split}
s_0:&\ q\to  q + \frac{a_0}{p-q-t},\ p\to  p+\frac{a_0}{p-q-t}, \\
s_1:&\ p\to  p-\frac{a_1}{q}, \\
s_2:&\ q\to  q +\frac{a_2}{p},\\
\pi_1:&\ q\to -p,\ p\to  -q,\\
\pi_2:&\ p\to  -p+q+t.
\end{split}
\end{equation}
Note that $\pi_1$ and $\pi_2$ change the sign of $\delta$.  In Section \ref{subsubsec:P4_intro} we
have shown the composition $\pi=\pi_1\pi_2$ as a B\"acklund transformation which preserves the
constraint $\delta=1$.
%

\subsubsection{d-P$(A_1^{(1)}/E_7^{(1)})$}\label{subsubsec:Weyl_d-A1}
\noindent Point configuration:
\begin{equation}\label{eqn:conf_dPA1}
(q_i,p_i)=\left(\frac{1}{\epsilon}, -a_1\epsilon\right)_2, \quad
\left(\frac{1}{\epsilon},\frac{2}{\epsilon^2}+t+(a_1-1)\epsilon\right)_6.
\end{equation}

\noindent Generators and actions on parameters:
\begin{equation}
 \begin{split}
s_0 &= s_{2H_1+H_2-E_3-E_4-E_5-E_6-E_7-E_8},\quad s_1 = s_{H_2-E_1-E_2},\\
s_1:&\ a_1\to -a_1\\
\pi:&\ a_1\to 1-a_1.
 \end{split}
\end{equation}

\noindent Actions on $q$, $p$:
\begin{equation}
\begin{split}
s_1 :&\ q\to q+\frac{a_1}{p},\\
\pi :&\ q\to-q,\ p\to -p+2q^2+t.
\end{split}
\end{equation}
\subsection{Lax pairs}\label{subsec:Lax_list}
In this section, we mainly use the parameters, $\kappa_1,\kappa_2, v_1,\ldots,v_8$, with
$\kappa_1^2\kappa_2^2=q \prod_{i=1}^8 v_i$ for multiplicative cases and
$2\kappa_1+2\kappa_2=\delta+\sum_{i=1}^8 v_i$ for additive cases.  $T_z$ denotes the shift operator
in $z$ such that $T_z: z \mapsto q z$ for multiplicative cases and $T_z: z \mapsto z+\delta$ for
additive cases.  Also, $T$ stands for the shift operator of the time evolution
$T=T_{\kappa_1}^{-1}T_{\kappa_2}$, and we write $\overline{f}=T(f)$ and $\underline{g}=T^{-1}(g)$.
We also include the list of points configuration characterizing the equation $L_1y(z)=0$ as a curve
of degree $(3,2)$ in $(f,g)$.
%
\subsubsection{$q$-P$(E_8^{(1)}/A_0^{(1)})$} \label{subsubsec:Lax_q-E8}
As given in \eqref{eqn:qpe8_L1} and \eqref{eqn:qpe8_L2}, the Lax pair for 
$q$-P$(E_8^{(1)}/A_0^{(1)})$ is
\begin{equation}
\begin{split}
&L_1=\dfrac{w(f,g)\left(\frac{z}{q}-\frac{\kk_1}{z}\right)\left\{\of-\of\left(\frac{z}{q}\right)\right\}}
{\left\{g - g\left(\frac{z}{q}\right)\right\}\left\{g-g\left(\frac{\kk_1}{z}\right)\right\}}
+ \dfrac{U\left(\frac{z}{q}\right)}{\left(\frac{z}{q}-\frac{q\kk_1}{z}\right)\left\{f-f(\frac{z}{q})\right\}}
\left[T_z^{-1} - \dfrac{g - g\left(\frac{q\kk_1}{z}\right)}{g - g\left(\frac{z}{q}\right)}\right]\\
&\quad +\dfrac{U\left(\frac{\kk_1}{z}\right)}{\left(z-\frac{\kk_1}{z}\right)\left\{f-f(z)\right\}}
\left[T_z - \dfrac{g-g(z)}{g-g\left(\frac{\kk_1}{z}\right)}\right],\\
&L_2=\left\{g-g\left(\frac{\kk_1}{z}\right)\right\}T_z
-\left\{g-g(z)\right\}
-\left(z-\frac{\kk_1}{z}\right)\left\{f-f(z)\right\}T,
\end{split}
\end{equation}
where $f(z)=z+\frac{\kk_1}{z}$, $g(z)=z+\frac{\kk_2}{z}$, $U(z)=z^{-4}\prod_{i=1}^8(z-v_i)$ and
$w(f,g)$ is a rational function in $f,g$ (independent of $z$) such as
\begin{equation}
w(f,g)\Big{|}_{g=g(z)}=q\frac{(1-\frac{\kk_2}{\kk_1})\frac{\kk_2}{\kk_1}}{z-\frac{\kk_2}{z}}
\left\{\frac{U(z)}{f-f(z)}-\frac{U\left(\frac{\kk_2}{z}\right)}{f-f\left(\frac{\kk_2}{z}\right)}\right\}.
\end{equation}
The linear equation $L_1y(z)=0$ is uniquely characterized as a curve of degree $(3,2)$ in $(f,g)$ passing through the 12 points:
\begin{equation}
\Bigl(f(u),g(u)\Bigr)_{u=v_1, \ldots, v_8, z, \frac{q\kk_1}{z}}, \quad 
\Bigl(f(u),\gamma_u\Bigr)_{u=z,\frac{z}{q}}, 
\end{equation}
where $\dfrac{\gamma_u-g\Bigl(\frac{\kk_1}{u}\Bigr)}{\gamma_u-g(u)}=\dfrac{y(u)}{y(qu)}$
for $u=z, \frac{z}{q}$.

%
\subsubsection{$q$-P$(E_7^{(1)}/A_1^{(1)})$} \label{subsubsec:Lax_q-E7}
\begin{equation}
\begin{split}
L_1=&
\left\{
\dfrac{q B_2\left(\frac{1}{cg}\right)\left(1-\frac{1}{c}\right)}{(1-z cg) (f-\frac{1}{cg})}
+\dfrac{B_1\left(\frac{1}{g}\right)\left(1-c\right)}{\left(1-\frac{z}{q}g\right)\left(f-\frac{1}{g}\right)}
\right\}
+\dfrac{B_1\left(\frac{z}{q}\right) }{f-\frac{z}{q}} \left(T_z^{-1}-\dfrac{q-zcg}{q-zg} \right)\\
&\hskip40pt 
+\dfrac{q B_2(z)}{f-z} \left(T_z-\dfrac{1-zg}{1-zcg} \right),\\
L_2= & (1-zcg)T_z -(1-zg) + z (z-f)gT ,
\end{split}
\end{equation}
where 
$B_1(z)=\frac{1}{z^2}\prod\limits_{i=1}^4\left(1-\frac{z}{v_i}\right)$, 
$B_2(z)=\frac{1}{z^2}\prod\limits_{i=5}^8\left(1-\frac{v_i}{\kappa_1} z\right)$
and $c=\frac{\kappa_2}{\kappa_1}$.\\[2mm]
\noindent Point configuration:
\begin{equation}
\Bigl(v_i,\frac{1}{v_i}\Bigr)_{i=1}^4, \quad
\Bigl(\frac{\kappa_1}{v_i},\frac{v_i}{\kappa_2}\Bigr)_{i=5}^8, \quad
\Bigl(\frac{z}{q}, \frac{q\kappa_1}{\kappa_2 z} \Bigr),\quad  \Bigl(z,\frac{1}{z}\Bigr), \quad
\bigl(z,\gamma_z\bigr), \quad \bigl(\frac{z}{q},\gamma_{\frac{z}{q}}\bigr),
\end{equation}
where $\gamma_u$ is given by
$\frac{1-u \gamma_u}{1-u \frac{\kappa_2}{\kappa_1} \gamma_u}=\frac{y(qu)}{y(u)}$
 ($u=z,\frac{z}{q}$).
%
\subsubsection{$q$-P$(E_6^{(1)}/A_2^{(1)})$} \label{subsubsec:Lax_q-E6}
\begin{equation}
\begin{split}
L_1=& \dfrac{\frac{z}{q} \prod\limits_{i=1}^4 (g v_i-1)}{g (f g-1) \left(\frac{z}{q}g - 1\right)}
-\dfrac{\prod\limits_{i=5}^6\left(\frac{g \kappa _2}{v_i}-1\right) \kappa _1^2}{f g q v_7 v_8}
+\dfrac{\prod\limits_{i=1}^4\left(v_i-\frac{z}{q}\right) }{f-\frac{z}{q}}\left\{\frac{g}{1-g \frac{z}{q}}-T_z^{-1}\right\}\\
&\hskip80pt 
+\dfrac{\prod\limits_{i=7}^8\left(\frac{\kappa_1}{v_i}-z\right) }{q (f-z)}\left\{\left(\frac{1}{g}-z\right) -T_z\right\},\\
L_2=& \left(1-\frac{f}{z}\right) T  + T_z - \left(\frac{1}{g}-z\right).\\
\end{split}
\end{equation}
\noindent Point configuration:
\begin{equation}
\Bigl(v_i,\frac{1}{v_i}\Bigr)_{i=1}^4, \quad
\Bigl(0,\frac{v_i}{\kappa_2}\Bigr)_{i=5}^6, \quad
\Bigl(\frac{\kappa_1}{v_i},0\Bigr)_{i=7}^8, \quad
\Bigl(\frac{z}{q}, 0\Bigr),\quad  \Bigl(z,\frac{1}{z}\Bigr), \quad
\bigl(z,\gamma_z\bigr), \quad \bigl(\frac{z}{q},\gamma_{\frac{z}{q}}\bigr),
\end{equation}
where $\gamma_u$ is given by
$\frac{1}{\gamma_u}-u=\frac{y(qu)}{y(u)}$ ($u=z,\frac{z}{q}$).
\begin{rem}\rm 
In order to obtain the Lax pair of $q$-P$(E_6^{(1)}/A_2^{(1)})$ by degeneration from
$q$-P$(E_7^{(1)}/A_2^{(1)})$, we need to apply the following gauge transformation
 \begin{equation}
  y(z) \to z^{\log_q(\frac{\varepsilon \kappa_1}{\kappa_2})}G_1(z)y(z),\quad G_1(z/q)=-zG_1(z),
 \end{equation}
together with the limiting procedure given in Section \ref{subsec:degeneration_qe7_to_qe6}.
\end{rem}
%
\subsubsection{$q$-P$(D_5^{(1)}/A_3^{(1)})$} \label{subsubsec:Lax_q-D5}
\begin{small}
\begin{equation}\label{eqn:L1L2_D5A3}
\begin{split}
L_1=& \left\{\dfrac{z \prod\limits_{i=1}^2(g v_i-1)}{qg}
-\dfrac{\prod\limits_{i=1}^4 v_i\, \prod\limits_{i=5}^6\left(g -\frac{v_i}{\kappa_2}\right) }{f g} \right\}
+\dfrac{v_1 v_2\prod\limits_{i=3}^4\left(\frac{z}{q}-v_i\right)}{f-\frac{z}{q}}\left(g -T_z^{-1}\right)
+\dfrac{\prod\limits_{i=7}^8\left(\frac{\kappa_1}{v_i}-z\right) }{q (f- z)}\left(T_z-\frac{1}{g}\right),\\
L_2=& \left(1-\frac{f}{z}\right)T  + T_z -\frac{1}{g}.
\end{split}
\end{equation}
\end{small}
\noindent Point configuration:
\begin{equation}
\begin{split}
&\Bigl(\infty,\frac{1}{v_i}\Bigr)_{i=1}^2, \quad
\Bigl(v_i,\infty\Bigr)_{i=3}^4, \quad
\Bigl(0,\frac{v_i}{\kappa_2}\Bigr)_{i=5}^6, \quad
\Bigl(\frac{\kappa_1}{v_i},0\Bigr)_{i=7}^8, \quad \\
&\Bigl(\frac{z}{q}, 0\Bigr),\quad  \Bigl(z,\infty\Bigr), \quad
\Bigl(z,\frac{y(qz)}{y(z)}\Bigr), \quad \Bigl(\frac{z}{q},\frac{y(z)}{y(\frac{z}{q})}\Bigr).
\end{split}
\end{equation}
\begin{rem}\rm
In order to take the degeneration limit from $q$-P$(E_6^{(1)}/A_2^{(1)})$ to
$q$-P$(D_5^{(1)}/A_3^{(1)})$ described in Section \ref{subsec:degeneration_qe6_to_qd5}, we
need to change variable $z \to z\varepsilon$.
\end{rem}

%
\subsubsection{$q$-P$(A_4^{(1)}/A_4^{(1)})$} \label{subsubsec:Lax_q-A4}
 \begin{equation}
L_1=\dfrac{\prod\limits_{i=5}^6\left(g- \frac{v_i}{\kappa_2}\right)\prod\limits_{i=1}^4v_i}{f g}
+\dfrac{z(g v_1-1)}{qg}
+\dfrac{\left(\frac{z}{q}-v_4\right) \prod\limits_{i=1}^3 v_i}{f-\frac{z}{q}}\left(g -T_z^{-1}\right)
+\dfrac{\prod\limits_{i=7}^8\left(z-\frac{\kappa_1}{v_i}\right)}{q (f-z)} \left(T_z-\frac{1}{g}\right).
\end{equation}
In this case and the further degenerations the $L_2$ operators are omitted, since they are the same as $q$-P($D_5^{(1)}/A_3^{(1)}$) case \eqref{eqn:L1L2_D5A3}. 
The extra four points are also the same as $q$-P($D_5^{(1)}/A_3^{(1)}$) case and will be omitted.

\noindent Point configuration:
\begin{equation}
\begin{split}
&\Bigl(\infty,\frac{1}{v_1}\Bigr), \quad
\Bigl(-v_2v_3\frac{1}{\epsilon},\frac{1}{\epsilon}\Bigr)_2, \quad
\Bigl(v_4,\infty\Bigr), \quad
\Bigl(0,\frac{v_i}{\kappa_2}\Bigr)_{i=5}^6, \quad
\Bigl(\frac{\kappa_1}{v_i},0\Bigr)_{i=7}^8.
\end{split}
\end{equation}
%
\subsubsection{$q$-P$(\esan^{(1)}/A_5^{(1)};a)$} \label{subsubsec:Lax_q-E3a}
\begin{equation}
L_1=\dfrac{\prod\limits_{i=5}^6\left(g-\frac{v_i}{\kappa_2}\right) \prod\limits_{i=1}^4v_i}{f g} + \frac{z}{q} v_1
+\dfrac{\left(\frac{z}{q}-v_4\right) \prod\limits_{i=1}^3 v_i}{f-\frac{z}{q}}\left(g -T_z^{-1}\right)
-\dfrac{\frac{\kappa_1}{v_8}\left(z-\frac{\kappa_1}{v_7}\right)}{q (f-z)} \left(T_z-\frac{1}{g}\right).
\end{equation}
\noindent Point configuration:
\begin{equation}
\begin{split}
&\Bigl(-\frac{\kappa_1}{v_1v_8\epsilon}, \epsilon\Bigr)_2, \quad
\Bigl(-v_2v_3\frac{1}{\epsilon},\frac{1}{\epsilon}\Bigr)_2, \quad
\Bigl(v_4,\infty\Bigr), \quad
\Bigl(0,\frac{v_i}{\kappa_2}\Bigr)_{i=5}^6, \quad
\Bigl(\frac{\kappa_1}{v_7},0\Bigr).
\end{split}
\end{equation}
%
\subsubsection{$q$-P$(\eni^{(1)}/A_6^{(1)};a)$} \label{subsubsec:Lax_q-E2a}
\begin{equation}
L_1=\dfrac{\left(g-\frac{v_5}{\kappa_2}\right)\prod\limits_{i=1}^4v_i }{f} + \frac{z}{q} v_1
+\dfrac{\left(\frac{z}{q}-v_4\right) \prod\limits_{i=1}^3 v_i}{f-\frac{z}{q}}\left(g -T_z^{-1}\right)
-\dfrac{\frac{\kappa_1}{v_8}\frac{z}{q}}{f-z}\left(T_z-\frac{1}{g}\right).
\end{equation}
\noindent Point configuration:
\begin{equation}
\begin{split}
&\Bigl(-\frac{\kappa_1}{v_1v_8\epsilon}, \epsilon\Bigr)_2, \quad
\Bigl(-v_2v_3\frac{1}{\epsilon},\frac{1}{\epsilon}\Bigr)_2, \quad
\Bigl(v_4,\infty\Bigr), \quad
\Bigl(0,\frac{v_5}{\kappa_2}\Bigr), \quad
\Bigl(-\frac{\kappa_1\kappa_2}{v_6v_7}\epsilon,\epsilon\Bigr)_2.
\end{split}
\end{equation}
%
\subsubsection{$q$-P$(\underset{|\alpha|^2=8}{A_{1}^{(1)}}/A_7^{(1)})$}\label{subsubsec:Lax_q-A1a} 
\begin{equation}
L_1=\dfrac{g \prod\limits_{i=1}^4v_i}{f} + \frac{z}{q} v_1 
+\dfrac{\left(\frac{z}{q}-v_4\right) \prod\limits_{i=1}^3 v_i}{f-\frac{z}{q}}\left(g -T_z^{-1}\right)
-\dfrac{\frac{\kappa_1}{v_8}\frac{z}{q}}{f-z}\left(T_z-\frac{1}{g}\right).
\end{equation}
\noindent Point configuration:
\begin{equation}
\begin{split}
&\Bigl(-\frac{\kappa_1}{v_1v_8\epsilon}, \epsilon\Bigr)_2, \quad
\Bigl(-v_2v_3\frac{1}{\epsilon},\frac{1}{\epsilon}\Bigr)_2, \quad
\Bigl(v_4,\infty\Bigr), \quad
\Bigl(\frac{\kappa_1\kappa_2^2}{v_5v_6v_7}\epsilon,\epsilon\Bigr)_3.
\end{split}
\end{equation}
%
\subsubsection{$q$-P$(\esan^{(1)}/A_5^{(1)};b)$} \label{subsubsec:Lax_q-E3b}
\begin{equation}
L_1=\dfrac{\left(g-\frac{v_5}{\kappa_2}\right)\prod\limits_{i=1}^4v_i}{f} + \dfrac{z (g v_1-1)}{qg}
+\dfrac{\left(\frac{z}{q}-v_4\right) \prod\limits_{i=1}^3 v_i}{f-\frac{z}{q}}\left(g -T_z^{-1}\right)
+\dfrac{\frac{z}{q} \left(z-\frac{\kappa_1}{v_8}\right)}{f-z} \left(T_z-\frac{1}{g}\right).
\end{equation}
\noindent Point configuration:
\begin{equation}
\begin{split}
&\Bigl(\infty,\frac{1}{v_1}\Bigr), \quad
\Bigl(-v_2v_3\frac{1}{\epsilon},\frac{1}{\epsilon}\Bigr)_2, \quad
\Bigl(v_4,\infty\Bigr), \quad
\Bigl(0,\frac{v_5}{\kappa_2}\Bigr), \quad
\Bigl(-\frac{\kappa_1\kappa_2}{v_6v_7}\epsilon,\epsilon\Bigr)_2, \quad
\Bigl(\frac{\kappa_1}{v_8},0\Bigr).
\end{split}
\end{equation}
%
\subsubsection{$q$-P$(\eni^{(1)}/A_6^{(1)};b)$} \label{subsubsec:Lax_q-E2b}
\begin{equation}
L_1=\dfrac{g \prod\limits_{i=1}^4v_i}{f}+\dfrac{z (g v_1-1)}{qg}
+\dfrac{\left(\frac{z}{q}-v_4\right) \prod\limits_{i=1}^3 v_i}{f-\frac{z}{q}}\left(g -T_z^{-1}\right)
+\dfrac{\frac{z}{q} \left(z-\frac{\kappa_1}{v_8}\right)}{f-z} \left(T_z-\frac{1}{g}\right).
\end{equation}
\noindent Point configuration:
\begin{equation}
\begin{split}
&\Bigl(\infty,\frac{1}{v_1}\Bigr), \quad
\Bigl(-v_2v_3\frac{1}{\epsilon},\frac{1}{\epsilon}\Bigr)_2, \quad
\Bigl(v_4,\infty\Bigr), \quad
\Bigl(\frac{\kappa_1\kappa_2^2}{v_5v_6v_7}\epsilon,\epsilon\Bigr)_3, \quad
\Bigl(\frac{\kappa_1}{v_8},0\Bigr).
\end{split}
\end{equation}
%
\subsubsection{$q$-P$(A_{1}^{(1)}/A_7^{(1)})$} \label{subsubsec:Lax_q-A1b}
\begin{equation}
L_1=\dfrac{g \prod\limits_{i=1}^4v_i}{f}-\dfrac{z}{qg}
+\dfrac{\left(\frac{z}{q}-v_4\right) \prod\limits_{i=1}^3 v_i}{f-\frac{z}{q}}\left(g -T_z^{-1}\right)
+\dfrac{\frac{z}{q} \left(z-\frac{\kappa_1}{v_8}\right)}{f-z}\left(T_z-\frac{1}{g}\right).
\end{equation}
\noindent Point configuration:
\begin{equation}
\begin{split}
&\Bigl(v_1v_2v_3\frac{1}{\epsilon},\frac{1}{\epsilon}\Bigr)_3, \quad
\Bigl(v_4,\infty\Bigr), \quad
\Bigl(\frac{\kappa_1\kappa_2^2}{v_5v_6v_7}\epsilon,\epsilon\Bigr)_3, \quad
\Bigl(\frac{\kappa_1}{v_8},0\Bigr).
\end{split}
\end{equation}
%
\subsubsection{d-P$(E_8^{(1)}/A_0^{(1)})$} \label{subsubsec:Lax_d-E8}
\begin{equation}
 \begin{split}
L_1& =
\frac{(2z-\delta-\kappa_1)w\Big\{\overline{f}-\overline{f}(z-\delta)\Big\}}
{\Big\{g-g(z-\delta)\Big\}\Big\{g-g(\kappa_1-z)\Big\}}\\
&+
\frac{U(z-\delta)}{(2z-2\delta-\kappa_1)\Big\{f-f(z-\delta)\Big\}}
\left[T_z^{-1} - \frac{g-g(\kappa_1+\delta-z)}{g-g(z-\delta)} 
\right]\\
& + \frac{U(\kappa_1-z)}{(2z-\kappa_1)\Big\{f-f(z)\Big\}}
\left[T_z - \frac{g-g(z)}{g-g(\kappa_1-z)}\right],\\
L_2 &=
\Big\{g-g(\kappa_1-z)\Big\}T_z - \Bigl\{g-g(z)\Bigr\} - (2z-\kappa_1)\Bigl\{f-f(z)\Bigr\}T^{-1},
\end{split}
\end{equation}
where $f(z) = z(z-\kappa_1)$, $g(z)=z(z-\kappa_2)$, $U(z)=\prod_{i=1}^8(z-v_i)$ and 
\begin{equation}
 w = \frac{(\kappa_1-\kappa_2-\delta)(\kappa_1-\kappa_2)U(t)}
{\Bigl\{\overline{f}-\overline{f}(t)\Bigr\}\Bigl\{f-f(t)\Bigr\}},\quad g=g(t).
\end{equation}
Point configuration:
\begin{equation}
\begin{split}
&\Bigl(f(v_i),g(v_i)\Bigr)_{i=1}^8, \quad  (f(z),g(z)), \quad
 \left(f(\kappa_1+\delta-z),\ g(\kappa_1+\delta-z)\right),\\
&\Bigl(f(z), \gamma_z\Bigr), \quad
\Bigl(f(\kappa_1+\delta-z), \gamma_{\kappa_1+\delta-z}\Bigr),
\end{split}
\end{equation}
where $\frac{\gamma_u-g(u)}{\gamma_u-g(\kappa_1-u)}=\frac{y(u+\delta)}{y(u)}$ ($u=z, \kappa_1+\delta-z$).
%
\subsubsection{d-P$(E_7^{(1)}/A_1^{(1)})$} \label{subsubsec:Lax_d-E7}
\begin{equation}
 \begin{split}
  L_1=& (\kappa_1-\kappa_2)\left\{
  \frac{(f+g-\kappa_1+\kappa_2)\prod\limits_{i=1}^4(g+v_i)}{(f+g)(g+z)}
- \frac{\prod\limits_{i=5}^8(g+\kappa_2-v_i)}{g-\kappa_1+\kappa_2+z+\delta}
\right\}\\
& + \frac{f+g-\kappa_1+\kappa_2}{z-f}\prod\limits_{i=1}^4(z-v_i)
\left(T_z^{-1} - \frac{g-\kappa_1+\kappa_2+z}{g+z}\right)\\
&+ \frac{f+g-\kappa_1+\kappa_2}{z-f+\delta}\prod\limits_{i=5}^8(z+\delta-\kappa_1+v_i)
\left(
T_z - \frac{g+z+\delta}{g-\kappa_1+\kappa_2+z+\delta}
\right),\\
L_2=& (g+z)T_z^{-1} + (-g+\kappa_1-\kappa_2-z) - (f-z)TT_z^{-1}.
 \end{split}
\end{equation}
Point configuration:
\begin{equation}
\begin{split}
&(v_i,-v_i)_{i=1}^4, \quad
(\kappa_1-v_i, v_i-\kappa_2)_{i=5}^8, \quad
 (z,\kappa_1-\kappa_2-z),\quad (z+\delta,-z-\delta),\\
&\Bigl(z, \gamma_z\Bigr),\quad
\Bigl(z+\delta, \gamma_{z+\delta}\Bigr),
\end{split}
\end{equation}
where $\frac{\gamma_u-\kappa_1+\kappa_2+u}{\gamma_u+u}=\frac{y(u-\delta)}{y(u)}$
($u=z, z+\delta$).
%

\subsubsection{d-P$(E_6^{(1)}/A_2^{(1)})$} \label{subsubsec:Lax_d-E6}
\begin{equation}
 \begin{split}
 L_1&= -\frac{\prod\limits_{i=1}^4(g+v_i)}{f+g} + (g+z)\prod\limits_{i=5}^6(g+\kappa_2-v_i)
+\frac{\prod\limits_{i=1}^4(z-v_i)}{f-z}\left\{-(g+z)T_z^{-1} + 1\right\},\\
&\hskip60pt +\frac{(g+z)\prod\limits_{i=7}^8(z+\delta-\kappa_1+v_i)}{f-z-\delta}\left\{(1+g+z)-T_z\right\},\\
L_2 & = (g+z)T_z^{-1} - \delta - (f-z)TT_z^{-1}.
 \end{split}
\end{equation}
Point configuration:
\begin{equation}
\begin{split}
&(v_i,-v_i)_{i=1}^4, \quad
(\infty, v_i-\kappa_2)_{i=5}^6, \quad
(\kappa_1-v_i, \infty)_{i=7}^8, \quad
 (z,\infty),\quad (z+\delta,-z-\delta),\\
&\Bigl(z, -z+\frac{y(z)}{y(z-\delta)}\Bigr),\quad
\Bigl(z+\delta, -z-\delta+\frac{y(z+\delta)}{y(z)}\Bigr).
\end{split}
\end{equation}

\subsubsection{d-P$(D_4^{(1)}/D_4^{(1)})$} \label{subsubsec:Lax_d-D4}
The following cases admit both the discrete flows and the continuous flows (i.e. Painlev\'e
differential equations). Both flows can be described as (i) deformations of a linear differential
equation and (ii) deformations of a linear difference equation.  We use two different coordinates
$(q,p)$ and $(f,g)=(q,qp)$ depending on the surfaces and the type of flows.  The point
configurations on $\mathbb{P}_q^1\times\mathbb{P}_p^1$ and/or $\mathbb{P}_f^1\times \mathbb{P}_g^1$
are shown schematically in Figure \ref{fig:degeneration_additive}.  The corresponding configurations
in blown-up space are given in Figure \ref{fig:config_add_blowup}.  For simplicity, we use the root
parameters $a_i$ instead of the parameters $\kappa_i, v_i$.

The continuous flow of the case d-P$(D_4^{(1)}/D_4^{(1)})$ is P$_{\rm VI}$ given by the Hamiltonian:
\begin{equation}\label{eqn:P6_Ham}
H=\frac{q(q-1)(q-t)}{t(t-1)}\Big\{p^2-\Big(\frac{a_0-1}{q-t}+\frac{a_3}{q-1}+\frac{a_4}{q}\Big) p\Big\}
+\frac{(q-t) a_2 (a_1+a_2)}{t(t-1)},
\end{equation}
where $a_0 + a_1 + 2a_2 + a_3 + a_4 =1$. In $(f,g)=(q, q p)$ coordinates, the eight points configuration is given by
\begin{equation}\label{eqn:P6_8pts}
(f_i,g_i)= (\infty,-a_2),\ (\infty,-a_1-a_2),\ \left(t (1+a_0 \epsilon),\frac{1}{\epsilon}\right)_2,\ (0,0),\ (0,a_4),\ 
\left(1+a_3 \epsilon,\frac{1}{\epsilon}\right)_2.
\end{equation}

\noindent(i) Differential Lax form:
\begin{equation}
\begin{split}
&{\cal L}_1=\frac{1}{x(x-1)}\Big\{{a_2(a_1+a_2)}+\frac{q(q-1)p}{x-q}-\frac{t(t-1)H}{x-t}\Big\} \\
&\hskip80pt +\Big\{\frac{1-a_0}{x-t}+\frac{1-a_3}{x-1}+\frac{1-a_4}{x}-\frac{1}{x-q}\Big\} {\partial_x}+{\partial_x}^2,\\
&{\cal L}_2=T_\alpha-\frac{1}{q-x}(x \partial_x-q p),\\
&{\cal B}={\partial_t}-\frac{t-q}{t(t-1)(x-q)}\Big(x(x-1){\partial_x}-q (q-1) p\Big),
\end{split}
\end{equation}
The curve ${\cal L}_1y=0$ is the unique curve of degree $(3,2)$ in $(f,g)$ passing through
\eqref{eqn:P6_8pts} and
\begin{equation}\label{eqn:fg_y_1}
\left(x+\epsilon,-\frac{x}{\epsilon}\right)_2,\quad 
\left(x+\epsilon,(x+\epsilon)\frac{y'(x+\epsilon)}{y(x+\epsilon)}\right)_2.
\end{equation}
Compatibility of ${\cal L}_1y={\cal L}_2y=0$ gives the discrete flow for $T_{\alpha}(=\pi_3 \pi_2 s_3 s_0 s_2 s_1 s_4 s_2)$:
\begin{equation}
\begin{split}
&\overline{a}_0=a_0-1, \ \overline{a}_2=a_2+1, \ \overline{a}_3=a_3-1,\\
&f\overline{f}=\frac{g t (g-a_4)}{(g+a_2) (g+a_1+a_2)},\quad
g+\overline{g}=a_0+a_3+a_4-2+\frac{t(a_0-1)}{\overline{f}-t}+\frac{a_3-1}{\overline{f}-1}.
\end{split}
\end{equation}
Compatibility of $L_1y=By=0$ gives the P$_{\rm VI}$ flow with Hamiltonian \eqref{eqn:P6_Ham}.

\noindent(ii) Difference Lax form: 
\begin{equation}
\begin{split}
&L_1=f(f-1)(f-t)\left(\frac{a_0}{f-t}+\frac{a_3}{f-1}-\frac{z+g-a_4}{f}\right)\\
&\hskip40pt +\frac{f(z-1+a_2)(z-a_0 - a_2 - a_3 - a_4)}{z-1 - g }(f-T_z^{-1}) + \frac{tz (z - a_4)}{z-g }(1-fT_z),\\
&L_2 = T_\beta T_z + \frac{1}{z-g}(1 - fT_z),\\
&B = (f-1)zT_z + \frac{(z+a_2)(z+a_1+a_2)}{z-g}(1 - fT_z) +t (t-1) \partial_t T_z.\\
\end{split}
\end{equation}
The curve $L_1y=0$ is the unique curve of degree $(2,3)$ in $(f,g)$ passing through
\eqref{eqn:P6_8pts} and
\begin{equation}\label{eqn:fg_y_2}
(\infty,z),\quad (0,z-1), \quad
\left(\frac{y(z)}{y(z+1)},z\right), \quad
\left(\frac{y(z-1)}{y(z)},z-1\right).
\end{equation}
Compatibility of $L_1y=L_2y=0$ gives the discrete flow for 
$T_{\beta}(=\pi_3 \pi_2 s_2 s_1 s_4 s_2 s_3 s_0)$:
\begin{equation}
\begin{split}
&\overline{a}_0=a_0+1, \ \overline{a}_2=a_2-1, \ \overline{a}_3=a_3+1,\\
&f\overline{f}=\frac{\overline{g} t \left(\overline{g}-a_4\right)}{\left(\overline{g}+a_2-1\right) \left(\overline{g}-a_0-a_2-a_3-a_4\right)}, \quad
g+\overline{g}=a_0+a_3+a_4+\frac{t a_0}{f-t}+\frac{a_3}{f-1}.\\
\end{split}
\end{equation}
Compatibility of $L_1y=By=0$ gives the P$_{\rm VI}$ flow with Hamiltonian \eqref{eqn:P6_Ham}.
\bigskip

\subsubsection{d-P$(A_3^{(1)}/D_5^{(1)})$}  \label{subsubsec:Lax_d-A3}
The corresponding continuous flow is the P$_{\rm V}$ equation given by the Hamiltonian:
\begin{equation}\label{eqn:P5_Ham}
H=\frac{1}{t}\Big\{q(q-1) p (p+t)-({a_1}+{a_3}) q p+a_1 p+{a_2} t q\Big\},
\end{equation}
with $a_0+a_1+a_2+a_3=1$. The eight points configuration in $(f,g)=(q,qp)$ coordinates is given as:
\begin{align}
&(f_i,g_i)=(\infty,-a_2),\  \left(\frac{1}{\epsilon},-\frac{t}{\epsilon}-a_0\right)_3,\ (0,0),\ (0,a_1), \ \left(1+a_3 \epsilon,\frac{1}{\epsilon}\right)_2. \label{eqn:P5_8pts}
\end{align}
\noindent(i) Differential Lax form:
\begin{equation}
\begin{split}
&{\cal L}_1=\frac{1}{x(x-1)}\Big\{\frac{q(q-1) p}{x-q} + t(a_2 x-H)\Big\}  
+\Big\{\frac{1-{a_1}}{x}+t+\frac{1-{a_3}}{x-1}-\frac{1}{x-q}\Big\}{\partial_x}+{\partial_x}^2,\\
&{\cal L}_2=T_\alpha-\frac{1}{x-q}(p -\partial_x),\\
&{\cal B}={\partial_t}-\frac{1}{t(x-q)}\Big(x(x-1){\partial_x}-q (q-1) p\Big).
\end{split}
\end{equation}
The linear equation ${\cal L}_1y=0$ is characterized as a curve of degree $(3,2)$ in $(f,g)$ passing through
the points \eqref{eqn:P5_8pts} and \eqref{eqn:fg_y_1}.
Compatibility of ${\cal L}_1y= {\cal L}_2y=0$ gives the discrete flow for $T_\alpha(=(\pi_1\pi_2)^2s_1s_3s_0s_2)$:
\begin{equation}
\begin{split}
&\overline{a}_0 = a_0+1,\ \overline{a}_1=a_1-1, \ \overline{a}_2=a_2+1, \ \overline{a}_3=a_3-1,\\
&q+\overline{q}=1-\frac{a_2}{p}-\frac{a_0}{p+t}, \quad
p+\overline{p}=-t+\frac{a_1-1}{\overline{q}}+\frac{a_3-1}{\overline{q}-1}.
\end{split}
\end{equation}
Compatibility of ${\cal L}_1y={\cal B}y=0$ gives the P$_{\rm V}$ flow with Hamiltonian \eqref{eqn:P5_Ham}.

\noindent(ii) Difference Lax form: 
\begin{equation}
\begin{split}
&L_1=\frac{t f (z+a_2-1)}{z-g-1}\left(f-{T_z}^{-1}\right)+\frac{z(z-a_1)}{z-g}\left(f T_z-1\right)\\
&\hskip40pt -(f-1) (z+g+t f-a_1)+a_3 f,\\
&L_2=T_\beta T_z+\frac{1}{z-g}(1-f T_z),\\
&B=\frac{z+a_2}{z-g}\left(1-f T_z\right)+\partial _t T_z+T_z.
\end{split}
\end{equation}
The curve $L_1y=0$ is the unique curve of degree $(2,3)$ in $(f,g)$ passing through
the points \eqref{eqn:P5_8pts} and \eqref{eqn:fg_y_2}.
Compatibility of $L_1y=L_2y=0$ gives the discrete flow for $T_\beta(=\pi_1\pi_2s_0s_1s_0s_3s_1)$:
\begin{equation}
\begin{split}
&\overline{a}_2=a_2 - 1, \quad \overline{a}_3=a_3 + 1,\\
&g+\overline{g}={a_1}+{a_3}-t f+\frac{a_3}{f-1},\quad
f\overline{f}=-\frac{(\overline{g}-{a_1}) \overline{g}}{t(\overline{g}+{a_2}-1)}.
\end{split}
\end{equation}
Compatibility of $L_1y=By=0$ gives the P$_{\rm V}$ flow with Hamiltonian \eqref{eqn:P5_Ham}.

\bigskip

\subsubsection{d-P$((2A_1)^{(1)}/D_6^{(1)})$} \label{subsubsec:Lax_d-2A1}
The corresponding continuous flow is P$_{\rm III}^{D_6^{(1)}}$ with the Hamiltonian:
\begin{equation}\label{eqn:P3D6_Ham}
H=\frac{1}{t}\Big\{p(p-1) q^2+(a_1+a_2)qp+t p-{a_2} q\Big\}.
\end{equation}
The eight points configuration in $(f,g)$ coordinates is given as:
\begin{equation}\label{eqn:P3D6_8pts}
(f_i,g_i)=\left(\frac{1}{\epsilon}, -a_2\right), \ \left(\frac{1}{\epsilon}, \frac{1}{\epsilon}-a_1\right)_3, \ (0,0),\ 
\left(\epsilon,-\frac{t}{\epsilon}+1-a_2-a_1\right)_3.
\end{equation}
\noindent(i) Differential Lax form: 
\begin{equation}
\begin{split}
&{\cal L}_1=\Big\{-\frac{{a_2}}{x}+\frac{p q}{x (x-q)}-\frac{t H}{x^2}\Big\} 
+\Big\{\frac{1+a_1+a_2}{x}-\frac{1}{x-q}+\frac{t}{x^2}-1\Big\} {\partial_x}+{\partial_x}^2,\\
&{\cal L}_2=T_{\alpha}-\frac{1}{x-q}(p -\partial_x),\\
&{\cal B} ={\partial_t}-\frac{q}{t(x-q)}(x\partial_x-qp).
\end{split}
\end{equation}
The curve ${\cal L}_1y=0$ is the unique curve of degree $(3,2)$ in $(f,g)$ passing through
the points \eqref{eqn:P3D6_8pts} and \eqref{eqn:fg_y_1}.
Compatibility of ${\cal L}_1y={\cal L}_2y=0$ gives the discrete flow for $T_{\alpha}(=(\pi_1\pi_2)^2s_2s_1)$:
\begin{equation}
\begin{split}
&\overline{a}_1=a_1+1, \ \overline{a}_2=a_2+1, \\
& q+\overline{q}=-\frac{a_2}{p}-\frac{a_1}{p-1}, \quad
p+\overline{p}=1-\frac{t}{\overline{q}^2}-\frac{a_1+a_2+1}{\overline{q}}.
\end{split}
\end{equation}
Compatibility of ${\cal L}_1y={\cal B}y=0$ gives the P$_{\rm III}^{D_6^{(1)}}$ flow with Hamiltonian \eqref{eqn:P3D6_Ham}.

\noindent(ii) Difference Lax form: 
\begin{equation}
\begin{split}
&L_1 = \frac{z+a_2-1}{z-g-1}f ({T_z}^{-1}-f)
+f^2 + f(1-a_1-a_2-g-z)-t + \frac{t z }{g-z}(fT_z-1),\\
&L_2=T_{\beta} T_z+\frac{1}{z-g}(1-fT_z),\\
&B=\frac{z+a_2}{z-g}\left(f T_z-1\right)+t\partial _t T_z+z T_z.
\end{split}
\end{equation}
The curve $L_1y=0$ is the unique curve of degree $(2,3)$ in $(f,g)$ passing through
the points \eqref{eqn:P3D6_8pts} and \eqref{eqn:fg_y_2}.
Compatibility of $L_1y=L_2y=0$ gives the discrete flow for $T_{\beta}(=s_2\pi_1\pi_2\pi_1)$:
\begin{equation}
\begin{split}
&\overline{a}_2=a_2-1, \quad g+\overline{g}=1+f-{a_1}-{a_2}-\frac{t}{f},\quad
f\overline{f}=-\frac{t\overline{g}}{\overline{g}+a_2-1}.
\end{split}
\end{equation}
Compatibility of $L_1, B$ gives the continuous P$_{\rm II}^{D_6^{(1)}}$ flow with the Hamiltonian \eqref{eqn:P3D6_Ham}.


\subsubsection{d-P$(\underset{|\alpha|^2=4}{A_{1}^{(1)}}/D_7^{(1)})$}\label{subsubsec:Lax_d-A1'}
  The corresponding continuous flow is P$_{\rm III}^{D_7^{(1)}}$ with Hamiltonian:
\begin{equation}\label{eqn:P3D7_Ham}
H=\frac{1}{t}\Big(p^2 q^2+q+p t+a_1 p q\Big).
\end{equation}
The eight points configuration in $(f,g)=(q,qp)$ coordinates is given as:
\begin{equation}\label{eqn:P3D7_8pts}
(f_i,g_i)=\left(-\frac{1}{\epsilon^2},\frac{1}{\epsilon}-\frac{a_1}{2}\right)_4,\ (0,0),\ \left(\epsilon,-\frac{t}{\epsilon}+1-a_1\right)_3.
\end{equation}

\noindent(i) Differential Lax form:
\begin{equation}
\begin{split}
&{\cal L}_1 =\Big\{\frac{1-p}{x}+\frac{p}{x-q}-\frac{t H}{x^2}\Big\} 
+\Big\{\frac{a_1+1}{x}-\frac{1}{x-q}+\frac{t}{x^2}\Big\} {\partial_x}+{\partial_x}^2,\\
&{\cal L}_2=T_{\alpha}-\frac{1}{x-q}(p -\partial_x),\\
&{\cal B} ={\partial_t}-\frac{q}{t(x-q)}(x\partial_x-q p).
\end{split}
\end{equation}
The curve ${\cal L}_1y=0$ is the unique curve of degree $(3,2)$ in $(f,g)$ passing through
the points \eqref{eqn:P3D7_8pts} and \eqref{eqn:fg_y_1}.
Compatibility of ${\cal L}_1y= {\cal L}_2y=0$ gives the discrete flow for 
$T_\alpha(=(\pi_1\pi_2)^2)$:
\begin{equation}
\begin{split}
&\overline{a_1}=a_1+2, \quad
q+\overline{q}=-\frac{1}{p^2}-\frac{a_1}{p}, \quad
p+\overline{p}=-\frac{t}{\overline{q}^2}-\frac{a_1+1}{\overline{q}}.
\end{split}
\end{equation}
Compatibility of ${\cal L}_1y={\cal B}y=0$ gives the P$_{\rm III}^{D_7^{(1)}}$ flow with Hamiltonian \eqref{eqn:P3D7_Ham}.

\noindent(ii) Difference Lax form: 
\begin{equation}
\begin{split}
&L_1=\frac{t z }{z-g}\left(1-f T_z\right)+\frac{f}{z-g-1}\left(f-{T_z}^{-1}\right)-f(z+g+a_1-1)-t,\\
&L_2=T_{\beta} T_z+\frac{1}{z-g}(1-f T_z),\\
&B=\frac{1}{z-g}(1-f T_z)+t\partial _t T_z+z T_z.
\end{split}
\end{equation}
The curve $L_1y=0$ is the unique curve of degree $(2,3)$ in $(f,g)$ passing through
the points \eqref{eqn:P3D7_8pts} and \eqref{eqn:fg_y_2}.
Compatibility of $L_1y=L_2y=0$ gives the discrete flow for 
$T_{\beta}(=\pi_2\pi_1)$:
\begin{equation}
\begin{split}
&\overline{a_1}=a_1-1, \quad
g+\overline{g}=1-a_1-\frac{t}{f}, \quad
f\overline{f}=t\overline{g}.\\
\end{split}
\end{equation}
Compatibility of $L_1y=By=0$ gives the P$_{\rm III}^{D_7^{(1)}}$ flow with Hamiltonian \eqref{eqn:P3D7_Ham}.

\bigskip

\subsubsection{d-P$(A_0^{(1)}/D_8^{(1)})$}\label{subsubsec:Lax_d-A0D8}
The corresponding continuous flow is P$_{\rm III}^{D_8^{(1)}}$ with the Hamiltonian
\begin{equation}\label{eqn:P3D8_Ham}
H=\frac{1}{t}\Bigl(p^2 q^2+pq+q+\frac{t}{q}\Bigr).
\end{equation}
The eight points configuration in $(f,g)$ coordinates is given as:
\begin{equation}\label{eqn:P3D8_8pts}
(f_i,g_i)=\left(-\frac{1}{\epsilon^2},-\frac{1}{\epsilon}-\frac{1}{2}\right)_4, \quad \left(-t\epsilon^2, \frac{1}{\epsilon}\right)_4. 
\end{equation}

\noindent(i) Differential Lax form:
\begin{equation}
\begin{split}
&{\cal L}_1 =\Big\{\frac{1-p}{x}+\frac{p}{x-q}-\frac{t H}{x^2}+\frac{t}{x^3}\Big\} 
+\Big\{\frac{2}{x}-\frac{1}{x-q}\Big\} {\partial_x}+{\partial_x}^2,\\
&{\cal B} ={\partial_t}-\frac{q}{t(x-q)}(x\partial_x-qp).
\end{split}
\end{equation}
The curve ${\cal L}_1=0$ is the unique curve of degree $(3,2)$ in $(f,g)$ passing through
the points \eqref{eqn:P3D8_8pts} and \eqref{eqn:fg_y_1}.
Compatibility of ${\cal L}_1y={\cal B}y=0$ gives the P$_{\rm III}^{D_8^{(1)}}$ flow with Hamiltonian \eqref{eqn:P3D8_Ham}.

\noindent(ii) Difference Lax form: 
\begin{equation}
\begin{split}
&L_1=\frac{t}{z-g}\left(T_z-f^{-1}\right)+\frac{1}{z-g-1}\left(T_z^{-1}-f\right)+g+z,\\
&B=\frac{1}{z-g}\left(1-f T_z\right)+t\partial _t T_z+z T_z.
\end{split}
\end{equation}
The curve $L_1y=0$ is the unique curve of degree $(2,3)$ in $(f,g)$ passing through
the points \eqref{eqn:P3D8_8pts} and \eqref{eqn:fg_y_2}.
Compatibility of $L_1y=By=0$ gives the P$_{\rm III}^{D_8^{(1)}}$ flow with Hamiltonian \eqref{eqn:P3D8_Ham}.
There is no discrete flow. 
\bigskip

\subsubsection{d-P$(A_2^{(1)}/E_6^{(1)})$}\label{subsubsec:Lax_d-A2}
The corresponding continuous flow is P$_{\rm IV}$ with Hamiltonian
\begin{equation}\label{eqn:P4_Ham}
H=q p (p-q-t) -{a_1} p-{a_2} q.
\end{equation}
The eight points configuration in $(f,g)=(q,qp)$ coordinates is given as:
\begin{equation}\label{eqn:P4_8pts}
(f_i,g_i) =\left(\infty, -a_2\right), \
\left(\frac{1}{\epsilon},\frac{1}{\epsilon^2}+\frac{t}{\epsilon}- a_0\right)_5,\ (0,0), \ (0,a_1).
\end{equation}

\noindent(i) Differential Lax form:
\begin{equation}
\begin{split}
&{\cal L}_1=\Big\{-{a_2}-\frac{H}{x}+\frac{p q}{x (x-q)}\Big\}
+\Big\{\frac{1-{a_1}}{x}-t-x-\frac{1}{x-q}\Big\} {\partial_x}+{\partial_x}^2,\\
&{\cal L}_2=T_{\alpha}-\frac{1}{x-q}(p -\partial_x),\\
&{\cal B} ={\partial_t}-\frac{1}{x-q}(x\partial_x-q p).
\end{split}
\end{equation}
The curve ${\cal L}_1y=0$ is the unique curve of degree $(3,2)$ in $(f,g)$ passing through
the points \eqref{eqn:P4_8pts} and \eqref{eqn:fg_y_1}.
Compatibility of ${\cal L}_1y={\cal L}_2y=0$ gives the discrete flow for $T_{\alpha}(=\pi_1\pi_2s_0s_2)$:
\begin{equation}
\begin{split}
&\overline{a}_1=a_1-1,\quad \overline{a}_2=a_2+1, \\
&q+\overline{q}=p-t-\frac{a_2}{p}, \quad
p+\overline{p}=\overline{q}+t+\frac{a_1-1}{\overline{q}}.
\end{split}
\end{equation}
Compatibility of ${\cal L}_1y= {\cal B}y=0$ gives the P$_{\rm IV}$ flow with Hamiltonian \eqref{eqn:P4_Ham}.

\noindent(ii) Difference Lax form: 
\begin{equation}
\begin{split}
&L_1=f\frac{a_2+z-1}{g-z+1}\left(f-T_z^{-1}\right)
-\frac{z \left(z-a_1\right)}{g-z} \left(1-f T_z\right)+a_1-g+f (f+t)-z,\\
&L_2=T_{\beta} T_z+\frac{1}{z-g}\left(1-f T_z\right),\\
&B=\frac{a_2+z}{z-g}\left(f T_z-1\right)+\partial _t T_z+T_z.
\end{split}
\end{equation}
The curve $L_1y=0$ is the unique curve of degree $(2,3)$ in $(f,g)$ passing through
the points \eqref{eqn:P4_8pts} and \eqref{eqn:fg_y_2}.
Compatibility of $L_1y=L_2y=0$ gives the discrete flow for 
$T_{\beta}(=\pi_1\pi_2s_1s_0)$:
\begin{equation}
\begin{split}
&\overline{a}_2=a_2-1, \quad
g+\overline{g}=f^2+tf+a_1, \quad
f\overline{f}=-\frac{\overline{g}(\overline{g}-a_1)}{\overline{g}+a_2-1}.\\
\end{split}
\end{equation}
Compatibility of $L_1y=By=0$ gives the P$_{\rm IV}$ flow with Hamiltonian \eqref{eqn:P4_Ham}.

\bigskip

\subsubsection{d-P$(A_1^{(1)}/E_7^{(1)})$}\label{subsubsec:Lax_d-A1}
The corresponding continuous flow is P$_{\rm II}$ with the Hamiltonian
\begin{equation}\label{eqn:P2_Ham}
H=\frac{p^2}{2}-\Big(q^2+\frac{t}{2}\Big) p-a q.
\end{equation}
The eight points configuration in $(q,p)$ coordinates is given as\footnote{Space of initial values can be realized by eight points configuration only in $(q,p)$ coordinates. More points are required in $(f,g)$ coordinates.}:
\begin{equation}\label{eqn:P2_8pts}
(q_i,p_i)=\left(\frac{1}{\epsilon}, -a\epsilon\right)_2, \quad
\left(\frac{1}{\epsilon},\frac{2}{\epsilon^2}+t+(a-1)\epsilon\right)_6.
\end{equation}

\noindent(i) Differential Lax form:
\begin{equation}
\begin{split}
&{\cal L}_1 =\Big\{\frac{p}{x-q}-2 H-2 a x\Big\}-\Big\{2x^2+t+\frac{1}{x-q}\Big\}{\partial_x}+{\partial_x}^2,\\
&{\cal L}_2=T_{\alpha}-\frac{1}{x-q}(p -\partial_x),\\
&{\cal B}={\partial_t}-\frac{1}{2(x-q)}(\partial_x-p).
\end{split}
\end{equation}
The curve ${\cal L}_1y=0$ is the unique curve of degree $(3,2)$ in $(q,p)$ passing through
the eight points \eqref{eqn:P2_8pts} and the extra four points:
\begin{equation}
(q,p)=\Bigl(x+\epsilon,-\frac{1}{\epsilon}\Bigr)_2, \quad
\Bigl(x+\epsilon, \frac{y'(x+\epsilon)}{y(x+\epsilon)}\Bigr)_2.
\end{equation}
Compatibility of ${\cal L}_1y={\cal L}_2y=0$ gives the discrete flow for $T_\alpha(=\pi s_1)$:
\begin{equation}
\begin{split}
&\overline{a}=a+1,\quad 
p+\overline{p}=2\overline{q}^2+t,\quad
q+\overline{q}=-\frac{a}{p}.
\end{split}
\end{equation}
Compatibility of ${\cal L}_1y={\cal B}y=0$ gives the P$_{\rm II}$ flow with the Hamiltonian \eqref{eqn:P2_Ham}.

\noindent(ii) Difference Lax form: We put $\varphi=p-2 q^2-t$.
\begin{equation}
\begin{split}
&L_2=-p+2 q T_{\alpha}+(z+1) T_z-{2 T_{\alpha}}{T_z}^{-1},\\
&L_3=(-a-z+1){T_z}^{-1}+\varphi T_{\alpha}{T_z}^{-1}-q z+z T_{\alpha},\\
&B_2=-p+2 q \partial _t+(z+1) T_z-{2 \partial _t}{T_z}^{-1},\\
&B_3=(-a-z+1){T_z}^{-1}+\varphi \partial _t {T_z}^{-1}-q
   z+z\partial _t.\\
\end{split}
\end{equation}
The linear difference equation $L_1y=0$ in $z$ obtained from $L_2y=L_3y=0$ is of third order in $T_z$:
\begin{equation}
\begin{split}
&L_1=2 (z+a-1) (z+\varphi q+1) T_z^{-1}+\Big(2q(z+a)+(\varphi p -2 a q)(z+\varphi q+1)\Big) \\
&+(z+1) (tz+t\varphi q-\varphi) T_z-(z+1) (z+2) (z+\varphi q) T_z^2.
\end{split}
\end{equation}
Compatibility of $L_2y=L_3y=0$ gives the discrete flow for $T_{\alpha}(=\pi s_1)$.
Compatibility of $B_2y=B_3y=0$ gives the P$_{\rm II}$ flow with Hamiltonian \eqref{eqn:P2_Ham}.

\bigskip

\subsubsection{d-P$(A_0^{(1)}/E_8^{(1)})$}\label{subsubsec:Lax_d-A0E8}
The corresponding continuous flow is P$_{\rm I}$ with the Hamiltonian
\begin{equation}\label{eqn:P1_Ham}
H=\frac{p^2}{2}-2 q^3-tq.
\end{equation} 
There is no discrete flow.

\noindent(i) Differential Lax form:
\begin{equation}
\begin{split}
&{\cal L}_1 =\Big\{-4 x^3-2 t x-2 H+\frac{p}{x-q}\Big\}-\frac{1}{x-q}{\partial_x}+{\partial_x}^2,\\
&{\cal B} ={\partial_t}-\frac{1}{2 (x-q)}(\partial_x-p).
\end{split}
\end{equation}
Compatibility of ${\cal L}_1y={\cal B}y=0$ gives the P$_{\rm I}$ flow with Hamiltonian \eqref{eqn:P1_Ham}.

\noindent(ii) Simple difference Lax form is not known.

\subsection{Hypergeometric solutions}\label{subsec:hyper_data}
%
\subsubsection{$e$-P$(E_8^{(1)}/A_0^{(1)})$}\label{subsubsec:hyper_e-E8}
\begin{center}
\begin{figure}[H]
\begin{center}
\includegraphics[scale=0.4,clip,viewport=60 70 753 450]{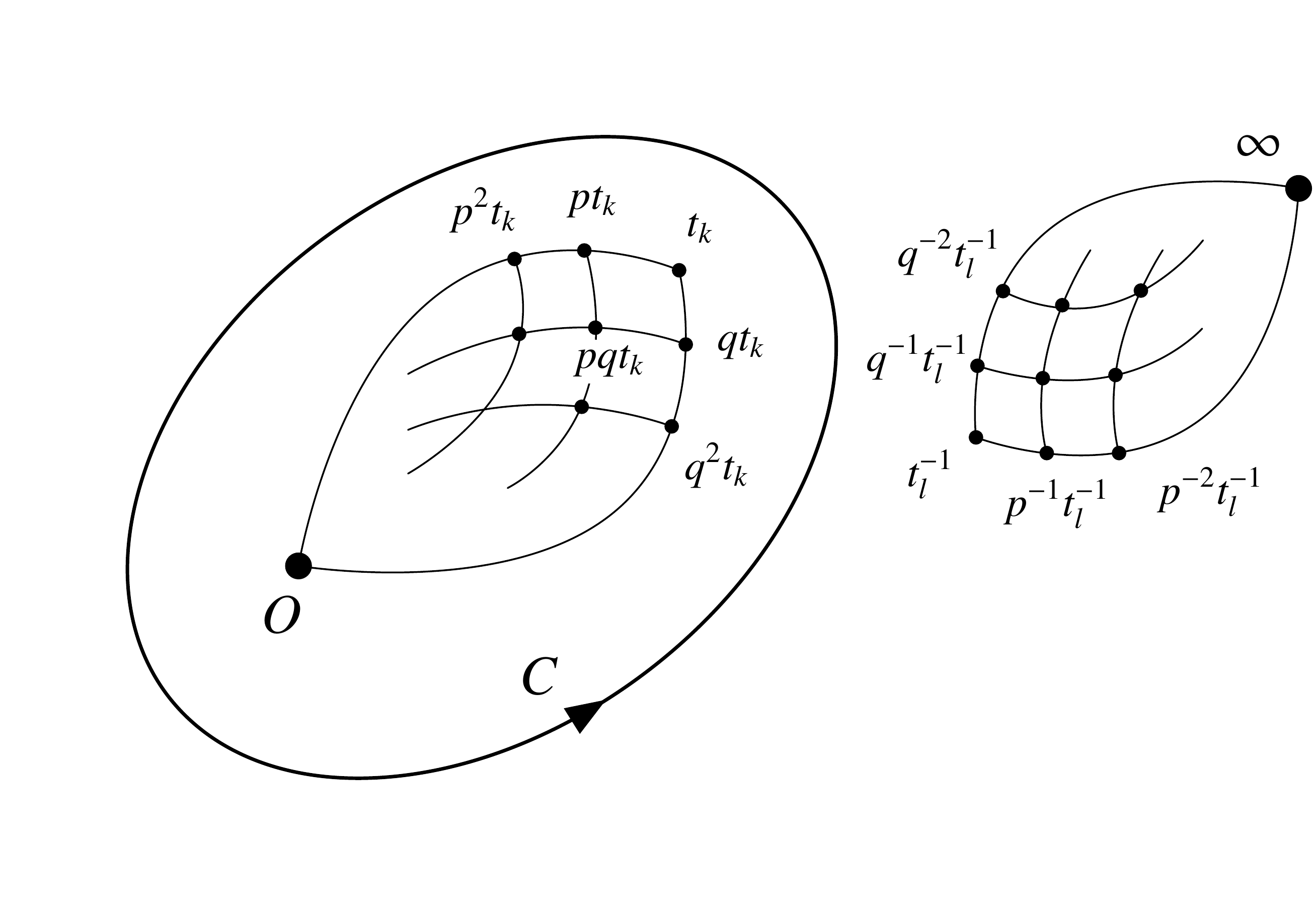}
\caption{Contour for the elliptic hypergeometric integral \eqref{eqn:elliptic_hyper_integral}. 
Contour encircles the sequence of poles accumulating to the origin but not those accumulating to the infinity.} \label{fig:contour_elliptic_hyper}
\end{center}
\end{figure} 
\end{center}

(i) Decoupling condition:
\begin{enumerate}
 \item ${\rm P}_1$, ${\rm P}_3$, ${\rm P}_5$, ${\rm P}_7$ are on a $(1,1)$ curve $C_1$:
\begin{equation}
 \kappa_1\kappa_2 = v_1v_3v_5v_7,\label{eqn:decoupling_param_e-e8}
\end{equation}
$(f,g)$ is also on $C_1$:
\begin{equation}
 \frac{f - f\left(\frac{\kappa_2}{t}\right)} {f - f(t)}
= \frac{f_a(t)}{f_a\left(\frac{\kappa_2}{t}\right)}
\prod_{j=1,3,5,7}\frac{\left[\frac{v_jt}{\kappa_2}\right]}{\left[\frac{v_j}{t}\right]},\quad \mbox{for}\ g=g(t),
\end{equation}
\begin{equation}
 \frac{g-g\left(\frac{\kappa_1}{s}\right)} {g-g(s)}
= \frac{g_a(s)}{g_a\left(\frac{\kappa_1}{s}\right)}
\prod_{j=1,3,5,7}\frac{\left[\frac{v_js}{\kappa_1}\right]}{\left[\frac{v_j}{s}\right]},\quad 
\mbox{for}\ f = f(s).
\end{equation}
 \item ${\rm P}_2'$, ${\rm P}_4'$, ${\rm P}_6'$, ${\rm P}_8'$ are on a $(1,1)$ curve $C_2$ where
${\rm P}_i'=(\overline{f}(v_i), g(v_i))$:
\begin{equation}
 \kappa_1\kappa_2 = qv_2v_4v_6v_8, \label{eqn:decoupling_param_e-e8_2}
\end{equation}
$(\overline{f},g)$ is also on $C_2$:
\begin{equation}
\frac{\overline{f}-\overline{f}\left(\frac{\kappa_2}{t}\right)} {\overline{f}-\overline{f}(t)}
= \frac{\overline{f_a}(t)}{\overline{f_a}\left(\frac{\kappa_2}{t}\right)}
\prod_{j=2,4,6,8}\frac{\left[\frac{v_jt}{\kappa_2}\right]}{\left[\frac{v_j}{t}\right]},\quad \mbox{for}\ g=g(t),
\end{equation}
and $(f,\underline{g})$ is on $\underline{C}_2$ which is a curve determined by 
$(f(v_i), \underline{g}(v_i))$ ($i=2,4,6,8$):
\begin{equation}
 \frac{\underline{g}-\underline{g}\left(\frac{\kappa_1}{s}\right)} {\underline{g}-\underline{g}(s)}
= \frac{\underline{g_a}(s)}{\underline{g_a}\left(\frac{\kappa_1}{s}\right)}
\prod_{j=2,4,6,8}
\frac{\left[\frac{v_js}{\kappa_1}\right]}{\left[\frac{v_j}{s}\right]},\quad \mbox{for}\ f = f(s),
\end{equation}
where 
\begin{equation}
\begin{split}
f_a(z)=\left[\frac{a}{z}\right]\left[\frac{\kappa_1}{az}\right],\quad
 g_a(z)=\left[\frac{a}{z}\right]\left[\frac{\kappa_2}{az}\right],\quad
f(z)=\frac{f_b(z)}{f_a(z)},\quad g(z)=\frac{g_b(z)}{g_a(z)},
\end{split}
\end{equation}
$[z]$ is the multiplicative theta function given in Section \ref{subsubsec:contiguity}, and $a$, $b$ are arbitrary.
\end{enumerate}
%
\noindent (ii) Linearized equation of the Riccati equation \eqref{eqn:hyper_linear_general}:
\begin{equation}\label{eqn:hyper_diffeq_eE8}
\begin{split}
&\hskip30pt U_1 (\overline{F} -F) + U_2 F + U_3 (\underline{F}-F)=0,\\
&\begin{array}{l}\medskip
 {\displaystyle 
U_1
=\frac{\left[\frac{v_1v_8}{\kappa_2}\right]\left[\frac{qv_1v_8}{\kappa_2}\right]}
{\left[\frac{\kappa_1}{\kappa_2}\right]\left[\frac{q\kappa_1}{\kappa_2}\right]}
\prod_{i=3,5,7}\left[\frac{v_1v_i}{\kappa_1}\right]\prod_{j=2,4,6}\left[\frac{v_jv_8}{\kappa_1}\right],
}\\\medskip
{\displaystyle 
U_2
=-\prod_{i=2,4,6}\left[\frac{v_1}{v_i}\right]\prod_{j=3,5,7}\left[\frac{v_j}{v_8}\right],
}\\
{\displaystyle 
U_3
=\frac{\left[\frac{v_1v_8}{\kappa_1}\right]\left[\frac{v_1v_8}{q\kappa_1}\right]}
{\left[\frac{q\kappa_1}{\kappa_2}\right]\left[\frac{q^2\kappa_1}{\kappa_2}\right]}
\prod_{i=3,5,7}\left[\frac{qv_1v_i}{\kappa_2}\right]\prod_{j=2,4,6}\left[\frac{qv_jv_8}{\kappa_2}\right].
}
\end{array}
\end{split}
\end{equation}
\noindent (iii) Elliptic hypergeometric integral \cite{Spiridonov:elliptic_hyper}:
\begin{equation}\label{eqn:elliptic_hyper_integral}
I(t_0,t_1,\ldots,t_7|p,q)
=\frac{(p;p)_\infty(q;q)_\infty}{4\pi\sqrt{-1}}
\int_{C}
\frac{\prod_{i=0}^{7}\Gamma(t_iz^{\pm1};p,q)}
{\Gamma(z^{\pm2};p,q)
}
\frac{dz}{z}, 
\end{equation}
\begin{equation}
t_0t_1\,\cdots\, t_7=p^2q^2,
\end{equation}
where 
\begin{equation}
\Gamma(z;p,q)=\frac{(pq/z;p,q)_\infty}{(z;p,q)_\infty},
\qquad (z;p,q)_\infty=\prod_{i,j=0}^{\infty}(1-p^iq^jz)
\quad(|p|,|q|<1),
\end{equation}
and each double sign indicates the product of two factors with different signs as
$\Gamma(az^{\pm1};p,q) = \Gamma(az;p,q)\Gamma(a/z;p,q)$.  Moreover, the integration contour $C$ in
\eqref{eqn:elliptic_hyper_integral} is a closed curve (or a cycle) which encircles in the positive direction the
sequence of poles
\begin{equation}
z=p^iq^jt_k\ (i,j\in\mathbb{N};k=0,\ldots,7),
\end{equation}
that accumulate to the origin as shown Figure \ref{fig:contour_elliptic_hyper}, under a certain
genericity condition for simplicity of poles. For instance, if $|t_k|<1$, then $C$ can be chosen as
the unit circle $|z|=1$.
%
\par\medskip

\noindent (iv) Linear difference equation for the hypergeometric function:
\begin{equation}\label{eqn:diffeq_elliptic_hyper}
\begin{split}
&\hskip40pt V_1(\overline{\Phi}-\Phi) + V_2\Phi  +V_3(\underline{\Phi}-\Phi)=0,\\
& \Phi(a_0,a_1,\ldots,a_7)=\Psi(u_0,u_1,\ldots,u_7)\\
& = \frac{
\prod_{r=1}^{6}
\Gamma(qu_r/u_0;p,q)
\Gamma(q/u_ru_7;p,q)
}
{\Gamma(q^2/u_0^2;p,q)
\Gamma(u_0/u_7;p,q)
\prod_{1\le r<s\le 6}\Gamma(u_ru_s;p,q)
}
I(pu_0,u_1,\ldots,u_6,pu_7|p,q), \\
&a_i = \frac{q}{u_0u_i}\ (i=0,\ldots,7),\quad q^2a_0^3 = a_1a_2\cdots a_7,\quad u_0u_1\cdots u_7=q^2,\\
&\overline{\Phi}=\Phi(a_0;a_1/q,qa_2,a_3,\ldots,a_7),\quad 
\underline{\Phi}=\Phi(a_0;qa_1,a_2/q,a_3,\ldots,a_7),\\
&\begin{array}{l}
{\displaystyle V_1=\frac{[a_2][a_0/a_2][qa_0/a_2]}{[a_2/a_1][qa_2/a_1]}\prod_{j=3}^7[qa_0/a_1a_j],
\quad V_2=[qa_0/a_1a_2]\prod_{j=3}^7[a_j],}\\
{\displaystyle V_3=\frac{[a_1][a_0/a_1][qa_0/a_1]}{[a_1/a_2][qa_1/a_2]}\prod_{j=3}^7[qa_0/a_2a_j],
\quad [z]=z^{-\frac{1}{2}}(z,p/z,p;p)_\infty}.
\end{array}
\end{split}
\end{equation}
%
\noindent (v) Contiguity relation \cite{Spiridonov:elliptic_hyper}:
\begin{equation}\label{eqn:elliptic_hyper_contiguity}
\begin{split}
& \overline{\Phi} - \Phi = 
-\frac{[qa_0][q^2a_0][qa_2/a_1][qa_0/a_1a_2]}{[qa_0/a_1][q^2a_0/a_1][a_0/a_2][qa_0/a_2]}\\
&\hskip60pt \times\prod_{i=3}^7 \frac{[a_i]}{[qa_0/a_i]}~\Phi(q^2a_0;a_1,qa_2,qa_3,\ldots,qa_7). 
\end{split}
\end{equation}
\begin{rem}\rm
When some of the parameters $a_i$ ($i=1,\ldots,6$) is $q^{-N}$ $(N\in\mathbb{N})$, 
the elliptic hypergeometric function $\Phi$ is expressed as a finite series 
\begin{equation}
\Phi={}_{12}V_{11}(a_0;a_1,\ldots,a_7)= \sum_{k=0}^N \frac{[q^{2k}a_0]}{[a_0]}\frac{[a_0]_k}{[q]_k}
\prod_{i=1}^7 \frac{[a_i]_k}{[qa_0/a_i]_k},
\end{equation}
where $[a]_k=[a][qa]\cdots [q^{k-1}a]$ \cite{KMNOY:10E9,Spiridonov:elliptic_hyper}.
\end{rem}
%
\noindent (vi) Hypergeometric solution \eqref{eqn:hypergeometric_solution_general}:
\begin{equation}
\begin{split}
 y&=\frac{f-f_1}{f-f_8}=
\frac{\left[\frac{v_1 v_8}{\kappa_2}\right]\prod_{i=3,5,7}\left[\frac{v_1 v_i}{\kappa_1}\right]}
{\left[\frac{\kappa_1}{\kappa_2}\right]\prod_{i=3,5,7}\left[\frac{v_8}{v_i}\right]} ~\frac{\overline{F}-F}{F}\\
&=\frac{\left[\frac{q v_1 v_8}{v_3 v_5}\right]\left[\frac{q^2 v_1 v_8}{v_3v_5}\right]
\prod_{i=2,4,6}\left[\frac{v_1}{v_i}\right] \prod_{i=3,5,7}\left[\frac{v_1v_i}{\kappa_1}\right] }
{\left[\frac{q v_1}{v_3}\right]\left[\frac{qv_1}{v_5}\right]\left[\frac{v_1v_8}{\kappa_1}\right]
\left[\frac{qv_1v_8}{\kappa_1}\right]\left[\frac{q v_1 v_8}{\kappa_2}\right]
\prod_{i=2,4,6}\left[\frac{q v_i v_8}{v_3 v_5}\right]
}~\frac{G}{F},\\
&F=\Phi(a_0,a_1,\cdots,a_7),\quad G = \Phi(q^2a_0;a_1,qa_2,qa_3,\ldots,qa_7).
\end{split}
\end{equation}
%
\noindent (vii) Identification of parameters:
\begin{equation}\label{eqn:identificattion_hyper_eE8}
 a_0 = \frac{v_1v_8}{v_3v_5},\ a_1=\frac{q\kappa_1}{v_3v_5},\ a_2=\frac{\kappa_2}{v_3v_5},\ a_3=\frac{v_1}{v_2},\ 
a_4=\frac{v_1}{v_4},\ a_5 = \frac{v_1}{v_6},\ a_6=\frac{v_8}{v_3},\ a_7=\frac{v_8}{v_5}.
\end{equation}
%
\subsubsection{$q$-P$(E_8^{(1)}/A_0^{(1)})$}\label{subsubsec:hyper_q-E8}
\noindent (i) Decoupling condition:
\begin{enumerate}
 \item ${\rm P}_1$, ${\rm P}_3$, ${\rm P}_5$, ${\rm P}_7$ are on a $(1,1)$ curve $C_1$:
\begin{equation}
 \kappa_1\kappa_2 = v_1v_3v_5v_7,\label{eqn:decoupling_param_q-e8}
\end{equation}
$(f,g)$ is also on $C_1$:
\begin{equation}
 \frac{f - f\left(\frac{\kappa_2}{t}\right)} {f - f(t)}
= \frac{t^4}{\kappa_2^2}\prod_{j=1,3,5,7}\frac{\frac{\kappa_2}{t}-v_j}{t-v_j},\quad \mbox{for}\ g=g(t),
\end{equation}
\begin{equation}
 \frac{g-g\left(\frac{\kappa_1}{s}\right)} {g-g(s)}
= \frac{s^4}{\kappa_1^2}\prod_{j=1,3,5,7}\frac{\frac{\kappa_1}{s}-v_j}{s-v_j},\quad 
\mbox{for}\ f = f(s).
\end{equation}
 \item ${\rm P}_2'$, ${\rm P}_4'$, ${\rm P}_6'$, ${\rm P}_8'$ are on a $(1,1)$ curve $C_2$ where
${\rm P}_i'=(\overline{f}(v_i), g(v_i))$:
\begin{equation}
 \kappa_1\kappa_2 = qv_2v_4v_6v_8, \label{eqn:decoupling_param_q-e8_2}
\end{equation}
$(\overline{f},g)$ is also on $C_2$:
\begin{equation}
\frac{\overline{f}-\overline{f}\left(\frac{\kappa_2}{t}\right)} {\overline{f}-\overline{f}(t)}
= \frac{t^4}{\kappa_2^2}\prod_{j=2,4,6,8}\frac{\frac{\kappa_2}{t}-v_j}{t-v_j},\quad \mbox{for}\ g=g(t),
\end{equation}
and $(f,\underline{g})$ is on $\underline{C}_2$:
\begin{equation}
 \frac{\underline{g}-\underline{g}\left(\frac{\kappa_1}{s}\right)} {\underline{g}-\underline{g}(s)}
= \frac{\kappa_1^2}{s^4} \prod_{j=2,4,6,8}
\frac{\frac{\kappa_1}{s}-v_j}{s-v_j},\quad \mbox{for}\ f = f(s),
\end{equation}
where $f(z)=z+\frac{\kappa_1}{z}$, $g(z)=z+\frac{\kappa_2}{z}$.
\end{enumerate}
%
\noindent(ii) Linearized equation of the Riccati equation \eqref{eqn:hyper_linear_general}:
\begin{equation}\label{eqn:hyper_diffeq_qE8}
\begin{split}
&\hskip60pt U_1 (\overline{F} -F) + U_2 F + U_3 (\underline{F}-F)=0,\\
&\begin{array}{l}\medskip
 {\displaystyle 
U_1
=\frac{(v_1v_8-\kappa_2)(qv_1v_8-\kappa_2)}{\kappa_1^2(\kappa_1-\kappa_2)(q\kappa_1-\kappa_2)}
\prod_{i=3,5,7}(v_1v_i-\kappa_1)\prod_{i=2,4,6}(v_iv_8-\kappa_1),
}\\\medskip
{\displaystyle 
U_2
=-v_1v_8\prod_{i=2,4,6}(v_1-v_i)\prod_{i=3,5,7}(v_i-v_8),
}\\
{\displaystyle 
U_3
=\frac{(v_1v_8-\kappa_1)(v_1v_8-q\kappa_1)}{q^2\kappa_2^2(q\kappa_1-\kappa_2)(q^2\kappa_1-\kappa_2)}
\prod_{i=3,5,7}(qv_1v_i-\kappa_2)\prod_{i=2,4,6}(qv_iv_8-\kappa_2).
}
\end{array}
\end{split}
\end{equation}
%
\noindent (iii) Linear difference equation for the hypergeometric function \cite{Gupta-Masson}:
\begin{equation}\label{eqn:10W9}
\begin{split}
& \Phi(a_0,a_1,\ldots,a_7)={}_{10}W_9(a_0;a_1,\ldots,a_7;q,q)
 + \frac{(qa_0,a_7/a_0;q)_\infty}{(a_0/a_7,qa_7^2/a_0;q)_\infty}\\
&\qquad \times\prod_{k=1}^6\frac{(a_k,qa_7/a_k;q)_\infty}{(qa_0/a_k,a_ka_7/a_0;q)_\infty}
{}_{10}W_9(a_7^2/a_0;a_1a_7/a_0,\ldots,a_6a_7/a_0,a_7;q,q) ,\\
&\hskip40pt q^2a_0^3 = a_1a_2\cdots a_7.
\end{split}
\end{equation}
\begin{equation}\label{eqn:diffeq_10W9}
\begin{split}
&V_1(\overline{\Phi}-\Phi) + V_2\Phi  +V_3(\underline{\Phi}-\Phi)=0,\\
&\begin{array}{l}
{\displaystyle V_1=\frac{(1-a_2)(1-a_0/a_2)(1-qa_0/a_2)}{(1-a_2/a_1)(1-qa_2/a_1)}\prod_{j=3}^7(1-qa_0/a_1a_j),}\\
{\displaystyle V_2=(1-qa_0/a_1a_2)\prod_{j=3}^7(1-a_j),}\\
{\displaystyle V_3=\frac{(1-a_1)(1-a_0/a_1)(1-qa_0/a_1)}{(1-a_1/a_2)(1-qa_1/a_2)}\prod_{j=3}^7(1-qa_0/a_2a_j),}\\
{\displaystyle \overline{\Phi}=\Phi(a_0;a_1/q,qa_2,a_3,\ldots,a_7),\quad 
\underline{\Phi}=\Phi(a_0;qa_1,a_2/q,a_3,\ldots,a_7).}
\end{array}
\end{split}
\end{equation}
%
\noindent(iv) Contiguity relation \cite{Gupta-Masson}:
\begin{equation}\label{eqn:10W9_contiguity}
\begin{split}
& \overline{\Phi} - \Phi = 
-\frac{a_1(1-qa_0)(1-q^2a_0)(1-qa_2/a_1)(1-qa_0/a_1a_2)}{(1-qa_0/a_1)(1-q^2a_0/a_1)(1-a_0/a_2)(1-qa_0/a_2)}\\
&\hskip60pt \times\prod_{i=3}^7 \frac{1-a_i}{1-qa_0/a_i}~\Phi(q^2a_0;a_1,qa_2,qa_3,\ldots,qa_7). 
\end{split}
\end{equation}
\noindent (v) Hypergeometric solution \eqref{eqn:hypergeometric_solution_general}:
\begin{equation}
\begin{split}
 y&=\frac{f-f_1}{f-f_8}
= \frac{v_8 (v_1 v_8-\kappa _2)\prod_{i=3,5,7}(v_1 v_i-\kappa_1)}
{v_1^2\kappa_1(\kappa_1-\kappa_2)\prod_{i=3,5,7}(v_8-v_i)} ~\frac{\overline{F}-F}{F}\\
&=\frac{q^2v_3v_5v_8^2 (q v_1 v_8-v_3 v_5) (q^2 v_1 v_8-v_3 v_5)}
{v_1\kappa_1 (q v_1-v_3) (q  v_1-v_5) (v_1 v_8-\kappa _1) (q v_1 v_8-\kappa _1) (q v_1 v_8-\kappa _2)}\\
&\quad \times 
\frac{\prod_{i=2,4,6}(v_1-v_i)   \prod_{i=3,5,7}(v_1 v_i-\kappa _1) }{\prod_{i=2,4,6}(q v_i v_8-v_3 v_5) }~\frac{G}{F},\\
&F = \Phi(a_0,a_1,\cdots,a_7),\quad G = \Phi(q^2a_0;a_1,qa_2,qa_3,\ldots,qa_7).
\end{split}
\end{equation}
%
\noindent (vi) Identification of parameters:
\begin{equation}\label{eqn:identificattion_hyper_qE8}
 a_0 = \frac{v_1v_8}{v_3v_5},\ a_1=\frac{q\kappa_1}{v_3v_5},\ a_2=\frac{\kappa_2}{v_3v_5},\ a_3=\frac{v_1}{v_2},\ 
a_4=\frac{v_1}{v_4},\ a_5 = \frac{v_1}{v_6},\ a_6=\frac{v_8}{v_3},\ a_7=\frac{v_8}{v_5}.
\end{equation}
\begin{rem}\rm
The difference equation \eqref{eqn:hyper_diffeq_qE8} is symmetric with respect to $v_3$, $v_5$,
$v_7$ and $v_2$, $v_4$, $v_6$. This means that the solution space retains this symmetry. However,
if we choose a particular solution, it is not always symmetric, as in the case of \eqref{eqn:10W9} with
\eqref{eqn:identificattion_hyper_qE8}. For example, exchanging $v_5$ and $v_7$ in this solution yields
another hypergeometric solution 
\begin{equation}
\widetilde\Phi=\Phi(\tilde{a}_0,\tilde{a}_1,\ldots,\tilde{a}_7),\quad
\begin{array}{c}
{\displaystyle \tilde{a}_0=qa_0^2/a_1a_2a_7,\ \tilde{a}_1=qa_0/a_2a_7,\ \tilde{a}_2=qa_0/a_1a_7,}\\
{\displaystyle \tilde{a}_3=a_3,\ldots,\tilde{a}_6=a_6,\ \tilde{a}_7=qa_0/a_1a_2 },
\end{array}
\end{equation}
of \eqref{eqn:hyper_diffeq_qE8} which is called a Bailey transform of the original solution \cite{Gasper-Rahman}.
\end{rem}
%
\subsubsection{$q$-P$(E_7^{(1)}/A_1^{(1)})$}\label{subsubsec:hyper_q-E7}
\noindent (i) Decoupling condition:
\begin{enumerate}
 \item ${\rm P}_1$, ${\rm P}_3$, ${\rm P}_5$, ${\rm P}_7$ are on a $(1,1)$ curve $C_1$:
\begin{equation}
 \kappa_1\kappa_2 = v_1v_3v_5v_7,\label{eqn:decoupling_param_q-e7}
\end{equation}
and $(f, g)$ is also on $C_1$:
\begin{equation}
 \frac{fg-\frac{\kappa_1}{\kappa_2}}{fg-1}= 
\frac{\left(g-\frac{v_5}{\kappa_2}\right)\left(g-\frac{v_7}{\kappa_2}\right)}
{\left(g-\frac{1}{v_1}\right)\left(g-\frac{1}{v_3}\right)},
\end{equation}
\begin{equation}
\frac{fg-\frac{\kappa_1}{\kappa_2}}{fg-1}= 
\frac{\left(f - \frac{\kappa_1}{v_5}\right)\left(f-\frac{\kappa_1}{v_7}\right)}{\left(f - v_1\right)\left(f - v_3\right)}.
\end{equation}
 \item ${\rm P}_2'$, ${\rm P}_4'$, ${\rm P}_6'$, ${\rm P}_8'$ are on a $(1,1)$ curve $C_2$ where
${\rm P}_i'=(\overline{f}(v_i), v(v_i))$:
\begin{equation}
 \kappa_1\kappa_2 = qv_2v_4v_6v_8, \label{eqn:decoupling_param_q-e7_2}
\end{equation} 
$(\overline{f}, g)$ is also on $C_2$:
\begin{equation}
 \frac{\overline{f} g-\frac{\kappa_1}{q\kappa_2}}{\overline{f} g-1}= 
\frac{\left(g-\frac{v_6}{\kappa_2}\right)\left(g-\frac{v_8}{\kappa_2}\right)}
{\left(g-\frac{1}{v_2}\right)\left(g-\frac{1}{v_4}\right)},
\end{equation}
and $(f, \underline{g})$ is on $\underline{C}_2$:
\begin{equation}
 \frac{f \underline{g}-\frac{q\kappa_1}{\kappa_2}}{f\underline{g}-1}= 
\frac{\left(f -\frac{\kappa_1}{\kappa_6}\right)\left(f -\frac{\kappa_1}{v_8}\right)}{(f - v_2)(f - v_4)}.
\end{equation}
\end{enumerate}
%
\noindent (ii) Linearized equation of the Riccati equation \eqref{eqn:hyper_linear_general}:
\begin{equation}\label{eqn:hyper_diffeq_qE7}
\begin{split}
&\hskip60pt U_1 (\overline{F} -F) + U_2 F + U_3 (\underline{F}-F)=0,\\
&\begin{array}{l}\medskip
 {\displaystyle 
U_1
=\frac{(v_1v_8-\kappa_2)(qv_1v_8-\kappa_2)}{\kappa_1(\kappa_1-\kappa_2)(q\kappa_1-\kappa_2)}
\prod_{j=2,4}(v_jv_8-\kappa_1)\prod_{i=5,7}(v_1v_i-\kappa_1)},\\
\medskip
{\displaystyle 
U_2 = v_1v_8\prod_{j=2,4}(v_1- v_j)\prod_{i=5,7}(v_i-v_8),}\\
{\displaystyle 
U_3
=\frac{(v_1v_8-\kappa_1)(v_1v_8-q\kappa_1)}{q\kappa_2(q\kappa_1-\kappa_2)(q^2\kappa_1-\kappa_2)}
\prod_{j=2,4}(qv_jv_8-\kappa_2) \prod_{i=5,7}(qv_1v_i-\kappa_2).
}
\end{array}
\end{split}
\end{equation}
%
\noindent (iii) Linear difference equation for the hypergeometric function \cite{Ismail-Rahman}:
\begin{equation}\label{eqn:8W7}
\Phi(a_0,a_1,\ldots,a_5)={}_{8}W_7(a_0;a_1,\ldots,a_5;q,q^2a_0^2/a_1a_2a_3a_4a_5),
\end{equation}
\begin{equation}\label{eqn:diffeq_8W7}
\begin{split}
&\qquad V_1(\overline{\Phi}-\Phi) + V_2\Phi  +V_3(\underline{\Phi}-\Phi)=0,\\
&\begin{array}{l}
{\displaystyle V_1=\frac{(1-a_2)(1-a_0/a_2)(1-qa_0/a_2)}{a_1(1-a_2/a_1)(1-qa_2/a_1)}\prod_{j=3,4,5}(1-qa_0/a_1a_j)},\\
{\displaystyle V_2=(qa_0^2/a_1a_2a_3a_4a_5)(1-qa_0/a_1a_2)\prod_{j=3,4,5}(1-a_j),}\\
{\displaystyle V_3=\frac{(1-a_1)(1-a_0/a_1)(1-qa_0/a_1)}{a_2(1-a_1/a_2)(1-qa_1/a_2)}\prod_{j=3,4,5}(1-qa_0/a_2a_j),}\\
{\displaystyle \overline{\Phi}=\Phi(a_0,a_1/q,qa_2,a_3,\ldots,a_5),\quad 
\underline{\Phi}=\Phi(a_0,qa_1,a_2/q,a_3,\ldots,a_5).}
\end{array}
\end{split}
\end{equation}
%
\noindent(iv) Contiguity relation \cite{Ismail-Rahman}:
\begin{equation}\label{eqn:8W7_contiguity}
\begin{split}
& \overline{\Phi} - \Phi = 
-\frac{a_1(1-qa_0)(1-q^2a_0)(1-qa_2/a_1)(1-qa_0/a_1a_2)}{(1-qa_0/a_1)(1-q^2a_0/a_1)(1-a_0/a_2)(1-qa_0/a_2)}\\
&\times(qa_0^2/a_1a_2a_3a_4a_5)\prod_{i=3}^5 \frac{1-a_i}{1-qa_0/a_i}~\Phi(q^2a_0;a_1,qa_2,qa_3,qa_4,qa_5). 
\end{split}
\end{equation}
%
\noindent (v) Hypergeometric solution \eqref{eqn:hypergeometric_solution_general}:
\begin{equation}
\begin{split}
 y&=\frac{f-f_1}{f-f_8}=
-\frac{v_8 (v_1 v_8-\kappa _2)(v_1 v_5-\kappa_1)(v_1 v_7-\kappa_1)}
{v_1\kappa_1(\kappa_1-\kappa_2)\prod_{i=5,7}(v_8- v_i)} ~\frac{\overline{F}-F}{F}\\
&=-\frac{qv_3v_8^2 (q v_1 v_8-v_3 v_5) (q^2 v_1 v_8-v_3 v_5)\prod_{i=2,4}(v_1-v_i)\prod_{i=5,7}(v_1 v_i-\kappa _1) }
{\kappa_1 (q v_1-v_3) (v_1 v_8-\kappa _1) (q v_1 v_8-\kappa _1) (q v_1 v_8-\kappa _2)\prod_{i=2,4}(q v_i v_8-v_3 v_5) }
~\frac{G}{F}.\\
&F=\Phi(a_0,a_1,\cdots,a_5),\quad G = \Phi(q^2a_0;a_1,qa_2,qa_3,qa_4,qa_5).
\end{split}
\end{equation}
%
\noindent (vi) Identification of parameters:
\begin{equation}\label{eqn:identificattion_hyper_qE7}
 a_0 = \frac{v_1v_8}{v_3v_5},\ a_1=\frac{q\kappa_1}{v_3v_5},\ a_2=\frac{\kappa_2}{v_3v_5},\ a_3=\frac{v_1}{v_2},\ 
a_4=\frac{v_1}{v_4},\ a_5=\frac{v_8}{v_5}.
\end{equation}
%
\subsubsection{$q$-P$(E_6^{(1)}/A_2^{(1)})$}\label{subsubsec:hyper_q-E6}
\noindent(i) Decoupling condition:
\begin{enumerate}
 \item ${\rm P}_1$, ${\rm P}_3$, ${\rm P}_5$, ${\rm P}_7$ are on a $(1,1)$ curve $C_1$:
\begin{equation}
 \kappa_1\kappa_2 = v_1v_3v_5v_7,\label{eqn:decoupling_param_q-e6}
\end{equation}
$(f, g)$ is also on $C_1$:
\begin{equation}
 \frac{fg-1}{f} = \frac{\left(g-\frac{1}{v_1}\right)\left(g-\frac{1}{v_3}\right)}{g-\frac{v_5}{\kappa_2}},
\end{equation}
\begin{equation}
 \frac{f g-1}{g} = \frac{(f -v_1)(f -v_3)}{f - \frac{\kappa_1}{v_7}}.
\end{equation}
 \item ${\rm P}_2'$, ${\rm P}_4'$, ${\rm P}_6'$, ${\rm P}_8'$ are on a $(1,1)$ curve $C_2$ where
${\rm P}_i'={\rm P}_i|_{\kappa_1\to \kappa_1/q}$:
\begin{equation}
 \kappa_1\kappa_2 = qv_2v_4v_6v_8, \label{eqn:decoupling_param_q-e6_2}
\end{equation} 
$(\overline{f}, g)$ is on $C_2$:
\begin{equation}
\frac{\overline{f}g-1}{\overline{f}} 
= \frac{\left(g-\frac{1}{v_2}\right)\left(g-\frac{1}{v_4}\right)}{g-\frac{v_6}{\kappa_2}},
\end{equation}
and $(f, \underline{g})$ is on $\underline{C}_2$:
\begin{equation}
\frac{f \underline{g}-1}{\underline{g}} = \frac{(f -v_2)(f -v_4)}{f - \frac{\kappa_1}{v_8}}.
\end{equation}
\end{enumerate}
%
\noindent (ii) Linearized equation of the Riccati equation \eqref{eqn:hyper_linear_general}:
\begin{equation}\label{eqn:hyper_diffeq_qE6}
\begin{split}
& U_1 (\overline{F} -F) + U_2 F + U_3 (\underline{F}-F)=0,\\
&\begin{array}{l}\medskip
 {\displaystyle 
U_1
=\left(1-\frac{\kappa_2}{v_3v_5}\right)\left(1-\frac{qv_2v_6}{\kappa_2}\right)\left(1-\frac{qv_4v_6}{\kappa_2}\right),}\\
\medskip
{\displaystyle 
U_2 = \left(1-\frac{v_1}{v_2}\right)\left(1-\frac{v_1}{v_4}\right)\left(1-\frac{v_7}{v_8}\right),}\\
{\displaystyle 
U_3
=q\frac{v_6}{v_5}\left(1-\frac{qv_1v_5}{\kappa_2}\right)\left(1-\frac{\kappa_1}{v_1v_8}\right)\left(1-\frac{v_1v_8}{q\kappa_1}\right).
}
\end{array}
\end{split}
\end{equation}
%
\noindent (iii) Linear difference equation for the hypergeometric function \cite{Gupta-Ismail-Masson}:
\begin{equation}\label{eqn:3phi2}
\Phi(a_1,a_2,a_3,b_1,b_2)={}_3\phi_2\left[\begin{array}{c}a_1,a_2,a_3\\b_1,b_2\end{array};q;\frac{b_1b_2}{a_1a_2a_3}\right],
\end{equation}
\begin{equation}\label{eqn:diffeq_3phi2}
\begin{split}
&V_1(\overline{\Phi}-\Phi) + V_2\Phi  +V_3(\underline{\Phi}-\Phi)=0,\\
&\begin{array}{l}
{\displaystyle V_1=\left(1-\frac{a_1}{b_2}\right)\left(1-\frac{a_1}{b_2}\right)(1-a_3),}\\
{\displaystyle V_2=(1-a_1)(1-a_2)\left(1-\frac{a_3}{b_2}\right),}\\
\smallskip
{\displaystyle V_3=\frac{qa_1a_2a_3}{b_1b_2}\left(1-\frac{b_2}{q}\right)\left(1-\frac{b_1}{a_3}\right)
\left(1-\frac{1}{b_2}\right),}\\
{\displaystyle \overline{\Phi}=\Phi(a_1,a_2,qa_3,b_1,qb_2),\quad \underline{\Phi}=\Phi(a_1,a_2,a_3/q,b_3,b_2/q)}.
\end{array}
\end{split}
\end{equation}
%
\noindent (iv) Contiguity relation \cite{Gupta-Ismail-Masson}:
\begin{equation}\label{eqn:3phi2_contiguity}
\begin{split}
& \overline{\Phi} - \Phi = 
\frac{b_1b_2\left(1-a_1\right)\left(1-a_2\right)
\left(1-\frac{b_2}{a_3}\right)}
{a_1a_2\left(1-b_1\right)\left(1-b_2\right)\left(1-qb_2\right)}
~\Phi(qa_1,qa_2,qa_3,qb_1,q^2b_2). 
\end{split}
\end{equation}
%
\noindent (v) Hypergeometric solution \eqref{eqn:hypergeometric_solution_general}:
\begin{equation}
\begin{split}
 y&=\frac{f-f_1}{f-f_8}
=  \frac{v_8(v_1v_7 - \kappa_1)}{\kappa_1(v_8-v_7)}
~\frac{\overline{F}-F}{F}
=\frac{q v_8^2\left(v_1v_7-\kappa_1\right)\left(v_1-v_2\right)\left(v_1-v_4\right)}
{v_7\left(qv_1-v_3\right)\left(v_1v_8-\kappa_1\right)\left(qv_1v_8-\kappa_1\right)}
~\frac{G}{F},\\
&F=\Phi(a_1,a_2,a_3,b_1,b_2),\quad G = \Phi(qa_1,qa_2,qa_3,qb_1,q^2b_2).
\end{split}
\end{equation}
%
\noindent(vi) Identification of parameters:
\begin{equation}\label{eqn:identificattion_hyper_qE6}
a_1=\frac{v_1}{v_2},\quad a_2=\frac{v_1}{v_4},\quad a_3=\frac{\kappa_2}{v_3v_5},\quad
b_1=\frac{qv_1}{v_3},\quad b_2 = \frac{v_1v_8}{\kappa_1}.
\end{equation}
%
\subsubsection{$q$-P$(D_5^{(1)}/A_3^{(1)})$}\label{subsubsec:hyper_q-D5}
\begin{rem}\rm
In order to take the degeneration limit from $q$-P$(E_6^{(1)}/A_2^{(1)})$ described in
\eqref{eqn:degeneration_of_qP} and \eqref{eqn:degeneration_points_1} consistently with the
decoupling condition, we first exchange P$_1$ and P$_3$ ($v_1$ and $v_3$), P$_2$ and P$_4$ ($v_2$
and $v_4$), and then take the limit 
\begin{displaymath}
 \kappa_1\to \epsilon\kappa_1,\quad \kappa_2\to \kappa_2/\epsilon,\quad
v_j\to \epsilon v_j\ (j=3,4),\quad v_k\to v_k/\epsilon\ (k=1,2),\quad \epsilon\to 0.
\end{displaymath}
\end{rem}
%
\noindent (i) Decoupling condition:
\begin{enumerate}
 \item ${\rm P}_1$, ${\rm P}_3$, ${\rm P}_5$, ${\rm P}_7$ are on a $(1,1)$ curve $C_1$:
\begin{equation}
 \kappa_1\kappa_2 = v_1v_3v_5v_7,\label{eqn:decoupling_param_q-d5}
\end{equation}
$(f, g)$ is also on $C_1$:
\begin{equation}
 f = v_3\frac{g-\frac{v_5}{\kappa_2}}{g-\frac{1}{v_1}},
\end{equation}
\begin{equation}
 g = \frac{1}{v_1}\frac{f - \frac{\kappa_1}{v_7}}{f -v_3}.
\end{equation}
 \item ${\rm P}_2'$, ${\rm P}_4'$, ${\rm P}_6'$, ${\rm P}_8'$ are on a $(1,1)$ curve $C_2$ where
${\rm P}_i'={\rm P}_i|_{\kappa_1\to \kappa_1/q}$:
\begin{equation}
 \kappa_1\kappa_2 = qv_2v_4v_6v_8, \label{eqn:decoupling_param_q-d5_2}
\end{equation} 
$(\overline{f}, g)$ is also on $C_2$
\begin{equation}
\overline{f}
= v_4\frac{g-\frac{v_6}{\kappa_2}}{g-\frac{1}{v_2}},
\end{equation}
and $(f, \underline{g})$ is on $\underline{C}_2$:
\begin{equation}
\underline{g}
=\frac{1}{v_2}\frac{f - \frac{\kappa_1}{v_8}}{f -v_4}.
\end{equation}
\end{enumerate}
%
\noindent (ii) Linearized equation of the Riccati equation \eqref{eqn:hyper_linear_general}:
\begin{equation}\label{eqn:hyper_diffeq_qD5}
\begin{split}
& U_1 (\overline{F} -F) + U_2 F + U_3 (\underline{F}-F)=0,\\
&\begin{array}{l}\medskip
 {\displaystyle 
U_1
=\left(1-\frac{\kappa_2}{v_1v_5}\right)\left(1-\frac{qv_2v_6}{\kappa_2}\right),}\\
\medskip
{\displaystyle 
U_2 = \left(1-\frac{v_3}{v_4}\right)\left(1-\frac{v_7}{v_8}\right),}\\
{\displaystyle 
U_3
=q\frac{v_6}{v_5}\left(1-\frac{\kappa_1}{v_3v_8}\right)\left(1-\frac{v_3v_8}{q\kappa_1}\right).
}
\end{array}
\end{split}
\end{equation}
%
\noindent(iii) Linear difference equation for the hypergeometric function \cite{Gupta-Ismail-Masson}:
\begin{equation}\label{eqn:2phi1}
\Phi(a_1,a_2,b_1,z)={}_2\phi_1\left[\begin{array}{c}a_1,a_2\\b_1\end{array};q;z\right],
\end{equation}
\begin{equation}\label{eqn:diffeq_2phi1}
\begin{split}
&V_1(\overline{\Phi}-\Phi) + V_2\Phi  +V_3(\underline{\Phi}-\Phi)=0,\\
&\begin{array}{l}\medskip
{\displaystyle V_1=(1-a_2)(a_1-b_1),}\\
\medskip
{\displaystyle V_2=(1-a_1)(a_2-b_1),}\\
\medskip
{\displaystyle V_3=\frac{(1-b_1)(q-b_1)}{z},}\\
{\displaystyle \overline{\Phi}=\Phi(a_1,qa_2,qb_1,z),\quad \underline{\Phi}=\Phi(a_1,a_2/q,b_1/q,z)}.
\end{array}
\end{split}
\end{equation}
%
\noindent (iv) Contiguity relation  \cite{Gupta-Ismail-Masson}:
\begin{equation}\label{eqn:2phi1_contiguity}
\begin{split}
& \overline{\Phi} - \Phi = 
\frac{(1-a_1)(a_2 - b_1)z}{(1-b_1)(1-qb_1)}~\Phi(qa_1,qa_2,q^2b_1,z). 
\end{split}
\end{equation}
%
\noindent (v) Hypergeometric solution \eqref{eqn:hypergeometric_solution_general}:
\begin{equation}
\begin{split}
 y&=\frac{f-f_3}{f-f_8}
=  \frac{v_8(v_3v_7-\kappa_1)}{\kappa_1(v_8-v_7)}
~\frac{\overline{F}-F}{F}
=\frac{v_3v_5v_8(v_3v_7-\kappa_1)(v_3-v_4)}{v_4v_6(v_3v_8-\kappa_1)(qv_3v_8-\kappa_1)}~\frac{G}{F},\\
&F=\Phi(a_1,a_2,b_1,z),\quad G = \Phi(qa_1,qa_2,q^2b_1,z).
\end{split}
\end{equation}
%
\noindent (vi) Identification of parameters:
\begin{equation}\label{eqn:identificattion_hyper_qD5}
a_1=\frac{v_3}{v_4},\quad a_2=\frac{v_3v_7}{\kappa_1},\quad b_1=\frac{v_3v_8}{\kappa_1},\quad z = \frac{v_5}{v_6}.
\end{equation}
%
\subsubsection{$q$-P$(A_4^{(1)}/A_4^{(1)})$}\label{subsubsec:hyper_q-A4}
\begin{rem}\rm
In order to take the degeneration limit from $q$-P$(D_5^{(1)}/A_3^{(1)})$ described in
\eqref{eqn:degeneration_of_qP} and \eqref{eqn:degeneration_points_1} consistently with the decoupling
condition, we first exchange P$_3$ and P$_4$ ($v_3$ and $v_4$), and then take the limit $v_2\to
v_2\epsilon$, $v_3\to v_3/\epsilon$, $\epsilon\to 0$.
\end{rem}
%
\noindent (i) Decoupling condition:
\begin{enumerate}
 \item ${\rm P}_1$, ${\rm P}_4$, ${\rm P}_5$, ${\rm P}_7$ are on a $(1,1)$ curve $C_1$:
\begin{equation}
 \kappa_1\kappa_2 = v_1v_4v_5v_7,\label{eqn:decoupling_param_q-a4}
\end{equation}
$(f, g)$ is also on $C_1$:
\begin{equation}
 f = v_4\frac{g-\frac{v_5}{\kappa_2}}{g-\frac{1}{v_1}},
\end{equation}
\begin{equation}
 g = \frac{1}{v_1}\frac{f - \frac{\kappa_1}{v_7}}{f -v_4}.
\end{equation}
\item ${\rm P}_2'$, ${\rm P}_3'$, ${\rm P}_6'$, ${\rm P}_8'$ are on a $(1,1)$ curve $C_2$ where
${\rm P}_i'={\rm P}_i|_{\kappa_1\to \kappa_1/q}$:
\begin{equation}
 \kappa_1\kappa_2 = qv_2v_3v_6v_8, \label{eqn:decoupling_param_q-a4_2}
\end{equation} 
$(\overline{f}, g)$ is also on $C_2$
\begin{equation}
\overline{f}
= -v_2v_3\left(g-\frac{v_6}{\kappa_2}\right),
\end{equation}
and $(f,\underline{g})$ is on $\underline{C}_2$:
\begin{equation}
\underline{g}
=-\frac{1}{v_2v_3}\left(f - \frac{\kappa_1}{v_8}\right).
\end{equation}
\end{enumerate}
%
\noindent (ii) Linearized equation of the Riccati equation \eqref{eqn:hyper_linear_general}:
\begin{equation}\label{eqn:hyper_diffeq_qA4}
\begin{split}
&\hskip40pt U_1 (\overline{F} -F) + U_2 F + U_3 (\underline{F}-F)=0,\\
& U_1=1-\frac{\kappa_2}{v_1v_5},\quad U_2 = 1-\frac{v_7}{v_8},\quad 
U_3 = q\frac{v_6}{v_5}\left(1-\frac{\kappa_1}{v_4v_8}\right)\left(1-\frac{v_4v_8}{q\kappa_1}\right).
\end{split}
\end{equation}
%
\noindent(iii) Linear difference equation for the hypergeometric function:
\begin{equation}\label{eqn:2phi1_A4}
\begin{split}
& \Phi(a,b,z)={}_2\phi_1\left[\begin{array}{c}a,0\\b\end{array};q;z\right],
\end{split}
\end{equation}
\begin{equation}\label{eqn:diffeq_2phi1_A4}
\begin{split}
&V_1(\overline{\Phi}-\Phi) + V_2\Phi  +V_3(\underline{\Phi}-\Phi)=0,\\
&\begin{array}{l}
{\displaystyle V_1=b(a-1),\quad V_2= a-b,\quad V_3=\frac{(b-1)(b-q)}{z},}\\
{\displaystyle \overline{\Phi}=\Phi(qa,qb,z),\quad \underline{\Phi}=\Phi(a/q,b/q,z)}.
\end{array}
\end{split}
\end{equation}
%
\noindent (iv) Contiguity relation:
\begin{equation}\label{eqn:2phi1_A4_contiguity}
\begin{split}
& \overline{\Phi} - \Phi = 
\frac{(a - b)z}{(1-b)(1-qb)}~\Phi(qa,q^2b,z). 
\end{split}
\end{equation}
%
\noindent (v) Hypergeometric solution \eqref{eqn:hypergeometric_solution_general}:
\begin{equation}
\begin{split}
 y&=\frac{f-f_4}{f-f_8}
=  \frac{v_8(v_4v_7-\kappa_1)}{\kappa_1(v_8-v_7)}
~\frac{\overline{F}-F}{F}
= -\frac{v_4v_5v_8(v_4v_7-\kappa_1)}{v_6(v_4v_8-\kappa_1)(qv_4v_8-\kappa_1)}~\frac{G}{F},\\
&F=\Phi(a,b,z),\quad G = \Phi(qa,q^2b,z).
\end{split}
\end{equation}
%
\noindent (vi) Identification of parameters:
\begin{equation}\label{eqn:identificattion_hyper_qA4}
a=\frac{v_4v_7}{\kappa_1},\quad b=\frac{v_4v_8}{\kappa_1},\quad z = \frac{v_5}{v_6}.
\end{equation}
%
\subsubsection{$q$-P$(E_3^{(1)}/A_5^{(1)};b)$}\label{subsubsec:hyper_q-E3b}
\begin{rem}\rm
In order to take the degeneration limit from $q$-P$(A_4^{(1)}/A_4^{(1)})$ described in
\eqref{eqn:degeneration_of_qP} and \eqref{eqn:degeneration_points_1} consistently with the decoupling
condition, we first exchange P$_5$ and P$_6$ ($v_5$ and $v_6$), and then take the limit $v_6\to
v_6\epsilon$, $v_7\to v_7/\epsilon$, $\epsilon\to 0$.
\end{rem}
\noindent (i) Decoupling condition:
\begin{enumerate}
 \item ${\rm P}_1$, ${\rm P}_4$, ${\rm P}_6$, ${\rm P}_7$ are on a $(1,1)$ curve $C_1$:
\begin{equation}
 \kappa_1\kappa_2 = v_1v_4v_6v_7,\label{eqn:decoupling_param_q-e3b}
\end{equation}
$(f, g)$ is also on $C_1$:
\begin{equation}
 f = v_4\frac{g}{g-\frac{1}{v_1}},
\end{equation}
\begin{equation}
 g = \frac{1}{v_1}\frac{f}{f -v_4}.
\end{equation}
 \item ${\rm P}_2'$, ${\rm P}_3'$, ${\rm P}_5'$, ${\rm P}_8'$ are on a $(1,1)$ curve $C_2$ where
${\rm P}_i'={\rm P}_i|_{\kappa_1\to \kappa_1/q}$:
\begin{equation}
 \kappa_1\kappa_2 = qv_2v_3v_5v_8, \label{eqn:decoupling_param_q-e3b_2}
\end{equation} 
$(\overline{f}, g)$ is also on $C_2$:
\begin{equation}
\overline{f}
= -v_2v_3\left(g-\frac{v_5}{\kappa_2}\right),
\end{equation}
and $(f,\underline{g})$ is on $\underline{C}_2$:
\begin{equation}
\underline{g}
=-\frac{1}{v_2v_3}\left(f - \frac{\kappa_1}{v_8}\right).
\end{equation}
\end{enumerate}
\noindent (ii) Linearized equation of the Riccati equation \eqref{eqn:hyper_linear_general}:
\begin{equation}\label{eqn:hyper_diffeq_e3b}
\begin{split}
&\hskip40pt U_1 (\overline{F} -F) + U_2 F + U_3 (\underline{F}-F)=0,\\
& 
U_1 = -\frac{\kappa_2}{v_1},\quad 
U_2 =-\frac{v_6v_7}{v_8},\quad 
U_3 = qv_5\left(1-\frac{\kappa_1}{v_4v_8}\right)\left(1-\frac{v_4v_8}{q\kappa_1}\right).
\end{split}
\end{equation}
\noindent (iii) Linear difference equation for the hypergeometric function:
\begin{equation}\label{eqn:hyper_e3b}
\begin{split}
& \Phi(b,z)={}_1\phi_1\left[\begin{array}{c}0\\b\end{array};q;z\right]\\
\end{split}
\end{equation}
\begin{equation}\label{eqn:diffeq_e3b}
\begin{split}
&V_1(\overline{\Phi}-\Phi) + V_2\Phi  +V_3(\underline{\Phi}-\Phi)=0,\\
&\begin{array}{l}
{\displaystyle V_1=bz,\quad V_2= z,\quad V_3=(b-1)(b-q),}\\
{\displaystyle \overline{\Phi}=\Phi(qb,qz),\quad \underline{\Phi}=\Phi(b/q,z/q)}.
\end{array}
\end{split}
\end{equation}
\noindent (iv) Contiguity relation:
\begin{equation}\label{eqn:hyper_e3b_contiguity}
\begin{split}
& \overline{\Phi} - \Phi = 
\frac{z}{(1-b)(1-qb)}~\Phi(q^2b,qz). 
\end{split}
\end{equation}
\noindent (v) Hypergeometric solution \eqref{eqn:hypergeometric_solution_general}:
\begin{equation}
\begin{split}
 y&=\frac{f-f_4}{f-f_8}
= -\frac{v_8v_4}{\kappa_1}~\frac{\overline{F}-F}{F}
= -\frac{v_4^2v_6v_7v_8}{v_5(v_4v_8-\kappa_1)(qv_4v_8-\kappa_1)}~\frac{G}{F},\\
&F=\Phi(b,z),\quad G = \Phi(q^2b,qz).
\end{split}
\end{equation}
\noindent (vi) Identification of parameters:
\begin{equation}\label{eqn:identificattion_hyper_qe3b}
b=\frac{v_4v_8}{\kappa_1},\quad z = \frac{v_4v_6v_7}{\kappa_1 v_5}.
\end{equation}
%
\subsubsection{$q$-P$(E_3^{(1)}/A_5^{(1)};a)$}\label{subsubsec:hyper_q-E3a}
\begin{rem}\rm
In order to take the degeneration limit from $q$-P$(A_4^{(1)}/A_3^{(1)})$ described in
\eqref{eqn:degeneration_of_qP} and \eqref{eqn:degeneration_points_1} consistently with the decoupling
condition, we first exchange P$_7$ and P$_8$ ($v_7$ and $v_8$), and then take the limit $v_8\to
v_8\epsilon$, $v_1\to v_1/\epsilon$, $\epsilon\to 0$.
\end{rem}
\begin{equation}
f \overline{f} 
= -v_2v_3v_4~ \frac{\prod\limits_{i=5}^6\Bigl(g-\dfrac{v_i}{\kappa_2}\Bigr)}{g}, \quad
g \underline{g} 
= \frac{\kappa_1}{v_1v_2v_3v_8}~\frac{f-\dfrac{\kappa_1}{v_7}}{f-v_4 }. 
\end{equation}
\noindent (i) Decoupling condition:
\begin{enumerate}
 \item ${\rm P}_1$, ${\rm P}_4$, ${\rm P}_5$, ${\rm P}_8$ are on a $(1,1)$ curve $C_1$:
\begin{equation}
 \kappa_1\kappa_2 = v_1v_4v_5v_8,\label{eqn:decoupling_param_q-e3a}
\end{equation}
$(f, g)$ is also on $C_1$:
\begin{equation}
 f = v_4\frac{g-\frac{v_5}{\kappa_2}}{g},
\end{equation}
\begin{equation}
 g = -\frac{\kappa_1}{v_1v_8}\frac{1}{f -v_4}.
\end{equation}
 \item ${\rm P}_2'$, ${\rm P}_3'$, ${\rm P}_6'$, ${\rm P}_7'$ are on a $(1,1)$ curve $C_2$ where
${\rm P}_i'={\rm P}_i|_{\kappa_1\to \kappa_1/q}$:
\begin{equation}
 \kappa_1\kappa_2 = qv_2v_3v_6v_7, \label{eqn:decoupling_param_q-e3a_2}
\end{equation} 
$(\overline{f}, g)$ is also on $C_2$:
\begin{equation}
\overline{f} = -v_2v_3\left(g-\frac{v_6}{\kappa_2}\right),
\end{equation}
and $(f,\underline{g})$ is on $\underline{C}_2$:
\begin{equation}
\underline{g} =-\frac{1}{v_2v_3}\left(f - \frac{\kappa_1}{v_7}\right).
\end{equation}
\end{enumerate}
\noindent (ii) Linearized equation of the Riccati equation \eqref{eqn:hyper_linear_general}:
\begin{equation}\label{eqn:hyper_diffeq_e3a}
\begin{split}
&\hskip40pt U_1 (\overline{F} -F) + U_2 F + U_3 (\underline{F}-F)=0,\\
& 
U_1=1,\quad 
U_2 = 1,\quad 
U_3 = q\frac{v_6}{v_5}\left(1-\frac{\kappa_1}{v_4v_7}\right)\left(1-\frac{v_4v_7}{q\kappa_1}\right).
\end{split}
\end{equation}
%
\noindent (iii) Linear difference equation for the hypergeometric function:
\begin{equation}\label{eqn:hyper_e3a}
\begin{split}
& \Phi(a,b,z)={}_2\phi_1\left[\begin{array}{c}0,0\\b\end{array};q;z\right],
\end{split}
\end{equation}
\begin{equation}\label{eqn:diffeq_e3_1}
\begin{split}
&V_1(\overline{\Phi}-\Phi) + V_2\Phi  +V_3(\underline{\Phi}-\Phi)=0,\\
&\begin{array}{l}
{\displaystyle V_1= -b,\quad V_2=-b,\quad V_3=\frac{(b-1)(b-q)}{z},}\\
{\displaystyle \overline{\Phi}=\Phi(qb,z),\quad \underline{\Phi}=\Phi(b/q,z)}.
\end{array}
\end{split}
\end{equation}
%
\noindent (iv) Contiguity relation: 
\begin{equation}\label{eqn:e3a_contiguity}
\begin{split}
& \overline{\Phi} - \Phi = 
-\frac{bz}{(1-b)(1-qb)}~\Phi(q^2b,z). 
\end{split}
\end{equation}
%
\noindent (v) Hypergeometric solution \eqref{eqn:hypergeometric_solution_general}:
\begin{equation}
\begin{split}
 y&=\frac{f-f_4}{f-f_7}
=-\frac{\overline{F}-F}{F}
= \frac{\kappa_1v_4v_5v_7}{v_6(v_4v_7-\kappa_1)(qv_4v_7-\kappa_1)}~\frac{G}{F},\\
&F=\Phi(b,z),\quad G = \Phi(q^2b,z).
\end{split}
\end{equation}
%
(vi) Identification of parameters:
\begin{equation}\label{eqn:identificattion_hyper_e3a}
b=\frac{v_4v_7}{\kappa_1},\quad z = \frac{v_5}{v_6}.
\end{equation}
%
\subsubsection{$q$-P$(E_2^{(1)}/A_6^{(1)};a)$}\label{subsubsec:hyper_q-E2a}
\begin{rem}\rm
In order to take the degeneration limit from $q$-P$(E_3^{(1)}/A_5^{(1)};a)$ described in
\eqref{eqn:degeneration_of_qP} and \eqref{eqn:degeneration_points_1} consistently with the decoupling
condition, we take the limit $v_6\to v_6\epsilon$, $v_7\to v_7/\epsilon$, $\epsilon\to 0$.
\end{rem}
\begin{equation}
f \overline{f} = -v_2v_3v_4\Bigl(g-\dfrac{v_5}{\kappa_2}\Bigr), \quad
g \underline{g}  = \frac{\kappa_1}{v_1v_2v_3v_8}~\frac{f}{f-v_4 }. 
\end{equation}
\noindent (i) Decoupling condition:
\begin{enumerate}
 \item ${\rm P}_1$, ${\rm P}_4$, ${\rm P}_5$, ${\rm P}_8$ are on a $(1,1)$ curve $C_1$:
\begin{equation}
 \kappa_1\kappa_2 = v_1v_4v_5v_8,\label{eqn:decoupling_param_q-e2_1}
\end{equation}
$(f, g)$ is also on $C_1$:
\begin{equation}
 f = v_4\frac{g-\frac{v_5}{\kappa_2}}{g},
\end{equation}
\begin{equation}
 g = -\frac{\kappa_1}{v_1v_8}\frac{1}{f -v_4}.
\end{equation}
 \item ${\rm P}_2'$, ${\rm P}_3'$, ${\rm P}_6'$, ${\rm P}_7'$ are on a $(1,1)$ curve $C_2$ where
${\rm P}_i'={\rm P}_i|_{\kappa_1\to \kappa_1/q}$:
\begin{equation}
 \kappa_1\kappa_2 = qv_2v_3v_6v_7, \label{eqn:decoupling_param_q-e2_2}
\end{equation} 
$(\overline{f}, g)$ is also on $C_2$:
\begin{equation}
\overline{f} = -v_2v_3 g,
\end{equation}
and $(f,\underline{g})$ is on $\underline{C}_2$:
\begin{equation}
\underline{g} =-\frac{1}{v_2v_3} f .
\end{equation}
\end{enumerate}
%
\noindent (ii) Linearized equation of the Riccati equation \eqref{eqn:hyper_linear_general}:
\begin{equation}\label{eqn:hyper_diffeq_e2a}
\begin{split}
&U_1 (\overline{F} -F) + U_2 F + U_3 (\underline{F}-F)=0,\\
& 
U_1=1,\quad 
U_2 = 1,\quad 
U_3 = -\frac{v_6v_4v_7}{\kappa_1 v_5}.
\end{split}
\end{equation}
\noindent (iii) Linear difference equation for the hypergeometric function:
\begin{equation}\label{eqn:hyper_2phi0}
\begin{split}
& \Phi(z)={}_2\phi_0\left[\begin{array}{c}0,0\\-\end{array};q;z\right],
\end{split}
\end{equation}
\begin{equation}\label{eqn:diffeq_2phi0}
\begin{split}
&z\overline{\Phi} + \Phi - \underline{\Phi}=0,\\
&\overline{\Phi}=\Phi(z/q),\quad \underline{\Phi}=\Phi(qz).
\end{split}
\end{equation}
\noindent (iv) Hypergeometric solution \eqref{eqn:hypergeometric_solution_general}
\begin{equation}
\begin{split}
 y&=\frac{f-f_4}{f}
=-\frac{\overline{F}-F}{F}
= \frac{\kappa_1 v_5}{qv_4v_6v_7}~\frac{G}{F},\\
&F=\Phi(z),\quad G = \Phi(z/q^2).
\end{split}
\end{equation}
\noindent (v) Identification of parameters:
\begin{equation}\label{eqn:identificattion_hyper_qe2a}
z=\frac{\kappa_1 v_5}{v_4v_6v_7}.
\end{equation}
%
\par\bigskip
\noindent In the following additive cases, we put $\delta=1$ for simplicity.
\subsubsection{d-P$(E_8^{(1)}/A_0^{(1)})$}\label{subsubsec:hyper_d-E8}
\noindent (i) Decoupling condition:
\begin{enumerate}
 \item ${\rm P}_1$, ${\rm P}_3$, ${\rm P}_5$, ${\rm P}_7$ are on a $(1,1)$ curve $C_1$:
\begin{equation}
 \kappa_1 + \kappa_2 = v_1 + v_3+ v_5+ v_7,\label{eqn:decoupling_param_d-e8}
\end{equation}
$(f,g)$ is also on $C_1$:
\begin{equation}
 \frac{f - f(\kappa_2-t)} {f - f(t)}
= 
\prod_{j=1,3,5,7}\frac{v_j+t-\kappa_2}{v_j-t},\quad \mbox{for}\ g=g(t),
\end{equation}
\begin{equation}
 \frac{g-g(\kappa_1-s)} {g-g(s)}
= 
\prod_{j=1,3,5,7}\frac{v_j+s-\kappa_1}{v_j-s},\quad 
\mbox{for}\ f = f(s).
\end{equation}
 \item ${\rm P}_2'$, ${\rm P}_4'$, ${\rm P}_6'$, ${\rm P}_8'$ are on a $(1,1)$ curve $C_2$ where
${\rm P}_i'=(\overline{f}(v_i), g(v_i))$:
\begin{equation}
 \kappa_1 + \kappa_2 = 1 + v_2 + v_4 + v_6 + v_8, \label{eqn:decoupling_param_d-e8_2}
\end{equation}
$(\overline{f},g)$ is also on $C_2$
\begin{equation}
\frac{\overline{f}-\overline{f}(\kappa_2-t)} {\overline{f}-\overline{f}(t)}
= \prod_{j=2,4,6,8}\frac{v_j+t-\kappa_2}{v_j - t},\quad \mbox{for}\ g=g(t),
\end{equation}
and $(f,\underline{g})$ is on $\underline{C}_2$:
\begin{equation}
 \frac{\underline{g}-\underline{g}(\kappa_1-s)} {\underline{g}-\underline{g}(s)}
= \prod_{j=2,4,6,8}
\frac{v_j+s-\kappa_1}{v_j - s},\quad \mbox{for}\ f = f(s),
\end{equation}
where 
\begin{equation}
\begin{split}
f(z)=z(z-\kappa_1),\quad g(z)=z(z-\kappa_2).
\end{split}
\end{equation}
\end{enumerate}
\noindent (ii) Linearized equation of the Riccati equation \eqref{eqn:hyper_linear_general}:
\begin{equation}\label{eqn:hyper_diffeq_dE8}
\begin{split}
&\hskip30pt U_1 (\overline{F} -F) + U_2 F + U_3 (\underline{F}-F)=0,\\
&\begin{array}{l}\medskip
 {\displaystyle 
U_1
=\frac{(v_1+v_8-\kappa_2)(1 + v_1+ v_8-\kappa_2)}
{(\kappa_1 - \kappa_2)(1+\kappa_1-\kappa_2)}
\prod_{i=3,5,7}(v_1+v_i-\kappa_1)\prod_{j=2,4,6}(v_j+v_8-\kappa_1),
}\\\medskip
{\displaystyle 
U_2
=-\prod_{i=2,4,6}(v_1-v_i)\prod_{j=3,5,7}(v_j-v_8),
}\\
{\displaystyle 
U_3
=\frac{(v_1+v_8-\kappa_1)(v_1+v_8-\kappa_1-1)}
{(1+\kappa_1-\kappa_2)(2+\kappa_1-\kappa_2)}
\prod_{i=3,5,7}(1+v_1+v_i-\kappa_2)\prod_{j=2,4,6}(1+v_j+v_8-\kappa_2)
}
\end{array}
\end{split}
\end{equation}
\par\medskip

\noindent (iii) Linear difference equation for the hypergeometric function:
\begin{equation}\label{eqn:diffeq_9F8}
\begin{split}
&\hskip40pt V_1(\overline{\Phi}-\Phi) + V_2\Phi  +V_3(\underline{\Phi}-\Phi)=0,\\
& \Phi(a_0,a_1,\ldots,a_7) = \phi(a_0;a_1,a_2,\ldots,a_7) + \widehat{\phi}(a_0;a_1,a_2,\ldots,a_7),\\
& \phi(a_0;a_1,a_2,\ldots,a_7) = 
{}_9F_8\left(\begin{array}{c} 
a_0, 1+\frac{a_0}{2},a_1,a_2,\ldots,a_7\\
\frac{a_0}{2},1+a_0-a_1,1+a_0-a_2,\ldots,1+a_0-a_7
\end{array}; 1\right),\\
& \widehat{\phi}(a_0;a_1,a_2,\ldots,a_7) 
= \frac{\Gamma(1+2a_7-a_0)\Gamma(a_0-a_7)\prod_{i=1}^6\Gamma(1+a_0-a_i)\Gamma(a_7+a_i-a_0)}
{\Gamma(1+a_0)\Gamma(a_7-a_0)\prod_{i=1}^6\Gamma(a_i)\Gamma(1+a_7-a_i)}\\
&\hskip40pt \times \phi(2a_7-a_0;a_1+a_7-a_0,\ldots,a_6+a_7-a_0,a_7),\\
&\hskip40pt 2 + 3a_0 = \sum_{i=1}^7 a_i,\\
&\overline{\Phi}=\Phi(a_0;a_1-1,a_2+1,a_3,\ldots,a_7),\quad 
\underline{\Phi}=\Phi(a_0;a_1+1,a_2-1,a_3,\ldots,a_7),\\
&\begin{array}{l}
{\displaystyle V_1=\frac{a_2(a_0-a_2)(1+a_0-a_2)}{(a_2-a_1)(1+a_2-a_1)}\prod_{j=3}^7(1+a_0-a_1-a_j)},\\
{\displaystyle V_2=(1+a_0-a_1-a_2)\prod_{j=3}^7 a_j,}\\
{\displaystyle V_3=\frac{a_1(a_0-a_1)(1+a_0-a_1)}{(a_1-a_2)(1+a_1-a_2)}\prod_{j=3}^7(1+a_0-a_2-a_j)}.
\end{array}
\end{split}
\end{equation}
\noindent (iv) Contiguity relation: 
\begin{equation}\label{eqn:9F8_contiguity}
\begin{split}
& \overline{\Phi} - \Phi = 
-\frac{(1+a_0)(2+a_0)(1+a_2-a_1)(1+a_0-a_1-a_2)}{(1+a_0-a_1)(2+a_0-a_1)(a_0-a_2)(1+a_0-a_2)}\\
&\hskip60pt \times\prod_{i=3}^7 \frac{a_i}{1+a_0-a_i}~\Phi(a_0+2;a_1,a_2+1,a_3+1,\ldots,a_7+1). 
\end{split}
\end{equation}
\noindent (v) Hypergeometric solution \eqref{eqn:hypergeometric_solution_general}:
\begin{equation}
\begin{split}
 y&=\frac{f-f_1}{f-f_8}=
\frac{(v_1+v_8-\kappa_2)\prod_{i=3,5,7}(v_1+v_i-\kappa_1)}
{(\kappa_1-\kappa_2)\prod_{i=3,5,7}(v_8-v_i)} ~\frac{\overline{F}-F}{F}\\
&=\frac{(1+v_1+v_8-v_3-v_5)(2+v_1+v_8-v_3-v_5)}
{(1+v_1-v_3)(1+v_1-v_5)(v_1+v_8-\kappa_1)(1+v_1+v_8-\kappa_1)(1+v_1+v_8-\kappa_2)}\\
&\times \frac{\prod_{i=2,4,6}(v_1-v_i) \prod_{i=3,5,7}(v_1+v_i-\kappa_1) }
{\prod_{i=2,4,6}(1+v_i+v_8-v_3-v_5) }~\frac{G}{F},\\
&F=\Phi(a_0,a_1,\cdots,a_7),\quad G = \Phi(a_0+2;a_1,a_2+1,a_3+1,\ldots,a_7+1).
\end{split}
\end{equation}
\noindent (vi) Identification of parameters:
\begin{equation}\label{eqn:identificattion_hyper_dE8}
\begin{split}
&a_0 = v_1+v_8-v_3-v_5,\ a_1= 1+\kappa_1-v_3-v_5,\ a_2=\kappa_2-v_3-v_5,\ a_3=v_1-v_2,\\
&a_4=v_1-v_4,\ a_5 = v_1-v_6,\ a_6=v_8-v_3,\ a_7=v_8-v_5. 
\end{split}
\end{equation}
%
\subsubsection{d-P$(E_7^{(1)}/A_1^{(1)})$}\label{subsubsec:hyper_d-E7}
\noindent (i) Decoupling condition:
\begin{enumerate}
 \item ${\rm P}_1$, ${\rm P}_3$, ${\rm P}_5$, ${\rm P}_7$ are on a $(1,1)$ curve $C_1$:
\begin{equation}
 \kappa_1+\kappa_2 = v_1+v_3+v_5+v_7,\label{eqn:decoupling_param_d-e7}
\end{equation}
and $(f, g)$ is also on $C_1$:
\begin{equation}
\frac{f+g-\kappa_1+\kappa_2}{f+g}=\frac{(g+\kappa_2-v_5)(g+\kappa_2-v_7)}{(g+v_1)(g+v_3)},\\
\end{equation}
\begin{equation}
\frac{f+g-\kappa_1+\kappa_2}{f+g}= \frac{(f-\kappa_1 + v_5)(f-\kappa_1 + v_7)}{(f-v_1)(f-v_3)}.
\end{equation}
 \item ${\rm P}_2'$, ${\rm P}_4'$, ${\rm P}_6'$, ${\rm P}_8'$ are on a $(1,1)$ curve $C_2$ where
${\rm P}_i'=(\overline{f}(v_i), v(v_i))$:
\begin{equation}
 \kappa_1+\kappa_2 = 1+v_2+v_4+v_6+v_8, \label{eqn:decoupling_param_d-e7_2}
\end{equation} 
$(\overline{f}, g)$ is also on $C_2$:
\begin{equation}
\begin{split}
\frac{\overline{f} + g - \kappa_1+\kappa_2+\delta}
{\of+g}=\frac{(g+\kappa_2-v_6)(g+\kappa_2-v_8)}{(g+v_2)(g+v_4)},
\end{split}
\end{equation}
and $(f, \underline{g})$ is on $\underline{C}_2$:
\begin{equation}
\frac{f + \underline{g} - \kappa_1+\kappa_2-\delta}{f+\underline{g}}
= \frac{(f-\kappa_1 + v_6)(f-\kappa_1 + v_8)}{(f-v_2)(f-v_4)}.
\end{equation}
\end{enumerate}
%
\noindent (ii) Linearized equation of the Riccati equation \eqref{eqn:hyper_linear_general}:
\begin{equation}\label{eqn:hyper_diffeq_dE7}
\begin{split}
&\hskip60pt U_1 (\overline{F} -F) + U_2 F + U_3 (\underline{F}-F)=0,\\
&\begin{array}{l}\medskip
 {\displaystyle 
U_1
=\frac{(v_1+v_8-\kappa_2)(1+v_1+v_8-\kappa_2)}{(\kappa_1-\kappa_2)(1+\kappa_1-\kappa_2)}
\prod_{j=2,4}(v_j+v_8-\kappa_1)\prod_{i=5,7}(v_1+v_i-\kappa_1),}\\
\medskip
{\displaystyle 
U_2 = \prod_{j=2,4}(v_1- v_j)\prod_{i=5,7}(v_i-v_8),}\\
{\displaystyle 
U_3
=\frac{(v_1+v_8-\kappa_1)(v_1+v_8-\kappa_1-1)}{(1+\kappa_1-\kappa_2)(2+\kappa_1-\kappa_2)}
\prod_{j=2,4}(1+v_j+v_8-\kappa_2) \prod_{i=5,7}(1+v_1+v_i-\kappa_2).
}
\end{array}
\end{split}
\end{equation}
\noindent (iii) Linear difference equation for the hypergeometric function \cite{Masson:9F8_7F6}:
\begin{equation}\label{eqn:7F6}
\Phi(a_0,a_1,\ldots,a_5)={}_{7}F_6\left(
\begin{array}{c}
 a_0,1+\frac{a_0}{2},a_1,\ldots,a_5\\
\frac{a_0}{2},1+a_0-a_1,\ldots,1+a_0-a_5
\end{array}; 1\right),
\end{equation}
\begin{equation}\label{eqn:diffeq_7F6}
\begin{split}
&\qquad V_1(\overline{\Phi}-\Phi) + V_2\Phi  +V_3(\underline{\Phi}-\Phi)=0,\\
&\begin{array}{l}
{\displaystyle V_1=\frac{a_2(a_0-a_2)(1+a_0-a_2)}{(a_2-a_1)(1+a_2-a_1)}\prod_{j=3,4,5}(1+a_0-a_1-a_j)},\\
{\displaystyle V_2=(1+a_0-a_1-a_2)\prod_{j=3,4,5}a_j,}\\
{\displaystyle V_3=\frac{a_1(a_0-a_1)(1+a_0-a_1)}{(a_1-a_2)(1+a_1-a_2)}\prod_{j=3,4,5}(1+a_0-a_2-a_j),}\\
{\displaystyle \overline{\Phi}=\Phi(a_0,a_1-1,a_2+1,a_3,\ldots,a_5),\quad 
\underline{\Phi}=\Phi(a_0;a_1+1,a_2-1,a_3,\ldots,a_5).}
\end{array}
\end{split}
\end{equation}
%
\noindent(iv) Contiguity relation: 
\begin{equation}\label{eqn:7W6_contiguity}
\begin{split}
& \overline{\Phi} - \Phi = 
-\frac{(1+a_0)(2+a_0)(1+a_2-a_1)(1+a_0-a_1-a_2)}{(1+a_0-a_1)(2+a_0-a_1)(a_0-a_2)(1+a_0-a_2)}\\
&\times\prod_{i=3}^5 \frac{a_i}{1+a_0-a_i}~\Phi(2+a_0;a_1,1+a_2,1+a_3,1+a_4,1+a_5). 
\end{split}
\end{equation}
%
\noindent (v) Hypergeometric solution \eqref{eqn:hypergeometric_solution_general}:
\begin{equation}
\begin{split}
 y&=\frac{f-f_1}{f-f_8}=
-\frac{(v_1 + v_8-\kappa _2)(v_1+ v_5-\kappa_1)(v_1+ v_7-\kappa_1)}
{(\kappa_1-\kappa_2)\prod_{i=5,7}(v_8 - v_i)} ~\frac{\overline{F}-F}{F}\\
&=-\frac{(1+v_1+v_8-v_3-v_5) (2+ v_1+v_8-v_3-v_5) }
{(1+ v_1-v_3) (v_1+v_8-\kappa _1) (1+ v_1+v_8-\kappa _1) (1+ v_1+v_8-\kappa _2)}
\\
&\qquad\times \frac{\prod_{i=2,4}(v_1-v_i)\prod_{i=5,7}(v_1 v_i-\kappa _1)}{\prod_{i=2,4}(1+ v_i+ v_8-v_3-v_5) }~\frac{G}{F}.\\
&F=\Phi(a_0,a_1,\cdots,a_5),\quad G = \Phi(2+a_0;a_1,1+a_2,1+a_3,1+a_4,1+a_5).
\end{split}
\end{equation}
%
\noindent (vi) Identification of parameters:
\begin{equation}\label{eqn:identificattion_hyper_dE7}
\begin{split}
&a_0 = v_1+v_8-v_3-v_5,\quad a_1= 1+\kappa_1-v_3-v_5,\quad a_2=\kappa_2-v_3-v_5,\\ 
&a_3=v_1-v_2,\quad a_4=v_1-v_4,\quad a_5=v_8-v_5. 
\end{split}
\end{equation}
%
\subsubsection{d-P$(E_6^{(1)}/A_2^{(1)})$}\label{subsubsec:hyper_d-E6}
\noindent(i) Decoupling condition:
\begin{enumerate}
 \item ${\rm P}_1$, ${\rm P}_3$, ${\rm P}_5$, ${\rm P}_7$ are on a $(1,1)$ curve $C_1$:
\begin{equation}
 \kappa_1+ \kappa_2 = v_1+ v_3+ v_5+ v_7,\label{eqn:decoupling_param_d-E6}
\end{equation}
$(f, g)$ is also on $C_1$:
\begin{equation}
f+g = \frac{(g+v_1)(g+v_3)}{g+\kappa_2-v_5},
\end{equation}
\begin{equation}
f+g = \frac{(f - v_1)(f - v_3)}{f -\kappa_1+v_7}.
\end{equation}
 \item ${\rm P}_2'$, ${\rm P}_4'$, ${\rm P}_6'$, ${\rm P}_8'$ are on a $(1,1)$ curve $C_2$ where
${\rm P}_i'={\rm P}_i|_{\kappa_1\to \kappa_1-1}$:
\begin{equation}
 \kappa_1+\kappa_2 = 1+v_2+v_4+v_6+v_8, \label{eqn:decoupling_param_d-E6_2}
\end{equation} 
$(\overline{f}, g)$ is on $C_2$:
\begin{equation}
\overline{f}+g
= \frac{(g + v_2)(g + v_4)}{g + \kappa_2 - v_6},
\end{equation}
and $(f, \underline{g})$ is on $\underline{C}_2$:
\begin{equation}
f+\underline{g}-1 = \frac{(f -v_2)(f -v_4)}{f -\kappa_1 + v_8}.
\end{equation}
\end{enumerate}
%
\noindent (ii) Linearized equation of the Riccati equation \eqref{eqn:hyper_linear_general}:
\begin{equation}\label{eqn:hyper_diffeq_d-E6}
\begin{split}
& U_1 (\overline{F} -F) + U_2 F + U_3 (\underline{F}-F)=0,\\
&\begin{array}{l}\medskip
 {\displaystyle 
U_1
=(\kappa_2-v_3-v_5)(1+v_2+v_6-\kappa_2)(1+v_4+v_6-\kappa_2),}\\
\medskip
{\displaystyle 
U_2 = (v_1-v_2)(v_1-v_4)(v_7-v_8),}\\
{\displaystyle 
U_3
=(v_1+v_5+1-\kappa_2)(\kappa_1-v_1-v_8)(v_1+v_8-1-\kappa_1).
}
\end{array}
\end{split}
\end{equation}
%
\noindent (iii) Linear difference equation for the hypergeometric function \cite{Gupta-Ismail-Masson:Continuous_Hahn}:
\begin{equation}\label{eqn:3F2}
\begin{split}
& \Phi(a_1,a_2,a_3,b_1,b_2)={}_3F_2\left[\begin{array}{c}a_1,a_2,a_3\\b_1,b_2\end{array};1\right],
\end{split}
\end{equation}
\begin{equation}\label{eqn:diffeq_3F2}
\begin{split}
&V_1(\overline{\Phi}-\Phi) + V_2\Phi  +V_3(\underline{\Phi}-\Phi)=0,\\
&\begin{array}{l}\medskip
{\displaystyle V_1=(a_1-b_2)(a_1-b_2)a_3,}\\
\medskip
{\displaystyle V_2=a_1a_2(a_3-b_2),}\\
\medskip
{\displaystyle V_3=b_2(1-b_2)(b_1-a_3),}\\
{\displaystyle \overline{\Phi}=\Phi(a_1,a_2,1+a_3,b_1,1+b_2),\quad \underline{\Phi}=\Phi(a_1,a_2,-1+a_3,b_3,-1+b_2)}.
\end{array}
\end{split}
\end{equation}
%
\noindent (iv) Contiguity relation \cite{Bailey:3F2,Gupta-Ismail-Masson:Continuous_Hahn,Wilson:3F2}:
\begin{equation}\label{eqn:3F2_contiguity}
\begin{split}
& \overline{\Phi} - \Phi = 
\frac{a_1a_2(b_2-a_3)}{b_1b_2(1+b_2)}
~\Phi(1+a_1,1+a_2,1+a_3,1+b_1,2+b_2). 
\end{split}
\end{equation}
%
\noindent (v) Hypergeometric solution \eqref{eqn:hypergeometric_solution_general}:
\begin{equation}
\begin{split}
 y&=\frac{f-f_1}{f-f_8}
=  \frac{(v_1+v_7 - \kappa_1)}{(v_8-v_7)}
~\frac{\overline{F}-F}{F}\\[2mm]
&=\frac{(v_1+v_7-\kappa_1)(v_1-v_2)(v_1-v_4)}
{(1+v_1-v_3)(v_1+v_8-\kappa_1)(1+v_1+v_8-\kappa_1)}
~\frac{G}{F},\\[2mm]
&F=\Phi(a_1,a_2,a_3,b_1,b_2),\quad G = \Phi(1+a_1,1+a_2,1+a_3,1+b_1,2+b_2).
\end{split}
\end{equation}
%
\noindent(vi) Identification of parameters:
\begin{equation}\label{eqn:identificattion_hyper_dE6}
a_1=v_1-v_2,\quad a_2=v_1-v_4,\quad a_3=\kappa_2-v_3-v_5,\quad
b_1=1+v_1-v_3,\quad b_2 = v_1+v_8-\kappa_1.
\end{equation}
%
\subsubsection{d-P$(D_4^{(1)}/D_4^{(1)})$}\label{subsubsec:hyper_d-D4}
\noindent(i) Decoupling condition:
\begin{enumerate}
 \item ${\rm P}_2$, ${\rm P}_3$, ${\rm P}_4$, ${\rm P}_5$ are on a $(1,1)$ curve $C_1$:
\begin{equation}
a_0 + a_1 + a_2=0,\label{eqn:decoupling_param_d-d4}
\end{equation}
and $(f,g)$ is on $C_1$
\begin{equation}
f=\frac{gt}{g+a_1+a_2},
\end{equation}
\begin{equation}
g = a_0 + \frac{a_0t}{f-t},
\end{equation}
 \item ${\rm P}_1'$, ${\rm P}_6'$, ${\rm P}_7'$, ${\rm P}_8'$ are on a $(1,1)$ curve $C_2$ where
${\rm P}_i'={\rm P}_i|_{{a_0\to a_0-1}\atop{a_3\to a_3-1}} = (\overline{x}_i,y_i)$:
\begin{equation}
a_2+a_3+a_4=1, \label{eqn:decoupling_param_d-d4_2}
\end{equation} 
$(\overline{f},g)$ is also on $C_2$:
\begin{equation}
\overline{f}=\frac{g-a_4}{g+a_2},
\end{equation}
and $(f,\underline{g})$ is on $\underline{C}_2$:
\begin{equation}
\underline{g}=\frac{a_3}{f-1}+a_3+a_4.
\end{equation}
\end{enumerate}
\noindent (ii) Linearized equation of the Riccati equation \eqref{eqn:hyper_linear_general}:
\begin{equation}\label{eqn:hyper_diffeq_D4}
\begin{split}
& U_1 (\overline{F} -F) + U_2 F + U_3 (\underline{F}-F)=0,\\
& U_1 = a_2(a_2-1)t,\quad U_2 = a_1(1-a_2-a_3),\quad U_3 = (1-a_1-a_2)a_3.
\end{split}
\end{equation}
\noindent(iii) Linear difference equation for the hypergeometric function \cite{Abramowitz-Stegun}:
\begin{equation}\label{eqn:2F1}
\Phi(\alpha_1,\alpha_2,\beta,z)={}_2F_1\left[\begin{array}{c}\alpha_1,\alpha_2\\\beta_1\end{array};z\right],
\end{equation}
\begin{equation}\label{eqn:diffeq_2F1}
\begin{split}
&V_1(\overline{\Phi}-\Phi) + V_2\Phi  +V_3(\underline{\Phi}-\Phi)=0,\\
& V_1=\frac{\beta(1-\beta)}{z},\quad V_2=\alpha_1(\alpha_2-\beta),\quad V_3=\alpha_2(\alpha_1-\beta),\\
& \overline{\Phi}=\Phi(\alpha_1,\alpha_2-1,\beta-1,z),\quad \underline{\Phi}=\Phi(\alpha_1,\alpha_2+1,\beta+1,z).
\end{split}
\end{equation}
\noindent (iv) Contiguity relation \cite{Abramowitz-Stegun}:
\begin{equation}\label{eqn:2F1_contiguity}
\begin{split}
& \overline{\Phi} - \Phi = 
\frac{\alpha_1(\alpha_2-\beta)z}{\beta(\beta-1)}~\Phi(\alpha_1+1,\alpha_2,\beta+1,z). 
\end{split}
\end{equation}
\noindent (v) Hypergeometric solution \eqref{eqn:hypergeometric_solution_general}:
\begin{equation}
\begin{split}
&f = \frac{a_2t}{a_1}\frac{\overline{F}-F}{F} 
= \frac{a_4}{1-a_2}\frac{G}{F},\\
&F=\Phi(\alpha_1,\alpha_2,\beta,z),\quad G = \Phi(\alpha_1+1,\alpha_2,\beta+1,z).
\end{split}
\end{equation}
\noindent (vi) Identification of parameters:
\begin{equation}\label{eqn:identificattion_hyper_D4}
\alpha_1 = a_1,\quad \alpha_2 = a_3,\quad \beta = 1-a_2,\quad z=\frac{1}{t}.
\end{equation}
%
\subsubsection{d-P$(A_3^{(1)}/D_5^{(1)})$}\label{subsubsec:hyper_d-A3}
\noindent(i) Decoupling condition:
\begin{enumerate}
 \item ${\rm P}_3$, ${\rm P}_4$, ${\rm P}_7$, ${\rm P}_8$ are on a $(1,1)$ curve $C_1$:
\begin{equation}
a_2 + a_3 =0,\label{eqn:decoupling_param_d-a3}
\end{equation}
and $(q,p)$ is on $C_1$
\begin{equation}
q = 1 - \frac{a_2}{p},
\end{equation}
\begin{equation}
p = \frac{a_2}{1-q}.
\end{equation}
 \item ${\rm P}_1'$, ${\rm P}_2'$, ${\rm P}_5'$, ${\rm P}_6'$ are on a $(1,1)$ curve $C_2$ where
${\rm P}_i'={\rm P}_i|_{{a_1\to a_1-1}\atop{a_3\to a_3-1}} = (\overline{x}_i,y_i)$:
\begin{equation}
a_0+a_1=1, \label{eqn:decoupling_param_d-a3_2}
\end{equation} 
$(\overline{q},p)$ is also on $C_2$:
\begin{equation}
\overline{q} = -\frac{a_0}{p+t},
\end{equation}
and $(q,\underline{p})$ is on $\underline{C}_2$:
\begin{equation}
\underline{p}= -t + \frac{a_1}{q}.
\end{equation}
\end{enumerate}
\noindent (ii) Riccati equation and linearized equation:
\begin{equation}
 \overline{q}= a_0\frac{-q+1}{tq-a_2-t},
\end{equation}
\begin{equation}
 q = -\frac{a_2+t}{t}~\frac{F-\underline{F}}{\underline{F}},
\end{equation}
\begin{equation}\label{eqn:hyper_diffeq_da3}
\begin{split}
& (t+a_2)(1+t+a_2)\overline{F} - (t+a_2)(1+t+a_0+a_2) F + a_0a_2\underline{F}=0.
\end{split}
\end{equation}
\noindent(iii) Linear difference equation for the hypergeometric function \cite{Abramowitz-Stegun}:
\begin{equation}\label{eqn:1F1}
\Phi(\alpha,\beta,z)=\frac{\Gamma(\alpha)}{\Gamma(\alpha+z)}{}_1F_1\left[\begin{array}{c}\alpha\\\beta\end{array};z\right],
\end{equation}
\begin{equation}\label{eqn:diffeq_1F1}
\begin{array}{c}\medskip
 {\displaystyle (z+\alpha)(z+\alpha-1)\overline{\Phi} - (z+\alpha-1)(z+2\alpha-\beta) \Phi 
+ (\alpha-\beta)(\alpha-1)\underline{\Phi}=0,}\\
{\displaystyle \overline{\Phi}=\Phi(\alpha+1,\beta,z),\quad \underline{\Phi}=\Phi(\alpha-1,\beta,z).}
\end{array}
\end{equation}
\noindent (iv) Contiguity relation:
\begin{equation}\label{eqn:1F1_contiguity}
\begin{split}
& \overline{\Phi} - \Phi = 
\frac{(\alpha-\beta)z}{\beta(z+\alpha)}~\Phi(\alpha,\beta+1,z). 
\end{split}
\end{equation}
\noindent (v) Hypergeometric solution:
\begin{equation}
\begin{split}
&q = -\frac{a_2+t}{t}~\frac{F-\underline{F}}{\underline{F}}
= \frac{1-a_0}{a_2-a_0+1}~\frac{{}_1F_1\left[\begin{array}{c}a_2\\ a_2-a_0+2\end{array};t\right]}
{{}_1F_1\left[\begin{array}{c}a_2\\a_2-a_0+1\end{array};t\right]}.
\end{split}
\end{equation}
\noindent (vi) Identification of parameters:
\begin{equation}\label{eqn:identificattion_hyper_dA3}
\alpha = a_2+1,\quad \quad \beta = a_2-a_0+1,\quad z=t.
\end{equation}
%
\subsubsection{d-P$(A_2^{(1)}/E_6^{(1)})$}\label{subsubsec:hyper_d-A2}
\noindent(i) Decoupling condition:
\begin{enumerate}
 \item ${\rm P}_5$, ${\rm P}_6$, ${\rm P}_7$, ${\rm P}_8$ are on a $(1,1)$ curve $C_1$:
\begin{equation}
a_0 =0,\label{eqn:decoupling_param_d-a2}
\end{equation}
and $(q,p)$ is on $C_1$
\begin{equation}
q = p - t,
\end{equation}
\begin{equation}
p = q + t.
\end{equation}
 \item ${\rm P}_1'$, ${\rm P}_2'$, ${\rm P}_3'$, ${\rm P}_4'$ are on a $(1,1)$ curve $C_2$ where
${\rm P}_i'={\rm P}_i|_{a_1\to a_1-1} = (\overline{x}_i,y_i)$:
\begin{equation}
a_1+a_2=1, \label{eqn:decoupling_param_d-a2_2}
\end{equation} 
$(\overline{q},p)$ is also on $C_2$:
\begin{equation}
\overline{q} = -\frac{a_2}{p},
\end{equation}
and $(q,\underline{p})$ is on $\underline{C}_2$:
\begin{equation}
\underline{p}= \frac{a_1}{q}.
\end{equation}
\end{enumerate}
\noindent (ii) Riccati equation and linearized equation:
\begin{equation}
 \overline{q}= -\frac{a_2}{q+t},
\end{equation}
\begin{equation}
 q = a_1\frac{\underline{F}}{F},
\end{equation}
\begin{equation}\label{eqn:hyper_diffeq_da2}
\overline{F}-tF - a_1\underline{F}=0.
\end{equation}
\noindent(iii) Linear difference equation for the hypergeometric function \cite{Abramowitz-Stegun}:
\begin{equation}\label{eqn:Hermite}
\Phi(\alpha,z)=
2^{\frac{\alpha}{2}}\sqrt{\pi}
\left[\frac{1}{\Gamma\left(\frac{1-\alpha}{2}\right)}\,
{}_1F_1\left[\begin{array}{c}\medskip\frac{-\alpha}{2}\\\frac{1}{2} \end{array};\frac{z^2}{2}\right] 
- 
\frac{\sqrt{2}t}{\Gamma\left(\frac{-\alpha}{2}\right)}\,{}_1F_1\left[
\begin{array}{c}\medskip\frac{1-\alpha}{2}\\\frac{3}{2}\end{array};\frac{z^2}{2}\right]
\right],
\end{equation}
\begin{equation}\label{eqn:diffeq_a2}
\begin{array}{c}\medskip
 {\displaystyle \overline{\Phi} - z \Phi 
+ \alpha \underline{\Phi}=0,}\\
{\displaystyle \overline{\Phi}=\Phi(\alpha+1,z),\quad \underline{\Phi}=\Phi(\alpha-1,z).}
\end{array}
\end{equation}
\noindent (iv) Hypergeometric solution:
\begin{equation}
\begin{split}
&q = a_1 ~\frac{\Phi(-a_1+1,t)}{\Phi(-a_1,t)}.
\end{split}
\end{equation}
\noindent (v) Identification of parameters:
\begin{equation}\label{eqn:identificattion_hyper_dA2}
\alpha = -a_1,\quad z=t.
\end{equation}
%
\subsubsection{d-P$(2A_1^{(1)}/D_6^{(1)})$}\label{subsubsec:hyper_d-2A1}
Equation \eqref{eqn:d-2A1} has no hypergeometric solution as constructed by the procedure used
above, since there is no $(1,1)$ curve preserved by the equation in the direction
$(\overline{a}_0,\overline{a}_1)=(a_0+1,a_1+1)$.  If we take the direction
$(\overline{a}_0,\overline{a}_1)=(a_0+1,a_1)$ or $(\overline{a}_0,\overline{a}_1)=(a_0,a_1+1)$, it
is known that the corresponding discrete Painlev\'e equation admits hypergeometric solutions
expressible in terms of the Bessel functions \cite{NSKGR:alt-dp2}.
\par
\end{document}